# Feynman Rules in the Two-Higgs Doublet Model Effective Field Theory

**Radovan Dermisek and Keith Hermanek**

*Physics Department, Indiana University,*
*Bloomington, Indiana 47405, USA*

*E-mail:* dermisek@iu.edu, khermane@iu.edu

Abstract: We derive a complete set of Feynman rules in the general two-Higgs doublet model effective field theory where the effects of additional new physics are parametrized by operators up to mass dimension-six. We calculate the physical Higgs spectrum, contributions to the couplings and masses of electroweak gauge bosons and fermions, and all contact interactions arising from dimension-six operators. We also present results in specific limits and types, which include the $CP$-conserving limit, alignment limit, and the four types of two-Higgs doublet models: type-I, -II, -X, and -Y. We discuss the differences between the two-Higgs doublet model effective field theory and the renormalizable two-Higgs doublet model or the standard model effective field theory. We create a FeynRules model package for calculating all Feynman rules in the general effective field theory and its specific limits and types.

## 1 Introduction

Effective field theories have been pivotal in providing a bridge to possible new physics, without knowing the details of a high-energy theory. The two-Higgs doublet model effective field theory (2HDM EFT) describes the effects of new physics, provided the low energy degrees of freedom are the known standard model (SM) fields and new Higgs bosons, with additional new physics entering at the scale $\Lambda$ substantially above the electroweak (EW) scale and the scale of new Higgs bosons. Much below this scale, heavy fields are integrated out, and their effects are parameterized by higher-dimensional operators, whose Wilson coefficients scale as $C_i \propto \Lambda^{4-d}$ for mass dimension $d > 4$ [1]. Various versions of the 2HDM

EFTs have been constructed at the mass-dimension six in Refs. [2–4] and most recently in Ref. [5].

The renormalizable two-Higgs doublet model (2HDM) [6] is one of the simplest extensions of the SM and has been widely studied, in both the general sense and in the context of specific types, type-I, -II, -X, and -Y models, resulting from restricting couplings of the Higgs doublets to SM fermions by $Z_2$ symmetries. In the effective field theory, the additional Higgs doublet results in a more expansive set of operators compared to the standard model effective field theory (SMEFT) [7]. While some operators are shared in all four types of models, due to $Z_2$ symmetries, many operators are forbidden in specific types. Additionally, assuming the conservation of charge-parity ($CP$) symmetry in the scalar sector also restricts the form of Wilson coefficients [2–4]. Ref. [5] provides the complete set of independent operators in general and in specific types without any additional assumptions.[1]

In this paper, we calculate the physical Higgs spectrum, contributions to the couplings and masses of electroweak gauge bosons and fermions after EWSB, and all contact interactions resulting from dimension-six operators in a general 2HDM EFT. We also present results in specific limits and types of 2HDM EFTs, which include the $CP$-conserving limit, alignment limit, and the four types of two-Higgs doublet models mentioned above. We discuss the differences between the 2HDM EFT and the renormalizable two-Higgs doublet model or the SMEFT. Previously in Ref. [4], a partial set of Feynman rules with the focus on those which modify existing interactions in the SM or 2HDM were obtained in the 2HDM EFT in the $CP$-conserving limit. While these are the most important for precision measurements, the 2HDM EFT provides a vast number of new contact interactions, not present either in the SM, SMEFT, or the 2HDM. We create a `FeynRules` [11, 12] model package publicly available in Ref. [13] for calculating all Feynman rules in the general 2HDM EFT, integrated with options for specifying $CP$-conservation in the scalar sector, the alignment limit, or the type of 2HDM. Because of the large number of possible interactions with very complex Feynman rules in the general theory, in Appendix C, we only list selected sets of Feynman rules in the $CP$-conserving type-II 2HDM EFT in the alignment limit. Even within this subset, the majority of interactions are new contact interactions not present in the SM, SMEFT, or the 2HDM.

We also present a novel method to obtain analytic formulas needed to diagonalize $4 \times 4$ matrices in the neutral scalar sector of the theory, which is needed unless the $CP$-symmetry is conserved, and has been previously done either numerically [14], or approximately [15]. These results are not specific to the 2HDM EFT and can be also used in studies of the renormalizable $CP$-violating 2HDM or other models.

Among the most important features of new contact interactions in the SMEFT [18] is the energy grows of the amplitudes for processes involving them (due to power counting [19]) which can often overcome the contributions from the corresponding SM background [20–26]. The same power counting holds for the 2HDM EFT, however, there are additional notable differences between interactions in the SMEFT and the 2HDM EFT. These include possible $\tan\beta$ enhancements [5, 26–29], where $\tan\beta$ is the ratio of the vacuum expectation values (VEVs) of the two Higgs doublets, additional momentum-dependent contributions to SM processes, and new interactions involving the additional Higgs scalars.

[1] In Ref. [5] the 2HDM EFT was also constructed in the Higgs basis, where the SM degrees of freedom are separated into one doublet, and the additional Higgs scalars are placed in another. Advantages of working in the Higgs basis include separating the effects which modify couplings and masses of the SM particles from ones which only contribute to scattering processes involving new Higgses, demonstrating clear scaling between processes involving SM particles and ones involving new Higgses, and identifying correlations between different operators in a given UV completion [8–10].

This paper is organized as follows: In Sec. 2, we describe all notations and conventions used in the general 2HDM EFT. In Sec. 3, we calculate the physical Higgs spectrum, contributions to the couplings and masses of electroweak gauge bosons and the fermions after EWSB. The main results are the Feynman rules in the 2HDM EFT presented in Sec. 4, where we also discuss the differences resulting in various limits and types of the 2HDM EFT, as well as the differences from the renormalizable 2HDM and SMEFT. We conclude in Sec. 5. In Appendices A and B, we provide supplementary details needed for Sec. 3. We also provide selected Feynman rules in the $CP$-conserving type-II 2HDM EFT in the alignment limit in Appendix C.

## 2 Conventions and Notations for the 2HDM EFT

We adopt all conventions and notations for parameters and effective operators of the 2HDM EFT summarized in Ref. [5]. When heavy degrees of freedom of a UV complete theory are integrated out at a high scale $\Lambda$, the effects of new physics are parametized by higher-dimensional operators, in addition to the renormalizable terms in the 2HDM. The complete Lagrangian up to dimension-six terms is

$$\mathcal{L} = \mathcal{L}^{(4)}_{2HDM} + \sum_i C_i^{(5)}\mathcal{O}_i^{(5)} + \sum_i C_i^{(6)}\mathcal{O}_i^{(6)} + \mathcal{O}\left(\frac{1}{\Lambda^3}\right), \tag{2.1}$$

where

$$\begin{aligned}\mathcal{L}^{(4)}_{2HDM} = &-\frac{1}{4}B_{\mu\nu}B^{\mu\nu} - \frac{1}{4}W^a_{\mu\nu}W^{a\mu\nu} - \frac{1}{4}G^a_{\mu\nu}G^{a\mu\nu} \\ &+ (D_\mu\Phi_1)^\dagger D^\mu\Phi_1 + (D_\mu\Phi_2)^\dagger D^\mu\Phi_2 + \Big(\eta(D_\mu\Phi_1)^\dagger D^\mu\Phi_2 + h.c.\Big) - V(\Phi_1,\Phi_2) \\ &+ i\bar{l}_L \not{D} l_L + i\bar{e}_R\not{D} e_R + i\bar{q}_L\not{D} q_L + i\bar{d}_R\not{D} d_R + i\bar{u}_R\not{D} u_R \\ &- \Big(y_e^{(1)}\bar{l}_L e_R\Phi_1 + y_e^{(2)}\bar{l}_L e_R\Phi_2 + y_d^{(1)}\bar{q}_L d_R\Phi_1 + y_d^{(2)}\bar{q}_L d_R\Phi_2 \\ &+ y_u^{(1)}\bar{q}_L u_R\cdot\Phi_1^\dagger + y_u^{(2)}\bar{q}_L u_R\cdot\Phi_2^\dagger + h.c.\Big),\end{aligned} \tag{2.2}$$

is the general renormalizable 2HDM Lagrangian (summarized in [30] without the mixed kinetic term). The lepton and quark doublets are defined as $l_L = (\nu_L, e_L)^T$ and $q_L = (u_L, d_L)^T$. The gauge covariant derivative acting on an object charged under $SU(3)\times SU(2)_L\times U(1)_Y$ is defined as

$$(D_\mu q)_{\alpha i} = \left[\delta_{\alpha\beta}\delta_{ij}\partial_\mu + \frac{ig}{2}\delta_{\alpha\beta}(\tau^a)_{ij}W^a_\mu + \frac{ig_s}{2}(\lambda^a)_{\alpha\beta}\delta_{ij}G^a_\mu + ig'Y_q B_\mu\delta_{\alpha\beta}\delta_{ij}\right]q_{\beta j}, \tag{2.3}$$

| | $l_L$ | $e_R$ | $q_L$ | $u_R$ | $d_R$ | $\Phi_1$ | $\Phi_2$ |
|---|---|---|---|---|---|---|---|
| $SU(3)$ | **1** | **1** | **3** | **3** | **3** | **1** | **1** |
| $SU(2)_L$ | **2** | **1** | **2** | **1** | **1** | **2** | **2** |
| $U(1)_Y$ | $-\frac{1}{2}$ | $-1$ | $\frac{1}{6}$ | $\frac{2}{3}$ | $-\frac{1}{3}$ | $\frac{1}{2}$ | $\frac{1}{2}$ |

**Table 1**. $SU(3)\times SU(2)_L\times U(1)_Y$ quantum numbers of standard model leptons, quarks and Higgs doublets. The electric charge generated after electroweak symmetry breaking is $Q = T^3 + Y$.

where $\tau^a/2$ and $\lambda^a/2$ are the generators of $SU(2)$ and $SU(3)$, respectively, with $\tau^a$ and $\lambda^a$ being the Pauli matrices and Gell-Mann matrices. $SU(2)$ and $SU(3)$ fundamental indices are written with $(ij)$ and $(\alpha\beta)$ where needed, respectively. The quantum numbers of the particle content are provided in Table 1. Field strength tensors of the gauge fields are defined as

$$\begin{aligned} B_{\mu\nu} &= \partial_\mu B_\nu - \partial_\nu B_\mu, \\ W^a_{\mu\nu} &= \partial_\mu W^a_\nu - \partial_\nu W^a_\mu - g\epsilon^{abc}W^b_\mu W^c_\nu, \\ G^a_{\mu\nu} &= \partial_\mu G^a_\nu - \partial_\nu G^a_\mu - g_s f^{abc}G^b_\mu G^c_\nu. \end{aligned} \tag{2.4}$$

The two Higgs fields $\Phi_{1,2}$ are complex $SU(2)$ doublets, each with hypercharge $Y_{1,2} = +1/2$, defined as [31]:

$$\Phi_1 = \begin{pmatrix} \Phi_1^+ \\ \Phi_1^0 \end{pmatrix} = \begin{pmatrix} \phi_1^+ \\ v_1 + \frac{1}{\sqrt{2}}(\rho_1 + ia_1) \end{pmatrix}, \qquad \Phi_2 = \begin{pmatrix} \Phi_2^+ \\ \Phi_2^0 \end{pmatrix} = \begin{pmatrix} \phi_2^+ \\ v_2 + \frac{1}{\sqrt{2}}(\rho_2 + ia_2) \end{pmatrix}, \tag{2.5}$$

where $\rho_1$, $\rho_2$, $a_1$, and $a_2$ are real scalar fields, and $\phi^\pm_{1,2}$ are charged scalar fields. After electroweak symmetry breaking (EWSB), the scalar doublets develop vacuum expectation values $\langle\Phi_1^0\rangle = v_1$ and $\langle\Phi_2^0\rangle = v_2$, where their ratio is parameterized by $v_2/v_1 = \tan\beta$, and $v = \sqrt{v_1^2 + v_2^2} = 174$ GeV.

The most general scalar potential with two Higgs doublets is

$$\begin{aligned} V(\Phi_1,\Phi_2) = {} & m_1^2\left(\Phi_1^\dagger\Phi_1\right) + m_2^2\left(\Phi_2^\dagger\Phi_2\right) + \left(m_{12}^2\Phi_1^\dagger\Phi_2 + h.c.\right) \\ & + \frac{1}{2}\lambda_1\left(\Phi_1^\dagger\Phi_1\right)^2 + \frac{1}{2}\lambda_2\left(\Phi_2^\dagger\Phi_2\right)^2 + \lambda_3\left(\Phi_1^\dagger\Phi_1\right)\left(\Phi_2^\dagger\Phi_2\right) + \lambda_4\left(\Phi_1^\dagger\Phi_2\right)\left(\Phi_2^\dagger\Phi_1\right) \\ & + \left(\frac{1}{2}\lambda_5\left(\Phi_1^\dagger\Phi_2\right)^2 + \lambda_6(\Phi_1^\dagger\Phi_1)\Phi_1^\dagger\Phi_2 + \lambda_7(\Phi_2^\dagger\Phi_2)\Phi_1^\dagger\Phi_2 + h.c.\right). \end{aligned} \tag{2.6}$$

The mixed kinetic terms in the dimension-four Lagrangian can be made diagonal via the non-unitary transformation,

$$(\Phi_1,\Phi_2) \to \left(\frac{\sqrt{\eta^*}\Phi_1 + \sqrt{\eta}\Phi_2}{2\sqrt{|\eta|(1+|\eta|)}} \pm \frac{\sqrt{\eta^*}\Phi_1 - \sqrt{\eta}\Phi_2}{2\sqrt{|\eta|(1-|\eta|)}}\right), \tag{2.7}$$

or if $\eta$ is present, we assume $\eta \ll 1$ at tree level. However, even if it is absent at tree level, as mentioned in Ref. [5], a general 2HDM can induce this kinetic term or not at loop level depending on the choice of renormalization scheme, with different schemes related by scale-dependent field redefinitions [32]. We choose to include it for completeness, which will affect the canonical definitions of the physical scalar and gauge fields in the main text. However, since it is expected that $\eta \ll 1$, we will neglect terms involving $|\eta|^2$ and $\eta \times \mathcal{O}(v^2/\Lambda^2)$.

Additional contributions due to dimension-five and -six operators are noted as $\mathcal{O}_i^{(5)}$ and $\mathcal{O}_i^{(6)}$, respectively, provided in Ref. [5]. The mass scale is contained inside their respective Wilson coefficients $C_i^{(5)}$ and $C_i^{(6)}$.

## 3 Canonical Field Redefinitions and Mass Eigenstates

### 3.1 Scalar Sector

In the effective theory, the scalar fields and masses are modified in the presence of dimension-six operators. The part of the Lagrangian containing the relevant kinetic terms is

$$
\begin{aligned}
\mathcal{L} \supset\ & (D_\mu\Phi_1)^\dagger D^\mu\Phi_1 + (D_\mu\Phi_2)^\dagger D^\mu\Phi_2 + \Big(\eta (D_\mu\Phi_1)^\dagger D^\mu\Phi_2 + h.c.\Big) \\
& + C_{\Phi\partial^2}^{(11)(11)} \partial_\mu(\Phi_1^\dagger\Phi_1)\partial^\mu(\Phi_1^\dagger\Phi_1) + C_{\Phi\partial^2}^{(22)(22)} \partial_\mu(\Phi_2^\dagger\Phi_2)\partial^\mu(\Phi_2^\dagger\Phi_2) \\
& + C_{\Phi\partial^2}^{(11)(22)} \partial_\mu(\Phi_1^\dagger\Phi_1)\partial^\mu(\Phi_2^\dagger\Phi_2) + C_{\Phi\partial^2}^{(21)(12)} \partial_\mu(\Phi_2^\dagger\Phi_1)\partial^\mu(\Phi_1^\dagger\Phi_2) \\
& + \Big( C_{\Phi\partial^2}^{(21)(21)} \partial_\mu(\Phi_2^\dagger\Phi_1)\partial^\mu(\Phi_2^\dagger\Phi_1) + C_{\Phi\partial^2}^{(21)(11)} \partial_\mu(\Phi_2^\dagger\Phi_1)\partial^\mu(\Phi_1^\dagger\Phi_1) \\
& + C_{\Phi\partial^2}^{(21)(22)} \partial_\mu(\Phi_2^\dagger\Phi_1)\partial^\mu(\Phi_2^\dagger\Phi_2) + h.c.\Big) \\
& + C_{\Phi D}^{(11)(11)} (\Phi_1^\dagger \overleftrightarrow{D}_\mu \Phi_1)(\Phi_1^\dagger \overleftrightarrow{D}^\mu \Phi_1) + C_{\Phi D}^{(22)(22)} (\Phi_2^\dagger \overleftrightarrow{D}_\mu \Phi_2)(\Phi_2^\dagger \overleftrightarrow{D}^\mu \Phi_2) \\
& + C_{\Phi D}^{(11)(22)} (\Phi_1^\dagger \overleftrightarrow{D}_\mu \Phi_1)(\Phi_2^\dagger \overleftrightarrow{D}^\mu \Phi_2) + C_{\Phi D}^{(21)(12)} (\Phi_2^\dagger \overleftrightarrow{D}_\mu \Phi_1)(\Phi_1^\dagger \overleftrightarrow{D}^\mu \Phi_2) \\
& + \Big( C_{\Phi D}^{(21)(21)} (\Phi_2^\dagger \overleftrightarrow{D}_\mu \Phi_1)(\Phi_2^\dagger \overleftrightarrow{D}^\mu \Phi_1) + C_{\Phi D}^{(21)(11)} (\Phi_2^\dagger \overleftrightarrow{D}_\mu \Phi_1)(\Phi_1^\dagger \overleftrightarrow{D}^\mu \Phi_1) \\
& + C_{\Phi D}^{(21)(22)} (\Phi_2^\dagger \overleftrightarrow{D}_\mu \Phi_1)(\Phi_2^\dagger \overleftrightarrow{D}^\mu \Phi_2) + h.c.\Big) ,
\end{aligned}
\tag{3.1}
$$

where the doublets are defined in Eq. (2.5). The covariant derivatives appear in both a symmetric combination,

$$
\partial_\mu(\Phi_{1,2}^\dagger\Phi_{1,2}) = \Phi_{1,2}^\dagger(D_\mu\Phi_{1,2}) + (D_\mu\Phi_{1,2})^\dagger\Phi_{1,2},
\tag{3.2}
$$

and an antisymmetric combination:

$$
\begin{aligned}
\Phi_{1,2}^\dagger \overleftrightarrow{D}_\mu \Phi_{1,2} &\equiv \Big(\Phi_{1,2}^\dagger(D_\mu\Phi_{1,2}) - (D_\mu\Phi_{1,2})^\dagger\Phi_{1,2}\Big), \\
\Phi_{1,2}^\dagger \overleftrightarrow{D}_\mu^a \Phi_{1,2} &\equiv \Big(\Phi_{1,2}^\dagger\tau^a(D_\mu\Phi_{1,2}) - (D_\mu\Phi_{1,2})^\dagger\tau^a\Phi_{1,2}\Big).
\end{aligned}
\tag{3.3}
$$

We collect kinetic terms from the unrotated $\rho_{1,2}$ and $a_{1,2}$, as well as charged fields:

$$
\begin{aligned}
\mathcal{L} \supset\ & \frac{1}{2}\begin{pmatrix}\partial_\mu\rho_1\\ \partial_\mu\rho_2\\ \partial_\mu a_1\\ \partial_\mu a_2\end{pmatrix}^T \begin{pmatrix} 1+A_1 & B & J & K \\ B & 1+A_2 & L & N \\ J & L & 1+A_1' & B' \\ K & N & B' & 1+A_2' \end{pmatrix} \begin{pmatrix}\partial^\mu\rho_1\\ \partial^\mu\rho_2\\ \partial^\mu a_1\\ \partial^\mu a_2\end{pmatrix} \\
& + \begin{pmatrix}\partial_\mu\phi_1^+ & \partial_\mu\phi_2^+\end{pmatrix} \begin{pmatrix} 1 & \eta^* \\ \eta & 1\end{pmatrix} \begin{pmatrix}\partial^\mu\phi_1^-\\ \partial^\mu\phi_2^-\end{pmatrix},
\end{aligned}
\tag{3.4}
$$

where

$$
\begin{aligned}
A_1 = v^2 \Big[ & 4c_\beta^2 C^{(11)(11)}_{\Phi\partial^2} + 2s_\beta^2 \, \mathrm{Re}[C^{(21)(21)}_{\Phi\partial^2}] + s_\beta^2 C^{(21)(12)}_{\Phi\partial^2} \\
& +4s_\beta c_\beta \, \mathrm{Re}[C^{(21)(11)}_{\Phi\partial^2}] + 2s_\beta^2 \mathrm{Re}[C^{(21)(21)}_{\Phi D}] - s_\beta^2 C^{(21)(12)}_{\Phi D} \Big] ,
\end{aligned} \tag{3.5}
$$

$$
\begin{aligned}
A_2 = v^2 \Big[ & 4s_\beta^2 C^{(22)(22)}_{\Phi\partial^2} + 2c_\beta^2 \, \mathrm{Re}[C^{(21)(21)}_{\Phi\partial^2}] + c_\beta^2 C^{(21)(12)}_{\Phi\partial^2} \\
& +4s_\beta c_\beta \mathrm{Re}[C^{(21)(22)}_{\Phi\partial^2}] + 2c_\beta^2 \mathrm{Re}[C^{(21)(21)}_{\Phi D}] - c_\beta^2 C^{(21)(12)}_{\Phi D} \Big] ,
\end{aligned} \tag{3.6}
$$

$$
\begin{aligned}
B = \mathrm{Re}[\eta] + v^2 \Big[ & 2s_\beta c_\beta C^{(11)(22)}_{\Phi\partial^2} + 2s_\beta c_\beta \mathrm{Re}[C^{(21)(21)}_{\Phi\partial^2}] + s_\beta c_\beta C^{(21)(12)}_{\Phi\partial^2} \\
& +2c_\beta^2 \mathrm{Re}[C^{(21)(11)}_{\Phi\partial^2}] + 2s_\beta^2 \, \mathrm{Re}[C^{(21)(22)}_{\Phi\partial^2}] - 2s_\beta c_\beta \mathrm{Re}[C^{(21)(21)}_{\Phi D}] + s_\beta c_\beta C^{(21)(12)}_{\Phi D} \Big] ,
\end{aligned} \tag{3.7}
$$

$$
\begin{aligned}
A_1' = v^2 \Big[ & 4c_\beta^2 C^{(11)(11)}_{\Phi\partial^2} + 2s_\beta^2 \mathrm{Re}[C^{(21)(21)}_{\Phi\partial^2}] + s_\beta^2 C^{(21)(12)}_{\Phi\partial^2} + 4s_\beta c_\beta \mathrm{Re}[C^{(21)(11)}_{\Phi\partial^2}] \\
& -4c_\beta^2 C^{(11)(11)}_{\Phi D} - 2s_\beta^2 \mathrm{Re}[C^{(21)(21)}_{\Phi D}] - s_\beta^2 C^{(21)(12)}_{\Phi D} - 4s_\beta c_\beta \mathrm{Re}[C^{(21)(11)}_{\Phi D}] \Big] ,
\end{aligned} \tag{3.8}
$$

$$
\begin{aligned}
A_2' = v^2 \Big[ & 4s_\beta^2 C^{(22)(22)}_{\Phi\partial^2} + 2c_\beta^2 \mathrm{Re}[C^{(21)(21)}_{\Phi\partial^2}] + c_\beta^2 C^{(21)(12)}_{\Phi\partial^2} + 4s_\beta c_\beta \mathrm{Re}[C^{(21)(22)}_{\Phi\partial^2}] \\
& -4s_\beta^2 C^{(22)(22)}_{\Phi D} - 2c_\beta^2 \mathrm{Re}[C^{(21)(21)}_{\Phi D}] - c_\beta^2 C^{(21)(12)}_{\Phi D} - 4s_\beta c_\beta \mathrm{Re}[C^{(21)(22)}_{\Phi D}] \Big] ,
\end{aligned} \tag{3.9}
$$

$$
\begin{aligned}
B' = \mathrm{Re}[\eta] + v^2 \Big[ & 2s_\beta c_\beta C^{(11)(22)}_{\Phi\partial^2} + 2s_\beta c_\beta \mathrm{Re}[C^{(21)(21)}_{\Phi\partial^2}] \\
& +s_\beta c_\beta C^{(21)(12)}_{\Phi\partial^2} + 2c_\beta^2 \mathrm{Re}[C^{(21)(11)}_{\Phi\partial^2}] + 2s_\beta^2 \mathrm{Re}[C^{(21)(22)}_{\Phi\partial^2}] \\
& -2s_\beta c_\beta C^{(11)(22)}_{\Phi D} - 2s_\beta c_\beta \mathrm{Re}[C^{(21)(21)}_{\Phi D}] \\
& -s_\beta c_\beta C^{(21)(12)}_{\Phi D} - 2c_\beta^2 \mathrm{Re}[C^{(21)(11)}_{\Phi D}] - 2s_\beta^2 \mathrm{Re}[C^{(21)(22)}_{\Phi D}] \Big] ,
\end{aligned} \tag{3.10}
$$

and

$$
\begin{aligned}
J &= -2v^2 \Big[ s_\beta^2 \, \mathrm{Im}[C^{(21)(21)}_{\Phi\partial^2}] + s_\beta c_\beta \, \mathrm{Im}[C^{(21)(11)}_{\Phi\partial^2}] \\
&\qquad +s_\beta^2 \, \mathrm{Im}[C^{(21)(21)}_{\Phi D}] + s_\beta c_\beta \, \mathrm{Im}[C^{(21)(11)}_{\Phi D}] \Big] , \\
N &= +2v^2 \Big[ c_\beta^2 \, \mathrm{Im}[C^{(21)(21)}_{\Phi\partial^2}] + s_\beta c_\beta \, \mathrm{Im}[C^{(21)(22)}_{\Phi\partial^2}] \\
&\qquad +c_\beta^2 \, \mathrm{Im}[C^{(21)(21)}_{\Phi D}] + s_\beta c_\beta \, \mathrm{Im}[C^{(21)(22)}_{\Phi D}] \Big] , \\
K &= - \mathrm{Im}[\eta] + 2v^2 \Big[ s_\beta c_\beta \, \mathrm{Im}[C^{(21)(21)}_{\Phi\partial^2}] + c_\beta^2 \, \mathrm{Im}[C^{(21)(11)}_{\Phi\partial^2}] \\
&\qquad -s_\beta c_\beta \, \mathrm{Im}[C^{(21)(21)}_{\Phi D}] - s_\beta^2 \, \mathrm{Im}[C^{(21)(22)}_{\Phi D}] \Big] , \\
L &= + \mathrm{Im}[\eta] - 2v^2 \Big[ s_\beta c_\beta \, \mathrm{Im}[C^{(21)(21)}_{\Phi\partial^2}] + s_\beta^2 \, \mathrm{Im}[C^{(21)(22)}_{\Phi\partial^2}] \\
&\qquad -s_\beta c_\beta \, \mathrm{Im}[C^{(21)(21)}_{\Phi D}] - c_\beta^2 \, \mathrm{Im}[C^{(21)(11)}_{\Phi D}] \Big] ,
\end{aligned} \tag{3.11}
$$

where for compactness, we define $\sin\beta \equiv s_\beta$, $\cos\beta \equiv c_\beta$, and $\tan\beta \equiv t_\beta$. Note that the entire $4 \times 4$ matrix in Eq. (3.4) is real and symmetric, and thus it can be written in terms

of block matrices

$$\mathcal{L} \supset \frac{1}{2}\begin{pmatrix}\partial_\mu\vec{\rho}\\ \partial_\mu\vec{a}\end{pmatrix}^T\begin{pmatrix}I_{2\times2}+A & D\\ D^T & I_{2\times2}+A'\end{pmatrix}\begin{pmatrix}\partial^\mu\vec{\rho}\\ \partial^\mu\vec{a}\end{pmatrix}, \tag{3.12}$$

where

$$A \equiv \begin{pmatrix}A_1 & B\\ B & A_2\end{pmatrix}, \qquad A' \equiv \begin{pmatrix}A_1' & B'\\ B' & A_2'\end{pmatrix}, \qquad D \equiv \begin{pmatrix}J & K\\ L & N\end{pmatrix}, \tag{3.13}$$

$\vec{\rho} = (\rho_1, \rho_2)^T$, and $\vec{a} = (a_1, a_2)^T$. If the theory were $CP$-conserving, $D = 0$ and the $\vec{\rho}$ and $\vec{a}$ fields would be decoupled from one another. One could then identify $\vec{\rho}$ containing the usual $CP$-even Higgs scalars $h$ and $H$ and $\vec{a}$ containing the usual $CP$-odd Higgs scalars $G$ and $A$.

In order to make the fields canonical, we need to rotate to the basis where the kinetic terms are diagonal and are normalized to 1. For the block matrix given in Eq. (3.12), this can be accomplished up to leading order $\mathcal{O}(1/\Lambda^2)$ by

$$\vec{\rho} \to \left(I_{2\times2} - \frac{A}{2}\right)\hat{\vec{\rho}} - \frac{D}{2}\hat{\vec{a}}, \qquad \hat{\vec{a}} \to -\frac{D^T}{2}\hat{\vec{\rho}} + \left(I_{2\times2} - \frac{A'}{2}\right)\hat{\vec{a}}, \tag{3.14}$$

where $\hat{\vec{\rho}} = (\hat{\rho}_1, \hat{\rho}_2)^T$ and $\hat{\vec{a}} = (\hat{a}_1, \hat{a}_2)^T$. In component form, the original $\rho_{1,2}$ and $a_{1,2}$ scalars transform as

$$\begin{aligned}
\rho_1 &\to \left(1 - \frac{A_1}{2}\right)\hat{\rho}_1 - \frac{B}{2}\hat{\rho}_2 - \frac{J}{2}\hat{a}_1 - \frac{K}{2}\hat{a}_2,\\
\rho_2 &\to -\frac{B}{2}\hat{\rho}_1 + \left(1 - \frac{A_2}{2}\right)\hat{\rho}_2 - \frac{L}{2}\hat{a}_1 - \frac{N}{2}\hat{a}_2,\\
a_1 &\to -\frac{J}{2}\hat{\rho}_1 - \frac{L}{2}\hat{\rho}_2 + \left(1 - \frac{A_1'}{2}\right)\hat{a}_1 - \frac{B'}{2}\hat{a}_2,\\
a_2 &\to -\frac{K}{2}\hat{\rho}_1 - \frac{N}{2}\hat{\rho}_2 - \frac{B'}{2}\hat{a}_1 + \left(1 - \frac{A_2'}{2}\right)\hat{a}_2,
\end{aligned} \tag{3.15}$$

and the charged fields transform as

$$\phi_1^+ \to \hat{\phi}_1^+ - \frac{\eta}{2}\hat{\phi}_2^+, \qquad \phi_2^+ \to -\frac{\eta^*}{2}\hat{\phi}_1^+ + \hat{\phi}_2^+. \tag{3.16}$$

The scalar masses are not only corrected by the dimension-six scalar terms, but also by the dimension-four potential due to the above field redefinitions. The Lagrangian containing all contributions to the scalar masses is

$$\begin{aligned}
\mathcal{L} \supset &- V(\Phi_1, \Phi_2) + C_\Phi^{(11)(11)(11)}(\Phi_1^\dagger\Phi_1)^3 + C_\Phi^{(11)(11)(22)}(\Phi_1^\dagger\Phi_1)^2(\Phi_2^\dagger\Phi_2)\\
&+ C_\Phi^{(11)(22)(22)}(\Phi_1^\dagger\Phi_1)(\Phi_2^\dagger\Phi_2)^2 + C_\Phi^{(22)(22)(22)}(\Phi_2^\dagger\Phi_2)^3\\
&+ C_\Phi^{(11)(21)(12)}(\Phi_1^\dagger\Phi_1)(\Phi_2^\dagger\Phi_1)(\Phi_1^\dagger\Phi_2) + C_\Phi^{(22)(21)(12)}(\Phi_2^\dagger\Phi_2)(\Phi_2^\dagger\Phi_1)(\Phi_1^\dagger\Phi_2)\\
&+ \Big(C_\Phi^{(11)(11)(21)}(\Phi_1^\dagger\Phi_1)^2(\Phi_2^\dagger\Phi_1) + C_\Phi^{(22)(22)(21)}(\Phi_2^\dagger\Phi_2)^2(\Phi_2^\dagger\Phi_1)\\
&+ C_\Phi^{(11)(21)(21)}(\Phi_1^\dagger\Phi_1)(\Phi_2^\dagger\Phi_1)^2 + C_\Phi^{(22)(21)(21)}(\Phi_2^\dagger\Phi_2)(\Phi_2^\dagger\Phi_1)^2\\
&+ C_\Phi^{(21)(21)(21)}(\Phi_2^\dagger\Phi_1)^3 + C_\Phi^{(21)(21)(12)}(\Phi_2^\dagger\Phi_1)^2(\Phi_1^\dagger\Phi_2)\\
&+ C_\Phi^{(11)(22)(21)}(\Phi_1^\dagger\Phi_1)(\Phi_2^\dagger\Phi_2)(\Phi_2^\dagger\Phi_1) + h.c.\Big),
\end{aligned} \tag{3.17}$$

where the scalar potential $V(\Phi_1, \Phi_2)$ is defined in Eq. (2.6). Focusing on the neutral scalars for a moment, the mass squared matrix for the $\rho$ and $a$ scalars is given by the $2 \times 2$ block matrix

$$\mathcal{L} \supset -\frac{1}{2}\begin{pmatrix} \vec{\rho} \\ \vec{a} \end{pmatrix}^T \begin{pmatrix} I_{2\times2} - \frac{A}{2} & -\frac{D}{2} \\ -\frac{D^T}{2} & I_{2\times2} - \frac{A'}{2} \end{pmatrix} \begin{pmatrix} M^2_{\rho\rho} & M^2_{\rho a} \\ M^2_{a\rho} & M^2_{aa} \end{pmatrix} \begin{pmatrix} I_{2\times2} - \frac{A}{2} & -\frac{D}{2} \\ -\frac{D^T}{2} & I_{2\times2} - \frac{A'}{2} \end{pmatrix} \begin{pmatrix} \vec{\rho} \\ \vec{a} \end{pmatrix}, \tag{3.18}$$

where $(M^2_{\rho a})^T = M^2_{a\rho}$ and, after rotating to the basis where the fields are canonical,

$$\begin{aligned}
&\mathcal{L} \supset -\frac{1}{2}\begin{pmatrix} \vec{\hat{\rho}} \\ \vec{\hat{a}} \end{pmatrix}^T \begin{pmatrix} \hat{M}^2_{\rho\rho} & \hat{M}^2_{\rho a} \\ (\hat{M}^2_{\rho a})^T & \hat{M}^2_{aa} \end{pmatrix} \begin{pmatrix} \vec{\hat{\rho}} \\ \vec{\hat{a}} \end{pmatrix}, \\
&\hat{M}^2_{\rho\rho} = M^2_{\rho\rho} - \frac{A}{2}M^2_{\rho\rho} - M^2_{\rho\rho}\frac{A}{2} - M^2_{\rho a}\frac{D^T}{2} - \frac{D}{2}(M^2_{\rho a})^T, \\
&\hat{M}^2_{\rho a} = M^2_{\rho a} - \frac{D}{2}M^2_{aa} - M^2_{\rho\rho}\frac{D}{2} - \frac{A}{2}M^2_{\rho a} - M^2_{\rho a}\frac{A'}{2}, \\
&\hat{M}^2_{aa} = M^2_{aa} - \frac{A'}{2}M^2_{aa} - M^2_{aa}\frac{A'}{2} - (M^2_{\rho a})^T\frac{D}{2} - \frac{D^T}{2}M^2_{\rho a},
\end{aligned} \tag{3.19}$$

to the leading order. All matrix elements in terms of Lagrangian parameters are given in Appendix A. Although the diagonalization of this matrix seems overwhelming since the theory now contains states of indefinite $CP$ raveled in a $4 \times 4$ matrix, we note that there is one mass eigenstate with zero eigenvalue, identified as the neutral Goldstone boson $G$, and its orthogonal eigenstate associated with the would-be $CP$-odd Higgs $A$ if $CP$ were conserved. These eigenstates are found by diagonalizing the bottom $\hat{M}^2_{aa}$ matrix via the rotation matrix

$$\begin{pmatrix} \hat{a}_1 \\ \hat{a}_2 \end{pmatrix} = \begin{pmatrix} \cos\hat{\beta} & -\sin\hat{\beta} \\ \sin\hat{\beta} & \cos\hat{\beta} \end{pmatrix} \begin{pmatrix} G \\ A \end{pmatrix}, \tag{3.20}$$

where the angle $\hat{\beta}$ is determined in Appendix B. After rearranging the components, we have

$$\mathcal{L} \supset -\frac{1}{2}\begin{pmatrix} G \\ A \\ \hat{\rho}_1 \\ \hat{\rho}_2 \end{pmatrix}^T \begin{pmatrix} m^2_G & 0 & 0 & 0 \\ 0 & m^2_A & \hat{M}^2_{A\rho_1} & \hat{M}^2_{A\rho_2} \\ 0 & \hat{M}^2_{A\rho_1} & \hat{M}^2_{\rho_1\rho_1} & \hat{M}^2_{\rho_1\rho_2} \\ 0 & \hat{M}^2_{A\rho_2} & \hat{M}^2_{\rho_2\rho_1} & \hat{M}^2_{\rho_2\rho_2} \end{pmatrix} \begin{pmatrix} G \\ A \\ \hat{\rho}_1 \\ \hat{\rho}_2 \end{pmatrix}, \tag{3.21}$$

where $m^2_A$ and $m^2_G = 0$ are also given in Appendix B and

$$\begin{aligned}
\hat{M}^2_{A\rho_1} &= -\sin\hat{\beta}\hat{M}^2_{\rho_1 a_1} + \cos\hat{\beta}\hat{M}^2_{\rho_1 a_2}, \\
\hat{M}^2_{A\rho_2} &= -\sin\hat{\beta}\hat{M}^2_{\rho_2 a_1} + \cos\hat{\beta}\hat{M}^2_{\rho_2 a_2}.
\end{aligned} \tag{3.22}$$

From here, one can further diagonalize to a basis of mass eigenstates $\{h_3, h_2, h_1\}$ indefinite in $CP$ via Euler angles [15] in the subspace spanned by $\{A, \hat{\rho}_1, \hat{\rho}_2\}$:

$$\begin{pmatrix} A \\ \hat{\rho}_1 \\ \hat{\rho}_2 \end{pmatrix} = R \begin{pmatrix} h_3 \\ h_2 \\ h_1 \end{pmatrix} = R_x(\hat{\alpha})R_y(\hat{\xi})R_z(\hat{\omega}) \begin{pmatrix} h_3 \\ h_2 \\ h_1 \end{pmatrix}, \tag{3.23}$$

where the rotation matrices are

$$R_x(\hat{\alpha}) = \begin{pmatrix} 1 & 0 & 0 \\ 0 & \cos\hat{\alpha} & -\sin\hat{\alpha} \\ 0 & \sin\hat{\alpha} & \cos\hat{\alpha} \end{pmatrix},$$
$$R_y(\hat{\xi}) = \begin{pmatrix} \cos\hat{\xi} & 0 & \sin\hat{\xi} \\ 0 & 1 & 0 \\ -\sin\hat{\xi} & 0 & \cos\hat{\xi} \end{pmatrix}, \qquad R_z(\hat{\omega}) = \begin{pmatrix} \cos\hat{\omega} & -\sin\hat{\omega} & 0 \\ \sin\hat{\omega} & \cos\hat{\omega} & 0 \\ 0 & 0 & 1 \end{pmatrix}. \tag{3.24}$$

To help simplify the problem while characterizing $CP$-violating effects, we can partially diagonalize the $2\times 2$ block matrix $\hat{M}^2_{\rho\rho}$ mixing the $\hat{\rho}_{1,2}$ fields by $R_x(\hat{\alpha})$:

$$\begin{pmatrix} A \\ \hat{\rho}_1 \\ \hat{\rho}_2 \end{pmatrix} = \begin{pmatrix} 1 & 0 & 0 \\ 0 & \cos\hat{\alpha} & -\sin\hat{\alpha} \\ 0 & \sin\hat{\alpha} & \cos\hat{\alpha} \end{pmatrix} \begin{pmatrix} A \\ H \\ h \end{pmatrix}, \tag{3.25}$$

to the basis spanned by the would-be $CP$-even scalars $H$ and $h$. Performing the rotation gives us

$$\mathcal{L} \supset -\frac{1}{2} \begin{pmatrix} A \\ H \\ h \end{pmatrix}^T \begin{pmatrix} m_A^2 & m_{AH}^2 & m_{Ah}^2 \\ m_{AH}^2 & m_H^2 & 0 \\ m_{Ah}^2 & 0 & m_h^2 \end{pmatrix} \begin{pmatrix} A \\ H \\ h \end{pmatrix}, \tag{3.26}$$

where $m_H^2 > m_h^2$ and

$$\begin{aligned} m_H^2 &= \cos^2\hat{\alpha}\hat{M}^2_{\rho_1\rho_1} + \sin 2\hat{\alpha}\hat{M}^2_{\rho_1\rho_2} + \sin^2\hat{\alpha}\hat{M}^2_{\rho_2\rho_2}, \\ m_h^2 &= \sin^2\hat{\alpha}\hat{M}^2_{\rho_1\rho_1} - \sin 2\hat{\alpha}\hat{M}^2_{\rho_1\rho_2} + \cos^2\hat{\alpha}\hat{M}^2_{\rho_2\rho_2}, \\ m_{AH}^2 &= \cos\hat{\alpha}\hat{M}^2_{A\rho_1} + \sin\hat{\alpha}\hat{M}^2_{A\rho_2}, \\ m_{Ah}^2 &= -\sin\hat{\alpha}\hat{M}^2_{A\rho_1} + \cos\hat{\alpha}\hat{M}^2_{A\rho_2}. \end{aligned} \tag{3.27}$$

The angle $\hat{\alpha}$ is calculated in Appendix B. The $CP$-violation is characterized by the off-diagonal elements $m_{AH}^2$ and $m_{Ah}^2$ of the above matrix [14, 15, 33] and in the limit that they approach zero, $\{A, H, h\}$ are the physical eigenstates. The diagonalization of the $3\times 3$ matrix has been done either numerically or explored in specific limits. Here, we present analytical results for the diagonalization of this $3\times 3$ matrix.

The remaining Euler angles which rotate the $\{A, H, h\}$ basis into $\{h_3, h_2, h_1\}$,

$$\begin{pmatrix} A \\ H \\ h \end{pmatrix} = R_y(\hat{\xi}) R_z(\hat{\omega}) \begin{pmatrix} h_3 \\ h_2 \\ h_1 \end{pmatrix}, \tag{3.28}$$

can be found in terms of the eigenvalues $\lambda^{(1,2,3)}$ of the $3\times 3$ matrix and matrix parameters following the method outlined in Ref. [16]:

$$\begin{aligned} \cos^2\hat{\xi} &= \frac{(m_{AH}^2)^2 + (m_{Ah}^2)^2 + (m_A^2 - \lambda^{(2)})(m_A^2 + \lambda^{(2)} - \lambda^{(3)} - \lambda^{(1)})}{(\lambda^{(1)} - \lambda^{(2)})(\lambda^{(2)} - \lambda^{(3)})}, \\ \cos^2\hat{\omega} &= \frac{m_A^2 - \lambda^{(2)} + \cos^2\hat{\xi}(\lambda^{(2)} - \lambda^{(1)})}{\cos^2\hat{\xi}(\lambda^{(3)} - \lambda^{(1)})}. \end{aligned} \tag{3.29}$$

The eigenvalues are found by solving the characteristic equation of the matrix in Eq. (3.26)

$$\lambda^3 + b\lambda^2 + c\lambda + d = 0, \tag{3.30}$$

where

$$\begin{aligned} b &= -|b| = -(m_A^2 + m_H^2 + m_h^2), \\ c &= m_H^2 m_h^2 + m_A^2(m_H^2 + m_h^2) - (m_{AH}^2)^2 - (m_{Ah}^2)^2, \\ d &= (m_{AH}^2)^2 m_h^2 + (m_{Ah}^2)^2 m_H^2 - m_A^2 m_H^2 m_h^2. \end{aligned} \tag{3.31}$$

Note that in the $CP$-conserving theory, $m_{AH}^2 = m_{Ah}^2 = 0$ and $\lambda^{(1,2,3)} = m_{h,H,A}^2$, respectively, implying $\hat{\xi} = \hat{\omega} = 0$. The cubic characteristic polynomial can be solved analytically [17] by reducing it to $y^3 + 3py + q = 0$ with

$$y = \lambda + \frac{b}{3}, \qquad p = -|p| = \frac{3c - b^2}{9}, \qquad q = -|q| = \frac{2b^3 - 9bc + 27d}{27}. \tag{3.32}$$

Given the fact that the eigenvalues are real, the discriminant $\Delta = q^2 + 4p^3 \leq 0$ is satisfied, where equality occurs if at least two solutions are identical. Then, the eigenvalues are

$$\begin{aligned} \lambda^{(1)} \equiv m_{h_1}^2 &= \frac{|b|}{3} - \frac{1}{2}\left(\zeta^{1/3} + \sigma^{1/3}\right) + i\frac{\sqrt{3}}{2}\left(\zeta^{1/3} - \sigma^{1/3}\right), \\ \lambda^{(2)} \equiv m_{h_2}^2 &= \frac{|b|}{3} - \frac{1}{2}\left(\zeta^{1/3} + \sigma^{1/3}\right) - i\frac{\sqrt{3}}{2}\left(\zeta^{1/3} - \sigma^{1/3}\right), \\ \lambda^{(3)} \equiv m_{h_3}^2 &= \frac{|b|}{3} + \zeta^{1/3} + \sigma^{1/3}, \end{aligned} \tag{3.33}$$

where $m_{h_1}^2 \leq m_{h_2}^2 \leq m_{h_3}^2$ and

$$\zeta \equiv \frac{1}{2}\left(|q| + i\sqrt{4|p|^3 - q^2}\right), \qquad \sigma \equiv \frac{1}{2}\left(|q| - i\sqrt{4|p|^3 - q^2}\right). \tag{3.34}$$

The reader may be concerned due to the presence of $i$ in the eigenvalues, however, note that these combinations are indeed real, as required for physical masses.[2] These formulas are exact, and could be used in studying the $CP$-violating 2HDM. In the 2HDM EFT, we need to expand to the leading order in $\mathcal{O}(1/\Lambda^2)$ for consistency by taking all of the parameters in the above formulas as $x \to x + \Delta x$ where the $\Delta x$ contains contributions from both the field redefinitions and the dimension-six operators.

The charged scalars have the following mass squared matrix

$$\mathcal{L} \supset -(\hat{\phi}_1^+ \; \hat{\phi}_2^+) \begin{pmatrix} M_{11}^{\pm 2} + \Delta M_{11}^{\pm 2} & M_{12}^{\pm 2} + \Delta M_{12}^{\pm 2} \\ M_{21}^{\pm 2} + \Delta M_{21}^{\pm 2} & M_{22}^{\pm 2} + \Delta M_{22}^{\pm 2} \end{pmatrix} \begin{pmatrix} \hat{\phi}_1^- \\ \hat{\phi}_2^- \end{pmatrix}, \tag{3.35}$$

[2] Because the real eigenvalues contain complex terms, one can alternatively parameterize the eigenvalues on the unit circle [17]: $\lambda^{(1)} = |b|/3 + 2|p|^{1/2}\cos((\theta + 2\pi)/3)$, $\lambda^{(2)} = |b|/3 + 2|p|^{1/2}\cos((\theta - 2\pi)/3)$, and $\lambda^{(3)} = |b|/3 + 2|p|^{1/2}\cos(\theta/3)$ with $\theta = \arccos(|q|/2|p|^{3/2}) \in [0, \pi/2]$. Although this parametrization is more attractive since it does not contain factors of $i$, it is less convenient when expanding the eigenvalues to the leading order $\mathcal{O}(1/\Lambda^2)$.

where the matrix elements are collected near the end of Appendix A. Diagonalizing the charged sector from the $\{\hat{\phi}_1^\pm, \hat{\phi}_2^\pm\}$ basis to the $\{G^\pm, H^\pm\}$ is straightforward and done in Appendix B.

In the general renormalizable 2HDM at tree level, the vacuum angle $\beta$ diagonalizes both the $\vec{a}$ and charged sectors. However, at the loop level or in the effective theory at mass dimension-six this is no longer true, and the diagonalization angles receive additional corrections: $\beta - \hat{\beta} \sim \mathcal{O}\left(v^4/\Lambda^2 m_A^2\right)$ and $\beta - \hat{\beta}^\pm \sim \mathcal{O}\left(v^4/\Lambda^2 m_{H^\pm}^2\right)$. In all the results we replace $\hat{\beta}$ and $\hat{\beta}^\pm$ by their relationship to the vacuum angle $\beta$ and Wilson coefficients. See Appendix B for details.

### 3.2 Gauge Sector

The part of the Lagrangian affecting the kinetic terms of the gauge fields is

$$
\begin{aligned}
\mathcal{L} \supset & -\frac{1}{4} B_{\mu\nu} B^{\mu\nu} - \frac{1}{4} W^a_{\mu\nu} W^{a\mu\nu} - \frac{1}{4} G^a_{\mu\nu} G^{a\mu\nu} \\
& + C^{(11)}_{\Phi B} (\Phi_1^\dagger \Phi_1) B_{\mu\nu} B^{\mu\nu} + C^{(22)}_{\Phi B} (\Phi_2^\dagger \Phi_2) B_{\mu\nu} B^{\mu\nu} + \left( C^{(21)}_{\Phi B} (\Phi_2^\dagger \Phi_1) B_{\mu\nu} B^{\mu\nu} + h.c. \right) \\
& + C^{(11)}_{\Phi W} (\Phi_1^\dagger \Phi_1) W^a_{\mu\nu} W^{a\mu\nu} + C^{(22)}_{\Phi W} (\Phi_2^\dagger \Phi_2) W^a_{\mu\nu} W^{a\mu\nu} + \left( C^{(21)}_{\Phi W} (\Phi_2^\dagger \Phi_1) W^a_{\mu\nu} W^{a\mu\nu} + h.c. \right) \\
& + C^{(11)}_{\Phi G} (\Phi_1^\dagger \Phi_1) G^a_{\mu\nu} G^{a\mu\nu} + C^{(22)}_{\Phi G} (\Phi_2^\dagger \Phi_2) G^a_{\mu\nu} G^{a\mu\nu} + \left( C^{(21)}_{\Phi G} (\Phi_2^\dagger \Phi_1) G^a_{\mu\nu} G^{a\mu\nu} + h.c. \right) \\
& + C^{(11)}_{\Phi WB} (\Phi_1^\dagger \tau^a \Phi_1) W^a_{\mu\nu} B^{\mu\nu} + C^{(22)}_{\Phi WB} (\Phi_2^\dagger \tau^a \Phi_2) W^a_{\mu\nu} B^{\mu\nu} + \left( C^{(21)}_{\Phi WB} (\Phi_2^\dagger \tau^a \Phi_1) W^a_{\mu\nu} B^{\mu\nu} + h.c. \right).
\end{aligned}
\tag{3.36}
$$

After EWSB, the kinetic terms are

$$
\begin{aligned}
\mathcal{L} \supset & -\frac{1}{4} \left[ 1 - 4 \left( v_1^2 C^{(11)}_{\Phi B} + v_2^2 C^{(22)}_{\Phi B} + 2 v_1 v_2 \ \mathrm{Re}[C^{(21)}_{\Phi B}] \right) \right] B_{\mu\nu} B^{\mu\nu} \\
& -\frac{1}{4} \left[ 1 - 4 \left( v_1^2 C^{(11)}_{\Phi W} + v_2^2 C^{(22)}_{\Phi W} + 2 v_1 v_2 \ \mathrm{Re}[C^{(21)}_{\Phi W}] \right) \right] W^a_{\mu\nu} W^{a\mu\nu} \\
& -\frac{1}{4} \left[ 1 - 4 \left( v_1^2 C^{(11)}_{\Phi G} + v_2^2 C^{(22)}_{\Phi G} + 2 v_1 v_2 \ \mathrm{Re}[C^{(21)}_{\Phi G}] \right) \right] G^a_{\mu\nu} G^{a\mu\nu} \\
& - \left[ v_1^2 C^{(11)}_{\Phi WB} + v_2^2 C^{(22)}_{\Phi WB} + 2 v_1 v_2 \ \mathrm{Re}[C^{(21)}_{\Phi WB}] \right] W^3_{\mu\nu} B^{\mu\nu}.
\end{aligned}
\tag{3.37}
$$

The charged fields and neutral fields of $W^a_\mu$ and $B_\mu$ can be separated:

$$
\begin{aligned}
\mathcal{L} \supset & -\frac{1}{4} \left( 1 + X_W \right) \left( W^1_{\mu\nu} W^{1\mu\nu} + W^2_{\mu\nu} W^{2\mu\nu} \right) - \frac{1}{4} \left( 1 + X_G \right) G^a_{\mu\nu} G^{a\mu\nu} \\
& - \frac{1}{4} \left( B_{\mu\nu} \ W^3_{\mu\nu} \right) \begin{pmatrix} 1 + X_B & X_{WB} \\ X_{WB} & 1 + X_W \end{pmatrix} \begin{pmatrix} B^{\mu\nu} \\ W^{3\mu\nu} \end{pmatrix},
\end{aligned}
\tag{3.38}
$$

where

$$
\begin{aligned}
X_B &= -4 \left( v_1^2 C^{(11)}_{\Phi B} + v_2^2 C^{(22)}_{\Phi B} + 2 v_1 v_2 \ \mathrm{Re}[C^{(21)}_{\Phi B}] \right), \\
X_W &= -4 \left( v_1^2 C^{(11)}_{\Phi W} + v_2^2 C^{(22)}_{\Phi W} + 2 v_1 v_2 \ \mathrm{Re}[C^{(21)}_{\Phi W}] \right), \\
X_G &= -4 \left( v_1^2 C^{(11)}_{\Phi G} + v_2^2 C^{(22)}_{\Phi G} + 2 v_1 v_2 \ \mathrm{Re}[C^{(21)}_{\Phi G}] \right), \\
X_{WB} &= +2 \left( v_1^2 C^{(11)}_{\Phi WB} + v_2^2 C^{(22)}_{\Phi WB} + 2 v_1 v_2 \ \mathrm{Re}[C^{(21)}_{\Phi WB}] \right).
\end{aligned}
\tag{3.39}
$$

The fields can be canonically normalized to the leading order by

$$B_\mu \to \left(1 - \frac{X_B}{2}\right) \hat{B}_\mu, \quad W^a_\mu \to \left(1 - \frac{X_W}{2}\right) \hat{W}^a_\mu, \quad G^a_\mu \to \left(1 - \frac{X_G}{2}\right) \hat{G}^a_\mu. \tag{3.40}$$

Note that to all orders, the vector fields scale as $Z_A A^{(a)}_\mu \to \hat{A}^{(a)}_\mu$ with $Z_A \equiv \sqrt{1 + X_A}$, for $A = B, W, G$. Through these field redefinitions, the structure of the covariant derivative is preserved:

$$(D_\mu q)_{\alpha i} \to (\hat{D}_\mu q)_{\alpha i} = \partial_\mu q_{\alpha i} + \frac{i\hat{g}}{2}(\tau^a)_{ij} q_{\alpha j} \hat{W}^a_\mu + \frac{i\hat{g}_s}{2}(\lambda^a)_{\alpha\beta} q_{\beta i} \hat{G}^a_\mu + i\hat{g}' Y_q \hat{B}_\mu q_{\alpha i}, \tag{3.41}$$

where the rescaled couplings are defined with carrots at the leading order:

$$\begin{aligned}
\hat{g} &= g \left[1 + 2\left(v_1^2 C^{(11)}_{\Phi W} + v_2^2 C^{(22)}_{\Phi W} + 2 v_1 v_2 \ \mathrm{Re}[C^{(21)}_{\Phi W}]\right)\right], \\
\hat{g}' &= g' \left[1 + 2\left(v_1^2 C^{(11)}_{\Phi B} + v_2^2 C^{(22)}_{\Phi B} + 2 v_1 v_2 \ \mathrm{Re}[C^{(21)}_{\Phi B}]\right)\right], \\
\hat{g}_s &= g_s \left[1 + 2\left(v_1^2 C^{(11)}_{\Phi G} + v_2^2 C^{(22)}_{\Phi G} + 2 v_1 v_2 \ \mathrm{Re}[C^{(21)}_{\Phi G}]\right)\right].
\end{aligned} \tag{3.42}$$

To all orders, the couplings are scaled as $g_A \to Z^{-1}_{g_A} \hat{g}_A = Z_A \hat{g}_A$.

The masses of gauge fields arise from terms with covariant derivatives:

$$\begin{aligned}
\mathcal{L} \supset & (D_\mu \Phi_1)^\dagger D^\mu \Phi_1 + (D_\mu \Phi_2)^\dagger D^\mu \Phi_2 + \left(\eta (D_\mu \Phi_1)^\dagger D^\mu \Phi_2 + h.c.\right) \\
& + C^{(11)(11)}_{\Phi D} (\Phi_1^\dagger \overleftrightarrow{D}_\mu \Phi_1)(\Phi_1^\dagger \overleftrightarrow{D}^\mu \Phi_1) + C^{(22)(22)}_{\Phi D} (\Phi_2^\dagger \overleftrightarrow{D}_\mu \Phi_2)(\Phi_2^\dagger \overleftrightarrow{D}^\mu \Phi_2) \\
& + C^{(11)(22)}_{\Phi D} (\Phi_1^\dagger \overleftrightarrow{D}_\mu \Phi_1)(\Phi_2^\dagger \overleftrightarrow{D}^\mu \Phi_2) + C^{(21)(12)}_{\Phi D} (\Phi_2^\dagger \overleftrightarrow{D}_\mu \Phi_1)(\Phi_1^\dagger \overleftrightarrow{D}^\mu \Phi_2) \\
& + \left(C^{(21)(21)}_{\Phi D} (\Phi_2^\dagger \overleftrightarrow{D}_\mu \Phi_1)(\Phi_2^\dagger \overleftrightarrow{D}^\mu \Phi_1) + C^{(21)(11)}_{\Phi D} (\Phi_2^\dagger \overleftrightarrow{D}_\mu \Phi_1)(\Phi_1^\dagger \overleftrightarrow{D}^\mu \Phi_1)\right. \\
& \left. + C^{(21)(22)}_{\Phi D} (\Phi_2^\dagger \overleftrightarrow{D}_\mu \Phi_1)(\Phi_2^\dagger \overleftrightarrow{D}^\mu \Phi_2) + h.c.\right).
\end{aligned} \tag{3.43}$$

After EWSB, the part of the Lagrangian containing the masses of gauge fields is

$$\begin{aligned}
\mathcal{L} \supset & \frac{g^2}{4}\left(v^2 + 2 v_2 v_2 \ \mathrm{Re}[\eta]\right)\left(W^1_\mu W^{1\mu} + W^2_\mu W^{2\mu}\right) \\
& + \left[v^2 + 2 v_1 v_2 \ \mathrm{Re}[\eta] - \left(v_1^4 C^{(11)(11)}_{\Phi D} + v_2^4 C^{(22)(22)}_{\Phi D} + v_1^2 v_2^3 C^{(21)(12)}_{\Phi D} + v_1^2 v_2^2 C^{(11)(22)}_{\Phi D}\right.\right. \\
& \left.\left. + 2 v_1^2 v_2^2 \ \mathrm{Re}[C^{(21)(21)}_{\Phi D}] + 2 v_1^3 v_2 \ \mathrm{Re}[C^{(21)(11)}_{\Phi D}] + 2 v_1 v_2^3 \ \mathrm{Re}[C^{(21)(22)}_{\Phi D}]\right)\right] \\
& \times \frac{1}{4} \left(B_\mu \ W^3_\mu\right) \begin{pmatrix} g'^2 & -g g' \\ -g g' & g^2 \end{pmatrix} \begin{pmatrix} B^\mu \\ W^{3\mu} \end{pmatrix}.
\end{aligned} \tag{3.44}$$

Defining $\hat{W}^\pm_\mu \equiv (\hat{W}^1_\mu \mp i \hat{W}^2_\mu)/\sqrt{2}$, after field redefinitions (Eqs. (3.40) and (3.42)), we obtain the mass squared for $W^\pm$ boson

$$\begin{aligned}
m_W^2 &= \frac{g^2}{2}\left(v^2 + 2 v_1 v_2 \ \mathrm{Re}[\eta]\right) \times \left[1 + 4\left(v_1^2 C^{(11)}_{\Phi W} + v_2^2 C^{(22)}_{\Phi W} + 2 v_1 v_2 \ \mathrm{Re}[C^{(21)}_{\Phi W}]\right)\right] \\
&= \frac{\hat{g}^2}{2}\left(v^2 + 2 v_1 v_2 \ \mathrm{Re}[\eta]\right).
\end{aligned} \tag{3.45}$$

The mass squared matrix for the neutral gauge bosons is then

$$\mathcal{L} \supset \frac{1}{2}\begin{pmatrix}\hat{B}_\mu & \hat{W}^3_\mu\end{pmatrix}\begin{pmatrix}M_{11}^2+\Delta M_{11}^2 & M_{12}^2+\Delta M_{12}^2 \\ M_{21}^2+\Delta M_{21}^2 & M_{22}^2+\Delta M_{22}^2\end{pmatrix}\begin{pmatrix}\hat{B}^\mu \\ \hat{W}^{3\mu}\end{pmatrix}, \tag{3.46}$$

where the matrix elements written in terms of Lagrangian parameters for this matrix are collected in Appendix A. The fields $\{\hat{B}_\mu, \hat{W}^3_\mu\}$ are rotated to the physical basis spanned by the photon and the $Z$ boson, $\{A_\mu, Z_\mu\}$ with mass eigenvalues

$$\begin{aligned}
&m_\gamma^2 = 0,\\
&m_Z^2 = \frac{1}{2}(\hat{g}^2+\hat{g}'^2)v^2\left[\left(1+X_{WB}\frac{2\hat{g}\hat{g}'}{\hat{g}^2+\hat{g}'^2}\right)(1+2s_\beta c_\beta\ \mathrm{Re}[\eta])\right.\\
&-v^2\left[c_\beta^4 C_{\Phi D}^{(11)(11)} + s_\beta^4 C_{\Phi D}^{(22)(22)} + s_\beta^2 c_\beta^2 C_{\Phi D}^{(21)(12)} + s_\beta^2 c_\beta^2 C_{\Phi D}^{(11)(22)}\right.\\
&\left.\left.+2s_\beta^2 c_\beta^2\ \mathrm{Re}[C_{\Phi D}^{(21)(21)}] + 2s_\beta c_\beta^3\ \mathrm{Re}[C_{\Phi D}^{(21)(11)}] + 2s_\beta^3 c_\beta\ \mathrm{Re}[C_{\Phi D}^{(21)(22)}]\right]\right].
\end{aligned} \tag{3.47}$$

The rotation matrix is given by

$$\begin{pmatrix}\hat{B}_\mu \\ \hat{W}^3_\mu\end{pmatrix} = \begin{pmatrix}\cos\hat{\theta}_W - \frac{X_{WB}}{2}\sin\hat{\theta}_W & -\sin\hat{\theta}_W - \frac{X_{WB}}{2}\cos\hat{\theta}_W \\ \sin\hat{\theta}_W - \frac{X_{WB}}{2}\cos\hat{\theta}_W & \cos\hat{\theta}_W + \frac{X_{WB}}{2}\sin\hat{\theta}_W\end{pmatrix}\begin{pmatrix}A_\mu \\ Z_\mu\end{pmatrix}. \tag{3.48}$$

This matrix is modified from the familiar weak mixing angle matrix in order to simultaneously enforce that the gauge kinetic terms are canonical (Eq. (3.38)) while diagonalizing to the physical basis after electroweak symmetry breaking.

After EWSB, the electroweak part of the gauge covariant derivative becomes

$$\begin{aligned}
\hat{D}_\mu q &= \partial_\mu q + \frac{i\hat{g}}{\sqrt{2}}(T^+\hat{W}^+_\mu + T^-\hat{W}^-_\mu)q\\
&+ i\left[\hat{g}\left(\sin\hat{\theta}_W - \frac{X_{WB}}{2}\cos\hat{\theta}_W\right)T^3 + \hat{g}'\left(\cos\hat{\theta}_W - \frac{X_{WB}}{2}\sin\hat{\theta}_W\right)Y_q\right]qA_\mu\\
&+ i\left[\hat{g}\left(\cos\hat{\theta}_W + \frac{X_{WB}}{2}\sin\hat{\theta}_W\right)T^3 - \hat{g}'\left(\sin\hat{\theta}_W + \frac{X_{WB}}{2}\cos\hat{\theta}_W\right)Y_q\right]qZ_\mu\\
&\equiv \partial_\mu q + \frac{i\hat{g}}{\sqrt{2}}(T^+\hat{W}^+_\mu + T^-\hat{W}^-_\mu)q + i\hat{e}QqA_\mu + i\hat{g}_Z\left[T^3 - \sin^2\hat{\theta}_W Q\right]qZ_\mu,
\end{aligned} \tag{3.49}$$

where $T^\pm = (\tau^1 \pm i\tau^2)/2$, the electric charge is $Q = T^3 + Y$, and

$$\hat{e} = \hat{g}\left(\sin\hat{\theta}_W - \frac{X_{WB}}{2}\cos\hat{\theta}_W\right) = \hat{g}'\left(\cos\hat{\theta}_W - \frac{X_{WB}}{2}\sin\hat{\theta}_W\right). \tag{3.50}$$

This condition gives us the modified weak angle, $\hat{\theta}_W$, in the effective theory:

$$\begin{aligned}
\tan\hat{\theta}_W &= \frac{\hat{g}'}{\hat{g}} + \frac{X_{WB}}{2}\left(1-\frac{\hat{g}'^2}{\hat{g}^2}\right),\\
\sin\hat{\theta}_W &= \frac{1}{\sqrt{\hat{g}'^2+\hat{g}^2}}\left(\hat{g}' + \frac{X_{WB}}{2}\frac{\hat{g}(\hat{g}^2-\hat{g}'^2)}{\hat{g}'^2+\hat{g}^2}\right),\\
\cos\hat{\theta}_W &= \frac{1}{\sqrt{\hat{g}'^2+\hat{g}^2}}\left(\hat{g} - \frac{X_{WB}}{2}\frac{\hat{g}'(\hat{g}^2-\hat{g}'^2)}{\hat{g}'^2+\hat{g}^2}\right).
\end{aligned} \tag{3.51}$$

Recall that in the SM $\tan\hat{\theta}_W = \hat{g}'/\hat{g}$. Therefore, the electromagnetic and $Z$ couplings can be written as

$$\hat{e} = \frac{\hat{g}\hat{g}'}{\sqrt{\hat{g}'^2+\hat{g}^2}}\left(1 - X_{WB}\frac{\hat{g}\hat{g}'}{\hat{g}'^2+\hat{g}^2}\right), \qquad \hat{g}_Z = \sqrt{\hat{g}'^2+\hat{g}^2}\left(1 + X_{WB}\frac{\hat{g}\hat{g}'}{\hat{g}'^2+\hat{g}^2}\right). \tag{3.52}$$

Since the main focus of the paper is identifying new interactions resulting from the 2HDM EFT, we do not discuss the non-linear gauge fixing procedure and ghost Lagrangian. A gauge-fixing prescription has been discussed before in the SM including dimension-six terms [18], and in the 2HDM in the context of the [$CP$-violating and conserving] Minimal Supersymmetric Standard Model [34], which can be straightforwardly applied here. In this case, the longitudinal components of the $Z$ and $W^\pm$ gauge bosons, $G$ and $G^\pm$, acquire a mass by the gauge-fixing procedure $m_G^2 \to \xi_Z m_Z^2$ and $m_{G^\pm}^2 \to \xi_W m_{W^\pm}^2$, where $\xi_{Z,W}$ are gauge-fixing parameters. In the FeynRules package, all gauge fixing parameters are set to 1.

### 3.3 Fermion Sector

The part of the Lagrangian contributing to the fermion masses is

$$\begin{aligned}
\mathcal{L} \supset &-\left(y_e^{(1)}\bar{l}_L e_R \Phi_1 + y_e^{(2)}\bar{l}_L e_R \Phi_2 + y_d^{(1)}\bar{q}_L d_R \Phi_1 + y_d^{(2)}\bar{q}_L d_R \Phi_2 + y_u^{(1)}\bar{q}_L u_R \cdot \Phi_1^\dagger + y_u^{(2)}\bar{q}_L u_R \cdot \Phi_2^\dagger\right)\\
&+ C_{\nu\nu\Phi}^{(11)}(\Phi_1\cdot l_L)^T\mathbf{C}(\Phi_1\cdot l_L) + C_{\nu\nu\Phi}^{(22)}(\Phi_2\cdot l_L)^T\mathbf{C}(\Phi_2\cdot l_L) + C_{\nu\nu\Phi}^{(12)}(\Phi_1\cdot l_L)^T\mathbf{C}(\Phi_2\cdot l_L)\\
&+ C_{l\Phi_1}^{(11)}\bar{l}_L e_R \Phi_1(\Phi_1^\dagger\Phi_1) + C_{l\Phi_1}^{(22)}\bar{l}_L e_R \Phi_1(\Phi_2^\dagger\Phi_2) + C_{l\Phi_1}^{(21)}\bar{l}_L e_R \Phi_1(\Phi_2^\dagger\Phi_1) + C_{l\Phi_1}^{(12)}\bar{l}_L e_R \Phi_1(\Phi_1^\dagger\Phi_2)\\
&+ C_{l\Phi_2}^{(22)}\bar{l}_L e_R \Phi_2(\Phi_2^\dagger\Phi_2) + C_{l\Phi_2}^{(11)}\bar{l}_L e_R \Phi_2(\Phi_1^\dagger\Phi_1) + C_{l\Phi_2}^{(21)}\bar{l}_L e_R \Phi_2(\Phi_2^\dagger\Phi_1) + C_{l\Phi_2}^{(12)}\bar{l}_L e_R \Phi_2(\Phi_1^\dagger\Phi_2)\\
&+ C_{d\Phi_1}^{(11)}\bar{q}_L d_R \Phi_1(\Phi_1^\dagger\Phi_1) + C_{d\Phi_1}^{(22)}\bar{q}_L d_R \Phi_1(\Phi_2^\dagger\Phi_2) + C_{d\Phi_1}^{(21)}\bar{q}_L d_R \Phi_1(\Phi_2^\dagger\Phi_1) + C_{d\Phi_1}^{(12)}\bar{q}_L d_R \Phi_1(\Phi_1^\dagger\Phi_2)\\
&+ C_{d\Phi_2}^{(22)}\bar{q}_L d_R \Phi_2(\Phi_2^\dagger\Phi_2) + C_{d\Phi_2}^{(11)}\bar{q}_L d_R \Phi_2(\Phi_1^\dagger\Phi_1) + C_{d\Phi_2}^{(21)}\bar{q}_L d_R \Phi_2(\Phi_2^\dagger\Phi_1) + C_{d\Phi_2}^{(12)}\bar{q}_L d_R \Phi_2(\Phi_1^\dagger\Phi_2)\\
&+ C_{u\Phi_1}^{(11)}\bar{q}_L u_R\cdot\Phi_1^\dagger(\Phi_1^\dagger\Phi_1) + C_{u\Phi_1}^{(22)}\bar{q}_L u_R\cdot\Phi_1^\dagger(\Phi_2^\dagger\Phi_2)\\
&+ C_{u\Phi_1}^{(21)}\bar{q}_L u_R\cdot\Phi_1^\dagger(\Phi_2^\dagger\Phi_1) + C_{u\Phi_1}^{(12)}\bar{q}_L u_R\cdot\Phi_1^\dagger(\Phi_1^\dagger\Phi_2)\\
&+ C_{u\Phi_2}^{(22)}\bar{q}_L u_R\cdot\Phi_2^\dagger(\Phi_2^\dagger\Phi_2) + C_{u\Phi_2}^{(11)}\bar{q}_L u_R\cdot\Phi_2^\dagger(\Phi_1^\dagger\Phi_1)\\
&+ C_{u\Phi_2}^{(21)}\bar{q}_L u_R\cdot\Phi_2^\dagger(\Phi_2^\dagger\Phi_1) + C_{u\Phi_2}^{(12)}\bar{q}_L u_R\cdot\Phi_2^\dagger(\Phi_1^\dagger\Phi_2) + h.c..
\end{aligned} \tag{3.53}$$

After each scalar doublet acquires a VEV, the Lagrangian with explicit flavor indices becomes

$$\mathcal{L} \supset -\bar{e}_{L,a}(M_e)_{ab}e_{R,b} - \bar{d}_{L,a}(M_d)_{ab}d_{R,b} - \bar{u}_{L,a}(M_u)_{ab}u_{R,b} - \frac{1}{2}\nu_{L,a}^T(M_\nu)_{ab}\mathbf{C}\nu_{L,b} + h.c., \tag{3.54}$$

where the mass matrices are

$$\begin{aligned}
(M_e)_{ab} = &\ v(y_{e,a}^{(1)}\cos\beta + y_{e,a}^{(2)}\sin\beta)\delta_{ab}\\
&- v^3\Big(\cos^3\beta(C_{l\Phi_1}^{(11)})_{ab} + \sin^2\beta\cos\beta(C_{l\Phi_1}^{(22)})_{ab} + \sin\beta\cos^2\beta((C_{l\Phi_1}^{(12)})_{ab} + (C_{l\Phi_1}^{(21)})_{ab})\\
&+ \sin^3\beta(C_{l\Phi_2}^{(22)})_{ab} + \sin\beta\cos^2\beta(C_{l\Phi_2}^{(11)})_{ab} + \sin^2\beta\cos\beta((C_{l\Phi_2}^{(12)})_{ab} + (C_{l\Phi_2}^{(21)})_{ab})\Big),
\end{aligned} \tag{3.55}$$

$$
\begin{aligned}
(M_d)_{ab} &= v(y^{(1)}_{d,a}\cos\beta + y^{(2)}_{d,a}\sin\beta)\delta_{ab} \\
&\quad - v^3\Big(\cos^3\beta(C^{(11)}_{d\Phi_1})_{ab} + \sin^2\beta\cos\beta(C^{(22)}_{d\Phi_1})_{ab} + \sin\beta\cos^2\beta((C^{(12)}_{d\Phi_1})_{ab} + (C^{(21)}_{d\Phi_1})_{ab}) \\
&\quad + \sin^3\beta(C^{(22)}_{d\Phi_2})_{ab} + \sin\beta\cos^2\beta(C^{(11)}_{d\Phi_2})_{ab} + \sin^2\beta\cos\beta((C^{(12)}_{d\Phi_2})_{ab} + (C^{(21)}_{d\Phi_2})_{ab})\Big)\,,
\end{aligned}
\tag{3.56}
$$

$$
\begin{aligned}
(M_u)_{ab} &= v(y^{(1)}_{u,a}\cos\beta + y^{(2)}_{u,a}\sin\beta)\delta_{ab} \\
&\quad - v^3\Big(\cos^3\beta(C^{(11)}_{u\Phi_1})_{ab} + \sin^2\beta\cos\beta(C^{(22)}_{u\Phi_1})_{ab} + \sin\beta\cos^2\beta((C^{(12)}_{u\Phi_1})_{ab} + (C^{(21)}_{u\Phi_1})_{ab}) \\
&\quad + \sin^3\beta(C^{(22)}_{u\Phi_2})_{ab} + \sin\beta\cos^2\beta(C^{(11)}_{u\Phi_2})_{ab} + \sin^2\beta\cos\beta((C^{(12)}_{u\Phi_2})_{ab} + (C^{(21)}_{u\Phi_2})_{ab})\Big)\,,
\end{aligned}
\tag{3.57}
$$

and the Majorana neutrino mass matrix generated by the dimension-five operators is

$$
(M_\nu)_{ab} = -2v^2\Big(\cos^2\beta(C^{(11)}_{\nu\nu\Phi})_{ab} + \sin^2\beta(C^{(22)}_{\nu\nu\Phi})_{ab} + \sin\beta\cos\beta(C^{(12)}_{\nu\nu\Phi})_{ab}\Big)\,. \tag{3.58}
$$

Note that we can also introduce $\overline{\nu}^c_L = \nu^T_L\mathbf{C}$, in which case the last term in the Lagrangian above can be written as $-\overline{\nu}^c_L M_\nu \nu_L/2 + h.c.$ (note that the Majorana condition is $\nu = \nu_L + \nu^c_L = \nu^c$).

Rotating the fields to the mass eigenstate basis can be done via bi-unitary transformations $\psi_L \to U_L\psi_L$ and $\psi_R \to U_R\psi_R$, whereby the unitary matrices $U_{L,R}$ are such that the diagonal entries of

$$
\begin{aligned}
\hat{M}_e &= U^{e\dagger}_L M_e U^e_R = \mathrm{diag}(m_e, m_\mu, m_\tau), \\
\hat{M}_d &= U^{d\dagger}_L M_d U^d_R = \mathrm{diag}(m_d, m_s, m_b), \\
\hat{M}_u &= U^{u\dagger}_L M_u U^u_R = \mathrm{diag}(m_u, m_c, m_t), \\
\hat{M}_\nu &= U^{\nu T}_L M_\nu U^\nu_L = \mathrm{diag}(m_{\nu_1}, m_{\nu_2}, m_{\nu_3}).
\end{aligned}
\tag{3.59}
$$

are real and positive. After field rotations, operators in the mass basis (with hats) are obtained as in Ref. [5]:

$$
\begin{aligned}
&(\hat{y}^{(1,2)}_{(e,d,u)})_{ab} = (U^{(e,d,u)\dagger}_L y^{(1,2)}_{(e,d,u)} U^{(e,d,u)}_R)_{ab}, \\
&(\hat{C}^{(11,22,12)}_{\nu\nu\Phi})_{ab} = (U^{\nu T}_L C^{(11,22,12)}_{\nu\nu\Phi} U^\nu_L)_{ab}, \\
&(\hat{C}^{(11,22,12,21)}_{(l,d,u)\Phi_{(1,2)}})_{ab} = (U^{(e,d,u)\dagger}_L C^{(11,22,12,21)}_{(l,d,u)\Phi_{(1,2)}} U^{(e,d,u)}_R)_{ab}, \\
&(\hat{C}_{l(B,W)\Phi_{(1,2)}})_{ab} = (U^{e\dagger}_L C_{l(B,W)\Phi_{(1,2)}} U^e_R)_{ab}, \\
&(\hat{C}_{d(B,W,G)\Phi_{(1,2)}})_{ab} = (U^{d\dagger}_L C_{d(B,W,G)\Phi_{(1,2)}} U^d_R)_{ab}, \\
&(\hat{C}_{u(B,W,G)\Phi_{(1,2)}})_{ab} = (U^{u\dagger}_L C_{u(B,W,G)\Phi_{(1,2)}} U^u_R)_{ab}, \\
&(\hat{C}^{(11,22,12)}_{\Phi(e,d,u)})_{ab} = (U^{(e,d,u)\dagger}_R C^{(11,22,12)}_{\Phi(e,d,u)} U^{(e,d,u)}_R)_{ab}, \\
&(\hat{C}^{(11,22,12)[1]}_{\Phi(l,q)})_{ab} = (U^{(e,d)\dagger}_L C^{(11,22,12)[1]}_{\Phi(l,q)} U^{(e,d)}_L)_{ab}, \\
&(\hat{C}^{(11,22,12)[3]}_{\Phi(l,q)})_{ab} = (U^{(e,d)\dagger}_L C^{(11,22,12)[3]}_{\Phi(l,q)} U^{(e,d)}_L)_{ab}, \\
&(\hat{C}^{(11,22,21)}_{\Phi ud})_{ab} = (U^{u\dagger}_R C^{(11,22,21)}_{\Phi ud} U^d_R)_{ab}.
\end{aligned}
\tag{3.60}
$$

Four-fermion operators are rotated just as in SMEFT and can be found in Ref. [18]. As a result of rotating to the mass eigenstate basis, the Cabbibo-Kobayashi-Maskawa (CKM) [35] and Pontecorvo–Maki–Nakagawa–Sakata (PMNS) [36, 37] matrices,

$$V \equiv U_L^{u\dagger} U_L^d, \qquad U \equiv U_L^{e\dagger} U_L^{\nu}, \tag{3.61}$$

appear in interactions mediated by the $W^{\pm}$ boson in the quark and lepton sectors, respectively.

All of the above definitions and conventions are implemented in the FeynRules package.

## 4 Main Results and Specific Types and Limits of the 2HDM EFT

The FeynRules package [13] can be used to obtain Feynman rules in the most general 2HDM EFT. However, specific versions or limits of the 2HDM are often considered, therefore, the FeynRules package can also provide a customized set of Feynman rules for four different types of 2HDMs, in addition to the $CP$-conserving limit and the alignment limit of the Higgs sector. In this section, we discuss these different types and limits. We then focus on the $CP$-conserving 2HDM EFT and compare it to a specific type of 2HDM EFT (type-II). Additionally, we compare the type-II 2HDM EFT to the SMEFT and the renormalizable type-II 2HDM. We present selected Feynman rules for the $CP$-conserving type-II 2HDM EFT in the alignment limit in Appendix C. Note that the results represent the effects of physics beyond the 2HDM and do not include loop effects from the renormalizable 2HDM (which can be staightfowardly added).

### 4.1 $CP$-Conserving 2HDM EFT

A $CP$-conserving Higgs sector can be accomplished by mandating that $m_{12}^2$ and couplings $\lambda_{5,6,7}$ in the scalar potential (Eq. (2.6)) are real. In addition, the imaginary parts of the dimension-six operators contributing to the Higgs potential, originating from $\phi^4 D^2$ (operators in Eq. (3.1)) and $\phi^6$ (operators in Eq. (3.17)) classes, must vanish. In this situation, the off-diagonal terms in the neutral scalar mass squared matrix $m_{AH}^2 = m_{Ah}^2 = 0$, as well as the matrix $D = 0$ from field redefinitions. The mass eigenstates become $h_1 \to h$, $h_2 \to H$, and $h_3 \to A$. This limit can be specified by the user in the FeynRules package.

### 4.2 Types of Two-Higgs Doublet Model EFTs

A discrete $Z_2$ symmetry imposed on the Higgs doublets, $\Phi_1 \to -\Phi_1$ and $\Phi_2 \to +\Phi_2$ restricts odd powers of either $\Phi_1$ and $\Phi_2$ in the renormalizable scalar potential [31] and the dimension-six operators contributing to it. As a consequence, this forbids terms involving $\lambda_6$ and $\lambda_7$ in the scalar potential (Eq. (2.6)) (and, if enforced exactly, possible effects from soft $Z_2$-breaking $m_{12}^2$). For generality, we include the soft-breaking term $m_{12}^2$ in our results, since this can allow for a heavy Higgs spectrum.

In addition to the scalar sector, the fermions can be charged under the $Z_2$ symmetry. We focus on four unique model types of a 2HDM: type-I, -II, -X, and -Y. Specifying how each of the doublets couple to SM fermions is provided in Table 2, where we use the convention that $\Phi_2$ always couples to the up-type sector. In any of the four models, the $Z_2$ symmetry prevents flavor changing neutral currents among fermionic species at the tree level. Under the $Z_2$ symmetry of $\Phi_1$ and $\Phi_2$, there are a total of 76 operators common in all 4 types of 2HDMs [5], and model-specific operators can be found in Tables VII, IX, X, and XI of Ref. [5] for type-I, -II, -X, and -Y, respectively. When working in a specific type of 2HDM, one can simply set the appropriate Wilson coefficients to zero in the general 2HDM. The specific type of 2HDM can be user-specified in the FeynRules package.

| Model | $u$ | $d$ | $e$ | $l_L$ | $e_R$ | $q_L$ | $u_R$ | $d_R$ | $\Phi_1$ | $\Phi_2$ |
|---|---|---|---|---|---|---|---|---|---|---|
| Type-I | $\Phi_2$ | $\Phi_2$ | $\Phi_2$ | $+$ | $+$ | $+$ | $+$ | $+$ | $-$ | $+$ |
| Type-II | $\Phi_2$ | $\Phi_1$ | $\Phi_1$ | $+$ | $-$ | $+$ | $+$ | $-$ | $-$ | $+$ |
| Type-X (Lepton-specific) | $\Phi_2$ | $\Phi_2$ | $\Phi_1$ | $+$ | $-$ | $+$ | $+$ | $+$ | $-$ | $+$ |
| Type-Y (Flipped) | $\Phi_2$ | $\Phi_1$ | $\Phi_2$ | $+$ | $+$ | $+$ | $+$ | $-$ | $-$ | $+$ |

**Table 2**. $Z_2$ charge assignments of leptons, quarks and Higgs doublets in each type of 2HDM. The middle column indicates which scalar doublet couples to each sector, where by convention, $\Phi_2$ always couples to the up-type sector, while $\Phi_1 \to -\Phi_1$ and $\Phi_2 \to +\Phi_2$.

### 4.3 Alignment Limit

Studies in a 2HDM often focus on evaluating the theory in the alignment limit when the difference between the vacuum angle and $\alpha$ (the angle which diagonalizes the $CP$-even scalars in the 2HDM) is $\beta - \alpha \to \pi/2$ [38]. The alignment limit enforces that the couplings of the light eigenstate $h$ are SM-like.[3] Simultaneously, the masses of the other scalar fields $H, A$, and $H^{\pm}$ are comparable. It is then convenient to work in the Higgs basis [31, 39–41], which separates SM degrees of freedom from additional scalars in the theory.

In the 2HDM EFT, no couplings of scalars are SM-like in general. Nevertheless, one can rotate to the Higgs basis and in the limit where $\beta - \hat{\alpha} \to \pi/2$, the SM degrees of freedom are approximately separated from new fields in two doublets. Using $\hat{\alpha}$ instead of $\alpha$ results in $h$ being contained completely in one doublet and $H$ in the other [5]. Furthermore, the Feynman rules simplify in this limit, and this is what we mean by “alignment limit” in the Appendix C. In the FeynRules package, one needs to specify if evaluating the alignment limit or not.

[3] Current constraints on the alignment are strong for type-II, -X, and -Y models at low and high values of $\tan\beta$, for example, at $\tan\beta \sim 1$, $\cos(\alpha - \beta) \in [-0.02, 0.08], [-0.1, 0.08]$, and $[-0.04, 0.08]$ at 95% C.L., respectively. A type-I 2HDM has the largest tolerance of $\cos(\alpha - \beta) \in [-0.17, 0.14]$ at $\tan\beta \sim 10$, allowing for the most freedom.

### 4.4 Discussion of Interactions

To illustrate the complex structure of the couplings of the $CP$-conserving 2HDM EFT, the Feynman rule for leptons coupling to the light $CP$-even Higgs eigenstate is given by

$$
\begin{aligned}
&+\frac{i}{2\sqrt{2}}\Big(\left(s_{\hat{\alpha}}\,(2-A_1)+c_{\hat{\alpha}}B\right)\Big(\mathcal{P}_R\hat{y}^{(1)}_{e,f_1f_2}+\mathcal{P}_L\hat{y}^{(1)*}_{e,f_2f_1}\Big)\\
&\qquad -\left(c_{\hat{\alpha}}\,(2-A_2)+s_{\hat{\alpha}}B\right)\Big(\mathcal{P}_R\hat{y}^{(2)}_{e,f_1f_2}+\mathcal{P}_L\hat{y}^{(2)*}_{e,f_2f_1}\Big)\Big)\\
&+\frac{iv^2}{\sqrt{2}}\Big(-3s_{\hat{\alpha}}c_\beta^2(\mathcal{P}_R\hat{C}^{(11)}_{l\Phi_1,f_1f_2}+\mathcal{P}_L\hat{C}^{(11)*}_{l\Phi_1,f_2f_1})\\
&\qquad +(2c_{\hat{\alpha}}s_\beta c_\beta-s_{\hat{\alpha}}s_\beta^2)(\mathcal{P}_R\hat{C}^{(22)}_{l\Phi_1,f_1f_2}+\mathcal{P}_L\hat{C}^{(22)*}_{l\Phi_1,f_2f_1})\\
&\qquad -(2s_{\hat{\alpha}}s_\beta c_\beta-c_{\hat{\alpha}}c_\beta^2)(\mathcal{P}_R\hat{C}^{(21)}_{l\Phi_1,f_1f_2}+\mathcal{P}_L\hat{C}^{(21)*}_{l\Phi_1,f_2f_1})\\
&\qquad -(2s_{\hat{\alpha}}s_\beta c_\beta-c_{\hat{\alpha}}c_\beta^2)(\mathcal{P}_R\hat{C}^{(12)}_{l\Phi_1,f_1f_2}+\mathcal{P}_L\hat{C}^{(12)*}_{l\Phi_1,f_2f_1})\\
&\qquad +3c_{\hat{\alpha}}s_\beta^2(\mathcal{P}_R\hat{C}^{(22)}_{l\Phi_2,f_1f_2}+\mathcal{P}_L\hat{C}^{(22)*}_{l\Phi_2,f_2f_1})\\
&\qquad -(2s_{\hat{\alpha}}s_\beta c_\beta-c_{\hat{\alpha}}c_\beta^2)(\mathcal{P}_R\hat{C}^{(11)}_{l\Phi_2,f_1f_2}+\mathcal{P}_L\hat{C}^{(11)*}_{l\Phi_2,f_2f_1})\\
&\qquad +(2c_{\hat{\alpha}}s_\beta c_\beta-s_{\hat{\alpha}}s_\beta^2)(\mathcal{P}_R\hat{C}^{(21)}_{l\Phi_2,f_1f_2}+\mathcal{P}_L\hat{C}^{(21)*}_{l\Phi_2,f_2f_1})\\
&\qquad +(2c_{\hat{\alpha}}s_\beta c_\beta-s_{\hat{\alpha}}s_\beta^2)(\mathcal{P}_R\hat{C}^{(12)}_{l\Phi_2,f_1f_2}+\mathcal{P}_L\hat{C}^{(12)*}_{l\Phi_2,f_2f_1})\Big)\\
&+\frac{iv}{\sqrt{2}}\not{p}_3\mathcal{P}_R c_{\hat{\alpha}-\beta}\Big(\hat{C}^{(12)}_{\Phi e,f_1f_2}-\hat{C}^{(12)*}_{\Phi e,f_2f_1}\Big)\\
&+\frac{iv}{\sqrt{2}}\not{p}_3\mathcal{P}_L c_{\hat{\alpha}-\beta}\Big(\hat{C}^{(12)[1]}_{\Phi l,f_1f_2}-\hat{C}^{(12)[1]*}_{\Phi l,f_2f_1}+\hat{C}^{(12)[3]}_{\Phi l,f_1f_2}-\hat{C}^{(12)[3]*}_{\Phi l,f_2f_1}\Big)\,,
\end{aligned}
\tag{4.1}
$$

where the indices $f_1$ and $f_2$ denote the flavor of lepton 1 and 2, respectively, and $p_3$ is the momentum of the light Higgs (see Appendix C for conventions). A similar expression can be found for the quarks. An interesting feature in the general theory is the momentum-enhanced contributions derived from covariant derivative operators for left- or right-handed lepton currents. Although the prefactor is $\cos(\hat{\alpha}-\beta)$ which would vanish if the alignment limit is strictly enforced, if one deviates from the exact limit $\beta-\hat{\alpha}=\pi/2$ from experiments, we see that these terms grow at higher energies and interfere with contributions from the mass operators themselves.

Moreover, since the 2HDM has in general two unique Yukawa couplings, one cannot rewrite the vertex in terms of the lepton mass $m_{e_{f_1}}\delta_{f_1f_2}$ (Eq. 3.55)), and same applies to the quark sector. Although one can trade $\psi^2\phi^3$ operators (operators in Eq. (3.53)) in favor for the physical mass, both complex Yukawa terms remain. If any of the four types of 2HDMs is enforced, terms with momentum dependence vanish due to the $Z_2$ symmetry and half of the mass operators and Yukawa terms as well, simplifying the vertex. In this case the Yukawa coupling can be completely removed in favor of the physical mass. The vertex simplifies greatly in the alignment limit of the type-II model:

$e^{f_2}$, $e^{f_1}$, $h$

$$
\begin{aligned}
&+\frac{i\delta_{f_1 f_2} m_{e_{f_1}}}{2\sqrt{2}v}\left(A_1 + Bt_\beta - 2\right)\\
&+i\sqrt{2}v^2 c_\beta \left(c_\beta^2\left(\mathcal{P}_L \hat{C}^{(11)*}_{l\Phi_1,f_2f_1} + \mathcal{P}_R \hat{C}^{(11)}_{l\Phi_1,f_1f_2}\right)\right.\\
&\qquad +s_\beta^2\left(\mathcal{P}_L \hat{C}^{(12)*}_{l\Phi_2,f_2f_1} + \mathcal{P}_R \hat{C}^{(12)}_{l\Phi_2,f_1f_2}\right)\\
&\qquad +s_\beta^2\left(\mathcal{P}_L \hat{C}^{(21)*}_{l\Phi_2,f_2f_1} + \mathcal{P}_R \hat{C}^{(21)}_{l\Phi_2,f_1f_2}\right)\\
&\qquad \left. +s_\beta^2\left(\mathcal{P}_L \hat{C}^{(22)*}_{l\Phi_1,f_2f_1} + \mathcal{P}_R \hat{C}^{(22)}_{l\Phi_1,f_1f_2}\right)\right),
\end{aligned}
\tag{4.2}
$$

and a similar simplified expression is expected in every type of 2HDM. There are many noticeable differences between the 2HDM EFT and the renormalizable theory. Due to field redefinitions of the scalar fields in the presence of higher-dimensional operators, each fermion coupling to the scalars $G, G^\pm, h, H, A$, and $H^\pm$ will receive contributions from scalar derivative operators $\phi^4 D^2$ (for example, the $A_1$, $A_2$, and $B$ terms in Eq. (4.1)). Additionally, vertices containing $G, A, G^\pm$, and $H^\pm$ are modified by $\phi^6$ operators affecting the diagonalization and rotation of the $CP-$odd and charged sectors given in Eq. (B.16) (for example, see the $\delta_{s_{\hat{\beta}}}$ term defined in Eq. (B.16) in the $\bar{e}^{f_1}e^{f_2}A$ vertex in Eq. (C.9)).

Interestingly, the renormalizable 2HDM does not have any scalars coupling to neutrinos, however, due to the presence of dimension-five terms, the neutrinos now couple to the SM and new scalars. The couplings of neutrinos to $G$ (also $h$ in the alignment limit) and $G^\pm$ can be written in terms of the physical neutrino masses (see for example the $\bar{\nu}^{f_1}e^{f_2}G^+$ vertex in Eq. (C.1)). Neutrino currents also receive contributions from left-handed lepton current operators $\psi^2\phi^2 D$ (Ref. [5]).

Examining the couplings of the photon, $Z$ and $W^\pm$ bosons to leptons and quarks in Appendices C.1 and C.2, respectively, we see that couplings are modified by EW dipole operators for all fermions and the chromomagnetic dipole operators only for the quarks. In addition, for right-handed charged currents of the quarks, there are model-specific operators such as $\mathcal{O}^{(11,22)}_{ud}$ allowed in type-I and type-X models, and only $\mathcal{O}^{(21)}_{ud}$ for type-II and -Y models.

Another interesting feature involving scalar-gauge boson interactions is found in three- and four-point vertices. In the renormalizable 2HDM, the theory admits up to four-point vertices from the kinetic terms in the Lagrangian. However, in the 2HDM EFT there are vertices which *do not* appear in the renormalizable theory at tree level that are derived from field redefinitions and higher-dimensional terms, such as $\gamma W^\pm H^\mp$. Due to the discrepancy between the vacuum angle $\beta$ and the diagonalization angle $\hat{\beta}^\pm$, the contributions to this vertex no longer cancel exactly as they would in the renormalizable 2HDM at tree level, and remain in addition to mixed boson operators $\mathcal{O}_{\Phi B\tilde{W}}$ (see Eq. (C.435)).[4] Additional vertices that do not appear in the renormalizable theory at tree level include ones with the $CP$-odd Higgs such as $A\gamma\gamma, AW^+W^-$, $AZZ$, and $Agg$ which are all possible through

[4] These vertices appear in Refs. [42] and [43] for example.

$\mathcal{O}^{(21)}_{\Phi(G,B,W),(\tilde{G},\tilde{B},\tilde{W})}$ operators. However, this operator is forbidden in any of the four types of 2HDM. Note however that $AZZ$ can also be generated by $\phi^4 D^2$ operators (see Eq. (C.428)).

As a consequence of the alignment limit being only approximate in the 2HDM EFT, as discussed earlier, other vertices such as $HW^+W^-(HZZ)$, that are usually proportional to $\cos(\alpha-\beta)$, or generally any vertex with at least one gauge boson *and* exactly 1 additional Higgs $H, A, H^\pm$ [38], will no longer vanish. The vertices with $H^\pm$ contain $\sin(\beta-\hat{\beta}^\pm)$ while the vertices with $H$ contain $\cos(\beta-\hat{\alpha})+\mathcal{O}(v^2/\Lambda^2)$, in addition to contributions from other operators (Eq. (C.429)). Similar arguments can be made for scalar-scalar-vector-vector vertices.

Comparing the 2HDM EFT to SMEFT, we see there are many new processes involving additional Higgs scalars $H, A$, and $H^\pm$ throughout Appendix C. There are new interactions of two fermions with up to three new scalars, gauge bosons with up to four new scalars[5], and SM Higgs and Goldstone bosons with up to five new Higgses (not listed in the Appendix C). Among these, there are mixed processes involving charged Goldstones and charged Higgses or $W^\pm$, for example, $\overline{\psi}\psi G^\pm H^\mp$ (Eq. (C.40)), which are *antisymmetric*; the Hermitian conjugate of this diagram is not the same due to operators with $\Phi_1^\dagger\Phi_2$ (or $\Phi_2^\dagger\Phi_1$ which is different). Thus, measuring the asymmetry of these interactions could probe the corresponding operators. In any of the four kinds of 2HDMs, this asymmetry originates from $\psi^2\phi^3$ operators $\mathcal{O}^{(12),(21)}_{\psi\Phi_{(1,2)}}$, whereas in the general theory, it also comes from $\psi^2\phi^2 D$ operators $\mathcal{O}^{(12)}_{\Phi\psi}$ for both left- and right-handed currents. In addition to new vertices, each vertex already appearing in the SMEFT receives more contributions due to the extra doublet in the theory.

Vertices containing any of the additional Higgs bosons compared to vertices where the new scalars are replaced by the SM Higgs or Goldstone bosons can be further enhanced or suppressed by various powers of $\tan\beta$ depending on which operators the vertices originate from. These effects can be straightforwardly understood in the Higgs basis, where the doublet approximately containing the SM degrees of freedom and the other doublet containing the new fields are accompanied by different factors of either $\cos\beta$ or $\sin\beta$. For a detailed discussion on the Higgs basis, see Sec. IV. of Ref. [5] (Wilson coefficients written in the Higgs basis are given in Appendices B2-B8 therein).

## 5 Conclusion

Given the 2HDM effective field theory as the low energy theory well below some scale $\Lambda$ with the SM particle content and additional Higgs degrees of freedom, we computed the physical Higgs spectrum and contributions to the couplings and masses of the gauge bosons and fermions from effective operators at mass dimension-six. We also introduced a new method to analytically diagonalize the $CP$-violating scalar sector in the 2HDM EFT, where the mass eigenstates are ones of indefinite $CP$-symmetry. This has general application towards many models where $CP$ violation can occur such as in the renormalizable 2HDM [15] or the MSSM [14, 33]. We found that in the scalar sector, the diagonalization angles are no longer the same as the vacuum angle $\beta$ due to additional corrections from canonical normalization, and $\beta-\hat{\beta} \sim \mathcal{O}\left(v^4/\Lambda^2 m_A^2\right)$ and $\beta-\hat{\beta}^{\pm} \sim \mathcal{O}\left(v^4/\Lambda^2 m_{H^\pm}^2\right)$ when diagonalizing either the neutral or charged scalar sector.

[5]Scalar-gauge vertices originate from the $X^2\phi^2$ and $\phi^4 D^2$ (with covariant derivatives) operators, and the number of gauge bosons can be distinguished from the operators: any six-, five-, or four-point vertex with 3 gauge bosons or more can only come from $X^2\phi^2$ operators, whereas six-point vertices with 2 vector bosons, five-point vertices with less than 2 vector bosons, and four-point vertices with one vector boson can only originate from $\phi^4 D^2$. Both operators can contribute to four-point vertices with 2 vector bosons.

We have shown that the general 2HDM EFT gives rise to many new contact interactions involving both the new Higgs bosons and the SM degrees of freedom: Interactions involving two fermions can include up to three new Higgses; gauge bosons with up to four new scalars; and the SM Higgs and Goldstone bosons with up to five new Higgses. We have presented selected Feynman rules for a $CP$-conserving type-II 2HDM EFT in the alignment limit to show the impact of dimension-six contributions for charged and neutral leptons, quarks, and scalar-gauge boson interactions. We highlighted differences between the 2HDM EFT and the SMEFT, as well as the 2HDM EFT and the renormalizable theory. We expect close similarities between the type-II model and other types. However, away from a specific type of 2HDM or the alignment limit, there are further interesting features not present in the SMEFT nor the renormalizable 2HDM, which include additional interactions involving the new Higgs spectrum, momentum-enhanced terms from left-handed or right-handed fermion current operators, vector bosons coupling to $CP$-odd Higgs such as $Agg$, and due to the additional charged scalar $H^{\pm}$, the 2HDM EFT admits non-Hermitian terms such as $\bar{\psi}\psi H^{\pm}G^{\mp}$ (measuring the asymmetry for such a process can isolate the contributing Wilson coefficient). As a consequence of $\beta \neq \hat{\beta}^{\pm}$ and $\beta \neq \hat{\beta}$, contributions to vertices involving these factors no longer cancel exactly as they would in the renormalizable 2HDM at tree level and remain in addition to contributions from other effective operators. Interactions proportional to $\cos(\hat{\alpha}-\beta)$ (for example in $HW^+W^-$ or $HZZ$) no longer vanish in the alignment limit since this limit is only approximate in the 2HDM EFT. The most-general 2HDM EFT allows an expansive set of new and interesting processes that can be studied to the reader's interest.

As a supplement, we prepared a FeynRules package [13] to compute the Feynman rules of the general 2HDM EFT, with specification of limits of the 2HDM EFT the user can implement, which include $CP$-conservation, any of the four types of 2HMDs listed earlier, and the alignment limit. Feynman rules can be printed out in LaTeX form. All results presented in the Appendix are made from this package in the variables and notation defined here and in [5].

## Acknowledgments

We would like to thank Taegyu Lee for testing and checking the final stages of our FeynRules model package.

## A Mass Squared Matrices for Scalars and Gauge Bosons

In this Appendix, we collect all matrix elements in terms of Lagrangian parameters needed to diagonalize the full $4 \times 4$ mass matrix in the neutral scalar sector, the $2 \times 2$ matrix for the charged scalars, and the $2 \times 2$ matrix for neutral gauge bosons. Given the Lagrangian in Eq. (3.17), the elements of the neutral scalar mass squared block matrices $\hat{M}^2_{\rho\rho}$, $\hat{M}^2_{aa}$, and $\hat{M}^2_{\rho a}$ in Eq. (3.19) can be written as

$$\begin{aligned}
\hat{M}^2_{\rho_1\rho_1} &\equiv M^2_{\rho_1\rho_1} + \Delta M^2_{\rho_1\rho_1}, \\
\hat{M}^2_{\rho_2\rho_2} &\equiv M^2_{\rho_2\rho_2} + \Delta M^2_{\rho_2\rho_2}, \\
\hat{M}^2_{\rho_1\rho_2} &= \hat{M}^2_{\rho_2\rho_1} \equiv M^2_{\rho_1\rho_2} + \Delta M^2_{\rho_1\rho_2},
\end{aligned} \tag{A.1}$$

$$\begin{aligned}
\hat{M}^2_{a_1a_1} &\equiv M^2_{a_1a_1} + \Delta M^2_{a_1a_1}, \\
\hat{M}^2_{a_2a_2} &\equiv M^2_{a_2a_2} + \Delta M^2_{a_2a_2}, \\
\hat{M}^2_{a_1a_2} &= \hat{M}^2_{a_2a_1} \equiv M^2_{a_1a_2} + \Delta M^2_{a_1a_2},
\end{aligned} \tag{A.2}$$

and

$$\begin{aligned}
\hat{M}^2_{\rho_1a_1} &\equiv M^2_{\rho_1a_1} + \Delta M^2_{\rho_1a_1}, \\
\hat{M}^2_{\rho_2a_2} &\equiv M^2_{\rho_2a_2} + \Delta M^2_{\rho_2a_2}, \\
\hat{M}^2_{\rho_1a_2} &\equiv M^2_{\rho_1a_2} + \Delta M^2_{\rho_1a_2}, \\
\hat{M}^2_{\rho_2a_1} &\equiv M^2_{\rho_2a_1} + \Delta M^2_{\rho_2a_1}.
\end{aligned} \tag{A.3}$$

All $M^2$ elements without the hat contain the standard dimension-four contributions from the potential:[6]

$$\begin{aligned}
M^2_{\rho_1\rho_1} &= m_1^2 + 3\lambda_1 v_1^2 + v_2^2(\lambda_3 + \lambda_4 + \mathrm{Re}[\lambda_5]) + 6v_1v_2\mathrm{Re}[\lambda_6], \\
M^2_{\rho_2\rho_2} &= m_2^2 + 3\lambda_2 v_2^2 + v_1^2(\lambda_3 + \lambda_4 + \mathrm{Re}[\lambda_5]) + 6v_1v_2\mathrm{Re}[\lambda_7], \\
M^2_{\rho_1\rho_2} &= M^2_{\rho_2\rho_1} = \ \mathrm{Re}[m_{12}^2] + 2v_1v_2(\lambda_3 + \lambda_4 + \mathrm{Re}[\lambda_5]) + 3v_1^2\mathrm{Re}[\lambda_6] + 3v_2^2\mathrm{Re}[\lambda_7],
\end{aligned} \tag{A.4}$$

$$\begin{aligned}
M^2_{a_1a_1} &= m_1^2 + \lambda_1 v_1^2 + v_2^2(\lambda_3 + \lambda_4 - \mathrm{Re}[\lambda_5]) + 2v_1v_2\mathrm{Re}[\lambda_6], \\
M^2_{a_2a_2} &= m_2^2 + \lambda_2 v_2^2 + v_1^2(\lambda_3 + \lambda_4 - \mathrm{Re}[\lambda_5]) + 2v_1v_2\mathrm{Re}[\lambda_7], \\
M^2_{a_1a_2} &= M^2_{a_2a_1} = \mathrm{Re}[m_{12}^2] + 2v_1v_2\mathrm{Re}[\lambda_5] + v_1^2\mathrm{Re}[\lambda_6] + v_2^2\mathrm{Re}[\lambda_7],
\end{aligned} \tag{A.5}$$

and

$$\begin{aligned}
M^2_{\rho_1a_1} &= +v_2^2\ \mathrm{Im}[\lambda_5] + 2v_1v_2\ \mathrm{Im}[\lambda_6], \\
M^2_{\rho_2a_2} &= -v_1^2\ \mathrm{Im}[\lambda_5] - 2v_1v_2\ \mathrm{Im}[\lambda_7], \\
M^2_{\rho_1a_2} &= -\mathrm{Im}[m_{12}^2] - 2v_1v_2\ \mathrm{Im}[\lambda_5] - 3v_1^2\ \mathrm{Im}[\lambda_6] - v_2^2\ \mathrm{Im}[\lambda_7], \\
M^2_{\rho_2a_1} &= +\mathrm{Im}[m_{12}^2] + 2v_1v_2\ \mathrm{Im}[\lambda_5] + v_1^2\ \mathrm{Im}[\lambda_6] + 3v_2^2\ \mathrm{Im}[\lambda_7].
\end{aligned} \tag{A.6}$$

[6] Note that one can trade off $m_{1,2}^2$ for $\mathrm{Re}[m_{12}^2]$ by using the vacuum criteria including dimension-six terms provided in Eq. (A1) and Eq. (A2) in Ref. [5] or $\mathrm{Im}[m_{12}^2]$ for $\mathrm{Im}[\lambda_5]$ using Eq. (A3) also in Ref. [5] to find other useful relations between terms.

Terms written as $\Delta M^2$ contain all contributions from dimension-six operators and field redefinitions given below

$$
\begin{aligned}
\Delta M^2_{\rho_1\rho_1} &= -A_1 M^2_{\rho_1\rho_1} - B M^2_{\rho_1\rho_2} - J M^2_{\rho_1 a_1} - K M^2_{\rho_1 a_2} \\
&- v^4 \Big[ 15 c_\beta^4 C_\Phi^{(11)(11)(11)} + 6 s_\beta^2 c_\beta^2 C_\Phi^{(11)(11)(22)} + s_\beta^4 C_\Phi^{(11)(22)(22)} + 20 s_\beta c_\beta^3 \ \mathrm{Re}[C_\Phi^{(11)(11)(21)}] \\
&+ 12 s_\beta^2 c_\beta^2 \ \mathrm{Re}[C_\Phi^{(11)(21)(21)}] + 2 s_\beta^4 \ \mathrm{Re}[C_\Phi^{(22)(21)(21)}] + 6 s_\beta^2 c_\beta^2 C_\Phi^{(11)(21)(12)} + s_\beta^4 C_\Phi^{(22)(21)(12)} \\
&+ 6 s_\beta^3 c_\beta \ \mathrm{Re}[C_\Phi^{(21)(21)(21)}] + 6 s_\beta^3 c_\beta \ \mathrm{Re}[C_\Phi^{(21)(21)(12)}] + 6 s_\beta^3 c_\beta \ \mathrm{Re}[C_\Phi^{(11)(22)(21)}] \Big] , \\
\Delta M^2_{\rho_2\rho_2} &= -A_2 M^2_{\rho_2\rho_2} - B M^2_{\rho_1\rho_2} - L M^2_{\rho_2 a_1} - N M^2_{\rho_2 a_2} \\
&- v^4 \Big[ 15 s_\beta^4 C_\Phi^{(22)(22)(22)} + 6 s_\beta^2 c_\beta^2 C_\Phi^{(11)(22)(22)} + c_\beta^4 C_\Phi^{(11)(11)(22)} + 20 s_\beta^3 c_\beta \ \mathrm{Re}[C_\Phi^{(22)(22)(21)}] \\
&+ 12 s_\beta^2 c_\beta^2 \ \mathrm{Re}[C_\Phi^{(22)(21)(21)}] + 2 c_\beta^4 \ \mathrm{Re}[C_\Phi^{(11)(21)(21)}] + 6 s_\beta^2 c_\beta^2 C_\Phi^{(22)(21)(12)} + c_\beta^4 C_\Phi^{(11)(21)(12)} \\
&+ 6 s_\beta c_\beta^3 \ \mathrm{Re}[C_\Phi^{(21)(21)(21)}] + 6 s_\beta c_\beta^3 \ \mathrm{Re}[C_\Phi^{(21)(21)(12)}] + 6 s_\beta c_\beta^3 \ \mathrm{Re}[C_\Phi^{(11)(22)(21)}] \Big] , \\
\Delta M^2_{\rho_1\rho_2} &= \Delta M^2_{\rho_2\rho_1} = -\frac{B}{2} (M^2_{\rho_1\rho_1} + M^2_{\rho_2\rho_2}) - \frac{1}{2} (A_1 + A_2) M^2_{\rho_1\rho_2} \\
&- \frac{1}{2} (J M^2_{\rho_2 a_1} + K M^2_{\rho_2 a_2} + L M^2_{\rho_1 a_1} + N M^2_{\rho_1 a_2}) \\
&- v^4 \Big[ 4 s_\beta c_\beta^3 C_\Phi^{(11)(11)(22)} + 4 s_\beta^3 c_\beta C_\Phi^{(11)(22)(22)} + 5 c_\beta^4 \ \mathrm{Re}[C_\Phi^{(11)(11)(21)}] + 5 s_\beta^4 \ \mathrm{Re}[C_\Phi^{(22)(22)(21)}] \\
&+ 8 s_\beta c_\beta^3 \ \mathrm{Re}[C_\Phi^{(11)(21)(21)}] + 8 s_\beta^3 c_\beta \ \mathrm{Re}[C_\Phi^{(22)(21)(21)}] + 4 s_\beta c_\beta^3 C_\Phi^{(11)(21)(12)} + 4 s_\beta^3 c_\beta C_\Phi^{(22)(21)(12)} \\
&+ 9 s_\beta^2 c_\beta^2 \ \mathrm{Re}[C_\Phi^{(21)(21)(21)}] + 9 s_\beta^2 c_\beta^2 \ \mathrm{Re}[C_\Phi^{(21)(21)(12)}] + 9 s_\beta^2 c_\beta^2 \ \mathrm{Re}[C_\Phi^{(11)(22)(21)}] \Big] ,
\end{aligned}
\tag{A.7}
$$

$$
\begin{aligned}
\Delta M^2_{a_1a_1} &= -A'_1 M^2_{a_1a_1} - B' M^2_{a_1a_2} - J M^2_{\rho_1 a_1} - L M^2_{\rho_2 a_1} \\
&\quad -v^4 \Big[ 3c_\beta^4 C_\Phi^{(11)(11)(11)} + 2s_\beta^2 c_\beta^2 C_\Phi^{(11)(11)(22)} + s_\beta^4 C_\Phi^{(11)(22)(22)} + 4 s_\beta c_\beta^3 \ \mathrm{Re}[C_\Phi^{(11)(11)(21)}] \\
&\qquad -2 s_\beta^4 \ \mathrm{Re}[C_\Phi^{(22)(21)(21)}] + 2 s_\beta^2 c_\beta^2 C_\Phi^{(11)(21)(12)} + s_\beta^4 C_\Phi^{(22)(21)(12)} \\
&\qquad -6 s_\beta^3 c_\beta \ \mathrm{Re}[C_\Phi^{(21)(21)(21)}] + 2 s_\beta^3 c_\beta \ \mathrm{Re}[C_\Phi^{(21)(21)(12)}] + 2 s_\beta^3 c_\beta \ \mathrm{Re}[C_\Phi^{(11)(22)(21)}] \Big] , \\
\Delta M^2_{a_2a_2} &= -A'_2 M^2_{a_2a_2} - B' M^2_{a_1a_2} - K M^2_{\rho_1 a_2} - N M^2_{\rho_2 a_2} \\
&\quad -v^4 \Big[ 3s_\beta^4 C_\Phi^{(22)(22)(22)} + 2s_\beta^2 c_\beta^2 C_\Phi^{(11)(22)(22)} + c_\beta^4 C_\Phi^{(11)(11)(22)} + 4 s_\beta^3 c_\beta \ \mathrm{Re}[C_\Phi^{(22)(22)(21)}] \\
&\qquad -2 c_\beta^4 \ \mathrm{Re}[C_\Phi^{(11)(21)(21)}] + 2 s_\beta^2 c_\beta^2 C_\Phi^{(22)(21)(12)} + c_\beta^4 C_\Phi^{(11)(21)(12)} \\
&\qquad -6 s_\beta c_\beta^3 \ \mathrm{Re}[C_\Phi^{(21)(21)(21)}] + 2 s_\beta c_\beta^3 \ \mathrm{Re}[C_\Phi^{(21)(21)(12)}] + 2 s_\beta c_\beta^3 \ \mathrm{Re}[C_\Phi^{(11)(22)(21)}] \Big] , \\
\Delta M^2_{a_1a_2} &= \Delta M^2_{a_2a_1} = -\frac{B'}{2} (M^2_{a_1a_1} + M^2_{a_2a_2}) - \frac{1}{2} (A'_1 + A'_2) M^2_{a_1a_2} \\
&\quad -\frac{1}{2} (J M^2_{\rho_1 a_2} + K M^2_{\rho_1 a_1} + L M^2_{\rho_2 a_2} + N M^2_{\rho_2 a_1}) \\
&\quad -v^4 \Big[ c_\beta^4 \ \mathrm{Re}[C_\Phi^{(11)(11)(21)}] + s_\beta^4 \ \mathrm{Re}[C_\Phi^{(22)(22)(21)}] + 4 s_\beta c_\beta^3 \ \mathrm{Re}[C_\Phi^{(11)(21)(21)}] \\
&\qquad +4 s_\beta^3 c_\beta \ \mathrm{Re}[C_\Phi^{(22)(21)(21)}] + 9 s_\beta^2 c_\beta^2 \ \mathrm{Re}[C_\Phi^{(21)(21)(21)}] + s_\beta^2 c_\beta^2 \ \mathrm{Re}[C_\Phi^{(21)(21)(12)}] \\
&\qquad + s_\beta^2 c_\beta^2 \ \mathrm{Re}[C_\Phi^{(11)(22)(21)}] \Big] ,
\end{aligned}
\tag{A.8}
$$

and

$$
\begin{aligned}
\Delta M^2_{\rho_1 a_1} =& -\frac{1}{2}\left(J M^2_{a_1 a_1} + K M^2_{a_2 a_1} + J M^2_{\rho_1 \rho_1} + L M^2_{\rho_1 \rho_2}\right) \\
& -\frac{1}{2}\left(A_1 M^2_{\rho_1 a_1} + B M^2_{\rho_2 a_1} + A_1' M^2_{\rho_1 a_1} + B' M^2_{\rho_1 a_2}\right) \\
& - 4v^4 \Big[2 s_\beta c_\beta^3 \mathrm{Im}[C_\Phi^{(11)(11)(21)}] + 3 s_\beta^2 c_\beta^2 \mathrm{Im}[C_\Phi^{(11)(21)(21)}] + s_\beta^4 \mathrm{Im}[C_\Phi^{(22)(21)(21)}] \\
& + 3 s_\beta^3 c_\beta \mathrm{Im}[C_\Phi^{(21)(21)(21)}] + s_\beta^3 c_\beta \mathrm{Im}[C_\Phi^{(21)(21)(12)}] + s_\beta^3 c_\beta \mathrm{Im}[C_\Phi^{(11)(22)(21)}]\Big]\,, \\
\Delta M^2_{\rho_2 a_2} =& -\frac{1}{2}\left(L M^2_{a_1 a_2} + N M^2_{a_2 a_2} + N M^2_{\rho_2 \rho_2} + K M^2_{\rho_2 \rho_1}\right) \\
& -\frac{1}{2}\left(B M^2_{\rho_1 a_2} + A_2 M^2_{\rho_2 a_2} + B' M^2_{\rho_2 a_1} + A_2' M^2_{\rho_2 a_2}\right) \\
& + 4v^4 \Big[2 s_\beta^3 c_\beta \mathrm{Im}[C_\Phi^{(22)(22)(21)}] + c_\beta^4 \mathrm{Im}[C_\Phi^{(11)(21)(21)}] + 3 s_\beta^2 c_\beta^2 \mathrm{Im}[C_\Phi^{(22)(21)(21)}] \\
& + 3 s_\beta c_\beta^3 \mathrm{Im}[C_\Phi^{(21)(21)(21)}] + s_\beta c_\beta^3 \mathrm{Im}[C_\Phi^{(21)(21)(12)}] + s_\beta c_\beta^3 \mathrm{Im}[C_\Phi^{(11)(22)(21)}]\Big] \\
\Delta M^2_{\rho_1 a_2} =& -\frac{1}{2}\left(J M^2_{a_1 a_2} + K M^2_{a_2 a_2} + K M^2_{\rho_1 \rho_1} + N M^2_{\rho_1 \rho_2}\right) \\
& -\frac{1}{2}\left(A_1 M^2_{\rho_1 a_2} + B M^2_{\rho_2 a_2} + B' M^2_{\rho_1 a_1} + A_2' M^2_{\rho_1 a_2}\right) \\
& + 2v^4 \Big[5 c_\beta^4 \mathrm{Im}[C_\Phi^{(11)(11)(21)}] + s_\beta^4 \mathrm{Im}[C_\Phi^{(22)(22)(21)}] + 8 s_\beta c_\beta^3 \mathrm{Im}[C_\Phi^{(11)(21)(21)}] \\
& + 4 s_\beta^3 c_\beta \mathrm{Im}[C_\Phi^{(22)(21)(21)}] + 9 s_\beta^2 c_\beta^2 \mathrm{Im}[C_\Phi^{(21)(21)(21)}] + 3 s_\beta^2 c_\beta^2 \mathrm{Im}[C_\Phi^{(21)(21)(12)}] \\
& + 3 s_\beta^2 c_\beta^2 \mathrm{Im}[C_\Phi^{(11)(22)(21)}]\Big]\,, \\
\Delta M^2_{\rho_2 a_1} =& -\frac{1}{2}\left(L M^2_{a_1 a_1} + N M^2_{a_2 a_1} + J M^2_{\rho_2 \rho_1} + L M^2_{\rho_2 \rho_2}\right) \\
& -\frac{1}{2}\left(B M^2_{\rho_1 a_1} + A_2 M^2_{\rho_2 a_1} + A_1' M^2_{\rho_2 a_1} + B' M^2_{\rho_2 a_2}\right) \\
& - 2v^4 \Big[c_\beta^4 \mathrm{Im}[C_\Phi^{(11)(11)(21)}] + 5 s_\beta^4 \mathrm{Im}[C_\Phi^{(22)(22)(21)}] + 4 s_\beta c_\beta^3 \mathrm{Im}[C_\Phi^{(11)(21)(21)}] \\
& + 8 s_\beta^3 c_\beta \mathrm{Im}[C_\Phi^{(22)(21)(21)}] + 9 s_\beta^2 c_\beta^2 \mathrm{Im}[C_\Phi^{(21)(21)(21)}] + 3 s_\beta^2 c_\beta^2 \mathrm{Im}[C_\Phi^{(21)(21)(12)}] \\
& + 3 s_\beta^2 c_\beta^2 \mathrm{Im}[C_\Phi^{(11)(22)(21)}]\Big]\,.
\end{aligned}
\tag{A.9}
$$

The matrix elements entering the charged scalar mass squared matrix in Eq. (3.35) are

$$
\begin{aligned}
M^{\pm 2}_{11} &= m_1^2 + \lambda_1 v_1^2 + v_2^2 \lambda_3 + 2 v_1 v_2 \ \mathrm{Re}[\lambda_6], \\
M^{\pm 2}_{22} &= m_2^2 + \lambda_2 v_2^2 + v_1^2 \lambda_3 + 2 v_1 v_2 \ \mathrm{Re}[\lambda_7], \\
M^{\pm 2}_{12} &= (M^{\pm 2}_{21})^* = (m_{12}^2)^* + v_1 v_2 (\lambda_4 + \lambda_5^*) + v_1^2 \lambda_6^* + v_2^2 \lambda_7^*,
\end{aligned}
\tag{A.10}
$$

and

$$
\begin{aligned}
\Delta M_{11}^{\pm 2} = -\text{Re}[\eta M_{12}^{\pm 2}] - v^4 \Big[ & 3c_\beta^4 C_\Phi^{(11)(11)(11)} + 2s_\beta^2 c_\beta^2 C_\Phi^{(11)(11)(22)} + s_\beta^4 C_\Phi^{(11)(22)(22)} \\
& + 4 s_\beta c_\beta^3 \text{ Re}[C_\Phi^{(11)(11)(21)}] + 2 s_\beta^2 c_\beta^2 \text{ Re}[C_\Phi^{(11)(21)(21)}] \\
& + s_\beta^2 c_\beta^2 C_\Phi^{(11)(21)(12)} + 2 s_\beta^3 c_\beta \text{ Re}[C_\Phi^{(11)(22)(21)}] \Big], \\
\Delta M_{22}^{\pm 2} = -\text{Re}[\eta M_{12}^{\pm 2}] - v^4 \Big[ & 3s_\beta^4 C_\Phi^{(22)(22)(22)} + 2s_\beta^2 c_\beta^2 C_\Phi^{(11)(22)(22)} + c_\beta^4 C_\Phi^{(11)(11)(22)} \\
& + 4 s_\beta^3 c_\beta \text{ Re}[C_\Phi^{(22)(22)(21)}] + 2 s_\beta^2 c_\beta^2 \text{ Re}[C_\Phi^{(22)(21)(21)}] \\
& + s_\beta^2 c_\beta^2 C_\Phi^{(22)(21)(12)} + 2 s_\beta c_\beta^3 \text{ Re}[C_\Phi^{(11)(22)(21)}] \Big], \\
\Delta M_{12}^{\pm 2} = (\Delta M_{21}^{\pm 2})^* = & -\frac{\eta^*}{2} (M_{11}^{\pm 2} + M_{22}^{\pm 2}) \\
& - v^4 \Big[ c_\beta^4 C_\Phi^{(11)(11)(21)} + s_\beta^4 C_\Phi^{(22)(22)(21)} + 2 s_\beta c_\beta^3 C_\Phi^{(11)(21)(21)} \\
& + 2 s_\beta^3 c_\beta C_\Phi^{(22)(21)(21)} + s_\beta c_\beta^3 C_\Phi^{(11)(21)(12)} + s_\beta^3 c_\beta C_\Phi^{(22)(21)(12)} \\
& + 3 s_\beta^2 c_\beta^2 C_\Phi^{(21)(21)(21)} + 3 s_\beta^2 c_\beta^2 C_\Phi^{(21)(21)(12)} + s_\beta^2 c_\beta^2 C_\Phi^{(11)(22)(21)} \Big].
\end{aligned}
\tag{A.11}
$$

Lastly, the matrix elements entering the mass squared matrix for the neutral gauge bosons in Eq. (3.46) from the Lagrangian in Eq. (3.44) are

$$
\begin{aligned}
M_{11}^2 &= \frac{g'^2}{2} (v^2 + 2 v_1 v_2 \text{ Re}[\eta]), \\
M_{12}^2 &= M_{21}^2 = -\frac{g g'}{2} (v^2 + 2 v_1 v_2 \text{ Re}[\eta]), \\
M_{22}^2 &= \frac{g^2}{2} (v^2 + 2 v_1 v_2 \text{ Re}[\eta]),
\end{aligned}
\tag{A.12}
$$

and

$$
\begin{aligned}
\Delta M_{11}^2 = -\frac{g'^2}{2} \Big( & v_1^4 C_{\Phi D}^{(11)(11)} + v_2^4 C_{\Phi D}^{(22)(22)} + v_1^2 v_2^2 C_{\Phi D}^{(21)(12)} + v_1^2 v_2^2 C_{\Phi D}^{(11)(22)} \\
& + 2 v_1^2 v_2^2 \text{ Re}[C_{\Phi D}^{(21)(21)}] + 2 v_1^3 v_2 \text{ Re}[C_{\Phi D}^{(21)(11)}] + 2 v_1 v_2^3 \text{ Re}[C_{\Phi D}^{(21)(22)}] \Big) \\
& - \frac{1}{2} (g'^2 X_B - g g' X_{WB}) (v^2 + 2 v_1 v_2 \text{ Re}[\eta]), \\
\Delta M_{22}^2 = -\frac{g^2}{2} \Big( & v_1^4 C_{\Phi D}^{(11)(11)} + v_2^4 C_{\Phi D}^{(22)(22)} + v_1^2 v_2^2 C_{\Phi D}^{(21)(12)} + v_1^2 v_2^2 C_{\Phi D}^{(11)(22)} \\
& + 2 v_1^2 v_2^2 \text{ Re}[C_{\Phi D}^{(21)(21)}] + 2 v_1^3 v_2 \text{ Re}[C_{\Phi D}^{(21)(11)}] + 2 v_1 v_2^3 \text{ Re}[C_{\Phi D}^{(21)(22)}] \Big) \\
& - \frac{1}{2} (g^2 X_W - g g' X_{WB}) (v^2 + 2 v_1 v_2 \text{ Re}[\eta]), \\
\Delta M_{12}^2 = \Delta M_{21}^2 = +\frac{g g'}{2} \Big( & v_1^4 C_{\Phi D}^{(11)(11)} + v_2^4 C_{\Phi D}^{(22)(22)} + v_1^2 v_2^2 C_{\Phi D}^{(21)(12)} + v_1^2 v_2^2 C_{\Phi D}^{(11)(22)} \\
& + 2 v_1^2 v_2^2 \text{ Re}[C_{\Phi D}^{(21)(21)}] + 2 v_1^3 v_2 \text{ Re}[C_{\Phi D}^{(21)(11)}] + 2 v_1 v_2^3 \text{ Re}[C_{\Phi D}^{(21)(22)}] \Big) \\
& + \frac{1}{4} (g g' (X_B + X_W) - (g^2 + g'^2) X_{WB}) (v^2 + 2 v_1 v_2 \text{ Re}[\eta]).
\end{aligned}
\tag{A.13}
$$

## B Diagonalization of the Scalar Sector

In this Appendix, we summarize the results of the diagonalization of the $2 \times 2$ matrices for $\hat{M}^2_{\rho\rho}$, $\hat{M}^2_{aa}$, and $\hat{M}^{\pm 2}$ not described in the main text. We also define terms used to simplify the Feynman rules.

The matrices $\hat{M}^2_{aa}$ and $\hat{M}^2_{\rho\rho}$ are real and symmetric, meanwhile the matrix $\hat{M}^{\pm 2}$ is Hermitian even in the presence of the generally-complex dimension-six effective operators. A $2 \times 2$ Hermitian matrix can be diagonalized by one angle:

$$\begin{pmatrix} \cos\chi & \sin\chi \\ -\sin\chi & \cos\chi \end{pmatrix} \begin{pmatrix} a & c \\ c^* & b \end{pmatrix} \begin{pmatrix} \cos\chi & -\sin\chi \\ \sin\chi & \cos\chi \end{pmatrix} = \begin{pmatrix} m_+^2 & 0 \\ 0 & m_-^2 \end{pmatrix}, \tag{B.1}$$

where $m^2_\pm = \frac{1}{2}\left[a + b \pm \sqrt{(a-b)^2 + 4|c|^2}\right]$. We also have

$$\begin{aligned} \tan 2\chi &= \frac{2|c|}{a-b}, \\ \sin 2\chi &= 2\sin\chi\cos\chi = \frac{2|c|}{\sqrt{(a-b)^2+4|c|^2}}. \end{aligned} \tag{B.2}$$

The individual scalars are rotated to the new basis in the following way

$$\begin{pmatrix} \hat{\rho}_1 \\ \hat{\rho}_2 \end{pmatrix} = \begin{pmatrix} \cos\hat{\alpha} & -\sin\hat{\alpha} \\ \sin\hat{\alpha} & \cos\hat{\alpha} \end{pmatrix} \begin{pmatrix} H \\ h \end{pmatrix}, \tag{B.3}$$

$$\begin{pmatrix} \hat{a}_1 \\ \hat{a}_2 \end{pmatrix} = \begin{pmatrix} \cos\hat{\beta} & -\sin\hat{\beta} \\ \sin\hat{\beta} & \cos\hat{\beta} \end{pmatrix} \begin{pmatrix} G \\ A \end{pmatrix}, \tag{B.4}$$

$$\begin{pmatrix} \hat{\phi}^+_1 \\ \hat{\phi}^+_2 \end{pmatrix} = \begin{pmatrix} \cos\hat{\beta}^\pm & -\sin\hat{\beta}^\pm \\ \sin\hat{\beta}^\pm & \cos\hat{\beta}^\pm \end{pmatrix} \begin{pmatrix} G^+ \\ H^+ \end{pmatrix}, \tag{B.5}$$

for the $\vec{\hat{\rho}}$, $\vec{\hat{a}}$, and charged eigenstates, respectively, with the following diagonalization angles in the effective theory at leading order:

$$\begin{aligned} \tan 2\hat{\alpha} &= \frac{2(M^2_{\rho_1\rho_2} + \Delta M^2_{\rho_1\rho_2})}{(M^2_{\rho_1\rho_1} - M^2_{\rho_2\rho_2}) + (\Delta M^2_{\rho_1\rho 1} - \Delta M^2_{\rho_2\rho_2})} \\ &\simeq \tan 2\alpha \left[1 - \frac{\Delta M^2_{\rho_1\rho_1} - \Delta M^2_{\rho_2\rho_2}}{M^2_{\rho_1\rho_1} - M^2_{\rho_2\rho_2}} + \frac{\Delta M^2_{\rho_1\rho_2}}{M^2_{\rho_1\rho_2}}\right], \end{aligned} \tag{B.6}$$

$$\begin{aligned} \tan 2\hat{\beta} &= \frac{2(M^2_{a_1a_2} + \Delta M^2_{a_1a_2})}{(M^2_{a_1a_1} - M^2_{a_2a_2}) + (\Delta M^2_{a_1a_1} - \Delta M^2_{a_2a_2})} \\ &\simeq \tan 2\beta \left[1 - \frac{\Delta M^2_{a_1a_1} - \Delta M^2_{a_2a_2}}{M^2_{a_1a_1} - M^2_{a_2a_2}} + \frac{\Delta M^2_{a_1a_2}}{M^2_{a_1a_2}}\right], \end{aligned} \tag{B.7}$$

$$\begin{aligned} \tan 2\hat{\beta}^\pm &= \frac{2|M^{\pm 2}_{12} + \Delta M^{\pm 2}_{12}|}{(M^{\pm 2}_{11} - M^{\pm 2}_{22}) + (\Delta M^{\pm 2}_{11} - \Delta M^{\pm 2}_{22})} \\ &\simeq \tan 2\beta \left[1 - \frac{\Delta M^{\pm 2}_{11} - \Delta M^{\pm 2}_{22}}{M^{\pm 2}_{11} - M^{\pm 2}_{22}} + \frac{|\Delta M^{\pm 2}_{12}|}{|M^{\pm 2}_{12}|}\ \mathrm{Re}[e^{i\phi}]\right], \end{aligned} \tag{B.8}$$

where $\phi = |\phi_{M_{12}^{\pm 2}} - \phi_{\Delta M_{12}^{\pm 2}}|$ is the phase difference between the dimension-four and -six contributions entering the off-diagonal matrix element. All of the above matrix elements can be found in Appendix A. The eigenvalues of the $\hat{M}^2_{\rho\rho}$, $\hat{M}^2_{aa}$, and $\hat{M}^{\pm 2}$ matrices are

$$
\begin{aligned}
m_H^2 &= \cos^2\hat{\alpha}(M^2_{\rho_1\rho_1} + \Delta M^2_{\rho_1\rho_1}) + \sin 2\hat{\alpha}(M^2_{\rho_1\rho_2} + \Delta M^2_{\rho_1\rho_2}) + \sin^2\hat{\alpha}(M^2_{\rho_2\rho_2} + \Delta M^2_{\rho_2\rho_2}), \\
m_h^2 &= \sin^2\hat{\alpha}(M^2_{\rho_1\rho_1} + \Delta M^2_{\rho_1\rho_1}) - \sin 2\hat{\alpha}(M^2_{\rho_1\rho_2} + \Delta M^2_{\rho_1\rho_2}) + \cos^2\hat{\alpha}(M^2_{\rho_2\rho_2} + \Delta M^2_{\rho_2\rho_2}),
\end{aligned}
\tag{B.9}
$$

$$
\begin{aligned}
m_G^2 &= \cos^2\hat{\beta}(M^2_{a_1a_1} + \Delta M^2_{a_1a_1}) + \sin 2\hat{\beta}(M^2_{a_1a_2} + \Delta M^2_{a_1a_2}) + \sin^2\hat{\beta}(M^2_{a_2a_2} + \Delta M^2_{a_2a_2}) \\
&\simeq 0, \\
m_A^2 &= \sin^2\hat{\beta}(M^2_{a_1a_1} + \Delta M^2_{a_1a_1}) - \sin 2\hat{\beta}(M^2_{a_1a_2} + \Delta M^2_{a_1a_2}) + \cos^2\hat{\beta}(M^2_{a_2a_2} + \Delta M^2_{a_2a_2}) \\
&\simeq M^2_{a_1a_1} + M^2_{a_2a_2} + \Delta M^2_{a_1a_1} + \Delta M^2_{a_2a_2},
\end{aligned}
\tag{B.10}
$$

$$
\begin{aligned}
m^2_{G^\pm} &= \cos^2\hat{\beta}^\pm(M_{11}^{\pm 2} + \Delta M_{11}^{\pm 2}) + \sin 2\hat{\beta}^\pm|M_{12}^{\pm 2} + \Delta M_{12}^{\pm 2}| + \sin^2\hat{\beta}^\pm(M_{22}^{\pm 2} + \Delta M_{22}^{\pm 2}) \\
&\simeq 0, \\
m^2_{H^\pm} &= \sin^2\hat{\beta}^\pm(M_{11}^{\pm 2} + \Delta M_{11}^{\pm 2}) - \sin 2\hat{\beta}^\pm|M_{12}^{\pm 2} + \Delta M_{12}^{\pm 2}| + \cos^2\hat{\beta}^\pm(M_{22}^{\pm 2} + \Delta M_{22}^{\pm 2}) \\
&\simeq M_{11}^{\pm 2} + M_{22}^{\pm 2} + \Delta M_{11}^{\pm 2} + \Delta M_{22}^{\pm 2}.
\end{aligned}
\tag{B.11}
$$

The matrices $\hat{M}^2_{aa}$ and $\hat{M}^{\pm 2}$, associated with the $G$ and $G^\pm$ bosons, admit one zero eigenvalue $m_-^2 = 0$ (at the dimension-four and dimension-six levels). In this case, equations for the diagonalization angle simplify to

$$
\begin{aligned}
\sin\chi &= \sqrt{\frac{b}{m_+^2}} = \sqrt{\frac{b}{a+b}}, \\
\cos\chi &= \sqrt{\frac{m_+^2 - b}{m_+^2}} = \sqrt{\frac{a}{a+b}}.
\end{aligned}
\tag{B.12}
$$

The angle $\chi$ for both $\hat{M}^2_{aa}$ and $\hat{M}^{\pm 2}$ is the same as the vacuum angle $\beta$ in the 2HDM at tree level. However, in the effective theory, $a \to a + \Delta a$ and $b \to b + \Delta b$, where $\Delta(a,b) \sim v^4/\Lambda^2$ are contributions from the dimension-six operators, in which case the angles are no longer equivalent. We find

$$
\begin{aligned}
\sin(\beta - \chi) &= \sin\beta\cos\chi - \sin\chi\cos\beta \\
&= \sqrt{\frac{b}{a+b}}\sqrt{\frac{a+\Delta a}{a+b+\Delta a+\Delta b}} - \sqrt{\frac{a}{a+b}}\sqrt{\frac{b+\Delta b}{a+b+\Delta a+\Delta b}} \\
&\simeq \frac{1}{2(a+b)}\left[\Delta a\sqrt{\frac{b}{a}} - \Delta b\sqrt{\frac{a}{b}}\right] = \beta - \chi \sim \mathcal{O}\left(\frac{v^4}{\Lambda^2 m_+^2}\right),
\end{aligned}
\tag{B.13}
$$

and

$$
\begin{aligned}
\cos(\beta - \chi) &= \cos\beta \cos\chi + \sin\beta \sin\chi \\
&\simeq 1 - \frac{1}{2}(\beta-\chi)^2 \sim 1 + \mathcal{O}\left(\frac{v^8}{\Lambda^4 m_+^4}\right) .
\end{aligned} \tag{B.14}
$$

We see the misalignment is additionally suppressed by $v^2/m_+^2$. Specifically, for the two matrices $\hat{M}_{aa}^2$ and $\hat{M}^{\pm 2}$, we have $\chi = \hat{\beta}, m_+^2 = m_A^2$ and $\chi = \hat{\beta}^{\pm}, m_+^2 = m_{H^\pm}^2$, respectively.[7] For convenience, we can also write

$$
\begin{aligned}
\sin\chi &\simeq \sin\beta\left[1 - \frac{\Delta a + \Delta b}{2(a+b)} + \frac{\Delta b}{2b}\right] \equiv \sin\beta\left[1 - \delta_{s_\chi}\right] , \\
\cos\chi &\simeq \cos\beta\left[1 - \frac{\Delta a + \Delta b}{2(a+b)} + \frac{\Delta a}{2a}\right] \equiv \cos\beta\left[1 - \delta_{c_\chi}\right] .
\end{aligned} \tag{B.15}
$$

Specifically, with the respective matrix elements from Eqs. (A.2) and (3.35), we have

$$
\begin{aligned}
\sin\hat{\beta} &\simeq \sin\beta\left[1 - \frac{\Delta M_{a_1a_1}^2 + \Delta M_{a_2a_2}^2}{M_{a_1a_1}^2 + M_{a_2a_2}^2} + \frac{\Delta M_{a_2a_2}^2}{M_{a_2a_2}^2}\right] = \sin\beta\left[1 - \delta_{s_{\hat{\beta}}}\right] , \\
\cos\hat{\beta} &\simeq \sin\beta\left[1 - \frac{\Delta M_{a_1a_1}^2 + \Delta M_{a_2a_2}^2}{M_{a_1a_1}^2 + M_{a_2a_2}^2} + \frac{\Delta M_{1_2a_1}^2}{M_{a_1a_1}^2}\right] = \sin\beta\left[1 - \delta_{c_{\hat{\beta}}}\right] , \\
\sin\hat{\beta}^{\pm} &\simeq \sin\beta\left[1 - \frac{\Delta M_{11}^{\pm 2} + \Delta M_{22}^{\pm 2}}{M_{11}^{\pm 2} + M_{22}^{\pm 2}} + \frac{\Delta M_{22}^{\pm 2}}{M_{22}^{\pm 2}}\right] = \sin\beta\left[1 - \delta_{s_{\hat{\beta}^\pm}}\right] , \\
\cos\hat{\beta}^{\pm} &\simeq \cos\beta\left[1 - \frac{\Delta M_{11}^{\pm 2} + \Delta M_{22}^{\pm 2}}{M_{11}^{\pm 2} + M_{22}^{\pm 2}} + \frac{\Delta M_{11}^{\pm 2}}{M_{11}^{\pm 2}}\right] = \cos\beta\left[1 - \delta_{c_{\hat{\beta}^\pm}}\right] ,
\end{aligned} \tag{B.16}
$$

where $\delta$'s simplify the Feynman rules in Appendix C.

In the $CP$-violating theory, we can separate the dimension-six contributions to rotation angles (Eq. (3.29)) in a similar way,

$$
\begin{aligned}
\sin\hat{\xi} &\simeq \sin\xi\left[1 - \delta_{s_{\hat{\xi}}}\right] , \qquad \cos\hat{\xi} \simeq \cos\xi\left[1 - \delta_{c_{\hat{\xi}}}\right] , \\
\sin\hat{\omega} &\simeq \sin\omega\left[1 - \delta_{s_{\hat{\omega}}}\right] , \qquad \cos\hat{\omega} \simeq \cos\omega\left[1 - \delta_{c_{\hat{\omega}}}\right] ,
\end{aligned} \tag{B.17}
$$

where the $\delta$'s can be obtained by expanding Eqs. (3.29) and (3.33) to leading order.

## C Selected Feynman Rules in the $CP$-Conserving Type-II 2HDM in the Alignment Limit

We present Feynman rules for the $CP$-Conserving type-II 2HDM in the alignment limit $\hat{\alpha} \to \beta - \pi/2$. For simplicity, we also set $\eta \to 0$. The subsection C.1 contains the complete set of Feynman rules for charged and neutral leptons coupling to gauge bosons and scalars. In subsection C.2, we present the complete set of Feynman rules for up-type quarks. To shorten the results, we omit any electroweak vertices involving only the down-type quarks (with the exception of the photon and $Z$ couplings) since they are identical to the charged lepton vertices upon replacement of $e \to d$ and $l \to q$ in the label of the Wilson coefficients and Feynman diagrams. We keep flavor indices explicit throughout the Appendix. This helps with matrix multiplication involving the PMNS or CKM matrices, $U$ or $V$, and the Wilson coefficients of the contributing operators. In subsection C.3, we present the complete set of Feynman rules for three- and four-point vertices involving electroweak gauge bosons and at least one scalar. The last subsection C.4 includes Feynman rules for vertices with two gluons and either one or two scalars.

[7] Modifications of the diagonalization angles where computed in detail at one-loop order in the 2HDM [42] and agree with our results when $\Delta(a,b)$ are matched to one-loop diagrams instead of contributions from dimension-six operators.

In order to keep the main results manageable, we do not present Feynman rules for charged-conjugated vertices such as $\overline{e}\nu$ and $\overline{d}u$. These can be obtained via complex conjugation of the Feynman rule (without an overall factor of $i$). Furthermore, we do not present five- and six-point vertices for scalar-gauge interactions, and any scalar self-interactions. To illustrate the structure of these vertices, we give one example of a five-point and six-point scalar-gauge vertex in Eqs. (C.568) and (C.569). We also omit any gauge self-interactions and four-fermion interactions, for which there is nothing new originating from the new scalars.

Momentum assignments for three-, four-, five-, and six-point diagrams are defined in the following way:

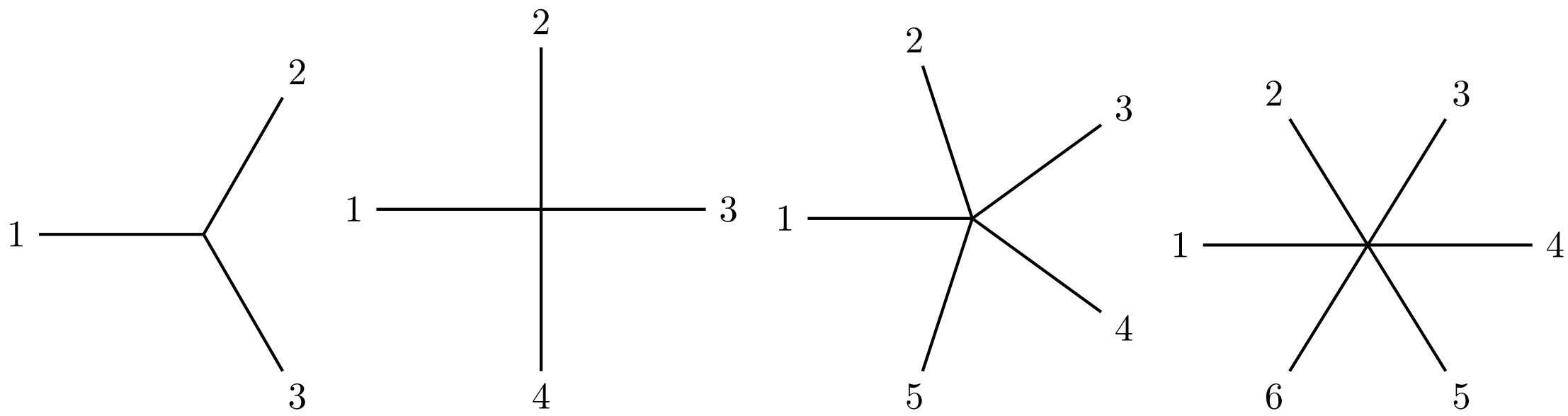

with all momenta incoming to the vertex.

For compactness, terms containing an angle are written as $\sin x \equiv s_x, \cos x \equiv c_x, \tan x \equiv t_x$, and $\cot x \equiv ct_x$. The indices appearing in a given Feynman rule follow the conventions in the FeynRules code [11, 12] and are the following: $f_i$ and $g_i$ represent the flavor index of the $i$-th particle, for example in $\overline{\psi}^{f_1}(\hat{C})_{f_1 f_2}\psi^{f_2}$ and its Hermitian conjugate $\overline{\psi}^{f_1}(\hat{C}^*)_{f_2 f_1}\psi^{f_2}$; $m_i$ represents the fundamental color index of the $i$-th particle, where $m_i = 1, \cdots, N$ for $N = 3$; $a_i$ and $b_i$ represent the adjoint color index of the $i$-th gluon, where $a_i, b_i = 1, \cdots, N^2 - 1$ for $N = 3$; $\mu_i$ represents the Lorentz index of the $i$-th particle (not to be confused with Greek letters *without* a number which are reserved for contracted indices in the expression).

For lepton-violating vertices arising from, for example, the Weinberg operators (Eq. (3.53)), external fermions are defined as $\psi^c = \mathbf{C}\overline{\psi}^T$ and $\overline{\psi}^c = \psi^T\mathbf{C}$, where $\mathbf{C} = i\gamma^2\gamma^0$ is the charge-conjugate matrix. Thus, the four-component Dirac spinors are given as $u(p) = \mathbf{C}\overline{v}(p)^T$ and $v(p) = \overline{u}(p)^T\mathbf{C}$ [18] in Eq. (C.15), for example.

Following the convention of the FeynRules code [11, 12], the neutral Goldstone boson

$G$ is written as $G^0$ and the $CP$-odd Higgs $A$ as $A^0$ in the Feynman rules below so that the code does not mix the labelings between the Goldstone boson and the gluon, and the $CP$-odd Higgs and the photon field.

### C.1 Charged Lepton and Neutrino Interactions

In this subsection, we present the complete set of Feynman rules for charged lepton and neutrino interactions.

$$\begin{aligned}&+\frac{iU^{*}_{f_2 f_1}}{v}\left(m_{\nu_{f_1}}\mathcal{P}_L+m_{e_{f_2}}\mathcal{P}_R\left(\delta_{c_{\hat{\beta}\pm}}-1\right)\right)\\&+2iv\not{p}_3\mathcal{P}_L U^{*}_{g_1 f_1}\left(c_\beta^2\hat{C}^{(11)[3]}_{\Phi l,g_1 f_2}+s_\beta^2\hat{C}^{(22)[3]}_{\Phi l,g_1 f_2}\right)\end{aligned}\tag{C.1}$$

$$\begin{aligned}&-\frac{\sqrt{2}\delta_{f_1 f_2}\gamma^5}{v}\left(m_{\nu_{f_1}}\right)\\&+\sqrt{2}vU_{g_2 f_2}U^{*}_{g_1 f_1}\not{p}_3\gamma^5\left(c_\beta^2\hat{C}^{(11)[1]}_{\Phi l,g_1 g_2}-c_\beta^2\hat{C}^{(11)[3]}_{\Phi l,g_1 g_2}+s_\beta^2\left(\hat{C}^{(22)[1]}_{\Phi l,g_1 g_2}-\hat{C}^{(22)[3]}_{\Phi l,g_1 g_2}\right)\right)\end{aligned}\tag{C.2}$$

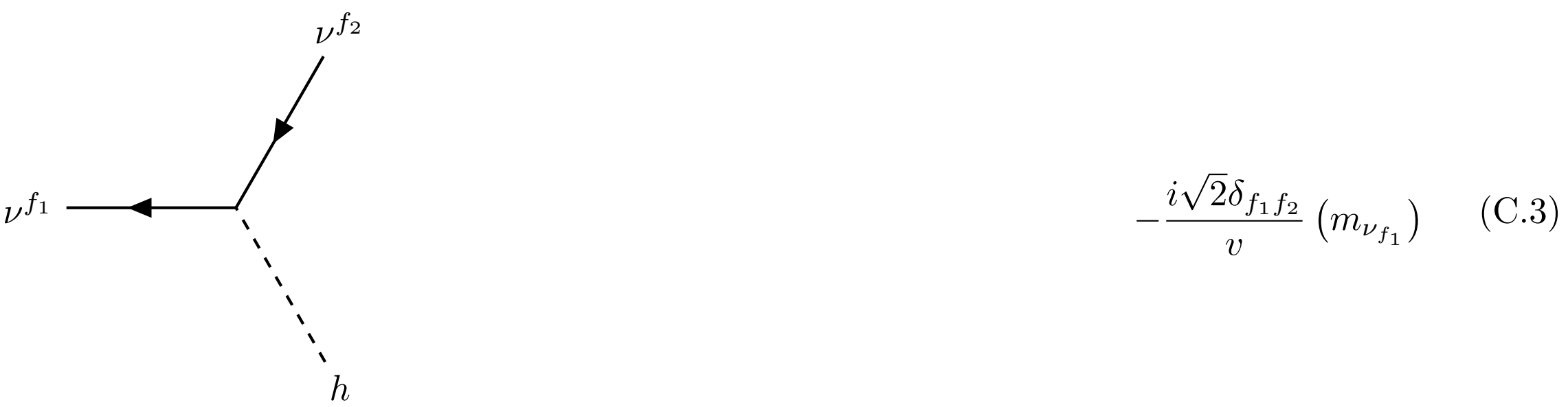

$$-\frac{i\sqrt{2}\delta_{f_1 f_2}}{v}\left(m_{\nu_{f_1}}\right)\tag{C.3}$$

$\nu^{f_2}$, $\nu^{f_1}$, $A^0$

$$
\begin{aligned}
&+\sqrt{2}vs_{2\beta}\left(\mathcal{P}_L\hat{C}^{(11)}_{\nu\nu\Phi,f_1f_2}-\mathcal{P}_L\hat{C}^{(22)}_{\nu\nu\Phi,f_1f_2}-\mathcal{P}_R\hat{C}^{(11)*}_{\nu\nu\Phi,f_1f_2}+\mathcal{P}_R\hat{C}^{(22)*}_{\nu\nu\Phi,f_1f_2}\right)\\
&+\sqrt{2}vs_\beta c_\beta U_{g_2f_2}U^*_{g_1f_1}\not{p}_3\gamma^5\left(-\hat{C}^{(11)[1]}_{\Phi l,g_1g_2}+\hat{C}^{(11)[3]}_{\Phi l,g_1g_2}+\hat{C}^{(22)[1]}_{\Phi l,g_1g_2}-\hat{C}^{(22)[3]}_{\Phi l,g_1g_2}\right)
\end{aligned}
\tag{C.4}
$$

$\nu^{f_2}$, $\nu^{f_1}$, $H$

$$
+i\sqrt{2}vs_{2\beta}\left(\mathcal{P}_L\hat{C}^{(11)}_{\nu\nu\Phi,f_1f_2}-\mathcal{P}_L\hat{C}^{(22)}_{\nu\nu\Phi,f_1f_2}+\mathcal{P}_R\hat{C}^{(11)*}_{\nu\nu\Phi,f_1f_2}-\mathcal{P}_R\hat{C}^{(22)*}_{\nu\nu\Phi,f_1f_2}\right)
\tag{C.5}
$$

$e^{f_2}$, $\nu^{f_1}$, $H^+$

$$
\begin{aligned}
&+2ivs_\beta c_\beta\mathcal{P}_LU^*_{f_2g_1}\left(\hat{C}^{(11)}_{\nu\nu\Phi,g_1f_1}-\hat{C}^{(22)}_{\nu\nu\Phi,g_1f_1}\right)\\
&-is_\beta\mathcal{P}_R\left(\delta_{s_{\hat{\beta}\pm}}-1\right)U^*_{g_1f_1}\left(\hat{y}^{(1)}_{e,g_1f_2}\right)\\
&-iv^2s_\beta\mathcal{P}_RU^*_{g_1f_1}\left(c^2_\beta\hat{C}^{(11)}_{l\Phi_1,g_1f_2}-c^2_\beta\hat{C}^{(12)}_{l\Phi_2,g_1f_2}\right.\\
&\qquad\left.-c^2_\beta\hat{C}^{(21)}_{l\Phi_2,g_1f_2}+s^2_\beta\hat{C}^{(22)}_{l\Phi_1,g_1f_2}\right)\\
&-ivs_{2\beta}\not{p}_3\mathcal{P}_LU^*_{g_1f_1}\left(\hat{C}^{(11)[3]}_{\Phi l,g_1f_2}-\hat{C}^{(22)[3]}_{\Phi l,g_1f_2}\right)
\end{aligned}
\tag{C.6}
$$

$e^{f_2}$, $e^{f_1}$, $G^0$

$$
\begin{aligned}
&-\frac{\delta_{f_1f_2}m_{e_{f_1}}\gamma^5}{2\sqrt{2}v}\left(A'_1+t_\beta B'+2\delta_{c_{\hat{\beta}}}-2\right)\\
&-\sqrt{2}v\not{p}_3\mathcal{P}_L\left(c^2_\beta\hat{C}^{(11)[1]}_{\Phi l,f_1f_2}+c^2_\beta\hat{C}^{(11)[3]}_{\Phi l,f_1f_2}+s^2_\beta\left(\hat{C}^{(22)[1]}_{\Phi l,f_1f_2}+\hat{C}^{(22)[3]}_{\Phi l,f_1f_2}\right)\right)\\
&-\sqrt{2}v\not{p}_3\mathcal{P}_R\left(c^2_\beta\hat{C}^{(11)}_{\Phi e,f_1f_2}+s^2_\beta\hat{C}^{(22)}_{\Phi e,f_1f_2}\right)
\end{aligned}
\tag{C.7}
$$

$e^{f_2}$ $e^{f_1}$ $h$

$$
\begin{aligned}
&+\frac{i\delta_{f_1 f_2} m_{e_{f_1}}}{2\sqrt{2}v}\left(A_1+Bt_\beta-2\right)\\
&+i\sqrt{2}v^2 c_\beta\left(c_\beta^2\left(\mathcal{P}_L\hat{C}^{(11)*}_{l\Phi_1,f_2f_1}+\mathcal{P}_R\hat{C}^{(11)}_{l\Phi_1,f_1f_2}\right)\right.\\
&\qquad+s_\beta^2\left(\mathcal{P}_L\hat{C}^{(12)*}_{l\Phi_2,f_2f_1}+\mathcal{P}_R\hat{C}^{(12)}_{l\Phi_2,f_1f_2}\right)\\
&\qquad+s_\beta^2\left(\mathcal{P}_L\hat{C}^{(21)*}_{l\Phi_2,f_2f_1}+\mathcal{P}_R\hat{C}^{(21)}_{l\Phi_2,f_1f_2}\right)\\
&\qquad\left.+s_\beta^2\left(\mathcal{P}_L\hat{C}^{(22)*}_{l\Phi_1,f_2f_1}+\mathcal{P}_R\hat{C}^{(22)}_{l\Phi_1,f_1f_2}\right)\right)
\end{aligned}
\tag{C.8}
$$

$e^{f_2}$ $e^{f_1}$ $A^0$

$$
\begin{aligned}
&+\frac{1}{2\sqrt{2}}\left(-\left(A_1'-2\right)s_\beta+c_\beta B'-2s_\beta\delta_{s_{\hat{\beta}}}\right)\left(\mathcal{P}_L\hat{y}^{(1)*}_{e,f_2f_1}-\mathcal{P}_R\hat{y}^{(1)}_{e,f_1f_2}\right)\\
&+\frac{v^2 s_\beta}{\sqrt{2}}\left(-c_\beta^2\left(\mathcal{P}_L\hat{C}^{(11)*}_{l\Phi_1,f_2f_1}-\mathcal{P}_R\hat{C}^{(11)}_{l\Phi_1,f_1f_2}\right)\right.\\
&\qquad+s_\beta^2\left(2ct_\beta^2+1\right)\left(\mathcal{P}_L\hat{C}^{(12)*}_{l\Phi_2,f_2f_1}-\mathcal{P}_R\hat{C}^{(12)}_{l\Phi_2,f_1f_2}\right)\\
&\qquad-s_\beta^2\left(\mathcal{P}_L\hat{C}^{(21)*}_{l\Phi_2,f_2f_1}-\mathcal{P}_R\hat{C}^{(21)}_{l\Phi_2,f_1f_2}\right)\\
&\qquad\left.-s_\beta^2\left(\mathcal{P}_L\hat{C}^{(22)*}_{l\Phi_1,f_2f_1}-\mathcal{P}_R\hat{C}^{(22)}_{l\Phi_1,f_1f_2}\right)\right)\\
&+\sqrt{2}vs_\beta c_\beta \not{p}_3\mathcal{P}_L\left(\hat{C}^{(11)[1]}_{\Phi l,f_1f_2}+\hat{C}^{(11)[3]}_{\Phi l,f_1f_2}-\hat{C}^{(22)[1]}_{\Phi l,f_1f_2}-\hat{C}^{(22)[3]}_{\Phi l,f_1f_2}\right)\\
&+\sqrt{2}vs_\beta c_\beta \not{p}_3\mathcal{P}_R\left(\hat{C}^{(11)}_{\Phi e,f_1f_2}-\hat{C}^{(22)}_{\Phi e,f_1f_2}\right)
\end{aligned}
\tag{C.9}
$$

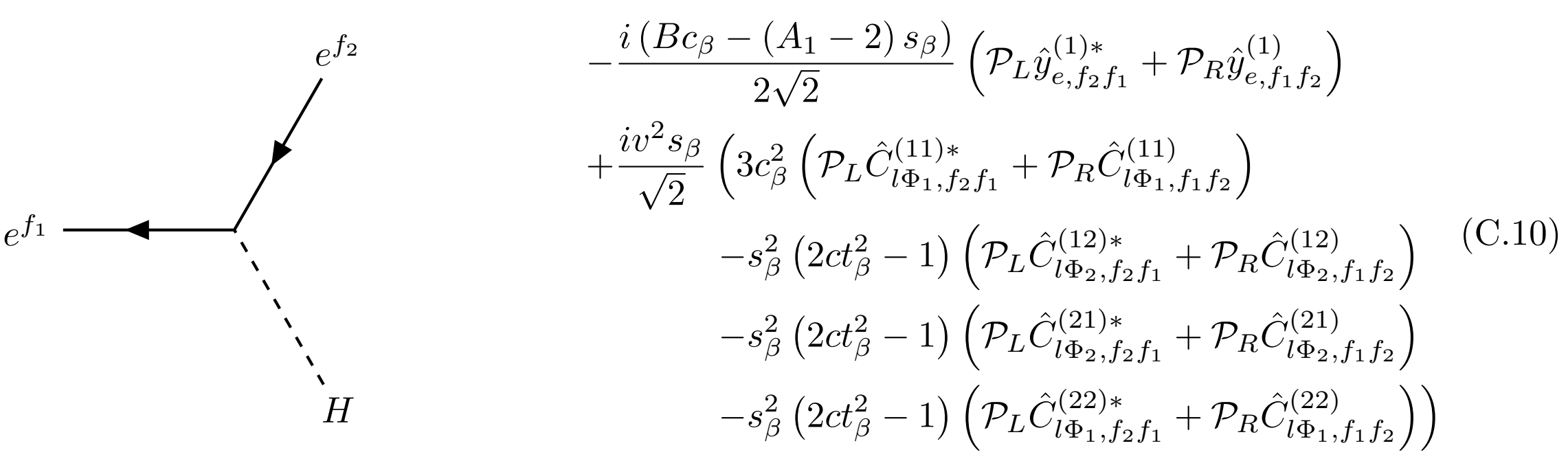

$$
\begin{aligned}
&-\frac{i\left(Bc_\beta-\left(A_1-2\right)s_\beta\right)}{2\sqrt{2}}\left(\mathcal{P}_L\hat{y}^{(1)*}_{e,f_2f_1}+\mathcal{P}_R\hat{y}^{(1)}_{e,f_1f_2}\right)\\
&+\frac{iv^2 s_\beta}{\sqrt{2}}\left(3c_\beta^2\left(\mathcal{P}_L\hat{C}^{(11)*}_{l\Phi_1,f_2f_1}+\mathcal{P}_R\hat{C}^{(11)}_{l\Phi_1,f_1f_2}\right)\right.\\
&\qquad-s_\beta^2\left(2ct_\beta^2-1\right)\left(\mathcal{P}_L\hat{C}^{(12)*}_{l\Phi_2,f_2f_1}+\mathcal{P}_R\hat{C}^{(12)}_{l\Phi_2,f_1f_2}\right)\\
&\qquad-s_\beta^2\left(2ct_\beta^2-1\right)\left(\mathcal{P}_L\hat{C}^{(21)*}_{l\Phi_2,f_2f_1}+\mathcal{P}_R\hat{C}^{(21)}_{l\Phi_2,f_1f_2}\right)\\
&\qquad\left.-s_\beta^2\left(2ct_\beta^2-1\right)\left(\mathcal{P}_L\hat{C}^{(22)*}_{l\Phi_1,f_2f_1}+\mathcal{P}_R\hat{C}^{(22)}_{l\Phi_1,f_1f_2}\right)\right)
\end{aligned}
\tag{C.10}
$$

$e^{f_2}$

$e^{f_1}$

$A_{\mu_3}$

$$
\begin{aligned}
&-\frac{i\hat{g}\delta_{f_1f_2}\hat{g}'\gamma^{\mu_3}}{\sqrt{\hat{g}'^2+\hat{g}^2}}\left(\frac{\hat{g}X_{WB}\hat{g}'}{\hat{g}'^2+\hat{g}^2}-1\right)\\
&+\frac{2vc_\beta p_{3\nu}}{\sqrt{\hat{g}'^2+\hat{g}^2}}\left(\hat{g}'\left(\hat{C}^{*}_{lW\Phi_1,f_2f_1}\sigma^{\mu_3\nu}\mathcal{P}_L+\hat{C}_{lW\Phi_1,f_1f_2}\sigma^{\mu_3\nu}\mathcal{P}_R\right)\right.\\
&\qquad\left.-\hat{g}\left(\hat{C}^{*}_{lB\Phi_1,f_2f_1}\sigma^{\mu_3\nu}\mathcal{P}_L+\hat{C}_{lB\Phi_1,f_1f_2}\sigma^{\mu_3\nu}\mathcal{P}_R\right)\right)
\end{aligned}
\tag{C.11}
$$

$e^{f_2}$

$\nu^{f_1}$

$W^+_{\mu_3}$

$$
\begin{aligned}
&-\frac{i\hat{g}U^*_{f_2f_1}}{\sqrt{2}}\left(\gamma^{\mu_3}\mathcal{P}_L\right)\\
&-2\sqrt{2}vc_\beta p_{3\nu}U^*_{g_1f_1}\sigma^{\mu_3\nu}\mathcal{P}_R\left(\hat{C}_{lW\Phi_1,g_1f_2}\right)\\
&-i\sqrt{2}\hat{g}v^2U^*_{g_1f_1}\gamma^{\mu_3}\mathcal{P}_L\left(c_\beta^2\hat{C}^{(11)[3]}_{\Phi l,g_1f_2}+s_\beta^2\hat{C}^{(22)[3]}_{\Phi l,g_1f_2}\right)
\end{aligned}
\tag{C.12}
$$

$\nu^{f_2}$

$\nu^{f_1}$

$Z_{\mu_3}$

$$
\begin{aligned}
&+\frac{i\delta_{f_1f_2}\gamma^{\mu_3}\gamma^5}{2\sqrt{\hat{g}'^2+\hat{g}^2}}\left(\hat{g}X_{WB}\hat{g}'+\hat{g}'^2+\hat{g}^2\right)\\
&-iv^2\sqrt{\hat{g}'^2+\hat{g}^2}U_{g_2f_2}U^*_{g_1f_1}\gamma^{\mu_3}\gamma^5\left(c_\beta^2\hat{C}^{(11)[1]}_{\Phi l,g_1g_2}-c_\beta^2\hat{C}^{(11)[3]}_{\Phi l,g_1g_2}\right.\\
&\qquad\left.+s_\beta^2\left(\hat{C}^{(22)[1]}_{\Phi l,g_1g_2}-\hat{C}^{(22)[3]}_{\Phi l,g_1g_2}\right)\right)
\end{aligned}
\tag{C.13}
$$

$e^{f_2}$

$e^{f_1}$

$Z_{\mu_3}$

$$
\begin{aligned}
&-\frac{i\delta_{f_1 f_2}}{2\left(\hat{g}'^2+\hat{g}^2\right)^{3/2}}\left(\left(\hat{g}'^2-\hat{g}^2\right)\left(-\hat{g}X_{WB}\hat{g}'+\hat{g}'^2+\hat{g}^2\right)\gamma^{\mu_3}\mathcal{P}_L\right.\\
&\qquad\left.+2\hat{g}'\left(\hat{g}'^3+\hat{g}^2\hat{g}'+\hat{g}^3X_{WB}\right)\gamma^{\mu_3}\mathcal{P}_R\right)\\
&+\frac{2vc_\beta p_{3\nu}}{\sqrt{\hat{g}'^2+\hat{g}^2}}\left(\hat{g}'\left(\hat{C}^*_{lB\Phi_1,f_2f_1}\sigma^{\mu_3\nu}\mathcal{P}_L+\hat{C}_{lB\Phi_1,f_1f_2}\sigma^{\mu_3\nu}\mathcal{P}_R\right)\right.\\
&\qquad\left.+\hat{g}\left(\hat{C}^*_{lW\Phi_1,f_2f_1}\sigma^{\mu_3\nu}\mathcal{P}_L+\hat{C}_{lW\Phi_1,f_1f_2}\sigma^{\mu_3\nu}\mathcal{P}_R\right)\right)\\
&+\frac{1}{2}iv^2\sqrt{\hat{g}'^2+\hat{g}^2}\gamma^{\mu_3}\mathcal{P}_L\left(2c_\beta^2\hat{C}^{(11)[1]}_{\Phi l,f_1f_2}+2c_\beta^2\hat{C}^{(11)[3]}_{\Phi l,f_1f_2}\right.\\
&\qquad\left.-\left(c_{2\beta}-1\right)\left(\hat{C}^{(22)[1]}_{\Phi l,f_1f_2}+\hat{C}^{(22)[3]}_{\Phi l,f_1f_2}\right)\right)\\
&+\frac{1}{2}iv^2\sqrt{\hat{g}'^2+\hat{g}^2}\gamma^{\mu_3}\mathcal{P}_R\left(\left(c_{2\beta}+1\right)\hat{C}^{(11)}_{\Phi e,f_1f_2}-\left(c_{2\beta}-1\right)\hat{C}^{(22)}_{\Phi e,f_1f_2}\right)
\end{aligned}
\tag{C.14}
$$

$e^{f_2}$

$(e^c)^{f_1}$ $G^+$

$G^+$

$$
-\frac{2im_{\nu_{g_1}}U^*_{f_1g_1}U^*_{f_2g_1}}{v^2}\left(\mathcal{P}_L\right)
\tag{C.15}
$$

$e^{f_2}$

$\nu^{f_1}$ $G^0$

$G^+$

$$
\begin{aligned}
&-\frac{m_{\nu_{f_1}}U^*_{f_2f_1}}{\sqrt{2}v^2}\left(\mathcal{P}_L\right)\\
&-\sqrt{2}\left(\not{p}_3\mathcal{P}_L-\not{p}_4\mathcal{P}_L\right)U^*_{g_1f_1}\left(c_\beta^2\hat{C}^{(11)[3]}_{\Phi l,g_1f_2}+s_\beta^2\hat{C}^{(22)[3]}_{\Phi l,g_1f_2}\right)
\end{aligned}
\tag{C.16}
$$

$e^{f_2}$, $\nu^{f_1}$, $G^+$, $h$

$$\begin{aligned}&+\frac{im_{\nu_{f_1}}U^*_{f_2f_1}}{\sqrt{2}v^2}\left(\mathcal{P}_L\right)\\&+i\sqrt{2}vc_\beta\mathcal{P}_R U^*_{g_1f_1}\left(c_\beta^2\hat{C}^{(11)}_{l\Phi_1,g_1f_2}+s_\beta^2\hat{C}^{(12)}_{l\Phi_2,g_1f_2}\right.\\&\qquad\qquad\left.+s_\beta^2\hat{C}^{(21)}_{l\Phi_2,g_1f_2}+s_\beta^2\hat{C}^{(22)}_{l\Phi_1,g_1f_2}\right)\\&+i\sqrt{2}\left(\not{p}_3\mathcal{P}_L-\not{p}_4\mathcal{P}_L\right)U^*_{g_1f_1}\left(c_\beta^2\hat{C}^{(11)[3]}_{\Phi l,g_1f_2}+s_\beta^2\hat{C}^{(22)[3]}_{\Phi l,g_1f_2}\right)\end{aligned}\tag{C.17}$$

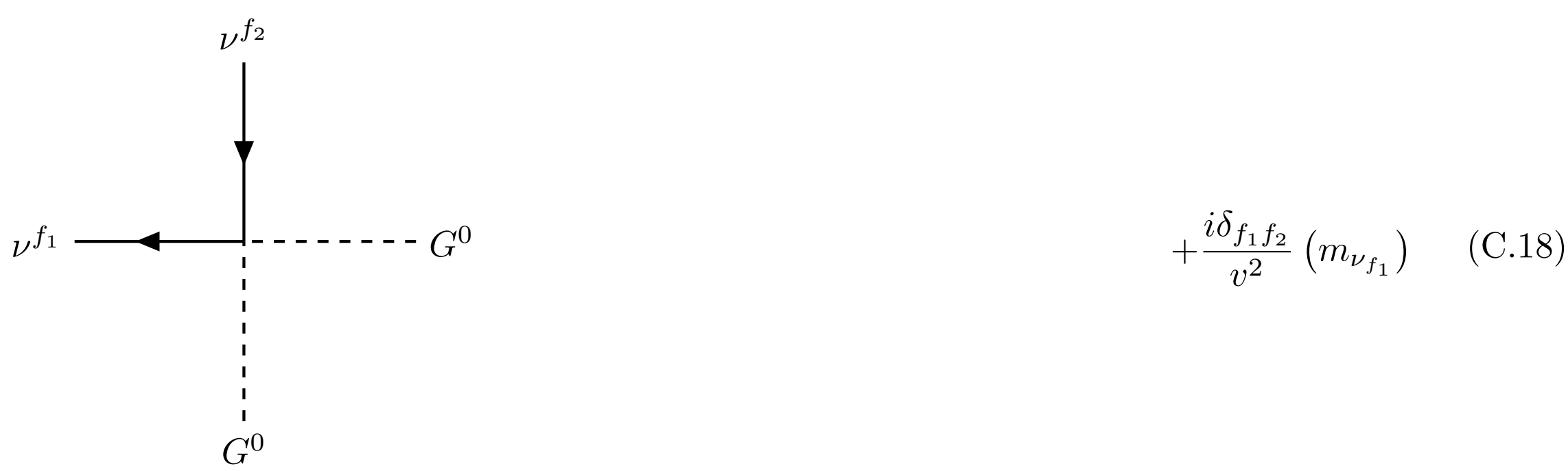

$$+\frac{i\delta_{f_1f_2}}{v^2}\left(m_{\nu_{f_1}}\right)\tag{C.18}$$

$\nu^{f_2}$, $\nu^{f_1}$, $G^0$, $h$

$$\begin{aligned}&-\frac{\delta_{f_1f_2}\gamma^5}{v^2}\left(m_{\nu_{f_1}}\right)\\&+U_{g_2f_2}U^*_{g_1f_1}\left(\not{p}_3\gamma^5-\not{p}_4\gamma^5\right)\left(c_\beta^2\hat{C}^{(11)[1]}_{\Phi l,g_1g_2}-c_\beta^2\hat{C}^{(11)[3]}_{\Phi l,g_1g_2}\right.\\&\qquad\qquad\left.+s_\beta^2\left(\hat{C}^{(22)[1]}_{\Phi l,g_1g_2}-\hat{C}^{(22)[3]}_{\Phi l,g_1g_2}\right)\right)\end{aligned}\tag{C.19}$$

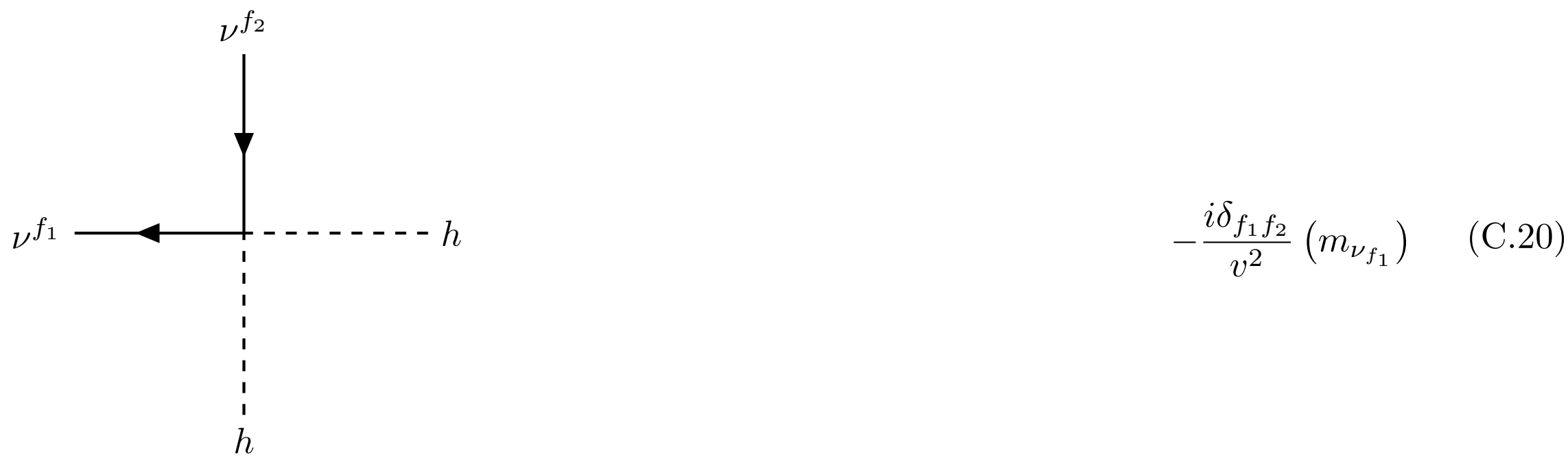

$$-\frac{i\delta_{f_1 f_2}}{v^2}\left(m_{\nu_{f_1}}\right) \tag{C.20}$$

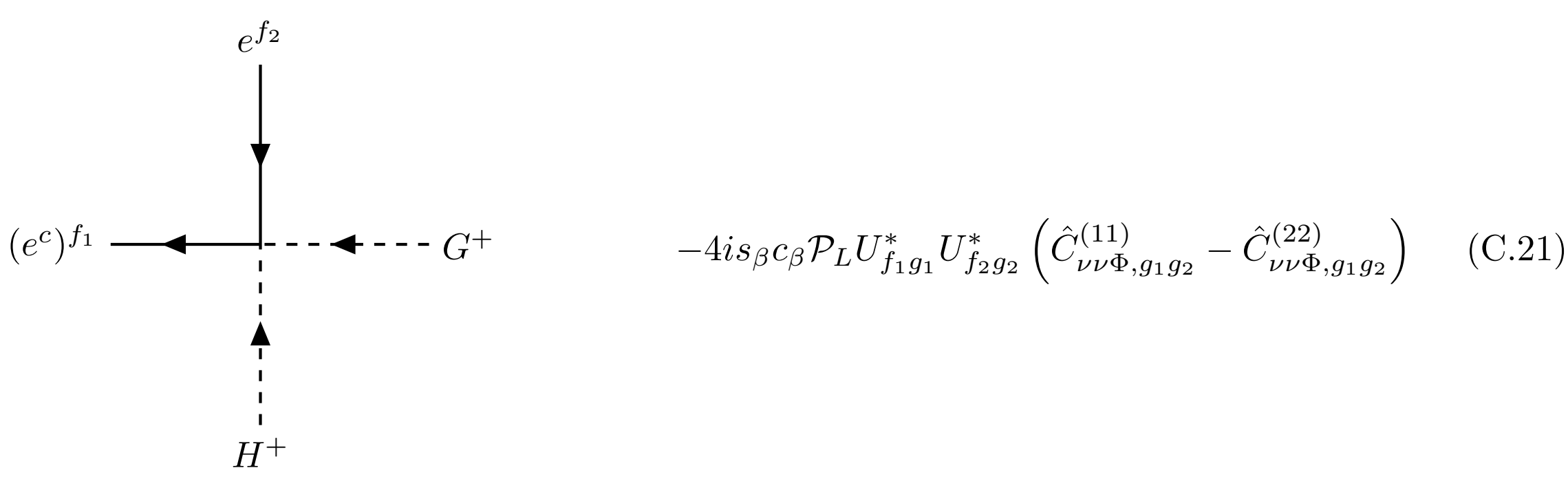

$$-4is_\beta c_\beta \mathcal{P}_L U^*_{f_1 g_1} U^*_{f_2 g_2}\left(\hat{C}^{(11)}_{\nu\nu\Phi,g_1 g_2} - \hat{C}^{(22)}_{\nu\nu\Phi,g_1 g_2}\right) \tag{C.21}$$

$e^{f_2}$, $\nu^{f_1}$, $A^0$, $G^+$

$$\begin{aligned}
&+\sqrt{2}s_\beta c_\beta \mathcal{P}_L U^*_{f_2 g_1}\left(\hat{C}^{(22)}_{\nu\nu\Phi,g_1 f_1} - \hat{C}^{(11)}_{\nu\nu\Phi,g_1 f_1}\right)\\
&-\frac{v s_\beta \mathcal{P}_R U^*_{g_1 f_1}}{\sqrt{2}}\left(s^2_\beta \hat{C}^{(12)}_{l\Phi_2,g_1 f_2} + c^2_\beta \hat{C}^{(12)}_{l\Phi_2,g_1 f_2}\right.\\
&\qquad\qquad \left. - s^2_\beta \hat{C}^{(21)}_{l\Phi_2,g_1 f_2} - c^2_\beta \hat{C}^{(21)}_{l\Phi_2,g_1 f_2}\right)\\
&+\sqrt{2}s_\beta c_\beta\left(\not{p}_3 \mathcal{P}_L - \not{p}_4 \mathcal{P}_L\right) U^*_{g_1 f_1}\left(\hat{C}^{(11)[3]}_{\Phi l,g_1 f_2} - \hat{C}^{(22)[3]}_{\Phi l,g_1 f_2}\right)
\end{aligned} \tag{C.22}$$

$e^{f_2}$, $\nu^{f_1}$, $G^+$, $H$

$$
\begin{aligned}
&-i\sqrt{2}s_\beta c_\beta \mathcal{P}_L U^*_{f_2 g_1}\left(\hat{C}^{(11)}_{\nu\nu\Phi,g_1f_1} - \hat{C}^{(22)}_{\nu\nu\Phi,g_1f_1}\right)\\
&+\frac{ivs_\beta \mathcal{P}_R U^*_{g_1f_1}}{\sqrt{2}}\Big(2c_\beta^2 \hat{C}^{(11)}_{l\Phi_1,g_1f_2} + s_\beta^2 \hat{C}^{(12)}_{l\Phi_2,g_1f_2}\\
&\qquad\qquad -c_\beta^2 \hat{C}^{(12)}_{l\Phi_2,g_1f_2} + s_\beta^2 \hat{C}^{(21)}_{l\Phi_2,g_1f_2}\\
&\qquad\qquad -c_\beta^2 \hat{C}^{(21)}_{l\Phi_2,g_1f_2} - 2c_\beta^2 \hat{C}^{(22)}_{l\Phi_1,g_1f_2}\Big)\\
&+i\sqrt{2}s_\beta c_\beta \left(\not p_3 \mathcal{P}_L - \not p_4 \mathcal{P}_L\right) U^*_{g_1f_1}\left(\hat{C}^{(11)[3]}_{\Phi l,g_1f_2} - \hat{C}^{(22)[3]}_{\Phi l,g_1f_2}\right)
\end{aligned}
\tag{C.23}
$$

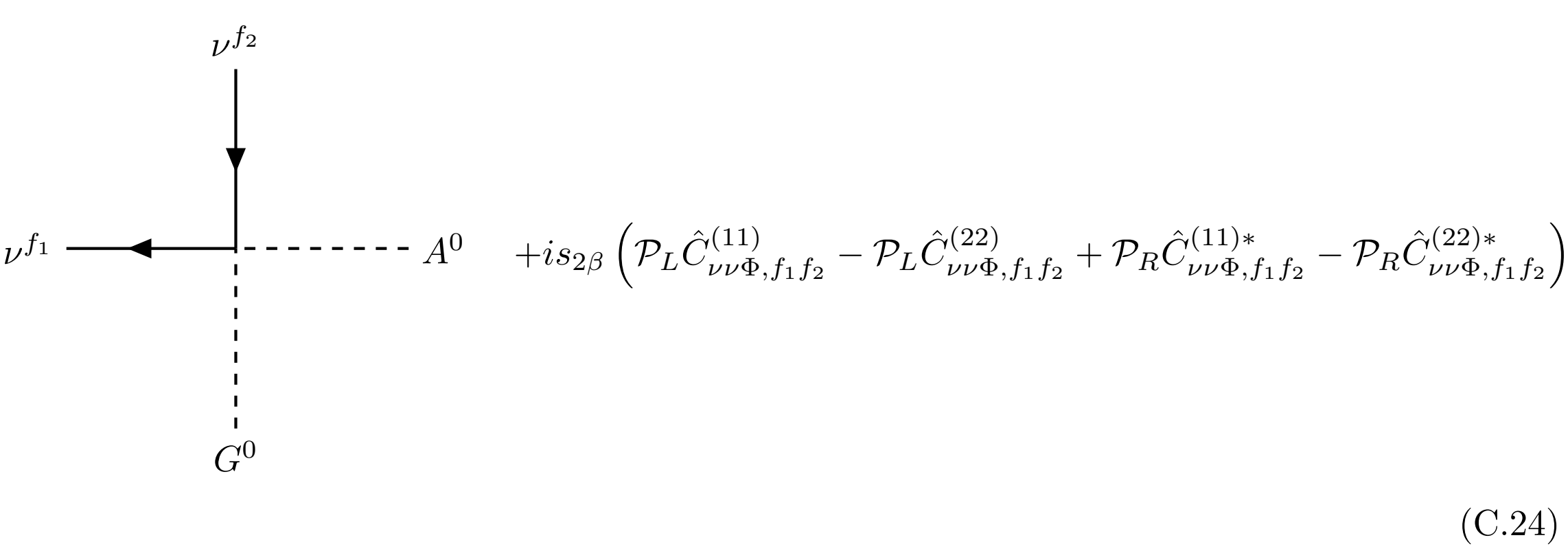

$$
+is_{2\beta}\left(\mathcal{P}_L \hat{C}^{(11)}_{\nu\nu\Phi,f_1f_2} - \mathcal{P}_L \hat{C}^{(22)}_{\nu\nu\Phi,f_1f_2} + \mathcal{P}_R \hat{C}^{(11)*}_{\nu\nu\Phi,f_1f_2} - \mathcal{P}_R \hat{C}^{(22)*}_{\nu\nu\Phi,f_1f_2}\right)
\tag{C.24}
$$

$\nu^{f_2}$, $\nu^{f_1}$, $G^0$, $H$

$$
\begin{aligned}
&+s_{2\beta}\left(-\mathcal{P}_L \hat{C}^{(11)}_{\nu\nu\Phi,f_1f_2} + \mathcal{P}_L \hat{C}^{(22)}_{\nu\nu\Phi,f_1f_2} + \mathcal{P}_R \hat{C}^{(11)*}_{\nu\nu\Phi,f_1f_2} - \mathcal{P}_R \hat{C}^{(22)*}_{\nu\nu\Phi,f_1f_2}\right)\\
&+s_\beta c_\beta U_{g_2f_2} U^*_{g_1f_1}\left(\not p_3 \gamma^5 - \not p_4 \gamma^5\right)\Big(\hat{C}^{(11)[1]}_{\Phi l,g_1g_2} - \hat{C}^{(11)[3]}_{\Phi l,g_1g_2}\\
&\qquad\qquad -\hat{C}^{(22)[1]}_{\Phi l,g_1g_2} + \hat{C}^{(22)[3]}_{\Phi l,g_1g_2}\Big)
\end{aligned}
\tag{C.25}
$$

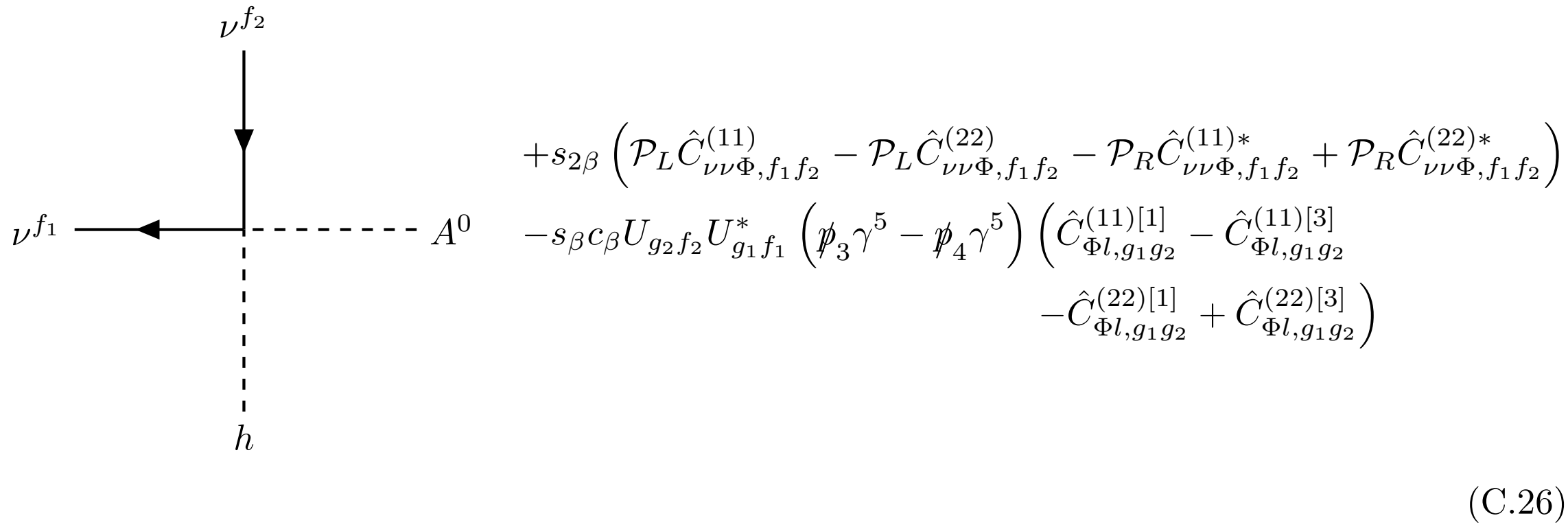

$$
\begin{aligned}
&+s_{2\beta}\left(\mathcal{P}_L\hat{C}^{(11)}_{\nu\nu\Phi,f_1f_2}-\mathcal{P}_L\hat{C}^{(22)}_{\nu\nu\Phi,f_1f_2}-\mathcal{P}_R\hat{C}^{(11)*}_{\nu\nu\Phi,f_1f_2}+\mathcal{P}_R\hat{C}^{(22)*}_{\nu\nu\Phi,f_1f_2}\right)\\
&-s_\beta c_\beta U_{g_2f_2}U^*_{g_1f_1}\left(\not{p}_3\gamma^5-\not{p}_4\gamma^5\right)\left(\hat{C}^{(11)[1]}_{\Phi l,g_1g_2}-\hat{C}^{(11)[3]}_{\Phi l,g_1g_2}\right.\\
&\qquad\left.-\hat{C}^{(22)[1]}_{\Phi l,g_1g_2}+\hat{C}^{(22)[3]}_{\Phi l,g_1g_2}\right)
\end{aligned}
\tag{C.26}
$$

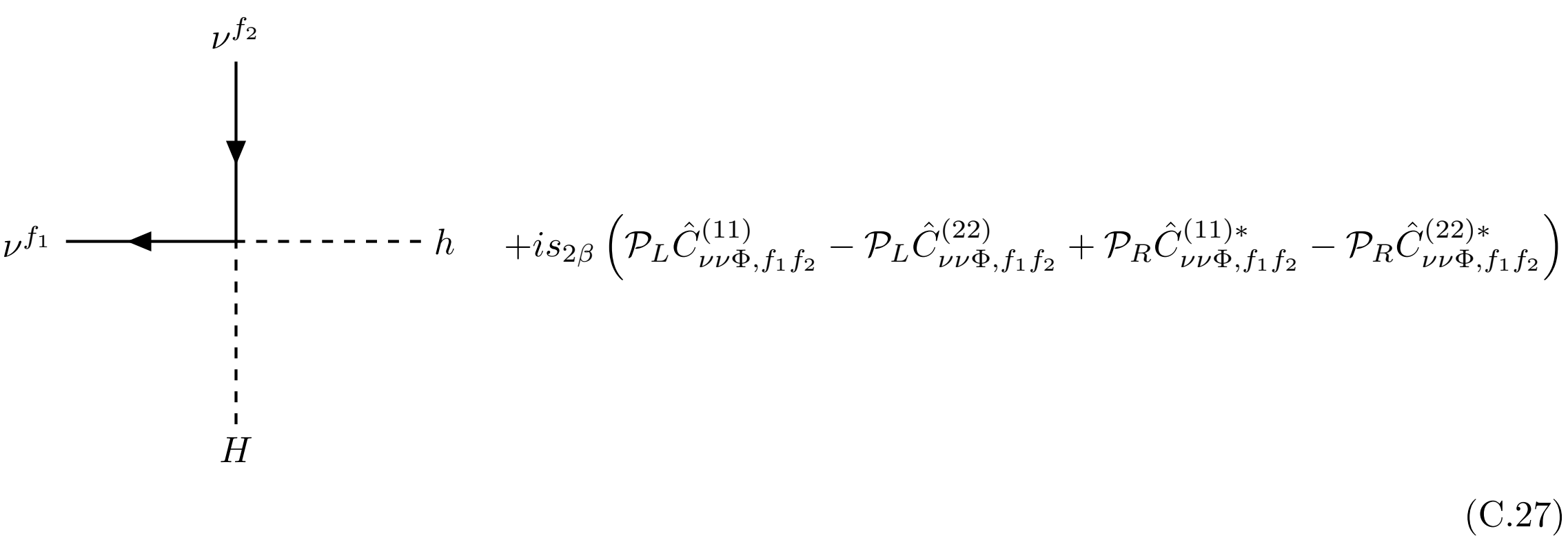

$$
+is_{2\beta}\left(\mathcal{P}_L\hat{C}^{(11)}_{\nu\nu\Phi,f_1f_2}-\mathcal{P}_L\hat{C}^{(22)}_{\nu\nu\Phi,f_1f_2}+\mathcal{P}_R\hat{C}^{(11)*}_{\nu\nu\Phi,f_1f_2}-\mathcal{P}_R\hat{C}^{(22)*}_{\nu\nu\Phi,f_1f_2}\right)
\tag{C.27}
$$

$$
\begin{aligned}
&+\sqrt{2}s_\beta c_\beta\mathcal{P}_L U^*_{f_2g_1}\left(\hat{C}^{(22)}_{\nu\nu\Phi,g_1f_1}-\hat{C}^{(11)}_{\nu\nu\Phi,g_1f_1}\right)\\
&+\sqrt{2}s_\beta c_\beta\left(\not{p}_3\mathcal{P}_L-\not{p}_4\mathcal{P}_L\right)U^*_{g_1f_1}\left(\hat{C}^{(11)[3]}_{\Phi l,g_1f_2}-\hat{C}^{(22)[3]}_{\Phi l,g_1f_2}\right)
\end{aligned}
\tag{C.28}
$$

$$
\begin{aligned}
&+i\sqrt{2}s_\beta c_\beta \mathcal{P}_L U^*_{f_2 g_1}\left(\hat{C}^{(11)}_{\nu\nu\Phi,g_1f_1} - \hat{C}^{(22)}_{\nu\nu\Phi,g_1f_1}\right)\\
&-i\sqrt{2}v s_\beta \mathcal{P}_R U^*_{g_1f_1}\left(c_\beta^2\hat{C}^{(11)}_{l\Phi_1,g_1f_2} - c_\beta^2\hat{C}^{(12)}_{l\Phi_2,g_1f_2}\right.\\
&\qquad\qquad\qquad\left. - c_\beta^2\hat{C}^{(21)}_{l\Phi_2,g_1f_2} + s_\beta^2\hat{C}^{(22)}_{l\Phi_1,g_1f_2}\right)\\
&+i\sqrt{2}s_\beta c_\beta\left(\not{p}_3\mathcal{P}_L - \not{p}_4\mathcal{P}_L\right)U^*_{g_1f_1}\left(\hat{C}^{(11)[3]}_{\Phi l,g_1f_2} - \hat{C}^{(22)[3]}_{\Phi l,g_1f_2}\right)
\end{aligned}
\tag{C.29}
$$

$$
+4i\mathcal{P}_L U^*_{f_1g_1}U^*_{f_2g_2}\left(s_\beta^2\hat{C}^{(11)}_{\nu\nu\Phi,g_1g_2} + c_\beta^2\hat{C}^{(22)}_{\nu\nu\Phi,g_1g_2}\right)
\tag{C.30}
$$

$$
\begin{aligned}
&+\sqrt{2}\mathcal{P}_L U^*_{f_2g_1}\left(s_\beta^2\hat{C}^{(11)}_{\nu\nu\Phi,g_1f_1} + c_\beta^2\hat{C}^{(22)}_{\nu\nu\Phi,g_1f_1}\right)\\
&-\frac{vc_\beta\mathcal{P}_R U^*_{g_1f_1}}{\sqrt{2}}\left(s_\beta^2\hat{C}^{(12)}_{l\Phi_2,g_1f_2} + c_\beta^2\hat{C}^{(12)}_{l\Phi_2,g_1f_2}\right.\\
&\qquad\qquad\qquad\left. - s_\beta^2\hat{C}^{(21)}_{l\Phi_2,g_1f_2} - c_\beta^2\hat{C}^{(21)}_{l\Phi_2,g_1f_2}\right)\\
&-\sqrt{2}\left(\not{p}_3\mathcal{P}_L - \not{p}_4\mathcal{P}_L\right)U^*_{g_1f_1}\left(s_\beta^2\hat{C}^{(11)[3]}_{\Phi l,g_1f_2} + c_\beta^2\hat{C}^{(22)[3]}_{\Phi l,g_1f_2}\right)
\end{aligned}
\tag{C.31}
$$

$$
\begin{aligned}
&+i\sqrt{2}\mathcal{P}_L U^*_{f_2 g_1}\left(s^2_\beta \hat{C}^{(11)}_{\nu\nu\Phi,g_1 f_1} + c^2_\beta \hat{C}^{(22)}_{\nu\nu\Phi,g_1 f_1}\right) \\
&-\frac{ivc_\beta \mathcal{P}_R U^*_{g_1 f_1}}{\sqrt{2}}\left(2s^2_\beta \hat{C}^{(11)}_{l\Phi_1,g_1 f_2} - s^2_\beta \hat{C}^{(12)}_{l\Phi_2,g_1 f_2}\right. \\
&\qquad\qquad\quad + c^2_\beta \hat{C}^{(12)}_{l\Phi_2,g_1 f_2} - s^2_\beta \hat{C}^{(21)}_{l\Phi_2,g_1 f_2} \\
&\qquad\qquad\quad \left. + c^2_\beta \hat{C}^{(21)}_{l\Phi_2,g_1 f_2} - 2s^2_\beta \hat{C}^{(22)}_{l\Phi_1,g_1 f_2}\right) \\
&+i\sqrt{2}\left(\not{p}_3 \mathcal{P}_L - \not{p}_4 \mathcal{P}_L\right) U^*_{g_1 f_1}\left(s^2_\beta \hat{C}^{(11)[3]}_{\Phi l,g_1 f_2} + c^2_\beta \hat{C}^{(22)[3]}_{\Phi l,g_1 f_2}\right)
\end{aligned}
\tag{C.32}
$$

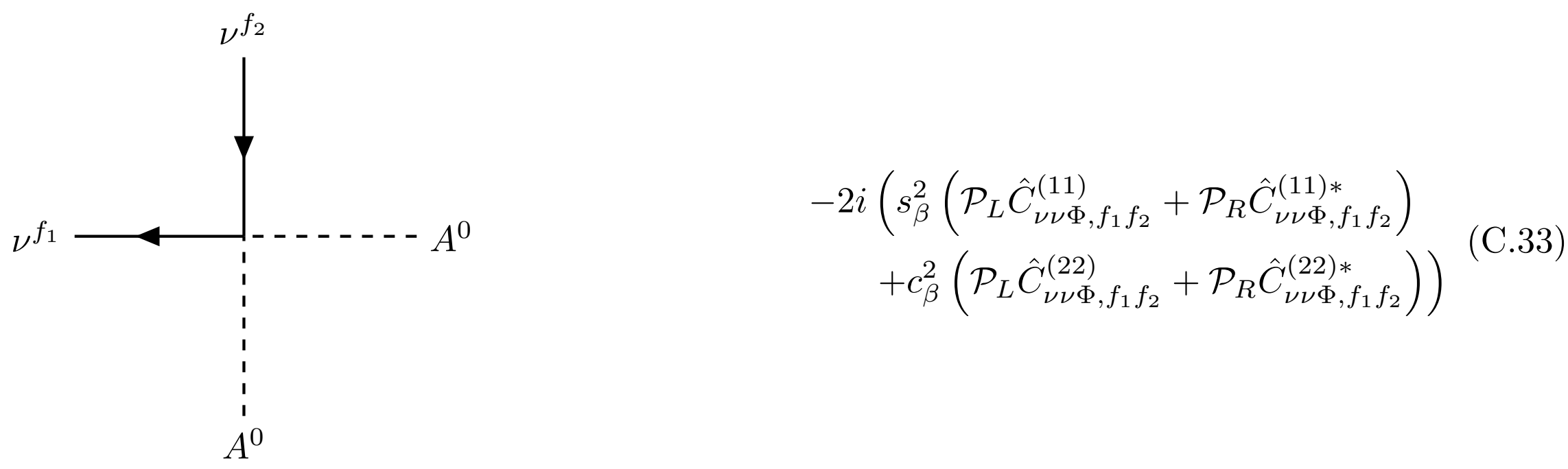

$$
\begin{aligned}
&-2i\left(s^2_\beta\left(\mathcal{P}_L \hat{C}^{(11)}_{\nu\nu\Phi,f_1 f_2} + \mathcal{P}_R \hat{C}^{(11)*}_{\nu\nu\Phi,f_1 f_2}\right)\right. \\
&\qquad \left. + c^2_\beta\left(\mathcal{P}_L \hat{C}^{(22)}_{\nu\nu\Phi,f_1 f_2} + \mathcal{P}_R \hat{C}^{(22)*}_{\nu\nu\Phi,f_1 f_2}\right)\right)
\end{aligned}
\tag{C.33}
$$

$$
\begin{aligned}
&+2\left(s^2_\beta\left(\mathcal{P}_L \hat{C}^{(11)}_{\nu\nu\Phi,f_1 f_2} - \mathcal{P}_R \hat{C}^{(11)*}_{\nu\nu\Phi,f_1 f_2}\right)\right. \\
&\qquad \left. + c^2_\beta\left(\mathcal{P}_L \hat{C}^{(22)}_{\nu\nu\Phi,f_1 f_2} - \mathcal{P}_R \hat{C}^{(22)*}_{\nu\nu\Phi,f_1 f_2}\right)\right) \\
&-U_{g_2 f_2} U^*_{g_1 f_1}\left(\not{p}_3 \gamma^5 - \not{p}_4 \gamma^5\right)\left(s^2_\beta\left(\hat{C}^{(11)[1]}_{\Phi l,g_1 g_2} - \hat{C}^{(11)[3]}_{\Phi l,g_1 g_2}\right)\right. \\
&\qquad\qquad\qquad\qquad \left. + c^2_\beta \hat{C}^{(22)[1]}_{\Phi l,g_1 g_2} - c^2_\beta \hat{C}^{(22)[3]}_{\Phi l,g_1 g_2}\right)
\end{aligned}
\tag{C.34}
$$

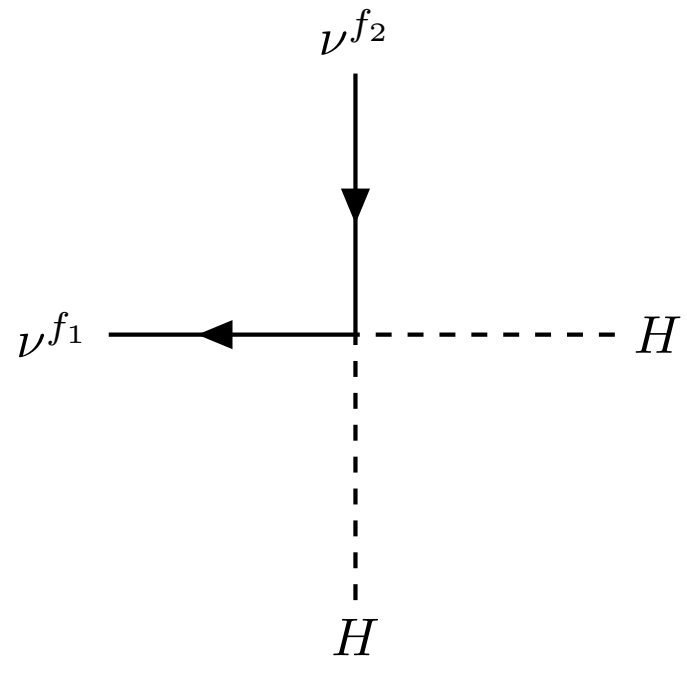

$$
\begin{aligned}
&+2i\left(s_\beta^2\left(\mathcal{P}_L\hat{C}^{(11)}_{\nu\nu\Phi,f_1f_2}+\mathcal{P}_R\hat{C}^{(11)*}_{\nu\nu\Phi,f_1f_2}\right)\right.\\
&\qquad\left.+c_\beta^2\left(\mathcal{P}_L\hat{C}^{(22)}_{\nu\nu\Phi,f_1f_2}+\mathcal{P}_R\hat{C}^{(22)*}_{\nu\nu\Phi,f_1f_2}\right)\right)
\end{aligned}
\tag{C.35}
$$

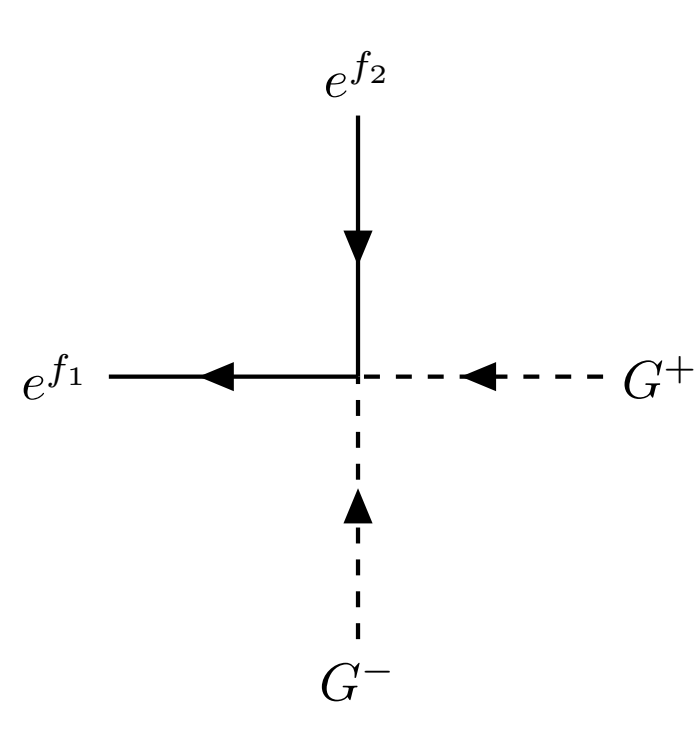

$$
\begin{aligned}
&+ivc_\beta\left(c_\beta^2\left(\mathcal{P}_L\hat{C}^{(11)*}_{l\Phi_1,f_2f_1}+\mathcal{P}_R\hat{C}^{(11)}_{l\Phi_1,f_1f_2}\right)\right.\\
&\qquad+s_\beta^2\left(\mathcal{P}_L\hat{C}^{(12)*}_{l\Phi_2,f_2f_1}+\mathcal{P}_R\hat{C}^{(12)}_{l\Phi_2,f_1f_2}\right)\\
&\qquad+s_\beta^2\left(\mathcal{P}_L\hat{C}^{(21)*}_{l\Phi_2,f_2f_1}+\mathcal{P}_R\hat{C}^{(21)}_{l\Phi_2,f_1f_2}\right)\\
&\qquad\left.+s_\beta^2\left(\mathcal{P}_L\hat{C}^{(22)*}_{l\Phi_1,f_2f_1}+\mathcal{P}_R\hat{C}^{(22)}_{l\Phi_1,f_1f_2}\right)\right)\\
&+i\left(\not{p}_3\mathcal{P}_L-\not{p}_4\mathcal{P}_L\right)\left(c_\beta^2\hat{C}^{(11)[1]}_{\Phi l,f_1f_2}-c_\beta^2\hat{C}^{(11)[3]}_{\Phi l,f_1f_2}\right.\\
&\qquad\qquad\left.+s_\beta^2\left(\hat{C}^{(22)[1]}_{\Phi l,f_1f_2}-\hat{C}^{(22)[3]}_{\Phi l,f_1f_2}\right)\right)\\
&+i\left(\not{p}_3\mathcal{P}_R-\not{p}_4\mathcal{P}_R\right)\left(c_\beta^2\hat{C}^{(11)}_{\Phi e,f_1f_2}+s_\beta^2\hat{C}^{(22)}_{\Phi e,f_1f_2}\right)
\end{aligned}
\tag{C.36}
$$

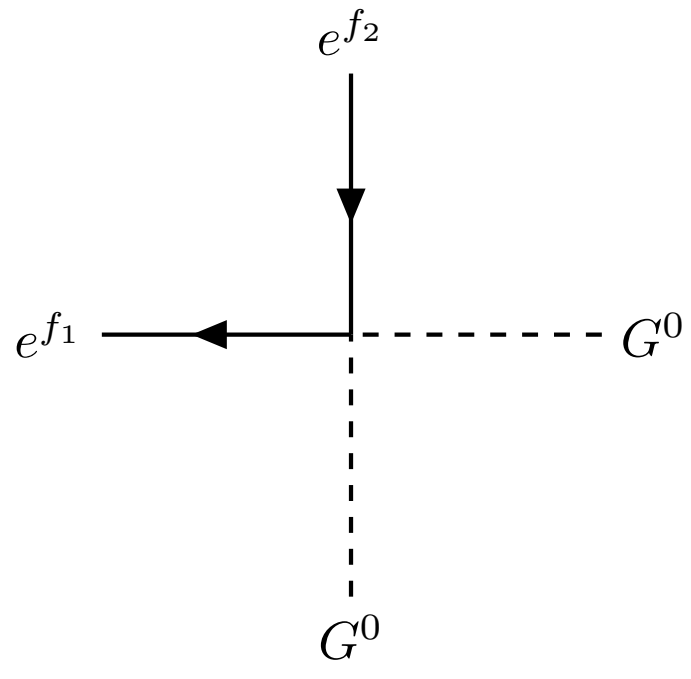

$$
\begin{aligned}
&+ivc_\beta\left(c_\beta^2\left(\mathcal{P}_L\hat{C}^{(11)*}_{l\Phi_1,f_2f_1}+\mathcal{P}_R\hat{C}^{(11)}_{l\Phi_1,f_1f_2}\right)\right.\\
&\qquad+s_\beta^2\left(\mathcal{P}_L\hat{C}^{(12)*}_{l\Phi_2,f_2f_1}+\mathcal{P}_R\hat{C}^{(12)}_{l\Phi_2,f_1f_2}\right)\\
&\qquad+s_\beta^2\left(\mathcal{P}_L\hat{C}^{(21)*}_{l\Phi_2,f_2f_1}+\mathcal{P}_R\hat{C}^{(21)}_{l\Phi_2,f_1f_2}\right)\\
&\qquad\left.+s_\beta^2\left(\mathcal{P}_L\hat{C}^{(22)*}_{l\Phi_1,f_2f_1}+\mathcal{P}_R\hat{C}^{(22)}_{l\Phi_1,f_1f_2}\right)\right)
\end{aligned}
\tag{C.37}
$$

$e^{f_2}$ $e^{f_1}$ $G^0$ $h$

$$
\begin{aligned}
&+vc_\beta\left(c_\beta^2\left(\mathcal{P}_L\hat{C}^{(11)*}_{l\Phi_1,f_2f_1}-\mathcal{P}_R\hat{C}^{(11)}_{l\Phi_1,f_1f_2}\right)\right.\\
&\qquad+s_\beta^2\left(\mathcal{P}_L\hat{C}^{(12)*}_{l\Phi_2,f_2f_1}-\mathcal{P}_R\hat{C}^{(12)}_{l\Phi_2,f_1f_2}\right)\\
&\qquad+s_\beta^2\left(\mathcal{P}_L\hat{C}^{(21)*}_{l\Phi_2,f_2f_1}-\mathcal{P}_R\hat{C}^{(21)}_{l\Phi_2,f_1f_2}\right)\\
&\qquad\left.+s_\beta^2\left(\mathcal{P}_L\hat{C}^{(22)*}_{l\Phi_1,f_2f_1}-\mathcal{P}_R\hat{C}^{(22)}_{l\Phi_1,f_1f_2}\right)\right)\\
&+\left(\not{p}_4\mathcal{P}_L-\not{p}_3\mathcal{P}_L\right)\left(c_\beta^2\hat{C}^{(11)[1]}_{\Phi l,f_1f_2}+c_\beta^2\hat{C}^{(11)[3]}_{\Phi l,f_1f_2}\right.\\
&\qquad\qquad\left.+s_\beta^2\left(\hat{C}^{(22)[1]}_{\Phi l,f_1f_2}+\hat{C}^{(22)[3]}_{\Phi l,f_1f_2}\right)\right)\\
&+\left(\not{p}_4\mathcal{P}_R-\not{p}_3\mathcal{P}_R\right)\left(c_\beta^2\hat{C}^{(11)}_{\Phi e,f_1f_2}+s_\beta^2\hat{C}^{(22)}_{\Phi e,f_1f_2}\right)
\end{aligned}
\tag{C.38}
$$

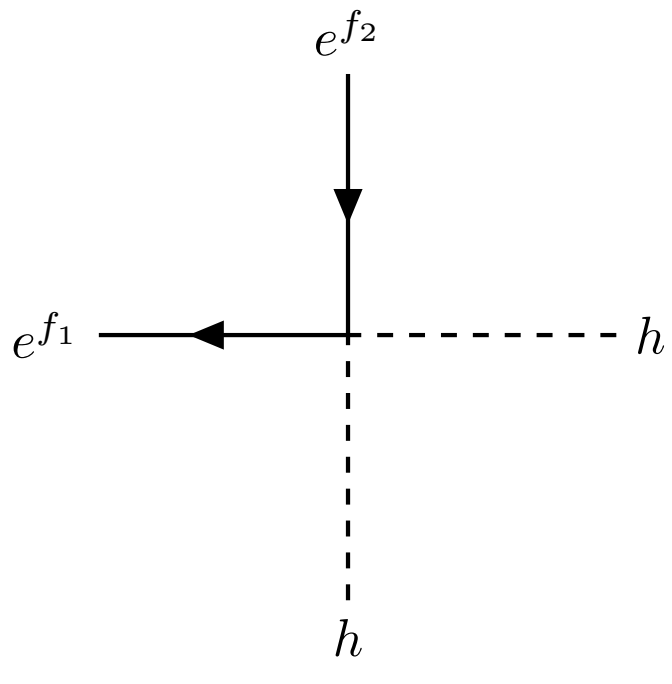

$$
\begin{aligned}
&+3ivc_\beta\left(c_\beta^2\left(\mathcal{P}_L\hat{C}^{(11)*}_{l\Phi_1,f_2f_1}+\mathcal{P}_R\hat{C}^{(11)}_{l\Phi_1,f_1f_2}\right)\right.\\
&\qquad+s_\beta^2\left(\mathcal{P}_L\hat{C}^{(12)*}_{l\Phi_2,f_2f_1}+\mathcal{P}_R\hat{C}^{(12)}_{l\Phi_2,f_1f_2}\right)\\
&\qquad+s_\beta^2\left(\mathcal{P}_L\hat{C}^{(21)*}_{l\Phi_2,f_2f_1}+\mathcal{P}_R\hat{C}^{(21)}_{l\Phi_2,f_1f_2}\right)\\
&\qquad\left.+s_\beta^2\left(\mathcal{P}_L\hat{C}^{(22)*}_{l\Phi_1,f_2f_1}+\mathcal{P}_R\hat{C}^{(22)}_{l\Phi_1,f_1f_2}\right)\right)
\end{aligned}
\tag{C.39}
$$

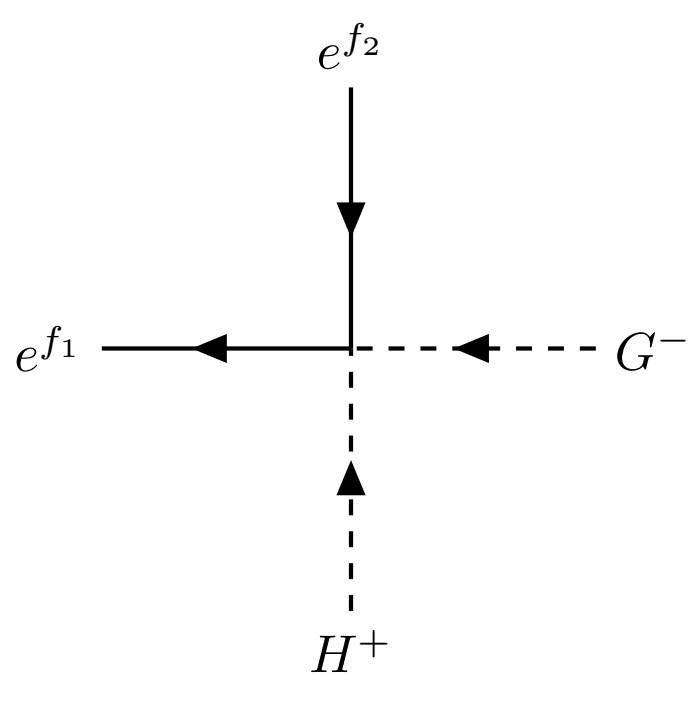

$$
\begin{aligned}
&-ivs_\beta\left(c_\beta^2\left(\mathcal{P}_L\hat{C}^{(11)*}_{l\Phi_1,f_2f_1}+\mathcal{P}_R\hat{C}^{(11)}_{l\Phi_1,f_1f_2}\right)\right.\\
&\qquad-c_\beta^2\left(\mathcal{P}_L\hat{C}^{(22)*}_{l\Phi_1,f_2f_1}+\mathcal{P}_R\hat{C}^{(22)}_{l\Phi_1,f_1f_2}\right)\\
&\qquad+s_\beta^2\mathcal{P}_L\hat{C}^{(12)*}_{l\Phi_2,f_2f_1}-c_\beta^2\mathcal{P}_L\hat{C}^{(21)*}_{l\Phi_2,f_2f_1}\\
&\qquad\left.-c_\beta^2\mathcal{P}_R\hat{C}^{(12)}_{l\Phi_2,f_1f_2}+s_\beta^2\mathcal{P}_R\hat{C}^{(21)}_{l\Phi_2,f_1f_2}\right)\\
&+\frac{1}{2}is_{2\beta}\left(\not{p}_3\mathcal{P}_L-\not{p}_4\mathcal{P}_L\right)\left(\hat{C}^{(11)[1]}_{\Phi l,f_1f_2}-\hat{C}^{(11)[3]}_{\Phi l,f_1f_2}-\hat{C}^{(22)[1]}_{\Phi l,f_1f_2}+\hat{C}^{(22)[3]}_{\Phi l,f_1f_2}\right)\\
&+\frac{1}{2}is_{2\beta}\left(\not{p}_3\mathcal{P}_R-\not{p}_4\mathcal{P}_R\right)\left(\hat{C}^{(11)}_{\Phi e,f_1f_2}-\hat{C}^{(22)}_{\Phi e,f_1f_2}\right)
\end{aligned}
\tag{C.40}
$$

$$
\begin{aligned}
&-ivs_\beta\left(c_\beta^2\left(\mathcal{P}_L\hat{C}^{(11)*}_{l\Phi_1,f_2f_1}+\mathcal{P}_R\hat{C}^{(11)}_{l\Phi_1,f_1f_2}\right)\right.\\
&\qquad+s_\beta^2\left(\mathcal{P}_L\hat{C}^{(12)*}_{l\Phi_2,f_2f_1}+\mathcal{P}_R\hat{C}^{(12)}_{l\Phi_2,f_1f_2}\right)\\
&\qquad-c_\beta^2\left(\mathcal{P}_L\hat{C}^{(21)*}_{l\Phi_2,f_2f_1}+\mathcal{P}_R\hat{C}^{(21)}_{l\Phi_2,f_1f_2}\right)\\
&\qquad\left.-c_\beta^2\left(\mathcal{P}_L\hat{C}^{(22)*}_{l\Phi_1,f_2f_1}+\mathcal{P}_R\hat{C}^{(22)}_{l\Phi_1,f_1f_2}\right)\right)
\end{aligned}
\tag{C.41}
$$

$$
\begin{aligned}
&+vs_\beta\left(-c_\beta^2\left(\mathcal{P}_L\hat{C}^{(11)*}_{l\Phi_1,f_2f_1}-\mathcal{P}_R\hat{C}^{(11)}_{l\Phi_1,f_1f_2}\right)\right.\\
&\qquad+s_\beta^2\left(2ct_\beta^2+1\right)\left(\mathcal{P}_L\hat{C}^{(12)*}_{l\Phi_2,f_2f_1}-\mathcal{P}_R\hat{C}^{(12)}_{l\Phi_2,f_1f_2}\right)\\
&\qquad-s_\beta^2\left(\mathcal{P}_L\hat{C}^{(21)*}_{l\Phi_2,f_2f_1}-\mathcal{P}_R\hat{C}^{(21)}_{l\Phi_2,f_1f_2}\right)\\
&\qquad\left.-s_\beta^2\left(\mathcal{P}_L\hat{C}^{(22)*}_{l\Phi_1,f_2f_1}-\mathcal{P}_R\hat{C}^{(22)}_{l\Phi_1,f_1f_2}\right)\right)\\
&+\frac{1}{2}s_{2\beta}\left(\not{p}_3\mathcal{P}_L-\not{p}_4\mathcal{P}_L\right)\left(\hat{C}^{(11)[1]}_{\Phi l,f_1f_2}+\hat{C}^{(11)[3]}_{\Phi l,f_1f_2}-\hat{C}^{(22)[1]}_{\Phi l,f_1f_2}-\hat{C}^{(22)[3]}_{\Phi l,f_1f_2}\right)\\
&+\frac{1}{2}s_{2\beta}\left(\not{p}_3\mathcal{P}_R-\not{p}_4\mathcal{P}_R\right)\left(\hat{C}^{(11)}_{\Phi e,f_1f_2}-\hat{C}^{(22)}_{\Phi e,f_1f_2}\right)
\end{aligned}
\tag{C.42}
$$

$$
\begin{aligned}
&+vs_\beta\left(c_\beta^2\left(\mathcal{P}_L\hat{C}^{(11)*}_{l\Phi_1,f_2f_1}-\mathcal{P}_R\hat{C}^{(11)}_{l\Phi_1,f_1f_2}\right)\right.\\
&\qquad+s_\beta^2\left(\mathcal{P}_L\hat{C}^{(12)*}_{l\Phi_2,f_2f_1}-\mathcal{P}_R\hat{C}^{(12)}_{l\Phi_2,f_1f_2}\right)\\
&\qquad-c_\beta^2\left(\mathcal{P}_L\hat{C}^{(21)*}_{l\Phi_2,f_2f_1}-\mathcal{P}_R\hat{C}^{(21)}_{l\Phi_2,f_1f_2}\right)\\
&\qquad\left.-c_\beta^2\left(\mathcal{P}_L\hat{C}^{(22)*}_{l\Phi_1,f_2f_1}-\mathcal{P}_R\hat{C}^{(22)}_{l\Phi_1,f_1f_2}\right)\right)\\
&-\frac{1}{2}s_{2\beta}\left(\not{p}_3\mathcal{P}_L-\not{p}_4\mathcal{P}_L\right)\left(\hat{C}^{(11)[1]}_{\Phi l,f_1f_2}+\hat{C}^{(11)[3]}_{\Phi l,f_1f_2}-\hat{C}^{(22)[1]}_{\Phi l,f_1f_2}-\hat{C}^{(22)[3]}_{\Phi l,f_1f_2}\right)\\
&-\frac{1}{2}s_{2\beta}\left(\not{p}_3\mathcal{P}_R-\not{p}_4\mathcal{P}_R\right)\left(\hat{C}^{(11)}_{\Phi e,f_1f_2}-\hat{C}^{(22)}_{\Phi e,f_1f_2}\right)
\end{aligned}
\tag{C.43}
$$

$e^{f_2}$, $e^{f_1}$, $h$, $H$

$$
\begin{aligned}
&+ivs_\beta \Big(3c_\beta^2 \left(\mathcal{P}_L \hat{C}^{(11)*}_{l\Phi_1,f_2f_1} + \mathcal{P}_R \hat{C}^{(11)}_{l\Phi_1,f_1f_2}\right) \\
&\qquad -s_\beta^2 \left(2ct_\beta^2 - 1\right)\left(\mathcal{P}_L \hat{C}^{(12)*}_{l\Phi_2,f_2f_1} + \mathcal{P}_R \hat{C}^{(12)}_{l\Phi_2,f_1f_2}\right) \\
&\qquad -s_\beta^2 \left(2ct_\beta^2 - 1\right)\left(\mathcal{P}_L \hat{C}^{(21)*}_{l\Phi_2,f_2f_1} + \mathcal{P}_R \hat{C}^{(21)}_{l\Phi_2,f_1f_2}\right) \\
&\qquad -s_\beta^2 \left(2ct_\beta^2 - 1\right)\left(\mathcal{P}_L \hat{C}^{(22)*}_{l\Phi_1,f_2f_1} + \mathcal{P}_R \hat{C}^{(22)}_{l\Phi_1,f_1f_2}\right)\Big)
\end{aligned}
\tag{C.44}
$$

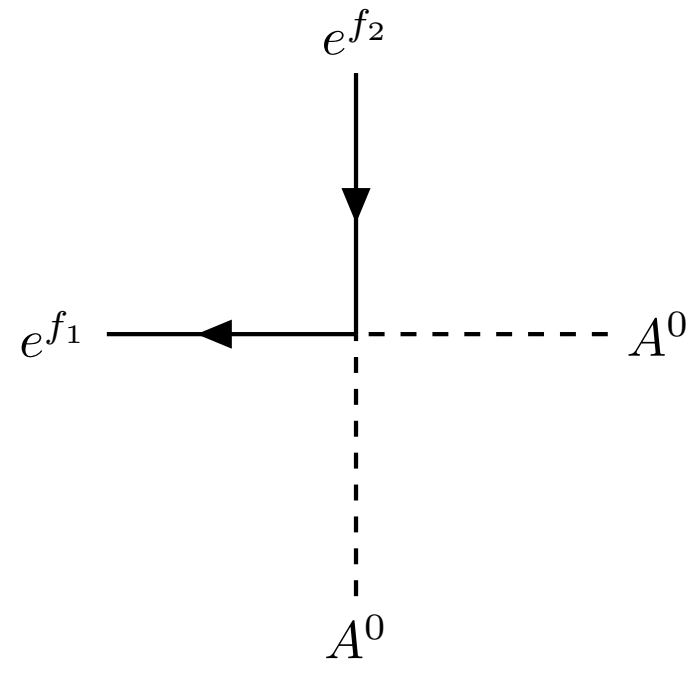

$$
\begin{aligned}
&+ivc_\beta \Big(s_\beta^2 \left(\mathcal{P}_L \hat{C}^{(11)*}_{l\Phi_1,f_2f_1} + \mathcal{P}_R \hat{C}^{(11)}_{l\Phi_1,f_1f_2}\right) \\
&\qquad -c_\beta^2 \left(2t_\beta^2 + 1\right)\left(\mathcal{P}_L \hat{C}^{(12)*}_{l\Phi_2,f_2f_1} + \mathcal{P}_R \hat{C}^{(12)}_{l\Phi_2,f_1f_2}\right) \\
&\qquad +c_\beta^2 \left(\mathcal{P}_L \hat{C}^{(21)*}_{l\Phi_2,f_2f_1} + \mathcal{P}_R \hat{C}^{(21)}_{l\Phi_2,f_1f_2}\right) \\
&\qquad +c_\beta^2 \left(\mathcal{P}_L \hat{C}^{(22)*}_{l\Phi_1,f_2f_1} + \mathcal{P}_R \hat{C}^{(22)}_{l\Phi_1,f_1f_2}\right)\Big)
\end{aligned}
\tag{C.45}
$$

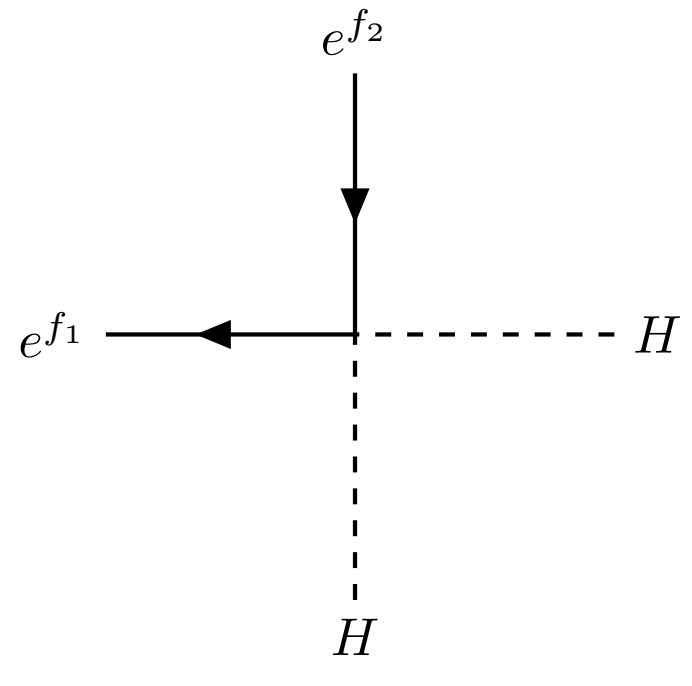

$$
\begin{aligned}
&+ivc_\beta \Big(3s_\beta^2 \left(\mathcal{P}_L \hat{C}^{(11)*}_{l\Phi_1,f_2f_1} + \mathcal{P}_R \hat{C}^{(11)}_{l\Phi_1,f_1f_2}\right) \\
&\qquad -c_\beta^2 \left(2t_\beta^2 - 1\right)\left(\mathcal{P}_L \hat{C}^{(12)*}_{l\Phi_2,f_2f_1} + \mathcal{P}_R \hat{C}^{(12)}_{l\Phi_2,f_1f_2}\right) \\
&\qquad -c_\beta^2 \left(2t_\beta^2 - 1\right)\left(\mathcal{P}_L \hat{C}^{(21)*}_{l\Phi_2,f_2f_1} + \mathcal{P}_R \hat{C}^{(21)}_{l\Phi_2,f_1f_2}\right) \\
&\qquad -c_\beta^2 \left(2t_\beta^2 - 1\right)\left(\mathcal{P}_L \hat{C}^{(22)*}_{l\Phi_1,f_2f_1} + \mathcal{P}_R \hat{C}^{(22)}_{l\Phi_1,f_1f_2}\right)\Big)
\end{aligned}
\tag{C.46}
$$

$e^{f_2}$

$e^{f_1}$ $H^+$

$H^-$

$$
\begin{aligned}
&+ivc_\beta\left(s_\beta^2\left(\mathcal{P}_L\hat{C}^{(11)*}_{l\Phi_1,f_2f_1}+\mathcal{P}_R\hat{C}^{(11)}_{l\Phi_1,f_1f_2}\right)\right.\\
&\qquad-s_\beta^2\left(\mathcal{P}_L\hat{C}^{(12)*}_{l\Phi_2,f_2f_1}+\mathcal{P}_R\hat{C}^{(12)}_{l\Phi_2,f_1f_2}\right)\\
&\qquad-s_\beta^2\left(\mathcal{P}_L\hat{C}^{(21)*}_{l\Phi_2,f_2f_1}+\mathcal{P}_R\hat{C}^{(21)}_{l\Phi_2,f_1f_2}\right)\\
&\qquad\left.+c_\beta^2\left(\mathcal{P}_L\hat{C}^{(22)*}_{l\Phi_1,f_2f_1}+\mathcal{P}_R\hat{C}^{(22)}_{l\Phi_1,f_1f_2}\right)\right)\\
&+i\left(\not{p}_3\mathcal{P}_L-\not{p}_4\mathcal{P}_L\right)\left(s_\beta^2\left(\hat{C}^{(11)[1]}_{\Phi l,f_1f_2}-\hat{C}^{(11)[3]}_{\Phi l,f_1f_2}\right)\right.\\
&\qquad\qquad\left.+c_\beta^2\hat{C}^{(22)[1]}_{\Phi l,f_1f_2}-c_\beta^2\hat{C}^{(22)[3]}_{\Phi l,f_1f_2}\right)\\
&+i\left(\not{p}_3\mathcal{P}_R-\not{p}_4\mathcal{P}_R\right)\left(s_\beta^2\hat{C}^{(11)}_{\Phi e,f_1f_2}+c_\beta^2\hat{C}^{(22)}_{\Phi e,f_1f_2}\right)
\end{aligned}
\tag{C.47}
$$

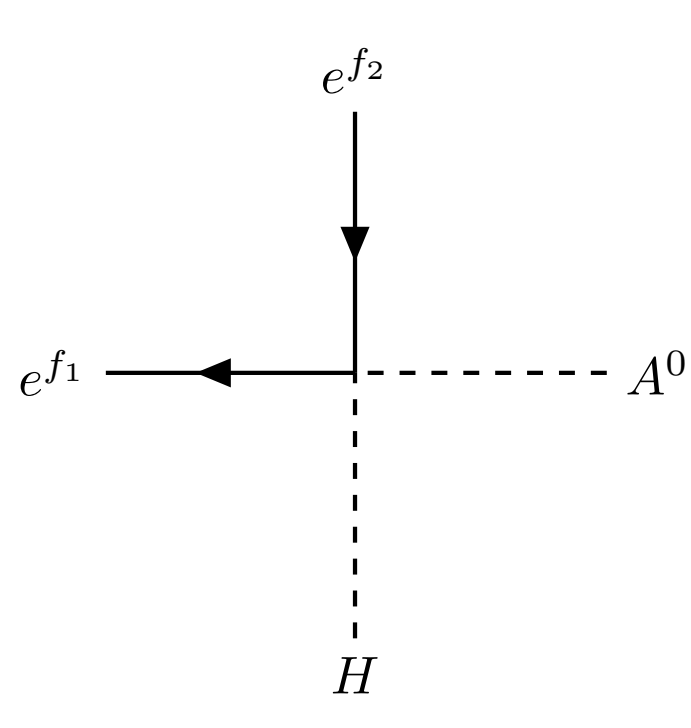

$$
\begin{aligned}
&+vc_\beta\left(-s_\beta^2\left(\mathcal{P}_L\hat{C}^{(11)*}_{l\Phi_1,f_2f_1}-\mathcal{P}_R\hat{C}^{(11)}_{l\Phi_1,f_1f_2}\right)\right.\\
&\qquad-c_\beta^2\left(\mathcal{P}_L\hat{C}^{(12)*}_{l\Phi_2,f_2f_1}-\mathcal{P}_R\hat{C}^{(12)}_{l\Phi_2,f_1f_2}\right)\\
&\qquad+s_\beta^2\left(\mathcal{P}_L\hat{C}^{(21)*}_{l\Phi_2,f_2f_1}-\mathcal{P}_R\hat{C}^{(21)}_{l\Phi_2,f_1f_2}\right)\\
&\qquad\left.+s_\beta^2\left(\mathcal{P}_L\hat{C}^{(22)*}_{l\Phi_1,f_2f_1}-\mathcal{P}_R\hat{C}^{(22)}_{l\Phi_1,f_1f_2}\right)\right)\\
&+\left(\not{p}_3\mathcal{P}_L-\not{p}_4\mathcal{P}_L\right)\left(s_\beta^2\left(\hat{C}^{(11)[1]}_{\Phi l,f_1f_2}+\hat{C}^{(11)[3]}_{\Phi l,f_1f_2}\right)\right.\\
&\qquad\qquad\left.+c_\beta^2\hat{C}^{(22)[1]}_{\Phi l,f_1f_2}+c_\beta^2\hat{C}^{(22)[3]}_{\Phi l,f_1f_2}\right)\\
&+\left(\not{p}_3\mathcal{P}_R-\not{p}_4\mathcal{P}_R\right)\left(s_\beta^2\hat{C}^{(11)}_{\Phi e,f_1f_2}+c_\beta^2\hat{C}^{(22)}_{\Phi e,f_1f_2}\right)
\end{aligned}
\tag{C.48}
$$

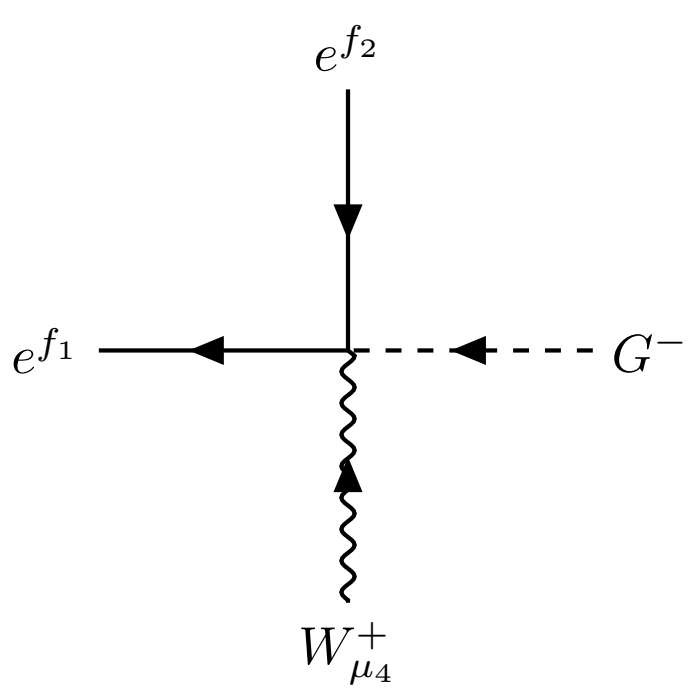

$$
\begin{aligned}
&-2\sqrt{2}c_\beta p_{4\nu}\sigma^{\mu_4\nu}\mathcal{P}_L\left(\hat{C}^{\;*}_{lW\Phi_1,f_2f_1}\right)\\
&-i\sqrt{2}\hat{g}v\gamma^{\mu_4}\mathcal{P}_L\left(c_\beta^2\hat{C}^{(11)[1]}_{\Phi l,f_1f_2}+s_\beta^2\hat{C}^{(22)[1]}_{\Phi l,f_1f_2}\right)\\
&-i\sqrt{2}\hat{g}v\gamma^{\mu_4}\mathcal{P}_R\left(c_\beta^2\hat{C}^{(11)}_{\Phi e,f_1f_2}+s_\beta^2\hat{C}^{(22)}_{\Phi e,f_1f_2}\right)
\end{aligned}
\tag{C.49}
$$

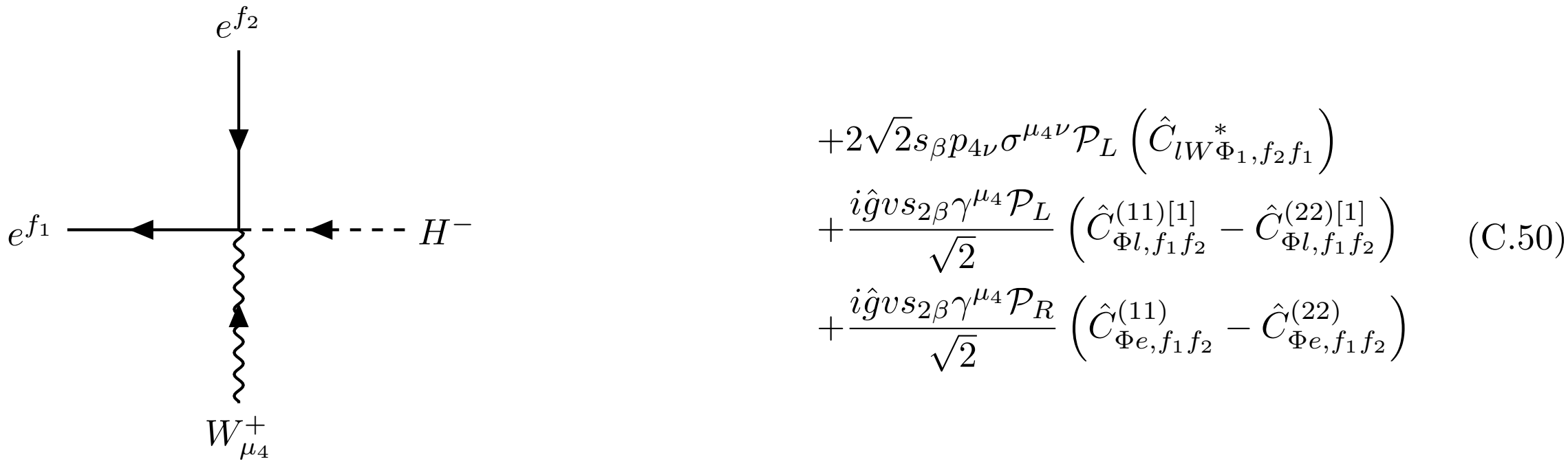

$$
\begin{aligned}
&+2\sqrt{2}s_\beta p_{4\nu}\sigma^{\mu_4\nu}\mathcal{P}_L\left(\hat{C}^*_{lW\Phi_1,f_2f_1}\right)\\
&+\frac{i\hat{g}vs_{2\beta}\gamma^{\mu_4}\mathcal{P}_L}{\sqrt{2}}\left(\hat{C}^{(11)[1]}_{\Phi l,f_1f_2}-\hat{C}^{(22)[1]}_{\Phi l,f_1f_2}\right)\\
&+\frac{i\hat{g}vs_{2\beta}\gamma^{\mu_4}\mathcal{P}_R}{\sqrt{2}}\left(\hat{C}^{(11)}_{\Phi e,f_1f_2}-\hat{C}^{(22)}_{\Phi e,f_1f_2}\right)
\end{aligned}
\tag{C.50}
$$

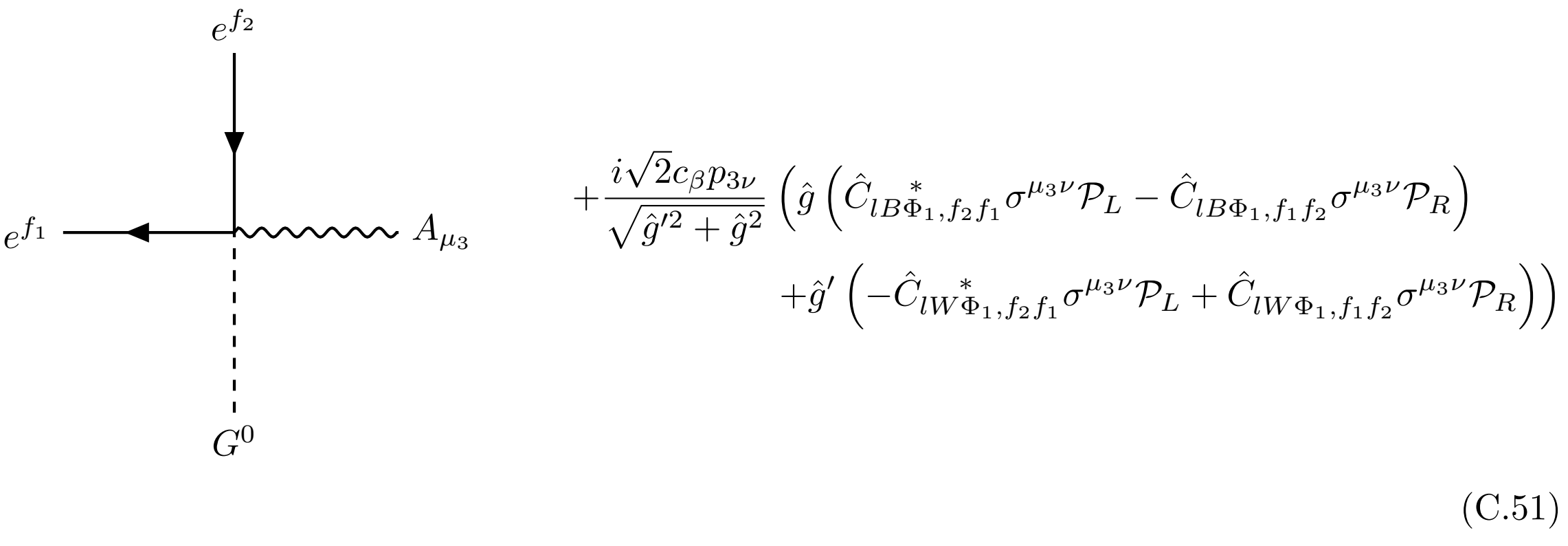

$$
\begin{aligned}
&+\frac{i\sqrt{2}c_\beta p_{3\nu}}{\sqrt{\hat{g}'^2+\hat{g}^2}}\left(\hat{g}\left(\hat{C}^*_{lB\Phi_1,f_2f_1}\sigma^{\mu_3\nu}\mathcal{P}_L-\hat{C}_{lB\Phi_1,f_1f_2}\sigma^{\mu_3\nu}\mathcal{P}_R\right)\right.\\
&\qquad\left.+\hat{g}'\left(-\hat{C}^*_{lW\Phi_1,f_2f_1}\sigma^{\mu_3\nu}\mathcal{P}_L+\hat{C}_{lW\Phi_1,f_1f_2}\sigma^{\mu_3\nu}\mathcal{P}_R\right)\right)
\end{aligned}
\tag{C.51}
$$

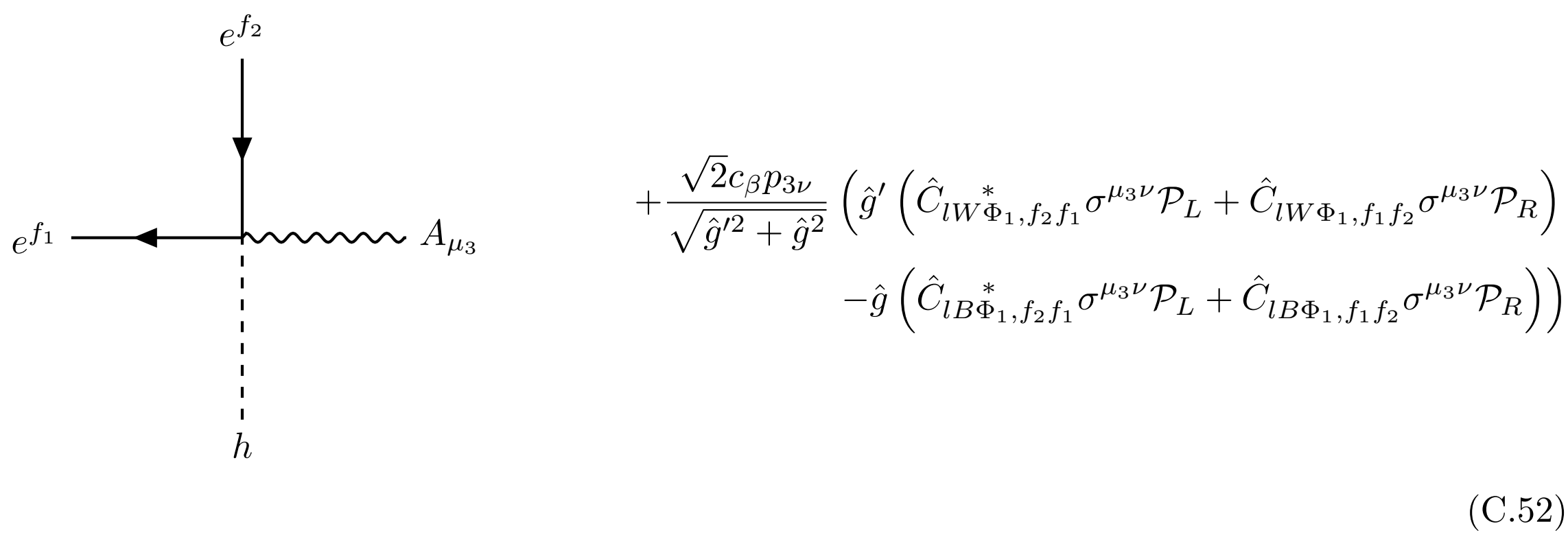

$$
\begin{aligned}
&+\frac{\sqrt{2}c_\beta p_{3\nu}}{\sqrt{\hat{g}'^2+\hat{g}^2}}\left(\hat{g}'\left(\hat{C}^*_{lW\Phi_1,f_2f_1}\sigma^{\mu_3\nu}\mathcal{P}_L+\hat{C}_{lW\Phi_1,f_1f_2}\sigma^{\mu_3\nu}\mathcal{P}_R\right)\right.\\
&\qquad\left.-\hat{g}\left(\hat{C}^*_{lB\Phi_1,f_2f_1}\sigma^{\mu_3\nu}\mathcal{P}_L+\hat{C}_{lB\Phi_1,f_1f_2}\sigma^{\mu_3\nu}\mathcal{P}_R\right)\right)
\end{aligned}
\tag{C.52}
$$

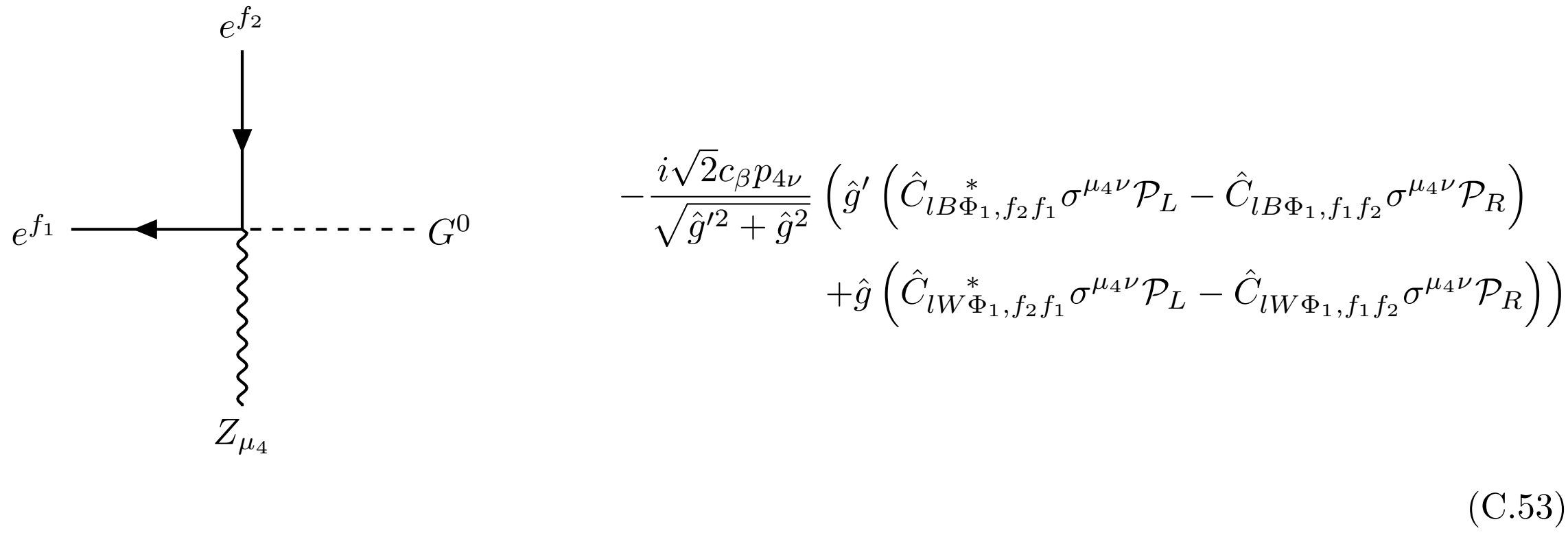

$$-\frac{i\sqrt{2}c_\beta p_{4\nu}}{\sqrt{\hat{g}'^2+\hat{g}^2}}\left(\hat{g}'\left(\hat{C}^*_{lB\Phi_1,f_2f_1}\sigma^{\mu_4\nu}\mathcal{P}_L-\hat{C}_{lB\Phi_1,f_1f_2}\sigma^{\mu_4\nu}\mathcal{P}_R\right)\right.$$
$$\left.+\hat{g}\left(\hat{C}^*_{lW\Phi_1,f_2f_1}\sigma^{\mu_4\nu}\mathcal{P}_L-\hat{C}_{lW\Phi_1,f_1f_2}\sigma^{\mu_4\nu}\mathcal{P}_R\right)\right) \tag{C.53}$$

$e^{f_2}$ $e^{f_1}$ $h$ $Z_{\mu_4}$

$$+\frac{\sqrt{2}c_\beta p_{4\nu}}{\sqrt{\hat{g}'^2+\hat{g}^2}}\left(\hat{g}'\left(\hat{C}^*_{lB\Phi_1,f_2f_1}\sigma^{\mu_4\nu}\mathcal{P}_L+\hat{C}_{lB\Phi_1,f_1f_2}\sigma^{\mu_4\nu}\mathcal{P}_R\right)\right.$$
$$\left.+\hat{g}\left(\hat{C}^*_{lW\Phi_1,f_2f_1}\sigma^{\mu_4\nu}\mathcal{P}_L+\hat{C}_{lW\Phi_1,f_1f_2}\sigma^{\mu_4\nu}\mathcal{P}_R\right)\right)$$
$$+\frac{iv\sqrt{\hat{g}'^2+\hat{g}^2}\gamma^{\mu_4}\mathcal{P}_L}{\sqrt{2}}\left(2c_\beta^2\hat{C}^{(11)[1]}_{\Phi l,f_1f_2}+2c_\beta^2\hat{C}^{(11)[3]}_{\Phi l,f_1f_2}\right.$$
$$\left.+2s_\beta^2\left(\hat{C}^{(22)[1]}_{\Phi l,f_1f_2}+\hat{C}^{(22)[3]}_{\Phi l,f_1f_2}\right)\right)$$
$$+\frac{iv\sqrt{\hat{g}'^2+\hat{g}^2}\gamma^{\mu_4}\mathcal{P}_R}{\sqrt{2}}\left(2c_\beta^2\hat{C}^{(11)}_{\Phi e,f_1f_2}+2s_\beta^2\hat{C}^{(22)}_{\Phi e,f_1f_2}\right) \tag{C.54}$$

$e^{f_2}$ $e^{f_1}$ $A_{\mu_3}$ $A^0$

$$+\frac{i\sqrt{2}s_\beta p_{3\nu}}{\sqrt{\hat{g}'^2+\hat{g}^2}}\left(\hat{g}'\left(\hat{C}^*_{lW\Phi_1,f_2f_1}\sigma^{\mu_3\nu}\mathcal{P}_L-\hat{C}_{lW\Phi_1,f_1f_2}\sigma^{\mu_3\nu}\mathcal{P}_R\right)\right.$$
$$\left.+\hat{g}\left(-\hat{C}^*_{lB\Phi_1,f_2f_1}\sigma^{\mu_3\nu}\mathcal{P}_L+\hat{C}_{lB\Phi_1,f_1f_2}\sigma^{\mu_3\nu}\mathcal{P}_R\right)\right) \tag{C.55}$$

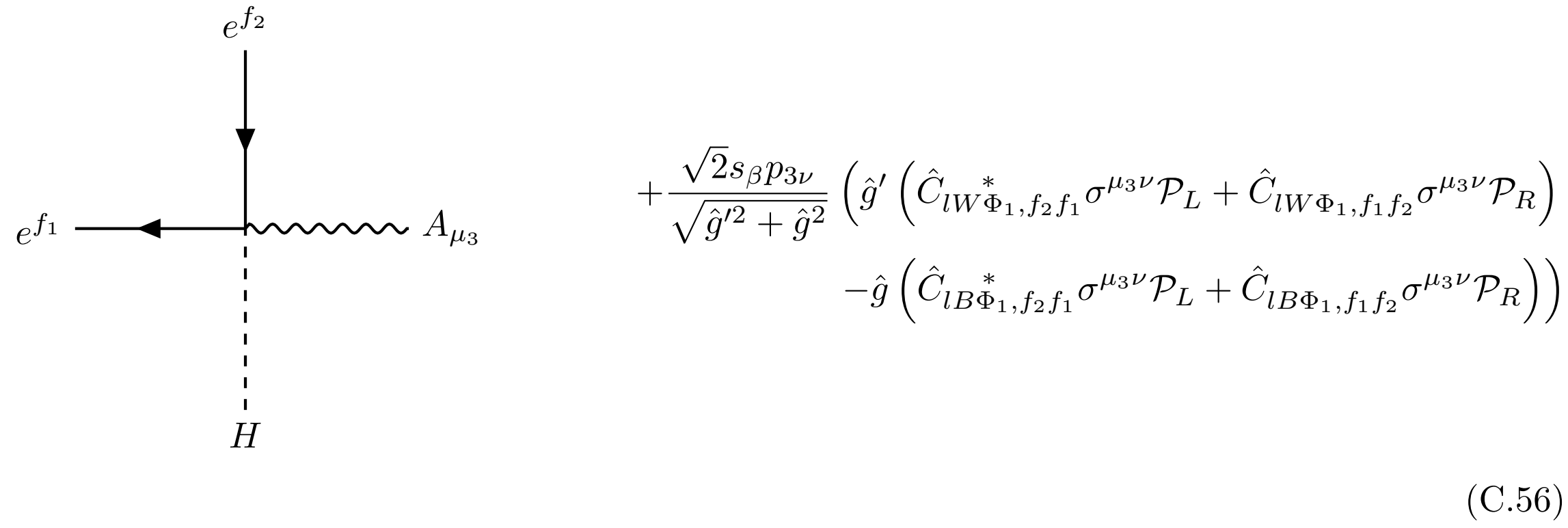

$$+\frac{\sqrt{2}s_\beta p_{3\nu}}{\sqrt{\hat{g}'^2+\hat{g}^2}}\left(\hat{g}'\left(\hat{C}^*_{lW\Phi_1,f_2f_1}\sigma^{\mu_3\nu}\mathcal{P}_L+\hat{C}_{lW\Phi_1,f_1f_2}\sigma^{\mu_3\nu}\mathcal{P}_R\right)\right.$$
$$\left.-\hat{g}\left(\hat{C}^*_{lB\Phi_1,f_2f_1}\sigma^{\mu_3\nu}\mathcal{P}_L+\hat{C}_{lB\Phi_1,f_1f_2}\sigma^{\mu_3\nu}\mathcal{P}_R\right)\right) \tag{C.56}$$

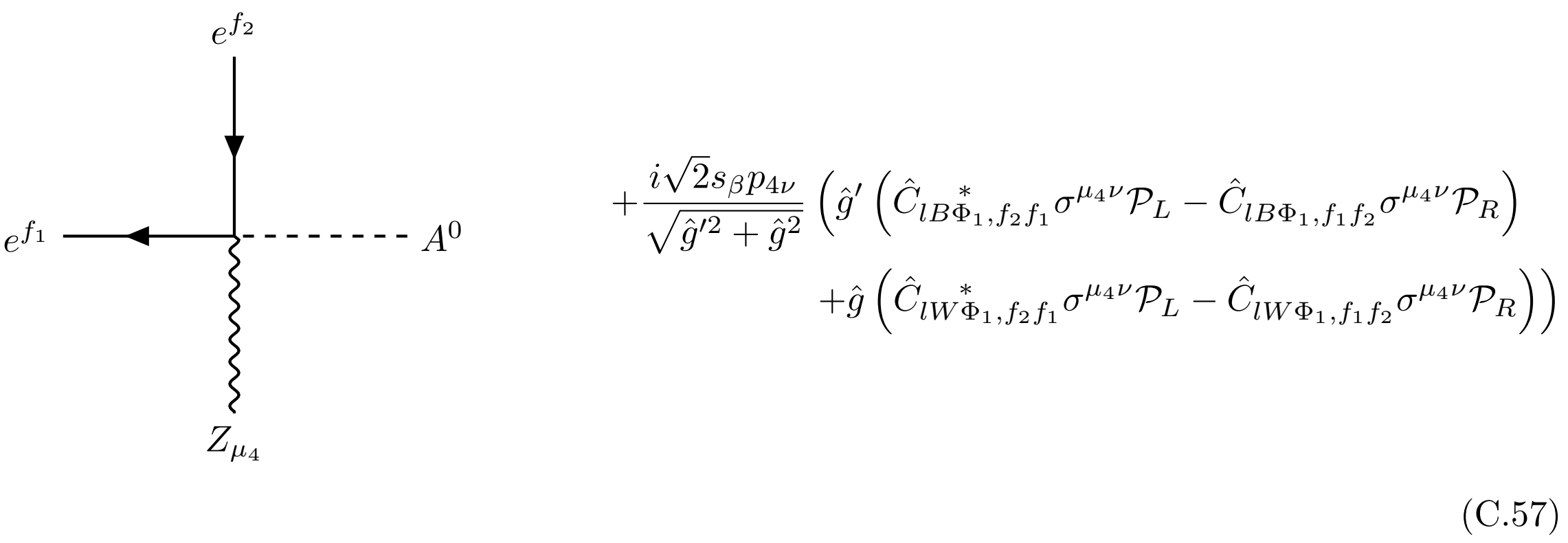

$$+\frac{i\sqrt{2}s_\beta p_{4\nu}}{\sqrt{\hat{g}'^2+\hat{g}^2}}\left(\hat{g}'\left(\hat{C}^*_{lB\Phi_1,f_2f_1}\sigma^{\mu_4\nu}\mathcal{P}_L-\hat{C}_{lB\Phi_1,f_1f_2}\sigma^{\mu_4\nu}\mathcal{P}_R\right)\right.$$
$$\left.+\hat{g}\left(\hat{C}^*_{lW\Phi_1,f_2f_1}\sigma^{\mu_4\nu}\mathcal{P}_L-\hat{C}_{lW\Phi_1,f_1f_2}\sigma^{\mu_4\nu}\mathcal{P}_R\right)\right) \tag{C.57}$$

$e^{f_2}$

$e^{f_1}$ $H$

$Z_{\mu_4}$

$$+\frac{\sqrt{2}s_\beta p_{4\nu}}{\sqrt{\hat{g}'^2+\hat{g}^2}}\left(\hat{g}'\left(\hat{C}^*_{lB\Phi_1,f_2f_1}\sigma^{\mu_4\nu}\mathcal{P}_L+\hat{C}_{lB\Phi_1,f_1f_2}\sigma^{\mu_4\nu}\mathcal{P}_R\right)\right.$$
$$\left.+\hat{g}\left(\hat{C}^*_{lW\Phi_1,f_2f_1}\sigma^{\mu_4\nu}\mathcal{P}_L+\hat{C}_{lW\Phi_1,f_1f_2}\sigma^{\mu_4\nu}\mathcal{P}_R\right)\right)$$
$$+i\sqrt{2}vs_\beta c_\beta\sqrt{\hat{g}'^2+\hat{g}^2}\gamma^{\mu_4}\mathcal{P}_L\left(\hat{C}^{(11)[1]}_{\Phi l,f_1f_2}+\hat{C}^{(11)[3]}_{\Phi l,f_1f_2}-\hat{C}^{(22)[1]}_{\Phi l,f_1f_2}-\hat{C}^{(22)[3]}_{\Phi l,f_1f_2}\right)$$
$$+i\sqrt{2}vs_\beta c_\beta\sqrt{\hat{g}'^2+\hat{g}^2}\gamma^{\mu_4}\mathcal{P}_R\left(\hat{C}^{(11)}_{\Phi e,f_1f_2}-\hat{C}^{(22)}_{\Phi e,f_1f_2}\right) \tag{C.58}$$

$$+2\hat{g}vc_\beta\left(\sigma^{\mu_3\mu_4}\mathcal{P}_L\hat{C}^*_{lW\Phi_1,f_2f_1}+\hat{C}_{lW\Phi_1,f_1f_2}\sigma^{\mu_3\mu_4}\mathcal{P}_R\right) \tag{C.59}$$

$$\begin{aligned}&-\frac{2c_\beta p_{3\nu}U^*_{g_1f_1}\sigma^{\mu_3\nu}\mathcal{P}_R}{\sqrt{\hat{g}'^2+\hat{g}^2}}\left(\hat{g}\hat{C}_{lB\Phi_1,g_1f_2}+\hat{g}'\hat{C}_{lW\Phi_1,g_1f_2}\right)\\&-\frac{2i\hat{g}v\hat{g}'U^*_{g_1f_1}\gamma^{\mu_3}\mathcal{P}_L}{\sqrt{\hat{g}'^2+\hat{g}^2}}\left(c_\beta^2\hat{C}^{(11)[3]}_{\Phi l,g_1f_2}+s_\beta^2\hat{C}^{(22)[3]}_{\Phi l,g_1f_2}\right)\end{aligned} \tag{C.60}$$

$$\begin{aligned}&+\frac{2c_\beta p_{4\nu}U^*_{g_1f_1}\sigma^{\mu_4\nu}\mathcal{P}_R}{\sqrt{\hat{g}'^2+\hat{g}^2}}\left(\hat{g}'\hat{C}_{lB\Phi_1,g_1f_2}-\hat{g}\hat{C}_{lW\Phi_1,g_1f_2}\right)\\&+\frac{2iv\hat{g}'^2U^*_{g_1f_1}\gamma^{\mu_4}\mathcal{P}_L}{\sqrt{\hat{g}'^2+\hat{g}^2}}\left(c_\beta^2\hat{C}^{(11)[3]}_{\Phi l,g_1f_2}+s_\beta^2\hat{C}^{(22)[3]}_{\Phi l,g_1f_2}\right)\end{aligned} \tag{C.61}$$

$$
\begin{aligned}
&+\frac{2s_\beta p_{3\nu} U^*_{g_1 f_1} \sigma^{\mu_3 \nu} \mathcal{P}_R}{\sqrt{\hat{g}'^2+\hat{g}^2}} \left( \hat{g} \hat{C}_{lB\Phi_1, g_1 f_2} + \hat{g}' \hat{C}_{lW\Phi_1, g_1 f_2} \right) \\
&+\frac{i\hat{g} v s_{2\beta} \hat{g}' U^*_{g_1 f_1} \gamma^{\mu_3} \mathcal{P}_L}{\sqrt{\hat{g}'^2+\hat{g}^2}} \left( \hat{C}^{(11)[3]}_{\Phi l, g_1 f_2} - \hat{C}^{(22)[3]}_{\Phi l, g_1 f_2} \right)
\end{aligned}
\tag{C.62}
$$

$$
\begin{aligned}
&+\frac{2s_\beta p_{4\nu} U^*_{g_1 f_1} \sigma^{\mu_4 \nu} \mathcal{P}_R}{\sqrt{\hat{g}'^2+\hat{g}^2}} \left( \hat{g} \hat{C}_{lW\Phi_1, g_1 f_2} - \hat{g}' \hat{C}_{lB\Phi_1, g_1 f_2} \right) \\
&-\frac{i v s_{2\beta} \hat{g}'^2 U^*_{g_1 f_1} \gamma^{\mu_4} \mathcal{P}_L}{\sqrt{\hat{g}'^2+\hat{g}^2}} \left( \hat{C}^{(11)[3]}_{\Phi l, g_1 f_2} - \hat{C}^{(22)[3]}_{\Phi l, g_1 f_2} \right)
\end{aligned}
\tag{C.63}
$$

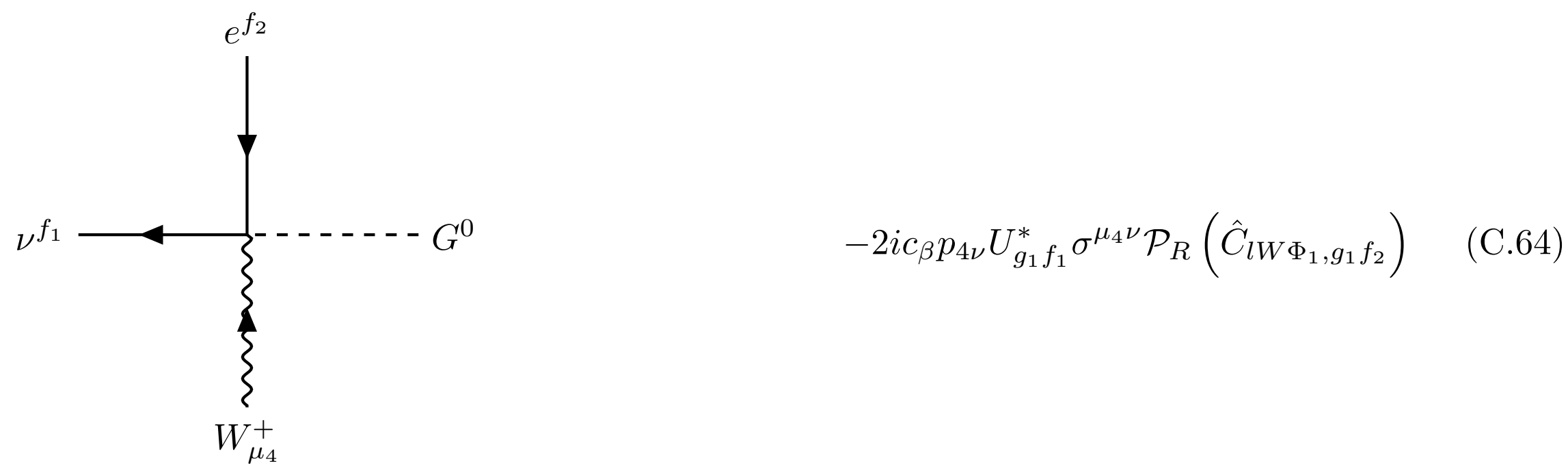

$$
-2ic_\beta p_{4\nu} U^*_{g_1 f_1} \sigma^{\mu_4 \nu} \mathcal{P}_R \left( \hat{C}_{lW\Phi_1, g_1 f_2} \right) \tag{C.64}
$$

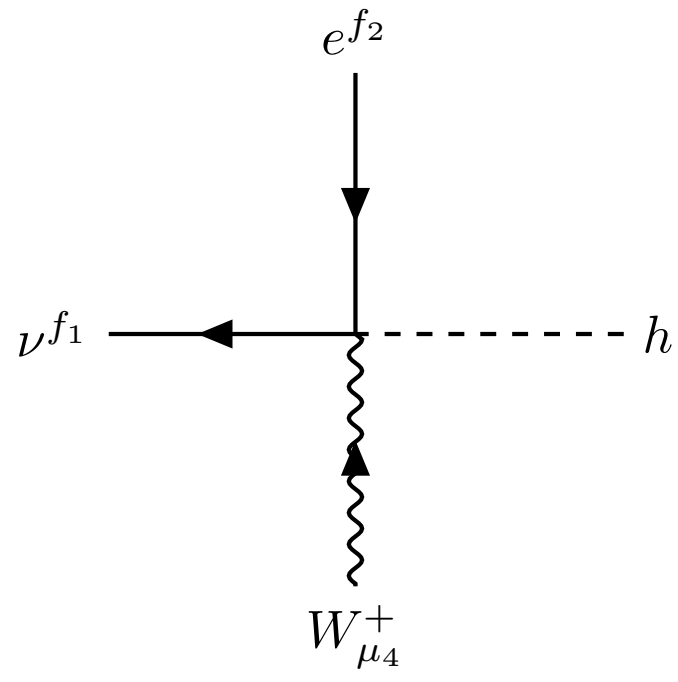

$$
\begin{aligned}
&-2c_\beta p_{4\nu} U^*_{g_1 f_1} \sigma^{\mu_4 \nu} \mathcal{P}_R \left( \hat{C}_{lW\Phi_1, g_1 f_2} \right) \\
&-2i\hat{g}v U^*_{g_1 f_1} \gamma^{\mu_4} \mathcal{P}_L \left( c^2_\beta \hat{C}^{(11)[3]}_{\Phi l, g_1 f_2} + s^2_\beta \hat{C}^{(22)[3]}_{\Phi l, g_1 f_2} \right)
\end{aligned}
\tag{C.65}
$$

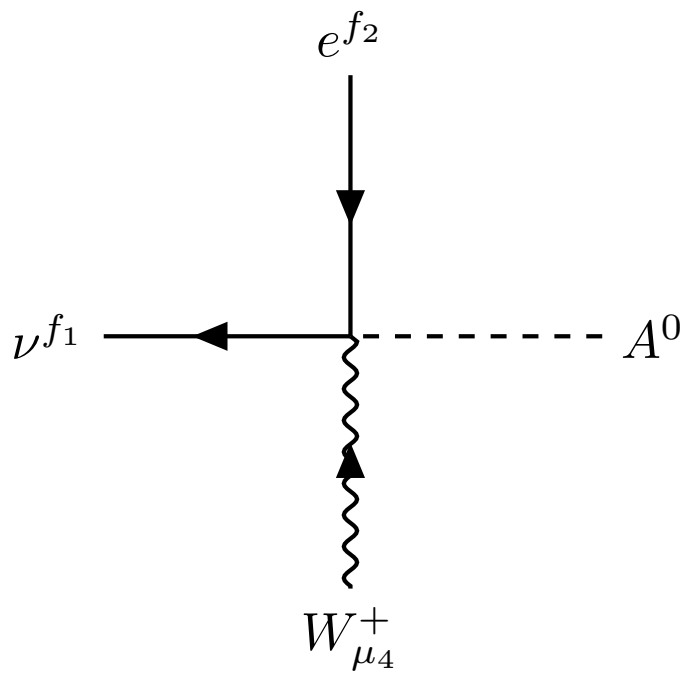

$$
+2is_\beta p_{4\nu} U^*_{g_1 f_1} \sigma^{\mu_4 \nu} \mathcal{P}_R \left( \hat{C}_{lW\Phi_1, g_1 f_2} \right)
\tag{C.66}
$$

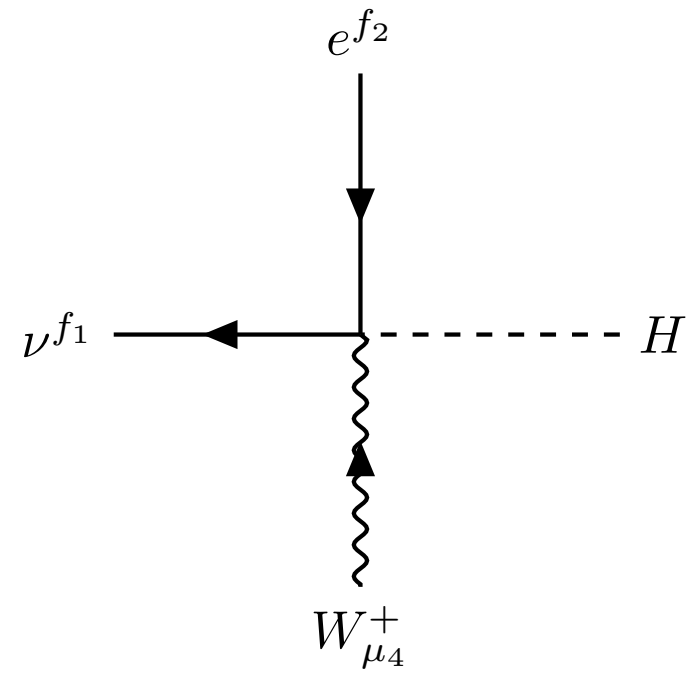

$$
\begin{aligned}
&-2s_\beta p_{4\nu} U^*_{g_1 f_1} \sigma^{\mu_4 \nu} \mathcal{P}_R \left( \hat{C}_{lW\Phi_1, g_1 f_2} \right) \\
&-i\hat{g}vs_{2\beta} U^*_{g_1 f_1} \gamma^{\mu_4} \mathcal{P}_L \left( \hat{C}^{(11)[3]}_{\Phi l, g_1 f_2} - \hat{C}^{(22)[3]}_{\Phi l, g_1 f_2} \right)
\end{aligned}
\tag{C.67}
$$

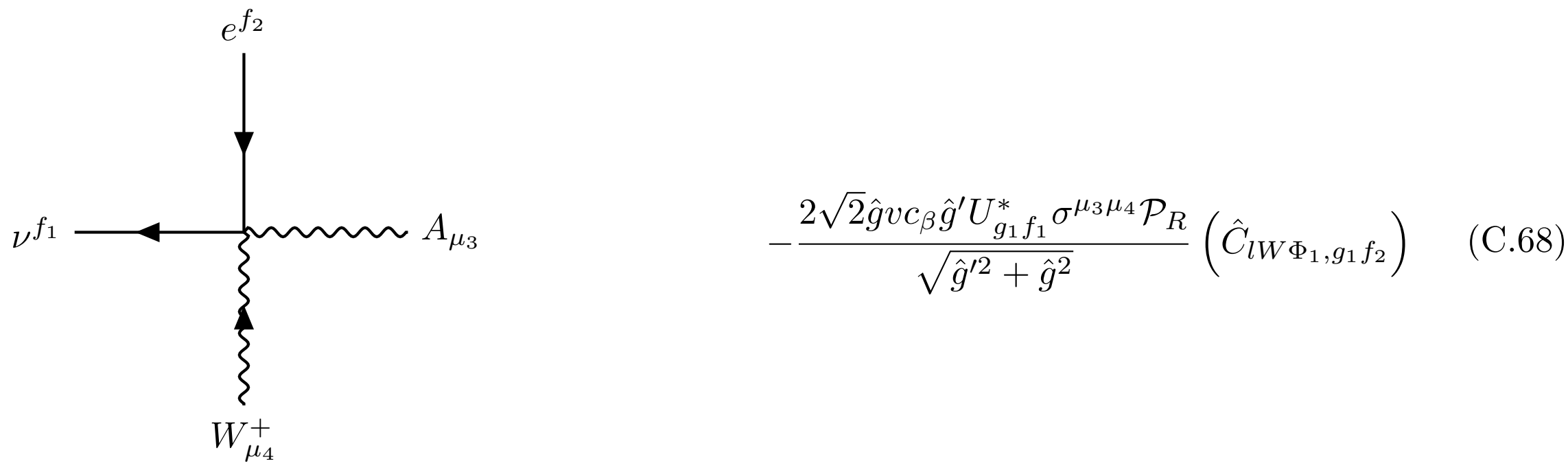

$$-\frac{2\sqrt{2}\hat{g}vc_\beta\hat{g}'U^*_{g_1f_1}\sigma^{\mu_3\mu_4}\mathcal{P}_R}{\sqrt{\hat{g}'^2+\hat{g}^2}}\left(\hat{C}_{lW\Phi_1,g_1f_2}\right) \tag{C.68}$$

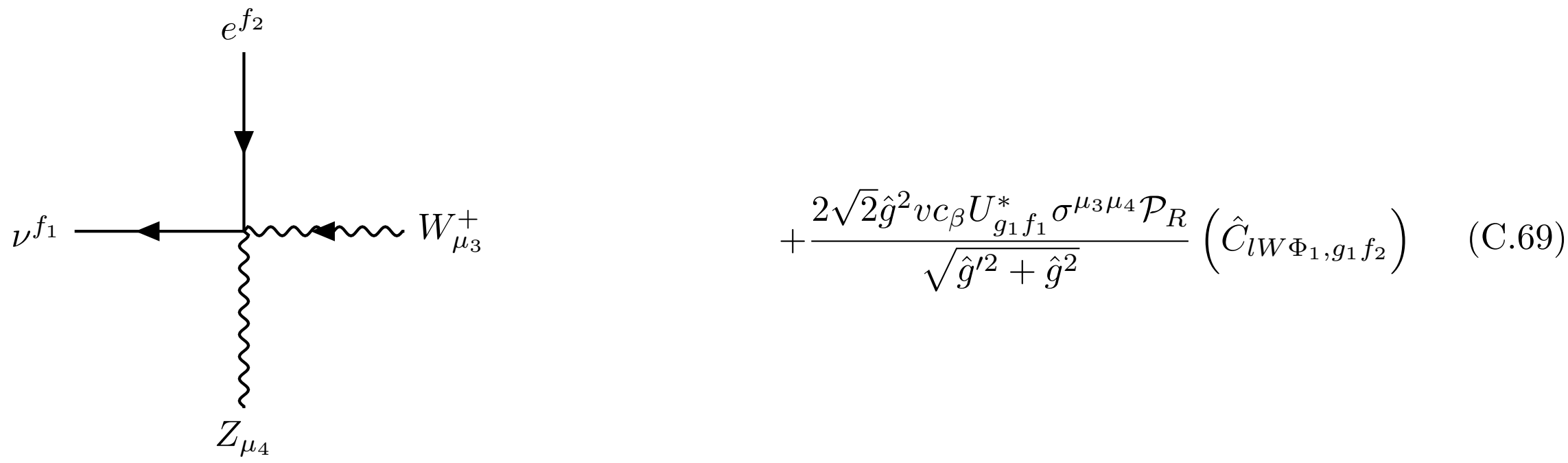

$$+\frac{2\sqrt{2}\hat{g}^2vc_\beta U^*_{g_1f_1}\sigma^{\mu_3\mu_4}\mathcal{P}_R}{\sqrt{\hat{g}'^2+\hat{g}^2}}\left(\hat{C}_{lW\Phi_1,g_1f_2}\right) \tag{C.69}$$

$\nu^{f_2}$

$\nu^{f_1}$ $G^+$

$G^-$

$$\begin{aligned}-iU_{g_2f_2}U^*_{g_1f_1}\left(\not{p}_3\gamma^5-\not{p}_4\gamma^5\right)\Big(&c_\beta^2\hat{C}^{(11)[1]}_{\Phi l,g_1g_2}+c_\beta^2\hat{C}^{(11)[3]}_{\Phi l,g_1g_2}\\&+s_\beta^2\left(\hat{C}^{(22)[1]}_{\Phi l,g_1g_2}+\hat{C}^{(22)[3]}_{\Phi l,g_1g_2}\right)\Big)\end{aligned} \tag{C.70}$$

$$-is_\beta c_\beta U_{g_2f_2}U^*_{g_1f_1}\left(\not{p}_3\gamma^5-\not{p}_4\gamma^5\right)\Big(\hat{C}^{(11)[1]}_{\Phi l,g_1g_2}+\hat{C}^{(11)[3]}_{\Phi l,g_1g_2} -\hat{C}^{(22)[1]}_{\Phi l,g_1g_2}-\hat{C}^{(22)[3]}_{\Phi l,g_1g_2}\Big) \tag{C.71}$$

$$-iU_{g_2f_2}U^*_{g_1f_1}\left(\not{p}_3\gamma^5-\not{p}_4\gamma^5\right)\Big(s^2_\beta\left(\hat{C}^{(11)[1]}_{\Phi l,g_1g_2}+\hat{C}^{(11)[3]}_{\Phi l,g_1g_2}\right) +c^2_\beta\hat{C}^{(22)[1]}_{\Phi l,g_1g_2}+c^2_\beta\hat{C}^{(22)[3]}_{\Phi l,g_1g_2}\Big) \tag{C.72}$$

$$+i\sqrt{2}\hat{g}vU_{g_2f_2}U^*_{g_1f_1}\gamma^{\mu_4}\gamma^5\left(c^2_\beta\hat{C}^{(11)[1]}_{\Phi l,g_1g_2}+s^2_\beta\hat{C}^{(22)[1]}_{\Phi l,g_1g_2}\right) \tag{C.73}$$

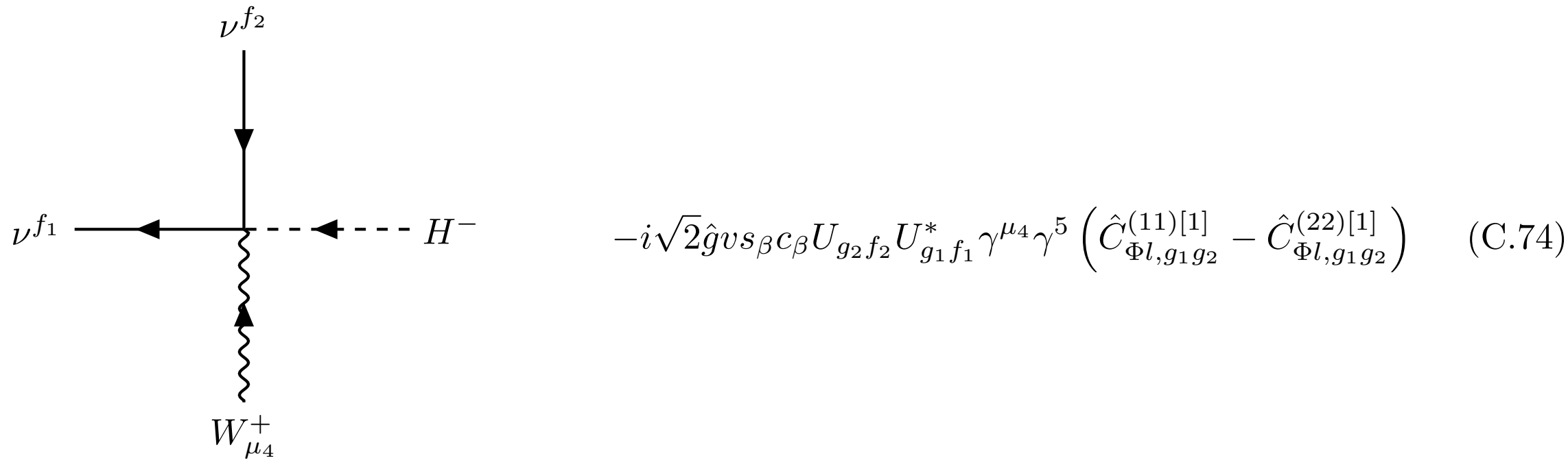

$$-i\sqrt{2}\hat{g}vs_\beta c_\beta U_{g_2 f_2} U^*_{g_1 f_1} \gamma^{\mu_4}\gamma^5 \left(\hat{C}^{(11)[1]}_{\Phi l,g_1 g_2} - \hat{C}^{(22)[1]}_{\Phi l,g_1 g_2}\right) \tag{C.74}$$

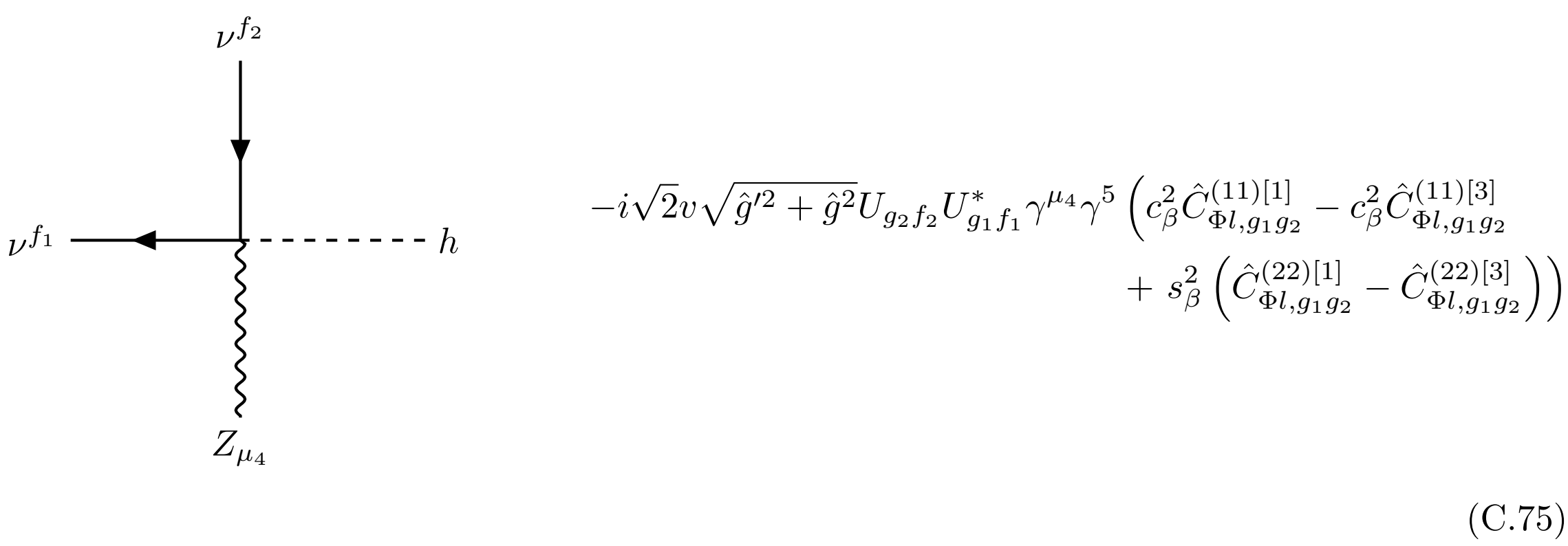

$$\begin{aligned} -i\sqrt{2}v\sqrt{\hat{g}'^2+\hat{g}^2}U_{g_2 f_2} U^*_{g_1 f_1} \gamma^{\mu_4}\gamma^5 \Big( & c^2_\beta \hat{C}^{(11)[1]}_{\Phi l,g_1 g_2} - c^2_\beta \hat{C}^{(11)[3]}_{\Phi l,g_1 g_2} \\ & + s^2_\beta \left(\hat{C}^{(22)[1]}_{\Phi l,g_1 g_2} - \hat{C}^{(22)[3]}_{\Phi l,g_1 g_2}\right)\Big) \end{aligned} \tag{C.75}$$

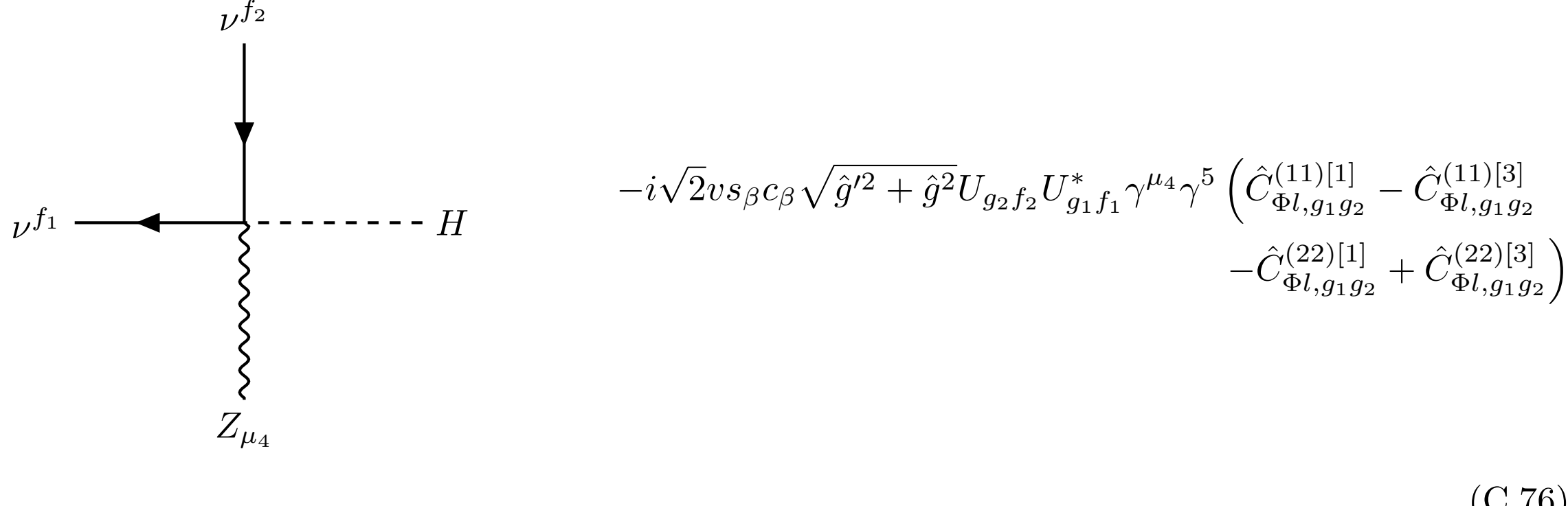

$$\begin{aligned} -i\sqrt{2}vs_\beta c_\beta\sqrt{\hat{g}'^2+\hat{g}^2}U_{g_2 f_2} U^*_{g_1 f_1} \gamma^{\mu_4}\gamma^5 \Big( & \hat{C}^{(11)[1]}_{\Phi l,g_1 g_2} - \hat{C}^{(11)[3]}_{\Phi l,g_1 g_2} \\ & -\hat{C}^{(22)[1]}_{\Phi l,g_1 g_2} + \hat{C}^{(22)[3]}_{\Phi l,g_1 g_2}\Big) \end{aligned} \tag{C.76}$$

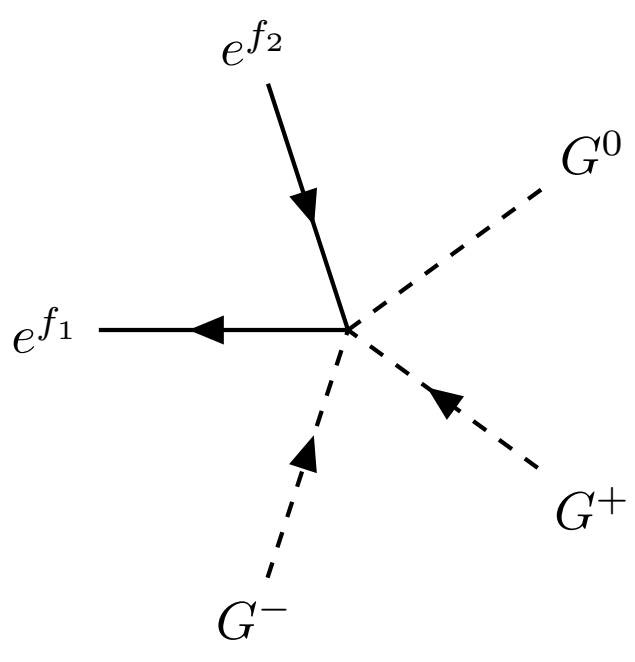

$$
\begin{aligned}
&+\frac{c_\beta}{\sqrt{2}}\Big(c_\beta^2\Big(\mathcal{P}_L\hat{C}^{(11)*}_{l\Phi_1,f_2f_1}-\mathcal{P}_R\hat{C}^{(11)}_{l\Phi_1,f_1f_2}\Big)\\
&\qquad+s_\beta^2\Big(\mathcal{P}_L\hat{C}^{(12)*}_{l\Phi_2,f_2f_1}-\mathcal{P}_R\hat{C}^{(12)}_{l\Phi_2,f_1f_2}\Big)\\
&\qquad+s_\beta^2\Big(\mathcal{P}_L\hat{C}^{(21)*}_{l\Phi_2,f_2f_1}-\mathcal{P}_R\hat{C}^{(21)}_{l\Phi_2,f_1f_2}\Big)\\
&\qquad+s_\beta^2\Big(\mathcal{P}_L\hat{C}^{(22)*}_{l\Phi_1,f_2f_1}-\mathcal{P}_R\hat{C}^{(22)}_{l\Phi_1,f_1f_2}\Big)\Big)
\end{aligned}
\tag{C.77}
$$

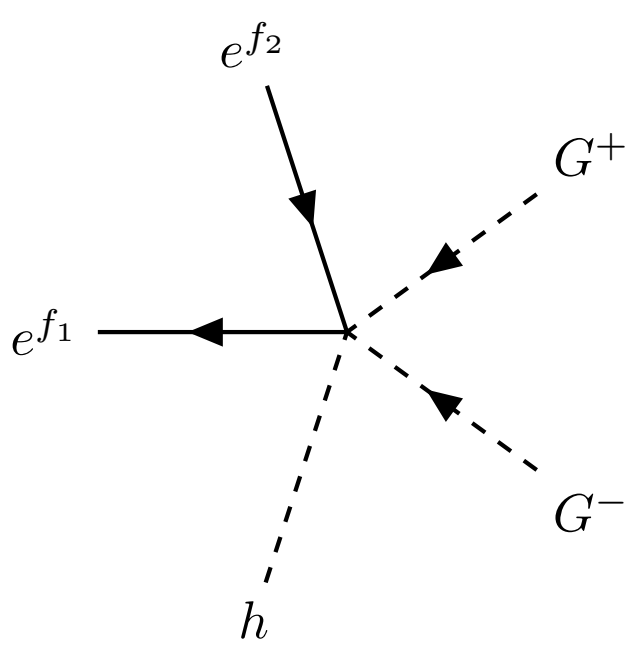

$$
\begin{aligned}
&+\frac{ic_\beta}{\sqrt{2}}\Big(c_\beta^2\Big(\mathcal{P}_L\hat{C}^{(11)*}_{l\Phi_1,f_2f_1}+\mathcal{P}_R\hat{C}^{(11)}_{l\Phi_1,f_1f_2}\Big)\\
&\qquad+s_\beta^2\Big(\mathcal{P}_L\hat{C}^{(12)*}_{l\Phi_2,f_2f_1}+\mathcal{P}_R\hat{C}^{(12)}_{l\Phi_2,f_1f_2}\Big)\\
&\qquad+s_\beta^2\Big(\mathcal{P}_L\hat{C}^{(21)*}_{l\Phi_2,f_2f_1}+\mathcal{P}_R\hat{C}^{(21)}_{l\Phi_2,f_1f_2}\Big)\\
&\qquad+s_\beta^2\Big(\mathcal{P}_L\hat{C}^{(22)*}_{l\Phi_1,f_2f_1}+\mathcal{P}_R\hat{C}^{(22)}_{l\Phi_1,f_1f_2}\Big)\Big)
\end{aligned}
\tag{C.78}
$$

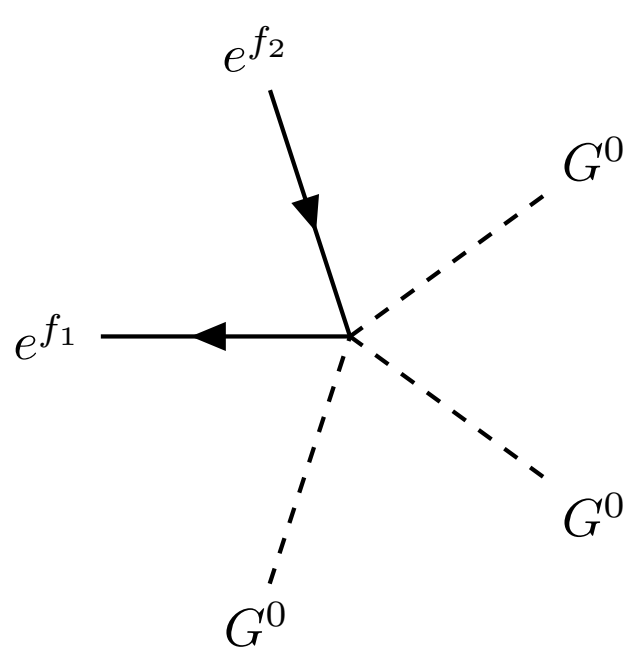

$$
\begin{aligned}
&+\frac{3c_\beta}{\sqrt{2}}\Big(c_\beta^2\Big(\mathcal{P}_L\hat{C}^{(11)*}_{l\Phi_1,f_2f_1}-\mathcal{P}_R\hat{C}^{(11)}_{l\Phi_1,f_1f_2}\Big)\\
&\qquad+s_\beta^2\Big(\mathcal{P}_L\hat{C}^{(12)*}_{l\Phi_2,f_2f_1}-\mathcal{P}_R\hat{C}^{(12)}_{l\Phi_2,f_1f_2}\Big)\\
&\qquad+s_\beta^2\Big(\mathcal{P}_L\hat{C}^{(21)*}_{l\Phi_2,f_2f_1}-\mathcal{P}_R\hat{C}^{(21)}_{l\Phi_2,f_1f_2}\Big)\\
&\qquad+s_\beta^2\Big(\mathcal{P}_L\hat{C}^{(22)*}_{l\Phi_1,f_2f_1}-\mathcal{P}_R\hat{C}^{(22)}_{l\Phi_1,f_1f_2}\Big)\Big)
\end{aligned}
\tag{C.79}
$$

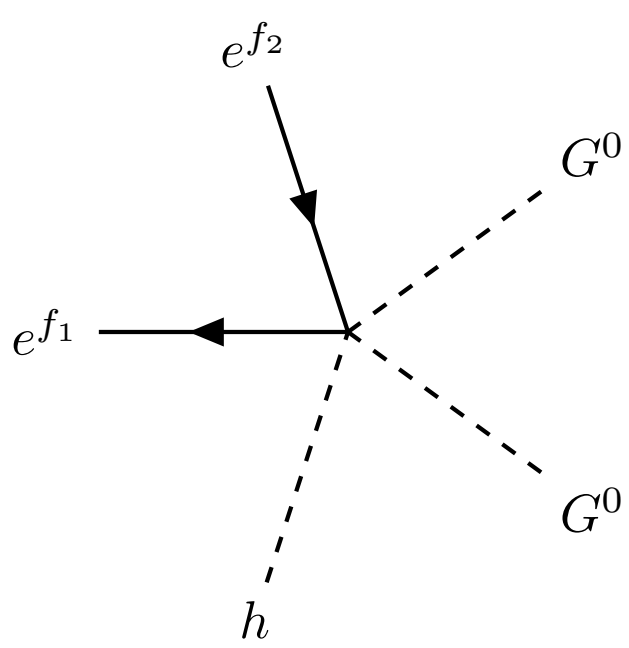

$$
\begin{aligned}
&+\frac{ic_\beta}{\sqrt{2}}\Big(c_\beta^2\Big(\mathcal{P}_L\hat{C}^{(11)*}_{l\Phi_1,f_2f_1}+\mathcal{P}_R\hat{C}^{(11)}_{l\Phi_1,f_1f_2}\Big)\\
&\qquad+s_\beta^2\Big(\mathcal{P}_L\hat{C}^{(12)*}_{l\Phi_2,f_2f_1}+\mathcal{P}_R\hat{C}^{(12)}_{l\Phi_2,f_1f_2}\Big)\\
&\qquad+s_\beta^2\Big(\mathcal{P}_L\hat{C}^{(21)*}_{l\Phi_2,f_2f_1}+\mathcal{P}_R\hat{C}^{(21)}_{l\Phi_2,f_1f_2}\Big)\\
&\qquad+s_\beta^2\Big(\mathcal{P}_L\hat{C}^{(22)*}_{l\Phi_1,f_2f_1}+\mathcal{P}_R\hat{C}^{(22)}_{l\Phi_1,f_1f_2}\Big)\Big)
\end{aligned}
\tag{C.80}
$$

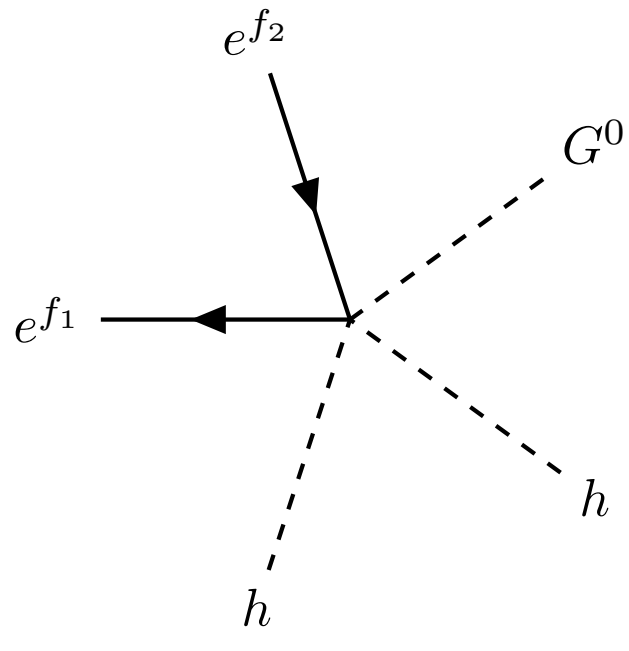

$$
\begin{aligned}
&+\frac{c_\beta}{\sqrt{2}}\left(c_\beta^2\left(\mathcal{P}_L\hat{C}^{(11)*}_{l\Phi_1,f_2f_1}-\mathcal{P}_R\hat{C}^{(11)}_{l\Phi_1,f_1f_2}\right)\right.\\
&\qquad+s_\beta^2\left(\mathcal{P}_L\hat{C}^{(12)*}_{l\Phi_2,f_2f_1}-\mathcal{P}_R\hat{C}^{(12)}_{l\Phi_2,f_1f_2}\right)\\
&\qquad+s_\beta^2\left(\mathcal{P}_L\hat{C}^{(21)*}_{l\Phi_2,f_2f_1}-\mathcal{P}_R\hat{C}^{(21)}_{l\Phi_2,f_1f_2}\right)\\
&\qquad\left.+s_\beta^2\left(\mathcal{P}_L\hat{C}^{(22)*}_{l\Phi_1,f_2f_1}-\mathcal{P}_R\hat{C}^{(22)}_{l\Phi_1,f_1f_2}\right)\right)
\end{aligned}
\tag{C.81}
$$

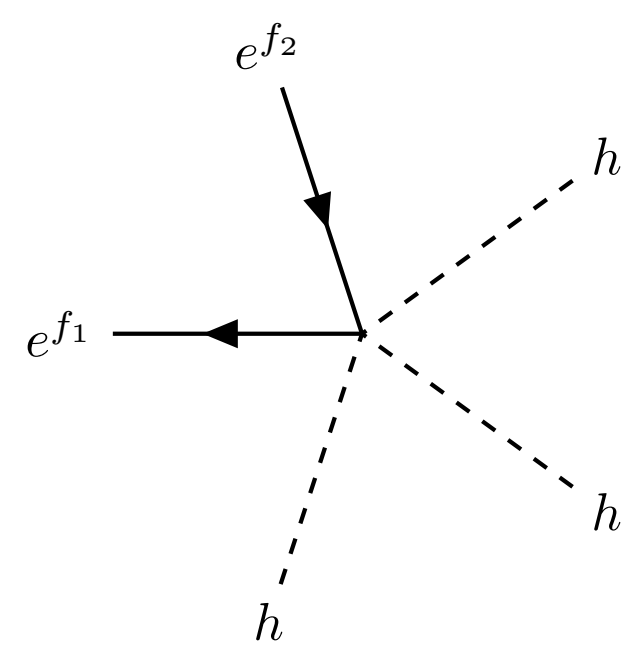

$$
\begin{aligned}
&+\frac{3ic_\beta}{\sqrt{2}}\left(c_\beta^2\left(\mathcal{P}_L\hat{C}^{(11)*}_{l\Phi_1,f_2f_1}+\mathcal{P}_R\hat{C}^{(11)}_{l\Phi_1,f_1f_2}\right)\right.\\
&\qquad+s_\beta^2\left(\mathcal{P}_L\hat{C}^{(12)*}_{l\Phi_2,f_2f_1}+\mathcal{P}_R\hat{C}^{(12)}_{l\Phi_2,f_1f_2}\right)\\
&\qquad+s_\beta^2\left(\mathcal{P}_L\hat{C}^{(21)*}_{l\Phi_2,f_2f_1}+\mathcal{P}_R\hat{C}^{(21)}_{l\Phi_2,f_1f_2}\right)\\
&\qquad\left.+s_\beta^2\left(\mathcal{P}_L\hat{C}^{(22)*}_{l\Phi_1,f_2f_1}+\mathcal{P}_R\hat{C}^{(22)}_{l\Phi_1,f_1f_2}\right)\right)
\end{aligned}
\tag{C.82}
$$

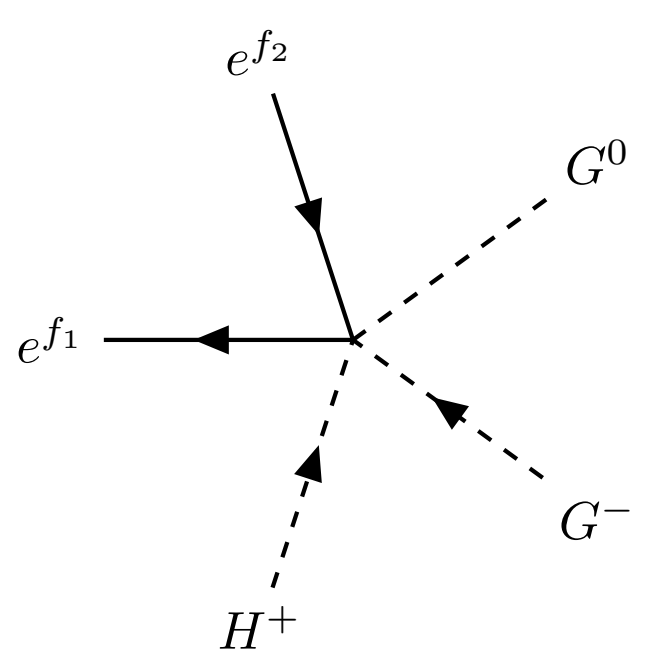

$$
\begin{aligned}
&+\frac{s_\beta}{\sqrt{2}}\left(-c_\beta^2\left(\mathcal{P}_L\hat{C}^{(11)*}_{l\Phi_1,f_2f_1}-\mathcal{P}_R\hat{C}^{(11)}_{l\Phi_1,f_1f_2}\right)\right.\\
&\qquad+c_\beta^2\left(\mathcal{P}_L\hat{C}^{(22)*}_{l\Phi_1,f_2f_1}-\mathcal{P}_R\hat{C}^{(22)}_{l\Phi_1,f_1f_2}\right)\\
&\qquad-\left(s_\beta^2\mathcal{P}_L\hat{C}^{(12)*}_{l\Phi_2,f_2f_1}+c_\beta^2\mathcal{P}_R\hat{C}^{(12)}_{l\Phi_2,f_1f_2}\right)\\
&\qquad\left.+\left(c_\beta^2\mathcal{P}_L\hat{C}^{(21)*}_{l\Phi_2,f_2f_1}+s_\beta^2\mathcal{P}_R\hat{C}^{(21)}_{l\Phi_2,f_1f_2}\right)\right)
\end{aligned}
\tag{C.83}
$$

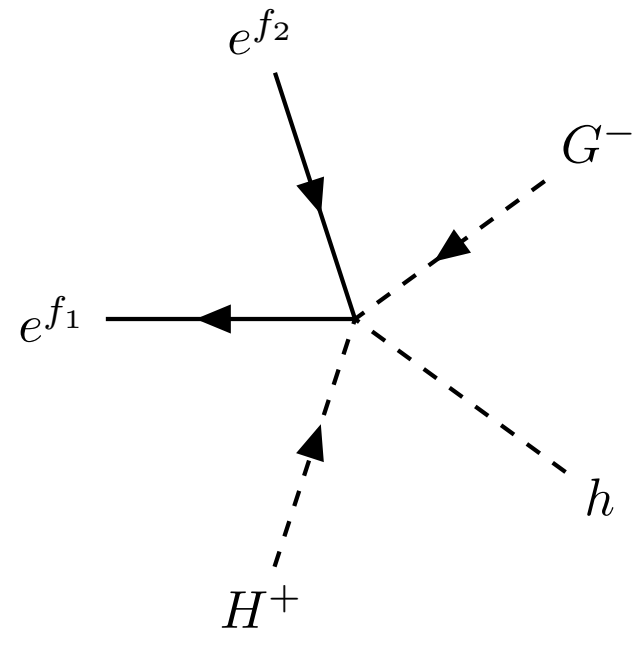

$$
\begin{aligned}
-\frac{is_\beta}{\sqrt{2}}\Big(c_\beta^2&\left(\mathcal{P}_L\hat{C}^{(11)*}_{l\Phi_1,f_2f_1}+\mathcal{P}_R\hat{C}^{(11)}_{l\Phi_1,f_1f_2}\right)\\
&-c_\beta^2\left(\mathcal{P}_L\hat{C}^{(22)*}_{l\Phi_1,f_2f_1}+\mathcal{P}_R\hat{C}^{(22)}_{l\Phi_1,f_1f_2}\right)\\
&+\left(s_\beta^2\mathcal{P}_L\hat{C}^{(12)*}_{l\Phi_2,f_2f_1}-c_\beta^2\mathcal{P}_R\hat{C}^{(12)}_{l\Phi_2,f_1f_2}\right)\\
&-\left(c_\beta^2\mathcal{P}_L\hat{C}^{(21)*}_{l\Phi_2,f_2f_1}-s_\beta^2\mathcal{P}_R\hat{C}^{(21)}_{l\Phi_2,f_1f_2}\right)\Big)
\end{aligned}
\tag{C.84}
$$

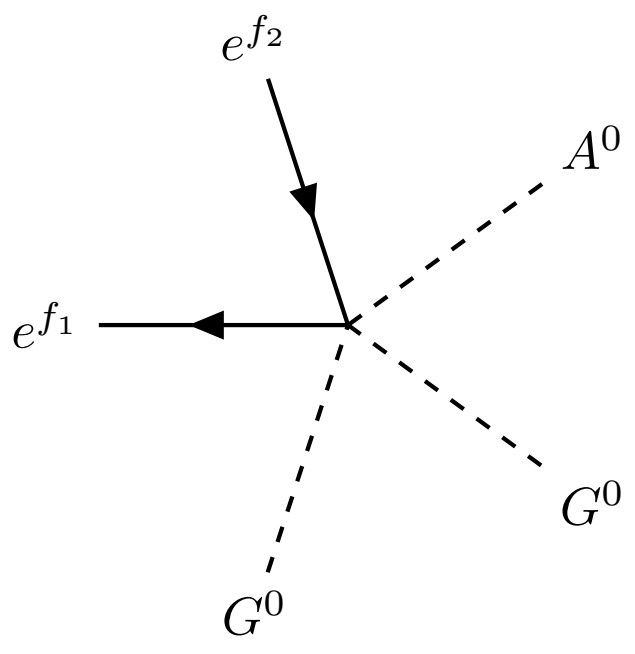

$$
\begin{aligned}
+\frac{s_\beta}{\sqrt{2}}\Big(-3c_\beta^2&\left(\mathcal{P}_L\hat{C}^{(11)*}_{l\Phi_1,f_2f_1}-\mathcal{P}_R\hat{C}^{(11)}_{l\Phi_1,f_1f_2}\right)\\
&+s_\beta^2\left(2ct_\beta^2-1\right)\left(\mathcal{P}_L\hat{C}^{(12)*}_{l\Phi_2,f_2f_1}-\mathcal{P}_R\hat{C}^{(12)}_{l\Phi_2,f_1f_2}\right)\\
&+s_\beta^2\left(2ct_\beta^2-1\right)\left(\mathcal{P}_L\hat{C}^{(21)*}_{l\Phi_2,f_2f_1}-\mathcal{P}_R\hat{C}^{(21)}_{l\Phi_2,f_1f_2}\right)\\
&+s_\beta^2\left(2ct_\beta^2-1\right)\left(\mathcal{P}_L\hat{C}^{(22)*}_{l\Phi_1,f_2f_1}-\mathcal{P}_R\hat{C}^{(22)}_{l\Phi_1,f_1f_2}\right)\Big)
\end{aligned}
\tag{C.85}
$$

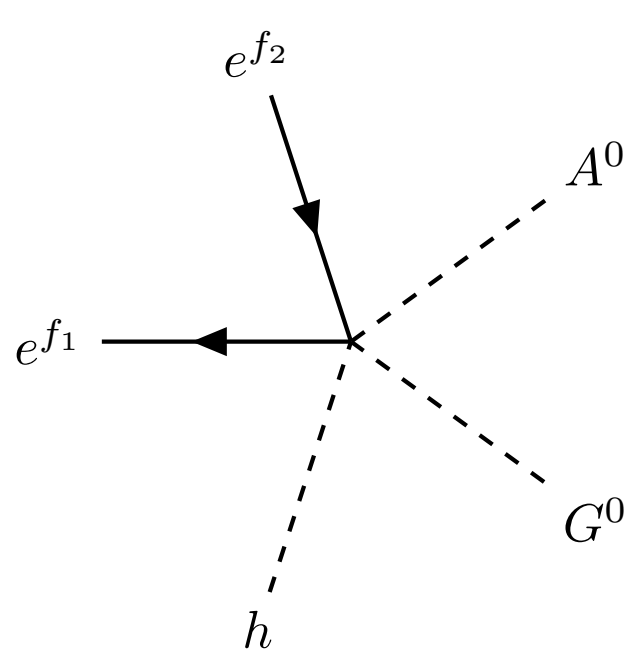

$$
\begin{aligned}
-\frac{is_\beta}{\sqrt{2}}\Big(c_\beta^2&\left(\mathcal{P}_L\hat{C}^{(11)*}_{l\Phi_1,f_2f_1}+\mathcal{P}_R\hat{C}^{(11)}_{l\Phi_1,f_1f_2}\right)\\
&+s_\beta^2\left(\mathcal{P}_L\hat{C}^{(12)*}_{l\Phi_2,f_2f_1}+\mathcal{P}_R\hat{C}^{(12)}_{l\Phi_2,f_1f_2}\right)\\
&-c_\beta^2\left(\mathcal{P}_L\hat{C}^{(21)*}_{l\Phi_2,f_2f_1}+\mathcal{P}_R\hat{C}^{(21)}_{l\Phi_2,f_1f_2}\right)\\
&-c_\beta^2\left(\mathcal{P}_L\hat{C}^{(22)*}_{l\Phi_1,f_2f_1}+\mathcal{P}_R\hat{C}^{(22)}_{l\Phi_1,f_1f_2}\right)\Big)
\end{aligned}
\tag{C.86}
$$

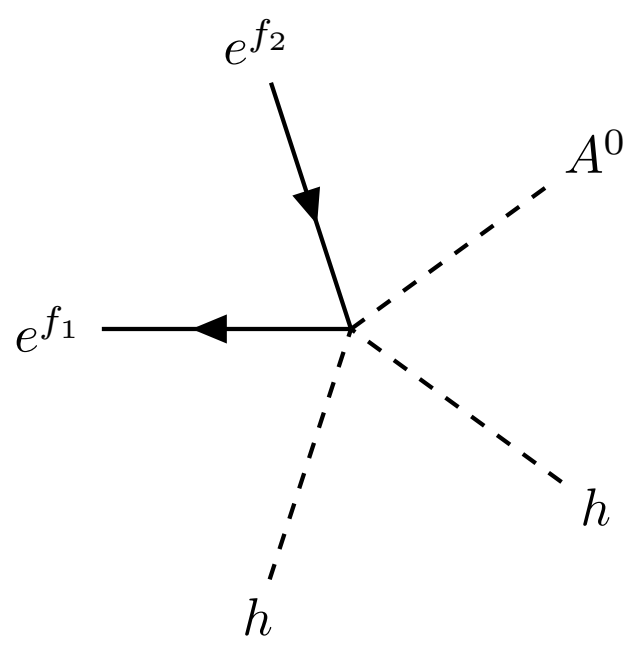

$$
\begin{aligned}
+\frac{s_\beta}{\sqrt{2}}\Big(-c_\beta^2&\left(\mathcal{P}_L\hat{C}^{(11)*}_{l\Phi_1,f_2f_1}-\mathcal{P}_R\hat{C}^{(11)}_{l\Phi_1,f_1f_2}\right)\\
&+s_\beta^2\left(2ct_\beta^2+1\right)\left(\mathcal{P}_L\hat{C}^{(12)*}_{l\Phi_2,f_2f_1}-\mathcal{P}_R\hat{C}^{(12)}_{l\Phi_2,f_1f_2}\right)\\
&-s_\beta^2\left(\mathcal{P}_L\hat{C}^{(21)*}_{l\Phi_2,f_2f_1}-\mathcal{P}_R\hat{C}^{(21)}_{l\Phi_2,f_1f_2}\right)\\
&-s_\beta^2\left(\mathcal{P}_L\hat{C}^{(22)*}_{l\Phi_1,f_2f_1}-\mathcal{P}_R\hat{C}^{(22)}_{l\Phi_1,f_1f_2}\right)\Big)
\end{aligned}
\tag{C.87}
$$

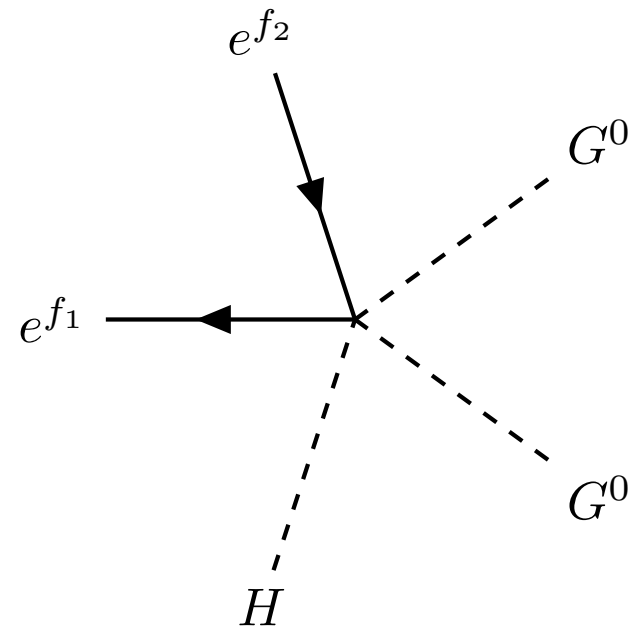

$$\begin{aligned}&+\frac{is_\beta}{\sqrt{2}}\left(c_\beta^2\left(\mathcal{P}_L\hat{C}^{(11)*}_{l\Phi_1,f_2f_1}+\mathcal{P}_R\hat{C}^{(11)}_{l\Phi_1,f_1f_2}\right)\right.\\&\qquad-s_\beta^2\left(2ct_\beta^2+1\right)\left(\mathcal{P}_L\hat{C}^{(12)*}_{l\Phi_2,f_2f_1}+\mathcal{P}_R\hat{C}^{(12)}_{l\Phi_2,f_1f_2}\right)\\&\qquad+s_\beta^2\left(\mathcal{P}_L\hat{C}^{(21)*}_{l\Phi_2,f_2f_1}+\mathcal{P}_R\hat{C}^{(21)}_{l\Phi_2,f_1f_2}\right)\\&\qquad\left.+s_\beta^2\left(\mathcal{P}_L\hat{C}^{(22)*}_{l\Phi_1,f_2f_1}+\mathcal{P}_R\hat{C}^{(22)}_{l\Phi_1,f_1f_2}\right)\right)\end{aligned}\tag{C.88}$$

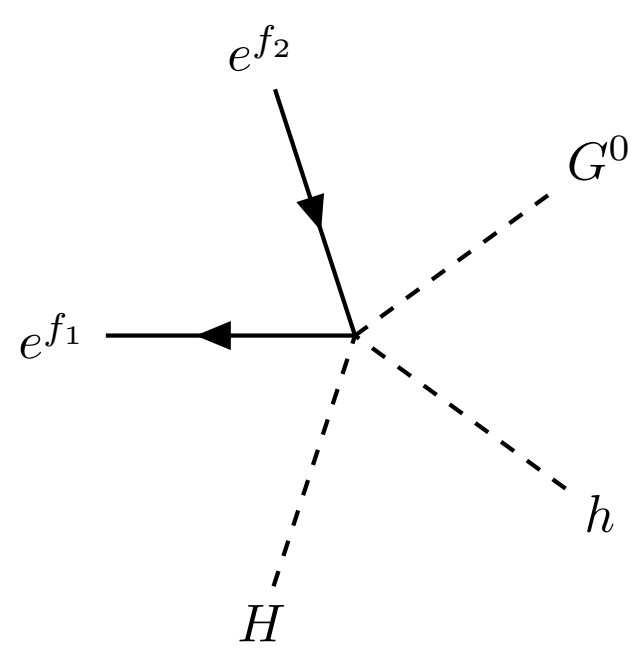

$$\begin{aligned}&+\frac{s_\beta}{\sqrt{2}}\left(c_\beta^2\left(\mathcal{P}_L\hat{C}^{(11)*}_{l\Phi_1,f_2f_1}-\mathcal{P}_R\hat{C}^{(11)}_{l\Phi_1,f_1f_2}\right)\right.\\&\qquad+s_\beta^2\left(\mathcal{P}_L\hat{C}^{(12)*}_{l\Phi_2,f_2f_1}-\mathcal{P}_R\hat{C}^{(12)}_{l\Phi_2,f_1f_2}\right)\\&\qquad-c_\beta^2\left(\mathcal{P}_L\hat{C}^{(21)*}_{l\Phi_2,f_2f_1}-\mathcal{P}_R\hat{C}^{(21)}_{l\Phi_2,f_1f_2}\right)\\&\qquad\left.-c_\beta^2\left(\mathcal{P}_L\hat{C}^{(22)*}_{l\Phi_1,f_2f_1}-\mathcal{P}_R\hat{C}^{(22)}_{l\Phi_1,f_1f_2}\right)\right)\end{aligned}\tag{C.89}$$

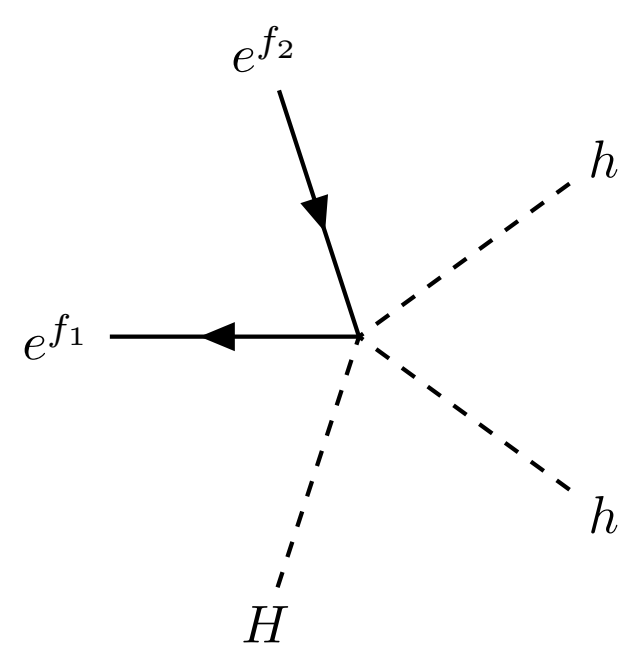

$$\begin{aligned}&+\frac{is_\beta}{\sqrt{2}}\left(3c_\beta^2\left(\mathcal{P}_L\hat{C}^{(11)*}_{l\Phi_1,f_2f_1}+\mathcal{P}_R\hat{C}^{(11)}_{l\Phi_1,f_1f_2}\right)\right.\\&\qquad-s_\beta^2\left(2ct_\beta^2-1\right)\left(\mathcal{P}_L\hat{C}^{(12)*}_{l\Phi_2,f_2f_1}+\mathcal{P}_R\hat{C}^{(12)}_{l\Phi_2,f_1f_2}\right)\\&\qquad-s_\beta^2\left(2ct_\beta^2-1\right)\left(\mathcal{P}_L\hat{C}^{(21)*}_{l\Phi_2,f_2f_1}+\mathcal{P}_R\hat{C}^{(21)}_{l\Phi_2,f_1f_2}\right)\\&\qquad\left.-s_\beta^2\left(2ct_\beta^2-1\right)\left(\mathcal{P}_L\hat{C}^{(22)*}_{l\Phi_1,f_2f_1}+\mathcal{P}_R\hat{C}^{(22)}_{l\Phi_1,f_1f_2}\right)\right)\end{aligned}\tag{C.90}$$

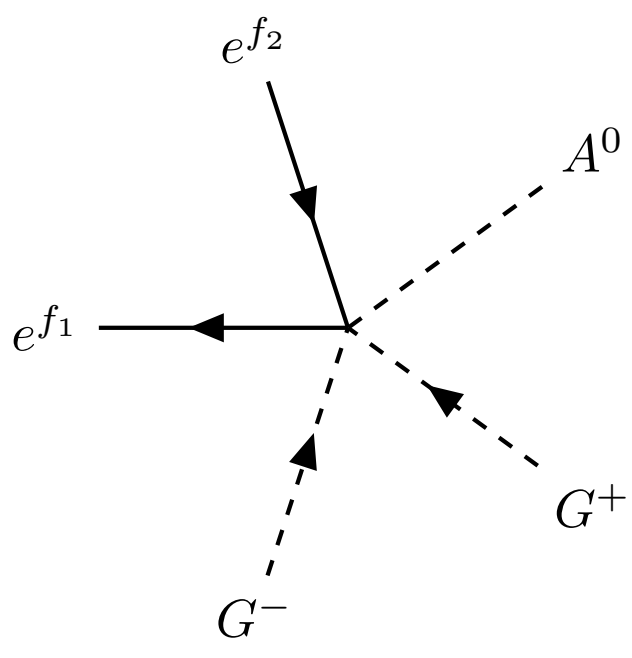

$$
\begin{aligned}
&+\frac{s_\beta}{\sqrt{2}}\Big(-c_\beta^2\Big(\mathcal{P}_L\hat{C}^{(11)*}_{l\Phi_1,f_2f_1}-\mathcal{P}_R\hat{C}^{(11)}_{l\Phi_1,f_1f_2}\Big)\\
&\qquad+c_\beta^2\Big(\mathcal{P}_L\hat{C}^{(12)*}_{l\Phi_2,f_2f_1}-\mathcal{P}_R\hat{C}^{(12)}_{l\Phi_2,f_1f_2}\Big)\\
&\qquad+c_\beta^2\Big(\mathcal{P}_L\hat{C}^{(21)*}_{l\Phi_2,f_2f_1}-\mathcal{P}_R\hat{C}^{(21)}_{l\Phi_2,f_1f_2}\Big)\\
&\qquad-s_\beta^2\Big(\mathcal{P}_L\hat{C}^{(22)*}_{l\Phi_1,f_2f_1}-\mathcal{P}_R\hat{C}^{(22)}_{l\Phi_1,f_1f_2}\Big)\Big)
\end{aligned}
\tag{C.91}
$$

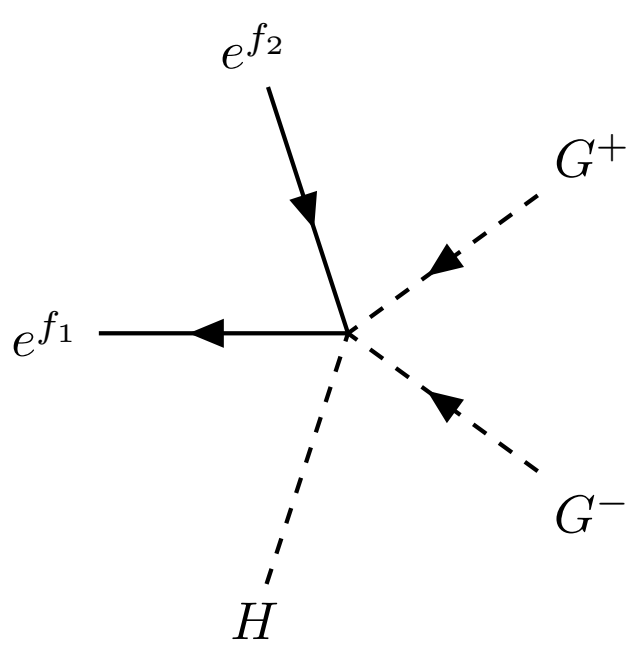

$$
\begin{aligned}
&+\frac{is_\beta}{\sqrt{2}}\Big(c_\beta^2\Big(\mathcal{P}_L\hat{C}^{(11)*}_{l\Phi_1,f_2f_1}+\mathcal{P}_R\hat{C}^{(11)}_{l\Phi_1,f_1f_2}\Big)\\
&\qquad-c_\beta^2\Big(\mathcal{P}_L\hat{C}^{(12)*}_{l\Phi_2,f_2f_1}+\mathcal{P}_R\hat{C}^{(12)}_{l\Phi_2,f_1f_2}\Big)\\
&\qquad-c_\beta^2\Big(\mathcal{P}_L\hat{C}^{(21)*}_{l\Phi_2,f_2f_1}+\mathcal{P}_R\hat{C}^{(21)}_{l\Phi_2,f_1f_2}\Big)\\
&\qquad+s_\beta^2\Big(\mathcal{P}_L\hat{C}^{(22)*}_{l\Phi_1,f_2f_1}+\mathcal{P}_R\hat{C}^{(22)}_{l\Phi_1,f_1f_2}\Big)\Big)
\end{aligned}
\tag{C.92}
$$

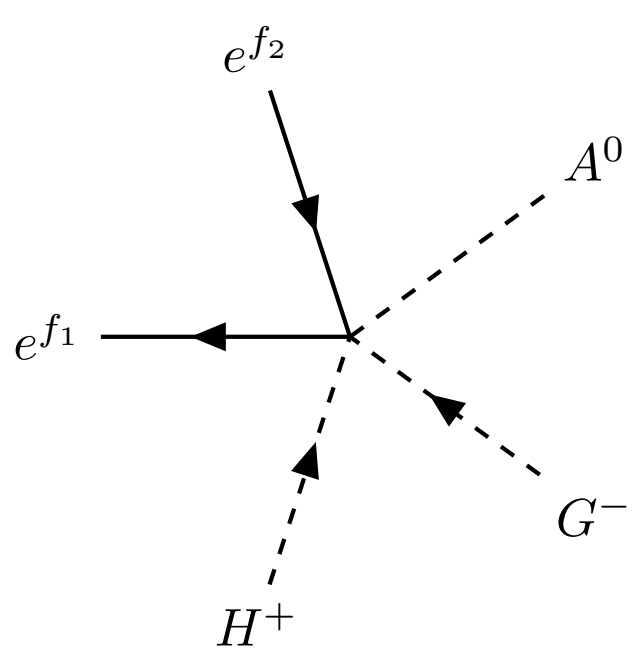

$$
\begin{aligned}
&+\frac{c_\beta}{\sqrt{2}}\Big(s_\beta^2\Big(\mathcal{P}_L\hat{C}^{(11)*}_{l\Phi_1,f_2f_1}-\mathcal{P}_R\hat{C}^{(11)}_{l\Phi_1,f_1f_2}\Big)\\
&\qquad-s_\beta^2\Big(\mathcal{P}_L\hat{C}^{(22)*}_{l\Phi_1,f_2f_1}-\mathcal{P}_R\hat{C}^{(22)}_{l\Phi_1,f_1f_2}\Big)\\
&\qquad-\Big(s_\beta^2\mathcal{P}_L\hat{C}^{(12)*}_{l\Phi_2,f_2f_1}+c_\beta^2\mathcal{P}_R\hat{C}^{(12)}_{l\Phi_2,f_1f_2}\Big)\\
&\qquad+\Big(c_\beta^2\mathcal{P}_L\hat{C}^{(21)*}_{l\Phi_2,f_2f_1}+s_\beta^2\mathcal{P}_R\hat{C}^{(21)}_{l\Phi_2,f_1f_2}\Big)\Big)
\end{aligned}
\tag{C.93}
$$

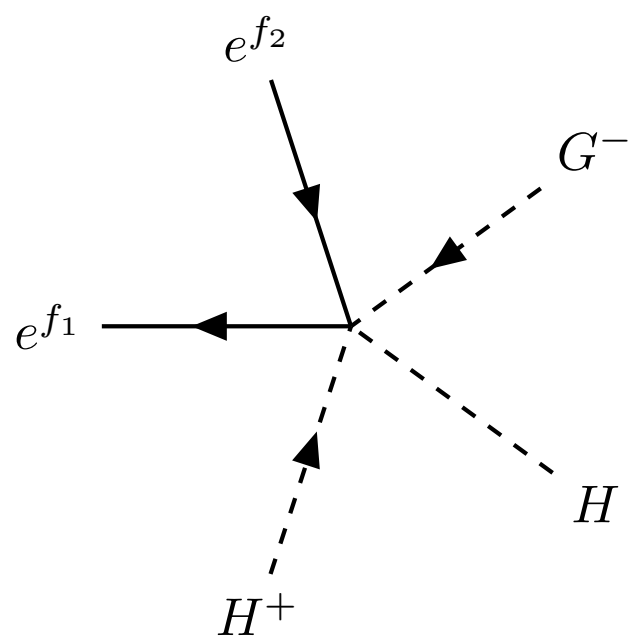

$$
\begin{aligned}
&-\frac{ic_\beta}{\sqrt{2}}\Big(s_\beta^2\Big(\mathcal{P}_L\hat{C}^{(11)*}_{l\Phi_1,f_2f_1}+\mathcal{P}_R\hat{C}^{(11)}_{l\Phi_1,f_1f_2}\Big)\\
&\qquad-s_\beta^2\Big(\mathcal{P}_L\hat{C}^{(22)*}_{l\Phi_1,f_2f_1}+\mathcal{P}_R\hat{C}^{(22)}_{l\Phi_1,f_1f_2}\Big)\\
&\qquad-\Big(s_\beta^2\mathcal{P}_L\hat{C}^{(12)*}_{l\Phi_2,f_2f_1}-c_\beta^2\mathcal{P}_R\hat{C}^{(12)}_{l\Phi_2,f_1f_2}\Big)\\
&\qquad+\Big(c_\beta^2\mathcal{P}_L\hat{C}^{(21)*}_{l\Phi_2,f_2f_1}-s_\beta^2\mathcal{P}_R\hat{C}^{(21)}_{l\Phi_2,f_1f_2}\Big)\Big)
\end{aligned}
\tag{C.94}
$$

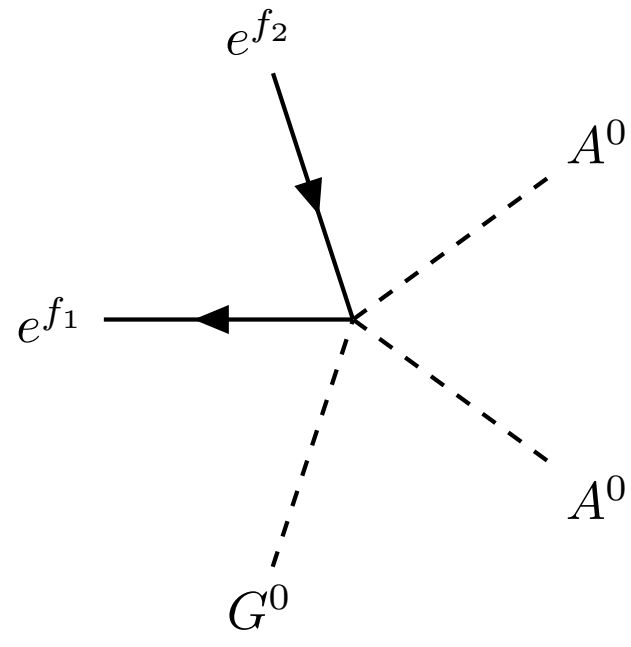

$$
\begin{aligned}
&+\frac{c_\beta}{\sqrt{2}}\Big(3s_\beta^2\left(\mathcal{P}_L\hat{C}^{(11)*}_{l\Phi_1,f_2f_1}-\mathcal{P}_R\hat{C}^{(11)}_{l\Phi_1,f_1f_2}\right)\\
&\qquad -c_\beta^2\left(2t_\beta^2-1\right)\left(\mathcal{P}_L\hat{C}^{(12)*}_{l\Phi_2,f_2f_1}-\mathcal{P}_R\hat{C}^{(12)}_{l\Phi_2,f_1f_2}\right)\\
&\qquad -c_\beta^2\left(2t_\beta^2-1\right)\left(\mathcal{P}_L\hat{C}^{(21)*}_{l\Phi_2,f_2f_1}-\mathcal{P}_R\hat{C}^{(21)}_{l\Phi_2,f_1f_2}\right)\\
&\qquad -c_\beta^2\left(2t_\beta^2-1\right)\left(\mathcal{P}_L\hat{C}^{(22)*}_{l\Phi_1,f_2f_1}-\mathcal{P}_R\hat{C}^{(22)}_{l\Phi_1,f_1f_2}\right)\Big)
\end{aligned}
\tag{C.95}
$$

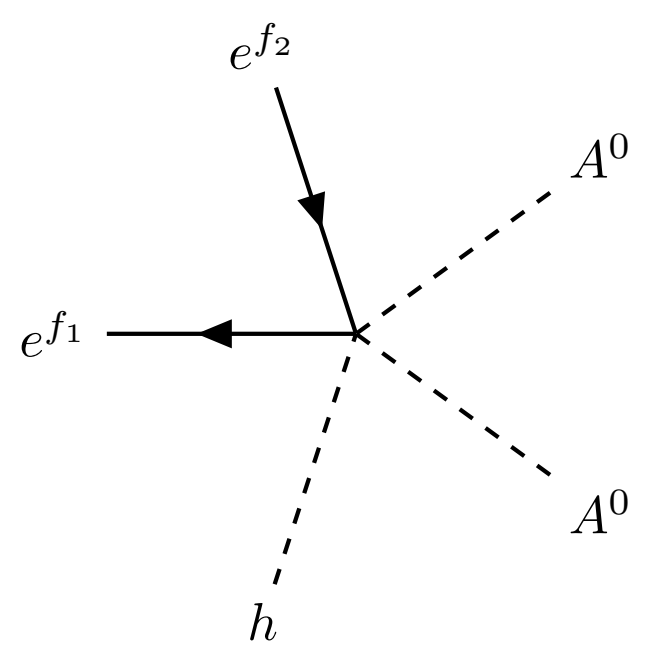

$$
\begin{aligned}
&+\frac{ic_\beta}{\sqrt{2}}\Big(s_\beta^2\left(\mathcal{P}_L\hat{C}^{(11)*}_{l\Phi_1,f_2f_1}+\mathcal{P}_R\hat{C}^{(11)}_{l\Phi_1,f_1f_2}\right)\\
&\qquad -c_\beta^2\left(2t_\beta^2+1\right)\left(\mathcal{P}_L\hat{C}^{(12)*}_{l\Phi_2,f_2f_1}+\mathcal{P}_R\hat{C}^{(12)}_{l\Phi_2,f_1f_2}\right)\\
&\qquad +c_\beta^2\left(\mathcal{P}_L\hat{C}^{(21)*}_{l\Phi_2,f_2f_1}+\mathcal{P}_R\hat{C}^{(21)}_{l\Phi_2,f_1f_2}\right)\\
&\qquad +c_\beta^2\left(\mathcal{P}_L\hat{C}^{(22)*}_{l\Phi_1,f_2f_1}+\mathcal{P}_R\hat{C}^{(22)}_{l\Phi_1,f_1f_2}\right)\Big)
\end{aligned}
\tag{C.96}
$$

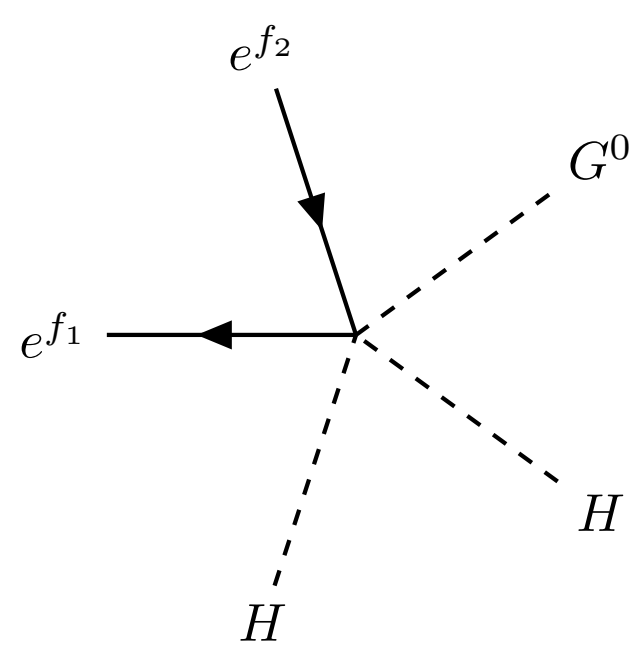

$$
\begin{aligned}
&+\frac{c_\beta}{\sqrt{2}}\Big(s_\beta^2\left(\mathcal{P}_L\hat{C}^{(11)*}_{l\Phi_1,f_2f_1}-\mathcal{P}_R\hat{C}^{(11)}_{l\Phi_1,f_1f_2}\right)\\
&\qquad -c_\beta^2\left(2t_\beta^2+1\right)\left(\mathcal{P}_L\hat{C}^{(12)*}_{l\Phi_2,f_2f_1}-\mathcal{P}_R\hat{C}^{(12)}_{l\Phi_2,f_1f_2}\right)\\
&\qquad +c_\beta^2\left(\mathcal{P}_L\hat{C}^{(21)*}_{l\Phi_2,f_2f_1}-\mathcal{P}_R\hat{C}^{(21)}_{l\Phi_2,f_1f_2}\right)\\
&\qquad +c_\beta^2\left(\mathcal{P}_L\hat{C}^{(22)*}_{l\Phi_1,f_2f_1}-\mathcal{P}_R\hat{C}^{(22)}_{l\Phi_1,f_1f_2}\right)\Big)
\end{aligned}
\tag{C.97}
$$

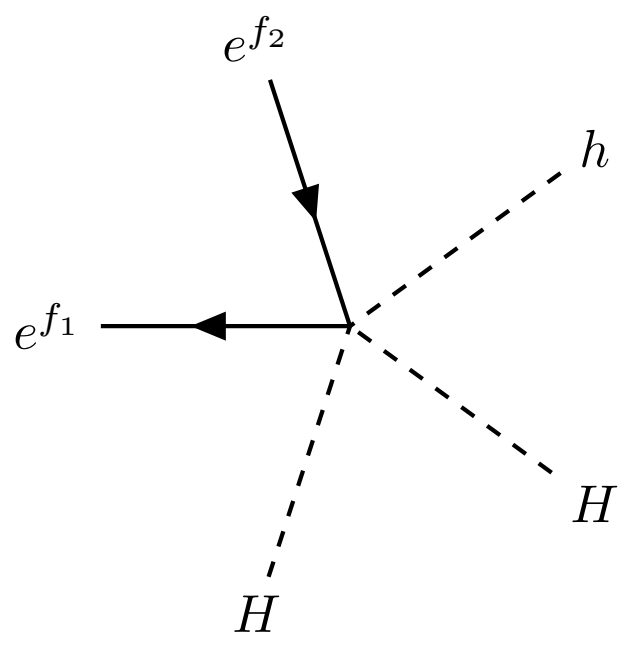

$$
\begin{aligned}
&+\frac{ic_\beta}{\sqrt{2}}\Big(3s_\beta^2\left(\mathcal{P}_L\hat{C}^{(11)*}_{l\Phi_1,f_2f_1}+\mathcal{P}_R\hat{C}^{(11)}_{l\Phi_1,f_1f_2}\right)\\
&\qquad -c_\beta^2\left(2t_\beta^2-1\right)\left(\mathcal{P}_L\hat{C}^{(12)*}_{l\Phi_2,f_2f_1}+\mathcal{P}_R\hat{C}^{(12)}_{l\Phi_2,f_1f_2}\right)\\
&\qquad -c_\beta^2\left(2t_\beta^2-1\right)\left(\mathcal{P}_L\hat{C}^{(21)*}_{l\Phi_2,f_2f_1}+\mathcal{P}_R\hat{C}^{(21)}_{l\Phi_2,f_1f_2}\right)\\
&\qquad -c_\beta^2\left(2t_\beta^2-1\right)\left(\mathcal{P}_L\hat{C}^{(22)*}_{l\Phi_1,f_2f_1}+\mathcal{P}_R\hat{C}^{(22)}_{l\Phi_1,f_1f_2}\right)\Big)
\end{aligned}
\tag{C.98}
$$

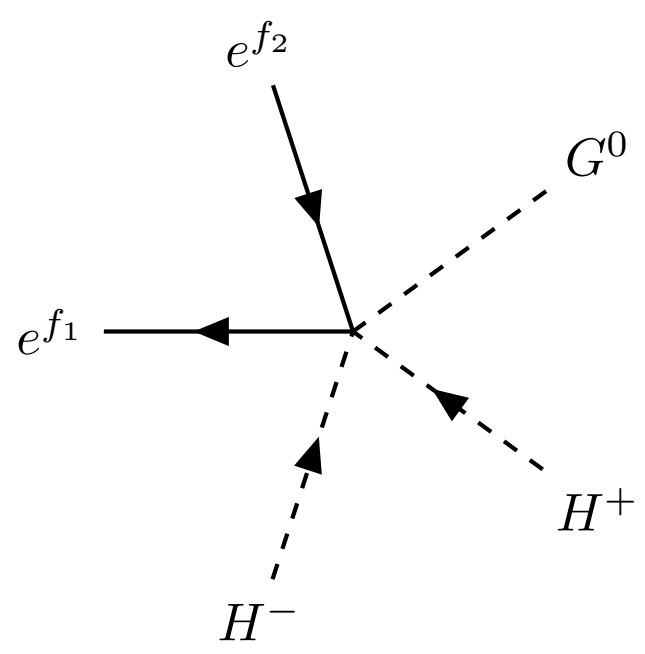

$$
\begin{aligned}
&+\frac{c_\beta}{\sqrt{2}}\Big(s_\beta^2\left(\mathcal{P}_L\hat{C}^{(11)*}_{l\Phi_1,f_2f_1}-\mathcal{P}_R\hat{C}^{(11)}_{l\Phi_1,f_1f_2}\right)\\
&\qquad -s_\beta^2\left(\mathcal{P}_L\hat{C}^{(12)*}_{l\Phi_2,f_2f_1}-\mathcal{P}_R\hat{C}^{(12)}_{l\Phi_2,f_1f_2}\right)\\
&\qquad -s_\beta^2\left(\mathcal{P}_L\hat{C}^{(21)*}_{l\Phi_2,f_2f_1}-\mathcal{P}_R\hat{C}^{(21)}_{l\Phi_2,f_1f_2}\right)\\
&\qquad +c_\beta^2\left(\mathcal{P}_L\hat{C}^{(22)*}_{l\Phi_1,f_2f_1}-\mathcal{P}_R\hat{C}^{(22)}_{l\Phi_1,f_1f_2}\right)\Big)
\end{aligned}
\tag{C.99}
$$

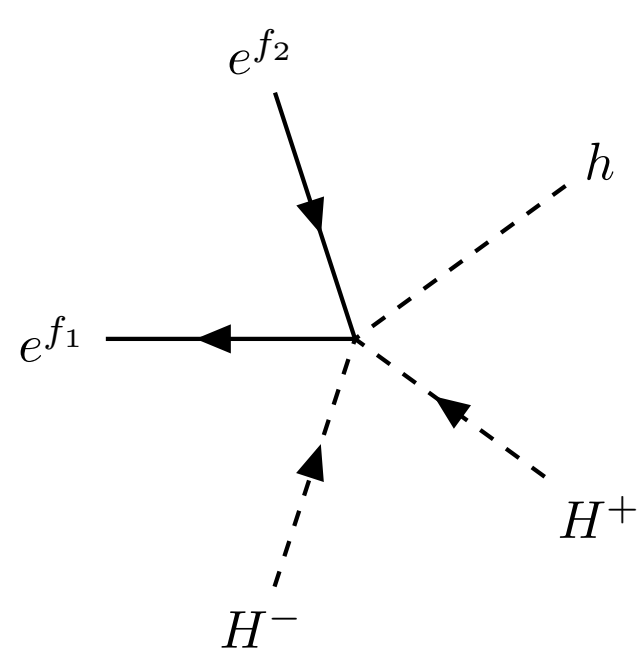

$$
\begin{aligned}
&+\frac{ic_\beta}{\sqrt{2}}\Big(s_\beta^2\left(\mathcal{P}_L\hat{C}^{(11)*}_{l\Phi_1,f_2f_1}+\mathcal{P}_R\hat{C}^{(11)}_{l\Phi_1,f_1f_2}\right)\\
&\qquad -s_\beta^2\left(\mathcal{P}_L\hat{C}^{(12)*}_{l\Phi_2,f_2f_1}+\mathcal{P}_R\hat{C}^{(12)}_{l\Phi_2,f_1f_2}\right)\\
&\qquad -s_\beta^2\left(\mathcal{P}_L\hat{C}^{(21)*}_{l\Phi_2,f_2f_1}+\mathcal{P}_R\hat{C}^{(21)}_{l\Phi_2,f_1f_2}\right)\\
&\qquad +c_\beta^2\left(\mathcal{P}_L\hat{C}^{(22)*}_{l\Phi_1,f_2f_1}+\mathcal{P}_R\hat{C}^{(22)}_{l\Phi_1,f_1f_2}\right)\Big)
\end{aligned}
\tag{C.100}
$$

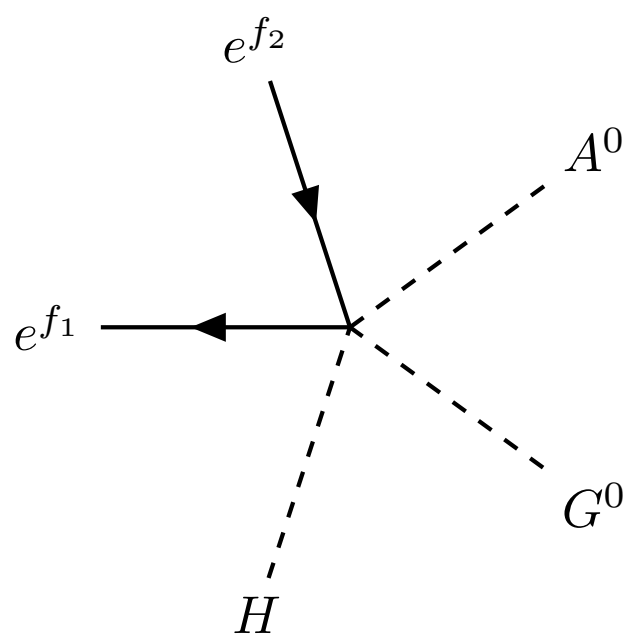

$$
\begin{aligned}
&-\frac{ic_\beta}{\sqrt{2}}\Big(s_\beta^2\left(\mathcal{P}_L\hat{C}^{(11)*}_{l\Phi_1,f_2f_1}+\mathcal{P}_R\hat{C}^{(11)}_{l\Phi_1,f_1f_2}\right)\\
&\qquad +c_\beta^2\left(\mathcal{P}_L\hat{C}^{(12)*}_{l\Phi_2,f_2f_1}+\mathcal{P}_R\hat{C}^{(12)}_{l\Phi_2,f_1f_2}\right)\\
&\qquad -s_\beta^2\left(\mathcal{P}_L\hat{C}^{(21)*}_{l\Phi_2,f_2f_1}+\mathcal{P}_R\hat{C}^{(21)}_{l\Phi_2,f_1f_2}\right)\\
&\qquad -s_\beta^2\left(\mathcal{P}_L\hat{C}^{(22)*}_{l\Phi_1,f_2f_1}+\mathcal{P}_R\hat{C}^{(22)}_{l\Phi_1,f_1f_2}\right)\Big)
\end{aligned}
\tag{C.101}
$$

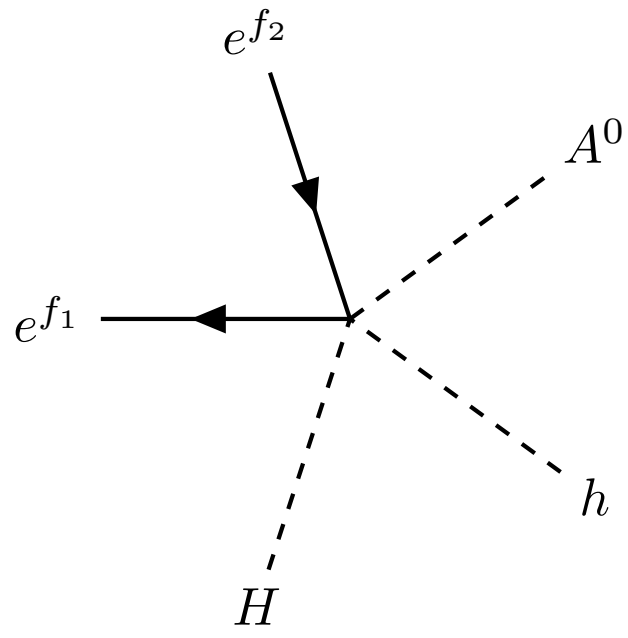

$$
\begin{aligned}
&+\frac{c_\beta}{\sqrt{2}}\left(-s_\beta^2\left(\mathcal{P}_L\hat{C}^{(11)*}_{l\Phi_1,f_2f_1}-\mathcal{P}_R\hat{C}^{(11)}_{l\Phi_1,f_1f_2}\right)\right.\\
&\qquad -c_\beta^2\left(\mathcal{P}_L\hat{C}^{(12)*}_{l\Phi_2,f_2f_1}-\mathcal{P}_R\hat{C}^{(12)}_{l\Phi_2,f_1f_2}\right)\\
&\qquad +s_\beta^2\left(\mathcal{P}_L\hat{C}^{(21)*}_{l\Phi_2,f_2f_1}-\mathcal{P}_R\hat{C}^{(21)}_{l\Phi_2,f_1f_2}\right)\\
&\qquad \left.+s_\beta^2\left(\mathcal{P}_L\hat{C}^{(22)*}_{l\Phi_1,f_2f_1}-\mathcal{P}_R\hat{C}^{(22)}_{l\Phi_1,f_1f_2}\right)\right)
\end{aligned}
\tag{C.102}
$$

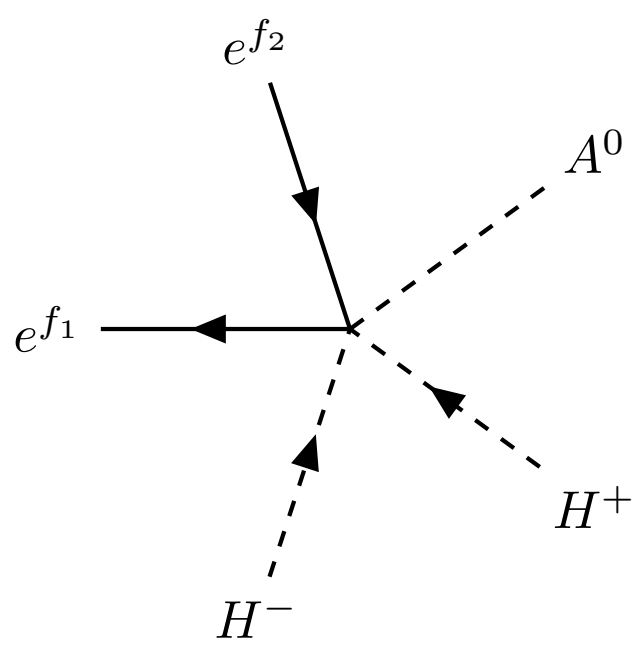

$$
\begin{aligned}
&+\frac{s_\beta}{\sqrt{2}}\left(-s_\beta^2\left(\mathcal{P}_L\hat{C}^{(11)*}_{l\Phi_1,f_2f_1}-\mathcal{P}_R\hat{C}^{(11)}_{l\Phi_1,f_1f_2}\right)\right.\\
&\qquad -c_\beta^2\left(\mathcal{P}_L\hat{C}^{(12)*}_{l\Phi_2,f_2f_1}-\mathcal{P}_R\hat{C}^{(12)}_{l\Phi_2,f_1f_2}\right)\\
&\qquad -c_\beta^2\left(\mathcal{P}_L\hat{C}^{(21)*}_{l\Phi_2,f_2f_1}-\mathcal{P}_R\hat{C}^{(21)}_{l\Phi_2,f_1f_2}\right)\\
&\qquad \left.-c_\beta^2\left(\mathcal{P}_L\hat{C}^{(22)*}_{l\Phi_1,f_2f_1}-\mathcal{P}_R\hat{C}^{(22)}_{l\Phi_1,f_1f_2}\right)\right)
\end{aligned}
\tag{C.103}
$$

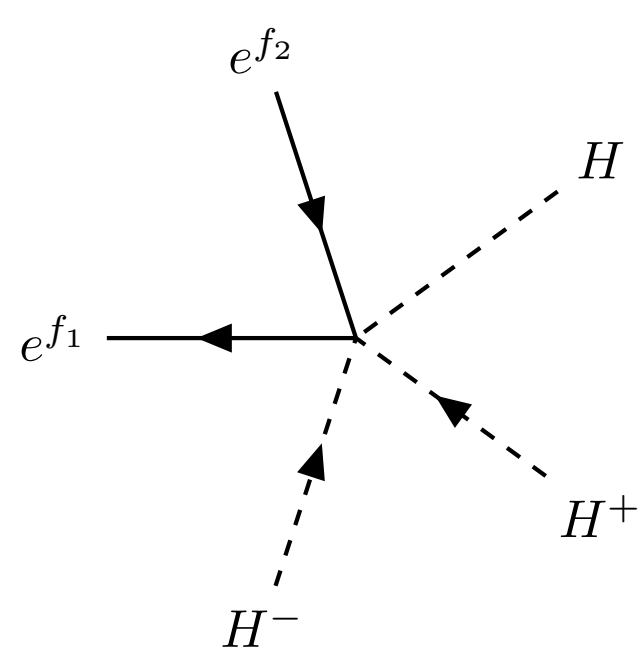

$$
\begin{aligned}
&+\frac{is_\beta}{\sqrt{2}}\left(s_\beta^2\left(\mathcal{P}_L\hat{C}^{(11)*}_{l\Phi_1,f_2f_1}+\mathcal{P}_R\hat{C}^{(11)}_{l\Phi_1,f_1f_2}\right)\right.\\
&\qquad +c_\beta^2\left(\mathcal{P}_L\hat{C}^{(12)*}_{l\Phi_2,f_2f_1}+\mathcal{P}_R\hat{C}^{(12)}_{l\Phi_2,f_1f_2}\right)\\
&\qquad +c_\beta^2\left(\mathcal{P}_L\hat{C}^{(21)*}_{l\Phi_2,f_2f_1}+\mathcal{P}_R\hat{C}^{(21)}_{l\Phi_2,f_1f_2}\right)\\
&\qquad \left.+c_\beta^2\left(\mathcal{P}_L\hat{C}^{(22)*}_{l\Phi_1,f_2f_1}+\mathcal{P}_R\hat{C}^{(22)}_{l\Phi_1,f_1f_2}\right)\right)
\end{aligned}
\tag{C.104}
$$

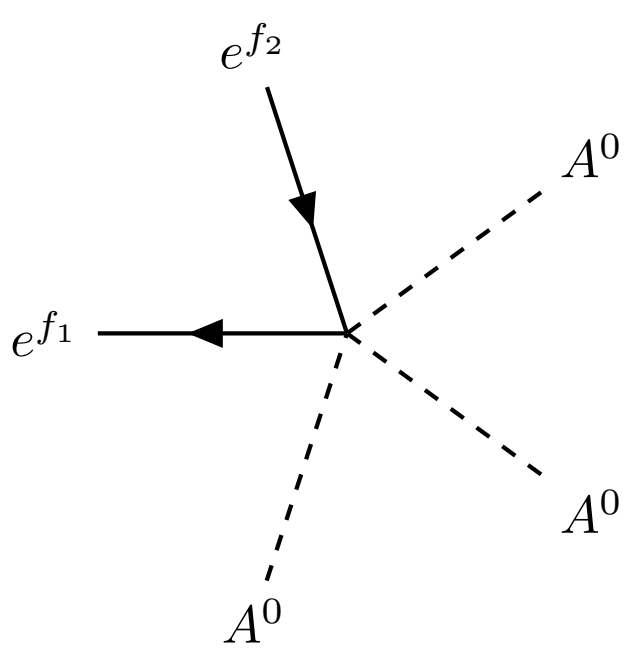

$$-\frac{3s_\beta}{\sqrt{2}}\left(s_\beta^2\left(\mathcal{P}_L\hat{C}^{(11)*}_{l\Phi_1,f_2f_1}-\mathcal{P}_R\hat{C}^{(11)}_{l\Phi_1,f_1f_2}\right)\right.$$
$$+c_\beta^2\left(\mathcal{P}_L\hat{C}^{(12)*}_{l\Phi_2,f_2f_1}-\mathcal{P}_R\hat{C}^{(12)}_{l\Phi_2,f_1f_2}\right)$$
$$+c_\beta^2\left(\mathcal{P}_L\hat{C}^{(21)*}_{l\Phi_2,f_2f_1}-\mathcal{P}_R\hat{C}^{(21)}_{l\Phi_2,f_1f_2}\right)$$
$$\left.+c_\beta^2\left(\mathcal{P}_L\hat{C}^{(22)*}_{l\Phi_1,f_2f_1}-\mathcal{P}_R\hat{C}^{(22)}_{l\Phi_1,f_1f_2}\right)\right) \tag{C.105}$$

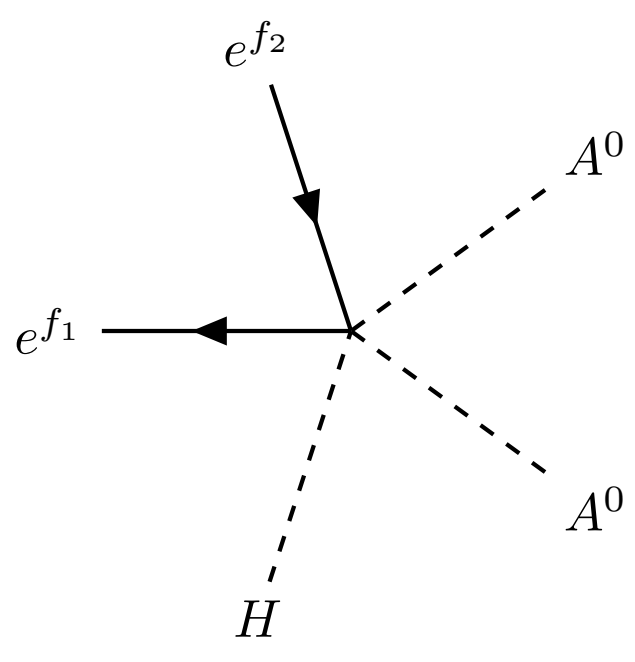

$$+\frac{is_\beta}{\sqrt{2}}\left(s_\beta^2\left(\mathcal{P}_L\hat{C}^{(11)*}_{l\Phi_1,f_2f_1}+\mathcal{P}_R\hat{C}^{(11)}_{l\Phi_1,f_1f_2}\right)\right.$$
$$+c_\beta^2\left(\mathcal{P}_L\hat{C}^{(12)*}_{l\Phi_2,f_2f_1}+\mathcal{P}_R\hat{C}^{(12)}_{l\Phi_2,f_1f_2}\right)$$
$$+c_\beta^2\left(\mathcal{P}_L\hat{C}^{(21)*}_{l\Phi_2,f_2f_1}+\mathcal{P}_R\hat{C}^{(21)}_{l\Phi_2,f_1f_2}\right)$$
$$\left.+c_\beta^2\left(\mathcal{P}_L\hat{C}^{(22)*}_{l\Phi_1,f_2f_1}+\mathcal{P}_R\hat{C}^{(22)}_{l\Phi_1,f_1f_2}\right)\right) \tag{C.106}$$

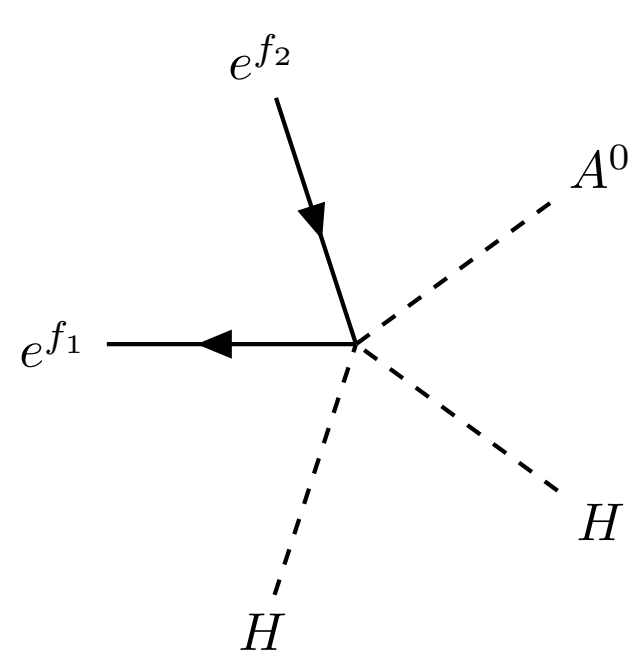

$$+\frac{s_\beta}{\sqrt{2}}\left(-s_\beta^2\left(\mathcal{P}_L\hat{C}^{(11)*}_{l\Phi_1,f_2f_1}-\mathcal{P}_R\hat{C}^{(11)}_{l\Phi_1,f_1f_2}\right)\right.$$
$$-c_\beta^2\left(\mathcal{P}_L\hat{C}^{(12)*}_{l\Phi_2,f_2f_1}-\mathcal{P}_R\hat{C}^{(12)}_{l\Phi_2,f_1f_2}\right)$$
$$-c_\beta^2\left(\mathcal{P}_L\hat{C}^{(21)*}_{l\Phi_2,f_2f_1}-\mathcal{P}_R\hat{C}^{(21)}_{l\Phi_2,f_1f_2}\right)$$
$$\left.-c_\beta^2\left(\mathcal{P}_L\hat{C}^{(22)*}_{l\Phi_1,f_2f_1}-\mathcal{P}_R\hat{C}^{(22)}_{l\Phi_1,f_1f_2}\right)\right) \tag{C.107}$$

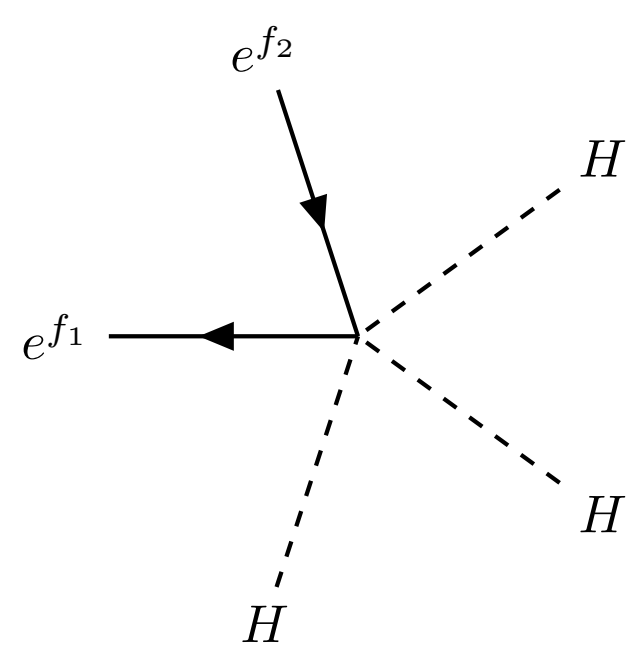

$$+\frac{3is_\beta}{\sqrt{2}}\left(s_\beta^2\left(\mathcal{P}_L\hat{C}^{(11)*}_{l\Phi_1,f_2f_1}+\mathcal{P}_R\hat{C}^{(11)}_{l\Phi_1,f_1f_2}\right)\right.$$
$$+c_\beta^2\left(\mathcal{P}_L\hat{C}^{(12)*}_{l\Phi_2,f_2f_1}+\mathcal{P}_R\hat{C}^{(12)}_{l\Phi_2,f_1f_2}\right)$$
$$+c_\beta^2\left(\mathcal{P}_L\hat{C}^{(21)*}_{l\Phi_2,f_2f_1}+\mathcal{P}_R\hat{C}^{(21)}_{l\Phi_2,f_1f_2}\right)$$
$$\left.+c_\beta^2\left(\mathcal{P}_L\hat{C}^{(22)*}_{l\Phi_1,f_2f_1}+\mathcal{P}_R\hat{C}^{(22)}_{l\Phi_1,f_1f_2}\right)\right) \tag{C.108}$$

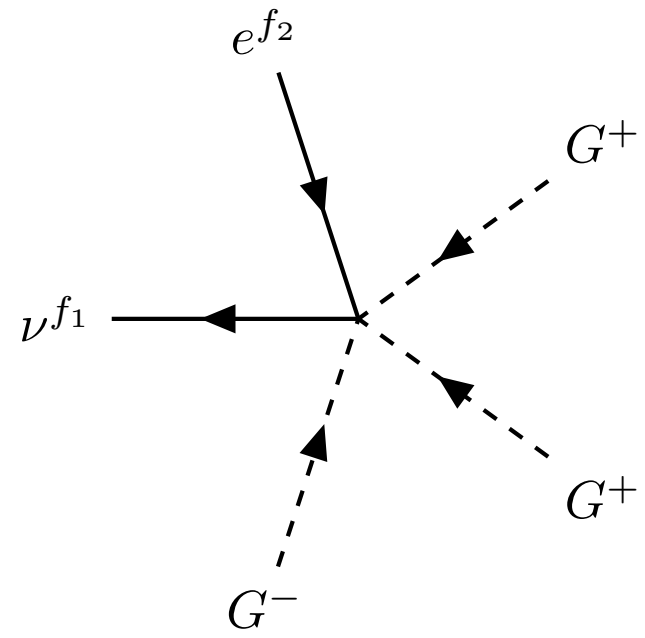

$$+2ic_\beta \mathcal{P}_R U^*_{g_1 f_1} \left( c^2_\beta \hat{C}^{(11)}_{l\Phi_1, g_1 f_2} + s^2_\beta \hat{C}^{(12)}_{l\Phi_2, g_1 f_2} + s^2_\beta \hat{C}^{(21)}_{l\Phi_2, g_1 f_2} + s^2_\beta \hat{C}^{(22)}_{l\Phi_1, g_1 f_2} \right) \tag{C.109}$$

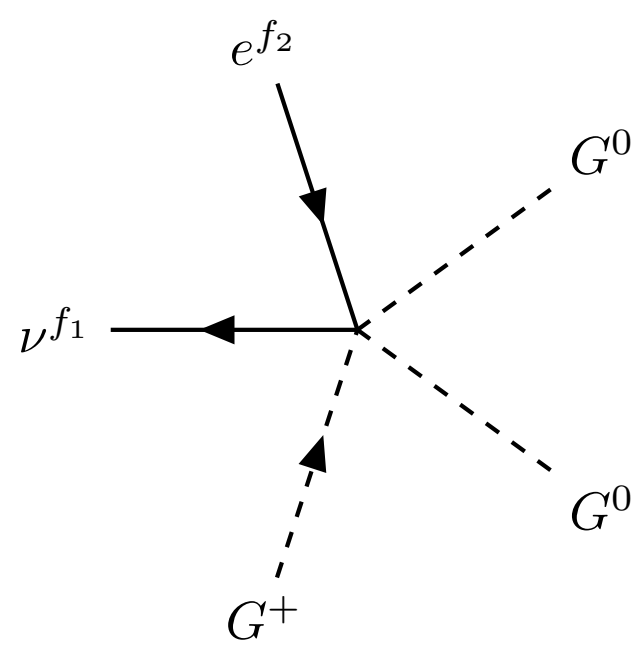

$$+ic_\beta \mathcal{P}_R U^*_{g_1 f_1} \left( c^2_\beta \hat{C}^{(11)}_{l\Phi_1, g_1 f_2} + s^2_\beta \hat{C}^{(12)}_{l\Phi_2, g_1 f_2} + s^2_\beta \hat{C}^{(21)}_{l\Phi_2, g_1 f_2} + s^2_\beta \hat{C}^{(22)}_{l\Phi_1, g_1 f_2} \right) \tag{C.110}$$

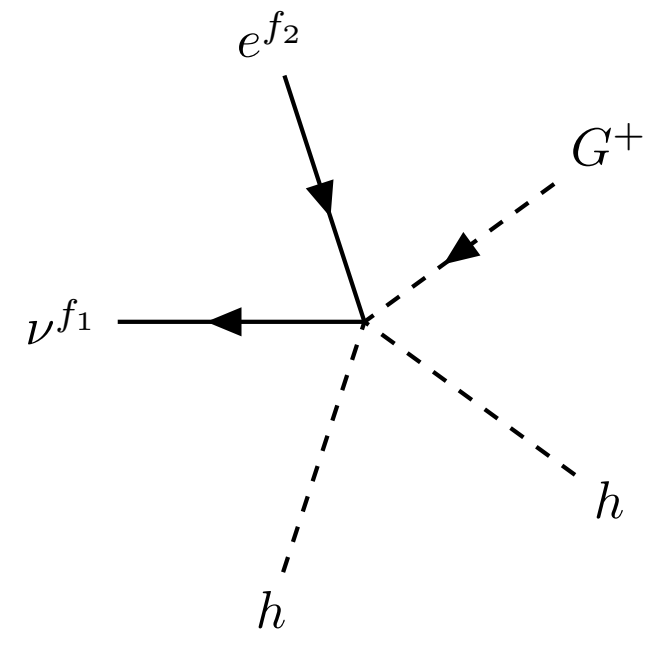

$$+ic_\beta \mathcal{P}_R U^*_{g_1 f_1} \left( c^2_\beta \hat{C}^{(11)}_{l\Phi_1, g_1 f_2} + s^2_\beta \hat{C}^{(12)}_{l\Phi_2, g_1 f_2} + s^2_\beta \hat{C}^{(21)}_{l\Phi_2, g_1 f_2} + s^2_\beta \hat{C}^{(22)}_{l\Phi_1, g_1 f_2} \right) \tag{C.111}$$

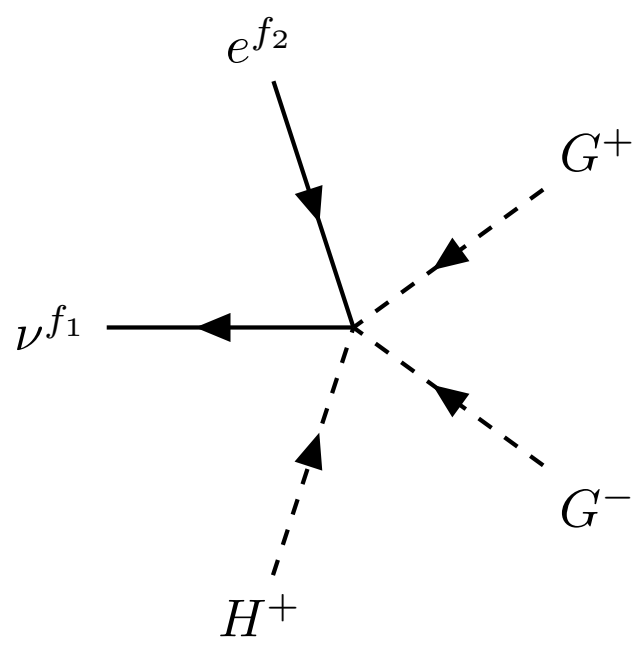

$$-is_\beta \mathcal{P}_R U^*_{g_1 f_1} \Big( 2c_\beta^2 \hat{C}^{(11)}_{l\Phi_1, g_1 f_2} - 2c_\beta^2 \hat{C}^{(12)}_{l\Phi_2, g_1 f_2} \\ + s_\beta^2 \hat{C}^{(21)}_{l\Phi_2, g_1 f_2} - c_\beta^2 \hat{C}^{(21)}_{l\Phi_2, g_1 f_2} \\ + s_\beta^2 \hat{C}^{(22)}_{l\Phi_1, g_1 f_2} - c_\beta^2 \hat{C}^{(22)}_{l\Phi_1, g_1 f_2} \Big) \tag{C.112}$$

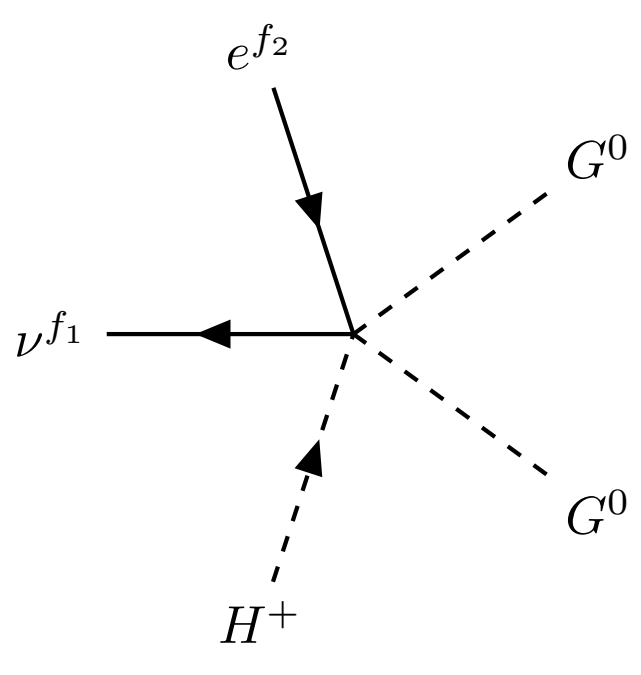

$$-is_\beta \mathcal{P}_R U^*_{g_1 f_1} \Big( c_\beta^2 \hat{C}^{(11)}_{l\Phi_1, g_1 f_2} - c_\beta^2 \hat{C}^{(12)}_{l\Phi_2, g_1 f_2} \\ - c_\beta^2 \hat{C}^{(21)}_{l\Phi_2, g_1 f_2} + s_\beta^2 \hat{C}^{(22)}_{l\Phi_1, g_1 f_2} \Big) \tag{C.113}$$

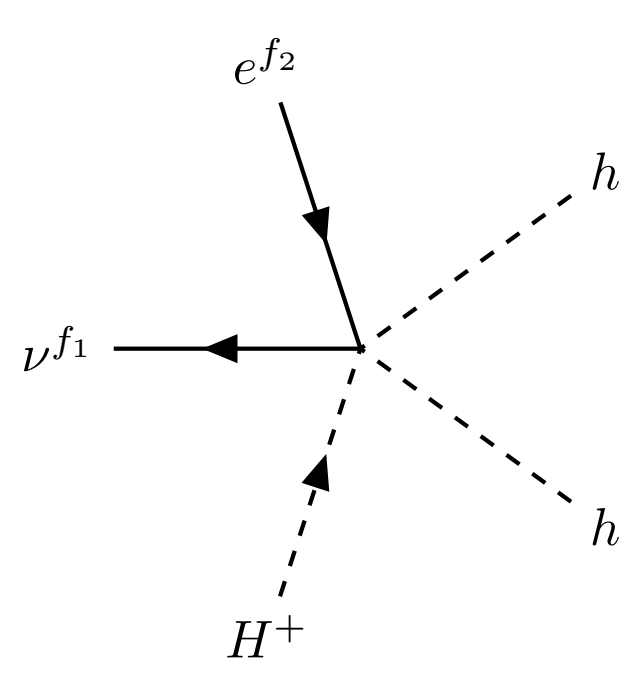

$$-is_\beta \mathcal{P}_R U^*_{g_1 f_1} \Big( c_\beta^2 \hat{C}^{(11)}_{l\Phi_1, g_1 f_2} - c_\beta^2 \hat{C}^{(12)}_{l\Phi_2, g_1 f_2} \\ - c_\beta^2 \hat{C}^{(21)}_{l\Phi_2, g_1 f_2} + s_\beta^2 \hat{C}^{(22)}_{l\Phi_1, g_1 f_2} \Big) \tag{C.114}$$

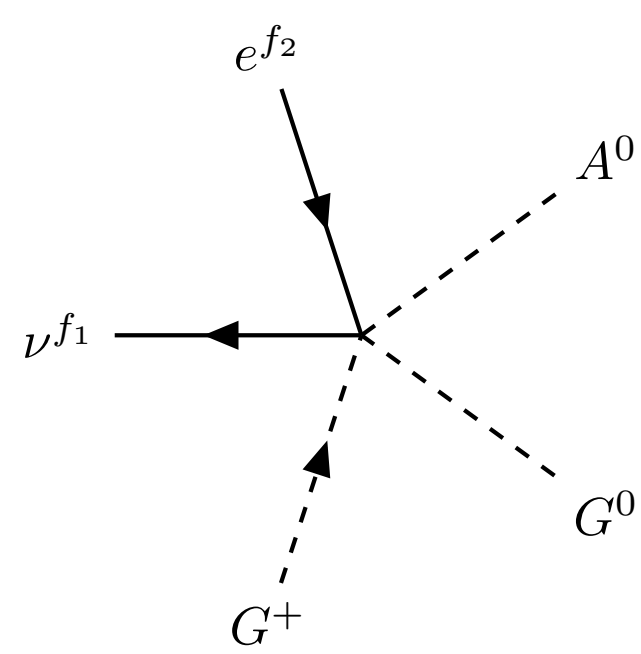

$$-\frac{1}{2} is_\beta \mathcal{P}_R U^*_{g_1 f_1} \Big( 2c_\beta^2 \hat{C}^{(11)}_{l\Phi_1, g_1 f_2} + s_\beta^2 \hat{C}^{(12)}_{l\Phi_2, g_1 f_2} \\ - c_\beta^2 \hat{C}^{(12)}_{l\Phi_2, g_1 f_2} + s_\beta^2 \hat{C}^{(21)}_{l\Phi_2, g_1 f_2} \\ - c_\beta^2 \hat{C}^{(21)}_{l\Phi_2, g_1 f_2} - 2c_\beta^2 \hat{C}^{(22)}_{l\Phi_1, g_1 f_2} \Big) \tag{C.115}$$

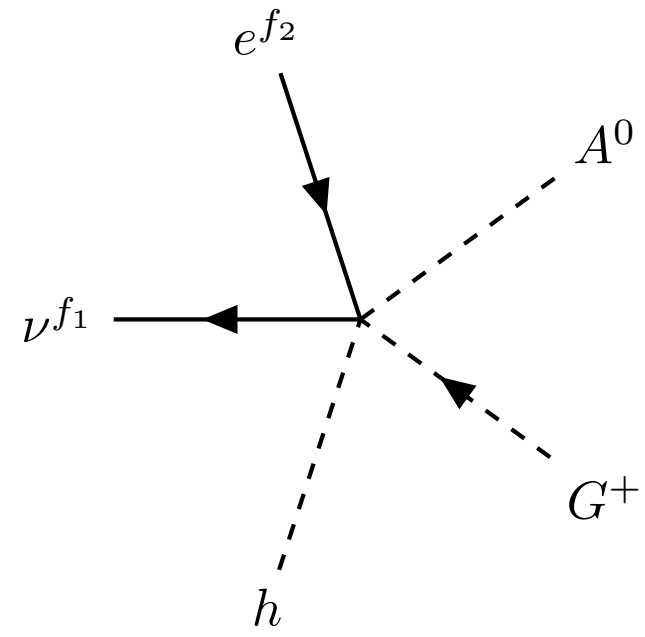

$$-\frac{1}{2}s_\beta \mathcal{P}_R U^*_{g_1f_1}\Big(s_\beta^2\hat{C}^{(12)}_{l\Phi_2,g_1f_2}+c_\beta^2\hat{C}^{(12)}_{l\Phi_2,g_1f_2} -s_\beta^2\hat{C}^{(21)}_{l\Phi_2,g_1f_2}-c_\beta^2\hat{C}^{(21)}_{l\Phi_2,g_1f_2}\Big) \tag{C.116}$$

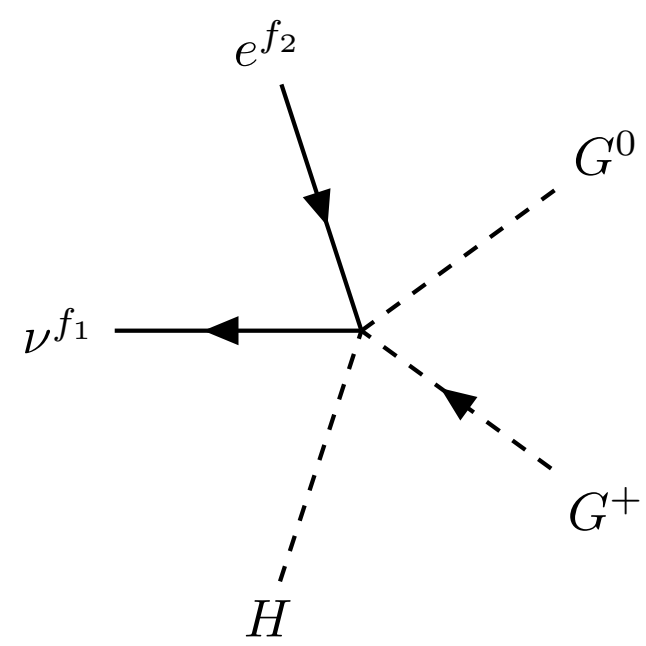

$$-\frac{1}{2}s_\beta \mathcal{P}_R U^*_{g_1f_1}\Big(s_\beta^2\hat{C}^{(12)}_{l\Phi_2,g_1f_2}+c_\beta^2\hat{C}^{(12)}_{l\Phi_2,g_1f_2} -s_\beta^2\hat{C}^{(21)}_{l\Phi_2,g_1f_2}-c_\beta^2\hat{C}^{(21)}_{l\Phi_2,g_1f_2}\Big) \tag{C.117}$$

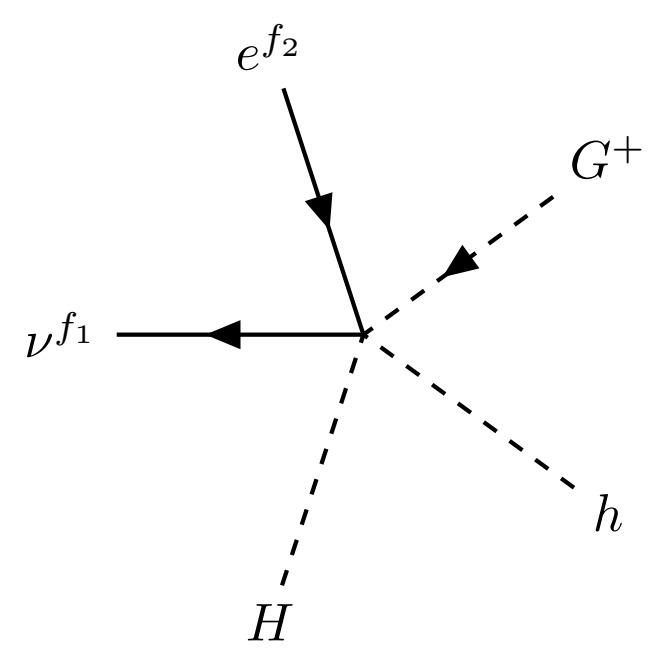

$$+\frac{1}{2}is_\beta \mathcal{P}_R U^*_{g_1f_1}\Big(2c_\beta^2\hat{C}^{(11)}_{l\Phi_1,g_1f_2}+s_\beta^2\hat{C}^{(12)}_{l\Phi_2,g_1f_2} -c_\beta^2\hat{C}^{(12)}_{l\Phi_2,g_1f_2}+s_\beta^2\hat{C}^{(21)}_{l\Phi_2,g_1f_2} -c_\beta^2\hat{C}^{(21)}_{l\Phi_2,g_1f_2}-2c_\beta^2\hat{C}^{(22)}_{l\Phi_1,g_1f_2}\Big) \tag{C.118}$$

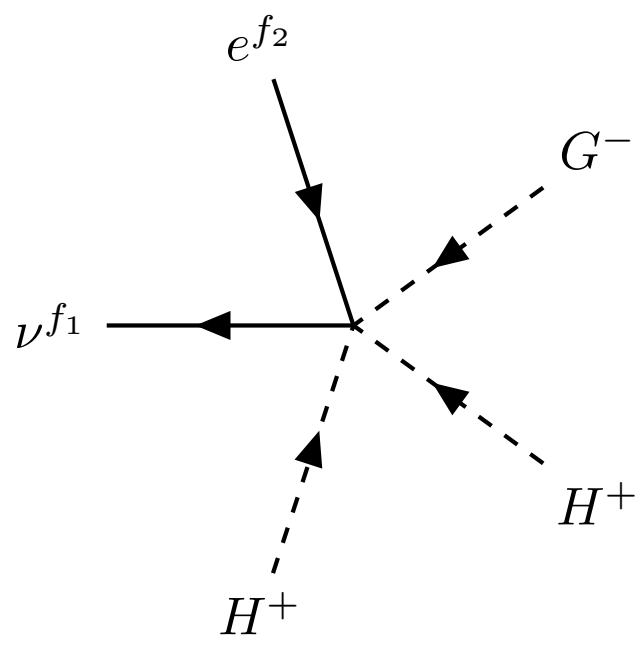

$$
\begin{aligned}
&+2ic_\beta\mathcal{P}_R U^*_{g_1 f_1}\Big(s_\beta^2\hat{C}^{(11)}_{l\Phi_1,g_1f_2}+c_\beta^2\hat{C}^{(12)}_{l\Phi_2,g_1f_2}\\
&\qquad -s_\beta^2\hat{C}^{(21)}_{l\Phi_2,g_1f_2}-s_\beta^2\hat{C}^{(22)}_{l\Phi_1,g_1f_2}\Big)
\end{aligned}
\tag{C.119}
$$

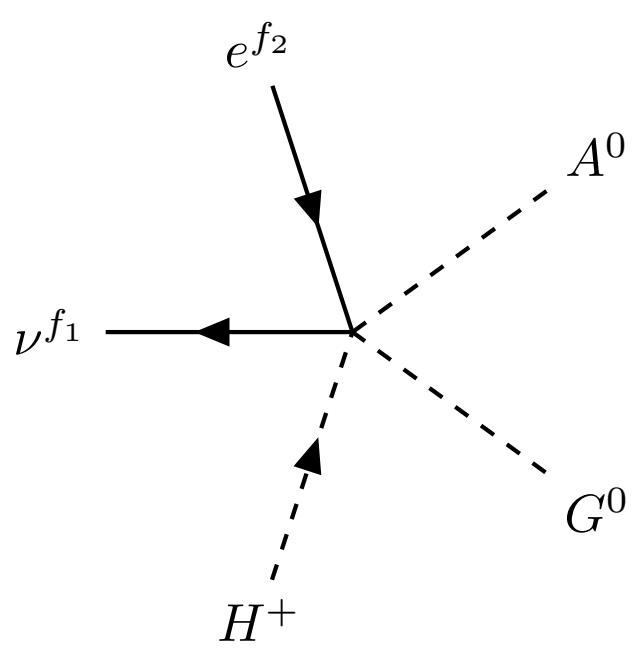

$$
\begin{aligned}
&+\frac{1}{2}ic_\beta\mathcal{P}_R U^*_{g_1 f_1}\Big(2s_\beta^2\hat{C}^{(11)}_{l\Phi_1,g_1f_2}-s_\beta^2\hat{C}^{(12)}_{l\Phi_2,g_1f_2}\\
&\qquad +c_\beta^2\hat{C}^{(12)}_{l\Phi_2,g_1f_2}-s_\beta^2\hat{C}^{(21)}_{l\Phi_2,g_1f_2}\\
&\qquad +c_\beta^2\hat{C}^{(21)}_{l\Phi_2,g_1f_2}-2s_\beta^2\hat{C}^{(22)}_{l\Phi_1,g_1f_2}\Big)
\end{aligned}
\tag{C.120}
$$

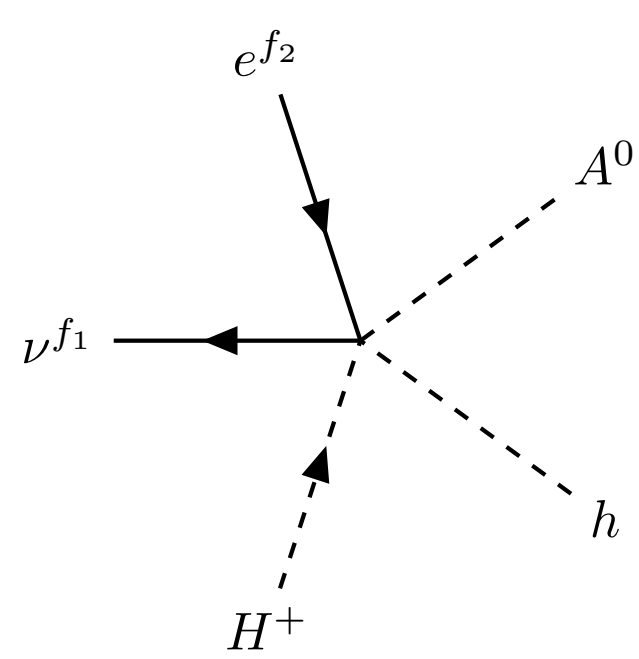

$$
\begin{aligned}
&-\frac{1}{2}c_\beta\mathcal{P}_R U^*_{g_1 f_1}\Big(s_\beta^2\hat{C}^{(12)}_{l\Phi_2,g_1f_2}+c_\beta^2\hat{C}^{(12)}_{l\Phi_2,g_1f_2}\\
&\qquad -s_\beta^2\hat{C}^{(21)}_{l\Phi_2,g_1f_2}-c_\beta^2\hat{C}^{(21)}_{l\Phi_2,g_1f_2}\Big)
\end{aligned}
\tag{C.121}
$$

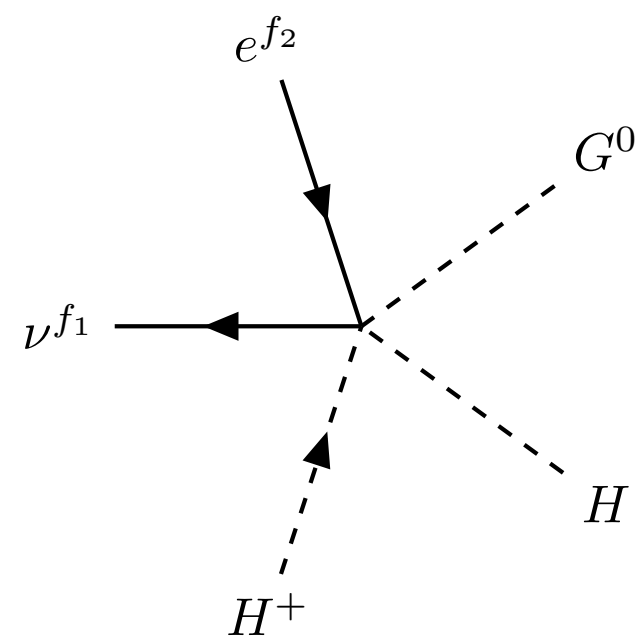

$$
\begin{aligned}
&-\frac{1}{2}c_\beta\mathcal{P}_R U^*_{g_1 f_1}\Big(s_\beta^2\hat{C}^{(12)}_{l\Phi_2,g_1f_2}+c_\beta^2\hat{C}^{(12)}_{l\Phi_2,g_1f_2}\\
&\qquad -s_\beta^2\hat{C}^{(21)}_{l\Phi_2,g_1f_2}-c_\beta^2\hat{C}^{(21)}_{l\Phi_2,g_1f_2}\Big)
\end{aligned}
\tag{C.122}
$$

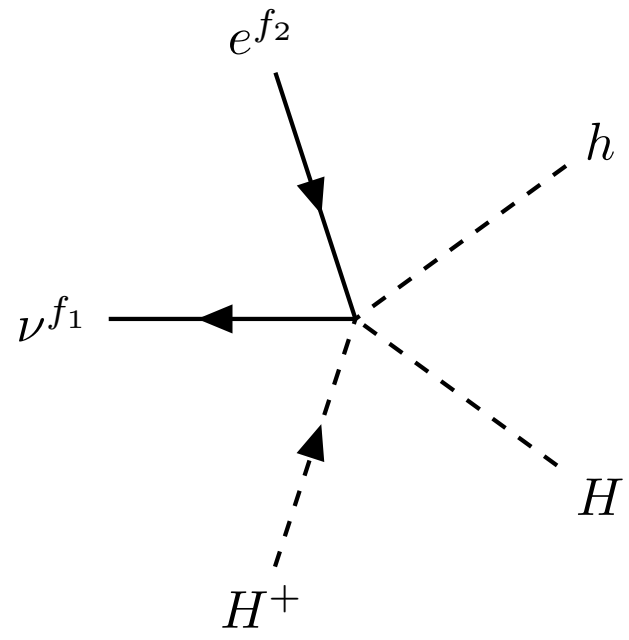

$$-\frac{1}{2} i c_\beta \mathcal{P}_R U^*_{g_1 f_1} \Big( 2 s_\beta^2 \hat{C}^{(11)}_{l\Phi_1,g_1f_2} - s_\beta^2 \hat{C}^{(12)}_{l\Phi_2,g_1f_2} + c_\beta^2 \hat{C}^{(12)}_{l\Phi_2,g_1f_2} - s_\beta^2 \hat{C}^{(21)}_{l\Phi_2,g_1f_2} + c_\beta^2 \hat{C}^{(21)}_{l\Phi_2,g_1f_2} - 2 s_\beta^2 \hat{C}^{(22)}_{l\Phi_1,g_1f_2} \Big) \tag{C.123}$$

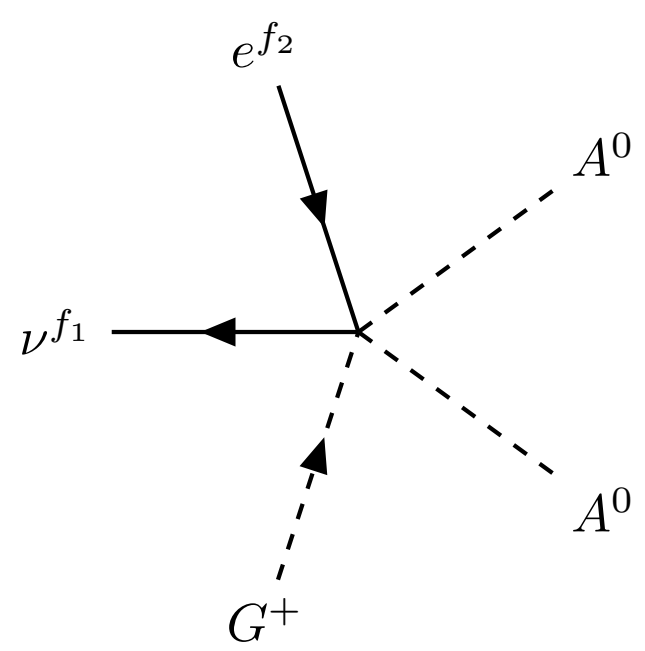

$$+ i c_\beta \mathcal{P}_R U^*_{g_1 f_1} \Big( s_\beta^2 \hat{C}^{(11)}_{l\Phi_1,g_1f_2} - s_\beta^2 \hat{C}^{(12)}_{l\Phi_2,g_1f_2} - s_\beta^2 \hat{C}^{(21)}_{l\Phi_2,g_1f_2} + c_\beta^2 \hat{C}^{(22)}_{l\Phi_1,g_1f_2} \Big) \tag{C.124}$$

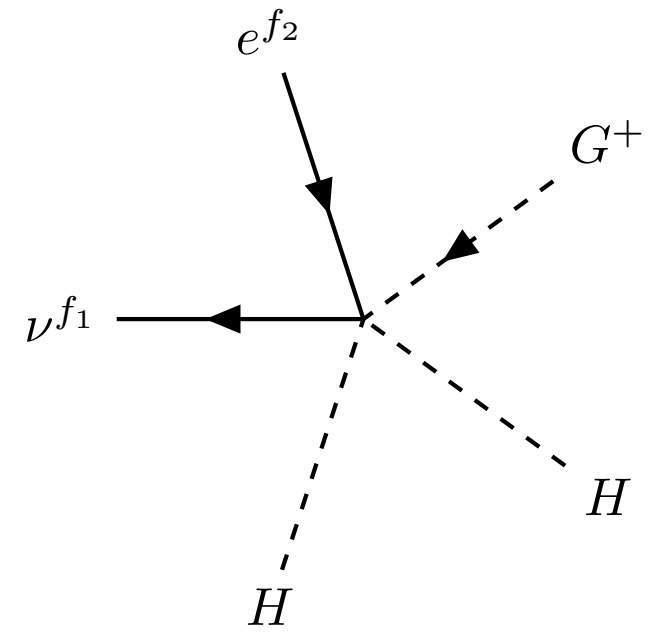

$$+ i c_\beta \mathcal{P}_R U^*_{g_1 f_1} \Big( s_\beta^2 \hat{C}^{(11)}_{l\Phi_1,g_1f_2} - s_\beta^2 \hat{C}^{(12)}_{l\Phi_2,g_1f_2} - s_\beta^2 \hat{C}^{(21)}_{l\Phi_2,g_1f_2} + c_\beta^2 \hat{C}^{(22)}_{l\Phi_1,g_1f_2} \Big) \tag{C.125}$$

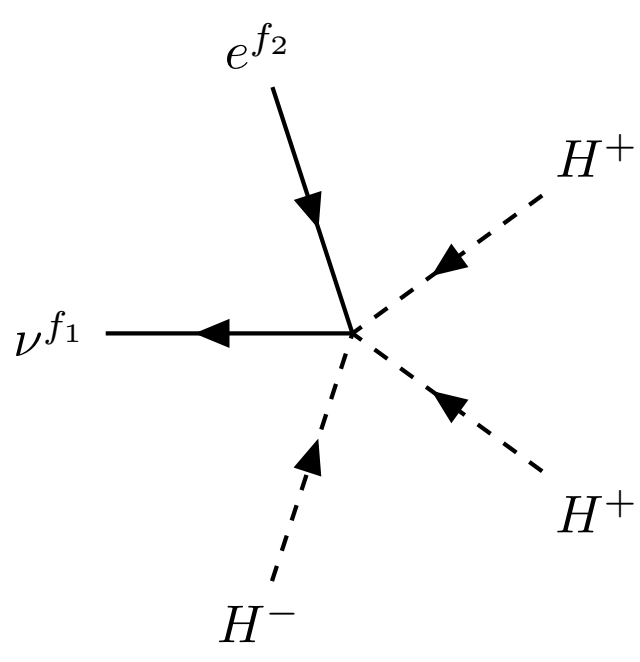

$$-2is_\beta\mathcal{P}_R U^*_{g_1f_1}\left(s^2_\beta\hat{C}^{(11)}_{l\Phi_1,g_1f_2}+c^2_\beta\hat{C}^{(12)}_{l\Phi_2,g_1f_2}\right.$$
$$\left.+c^2_\beta\hat{C}^{(21)}_{l\Phi_2,g_1f_2}+c^2_\beta\hat{C}^{(22)}_{l\Phi_1,g_1f_2}\right) \tag{C.126}$$

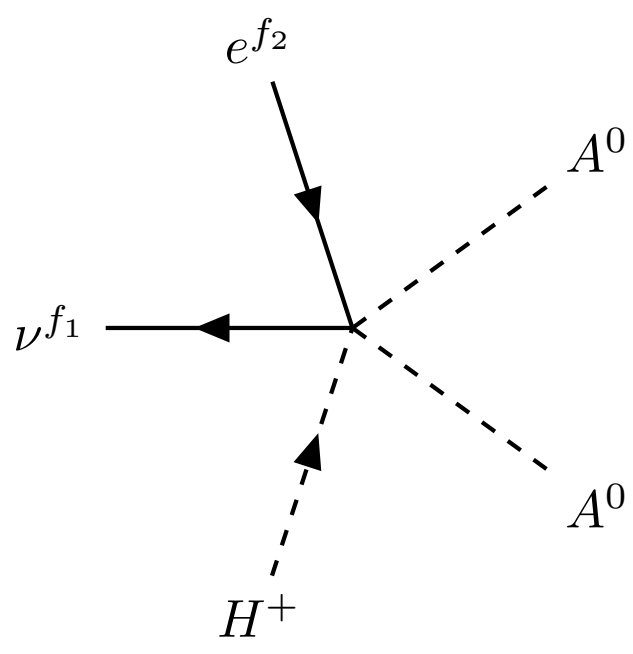

$$-is_\beta\mathcal{P}_R U^*_{g_1f_1}\left(s^2_\beta\hat{C}^{(11)}_{l\Phi_1,g_1f_2}+c^2_\beta\hat{C}^{(12)}_{l\Phi_2,g_1f_2}\right.$$
$$\left.+c^2_\beta\hat{C}^{(21)}_{l\Phi_2,g_1f_2}+c^2_\beta\hat{C}^{(22)}_{l\Phi_1,g_1f_2}\right) \tag{C.127}$$

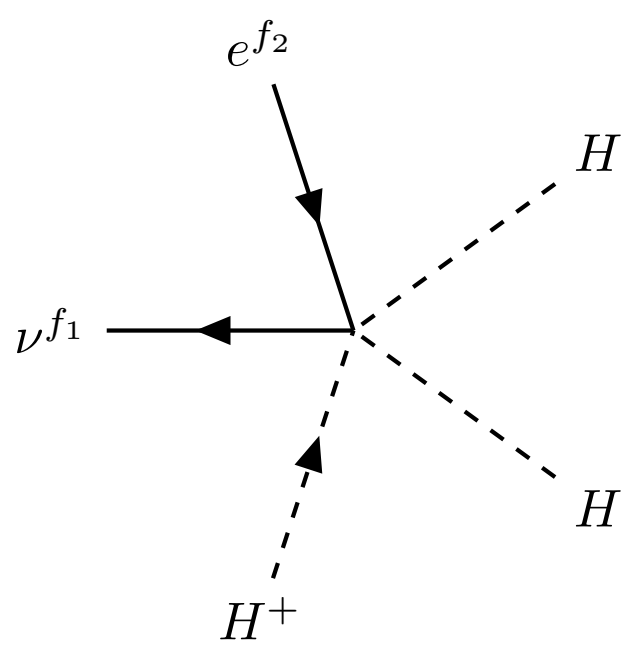

$$-is_\beta\mathcal{P}_R U^*_{g_1f_1}\left(s^2_\beta\hat{C}^{(11)}_{l\Phi_1,g_1f_2}+c^2_\beta\hat{C}^{(12)}_{l\Phi_2,g_1f_2}\right.$$
$$\left.+c^2_\beta\hat{C}^{(21)}_{l\Phi_2,g_1f_2}+c^2_\beta\hat{C}^{(22)}_{l\Phi_1,g_1f_2}\right) \tag{C.128}$$

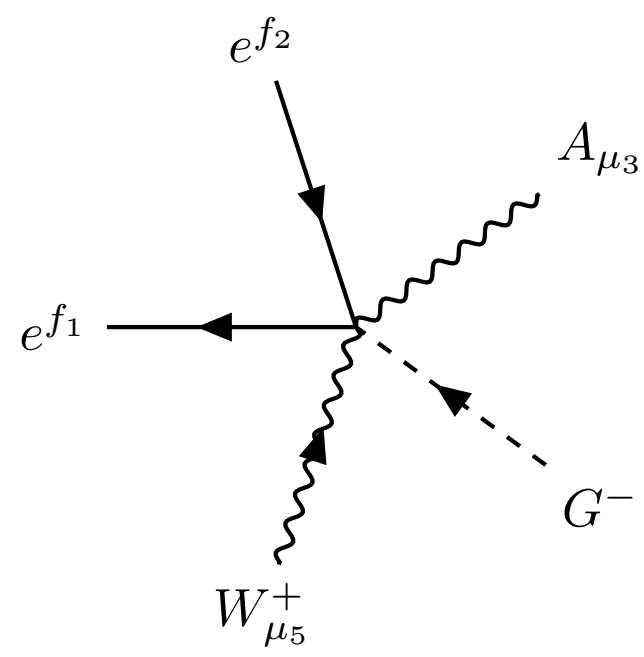

$$-\frac{2\sqrt{2}\hat{g}c_\beta\hat{g}'\sigma^{\mu_3\mu_5}\mathcal{P}_L}{\sqrt{\hat{g}'^2+\hat{g}^2}}\left(\hat{C}_{lW\Phi_1,f_2f_1*}\right) \tag{C.129}$$

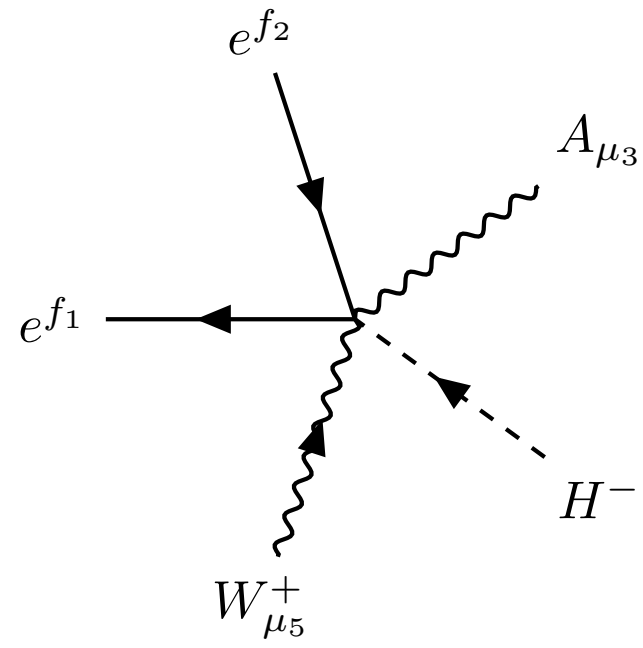

$$+\frac{2\sqrt{2}\hat{g}s_\beta\hat{g}'\sigma^{\mu_3\mu_5}\mathcal{P}_L}{\sqrt{\hat{g}'^2+\hat{g}^2}}\left(\hat{C}_{lW\Phi_1,f_2f_1*}\right) \tag{C.130}$$

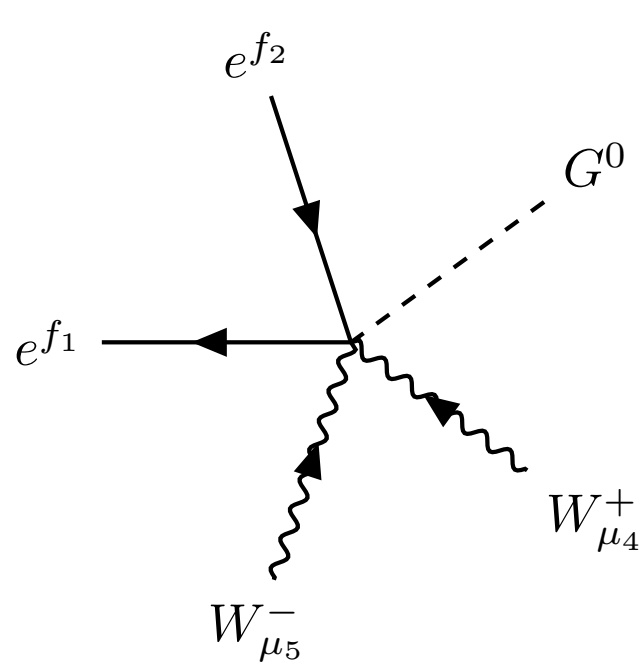

$$-i\sqrt{2}\hat{g}c_\beta\left(\sigma^{\mu_4\mu_5}\mathcal{P}_L\hat{C}^*_{lW\Phi_1,f_2f_1}-\hat{C}_{lW\Phi_1,f_1f_2}\sigma^{\mu_4\mu_5}\mathcal{P}_R\right) \tag{C.131}$$

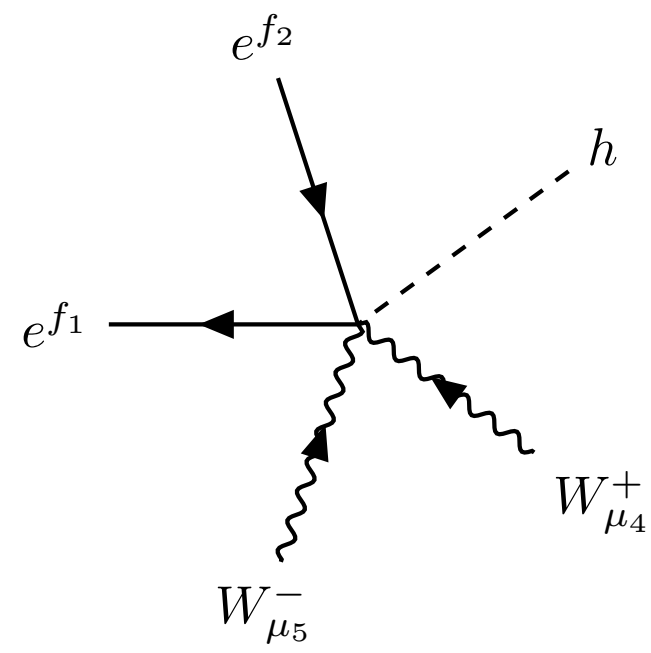

$$+\sqrt{2}\hat{g}c_\beta\left(\sigma^{\mu_4\mu_5}\mathcal{P}_L\hat{C}^*_{lW\Phi_1,f_2f_1}+\hat{C}_{lW\Phi_1,f_1f_2}\sigma^{\mu_4\mu_5}\mathcal{P}_R\right) \tag{C.132}$$

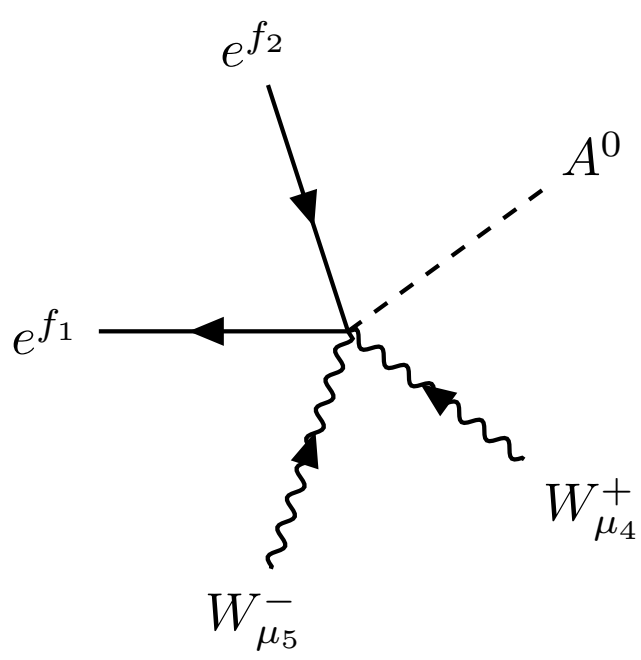

$$+i\sqrt{2}\hat{g}s_\beta\left(\sigma^{\mu_4\mu_5}\mathcal{P}_L\hat{C}^*_{lW\Phi_1,f_2f_1}-\hat{C}_{lW\Phi_1,f_1f_2}\sigma^{\mu_4\mu_5}\mathcal{P}_R\right) \tag{C.133}$$

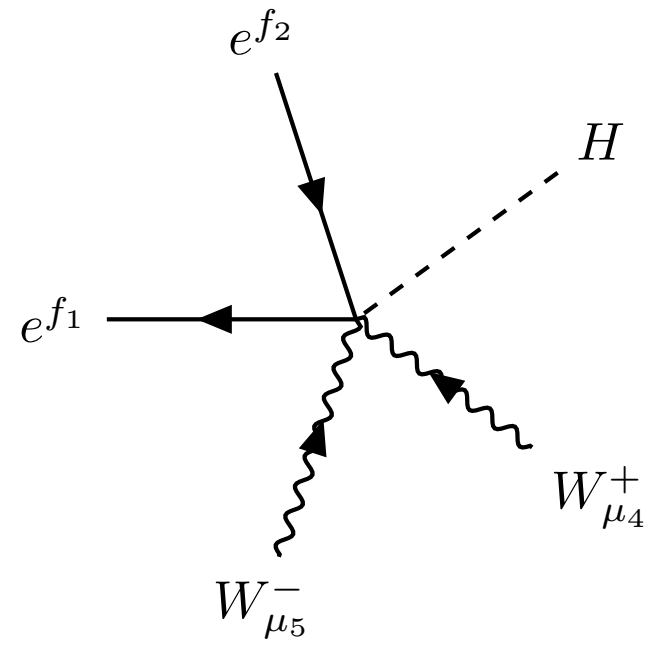

$$+\sqrt{2}\hat{g}s_\beta\left(\sigma^{\mu_4\mu_5}\mathcal{P}_L\hat{C}^*_{lW\Phi_1,f_2f_1}+\hat{C}_{lW\Phi_1,f_1f_2}\sigma^{\mu_4\mu_5}\mathcal{P}_R\right) \tag{C.134}$$

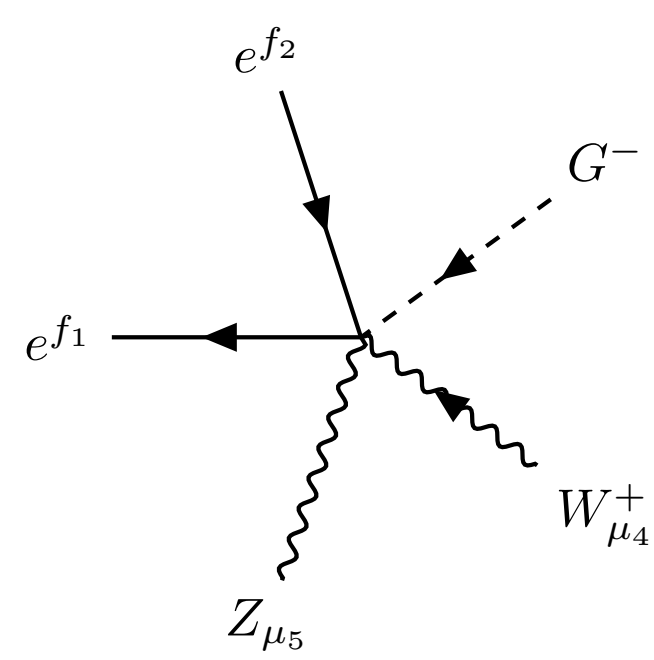

$$+\frac{2\sqrt{2}\hat{g}^2c_\beta\sigma^{\mu_4\mu_5}\mathcal{P}_L}{\sqrt{\hat{g}'^2+\hat{g}^2}}\left(\hat{C}^*_{lW\Phi_1,f_2f_1}\right) \tag{C.135}$$

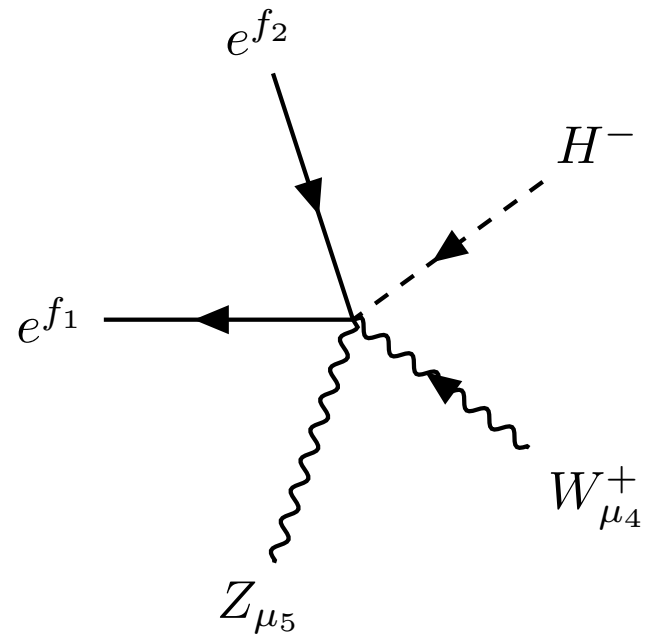

$$-\frac{2\sqrt{2}\hat{g}^2s_\beta\sigma^{\mu_4\mu_5}\mathcal{P}_L}{\sqrt{\hat{g}'^2+\hat{g}^2}}\left(\hat{C}^*_{lW\Phi_1,f_2f_1}\right) \tag{C.136}$$

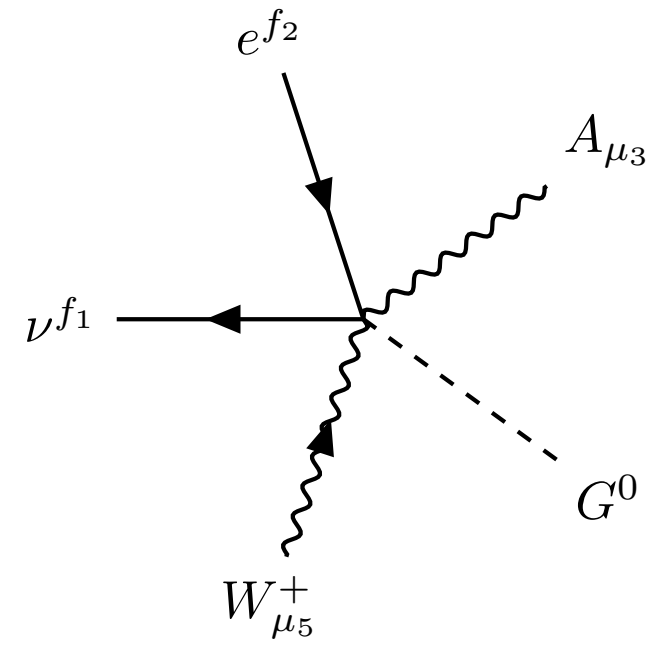

$$-\frac{2i\hat{g}c_\beta\hat{g}'U^*_{g_1f_1}\sigma^{\mu_3\mu_5}\mathcal{P}_R}{\sqrt{\hat{g}'^2+\hat{g}^2}}\left(\hat{C}_{lW\Phi_1,g_1f_2}\right) \tag{C.137}$$

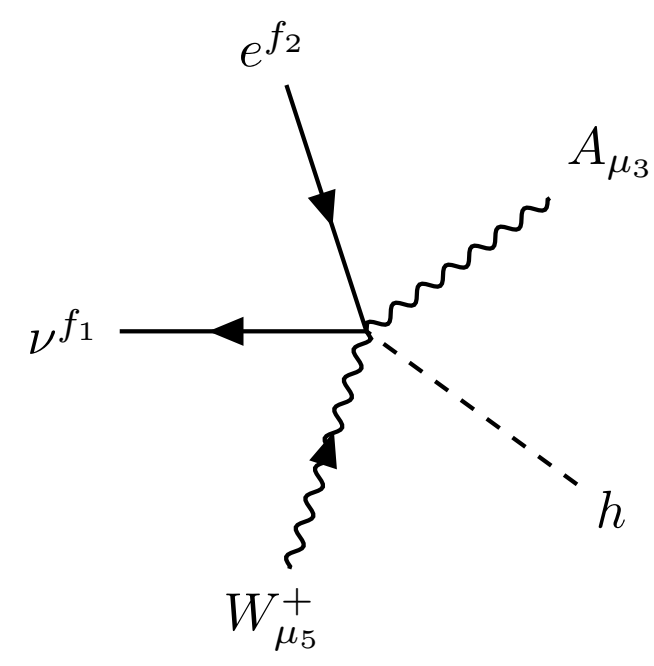

$$-\frac{2\hat{g}c_\beta\hat{g}'U^*_{g_1f_1}\sigma^{\mu_3\mu_5}\mathcal{P}_R}{\sqrt{\hat{g}'^2+\hat{g}^2}}\left(\hat{C}_{lW\Phi_1,g_1f_2}\right) \tag{C.138}$$

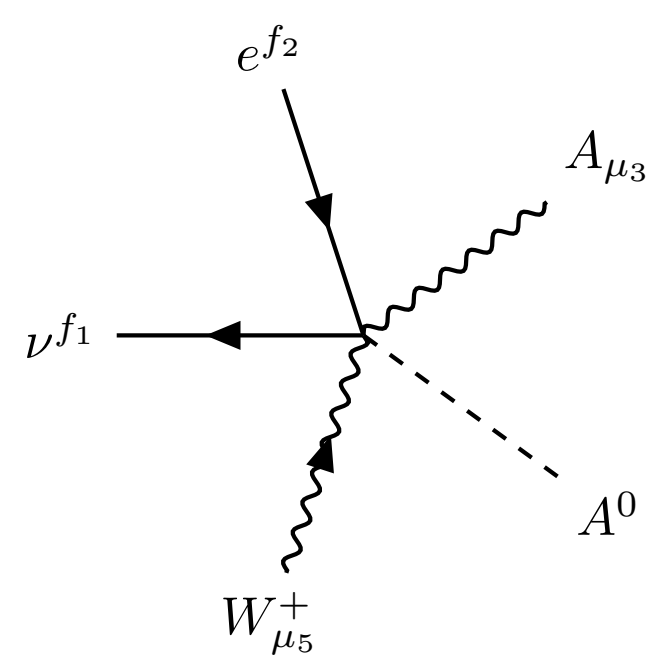

$$+\frac{2i\hat{g}s_\beta\hat{g}'U^*_{g_1f_1}\sigma^{\mu_3\mu_5}\mathcal{P}_R}{\sqrt{\hat{g}'^2+\hat{g}^2}}\left(\hat{C}_{lW\Phi_1,g_1f_2}\right) \tag{C.139}$$

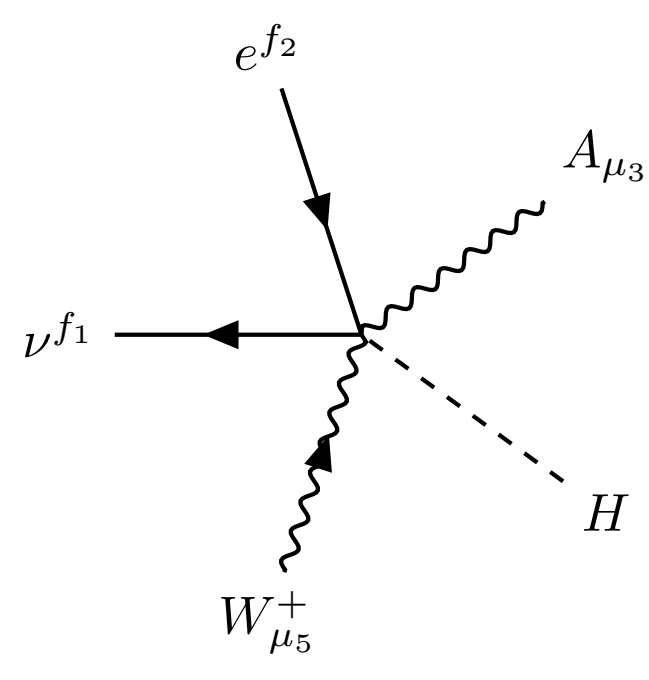

$$-\frac{2\hat{g}s_\beta\hat{g}'U^*_{g_1f_1}\sigma^{\mu_3\mu_5}\mathcal{P}_R}{\sqrt{\hat{g}'^2+\hat{g}^2}}\left(\hat{C}_{lW\Phi_1,g_1f_2}\right) \tag{C.140}$$

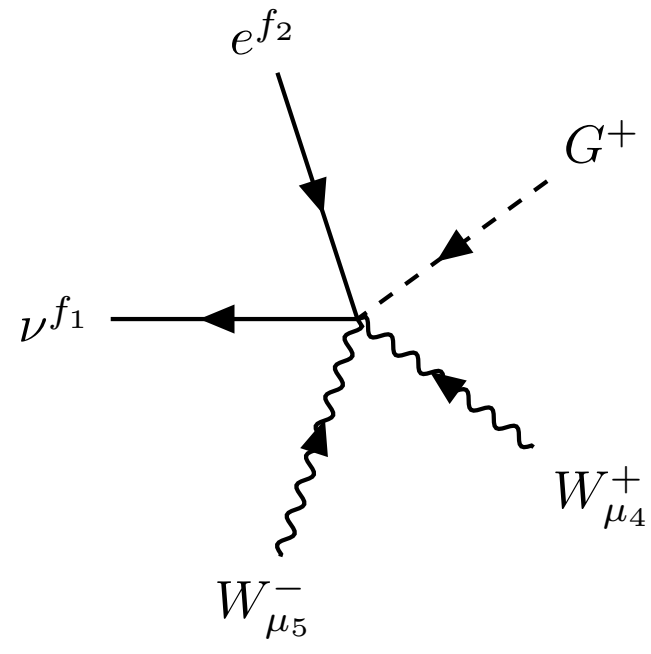

$$-2\hat{g}c_\beta U^*_{g_1f_1}\sigma^{\mu_4\mu_5}\mathcal{P}_R\left(\hat{C}_{lW\Phi_1,g_1f_2}\right) \tag{C.141}$$

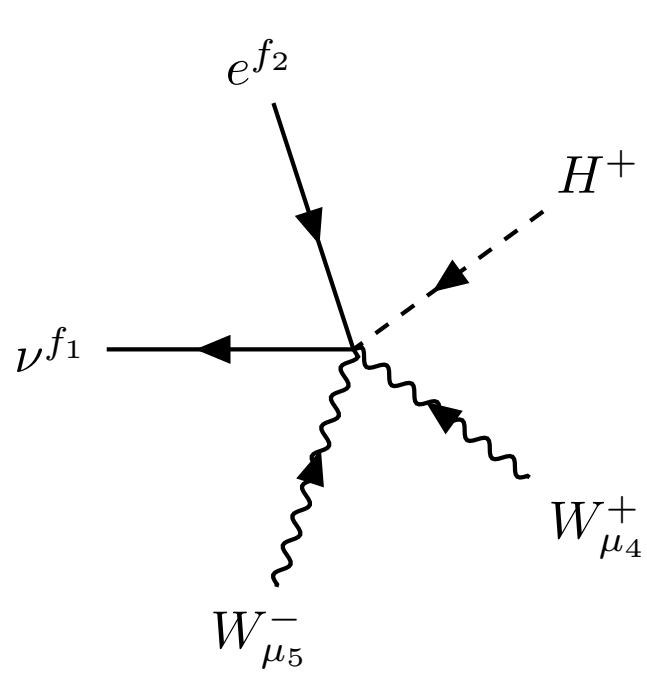

$$+2\hat{g}s_\beta U^*_{g_1f_1}\sigma^{\mu_4\mu_5}\mathcal{P}_R\left(\hat{C}_{lW\Phi_1,g_1f_2}\right) \tag{C.142}$$

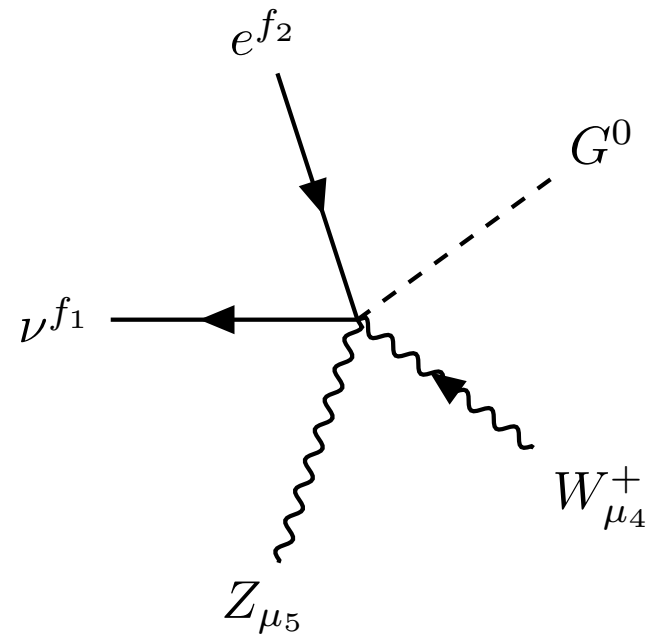

$$+\frac{2i\hat{g}^2c_\beta U^*_{g_1f_1}\sigma^{\mu_4\mu_5}\mathcal{P}_R}{\sqrt{\hat{g}'^2+\hat{g}^2}}\left(\hat{C}_{lW\Phi_1,g_1f_2}\right) \tag{C.143}$$

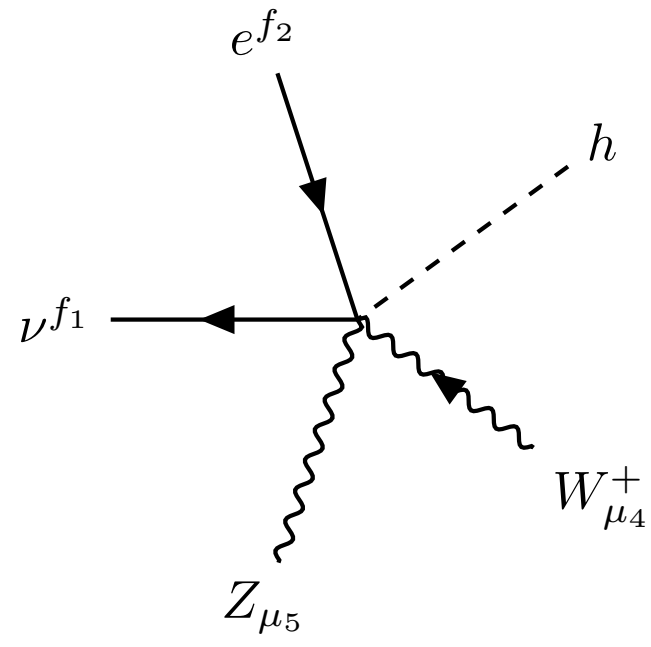

$$+\frac{2\hat{g}^2 c_\beta U^*_{g_1 f_1}\sigma^{\mu_4\mu_5}\mathcal{P}_R}{\sqrt{\hat{g}'^2+\hat{g}^2}}\left(\hat{C}_{lW\Phi_1,g_1 f_2}\right) \tag{C.144}$$

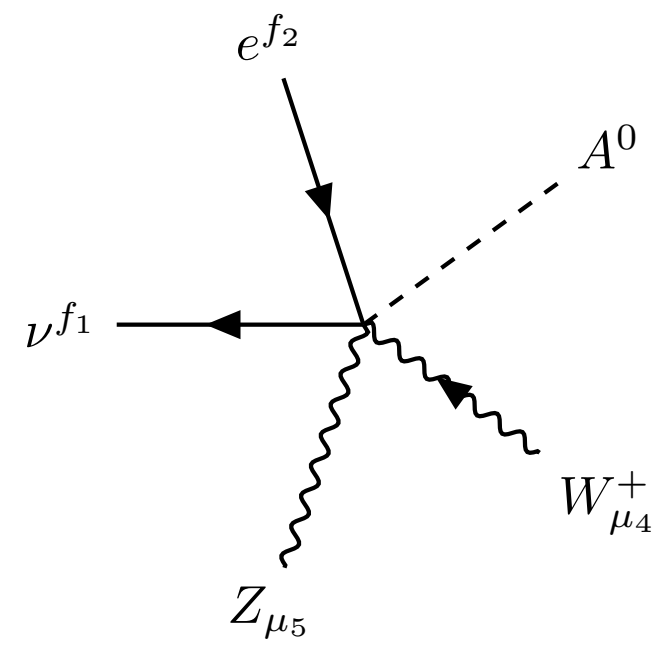

$$-\frac{2i\hat{g}^2 s_\beta U^*_{g_1 f_1}\sigma^{\mu_4\mu_5}\mathcal{P}_R}{\sqrt{\hat{g}'^2+\hat{g}^2}}\left(\hat{C}_{lW\Phi_1,g_1 f_2}\right) \tag{C.145}$$

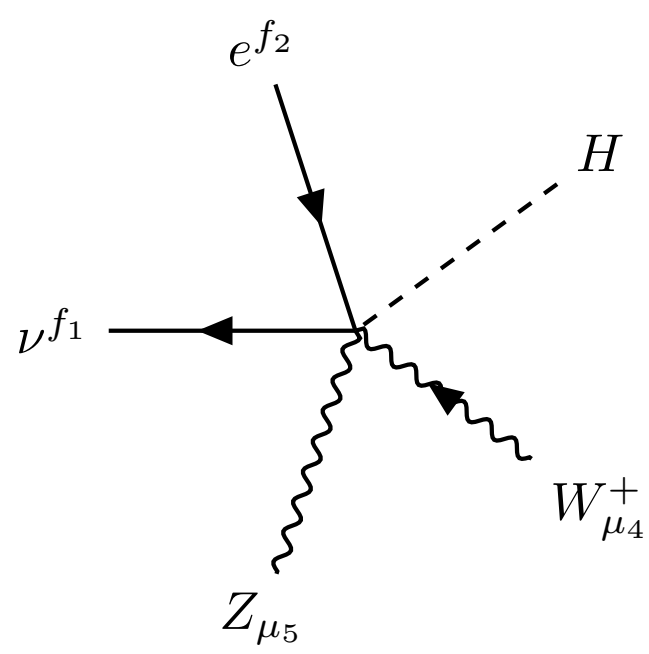

$$+\frac{2\hat{g}^2 s_\beta U^*_{g_1 f_1}\sigma^{\mu_4\mu_5}\mathcal{P}_R}{\sqrt{\hat{g}'^2+\hat{g}^2}}\left(\hat{C}_{lW\Phi_1,g_1 f_2}\right) \tag{C.146}$$

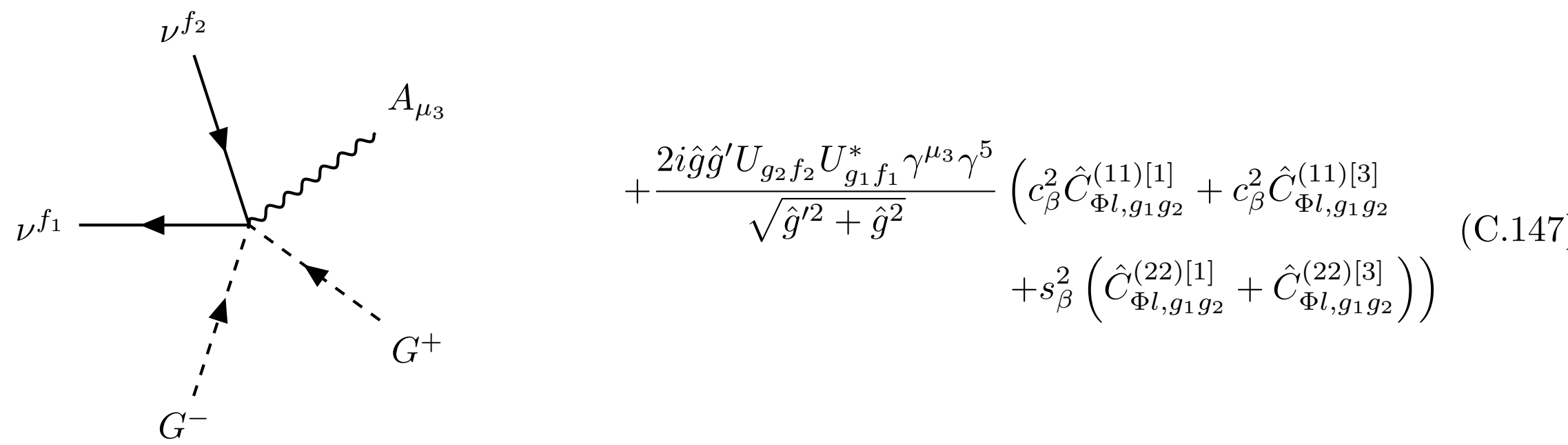

$$+\frac{2i\hat{g}\hat{g}' U_{g_2 f_2} U^*_{g_1 f_1} \gamma^{\mu_3}\gamma^5}{\sqrt{\hat{g}'^2+\hat{g}^2}} \Big( c_\beta^2 \hat{C}^{(11)[1]}_{\Phi l, g_1 g_2} + c_\beta^2 \hat{C}^{(11)[3]}_{\Phi l, g_1 g_2} + s_\beta^2 \Big( \hat{C}^{(22)[1]}_{\Phi l, g_1 g_2} + \hat{C}^{(22)[3]}_{\Phi l, g_1 g_2} \Big) \Big) \tag{C.147}$$

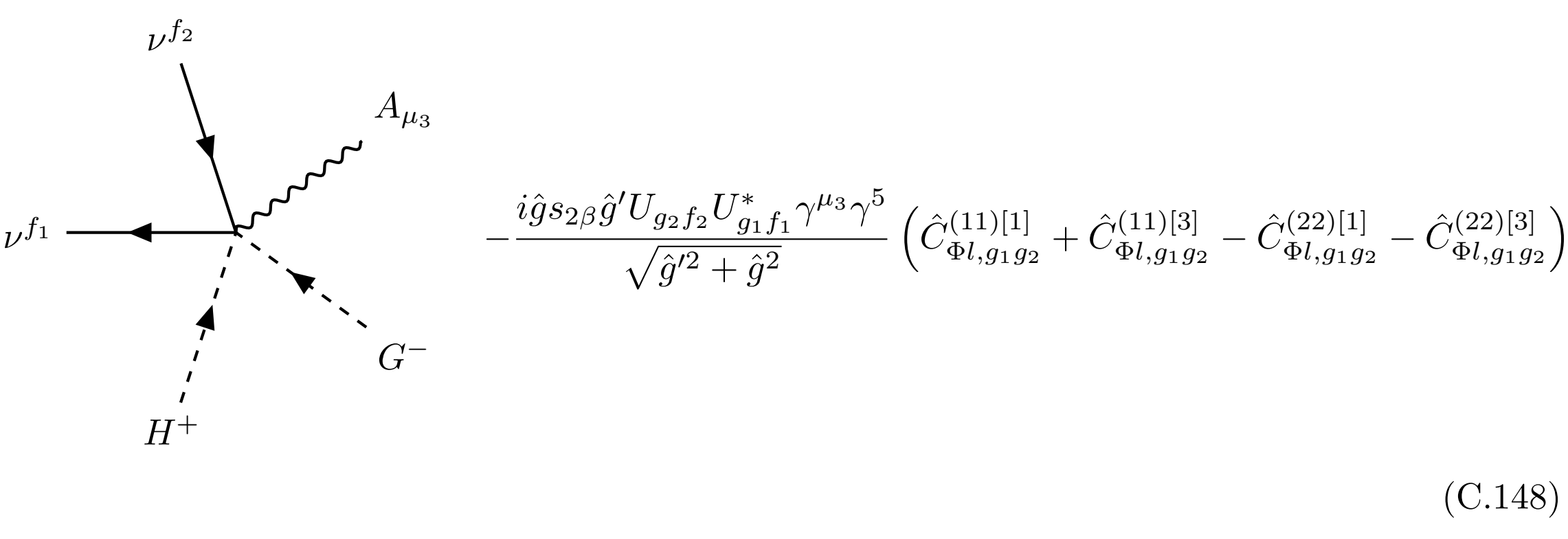

$$-\frac{i\hat{g} s_{2\beta} \hat{g}' U_{g_2 f_2} U^*_{g_1 f_1} \gamma^{\mu_3}\gamma^5}{\sqrt{\hat{g}'^2+\hat{g}^2}} \Big( \hat{C}^{(11)[1]}_{\Phi l, g_1 g_2} + \hat{C}^{(11)[3]}_{\Phi l, g_1 g_2} - \hat{C}^{(22)[1]}_{\Phi l, g_1 g_2} - \hat{C}^{(22)[3]}_{\Phi l, g_1 g_2} \Big) \tag{C.148}$$

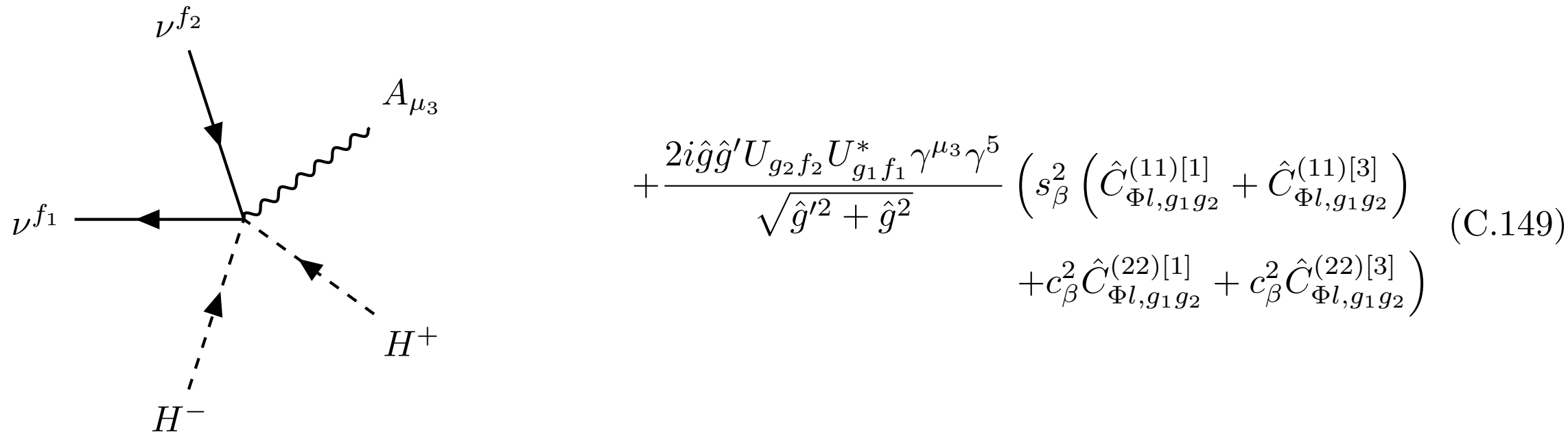

$$+\frac{2i\hat{g}\hat{g}' U_{g_2 f_2} U^*_{g_1 f_1} \gamma^{\mu_3}\gamma^5}{\sqrt{\hat{g}'^2+\hat{g}^2}} \Big( s_\beta^2 \Big( \hat{C}^{(11)[1]}_{\Phi l, g_1 g_2} + \hat{C}^{(11)[3]}_{\Phi l, g_1 g_2} \Big) + c_\beta^2 \hat{C}^{(22)[1]}_{\Phi l, g_1 g_2} + c_\beta^2 \hat{C}^{(22)[3]}_{\Phi l, g_1 g_2} \Big) \tag{C.149}$$

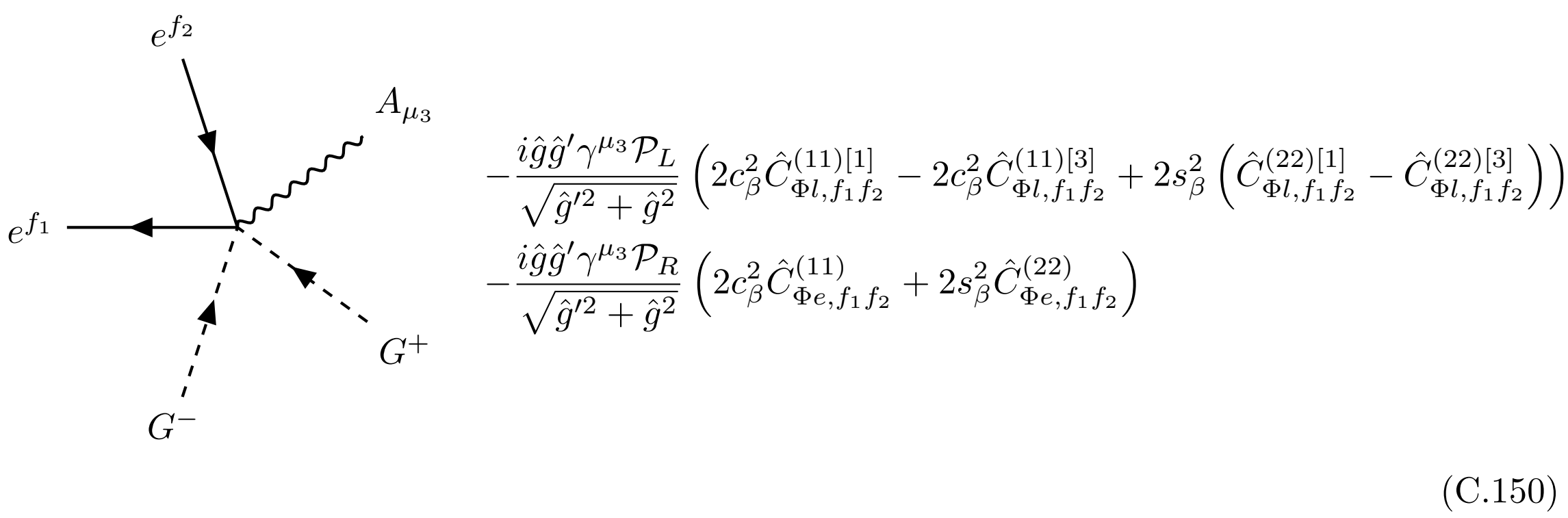

$$
\begin{aligned}
&-\frac{i\hat{g}\hat{g}'\gamma^{\mu_3}\mathcal{P}_L}{\sqrt{\hat{g}'^2+\hat{g}^2}}\left(2c_\beta^2\hat{C}^{(11)[1]}_{\Phi l,f_1f_2}-2c_\beta^2\hat{C}^{(11)[3]}_{\Phi l,f_1f_2}+2s_\beta^2\left(\hat{C}^{(22)[1]}_{\Phi l,f_1f_2}-\hat{C}^{(22)[3]}_{\Phi l,f_1f_2}\right)\right)\\
&-\frac{i\hat{g}\hat{g}'\gamma^{\mu_3}\mathcal{P}_R}{\sqrt{\hat{g}'^2+\hat{g}^2}}\left(2c_\beta^2\hat{C}^{(11)}_{\Phi e,f_1f_2}+2s_\beta^2\hat{C}^{(22)}_{\Phi e,f_1f_2}\right)
\end{aligned}
\tag{C.150}
$$

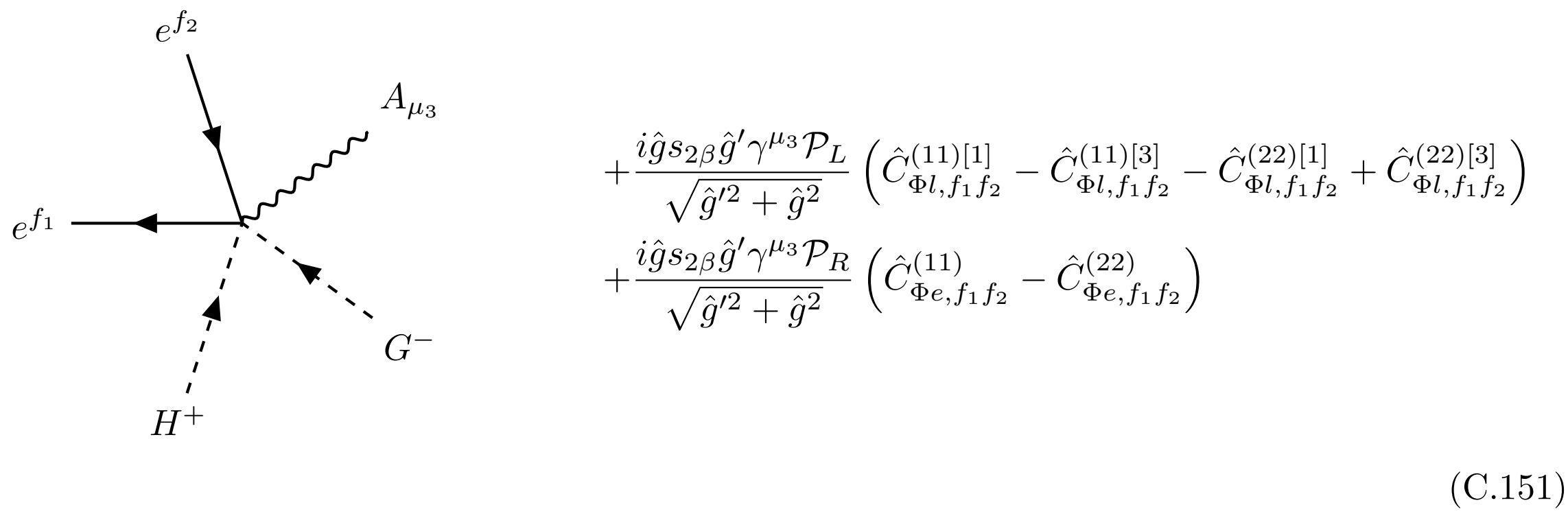

$$
\begin{aligned}
&+\frac{i\hat{g}s_{2\beta}\hat{g}'\gamma^{\mu_3}\mathcal{P}_L}{\sqrt{\hat{g}'^2+\hat{g}^2}}\left(\hat{C}^{(11)[1]}_{\Phi l,f_1f_2}-\hat{C}^{(11)[3]}_{\Phi l,f_1f_2}-\hat{C}^{(22)[1]}_{\Phi l,f_1f_2}+\hat{C}^{(22)[3]}_{\Phi l,f_1f_2}\right)\\
&+\frac{i\hat{g}s_{2\beta}\hat{g}'\gamma^{\mu_3}\mathcal{P}_R}{\sqrt{\hat{g}'^2+\hat{g}^2}}\left(\hat{C}^{(11)}_{\Phi e,f_1f_2}-\hat{C}^{(22)}_{\Phi e,f_1f_2}\right)
\end{aligned}
\tag{C.151}
$$

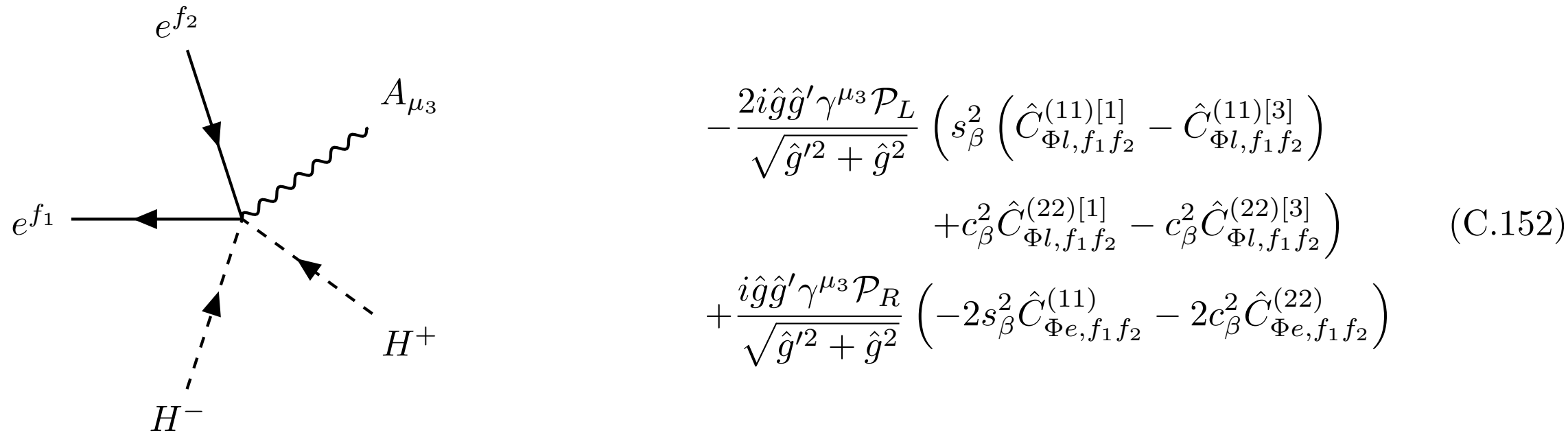

$$
\begin{aligned}
&-\frac{2i\hat{g}\hat{g}'\gamma^{\mu_3}\mathcal{P}_L}{\sqrt{\hat{g}'^2+\hat{g}^2}}\left(s_\beta^2\left(\hat{C}^{(11)[1]}_{\Phi l,f_1f_2}-\hat{C}^{(11)[3]}_{\Phi l,f_1f_2}\right)\right.\\
&\qquad\left.+c_\beta^2\hat{C}^{(22)[1]}_{\Phi l,f_1f_2}-c_\beta^2\hat{C}^{(22)[3]}_{\Phi l,f_1f_2}\right)\\
&+\frac{i\hat{g}\hat{g}'\gamma^{\mu_3}\mathcal{P}_R}{\sqrt{\hat{g}'^2+\hat{g}^2}}\left(-2s_\beta^2\hat{C}^{(11)}_{\Phi e,f_1f_2}-2c_\beta^2\hat{C}^{(22)}_{\Phi e,f_1f_2}\right)
\end{aligned}
\tag{C.152}
$$

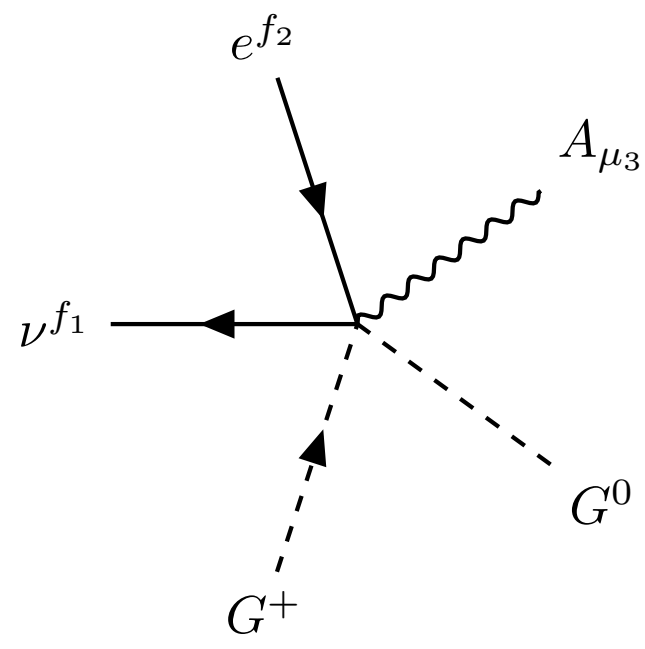

$$-\frac{\sqrt{2}\hat{g}\hat{g}' U^*_{g_1 f_1}\gamma^{\mu_3}\mathcal{P}_L}{\sqrt{\hat{g}'^2+\hat{g}^2}}\left(c_\beta^2 \hat{C}^{(11)[3]}_{\Phi l,g_1 f_2} + s_\beta^2 \hat{C}^{(22)[3]}_{\Phi l,g_1 f_2}\right) \tag{C.153}$$

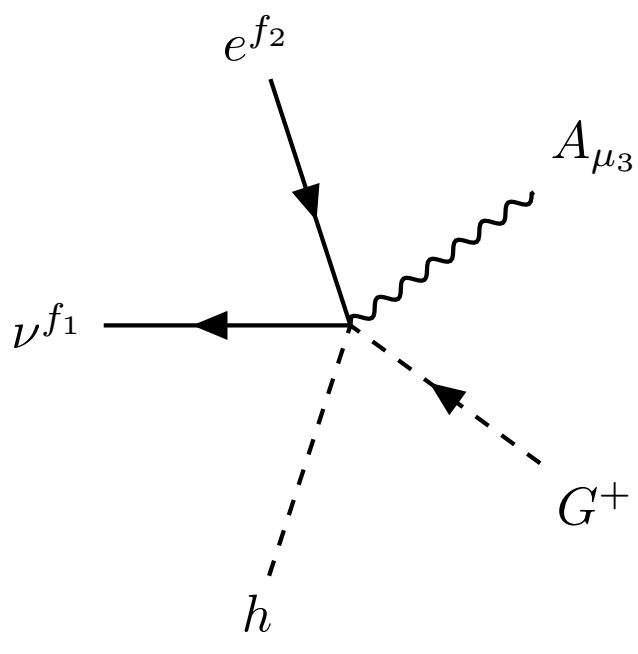

$$-\frac{i\sqrt{2}\hat{g}\hat{g}' U^*_{g_1 f_1}\gamma^{\mu_3}\mathcal{P}_L}{\sqrt{\hat{g}'^2+\hat{g}^2}}\left(c_\beta^2 \hat{C}^{(11)[3]}_{\Phi l,g_1 f_2} + s_\beta^2 \hat{C}^{(22)[3]}_{\Phi l,g_1 f_2}\right) \tag{C.154}$$

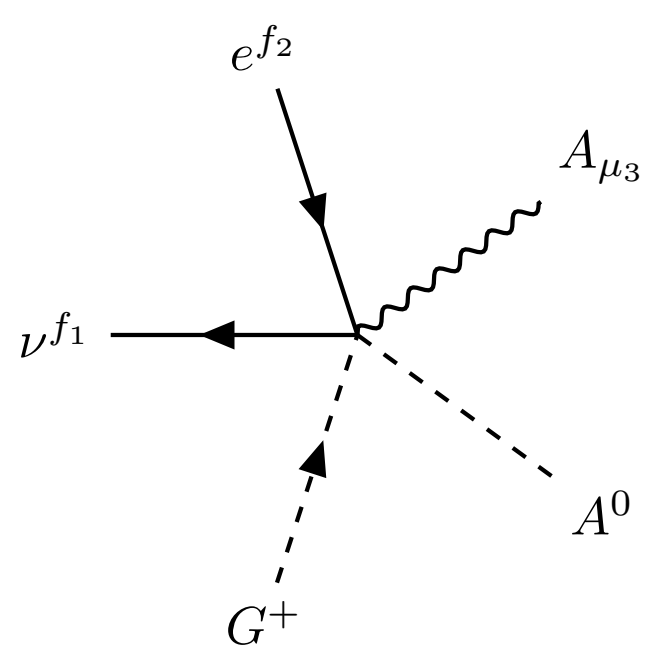

$$+\frac{\sqrt{2}\hat{g}s_\beta c_\beta\hat{g}' U^*_{g_1 f_1}\gamma^{\mu_3}\mathcal{P}_L}{\sqrt{\hat{g}'^2+\hat{g}^2}}\left(\hat{C}^{(11)[3]}_{\Phi l,g_1 f_2} - \hat{C}^{(22)[3]}_{\Phi l,g_1 f_2}\right) \tag{C.155}$$

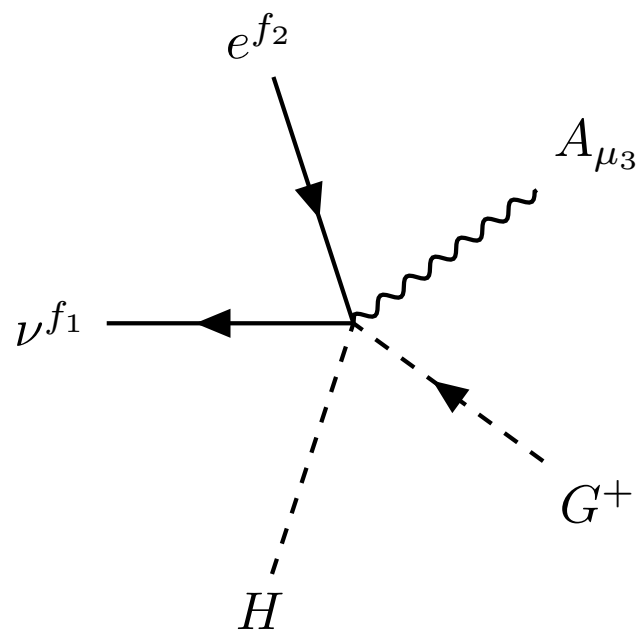

$$-\frac{i\sqrt{2}\hat{g}s_\beta c_\beta\hat{g}' U^*_{g_1 f_1}\gamma^{\mu_3}\mathcal{P}_L}{\sqrt{\hat{g}'^2+\hat{g}^2}}\left(\hat{C}^{(11)[3]}_{\Phi l,g_1 f_2} - \hat{C}^{(22)[3]}_{\Phi l,g_1 f_2}\right) \tag{C.156}$$

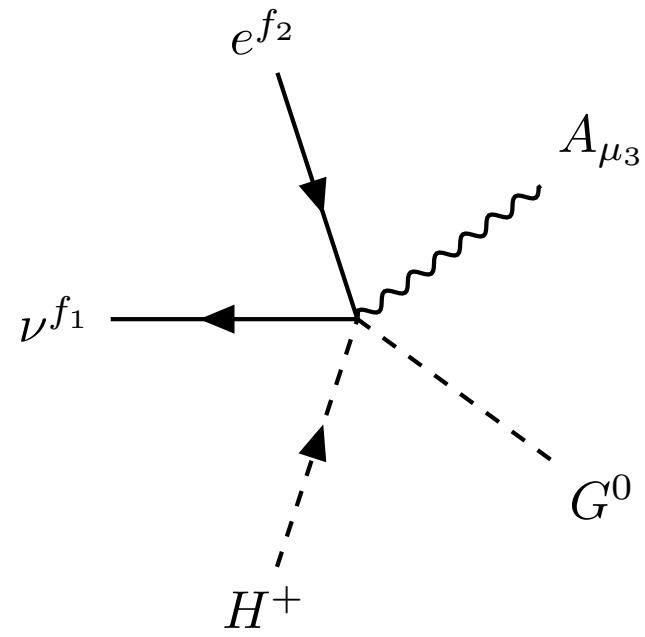

$$+\frac{\sqrt{2}\hat{g}s_\beta c_\beta \hat{g}' U^*_{g_1 f_1}\gamma^{\mu_3}\mathcal{P}_L}{\sqrt{\hat{g}'^2+\hat{g}^2}}\left(\hat{C}^{(11)[3]}_{\Phi l,g_1 f_2}-\hat{C}^{(22)[3]}_{\Phi l,g_1 f_2}\right) \tag{C.157}$$

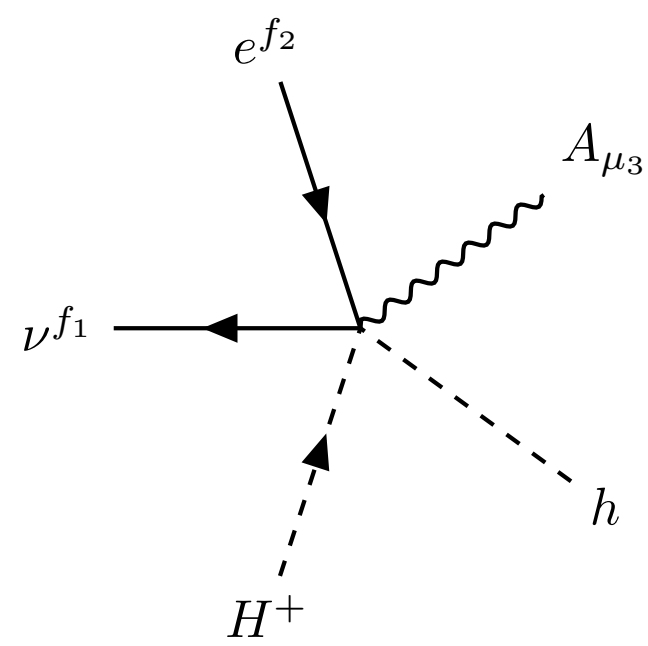

$$+\frac{i\sqrt{2}\hat{g}s_\beta c_\beta \hat{g}' U^*_{g_1 f_1}\gamma^{\mu_3}\mathcal{P}_L}{\sqrt{\hat{g}'^2+\hat{g}^2}}\left(\hat{C}^{(11)[3]}_{\Phi l,g_1 f_2}-\hat{C}^{(22)[3]}_{\Phi l,g_1 f_2}\right) \tag{C.158}$$

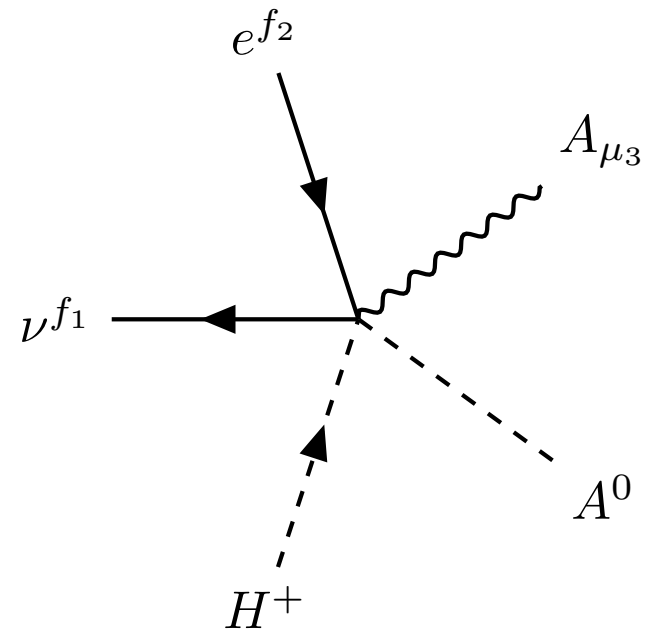

$$-\frac{\sqrt{2}\hat{g}\hat{g}' U^*_{g_1 f_1}\gamma^{\mu_3}\mathcal{P}_L}{\sqrt{\hat{g}'^2+\hat{g}^2}}\left(s^2_\beta\hat{C}^{(11)[3]}_{\Phi l,g_1 f_2}+c^2_\beta\hat{C}^{(22)[3]}_{\Phi l,g_1 f_2}\right) \tag{C.159}$$

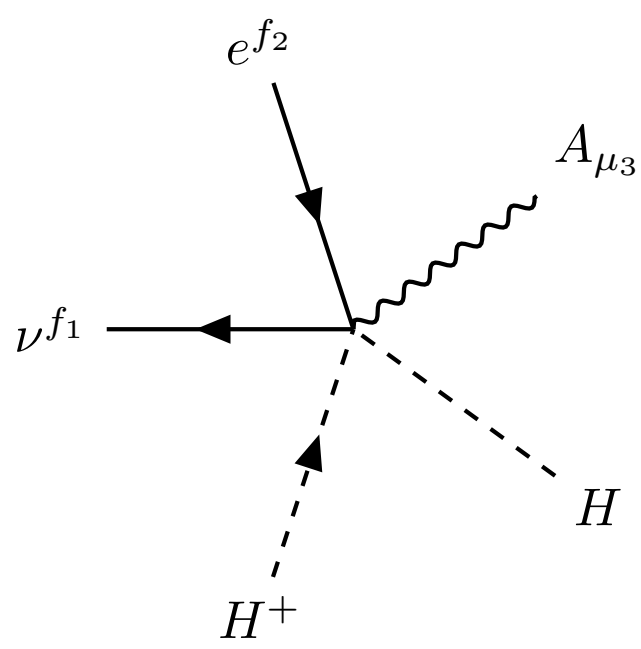

$$+\frac{i\sqrt{2}\hat{g}\hat{g}' U^*_{g_1 f_1} \gamma^{\mu_3} \mathcal{P}_L}{\sqrt{\hat{g}'^2+\hat{g}^2}} \left( s_\beta^2 \hat{C}^{(11)[3]}_{\Phi l, g_1 f_2} + c_\beta^2 \hat{C}^{(22)[3]}_{\Phi l, g_1 f_2} \right) \tag{C.160}$$

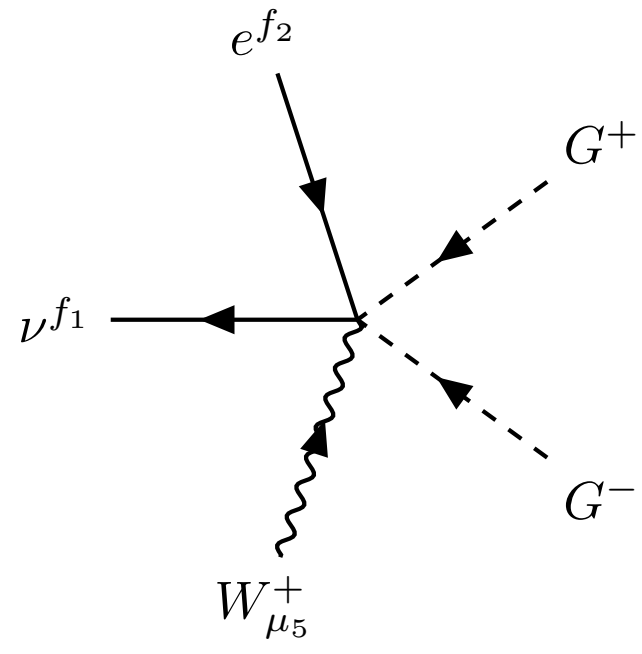

$$-i\sqrt{2}\hat{g} U^*_{g_1 f_1} \gamma^{\mu_5} \mathcal{P}_L \left( c_\beta^2 \hat{C}^{(11)[3]}_{\Phi l, g_1 f_2} + s_\beta^2 \hat{C}^{(22)[3]}_{\Phi l, g_1 f_2} \right) \tag{C.161}$$

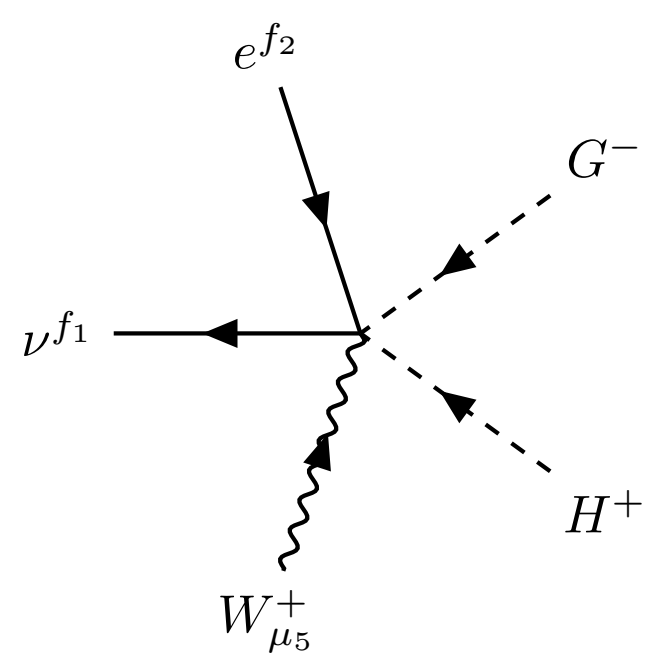

$$+i\sqrt{2}\hat{g} s_\beta c_\beta U^*_{g_1 f_1} \gamma^{\mu_5} \mathcal{P}_L \left( \hat{C}^{(11)[3]}_{\Phi l, g_1 f_2} - \hat{C}^{(22)[3]}_{\Phi l, g_1 f_2} \right) \tag{C.162}$$

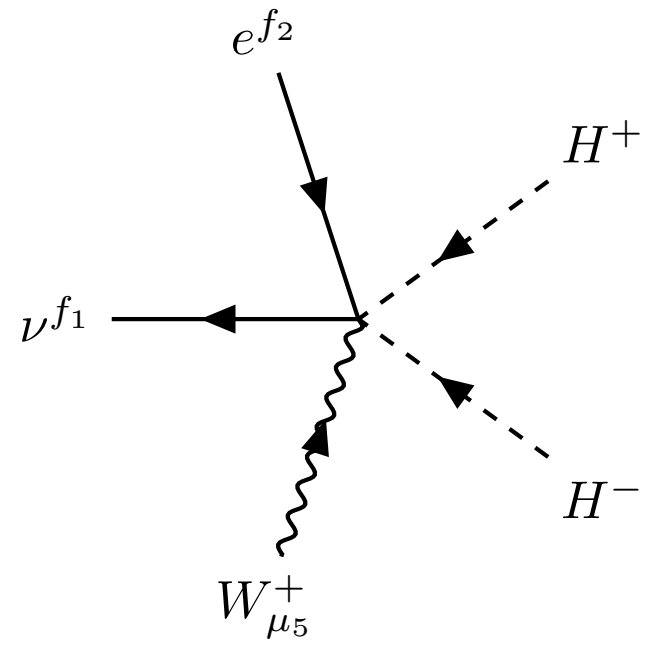

$$-i\sqrt{2}\hat{g} U^*_{g_1 f_1} \gamma^{\mu_5} \mathcal{P}_L \left( s_\beta^2 \hat{C}^{(11)[3]}_{\Phi l, g_1 f_2} + c_\beta^2 \hat{C}^{(22)[3]}_{\Phi l, g_1 f_2} \right) \tag{C.163}$$

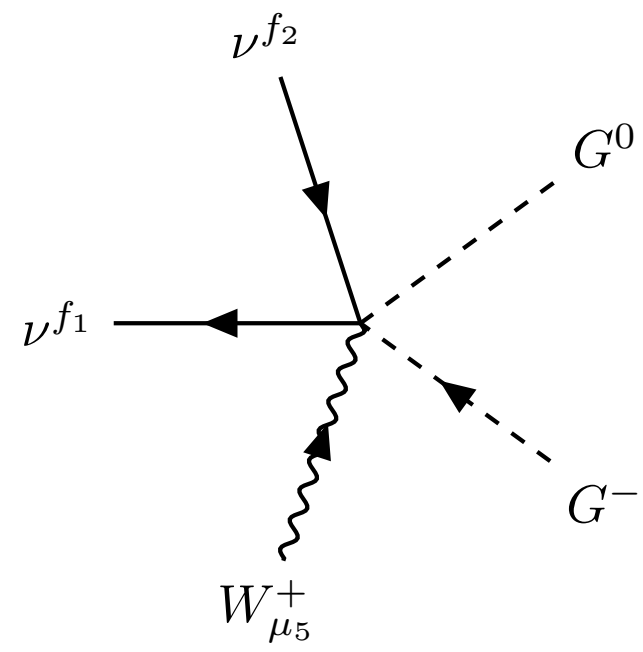

$$-\hat{g}U_{g_2 f_2}U^*_{g_1 f_1}\gamma^{\mu_5}\gamma^5\left(c^2_\beta \hat{C}^{(11)[1]}_{\Phi l,g_1 g_2} + s^2_\beta \hat{C}^{(22)[1]}_{\Phi l,g_1 g_2}\right) \qquad \text{(C.164)}$$

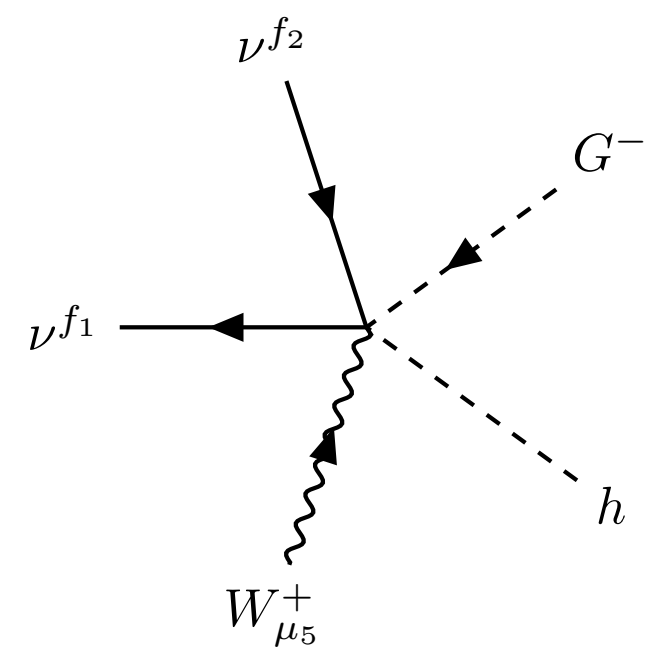

$$+i\hat{g}U_{g_2 f_2}U^*_{g_1 f_1}\gamma^{\mu_5}\gamma^5\left(c^2_\beta \hat{C}^{(11)[1]}_{\Phi l,g_1 g_2} + s^2_\beta \hat{C}^{(22)[1]}_{\Phi l,g_1 g_2}\right) \qquad \text{(C.165)}$$

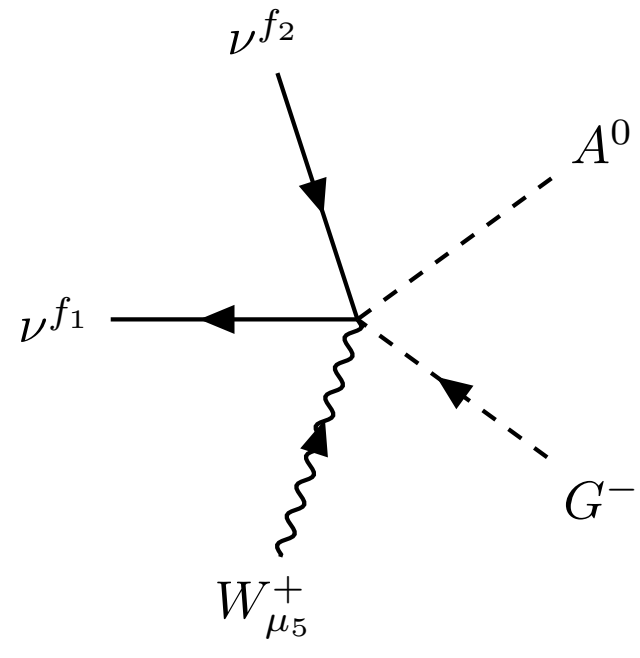

$$+\hat{g}s_\beta c_\beta U_{g_2 f_2}U^*_{g_1 f_1}\gamma^{\mu_5}\gamma^5\left(\hat{C}^{(11)[1]}_{\Phi l,g_1 g_2} - \hat{C}^{(22)[1]}_{\Phi l,g_1 g_2}\right) \qquad \text{(C.166)}$$

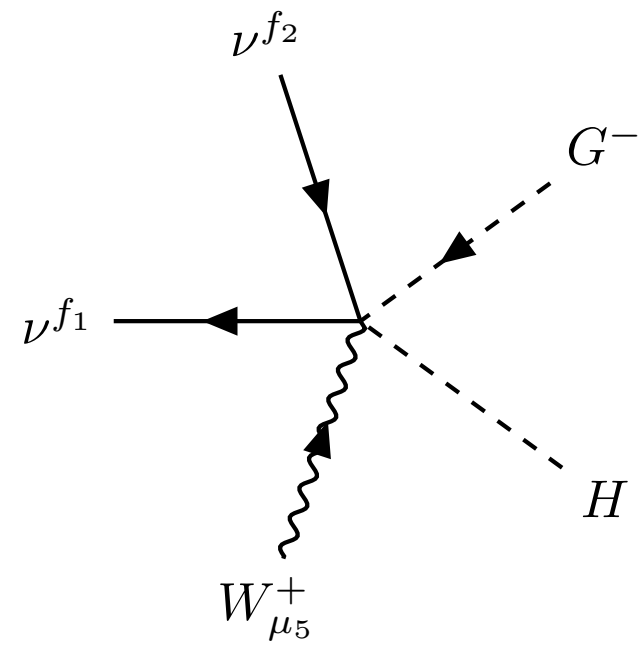

$$+i\hat{g}s_\beta c_\beta U_{g_2 f_2} U^*_{g_1 f_1} \gamma^{\mu_5}\gamma^5 \left(\hat{C}^{(11)[1]}_{\Phi l,g_1 g_2} - \hat{C}^{(22)[1]}_{\Phi l,g_1 g_2}\right) \qquad \text{(C.167)}$$

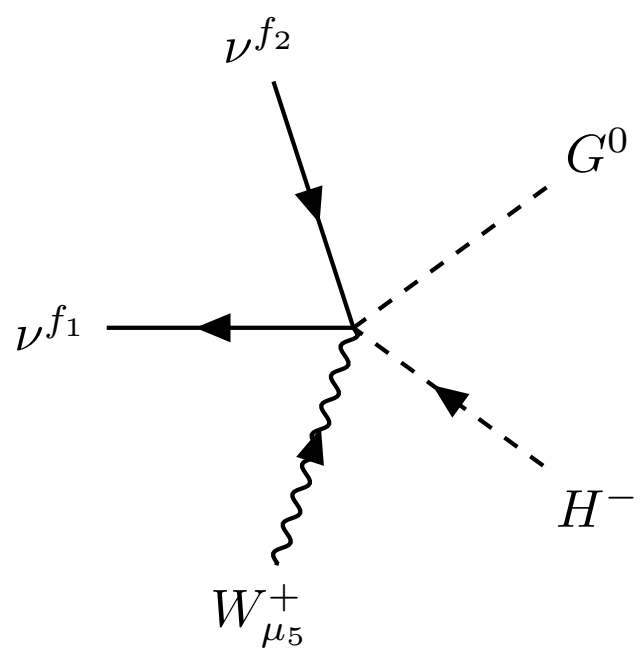

$$+\hat{g}s_\beta c_\beta U_{g_2 f_2} U^*_{g_1 f_1} \gamma^{\mu_5}\gamma^5 \left(\hat{C}^{(11)[1]}_{\Phi l,g_1 g_2} - \hat{C}^{(22)[1]}_{\Phi l,g_1 g_2}\right) \qquad \text{(C.168)}$$

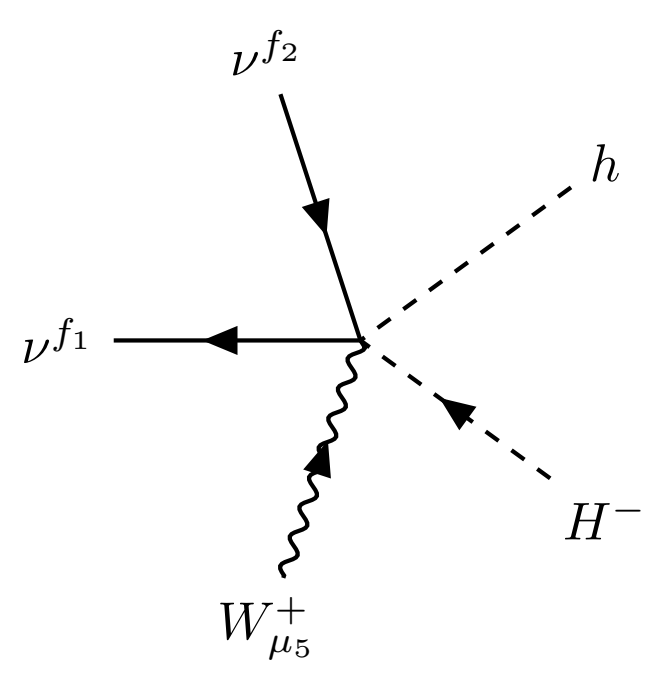

$$-i\hat{g}s_\beta c_\beta U_{g_2 f_2} U^*_{g_1 f_1} \gamma^{\mu_5}\gamma^5 \left(\hat{C}^{(11)[1]}_{\Phi l,g_1 g_2} - \hat{C}^{(22)[1]}_{\Phi l,g_1 g_2}\right) \qquad \text{(C.169)}$$

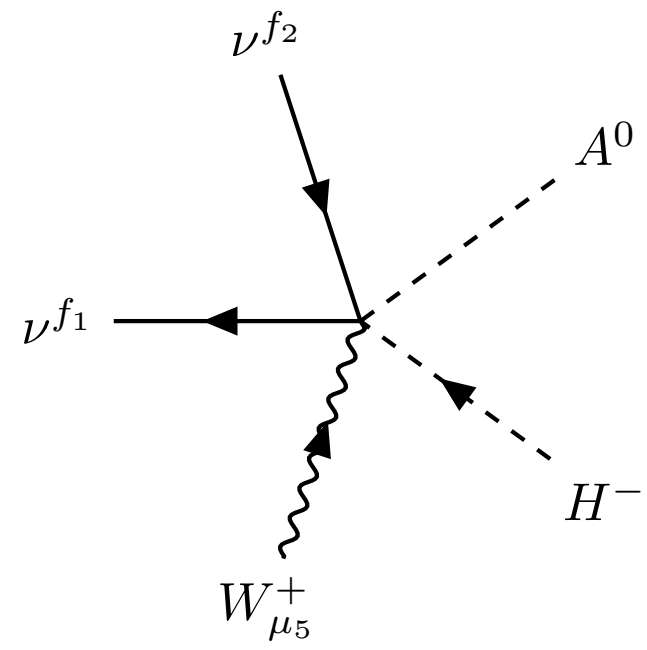

$$-\hat{g} U_{g_2 f_2} U^*_{g_1 f_1} \gamma^{\mu_5}\gamma^5 \left(s^2_\beta \hat{C}^{(11)[1]}_{\Phi l,g_1 g_2} + c^2_\beta \hat{C}^{(22)[1]}_{\Phi l,g_1 g_2}\right) \qquad \text{(C.170)}$$

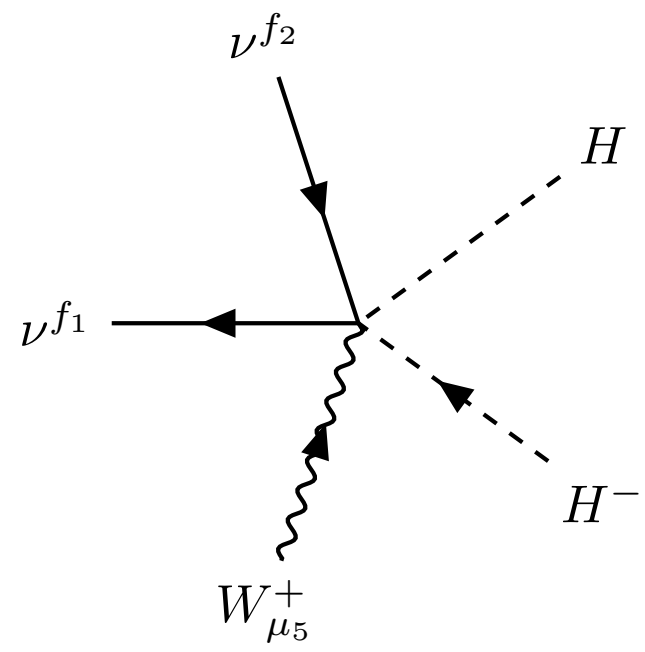

$$-i\hat{g}U_{g_2f_2}U^*_{g_1f_1}\gamma^{\mu_5}\gamma^5\left(s_\beta^2\hat{C}^{(11)[1]}_{\Phi l,g_1g_2}+c_\beta^2\hat{C}^{(22)[1]}_{\Phi l,g_1g_2}\right) \tag{C.171}$$

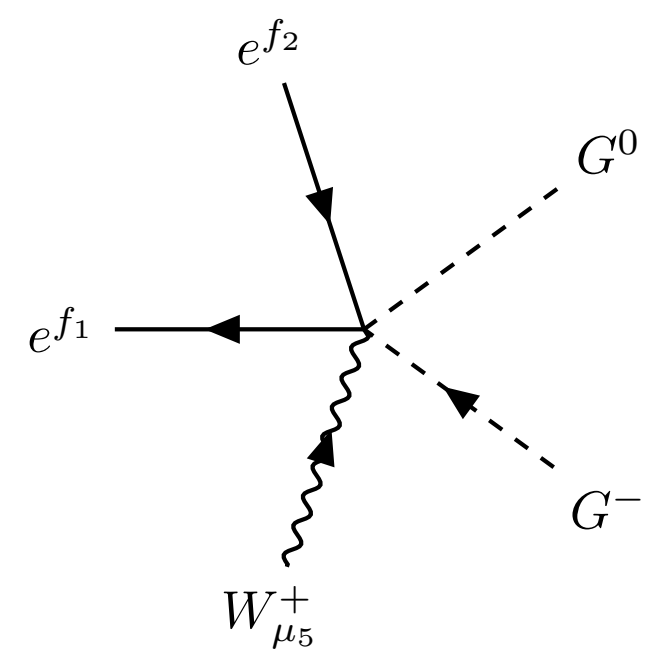

$$\begin{aligned}&+\hat{g}\gamma^{\mu_5}\mathcal{P}_L\left(c_\beta^2\hat{C}^{(11)[1]}_{\Phi l,f_1f_2}+s_\beta^2\hat{C}^{(22)[1]}_{\Phi l,f_1f_2}\right)\\&+\hat{g}\gamma^{\mu_5}\mathcal{P}_R\left(c_\beta^2\hat{C}^{(11)}_{\Phi e,f_1f_2}+s_\beta^2\hat{C}^{(22)}_{\Phi e,f_1f_2}\right)\end{aligned} \tag{C.172}$$

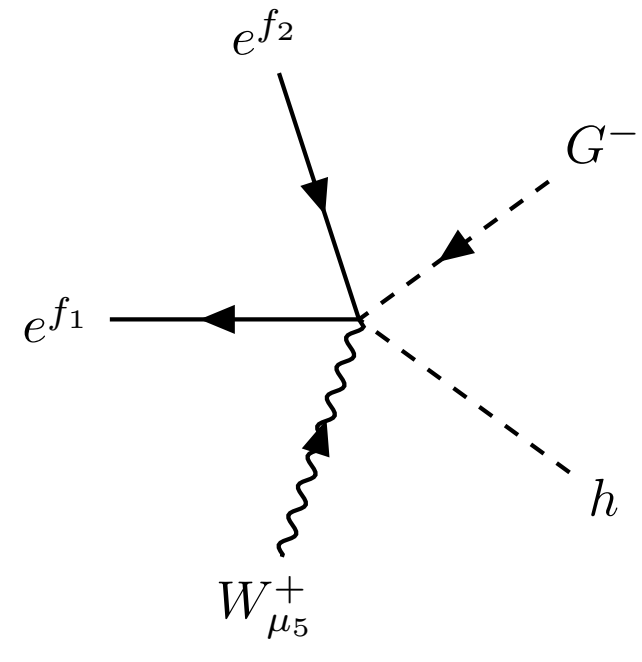

$$\begin{aligned}&-i\hat{g}\gamma^{\mu_5}\mathcal{P}_L\left(c_\beta^2\hat{C}^{(11)[1]}_{\Phi l,f_1f_2}+s_\beta^2\hat{C}^{(22)[1]}_{\Phi l,f_1f_2}\right)\\&-i\hat{g}\gamma^{\mu_5}\mathcal{P}_R\left(c_\beta^2\hat{C}^{(11)}_{\Phi e,f_1f_2}+s_\beta^2\hat{C}^{(22)}_{\Phi e,f_1f_2}\right)\end{aligned} \tag{C.173}$$

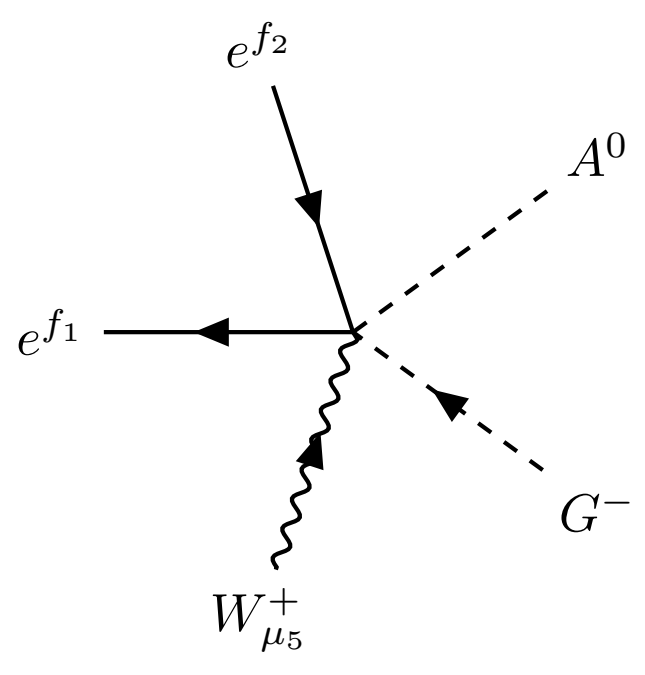

$$\begin{aligned}&+\frac{1}{2}\hat{g}s_{2\beta}\gamma^{\mu_5}\mathcal{P}_L\left(\hat{C}^{(22)[1]}_{\Phi l,f_1f_2}-\hat{C}^{(11)[1]}_{\Phi l,f_1f_2}\right)\\&+\frac{1}{2}\hat{g}s_{2\beta}\gamma^{\mu_5}\mathcal{P}_R\left(\hat{C}^{(22)}_{\Phi e,f_1f_2}-\hat{C}^{(11)}_{\Phi e,f_1f_2}\right)\end{aligned}\tag{C.174}$$

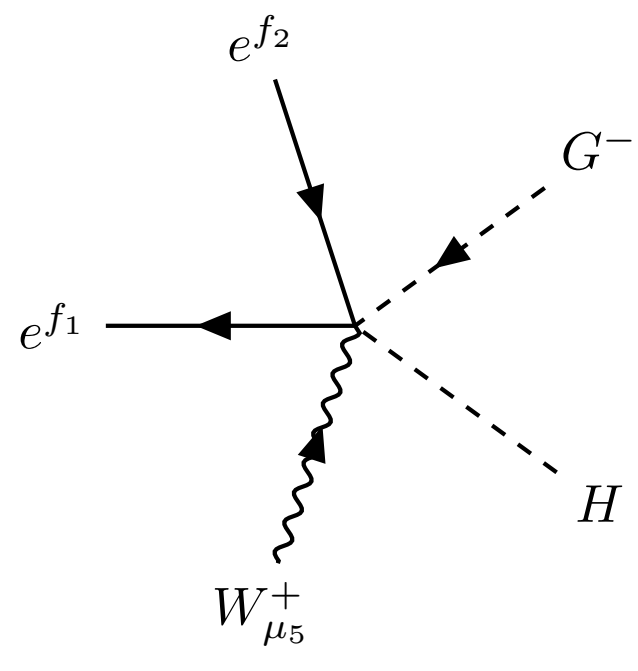

$$\begin{aligned}&-\frac{1}{2}i\hat{g}s_{2\beta}\gamma^{\mu_5}\mathcal{P}_L\left(\hat{C}^{(11)[1]}_{\Phi l,f_1f_2}-\hat{C}^{(22)[1]}_{\Phi l,f_1f_2}\right)\\&-\frac{1}{2}i\hat{g}s_{2\beta}\gamma^{\mu_5}\mathcal{P}_R\left(\hat{C}^{(11)}_{\Phi e,f_1f_2}-\hat{C}^{(22)}_{\Phi e,f_1f_2}\right)\end{aligned}\tag{C.175}$$

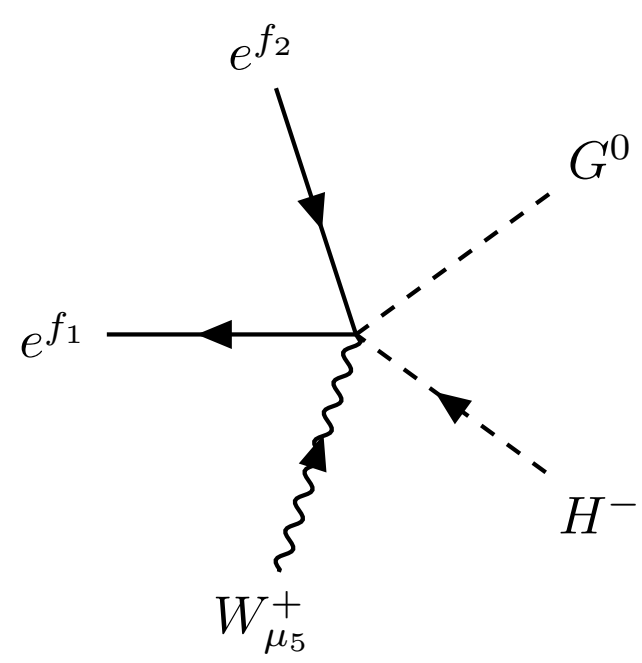

$$\begin{aligned}&+\frac{1}{2}\hat{g}s_{2\beta}\gamma^{\mu_5}\mathcal{P}_L\left(\hat{C}^{(22)[1]}_{\Phi l,f_1f_2}-\hat{C}^{(11)[1]}_{\Phi l,f_1f_2}\right)\\&+\frac{1}{2}\hat{g}s_{2\beta}\gamma^{\mu_5}\mathcal{P}_R\left(\hat{C}^{(22)}_{\Phi e,f_1f_2}-\hat{C}^{(11)}_{\Phi e,f_1f_2}\right)\end{aligned}\tag{C.176}$$

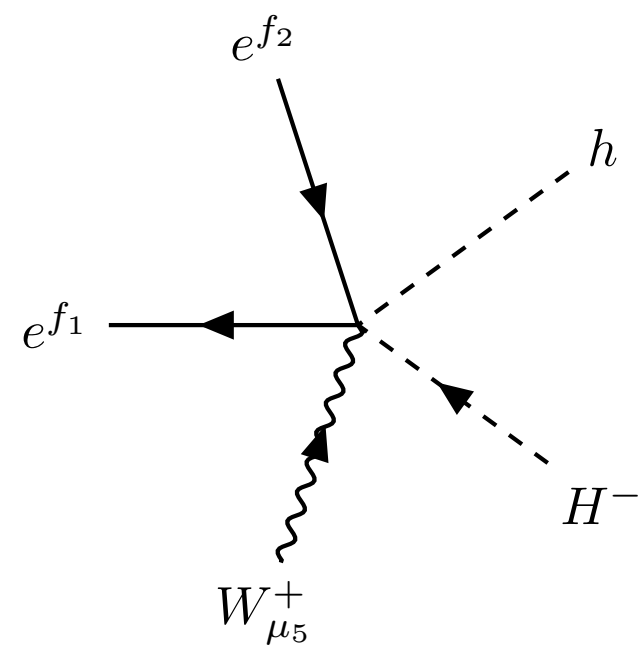

$$\begin{aligned}&+\frac{1}{2}i\hat{g}s_{2\beta}\gamma^{\mu_5}\mathcal{P}_L\left(\hat{C}^{(11)[1]}_{\Phi l,f_1f_2}-\hat{C}^{(22)[1]}_{\Phi l,f_1f_2}\right)\\&+\frac{1}{2}i\hat{g}s_{2\beta}\gamma^{\mu_5}\mathcal{P}_R\left(\hat{C}^{(11)}_{\Phi e,f_1f_2}-\hat{C}^{(22)}_{\Phi e,f_1f_2}\right)\end{aligned}\tag{C.177}$$

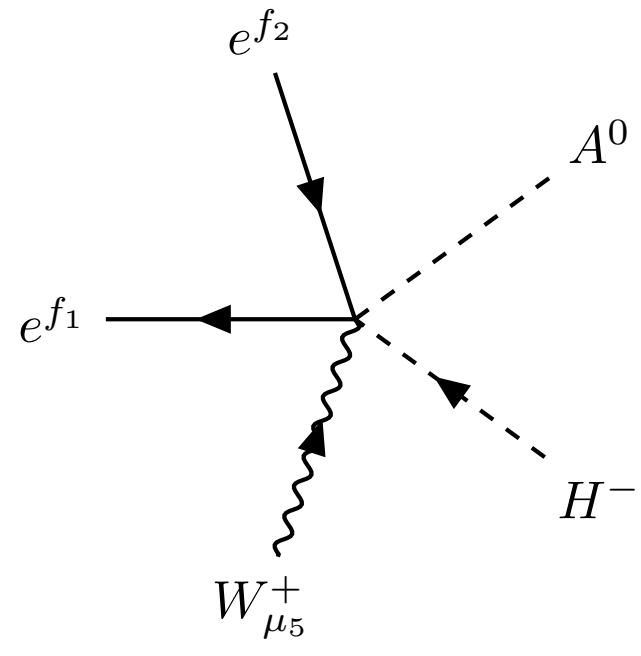

$$+\hat{g}\gamma^{\mu_5}\mathcal{P}_L\left(s_\beta^2\hat{C}^{(11)[1]}_{\Phi l,f_1f_2}+c_\beta^2\hat{C}^{(22)[1]}_{\Phi l,f_1f_2}\right)$$
$$+\hat{g}\gamma^{\mu_5}\mathcal{P}_R\left(s_\beta^2\hat{C}^{(11)}_{\Phi e,f_1f_2}+c_\beta^2\hat{C}^{(22)}_{\Phi e,f_1f_2}\right) \tag{C.178}$$

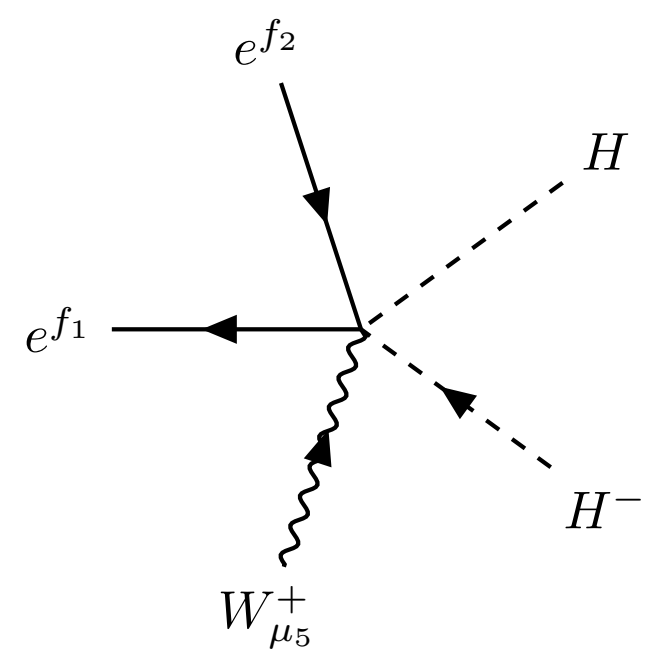

$$+i\hat{g}\gamma^{\mu_5}\mathcal{P}_L\left(s_\beta^2\hat{C}^{(11)[1]}_{\Phi l,f_1f_2}+c_\beta^2\hat{C}^{(22)[1]}_{\Phi l,f_1f_2}\right)$$
$$+i\hat{g}\gamma^{\mu_5}\mathcal{P}_R\left(s_\beta^2\hat{C}^{(11)}_{\Phi e,f_1f_2}+c_\beta^2\hat{C}^{(22)}_{\Phi e,f_1f_2}\right) \tag{C.179}$$

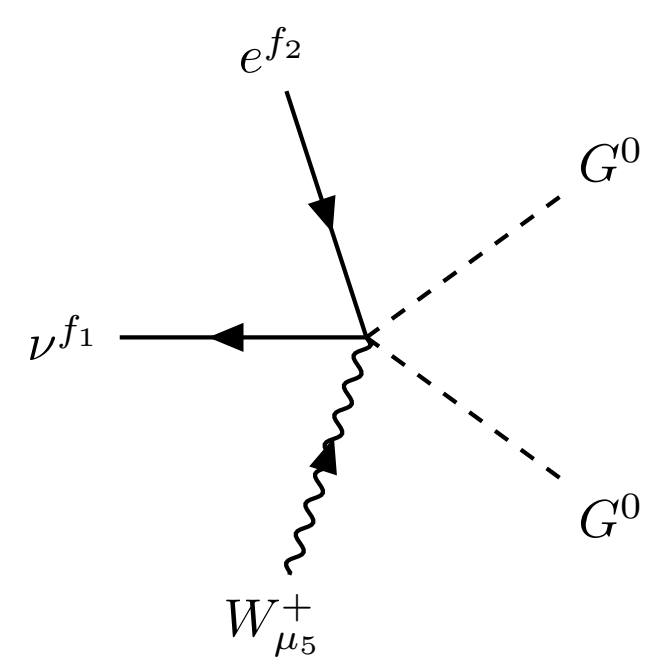

$$-i\sqrt{2}\hat{g}U^*_{g_1f_1}\gamma^{\mu_5}\mathcal{P}_L\left(c_\beta^2\hat{C}^{(11)[3]}_{\Phi l,g_1f_2}+s_\beta^2\hat{C}^{(22)[3]}_{\Phi l,g_1f_2}\right) \tag{C.180}$$

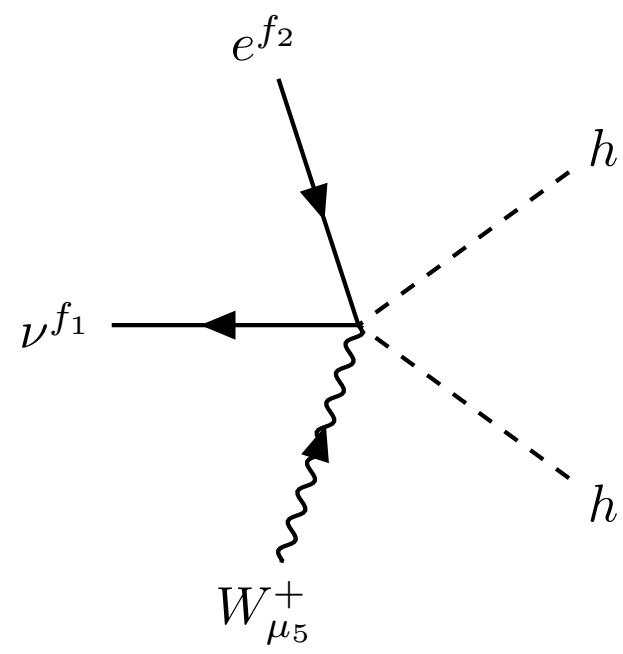

$$-i\sqrt{2}\hat{g}U^*_{g_1f_1}\gamma^{\mu_5}\mathcal{P}_L\left(c^2_\beta\hat{C}^{(11)[3]}_{\Phi l,g_1f_2}+s^2_\beta\hat{C}^{(22)[3]}_{\Phi l,g_1f_2}\right) \tag{C.181}$$

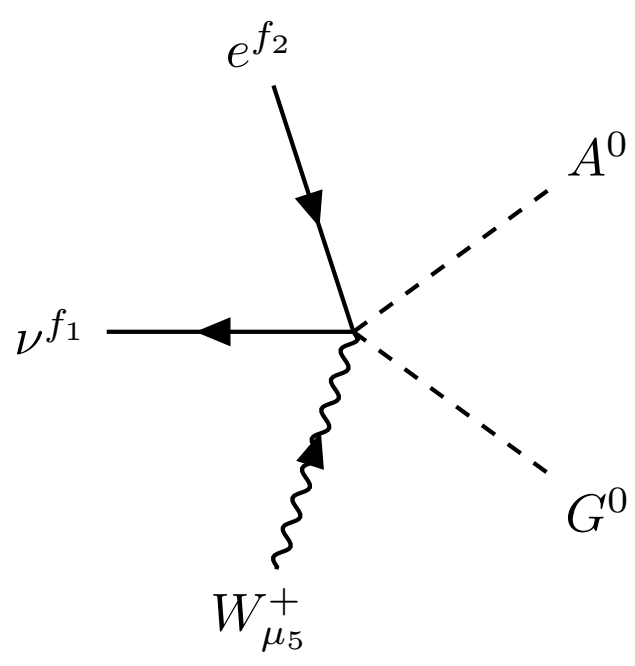

$$+i\sqrt{2}\hat{g}s_\beta c_\beta U^*_{g_1f_1}\gamma^{\mu_5}\mathcal{P}_L\left(\hat{C}^{(11)[3]}_{\Phi l,g_1f_2}-\hat{C}^{(22)[3]}_{\Phi l,g_1f_2}\right) \tag{C.182}$$

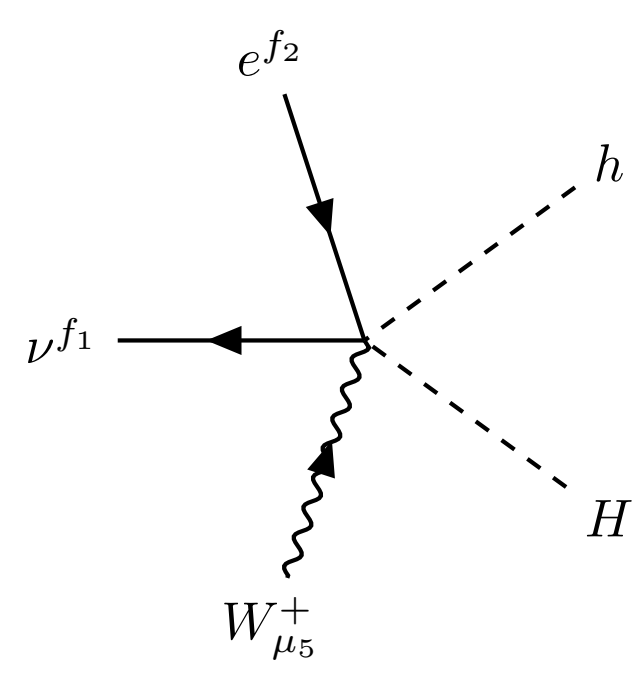

$$-i\sqrt{2}\hat{g}s_\beta c_\beta U^*_{g_1f_1}\gamma^{\mu_5}\mathcal{P}_L\left(\hat{C}^{(11)[3]}_{\Phi l,g_1f_2}-\hat{C}^{(22)[3]}_{\Phi l,g_1f_2}\right) \tag{C.183}$$

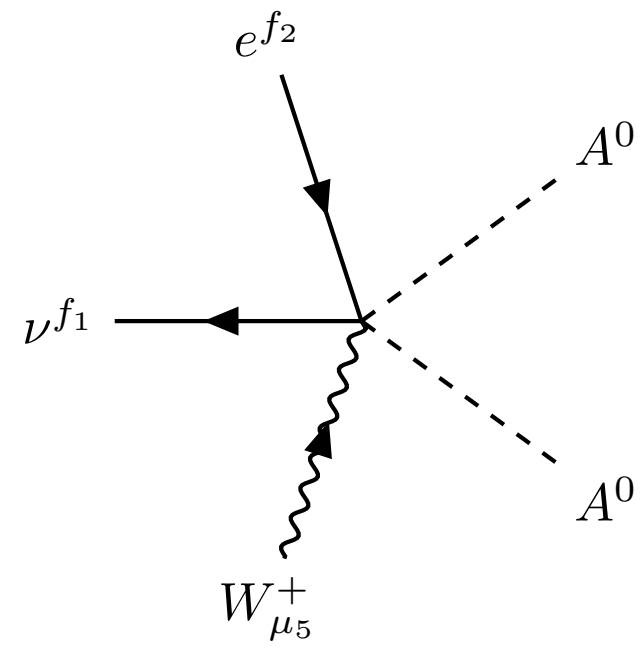

$$-i\sqrt{2}\hat{g}U^*_{g_1f_1}\gamma^{\mu_5}\mathcal{P}_L\left(s^2_\beta\hat{C}^{(11)[3]}_{\Phi l,g_1f_2}+c^2_\beta\hat{C}^{(22)[3]}_{\Phi l,g_1f_2}\right) \tag{C.184}$$

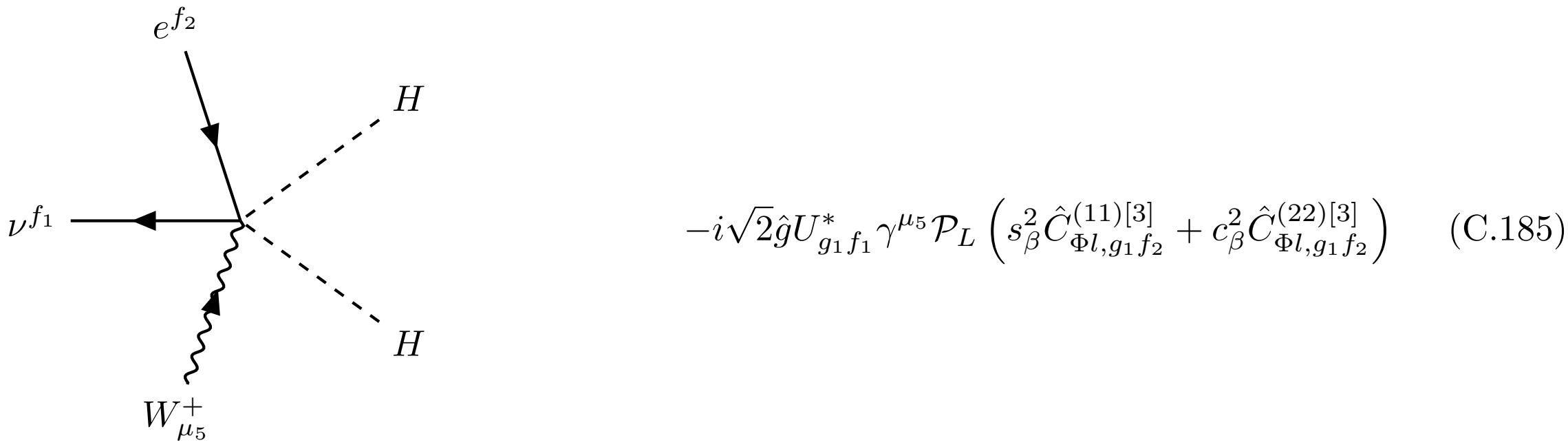

$$-i\sqrt{2}\hat{g}U^*_{g_1f_1}\gamma^{\mu_5}\mathcal{P}_L\left(s_\beta^2\hat{C}^{(11)[3]}_{\Phi l,g_1f_2}+c_\beta^2\hat{C}^{(22)[3]}_{\Phi l,g_1f_2}\right) \tag{C.185}$$

$$-\frac{i\left(\hat{g}'^2-\hat{g}^2\right)U_{g_2f_2}U^*_{g_1f_1}\gamma^{\mu_5}\gamma^5}{\sqrt{\hat{g}'^2+\hat{g}^2}}\Big(c_\beta^2\hat{C}^{(11)[1]}_{\Phi l,g_1g_2}+c_\beta^2\hat{C}^{(11)[3]}_{\Phi l,g_1g_2} +s_\beta^2\left(\hat{C}^{(22)[1]}_{\Phi l,g_1g_2}+\hat{C}^{(22)[3]}_{\Phi l,g_1g_2}\right)\Big) \tag{C.186}$$

$$+\frac{is_\beta c_\beta\left(\hat{g}'^2-\hat{g}^2\right)U_{g_2f_2}U^*_{g_1f_1}\gamma^{\mu_5}\gamma^5}{\sqrt{\hat{g}'^2+\hat{g}^2}}\left(\hat{C}^{(11)[1]}_{\Phi l,g_1g_2}+\hat{C}^{(11)[3]}_{\Phi l,g_1g_2}-\hat{C}^{(22)[1]}_{\Phi l,g_1g_2}-\hat{C}^{(22)[3]}_{\Phi l,g_1g_2}\right) \tag{C.187}$$

$$
-\frac{i\left(\hat{g}'^2-\hat{g}^2\right)U_{g_2f_2}U^*_{g_1f_1}\gamma^{\mu_5}\gamma^5}{\sqrt{\hat{g}'^2+\hat{g}^2}}\left(s_\beta^2\left(\hat{C}^{(11)[1]}_{\Phi l,g_1g_2}+\hat{C}^{(11)[3]}_{\Phi l,g_1g_2}\right)\right.
$$
$$
\left.+c_\beta^2\hat{C}^{(22)[1]}_{\Phi l,g_1g_2}+c_\beta^2\hat{C}^{(22)[3]}_{\Phi l,g_1g_2}\right) \tag{C.188}
$$

$$
+\frac{i\left(\hat{g}'^2-\hat{g}^2\right)\gamma^{\mu_5}\mathcal{P}_L}{2\sqrt{\hat{g}'^2+\hat{g}^2}}\left(2c_\beta^2\hat{C}^{(11)[1]}_{\Phi l,f_1f_2}-2c_\beta^2\hat{C}^{(11)[3]}_{\Phi l,f_1f_2}+2s_\beta^2\left(\hat{C}^{(22)[1]}_{\Phi l,f_1f_2}-\hat{C}^{(22)[3]}_{\Phi l,f_1f_2}\right)\right)
$$
$$
+\frac{i\left(\hat{g}'^2-\hat{g}^2\right)\gamma^{\mu_5}\mathcal{P}_R}{2\sqrt{\hat{g}'^2+\hat{g}^2}}\left(2c_\beta^2\hat{C}^{(11)}_{\Phi e,f_1f_2}+2s_\beta^2\hat{C}^{(22)}_{\Phi e,f_1f_2}\right) \tag{C.189}
$$

$$
-\frac{is_\beta c_\beta\left(\hat{g}'^2-\hat{g}^2\right)\gamma^{\mu_5}\mathcal{P}_L}{\sqrt{\hat{g}'^2+\hat{g}^2}}\left(\hat{C}^{(11)[1]}_{\Phi l,f_1f_2}-\hat{C}^{(11)[3]}_{\Phi l,f_1f_2}-\hat{C}^{(22)[1]}_{\Phi l,f_1f_2}+\hat{C}^{(22)[3]}_{\Phi l,f_1f_2}\right)
$$
$$
-\frac{is_\beta c_\beta\left(\hat{g}'^2-\hat{g}^2\right)\gamma^{\mu_5}\mathcal{P}_R}{\sqrt{\hat{g}'^2+\hat{g}^2}}\left(\hat{C}^{(11)}_{\Phi e,f_1f_2}-\hat{C}^{(22)}_{\Phi e,f_1f_2}\right) \tag{C.190}
$$

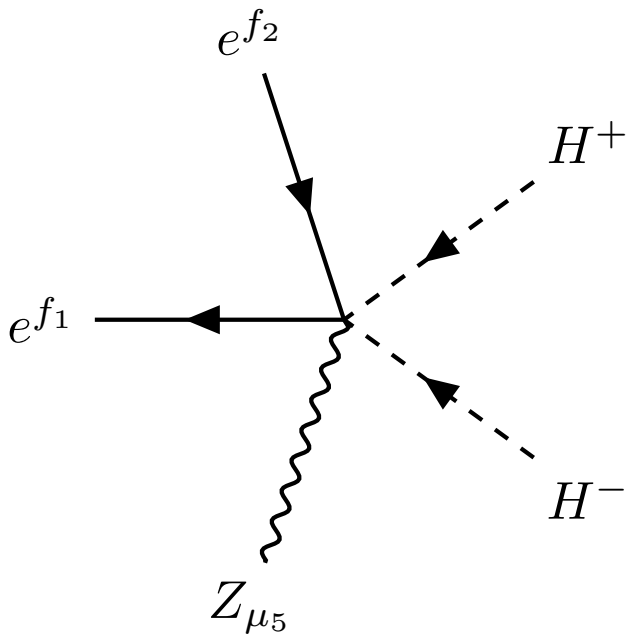

$$
\begin{aligned}
&+\frac{i\left(\hat{g}'^2-\hat{g}^2\right)\gamma^{\mu_5}\mathcal{P}_L}{\sqrt{\hat{g}'^2+\hat{g}^2}}\left(s_\beta^2\left(\hat{C}^{(11)[1]}_{\Phi l,f_1f_2}-\hat{C}^{(11)[3]}_{\Phi l,f_1f_2}\right)\right.\\
&\qquad\left.+c_\beta^2\hat{C}^{(22)[1]}_{\Phi l,f_1f_2}-c_\beta^2\hat{C}^{(22)[3]}_{\Phi l,f_1f_2}\right)\\
&+\frac{i\left(\hat{g}'^2-\hat{g}^2\right)\gamma^{\mu_5}\mathcal{P}_R}{2\sqrt{\hat{g}'^2+\hat{g}^2}}\left(2s_\beta^2\hat{C}^{(11)}_{\Phi e,f_1f_2}+2c_\beta^2\hat{C}^{(22)}_{\Phi e,f_1f_2}\right)
\end{aligned}
\tag{C.191}
$$

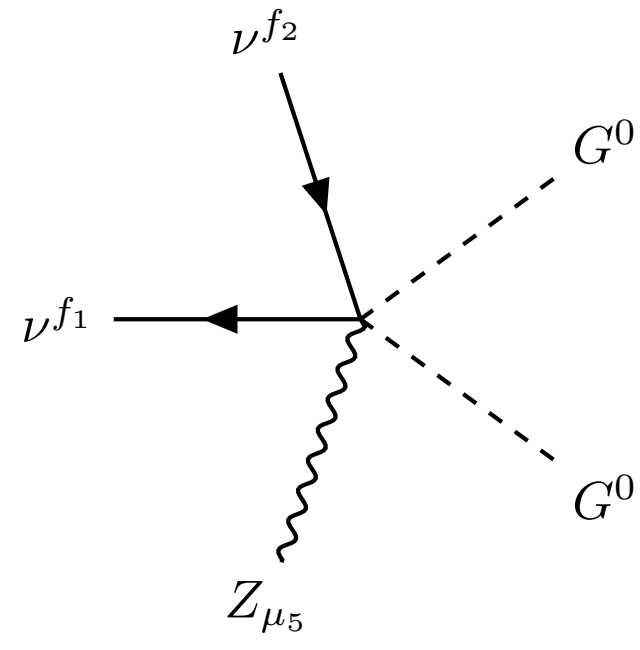

$$
\begin{aligned}
&-i\sqrt{\hat{g}'^2+\hat{g}^2}U_{g_2f_2}U^*_{g_1f_1}\gamma^{\mu_5}\gamma^5\left(c_\beta^2\hat{C}^{(11)[1]}_{\Phi l,g_1g_2}-c_\beta^2\hat{C}^{(11)[3]}_{\Phi l,g_1g_2}\right.\\
&\qquad\left.+s_\beta^2\left(\hat{C}^{(22)[1]}_{\Phi l,g_1g_2}-\hat{C}^{(22)[3]}_{\Phi l,g_1g_2}\right)\right)
\end{aligned}
\tag{C.192}
$$

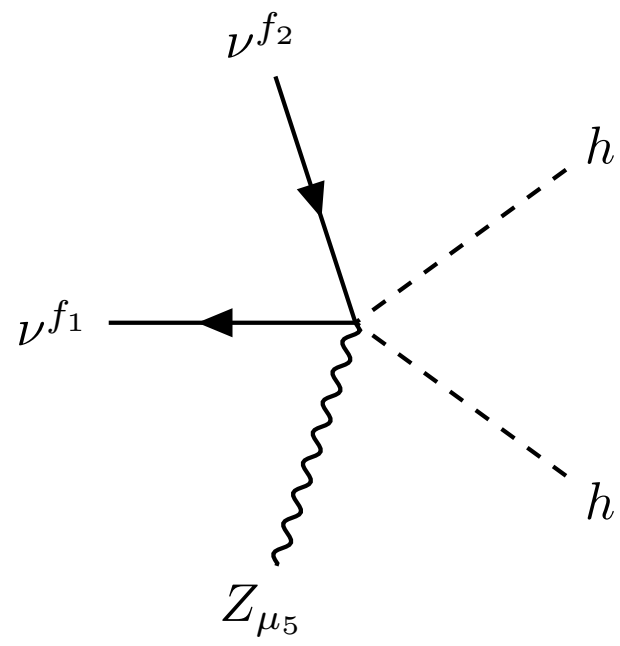

$$
\begin{aligned}
&-i\sqrt{\hat{g}'^2+\hat{g}^2}U_{g_2f_2}U^*_{g_1f_1}\gamma^{\mu_5}\gamma^5\left(c_\beta^2\hat{C}^{(11)[1]}_{\Phi l,g_1g_2}-c_\beta^2\hat{C}^{(11)[3]}_{\Phi l,g_1g_2}\right.\\
&\qquad\left.+s_\beta^2\left(\hat{C}^{(22)[1]}_{\Phi l,g_1g_2}-\hat{C}^{(22)[3]}_{\Phi l,g_1g_2}\right)\right)
\end{aligned}
\tag{C.193}
$$

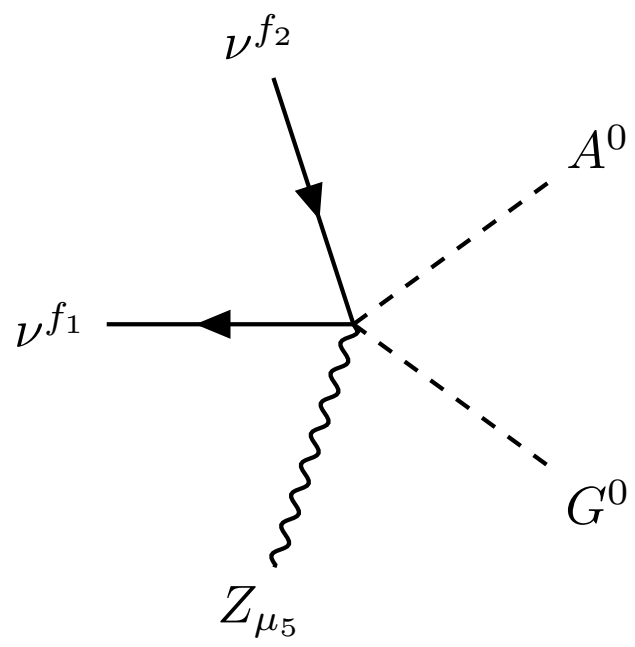

$$
+is_\beta c_\beta\sqrt{\hat{g}'^2+\hat{g}^2}U_{g_2f_2}U^*_{g_1f_1}\gamma^{\mu_5}\gamma^5\left(\hat{C}^{(11)[1]}_{\Phi l,g_1g_2}-\hat{C}^{(11)[3]}_{\Phi l,g_1g_2}-\hat{C}^{(22)[1]}_{\Phi l,g_1g_2}+\hat{C}^{(22)[3]}_{\Phi l,g_1g_2}\right)
\tag{C.194}
$$

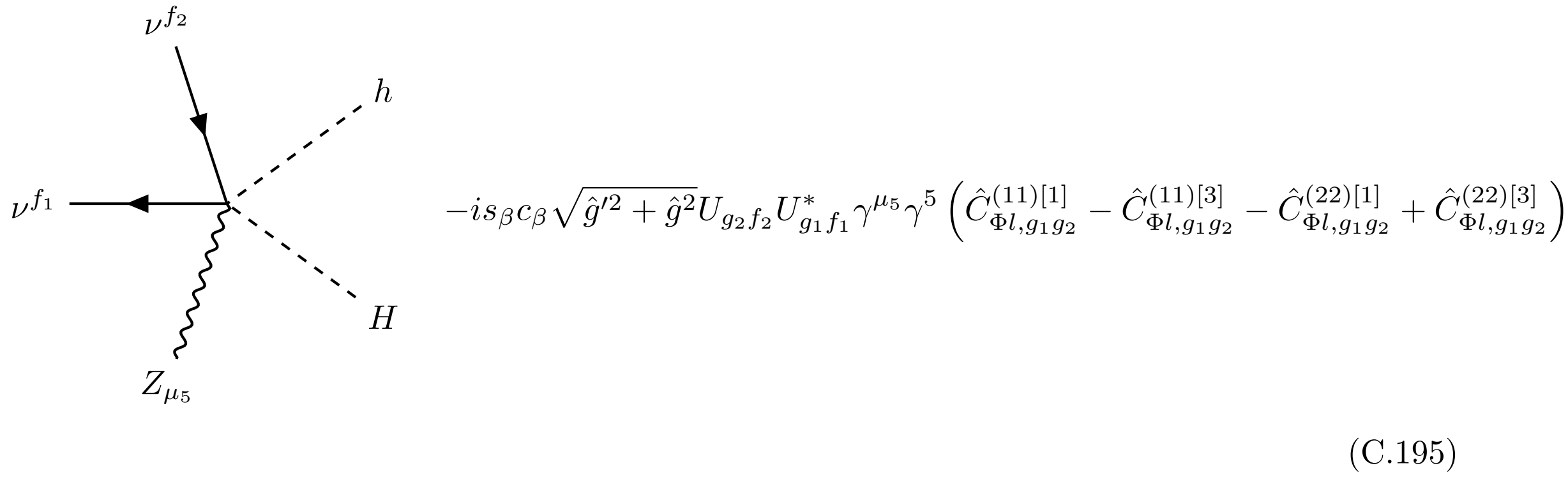

$$-is_\beta c_\beta \sqrt{\hat{g}'^2+\hat{g}^2}U_{g_2 f_2}U^*_{g_1 f_1}\gamma^{\mu_5}\gamma^5\left(\hat{C}^{(11)[1]}_{\Phi l,g_1g_2}-\hat{C}^{(11)[3]}_{\Phi l,g_1g_2}-\hat{C}^{(22)[1]}_{\Phi l,g_1g_2}+\hat{C}^{(22)[3]}_{\Phi l,g_1g_2}\right) \tag{C.195}$$

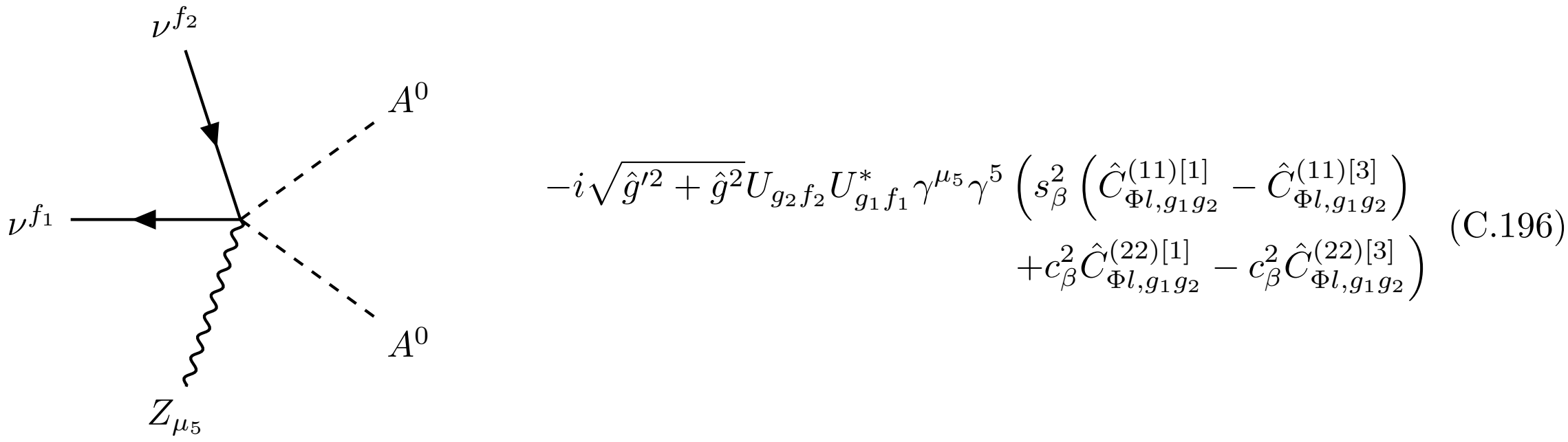

$$\begin{aligned}-i\sqrt{\hat{g}'^2+\hat{g}^2}U_{g_2 f_2}U^*_{g_1 f_1}\gamma^{\mu_5}\gamma^5\Big(s^2_\beta\left(\hat{C}^{(11)[1]}_{\Phi l,g_1g_2}-\hat{C}^{(11)[3]}_{\Phi l,g_1g_2}\right)\\ +c^2_\beta\hat{C}^{(22)[1]}_{\Phi l,g_1g_2}-c^2_\beta\hat{C}^{(22)[3]}_{\Phi l,g_1g_2}\Big)\end{aligned} \tag{C.196}$$

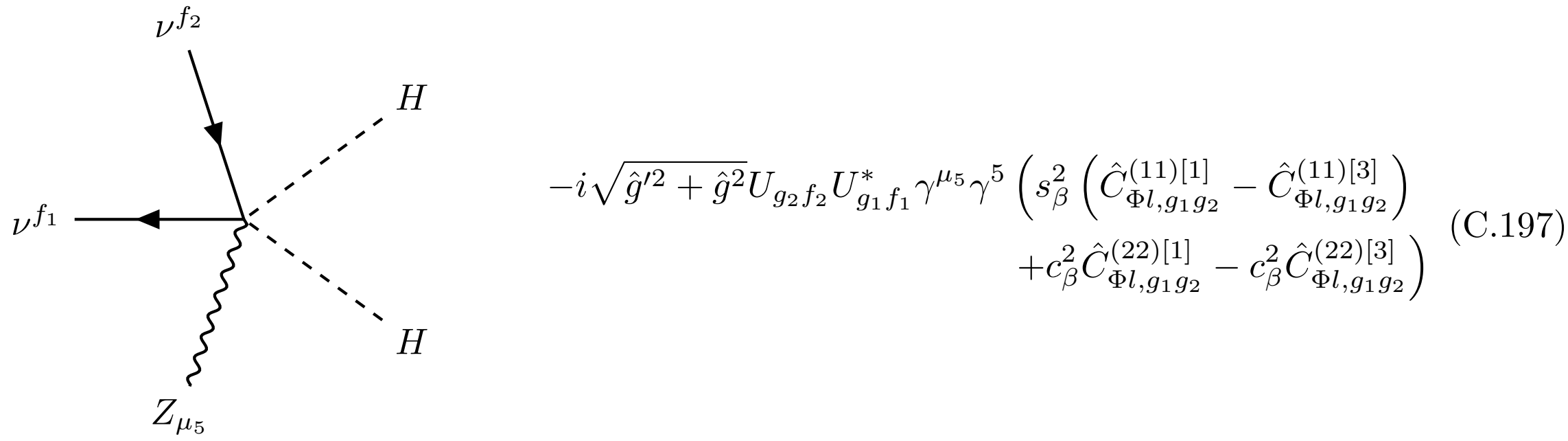

$$\begin{aligned}-i\sqrt{\hat{g}'^2+\hat{g}^2}U_{g_2 f_2}U^*_{g_1 f_1}\gamma^{\mu_5}\gamma^5\Big(s^2_\beta\left(\hat{C}^{(11)[1]}_{\Phi l,g_1g_2}-\hat{C}^{(11)[3]}_{\Phi l,g_1g_2}\right)\\ +c^2_\beta\hat{C}^{(22)[1]}_{\Phi l,g_1g_2}-c^2_\beta\hat{C}^{(22)[3]}_{\Phi l,g_1g_2}\Big)\end{aligned} \tag{C.197}$$

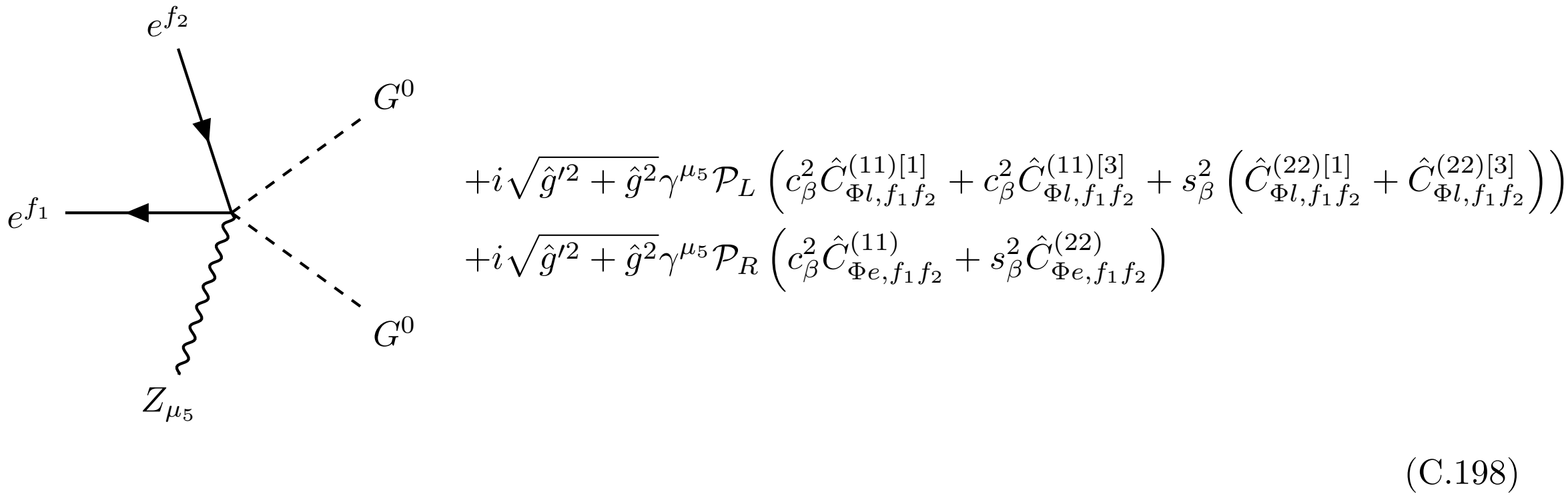

$$
\begin{aligned}
&+i\sqrt{\hat{g}'^2+\hat{g}^2}\,\gamma^{\mu_5}\mathcal{P}_L\left(c_\beta^2\hat{C}^{(11)[1]}_{\Phi l,f_1f_2}+c_\beta^2\hat{C}^{(11)[3]}_{\Phi l,f_1f_2}+s_\beta^2\left(\hat{C}^{(22)[1]}_{\Phi l,f_1f_2}+\hat{C}^{(22)[3]}_{\Phi l,f_1f_2}\right)\right)\\
&+i\sqrt{\hat{g}'^2+\hat{g}^2}\,\gamma^{\mu_5}\mathcal{P}_R\left(c_\beta^2\hat{C}^{(11)}_{\Phi e,f_1f_2}+s_\beta^2\hat{C}^{(22)}_{\Phi e,f_1f_2}\right)
\end{aligned}
\tag{C.198}
$$

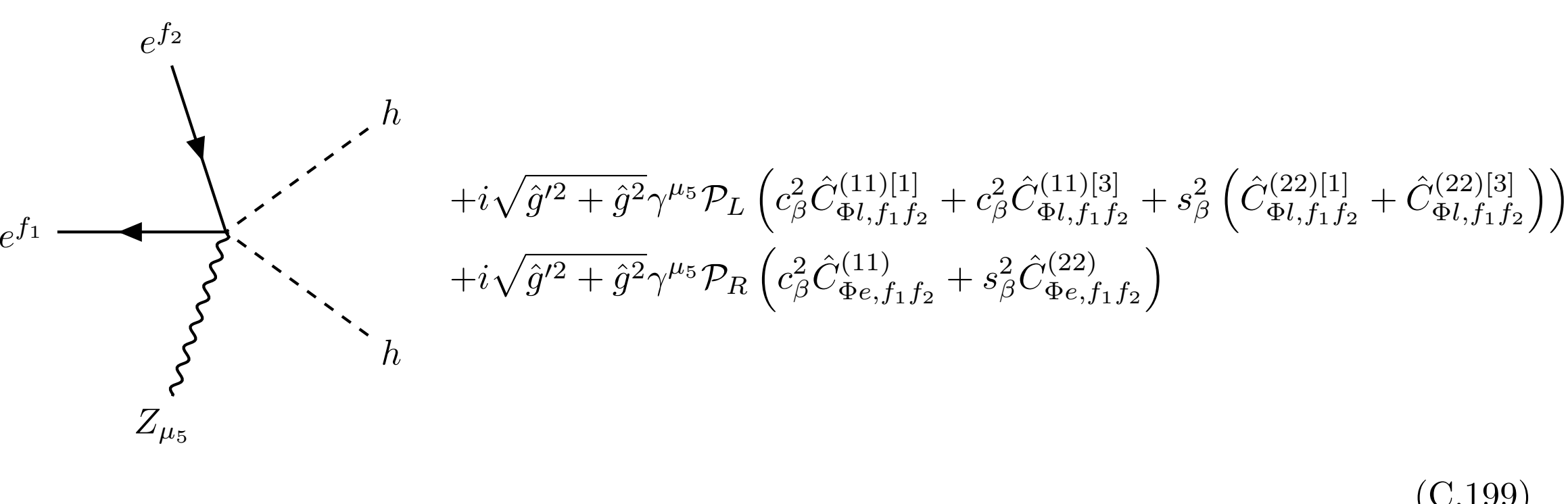

$$
\begin{aligned}
&+i\sqrt{\hat{g}'^2+\hat{g}^2}\,\gamma^{\mu_5}\mathcal{P}_L\left(c_\beta^2\hat{C}^{(11)[1]}_{\Phi l,f_1f_2}+c_\beta^2\hat{C}^{(11)[3]}_{\Phi l,f_1f_2}+s_\beta^2\left(\hat{C}^{(22)[1]}_{\Phi l,f_1f_2}+\hat{C}^{(22)[3]}_{\Phi l,f_1f_2}\right)\right)\\
&+i\sqrt{\hat{g}'^2+\hat{g}^2}\,\gamma^{\mu_5}\mathcal{P}_R\left(c_\beta^2\hat{C}^{(11)}_{\Phi e,f_1f_2}+s_\beta^2\hat{C}^{(22)}_{\Phi e,f_1f_2}\right)
\end{aligned}
\tag{C.199}
$$

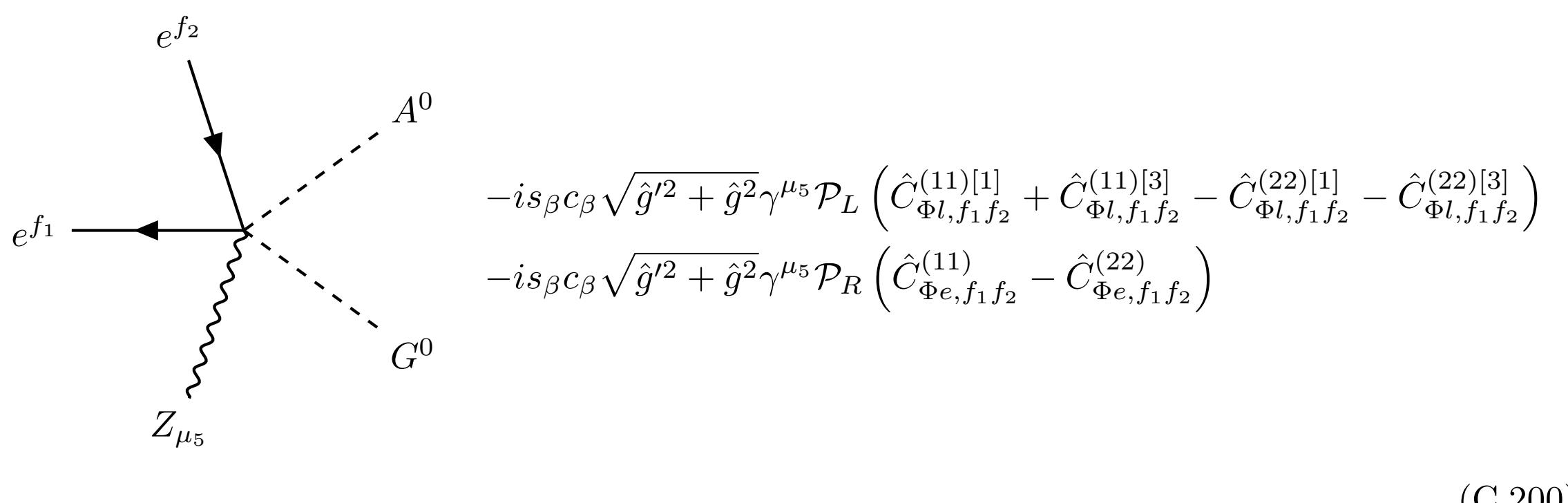

$$
\begin{aligned}
&-is_\beta c_\beta\sqrt{\hat{g}'^2+\hat{g}^2}\,\gamma^{\mu_5}\mathcal{P}_L\left(\hat{C}^{(11)[1]}_{\Phi l,f_1f_2}+\hat{C}^{(11)[3]}_{\Phi l,f_1f_2}-\hat{C}^{(22)[1]}_{\Phi l,f_1f_2}-\hat{C}^{(22)[3]}_{\Phi l,f_1f_2}\right)\\
&-is_\beta c_\beta\sqrt{\hat{g}'^2+\hat{g}^2}\,\gamma^{\mu_5}\mathcal{P}_R\left(\hat{C}^{(11)}_{\Phi e,f_1f_2}-\hat{C}^{(22)}_{\Phi e,f_1f_2}\right)
\end{aligned}
\tag{C.200}
$$

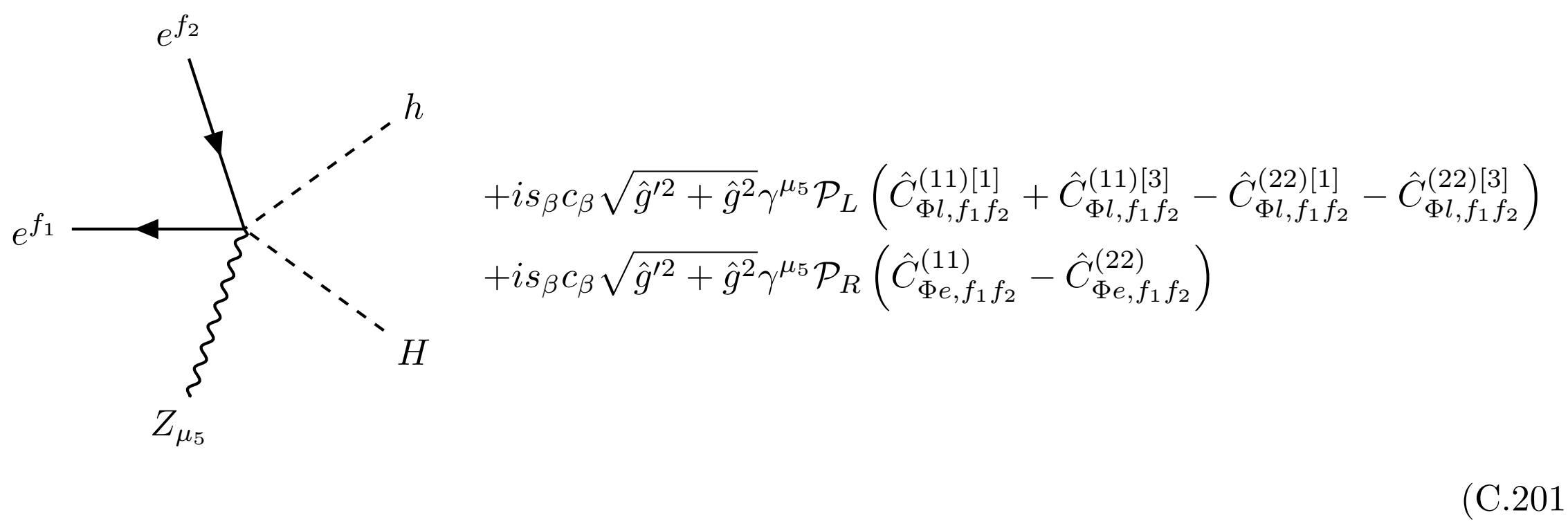

$$+is_\beta c_\beta\sqrt{\hat{g}'^2+\hat{g}^2}\gamma^{\mu_5}\mathcal{P}_L\left(\hat{C}^{(11)[1]}_{\Phi l,f_1f_2}+\hat{C}^{(11)[3]}_{\Phi l,f_1f_2}-\hat{C}^{(22)[1]}_{\Phi l,f_1f_2}-\hat{C}^{(22)[3]}_{\Phi l,f_1f_2}\right)$$
$$+is_\beta c_\beta\sqrt{\hat{g}'^2+\hat{g}^2}\gamma^{\mu_5}\mathcal{P}_R\left(\hat{C}^{(11)}_{\Phi e,f_1f_2}-\hat{C}^{(22)}_{\Phi e,f_1f_2}\right)$$

(C.201)

$$\begin{aligned}&+i\sqrt{\hat{g}'^2+\hat{g}^2}\gamma^{\mu_5}\mathcal{P}_L\left(s^2_\beta\left(\hat{C}^{(11)[1]}_{\Phi l,f_1f_2}+\hat{C}^{(11)[3]}_{\Phi l,f_1f_2}\right)\right.\\&\qquad\left.+c^2_\beta\hat{C}^{(22)[1]}_{\Phi l,f_1f_2}+c^2_\beta\hat{C}^{(22)[3]}_{\Phi l,f_1f_2}\right)\\&+i\sqrt{\hat{g}'^2+\hat{g}^2}\gamma^{\mu_5}\mathcal{P}_R\left(s^2_\beta\hat{C}^{(11)}_{\Phi e,f_1f_2}+c^2_\beta\hat{C}^{(22)}_{\Phi e,f_1f_2}\right)\end{aligned}\tag{C.202}$$

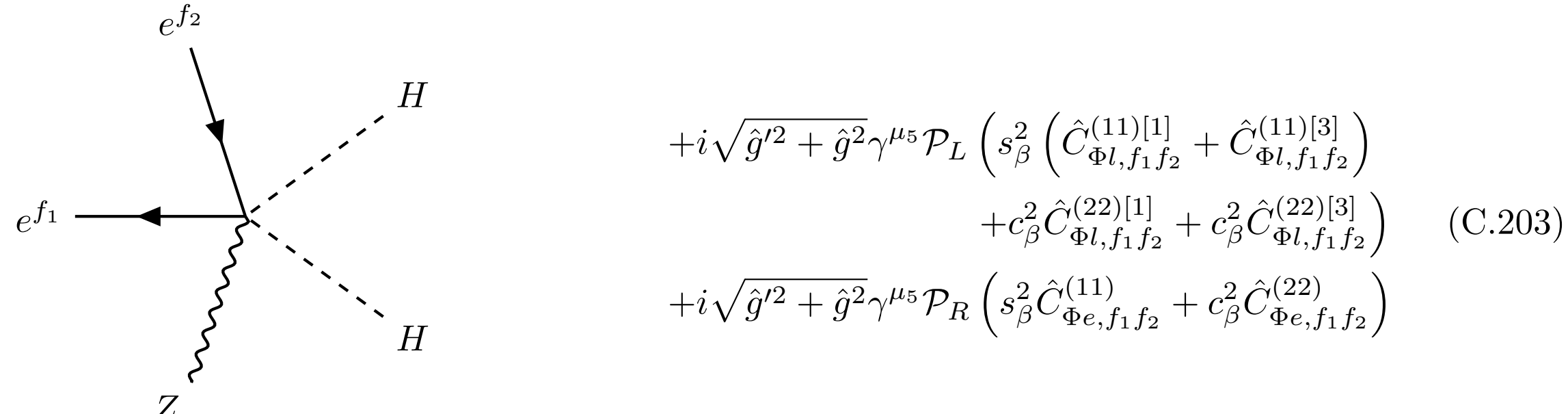

$$\begin{aligned}&+i\sqrt{\hat{g}'^2+\hat{g}^2}\gamma^{\mu_5}\mathcal{P}_L\left(s^2_\beta\left(\hat{C}^{(11)[1]}_{\Phi l,f_1f_2}+\hat{C}^{(11)[3]}_{\Phi l,f_1f_2}\right)\right.\\&\qquad\left.+c^2_\beta\hat{C}^{(22)[1]}_{\Phi l,f_1f_2}+c^2_\beta\hat{C}^{(22)[3]}_{\Phi l,f_1f_2}\right)\\&+i\sqrt{\hat{g}'^2+\hat{g}^2}\gamma^{\mu_5}\mathcal{P}_R\left(s^2_\beta\hat{C}^{(11)}_{\Phi e,f_1f_2}+c^2_\beta\hat{C}^{(22)}_{\Phi e,f_1f_2}\right)\end{aligned}\tag{C.203}$$

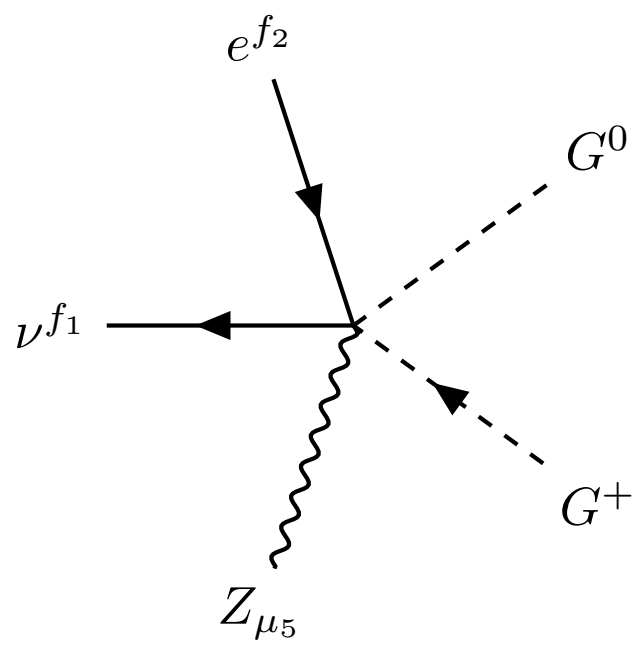

$$+\frac{\sqrt{2}\hat{g}'^2 U^*_{g_1 f_1}\gamma^{\mu_5}\mathcal{P}_L}{\sqrt{\hat{g}'^2+\hat{g}^2}}\left(c_\beta^2 \hat{C}^{(11)[3]}_{\Phi l,g_1 f_2} + s_\beta^2 \hat{C}^{(22)[3]}_{\Phi l,g_1 f_2}\right) \tag{C.204}$$

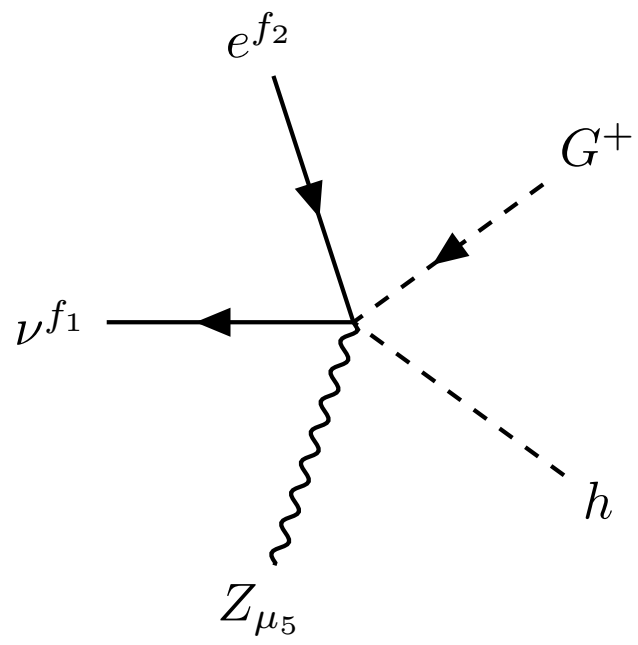

$$+\frac{i\sqrt{2}\hat{g}'^2 U^*_{g_1 f_1}\gamma^{\mu_5}\mathcal{P}_L}{\sqrt{\hat{g}'^2+\hat{g}^2}}\left(c_\beta^2 \hat{C}^{(11)[3]}_{\Phi l,g_1 f_2} + s_\beta^2 \hat{C}^{(22)[3]}_{\Phi l,g_1 f_2}\right) \tag{C.205}$$

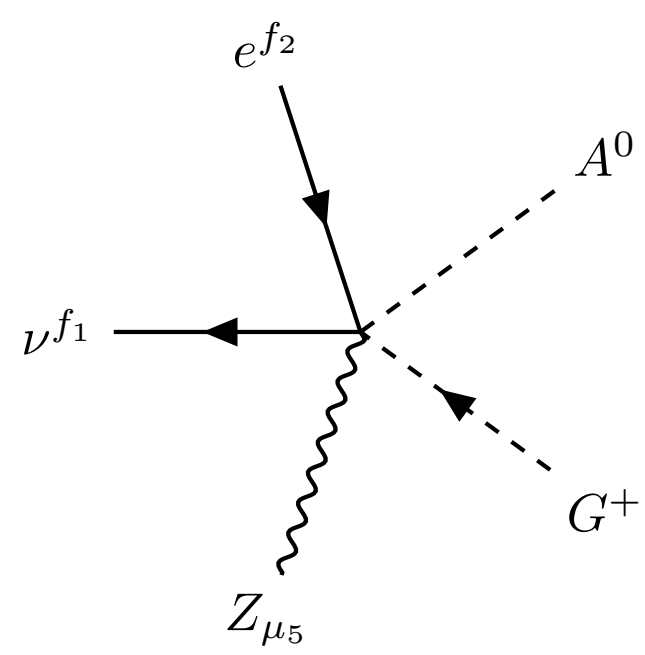

$$+\frac{\sqrt{2}s_\beta c_\beta \hat{g}'^2 U^*_{g_1 f_1}\gamma^{\mu_5}\mathcal{P}_L}{\sqrt{\hat{g}'^2+\hat{g}^2}}\left(\hat{C}^{(22)[3]}_{\Phi l,g_1 f_2} - \hat{C}^{(11)[3]}_{\Phi l,g_1 f_2}\right) \tag{C.206}$$

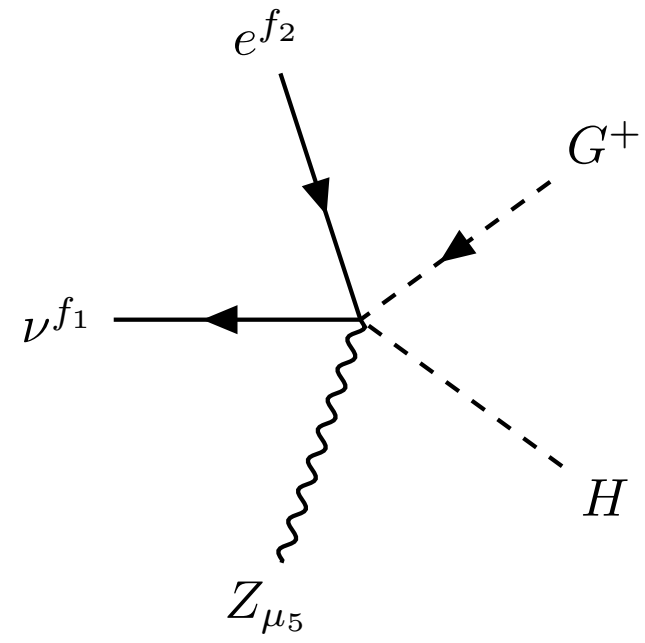

$$+\frac{i\sqrt{2}s_\beta c_\beta \hat{g}'^2 U^*_{g_1 f_1}\gamma^{\mu_5}\mathcal{P}_L}{\sqrt{\hat{g}'^2+\hat{g}^2}}\left(\hat{C}^{(11)[3]}_{\Phi l,g_1 f_2} - \hat{C}^{(22)[3]}_{\Phi l,g_1 f_2}\right) \tag{C.207}$$

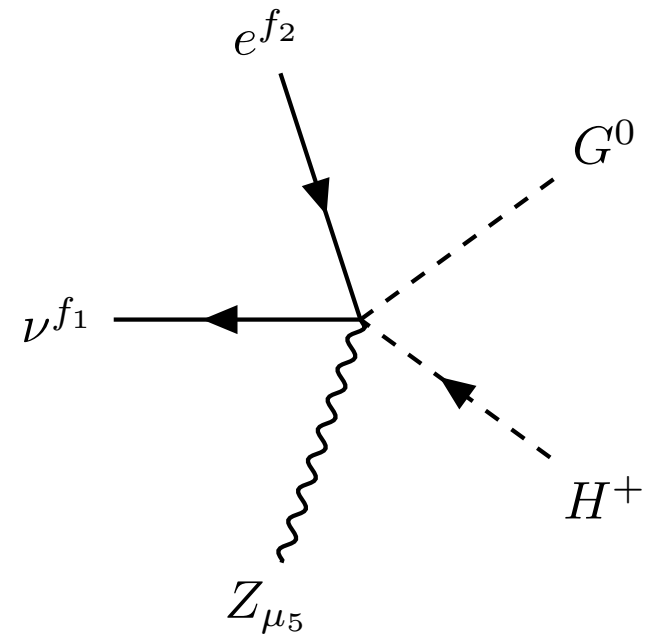

$$+\frac{\sqrt{2}s_\beta c_\beta \hat{g}'^2 U^*_{g_1 f_1}\gamma^{\mu_5}\mathcal{P}_L}{\sqrt{\hat{g}'^2+\hat{g}^2}}\left(\hat{C}^{(22)[3]}_{\Phi l,g_1 f_2}-\hat{C}^{(11)[3]}_{\Phi l,g_1 f_2}\right) \quad \text{(C.208)}$$

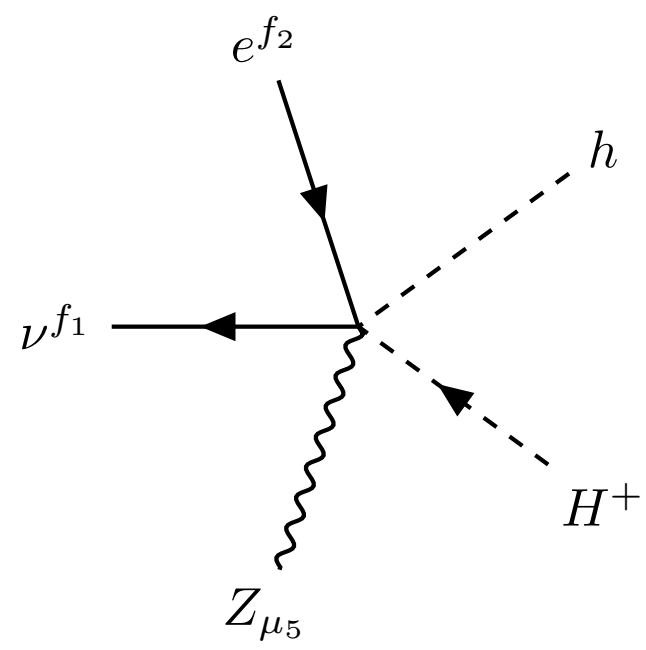

$$-\frac{i\sqrt{2}s_\beta c_\beta \hat{g}'^2 U^*_{g_1 f_1}\gamma^{\mu_5}\mathcal{P}_L}{\sqrt{\hat{g}'^2+\hat{g}^2}}\left(\hat{C}^{(11)[3]}_{\Phi l,g_1 f_2}-\hat{C}^{(22)[3]}_{\Phi l,g_1 f_2}\right) \quad \text{(C.209)}$$

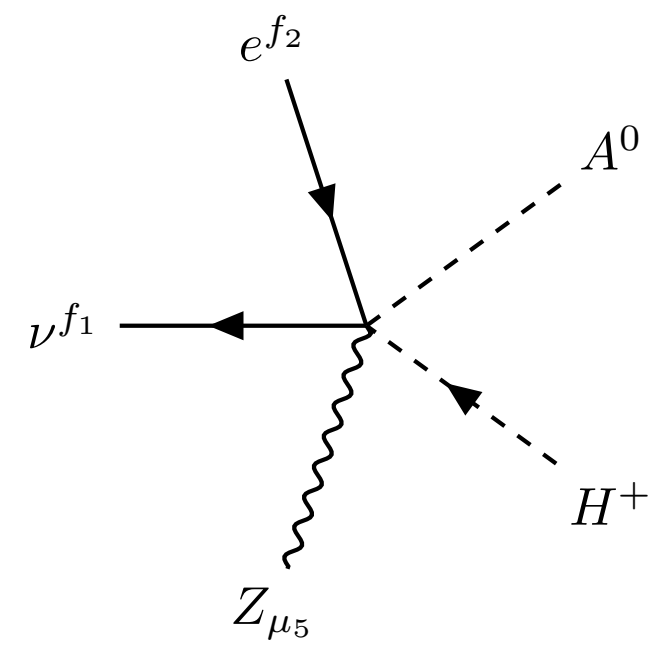

$$+\frac{\sqrt{2}\hat{g}'^2 U^*_{g_1 f_1}\gamma^{\mu_5}\mathcal{P}_L}{\sqrt{\hat{g}'^2+\hat{g}^2}}\left(s^2_\beta\hat{C}^{(11)[3]}_{\Phi l,g_1 f_2}+c^2_\beta\hat{C}^{(22)[3]}_{\Phi l,g_1 f_2}\right) \quad \text{(C.210)}$$

$e^{f_2}$, $H$, $\nu^{f_1}$, $H^+$, $Z_{\mu_5}$

$$
-\frac{i\sqrt{2}\hat{g}'^2 U^*_{g_1 f_1}\gamma^{\mu_5}\mathcal{P}_L}{\sqrt{\hat{g}'^2+\hat{g}^2}}\left(s_\beta^2 \hat{C}^{(11)[3]}_{\Phi l, g_1 f_2} + c_\beta^2 \hat{C}^{(22)[3]}_{\Phi l, g_1 f_2}\right) \tag{C.211}
$$

## C.2 Quark Interactions

In this subsection, we present the complete set of Feynman rules for up-type quarks. We omit any electroweak vertices involving only the down-type quarks with the exception of the photon and $Z$ couplings. We do this because the vertices are identical to the charged lepton vertices upon replacement of $e \to d$ and $l \to q$ in the label of the Wilson coefficients and Feynman diagrams.

$u^{f_2}$, $u^{f_1}$, $G^0$

$$
\begin{aligned}
&+\frac{\delta_{f_1 f_2} m_{u_{f_1}}\gamma^5}{2\sqrt{2}v}\left(A_1' + t_\beta B' + 2\delta_{c_{\hat{\beta}}} - 2\right)\\
&+\sqrt{2}v V_{f_1 g_1} \not{p}_3 \mathcal{P}_L V^*_{f_2 g_2}\left(-c_\beta^2 \hat{C}^{(11)[1]}_{\Phi q, g_1 g_2} + c_\beta^2 \hat{C}^{(11)[3]}_{\Phi q, g_1 g_2} + s_\beta^2\left(\hat{C}^{(22)[3]}_{\Phi q, g_1 g_2} - \hat{C}^{(22)[1]}_{\Phi q, g_1 g_2}\right)\right)\\
&-\sqrt{2}v \not{p}_3 \mathcal{P}_R\left(c_\beta^2 \hat{C}^{(11)}_{\Phi u, f_1 f_2} + s_\beta^2 \hat{C}^{(22)}_{\Phi u, f_1 f_2}\right)
\end{aligned} \tag{C.212}
$$

$u^{f_2}$, $u^{f_1}$, $h$

$$
\begin{aligned}
&+\frac{i\delta_{f_1 f_2} m_{u_{f_1}}}{2\sqrt{2}v}\left(A_1 + B t_\beta - 2\right)\\
&+i\sqrt{2}v^2 s_\beta\left(c_\beta^2\left(\mathcal{P}_L \hat{C}^{(11)*}_{u\Phi_2, f_2 f_1} + \mathcal{P}_R \hat{C}^{(11)}_{u\Phi_2, f_1 f_2}\right)\right.\\
&\qquad\qquad +c_\beta^2\left(\mathcal{P}_L \hat{C}^{(12)*}_{u\Phi_1, f_2 f_1} + \mathcal{P}_R \hat{C}^{(12)}_{u\Phi_1, f_1 f_2}\right)\\
&\qquad\qquad +c_\beta^2\left(\mathcal{P}_L \hat{C}^{(21)*}_{u\Phi_1, f_2 f_1} + \mathcal{P}_R \hat{C}^{(21)}_{u\Phi_1, f_1 f_2}\right)\\
&\qquad\qquad \left.+s_\beta^2\left(\mathcal{P}_L \hat{C}^{(22)*}_{u\Phi_2, f_2 f_1} + \mathcal{P}_R \hat{C}^{(22)}_{u\Phi_2, f_1 f_2}\right)\right)
\end{aligned} \tag{C.213}
$$

$u^{f_1}$, $u^{f_2}$, $A^0$

$$
\begin{aligned}
&-\frac{\left(c_\beta\left(A_2'+2\delta_{c_{\hat\beta}}-2\right)-s_\beta B'\right)}{2\sqrt{2}}\left(\mathcal{P}_L\hat{y}^{(2)*}_{u,f_2f_1}-\mathcal{P}_R\hat{y}^{(2)}_{u,f_1f_2}\right)\\
&+\frac{v^2c_\beta}{\sqrt{2}}\Big(-c_\beta^2\left(\mathcal{P}_L\hat{C}^{(11)*}_{u\Phi_2,f_2f_1}-\mathcal{P}_R\hat{C}^{(11)}_{u\Phi_2,f_1f_2}\right)\\
&\qquad+\left(2s_\beta^2+c_\beta^2\right)\left(\mathcal{P}_L\hat{C}^{(12)*}_{u\Phi_1,f_2f_1}-\mathcal{P}_R\hat{C}^{(12)}_{u\Phi_1,f_1f_2}\right)\\
&\qquad-c_\beta^2\left(\mathcal{P}_L\hat{C}^{(21)*}_{u\Phi_1,f_2f_1}-\mathcal{P}_R\hat{C}^{(21)}_{u\Phi_1,f_1f_2}\right)\\
&\qquad-s_\beta^2\left(\mathcal{P}_L\hat{C}^{(22)*}_{u\Phi_2,f_2f_1}-\mathcal{P}_R\hat{C}^{(22)}_{u\Phi_2,f_1f_2}\right)\Big)\\
&+\sqrt{2}vs_\beta c_\beta V_{f_1g_1}\not{p}_3\mathcal{P}_LV^*_{f_2g_2}\left(\hat{C}^{(11)[1]}_{\Phi q,g_1g_2}-\hat{C}^{(11)[3]}_{\Phi q,g_1g_2}-\hat{C}^{(22)[1]}_{\Phi q,g_1g_2}+\hat{C}^{(22)[3]}_{\Phi q,g_1g_2}\right)\\
&+\sqrt{2}vs_\beta c_\beta\not{p}_3\mathcal{P}_R\left(\hat{C}^{(11)}_{\Phi u,f_1f_2}-\hat{C}^{(22)}_{\Phi u,f_1f_2}\right)
\end{aligned}
\tag{C.214}
$$

$u^{f_1}$, $u^{f_2}$, $H$

$$
\begin{aligned}
&-\frac{i\left(\left(A_2-2\right)c_\beta-Bs_\beta\right)}{2\sqrt{2}}\left(\mathcal{P}_L\hat{y}^{(2)*}_{u,f_2f_1}+\mathcal{P}_R\hat{y}^{(2)}_{u,f_1f_2}\right)\\
&-\frac{iv^2c_\beta}{\sqrt{2}}\Big(-\left(2s_\beta^2-c_\beta^2\right)\left(\mathcal{P}_L\hat{C}^{(11)*}_{u\Phi_2,f_2f_1}+\mathcal{P}_R\hat{C}^{(11)}_{u\Phi_2,f_1f_2}\right)\\
&\qquad-\left(2s_\beta^2-c_\beta^2\right)\left(\mathcal{P}_L\hat{C}^{(12)*}_{u\Phi_1,f_2f_1}+\mathcal{P}_R\hat{C}^{(12)}_{u\Phi_1,f_1f_2}\right)\\
&\qquad-\left(2s_\beta^2-c_\beta^2\right)\left(\mathcal{P}_L\hat{C}^{(21)*}_{u\Phi_1,f_2f_1}+\mathcal{P}_R\hat{C}^{(21)}_{u\Phi_1,f_1f_2}\right)\\
&\qquad+3s_\beta^2\left(\mathcal{P}_L\hat{C}^{(22)*}_{u\Phi_2,f_2f_1}+\mathcal{P}_R\hat{C}^{(22)}_{u\Phi_2,f_1f_2}\right)\Big)
\end{aligned}
\tag{C.215}
$$

$u^{f_1}$, $d^{f_2}$, $G^+$

$$
\begin{aligned}
&-\frac{i}{v}\left(\delta_{c_{\hat\beta\pm}}-1\right)\left(m_{u_{f_1}}\mathcal{P}_L-m_{d_{f_2}}\mathcal{P}_R\right)\left(V_{f_1f_2}\right)\\
&-2ivs_\beta c_\beta\not{p}_3\mathcal{P}_R\left(\hat{C}^{(21)}_{\Phi ud,f_1f_2}\right)\\
&+2ivV_{f_1g_1}\not{p}_3\mathcal{P}_L\left(c_\beta^2\hat{C}^{(11)[3]}_{\Phi q,g_1f_2}+s_\beta^2\hat{C}^{(22)[3]}_{\Phi q,g_1f_2}\right)
\end{aligned}
\tag{C.216}
$$

$d^{f_2}$, $u^{f_1}$, $H^+$

$$
\begin{aligned}
&-i\left(s_\beta \mathcal{P}_R V_{f_1 g_1}\left(\delta_{s_{\hat{\beta}\pm}}-1\right)\hat{y}^{(1)}_{d,g_1 f_2}+c_\beta\left(\delta_{c_{\hat{\beta}\pm}}-1\right)\mathcal{P}_L V_{g_1 f_2}\hat{y}^{(2)*}_{u,g_1 f_1}\right)\\
&-iv^2\left(c_\beta^3 \mathcal{P}_L V_{g_1 f_2}\hat{C}^{(11)*}_{u\Phi_2,g_1 f_1}-s_\beta^2 c_\beta \mathcal{P}_L V_{g_1 f_2}\hat{C}^{(12)*}_{u\Phi_1,g_1 f_1}\right.\\
&\qquad -s_\beta^2 c_\beta \mathcal{P}_L V_{g_1 f_2}\hat{C}^{(21)*}_{u\Phi_1,g_1 f_1}+s_\beta^2 c_\beta \mathcal{P}_L V_{g_1 f_2}\hat{C}^{(22)*}_{u\Phi_2,g_1 f_1}\\
&\qquad +s_\beta c_\beta^2 \mathcal{P}_R V_{f_1 g_1}\hat{C}^{(11)}_{d\Phi_1,g_1 f_2}-s_\beta c_\beta^2 \mathcal{P}_R V_{f_1 g_1}\hat{C}^{(12)}_{d\Phi_2,g_1 f_2}\\
&\qquad \left.-s_\beta c_\beta^2 \mathcal{P}_R V_{f_1 g_1}\hat{C}^{(21)}_{d\Phi_2,g_1 f_2}+s_\beta^3 \mathcal{P}_R V_{f_1 g_1}\hat{C}^{(22)}_{d\Phi_1,g_1 f_2}\right)\\
&-ivc_{2\beta}\not{p}_3 \mathcal{P}_R\left(\hat{C}^{(21)}_{\Phi ud,f_1 f_2}\right)\\
&-ivs_{2\beta}V_{f_1 g_1}\not{p}_3 \mathcal{P}_L\left(\hat{C}^{(11)[3]}_{\Phi q,g_1 f_2}-\hat{C}^{(22)[3]}_{\Phi q,g_1 f_2}\right)
\end{aligned}
\tag{C.217}
$$

$u^{f_2}$, $u^{f_1}$, $A_{\mu_3}$

$$
\begin{aligned}
&+\frac{2i\hat{g}\delta_{f_1 f_2}\hat{g}'\gamma^{\mu_3}}{3\sqrt{\hat{g}'^2+\hat{g}^2}}\left(\frac{\hat{g}X_{WB}\hat{g}'}{\hat{g}'^2+\hat{g}^2}-1\right)\\
&-\frac{2vs_\beta p_{3\nu}}{\sqrt{\hat{g}'^2+\hat{g}^2}}\left(\hat{g}'\left(\hat{C}^{*}_{uW\Phi_2,f_2 f_1}\sigma^{\mu_3\nu}\mathcal{P}_L+\hat{C}_{uW\Phi_2,f_1 f_2}\sigma^{\mu_3\nu}\mathcal{P}_R\right)\right.\\
&\qquad\qquad \left.+\hat{g}\left(\hat{C}^{*}_{uB\Phi_2,f_2 f_1}\sigma^{\mu_3\nu}\mathcal{P}_L+\hat{C}_{uB\Phi_2,f_1 f_2}\sigma^{\mu_3\nu}\mathcal{P}_R\right)\right)
\end{aligned}
\tag{C.218}
$$

$d^{f_2}$, $d^{f_1}$, $A_{\mu_3}$

$$
\begin{aligned}
&-\frac{i\hat{g}\delta_{f_1 f_2}\hat{g}'\gamma^{\mu_3}}{3\sqrt{\hat{g}'^2+\hat{g}^2}}\left(\frac{\hat{g}X_{WB}\hat{g}'}{\hat{g}'^2+\hat{g}^2}-1\right)\\
&+\frac{2vc_\beta p_{3\nu}}{\sqrt{\hat{g}'^2+\hat{g}^2}}\left(\hat{g}'\left(\hat{C}^{*}_{dW\Phi_1,f_2 f_1}\sigma^{\mu_3\nu}\mathcal{P}_L+\hat{C}_{dW\Phi_1,f_1 f_2}\sigma^{\mu_3\nu}\mathcal{P}_R\right)\right.\\
&\qquad\qquad \left.-\hat{g}\left(\hat{C}^{*}_{dB\Phi_1,f_2 f_1}\sigma^{\mu_3\nu}\mathcal{P}_L+\hat{C}_{dB\Phi_1,f_1 f_2}\sigma^{\mu_3\nu}\mathcal{P}_R\right)\right)
\end{aligned}
\tag{C.219}
$$

$u^{f_2}$, $u^{f_1}$, $g^{a_3}_{\mu_3}$

$$
\begin{aligned}
&-i\hat{g}_s\delta_{f_1 f_2}\gamma^{\mu_3}\left(T^{a_3}_{m_1 m_2}\right)\\
&-4vs_\beta p_{3\nu}T^{a_3}_{m_1 m_2}\left(\hat{C}^{*}_{uG\Phi_2,f_2 f_1}\sigma^{\mu_3\nu}\mathcal{P}_L+\hat{C}_{uG\Phi_2,f_1 f_2}\sigma^{\mu_3\nu}\mathcal{P}_R\right)
\end{aligned}
\tag{C.220}
$$

$d^{f_2}$, $d^{f_1}$, $g^{a_3}_{\mu_3}$

$$
\begin{aligned}
&-i\hat{g}_s\delta_{f_1f_2}\gamma^{\mu_3}\left(T^{a_3}_{m_1m_2}\right)\\
&-4vc_\beta p_{3\nu}T^{a_3}_{m_1m_2}\left(\hat{C}^{*}_{dG\Phi_1,f_2f_1}\sigma^{\mu_3\nu}\mathcal{P}_L+\hat{C}_{dG\Phi_1,f_1f_2}\sigma^{\mu_3\nu}\mathcal{P}_R\right)
\end{aligned}
\tag{C.221}
$$

$d^{f_2}$, $u^{f_1}$, $W^+_{\mu_3}$

$$
\begin{aligned}
&-\frac{i\hat{g}V_{f_1f_2}}{\sqrt{2}}\left(\gamma^{\mu_3}\mathcal{P}_L\right)\\
&-2\sqrt{2}vp_{3\nu}\left(s_\beta V_{g_1f_2}\sigma^{\mu_3\nu}\mathcal{P}_L\hat{C}^{*}_{uW\Phi_2,g_1f_1}+c_\beta V_{f_1g_1}\hat{C}_{dW\Phi_1,g_1f_2}\sigma^{\mu_3\nu}\mathcal{P}_R\right)\\
&+i\sqrt{2}\hat{g}v^2s_\beta c_\beta\gamma^{\mu_3}\mathcal{P}_R\left(\hat{C}^{(21)}_{\Phi ud,f_1f_2}\right)\\
&-i\sqrt{2}\hat{g}v^2V_{f_1g_1}\gamma^{\mu_3}\mathcal{P}_L\left(c_\beta^2\hat{C}^{(11)[3]}_{\Phi q,g_1f_2}+s_\beta^2\hat{C}^{(22)[3]}_{\Phi q,g_1f_2}\right)
\end{aligned}
\tag{C.222}
$$

$u^{f_2}$, $u^{f_1}$, $Z_{\mu_3}$

$$
\begin{aligned}
&+\frac{i\delta_{f_1f_2}}{6\left(\hat{g}'^2+\hat{g}^2\right)^{3/2}}\left(\left(-3\hat{g}X_{WB}\hat{g}'^3+\hat{g}^3X_{WB}\hat{g}'+\hat{g}'^4-2\hat{g}^2\hat{g}'^2-3\hat{g}^4\right)\gamma^{\mu_3}\mathcal{P}_L\right.\\
&\qquad\left.+4\hat{g}'\left(\hat{g}'^3+\hat{g}^2\hat{g}'+\hat{g}^3X_{WB}\right)\gamma^{\mu_3}\mathcal{P}_R\right)\\
&+\frac{2vs_\beta p_{3\nu}}{\sqrt{\hat{g}'^2+\hat{g}^2}}\left(\hat{g}'\left(\hat{C}^{*}_{uB\Phi_2,f_2f_1}\sigma^{\mu_3\nu}\mathcal{P}_L+\hat{C}_{uB\Phi_2,f_1f_2}\sigma^{\mu_3\nu}\mathcal{P}_R\right)\right.\\
&\qquad\left.-\hat{g}\left(\hat{C}^{*}_{uW\Phi_2,f_2f_1}\sigma^{\mu_3\nu}\mathcal{P}_L+\hat{C}_{uW\Phi_2,f_1f_2}\sigma^{\mu_3\nu}\mathcal{P}_R\right)\right)\\
&+iv^2\sqrt{\hat{g}'^2+\hat{g}^2}V_{f_1g_1}V_{f_2g_2*}\gamma^{\mu_3}\mathcal{P}_L\left(c_\beta^2\hat{C}^{(11)[1]}_{\Phi q,g_1g_2}-c_\beta^2\hat{C}^{(11)[3]}_{\Phi q,g_1g_2}\right.\\
&\qquad\left.+s_\beta^2\left(\hat{C}^{(22)[1]}_{\Phi q,g_1g_2}-\hat{C}^{(22)[3]}_{\Phi q,g_1g_2}\right)\right)\\
&+iv^2\sqrt{\hat{g}'^2+\hat{g}^2}\gamma^{\mu_3}\mathcal{P}_R\left(c_\beta^2\hat{C}^{(11)}_{\Phi u,f_1f_2}+s_\beta^2\hat{C}^{(22)}_{\Phi u,f_1f_2}\right)
\end{aligned}
\tag{C.223}
$$

$d^{f_2}$ $d^{f_1}$ $Z_{\mu_3}$

$$
\begin{aligned}
&+\frac{i\delta_{f_1f_2}}{6\left(\hat{g}'^2+\hat{g}^2\right)^{3/2}}\left(\left(3\hat{g}X_{WB}\hat{g}'^3+\hat{g}^3X_{WB}\hat{g}'+\hat{g}'^4+4\hat{g}^2\hat{g}'^2+3\hat{g}^4\right)\gamma^{\mu_3}\mathcal{P}_L\right.\\
&\qquad\qquad\left.-2\hat{g}'\left(\hat{g}'^3+\hat{g}^2\hat{g}'+\hat{g}^3X_{WB}\right)\gamma^{\mu_3}\mathcal{P}_R\right)\\
&+\frac{2vc_\beta p_{3\nu}}{\sqrt{\hat{g}'^2+\hat{g}^2}}\left(\hat{g}'\left(\hat{C}^{*}_{dB\Phi_1,f_2f_1}\sigma^{\mu_3\nu}\mathcal{P}_L+\hat{C}_{dB\Phi_1,f_1f_2}\sigma^{\mu_3\nu}\mathcal{P}_R\right)\right.\\
&\qquad\left.+\hat{g}\left(\hat{C}^{*}_{dW\Phi_1,f_2f_1}\sigma^{\mu_3\nu}\mathcal{P}_L+\hat{C}_{dW\Phi_1,f_1f_2}\sigma^{\mu_3\nu}\mathcal{P}_R\right)\right)\\
&+iv^2\sqrt{\hat{g}'^2+\hat{g}^2}\gamma^{\mu_3}\mathcal{P}_L\left(c_\beta^2\hat{C}^{(11)[1]}_{\Phi q,f_1f_2}+c_\beta^2\hat{C}^{(11)[3]}_{\Phi q,f_1f_2}+s_\beta^2\left(\hat{C}^{(22)[1]}_{\Phi q,f_1f_2}+\hat{C}^{(22)[3]}_{\Phi q,f_1f_2}\right)\right)\\
&+iv^2\sqrt{\hat{g}'^2+\hat{g}^2}\gamma^{\mu_3}\mathcal{P}_R\left(c_\beta^2\hat{C}^{(11)}_{\Phi d,f_1f_2}+s_\beta^2\hat{C}^{(22)}_{\Phi d,f_1f_2}\right)
\end{aligned}
\tag{C.224}
$$

$u^{f_2}$ $u^{f_1}$ $A^0$ $A^0$

$$
\begin{aligned}
-ivs_\beta\Big(&-s_\beta^2\left(\mathcal{P}_L\hat{C}^{(11)*}_{u\Phi_2,f_2f_1}+\mathcal{P}_R\hat{C}^{(11)}_{u\Phi_2,f_1f_2}\right)\\
&+\left(2c_\beta^2+s_\beta^2\right)\left(\mathcal{P}_L\hat{C}^{(12)*}_{u\Phi_1,f_2f_1}+\mathcal{P}_R\hat{C}^{(12)}_{u\Phi_1,f_1f_2}\right)\\
&-s_\beta^2\left(\mathcal{P}_L\hat{C}^{(21)*}_{u\Phi_1,f_2f_1}+\mathcal{P}_R\hat{C}^{(21)}_{u\Phi_1,f_1f_2}\right)\\
&-c_\beta^2\left(\mathcal{P}_L\hat{C}^{(22)*}_{u\Phi_2,f_2f_1}+\mathcal{P}_R\hat{C}^{(22)}_{u\Phi_2,f_1f_2}\right)\Big)
\end{aligned}
\tag{C.225}
$$

$u^{f_2}$ $u^{f_1}$ $H$ $H$

$$
\begin{aligned}
+ivs_\beta\Big(&-\left(2c_\beta^2-s_\beta^2\right)\left(\mathcal{P}_L\hat{C}^{(11)*}_{u\Phi_2,f_2f_1}+\mathcal{P}_R\hat{C}^{(11)}_{u\Phi_2,f_1f_2}\right)\\
&-\left(2c_\beta^2-s_\beta^2\right)\left(\mathcal{P}_L\hat{C}^{(12)*}_{u\Phi_1,f_2f_1}+\mathcal{P}_R\hat{C}^{(12)}_{u\Phi_1,f_1f_2}\right)\\
&-\left(2c_\beta^2-s_\beta^2\right)\left(\mathcal{P}_L\hat{C}^{(21)*}_{u\Phi_1,f_2f_1}+\mathcal{P}_R\hat{C}^{(21)}_{u\Phi_1,f_1f_2}\right)\\
&+3c_\beta^2\left(\mathcal{P}_L\hat{C}^{(22)*}_{u\Phi_2,f_2f_1}+\mathcal{P}_R\hat{C}^{(22)}_{u\Phi_2,f_1f_2}\right)\Big)
\end{aligned}
\tag{C.226}
$$

$$
\begin{aligned}
&+ivs_\beta\left(s_\beta^2\left(\mathcal{P}_L\hat{C}^{(11)*}_{u\Phi_2,f_2f_1}+\mathcal{P}_R\hat{C}^{(11)}_{u\Phi_2,f_1f_2}\right)\right.\\
&\qquad -c_\beta^2\left(\mathcal{P}_L\hat{C}^{(12)*}_{u\Phi_1,f_2f_1}+\mathcal{P}_R\hat{C}^{(12)}_{u\Phi_1,f_1f_2}\right)\\
&\qquad -c_\beta^2\left(\mathcal{P}_L\hat{C}^{(21)*}_{u\Phi_1,f_2f_1}+\mathcal{P}_R\hat{C}^{(21)}_{u\Phi_1,f_1f_2}\right)\\
&\qquad \left.+c_\beta^2\left(\mathcal{P}_L\hat{C}^{(22)*}_{u\Phi_2,f_2f_1}+\mathcal{P}_R\hat{C}^{(22)}_{u\Phi_2,f_1f_2}\right)\right)\\
&+iV_{f_1g_1}V^*_{f_2g_2}\left(\not{p}_3\mathcal{P}_L-\not{p}_4\mathcal{P}_L\right)\left(s_\beta^2\left(\hat{C}^{(11)[1]}_{\Phi q,g_1g_2}+\hat{C}^{(11)[3]}_{\Phi q,g_1g_2}\right)\right.\\
&\qquad\qquad\left.+c_\beta^2\hat{C}^{(22)[1]}_{\Phi q,g_1g_2}+c_\beta^2\hat{C}^{(22)[3]}_{\Phi q,g_1g_2}\right)\\
&+i\left(\not{p}_3\mathcal{P}_R-\not{p}_4\mathcal{P}_R\right)\left(s_\beta^2\hat{C}^{(11)}_{\Phi u,f_1f_2}+c_\beta^2\hat{C}^{(22)}_{\Phi u,f_1f_2}\right)
\end{aligned}
\tag{C.227}
$$

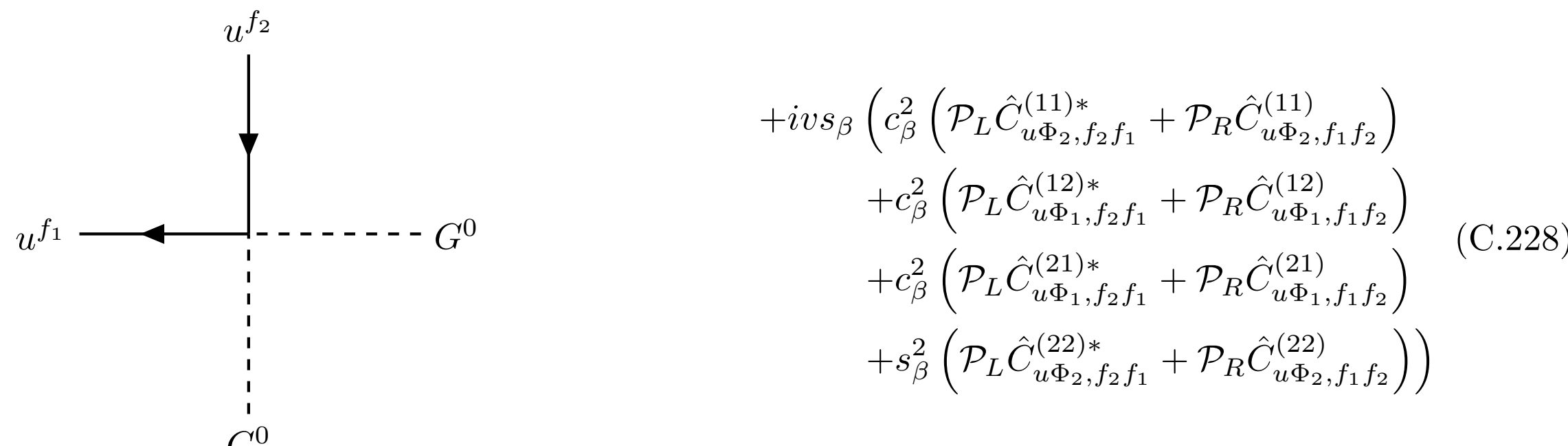

$$
\begin{aligned}
&+ivs_\beta\left(c_\beta^2\left(\mathcal{P}_L\hat{C}^{(11)*}_{u\Phi_2,f_2f_1}+\mathcal{P}_R\hat{C}^{(11)}_{u\Phi_2,f_1f_2}\right)\right.\\
&\qquad +c_\beta^2\left(\mathcal{P}_L\hat{C}^{(12)*}_{u\Phi_1,f_2f_1}+\mathcal{P}_R\hat{C}^{(12)}_{u\Phi_1,f_1f_2}\right)\\
&\qquad +c_\beta^2\left(\mathcal{P}_L\hat{C}^{(21)*}_{u\Phi_1,f_2f_1}+\mathcal{P}_R\hat{C}^{(21)}_{u\Phi_1,f_1f_2}\right)\\
&\qquad \left.+s_\beta^2\left(\mathcal{P}_L\hat{C}^{(22)*}_{u\Phi_2,f_2f_1}+\mathcal{P}_R\hat{C}^{(22)}_{u\Phi_2,f_1f_2}\right)\right)
\end{aligned}
\tag{C.228}
$$

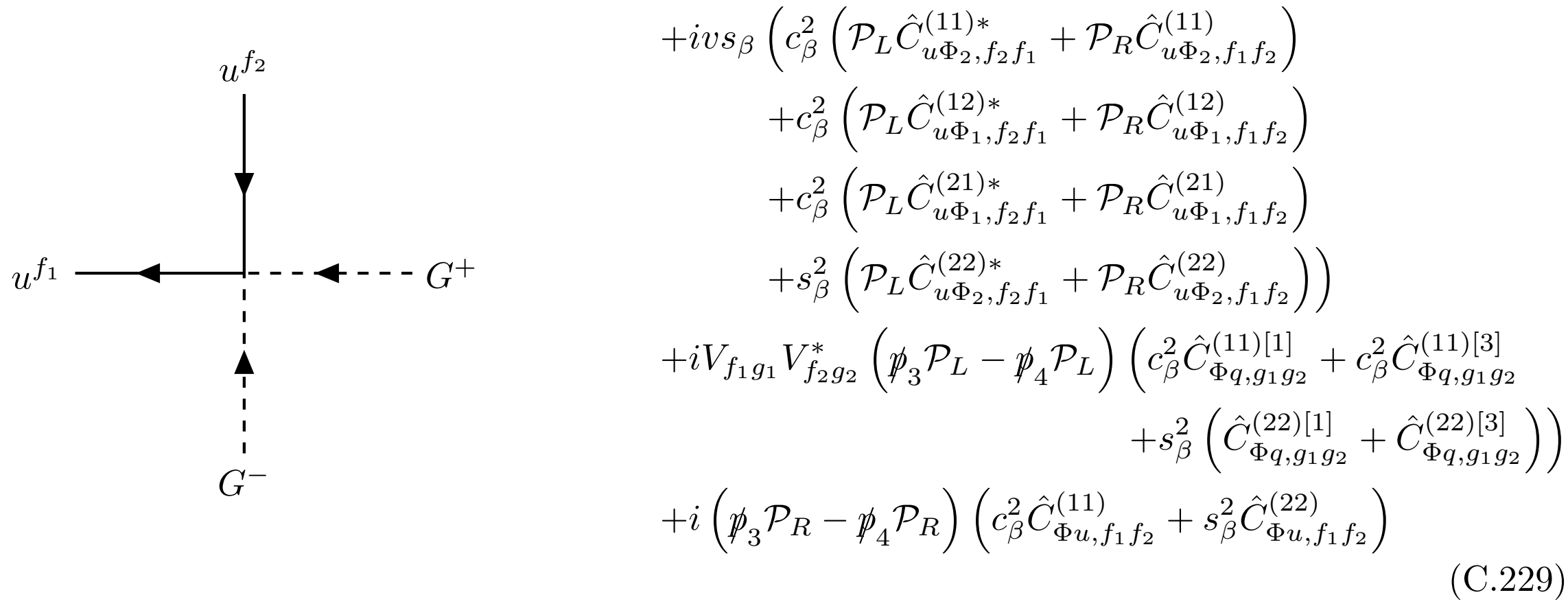

$$
\begin{aligned}
&+ivs_\beta\left(c_\beta^2\left(\mathcal{P}_L\hat{C}^{(11)*}_{u\Phi_2,f_2f_1}+\mathcal{P}_R\hat{C}^{(11)}_{u\Phi_2,f_1f_2}\right)\right.\\
&\qquad +c_\beta^2\left(\mathcal{P}_L\hat{C}^{(12)*}_{u\Phi_1,f_2f_1}+\mathcal{P}_R\hat{C}^{(12)}_{u\Phi_1,f_1f_2}\right)\\
&\qquad +c_\beta^2\left(\mathcal{P}_L\hat{C}^{(21)*}_{u\Phi_1,f_2f_1}+\mathcal{P}_R\hat{C}^{(21)}_{u\Phi_1,f_1f_2}\right)\\
&\qquad \left.+s_\beta^2\left(\mathcal{P}_L\hat{C}^{(22)*}_{u\Phi_2,f_2f_1}+\mathcal{P}_R\hat{C}^{(22)}_{u\Phi_2,f_1f_2}\right)\right)\\
&+iV_{f_1g_1}V^*_{f_2g_2}\left(\not{p}_3\mathcal{P}_L-\not{p}_4\mathcal{P}_L\right)\left(c_\beta^2\hat{C}^{(11)[1]}_{\Phi q,g_1g_2}+c_\beta^2\hat{C}^{(11)[3]}_{\Phi q,g_1g_2}\right.\\
&\qquad\qquad\left.+s_\beta^2\left(\hat{C}^{(22)[1]}_{\Phi q,g_1g_2}+\hat{C}^{(22)[3]}_{\Phi q,g_1g_2}\right)\right)\\
&+i\left(\not{p}_3\mathcal{P}_R-\not{p}_4\mathcal{P}_R\right)\left(c_\beta^2\hat{C}^{(11)}_{\Phi u,f_1f_2}+s_\beta^2\hat{C}^{(22)}_{\Phi u,f_1f_2}\right)
\end{aligned}
\tag{C.229}
$$

$$
\begin{aligned}
&+vs_\beta\Big(-c_\beta^2\left(\mathcal{P}_L\hat{C}^{(11)*}_{u\Phi_2,f_2f_1}-\mathcal{P}_R\hat{C}^{(11)}_{u\Phi_2,f_1f_2}\right)\\
&\qquad -c_\beta^2\left(\mathcal{P}_L\hat{C}^{(12)*}_{u\Phi_1,f_2f_1}-\mathcal{P}_R\hat{C}^{(12)}_{u\Phi_1,f_1f_2}\right)\\
&\qquad -c_\beta^2\left(\mathcal{P}_L\hat{C}^{(21)*}_{u\Phi_1,f_2f_1}-\mathcal{P}_R\hat{C}^{(21)}_{u\Phi_1,f_1f_2}\right)\\
&\qquad -s_\beta^2\left(\mathcal{P}_L\hat{C}^{(22)*}_{u\Phi_2,f_2f_1}-\mathcal{P}_R\hat{C}^{(22)}_{u\Phi_2,f_1f_2}\right)\Big)\\
&-V_{f_1g_1}V^*_{f_2g_2}\left(\not{p}_3\mathcal{P}_L-\not{p}_4\mathcal{P}_L\right)\Big(c_\beta^2\hat{C}^{(11)[1]}_{\Phi q,g_1g_2}-c_\beta^2\hat{C}^{(11)[3]}_{\Phi q,g_1g_2}\\
&\qquad\qquad +s_\beta^2\left(\hat{C}^{(22)[1]}_{\Phi q,g_1g_2}-\hat{C}^{(22)[3]}_{\Phi q,g_1g_2}\right)\Big)\\
&-\left(\not{p}_3\mathcal{P}_R-\not{p}_4\mathcal{P}_R\right)\left(c_\beta^2\hat{C}^{(11)}_{\Phi u,f_1f_2}+s_\beta^2\hat{C}^{(22)}_{\Phi u,f_1f_2}\right)
\end{aligned}
\tag{C.230}
$$

$$
\begin{aligned}
&+3ivs_\beta\Big(c_\beta^2\left(\mathcal{P}_L\hat{C}^{(11)*}_{u\Phi_2,f_2f_1}+\mathcal{P}_R\hat{C}^{(11)}_{u\Phi_2,f_1f_2}\right)\\
&\qquad +c_\beta^2\left(\mathcal{P}_L\hat{C}^{(12)*}_{u\Phi_1,f_2f_1}+\mathcal{P}_R\hat{C}^{(12)}_{u\Phi_1,f_1f_2}\right)\\
&\qquad +c_\beta^2\left(\mathcal{P}_L\hat{C}^{(21)*}_{u\Phi_1,f_2f_1}+\mathcal{P}_R\hat{C}^{(21)}_{u\Phi_1,f_1f_2}\right)\\
&\qquad +s_\beta^2\left(\mathcal{P}_L\hat{C}^{(22)*}_{u\Phi_2,f_2f_1}+\mathcal{P}_R\hat{C}^{(22)}_{u\Phi_2,f_1f_2}\right)\Big)
\end{aligned}
\tag{C.231}
$$

$$
\begin{aligned}
&+vc_\beta\Big(-c_\beta^2\left(\mathcal{P}_L\hat{C}^{(11)*}_{u\Phi_2,f_2f_1}-\mathcal{P}_R\hat{C}^{(11)}_{u\Phi_2,f_1f_2}\right)\\
&\qquad +\left(2s_\beta^2+c_\beta^2\right)\left(\mathcal{P}_L\hat{C}^{(12)*}_{u\Phi_1,f_2f_1}-\mathcal{P}_R\hat{C}^{(12)}_{u\Phi_1,f_1f_2}\right)\\
&\qquad -c_\beta^2\left(\mathcal{P}_L\hat{C}^{(21)*}_{u\Phi_1,f_2f_1}-\mathcal{P}_R\hat{C}^{(21)}_{u\Phi_1,f_1f_2}\right)\\
&\qquad -s_\beta^2\left(\mathcal{P}_L\hat{C}^{(22)*}_{u\Phi_2,f_2f_1}-\mathcal{P}_R\hat{C}^{(22)}_{u\Phi_2,f_1f_2}\right)\Big)\\
&+s_\beta c_\beta V_{f_1g_1}V^*_{f_2g_2}\left(\not{p}_3\mathcal{P}_L-\not{p}_4\mathcal{P}_L\right)\Big(\hat{C}^{(11)[1]}_{\Phi q,g_1g_2}-\hat{C}^{(11)[3]}_{\Phi q,g_1g_2}\\
&\qquad\qquad -\hat{C}^{(22)[1]}_{\Phi q,g_1g_2}+\hat{C}^{(22)[3]}_{\Phi q,g_1g_2}\Big)\\
&+s_\beta c_\beta\left(\not{p}_3\mathcal{P}_R-\not{p}_4\mathcal{P}_R\right)\left(\hat{C}^{(11)}_{\Phi u,f_1f_2}-\hat{C}^{(22)}_{\Phi u,f_1f_2}\right)
\end{aligned}
\tag{C.232}
$$

$$
\begin{aligned}
-ivc_\beta \Big( &-\left(2s_\beta^2 - c_\beta^2\right)\left(\mathcal{P}_L \hat{C}^{(11)*}_{u\Phi_2,f_2f_1} + \mathcal{P}_R \hat{C}^{(11)}_{u\Phi_2,f_1f_2}\right)\\
&-\left(2s_\beta^2 - c_\beta^2\right)\left(\mathcal{P}_L \hat{C}^{(12)*}_{u\Phi_1,f_2f_1} + \mathcal{P}_R \hat{C}^{(12)}_{u\Phi_1,f_1f_2}\right)\\
&-\left(2s_\beta^2 - c_\beta^2\right)\left(\mathcal{P}_L \hat{C}^{(21)*}_{u\Phi_1,f_2f_1} + \mathcal{P}_R \hat{C}^{(21)}_{u\Phi_1,f_1f_2}\right)\\
&+3s_\beta^2\left(\mathcal{P}_L \hat{C}^{(22)*}_{u\Phi_2,f_2f_1} + \mathcal{P}_R \hat{C}^{(22)}_{u\Phi_2,f_1f_2}\right)\Big)
\end{aligned} \tag{C.233}
$$

$$
\begin{aligned}
&+ivc_\beta \Big(-s_\beta^2\left(\mathcal{P}_L \hat{C}^{(11)*}_{u\Phi_2,f_2f_1} + \mathcal{P}_R \hat{C}^{(11)}_{u\Phi_2,f_1f_2}\right)\\
&\qquad +s_\beta^2\left(\mathcal{P}_L \hat{C}^{(22)*}_{u\Phi_2,f_2f_1} + \mathcal{P}_R \hat{C}^{(22)}_{u\Phi_2,f_1f_2}\right)\\
&\qquad -s_\beta^2 \mathcal{P}_L \hat{C}^{(12)*}_{u\Phi_1,f_2f_1} + c_\beta^2 \mathcal{P}_L \hat{C}^{(21)*}_{u\Phi_1,f_2f_1}\\
&\qquad +c_\beta^2 \mathcal{P}_R \hat{C}^{(12)}_{u\Phi_1,f_1f_2} - s_\beta^2 \mathcal{P}_R \hat{C}^{(21)}_{u\Phi_1,f_1f_2}\Big)\\
&+\frac{1}{2} i s_{2\beta} V_{f_1g_1} V^*_{f_2g_2}\left(\not{p}_3 \mathcal{P}_L - \not{p}_4 \mathcal{P}_L\right)\Big(\hat{C}^{(11)[1]}_{\Phi q,g_1g_2} + \hat{C}^{(11)[3]}_{\Phi q,g_1g_2}\\
&\qquad\qquad -\hat{C}^{(22)[1]}_{\Phi q,g_1g_2} - \hat{C}^{(22)[3]}_{\Phi q,g_1g_2}\Big)\\
&+\frac{1}{2} i s_{2\beta}\left(\not{p}_3 \mathcal{P}_R - \not{p}_4 \mathcal{P}_R\right)\left(\hat{C}^{(11)}_{\Phi u,f_1f_2} - \hat{C}^{(22)}_{\Phi u,f_1f_2}\right)
\end{aligned} \tag{C.234}
$$

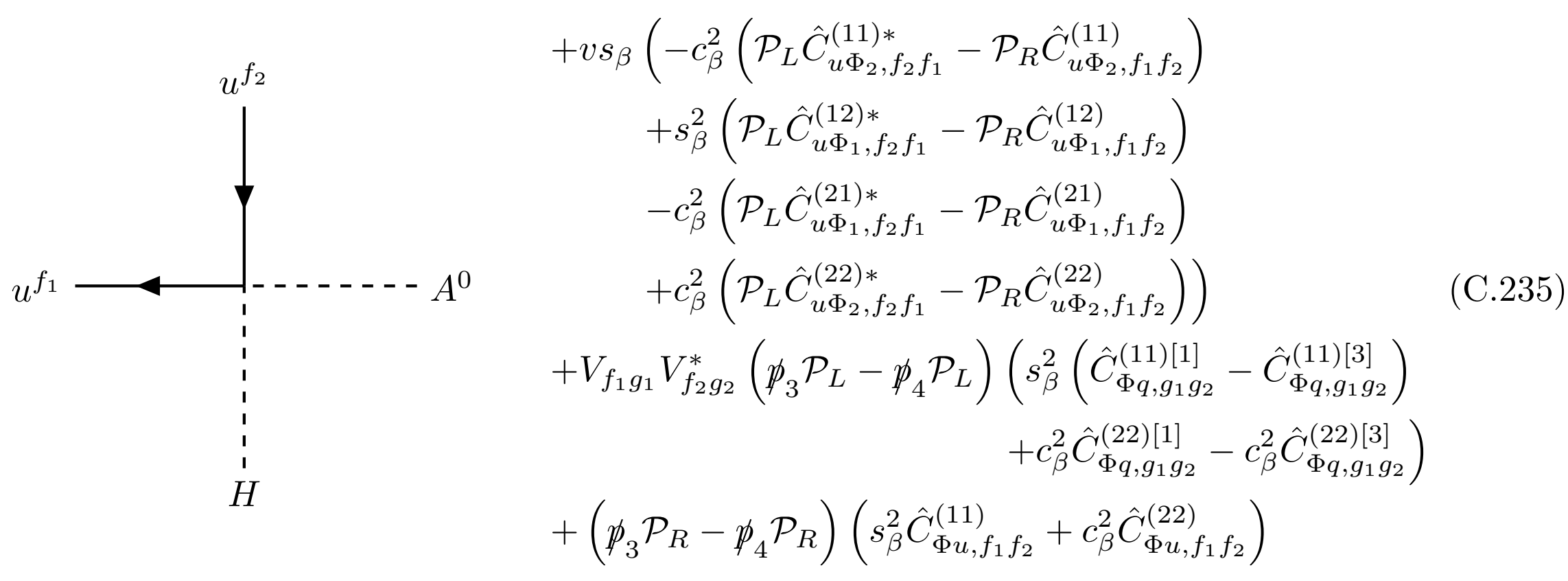

$$
\begin{aligned}
&+vs_\beta \Big(-c_\beta^2\left(\mathcal{P}_L \hat{C}^{(11)*}_{u\Phi_2,f_2f_1} - \mathcal{P}_R \hat{C}^{(11)}_{u\Phi_2,f_1f_2}\right)\\
&\qquad +s_\beta^2\left(\mathcal{P}_L \hat{C}^{(12)*}_{u\Phi_1,f_2f_1} - \mathcal{P}_R \hat{C}^{(12)}_{u\Phi_1,f_1f_2}\right)\\
&\qquad -c_\beta^2\left(\mathcal{P}_L \hat{C}^{(21)*}_{u\Phi_1,f_2f_1} - \mathcal{P}_R \hat{C}^{(21)}_{u\Phi_1,f_1f_2}\right)\\
&\qquad +c_\beta^2\left(\mathcal{P}_L \hat{C}^{(22)*}_{u\Phi_2,f_2f_1} - \mathcal{P}_R \hat{C}^{(22)}_{u\Phi_2,f_1f_2}\right)\Big)\\
&+V_{f_1g_1} V^*_{f_2g_2}\left(\not{p}_3 \mathcal{P}_L - \not{p}_4 \mathcal{P}_L\right)\Big(s_\beta^2\left(\hat{C}^{(11)[1]}_{\Phi q,g_1g_2} - \hat{C}^{(11)[3]}_{\Phi q,g_1g_2}\right)\\
&\qquad\qquad +c_\beta^2 \hat{C}^{(22)[1]}_{\Phi q,g_1g_2} - c_\beta^2 \hat{C}^{(22)[3]}_{\Phi q,g_1g_2}\Big)\\
&+\left(\not{p}_3 \mathcal{P}_R - \not{p}_4 \mathcal{P}_R\right)\left(s_\beta^2 \hat{C}^{(11)}_{\Phi u,f_1f_2} + c_\beta^2 \hat{C}^{(22)}_{\Phi u,f_1f_2}\right)
\end{aligned} \tag{C.235}
$$

$u^{f_2}$

$u^{f_1}$ $A^0$

$G^0$

$$
\begin{aligned}
&+ivc_\beta \Big(-s_\beta^2 \left(\mathcal{P}_L \hat{C}^{(11)*}_{u\Phi_2,f_2f_1} + \mathcal{P}_R \hat{C}^{(11)}_{u\Phi_2,f_1f_2}\right) \\
&\qquad +c_\beta^2 \left(\mathcal{P}_L \hat{C}^{(12)*}_{u\Phi_1,f_2f_1} + \mathcal{P}_R \hat{C}^{(12)}_{u\Phi_1,f_1f_2}\right) \\
&\qquad -s_\beta^2 \left(\mathcal{P}_L \hat{C}^{(21)*}_{u\Phi_1,f_2f_1} + \mathcal{P}_R \hat{C}^{(21)}_{u\Phi_1,f_1f_2}\right) \\
&\qquad +s_\beta^2 \left(\mathcal{P}_L \hat{C}^{(22)*}_{u\Phi_2,f_2f_1} + \mathcal{P}_R \hat{C}^{(22)}_{u\Phi_2,f_1f_2}\right)\Big)
\end{aligned} \tag{C.236}
$$

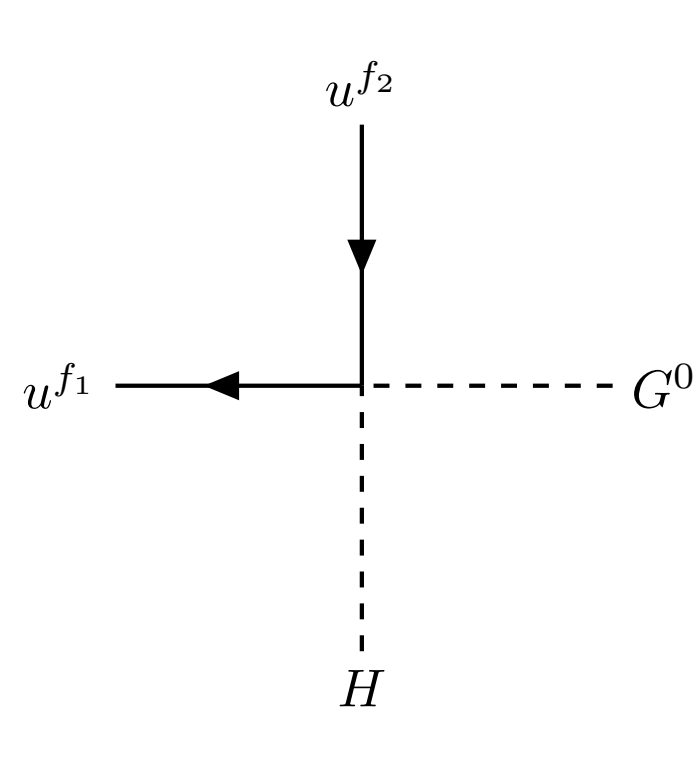

$$
\begin{aligned}
&+vc_\beta \Big(-s_\beta^2 \left(\mathcal{P}_L \hat{C}^{(11)*}_{u\Phi_2,f_2f_1} - \mathcal{P}_R \hat{C}^{(11)}_{u\Phi_2,f_1f_2}\right) \\
&\qquad +c_\beta^2 \left(\mathcal{P}_L \hat{C}^{(12)*}_{u\Phi_1,f_2f_1} - \mathcal{P}_R \hat{C}^{(12)}_{u\Phi_1,f_1f_2}\right) \\
&\qquad -s_\beta^2 \left(\mathcal{P}_L \hat{C}^{(21)*}_{u\Phi_1,f_2f_1} - \mathcal{P}_R \hat{C}^{(21)}_{u\Phi_1,f_1f_2}\right) \\
&\qquad +s_\beta^2 \left(\mathcal{P}_L \hat{C}^{(22)*}_{u\Phi_2,f_2f_1} - \mathcal{P}_R \hat{C}^{(22)}_{u\Phi_2,f_1f_2}\right)\Big) \\
&-s_\beta c_\beta V_{f_1g_1} V_{f_2g_2*} \left(\not{p}_3 \mathcal{P}_L - \not{p}_4 \mathcal{P}_L\right) \Big(\hat{C}^{(11)[1]}_{\Phi q,g_1g_2} - \hat{C}^{(11)[3]}_{\Phi q,g_1g_2} \\
&\qquad\qquad -\hat{C}^{(22)[1]}_{\Phi q,g_1g_2} + \hat{C}^{(22)[3]}_{\Phi q,g_1g_2}\Big) \\
&-s_\beta c_\beta \left(\not{p}_3 \mathcal{P}_R - \not{p}_4 \mathcal{P}_R\right) \left(\hat{C}^{(11)}_{\Phi u,f_1f_2} - \hat{C}^{(22)}_{\Phi u,f_1f_2}\right)
\end{aligned} \tag{C.237}
$$

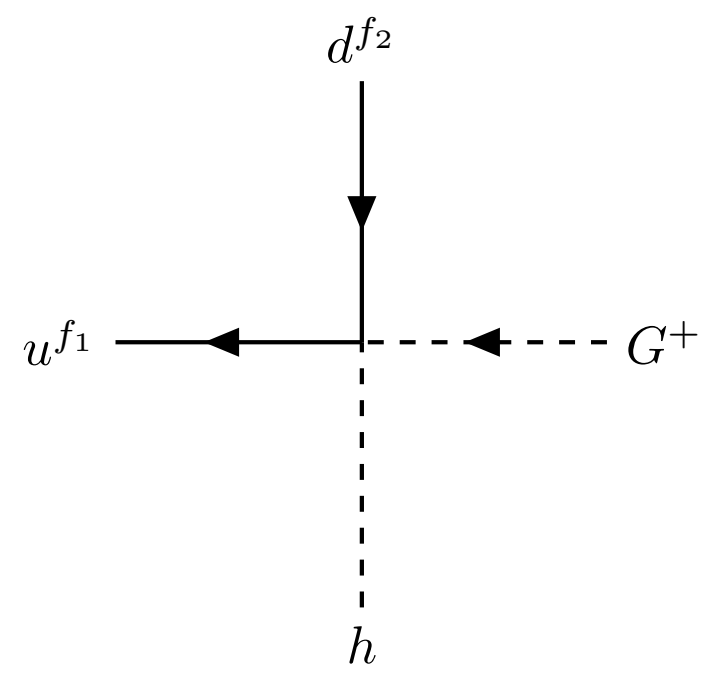

$$
\begin{aligned}
&-i\sqrt{2}v \Big(s_\beta c_\beta^2 \mathcal{P}_L V_{g_1f_2} \hat{C}^{(11)*}_{u\Phi_2,g_1f_1} + s_\beta c_\beta^2 \mathcal{P}_L V_{g_1f_2} \hat{C}^{(12)*}_{u\Phi_1,g_1f_1} \\
&\qquad +s_\beta c_\beta^2 \mathcal{P}_L V_{g_1f_2} \hat{C}^{(21)*}_{u\Phi_1,g_1f_1} + s_\beta^3 \mathcal{P}_L V_{g_1f_2} \hat{C}^{(22)*}_{u\Phi_2,g_1f_1} \\
&\qquad -c_\beta^3 \mathcal{P}_R V_{f_1g_1} \hat{C}^{(11)}_{d\Phi_1,g_1f_2} - s_\beta^2 c_\beta \mathcal{P}_R V_{f_1g_1} \hat{C}^{(12)}_{d\Phi_2,g_1f_2} \\
&\qquad -s_\beta^2 c_\beta \mathcal{P}_R V_{f_1g_1} \hat{C}^{(21)}_{d\Phi_2,g_1f_2} - s_\beta^2 c_\beta \mathcal{P}_R V_{f_1g_1} \hat{C}^{(22)}_{d\Phi_1,g_1f_2}\Big) \\
&-i\sqrt{2} s_\beta c_\beta \left(\not{p}_3 \mathcal{P}_R - \not{p}_4 \mathcal{P}_R\right) \left(\hat{C}^{(21)}_{\Phi ud,f_1f_2}\right) \\
&+i\sqrt{2} V_{f_1g_1} \left(\not{p}_3 \mathcal{P}_L - \not{p}_4 \mathcal{P}_L\right) \left(c_\beta^2 \hat{C}^{(11)[3]}_{\Phi q,g_1f_2} + s_\beta^2 \hat{C}^{(22)[3]}_{\Phi q,g_1f_2}\right)
\end{aligned} \tag{C.238}
$$

$$
\begin{aligned}
&+\frac{iv}{\sqrt{2}}\Big(-2s_\beta^2 c_\beta \mathcal{P}_L V_{g_1f_2}\hat{C}^{(11)*}_{u\Phi_2,g_1f_1} + c_\beta^3 \mathcal{P}_L V_{g_1f_2}\hat{C}^{(12)*}_{u\Phi_1,g_1f_1} \\
&\qquad -s_\beta^2 c_\beta \mathcal{P}_L V_{g_1f_2}\hat{C}^{(12)*}_{u\Phi_1,g_1f_1} + c_\beta^3 \mathcal{P}_L V_{g_1f_2}\hat{C}^{(21)*}_{u\Phi_1,g_1f_1} \\
&\qquad -s_\beta^2 c_\beta \mathcal{P}_L V_{g_1f_2}\hat{C}^{(21)*}_{u\Phi_1,g_1f_1} + 2s_\beta^2 c_\beta \mathcal{P}_L V_{g_1f_2}\hat{C}^{(22)*}_{u\Phi_2,g_1f_1} \\
&\qquad +2s_\beta c_\beta^2 \mathcal{P}_R V_{f_1g_1}\hat{C}^{(11)}_{d\Phi_1,g_1f_2} + s_\beta^3 \mathcal{P}_R V_{f_1g_1}\hat{C}^{(12)}_{d\Phi_2,g_1f_2} \\
&\qquad -s_\beta c_\beta^2 \mathcal{P}_R V_{f_1g_1}\hat{C}^{(12)}_{d\Phi_2,g_1f_2} + s_\beta^3 \mathcal{P}_R V_{f_1g_1}\hat{C}^{(21)}_{d\Phi_2,g_1f_2} \\
&\qquad -s_\beta c_\beta^2 \mathcal{P}_R V_{f_1g_1}\hat{C}^{(21)}_{d\Phi_2,g_1f_2} - 2s_\beta c_\beta^2 \mathcal{P}_R V_{f_1g_1}\hat{C}^{(22)}_{d\Phi_1,g_1f_2}\Big) \\
&+\frac{ic_{2\beta}\left(\not{p}_3\mathcal{P}_R - \not{p}_4\mathcal{P}_R\right)}{\sqrt{2}}\left(\hat{C}^{(21)}_{\Phi ud,f_1f_2}\right) \\
&+\frac{is_{2\beta}V_{f_1g_1}\left(\not{p}_3\mathcal{P}_L - \not{p}_4\mathcal{P}_L\right)}{\sqrt{2}}\left(\hat{C}^{(11)[3]}_{\Phi q,g_1f_2} - \hat{C}^{(22)[3]}_{\Phi q,g_1f_2}\right)
\end{aligned}
\tag{C.239}
$$

$$
\begin{aligned}
&-i\sqrt{2}v\Big(c_\beta^3 \mathcal{P}_L V_{g_1f_2}\hat{C}^{(11)*}_{u\Phi_2,g_1f_1} - s_\beta^2 c_\beta \mathcal{P}_L V_{g_1f_2}\hat{C}^{(12)*}_{u\Phi_1,g_1f_1} \\
&\qquad -s_\beta^2 c_\beta \mathcal{P}_L V_{g_1f_2}\hat{C}^{(21)*}_{u\Phi_1,g_1f_1} + s_\beta^2 c_\beta \mathcal{P}_L V_{g_1f_2}\hat{C}^{(22)*}_{u\Phi_2,g_1f_1} \\
&\qquad +s_\beta c_\beta^2 \mathcal{P}_R V_{f_1g_1}\hat{C}^{(11)}_{d\Phi_1,g_1f_2} - s_\beta c_\beta^2 \mathcal{P}_R V_{f_1g_1}\hat{C}^{(12)}_{d\Phi_2,g_1f_2} \\
&\qquad -s_\beta c_\beta^2 \mathcal{P}_R V_{f_1g_1}\hat{C}^{(21)}_{d\Phi_2,g_1f_2} + s_\beta^3 \mathcal{P}_R V_{f_1g_1}\hat{C}^{(22)}_{d\Phi_1,g_1f_2}\Big) \\
&+\frac{ic_{2\beta}\left(\not{p}_3\mathcal{P}_R - \not{p}_4\mathcal{P}_R\right)}{\sqrt{2}}\left(\hat{C}^{(21)}_{\Phi ud,f_1f_2}\right) \\
&+\frac{is_{2\beta}V_{f_1g_1}\left(\not{p}_3\mathcal{P}_L - \not{p}_4\mathcal{P}_L\right)}{\sqrt{2}}\left(\hat{C}^{(11)[3]}_{\Phi q,g_1f_2} - \hat{C}^{(22)[3]}_{\Phi q,g_1f_2}\right)
\end{aligned}
\tag{C.240}
$$

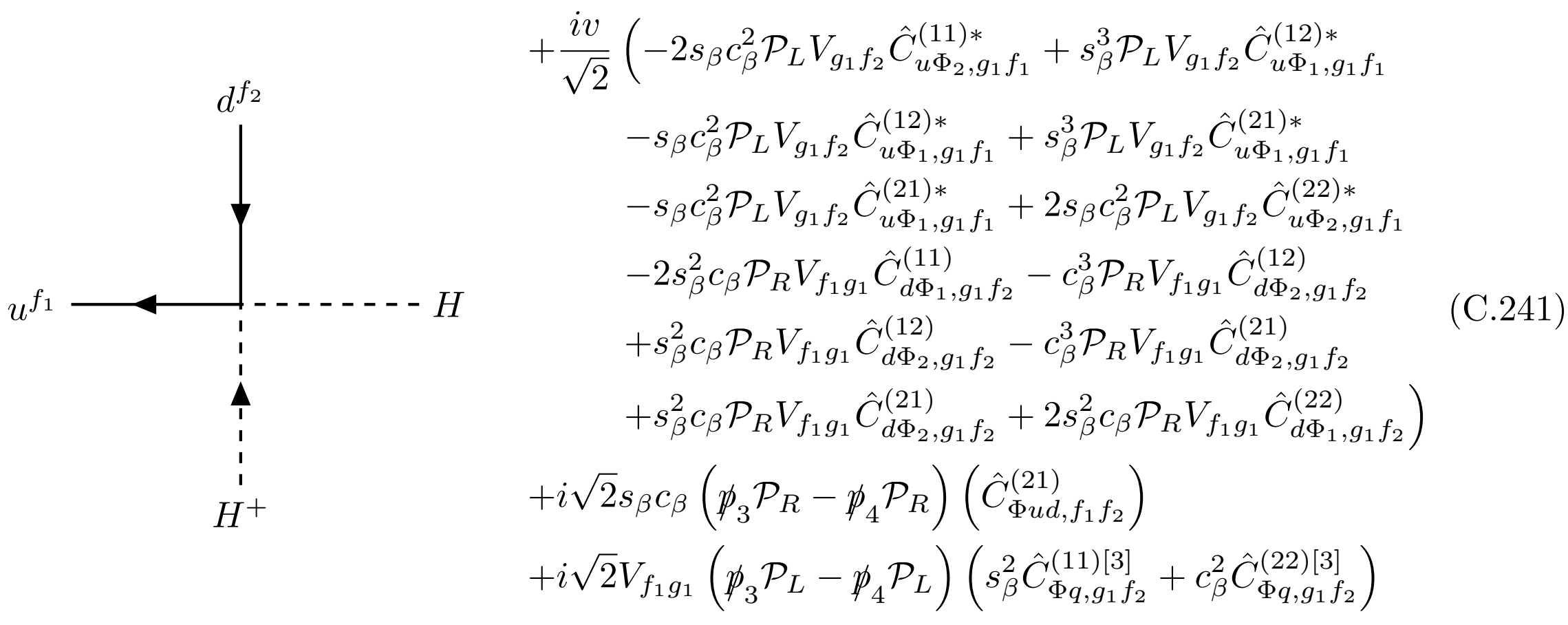

$$
\begin{aligned}
&+\frac{iv}{\sqrt{2}}\Big(-2s_\beta c_\beta^2 \mathcal{P}_L V_{g_1f_2}\hat{C}^{(11)*}_{u\Phi_2,g_1f_1} + s_\beta^3 \mathcal{P}_L V_{g_1f_2}\hat{C}^{(12)*}_{u\Phi_1,g_1f_1} \\
&\qquad -s_\beta c_\beta^2 \mathcal{P}_L V_{g_1f_2}\hat{C}^{(12)*}_{u\Phi_1,g_1f_1} + s_\beta^3 \mathcal{P}_L V_{g_1f_2}\hat{C}^{(21)*}_{u\Phi_1,g_1f_1} \\
&\qquad -s_\beta c_\beta^2 \mathcal{P}_L V_{g_1f_2}\hat{C}^{(21)*}_{u\Phi_1,g_1f_1} + 2s_\beta c_\beta^2 \mathcal{P}_L V_{g_1f_2}\hat{C}^{(22)*}_{u\Phi_2,g_1f_1} \\
&\qquad -2s_\beta^2 c_\beta \mathcal{P}_R V_{f_1g_1}\hat{C}^{(11)}_{d\Phi_1,g_1f_2} - c_\beta^3 \mathcal{P}_R V_{f_1g_1}\hat{C}^{(12)}_{d\Phi_2,g_1f_2} \\
&\qquad +s_\beta^2 c_\beta \mathcal{P}_R V_{f_1g_1}\hat{C}^{(12)}_{d\Phi_2,g_1f_2} - c_\beta^3 \mathcal{P}_R V_{f_1g_1}\hat{C}^{(21)}_{d\Phi_2,g_1f_2} \\
&\qquad +s_\beta^2 c_\beta \mathcal{P}_R V_{f_1g_1}\hat{C}^{(21)}_{d\Phi_2,g_1f_2} + 2s_\beta^2 c_\beta \mathcal{P}_R V_{f_1g_1}\hat{C}^{(22)}_{d\Phi_1,g_1f_2}\Big) \\
&+i\sqrt{2}s_\beta c_\beta\left(\not{p}_3\mathcal{P}_R - \not{p}_4\mathcal{P}_R\right)\left(\hat{C}^{(21)}_{\Phi ud,f_1f_2}\right) \\
&+i\sqrt{2}V_{f_1g_1}\left(\not{p}_3\mathcal{P}_L - \not{p}_4\mathcal{P}_L\right)\left(s_\beta^2\hat{C}^{(11)[3]}_{\Phi q,g_1f_2} + c_\beta^2\hat{C}^{(22)[3]}_{\Phi q,g_1f_2}\right)
\end{aligned}
\tag{C.241}
$$

$$
\begin{aligned}
&-\frac{v}{\sqrt{2}}\Big(c_\beta^3\mathcal{P}_L V_{g_1f_2}\hat{C}^{(12)*}_{u\Phi_1,g_1f_1}+s_\beta^2c_\beta\mathcal{P}_L V_{g_1f_2}\hat{C}^{(12)*}_{u\Phi_1,g_1f_1}\\
&\qquad -c_\beta^3\mathcal{P}_L V_{g_1f_2}\hat{C}^{(21)*}_{u\Phi_1,g_1f_1}-s_\beta^2c_\beta\mathcal{P}_L V_{g_1f_2}\hat{C}^{(21)*}_{u\Phi_1,g_1f_1}\\
&\qquad +s_\beta^3\mathcal{P}_R V_{f_1g_1}\hat{C}^{(12)}_{d\Phi_2,g_1f_2}+s_\beta c_\beta^2\mathcal{P}_R V_{f_1g_1}\hat{C}^{(12)}_{d\Phi_2,g_1f_2}\\
&\qquad -s_\beta^3\mathcal{P}_R V_{f_1g_1}\hat{C}^{(21)}_{d\Phi_2,g_1f_2}-s_\beta c_\beta^2\mathcal{P}_R V_{f_1g_1}\hat{C}^{(21)}_{d\Phi_2,g_1f_2}\Big)\\
&+\frac{c_{2\beta}\left(\not{p}_4\mathcal{P}_R-\not{p}_3\mathcal{P}_R\right)}{\sqrt{2}}\left(\hat{C}^{(21)}_{\Phi ud,f_1f_2}\right)\\
&+\frac{s_{2\beta}V_{f_1g_1}\left(\not{p}_3\mathcal{P}_L-\not{p}_4\mathcal{P}_L\right)}{\sqrt{2}}\left(\hat{C}^{(11)[3]}_{\Phi q,g_1f_2}-\hat{C}^{(22)[3]}_{\Phi q,g_1f_2}\right)
\end{aligned}
\tag{C.242}
$$

$$
\begin{aligned}
&+\frac{v}{\sqrt{2}}\Big(s_\beta^3\mathcal{P}_L V_{g_1f_2}\hat{C}^{(12)*}_{u\Phi_1,g_1f_1}+s_\beta c_\beta^2\mathcal{P}_L V_{g_1f_2}\hat{C}^{(12)*}_{u\Phi_1,g_1f_1}\\
&\qquad -s_\beta^3\mathcal{P}_L V_{g_1f_2}\hat{C}^{(21)*}_{u\Phi_1,g_1f_1}-s_\beta c_\beta^2\mathcal{P}_L V_{g_1f_2}\hat{C}^{(21)*}_{u\Phi_1,g_1f_1}\\
&\qquad -c_\beta^3\mathcal{P}_R V_{f_1g_1}\hat{C}^{(12)}_{d\Phi_2,g_1f_2}-s_\beta^2c_\beta\mathcal{P}_R V_{f_1g_1}\hat{C}^{(12)}_{d\Phi_2,g_1f_2}\\
&\qquad +c_\beta^3\mathcal{P}_R V_{f_1g_1}\hat{C}^{(21)}_{d\Phi_2,g_1f_2}+s_\beta^2c_\beta\mathcal{P}_R V_{f_1g_1}\hat{C}^{(21)}_{d\Phi_2,g_1f_2}\Big)\\
&+\sqrt{2}s_\beta c_\beta\left(\not{p}_3\mathcal{P}_R-\not{p}_4\mathcal{P}_R\right)\left(\hat{C}^{(21)}_{\Phi ud,f_1f_2}\right)\\
&-\sqrt{2}V_{f_1g_1}\left(\not{p}_3\mathcal{P}_L-\not{p}_4\mathcal{P}_L\right)\left(s_\beta^2\hat{C}^{(11)[3]}_{\Phi q,g_1f_2}+c_\beta^2\hat{C}^{(22)[3]}_{\Phi q,g_1f_2}\right)
\end{aligned}
\tag{C.243}
$$

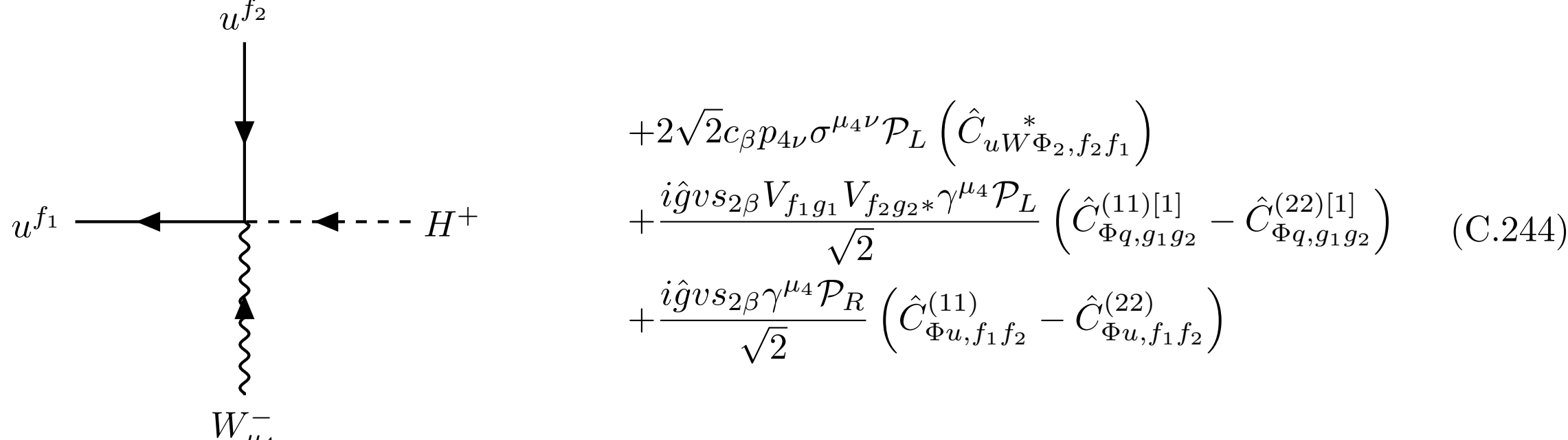

$$
\begin{aligned}
&+2\sqrt{2}c_\beta p_{4\nu}\sigma^{\mu_4\nu}\mathcal{P}_L\left(\hat{C}^{*}_{uW\Phi_2,f_2f_1}\right)\\
&+\frac{i\hat{g}vs_{2\beta}V_{f_1g_1}V_{f_2g_2*}\gamma^{\mu_4}\mathcal{P}_L}{\sqrt{2}}\left(\hat{C}^{(11)[1]}_{\Phi q,g_1g_2}-\hat{C}^{(22)[1]}_{\Phi q,g_1g_2}\right)\\
&+\frac{i\hat{g}vs_{2\beta}\gamma^{\mu_4}\mathcal{P}_R}{\sqrt{2}}\left(\hat{C}^{(11)}_{\Phi u,f_1f_2}-\hat{C}^{(22)}_{\Phi u,f_1f_2}\right)
\end{aligned}
\tag{C.244}
$$

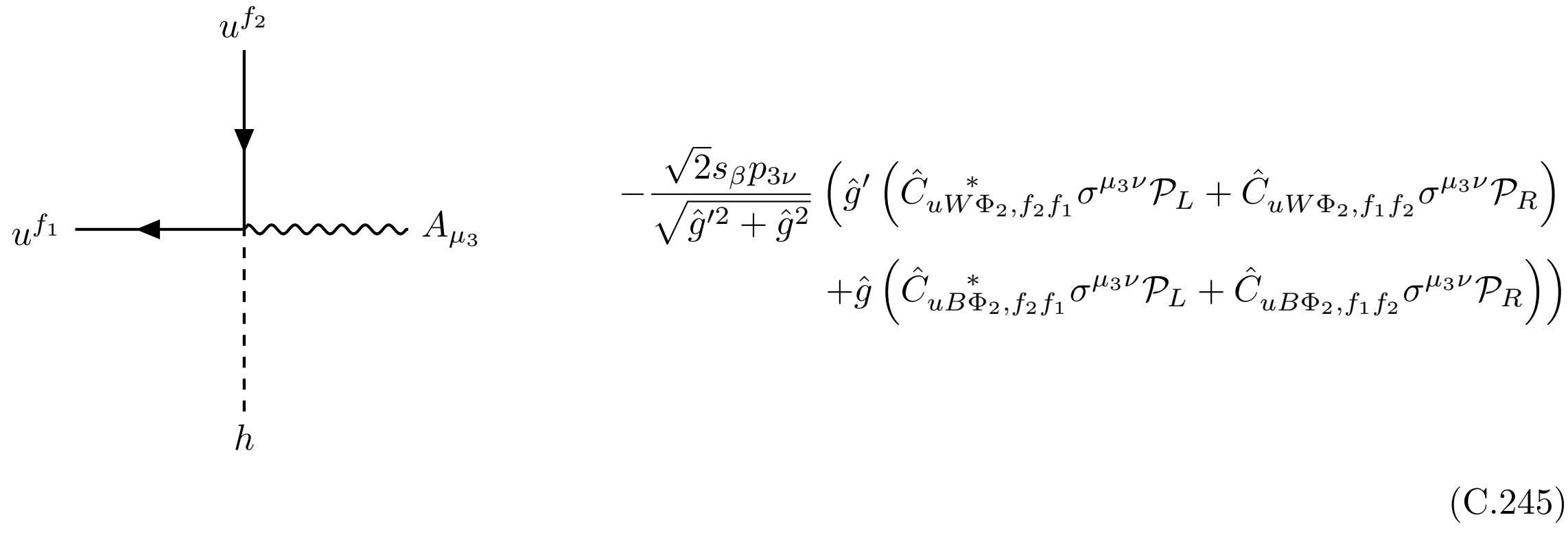

$$-\frac{\sqrt{2}s_\beta p_{3\nu}}{\sqrt{\hat{g}'^2+\hat{g}^2}}\left(\hat{g}'\left(\hat{C}^*_{uW\Phi_2,f_2f_1}\sigma^{\mu_3\nu}\mathcal{P}_L+\hat{C}_{uW\Phi_2,f_1f_2}\sigma^{\mu_3\nu}\mathcal{P}_R\right)\right.$$
$$\left.+\hat{g}\left(\hat{C}^*_{uB\Phi_2,f_2f_1}\sigma^{\mu_3\nu}\mathcal{P}_L+\hat{C}_{uB\Phi_2,f_1f_2}\sigma^{\mu_3\nu}\mathcal{P}_R\right)\right) \tag{C.245}$$

$$+\frac{\sqrt{2}s_\beta p_{4\nu}}{\sqrt{\hat{g}'^2+\hat{g}^2}}\left(\hat{g}'\left(\hat{C}^*_{uB\Phi_2,f_2f_1}\sigma^{\mu_4\nu}\mathcal{P}_L+\hat{C}_{uB\Phi_2,f_1f_2}\sigma^{\mu_4\nu}\mathcal{P}_R\right)\right.$$
$$\left.-\hat{g}\left(\hat{C}^*_{uW\Phi_2,f_2f_1}\sigma^{\mu_4\nu}\mathcal{P}_L+\hat{C}_{uW\Phi_2,f_1f_2}\sigma^{\mu_4\nu}\mathcal{P}_R\right)\right)$$
$$+i\sqrt{2}v\sqrt{\hat{g}'^2+\hat{g}^2}V_{f_1g_1}V^*_{f_2g_2}\gamma^{\mu_4}\mathcal{P}_L\left(c^2_\beta\hat{C}^{(11)[1]}_{\Phi q,g_1g_2}-c^2_\beta\hat{C}^{(11)[3]}_{\Phi q,g_1g_2}\right.$$
$$\left.+s^2_\beta\left(\hat{C}^{(22)[1]}_{\Phi q,g_1g_2}-\hat{C}^{(22)[3]}_{\Phi q,g_1g_2}\right)\right)$$
$$+i\sqrt{2}v\sqrt{\hat{g}'^2+\hat{g}^2}\gamma^{\mu_4}\mathcal{P}_R\left(c^2_\beta\hat{C}^{(11)}_{\Phi u,f_1f_2}+s^2_\beta\hat{C}^{(22)}_{\Phi u,f_1f_2}\right) \tag{C.246}$$

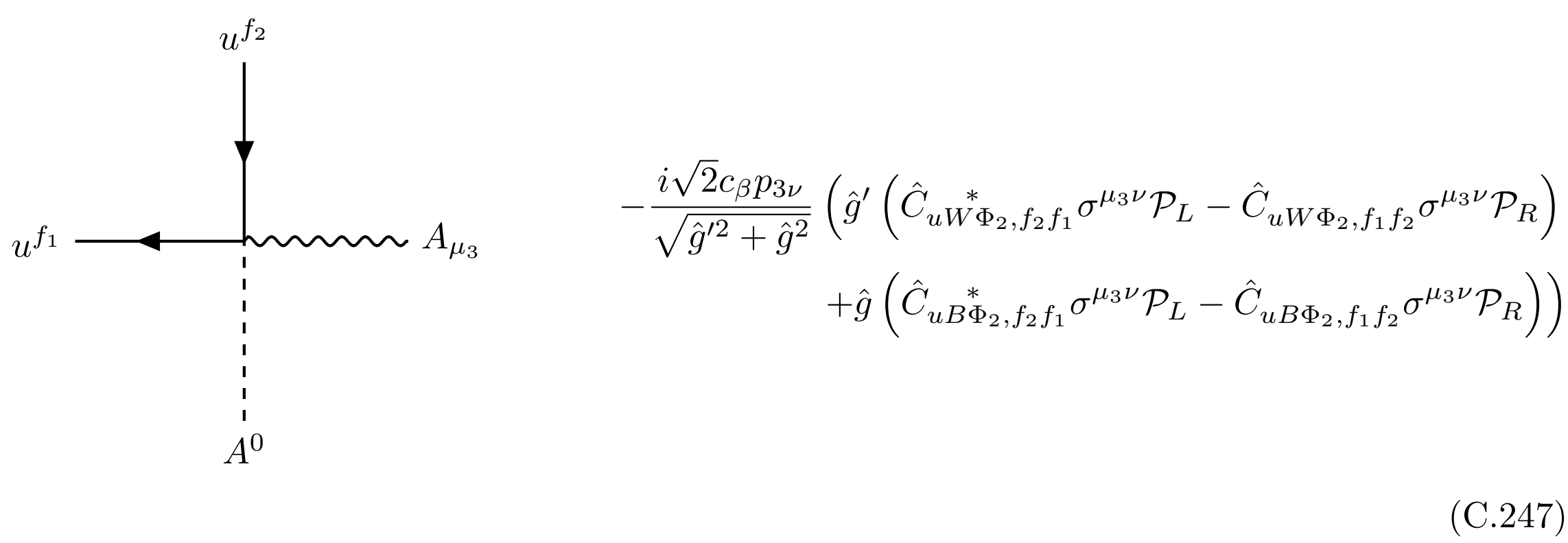

$$-\frac{i\sqrt{2}c_\beta p_{3\nu}}{\sqrt{\hat{g}'^2+\hat{g}^2}}\left(\hat{g}'\left(\hat{C}^*_{uW\Phi_2,f_2f_1}\sigma^{\mu_3\nu}\mathcal{P}_L-\hat{C}_{uW\Phi_2,f_1f_2}\sigma^{\mu_3\nu}\mathcal{P}_R\right)\right.$$
$$\left.+\hat{g}\left(\hat{C}^*_{uB\Phi_2,f_2f_1}\sigma^{\mu_3\nu}\mathcal{P}_L-\hat{C}_{uB\Phi_2,f_1f_2}\sigma^{\mu_3\nu}\mathcal{P}_R\right)\right) \tag{C.247}$$

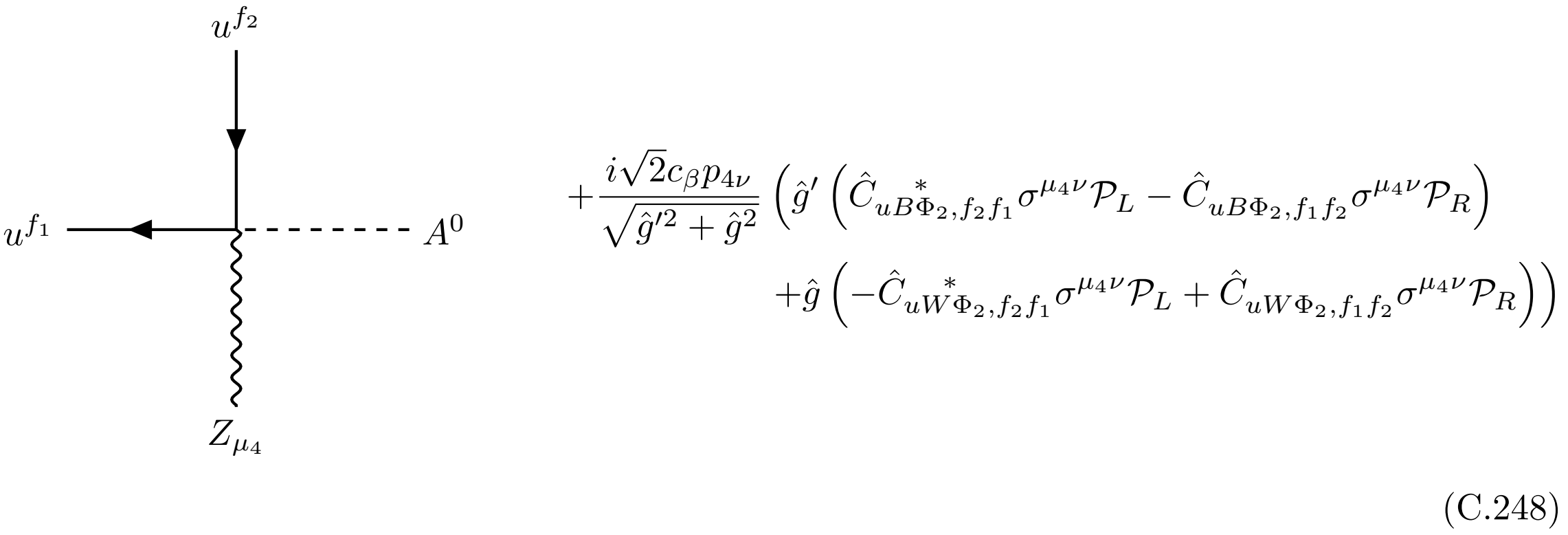

$$+\frac{i\sqrt{2}c_\beta p_{4\nu}}{\sqrt{\hat{g}'^2+\hat{g}^2}}\left(\hat{g}'\left(\hat{C}^*_{uB\Phi_2,f_2f_1}\sigma^{\mu_4\nu}\mathcal{P}_L-\hat{C}_{uB\Phi_2,f_1f_2}\sigma^{\mu_4\nu}\mathcal{P}_R\right)\right.$$
$$\left.+\hat{g}\left(-\hat{C}^*_{uW\Phi_2,f_2f_1}\sigma^{\mu_4\nu}\mathcal{P}_L+\hat{C}_{uW\Phi_2,f_1f_2}\sigma^{\mu_4\nu}\mathcal{P}_R\right)\right) \tag{C.248}$$

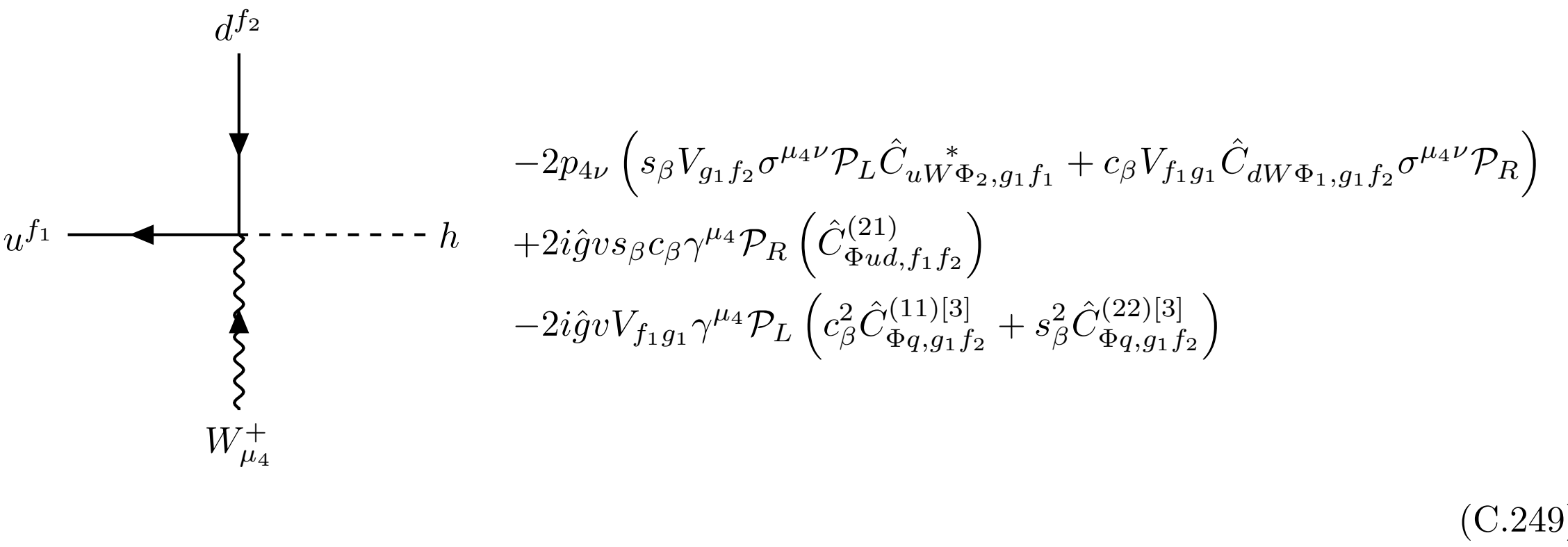

$$-2p_{4\nu}\left(s_\beta V_{g_1f_2}\sigma^{\mu_4\nu}\mathcal{P}_L\hat{C}^*_{uW\Phi_2,g_1f_1}+c_\beta V_{f_1g_1}\hat{C}_{dW\Phi_1,g_1f_2}\sigma^{\mu_4\nu}\mathcal{P}_R\right)$$
$$+2i\hat{g}vs_\beta c_\beta\gamma^{\mu_4}\mathcal{P}_R\left(\hat{C}^{(21)}_{\Phi ud,f_1f_2}\right)$$
$$-2i\hat{g}vV_{f_1g_1}\gamma^{\mu_4}\mathcal{P}_L\left(c^2_\beta\hat{C}^{(11)[3]}_{\Phi q,g_1f_2}+s^2_\beta\hat{C}^{(22)[3]}_{\Phi q,g_1f_2}\right) \tag{C.249}$$

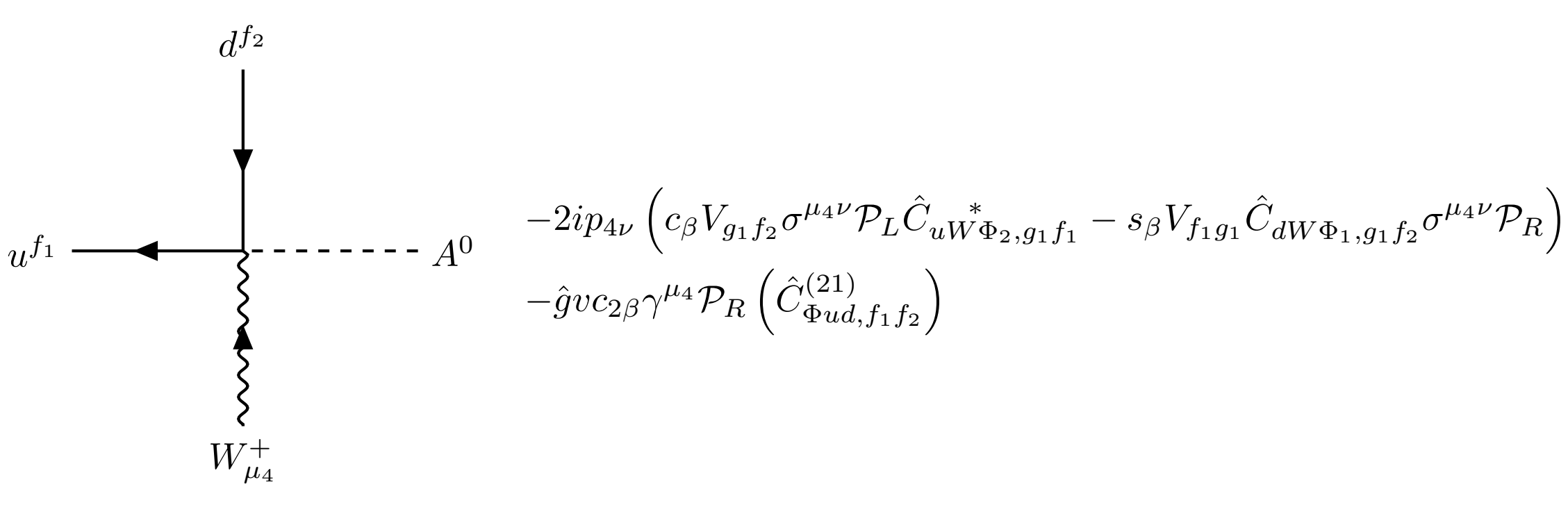

$$-2ip_{4\nu}\left(c_\beta V_{g_1f_2}\sigma^{\mu_4\nu}\mathcal{P}_L\hat{C}^*_{uW\Phi_2,g_1f_1}-s_\beta V_{f_1g_1}\hat{C}_{dW\Phi_1,g_1f_2}\sigma^{\mu_4\nu}\mathcal{P}_R\right)$$
$$-\hat{g}vc_{2\beta}\gamma^{\mu_4}\mathcal{P}_R\left(\hat{C}^{(21)}_{\Phi ud,f_1f_2}\right) \tag{C.250}$$

$d^{f_2}$, $u^{f_1}$, $A_{\mu_3}$, $H^+$

$$
\begin{aligned}
&+\frac{2p_{3\nu}}{\sqrt{\hat{g}'^2+\hat{g}^2}}\Big(-c_\beta\hat{g}'V_{g_1f_2}\sigma^{\mu_3\nu}\mathcal{P}_L\hat{C}^*_{uW\Phi_2,g_1f_1}\\
&\qquad+\hat{g}c_\beta V_{g_1f_2}\sigma^{\mu_3\nu}\mathcal{P}_L\hat{C}^*_{uB\Phi_2,g_1f_1}\\
&\qquad+\hat{g}s_\beta V_{f_1g_1}\hat{C}_{dB\Phi_1,g_1f_2}\sigma^{\mu_3\nu}\mathcal{P}_R\\
&\qquad+s_\beta\hat{g}'V_{f_1g_1}\hat{C}_{dW\Phi_1,g_1f_2}\sigma^{\mu_3\nu}\mathcal{P}_R\Big)\\
&+\frac{i\hat{g}vc_{2\beta}\hat{g}'\gamma^{\mu_3}\mathcal{P}_R}{\sqrt{\hat{g}'^2+\hat{g}^2}}\left(\hat{C}^{(21)}_{\Phi ud,f_1f_2}\right)\\
&+\frac{i\hat{g}vs_{2\beta}\hat{g}'V_{f_1g_1}\gamma^{\mu_3}\mathcal{P}_L}{\sqrt{\hat{g}'^2+\hat{g}^2}}\left(\hat{C}^{(11)[3]}_{\Phi q,g_1f_2}-\hat{C}^{(22)[3]}_{\Phi q,g_1f_2}\right)
\end{aligned}
\tag{C.251}
$$

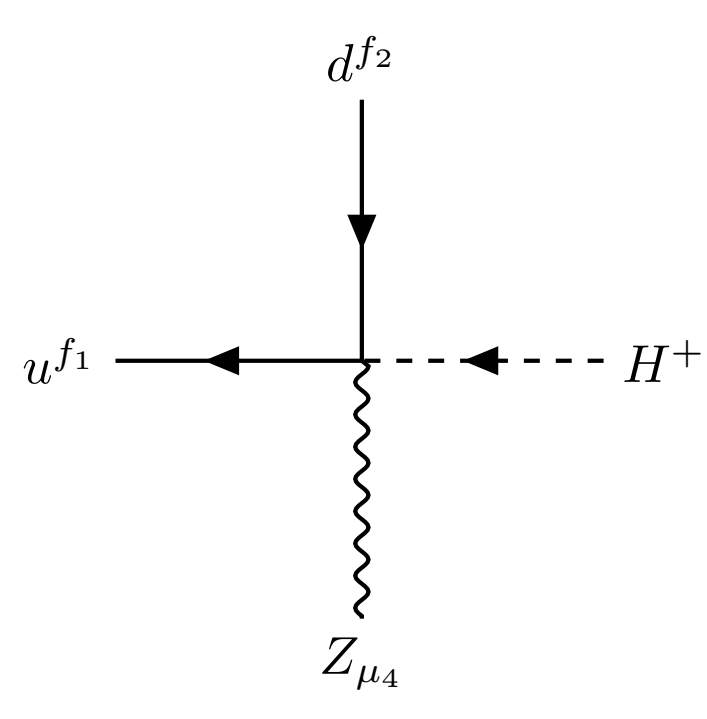

$$
\begin{aligned}
&-\frac{2p_{4\nu}}{\sqrt{\hat{g}'^2+\hat{g}^2}}\Big(c_\beta\hat{g}'V_{g_1f_2}\sigma^{\mu_4\nu}\mathcal{P}_L\hat{C}^*_{uB\Phi_2,g_1f_1}\\
&\qquad+\hat{g}c_\beta V_{g_1f_2}\sigma^{\mu_4\nu}\mathcal{P}_L\hat{C}^*_{uW\Phi_2,g_1f_1}\\
&\qquad+s_\beta\hat{g}'V_{f_1g_1}\hat{C}_{dB\Phi_1,g_1f_2}\sigma^{\mu_4\nu}\mathcal{P}_R\\
&\qquad-\hat{g}s_\beta V_{f_1g_1}\hat{C}_{dW\Phi_1,g_1f_2}\sigma^{\mu_4\nu}\mathcal{P}_R\Big)\\
&+\frac{i\hat{g}^2vc_{2\beta}\gamma^{\mu_4}\mathcal{P}_R}{\sqrt{\hat{g}'^2+\hat{g}^2}}\left(\hat{C}^{(21)}_{\Phi ud,f_1f_2}\right)\\
&-\frac{ivs_{2\beta}\hat{g}'^2V_{f_1g_1}\gamma^{\mu_4}\mathcal{P}_L}{\sqrt{\hat{g}'^2+\hat{g}^2}}\left(\hat{C}^{(11)[3]}_{\Phi q,g_1f_2}-\hat{C}^{(22)[3]}_{\Phi q,g_1f_2}\right)
\end{aligned}
\tag{C.252}
$$

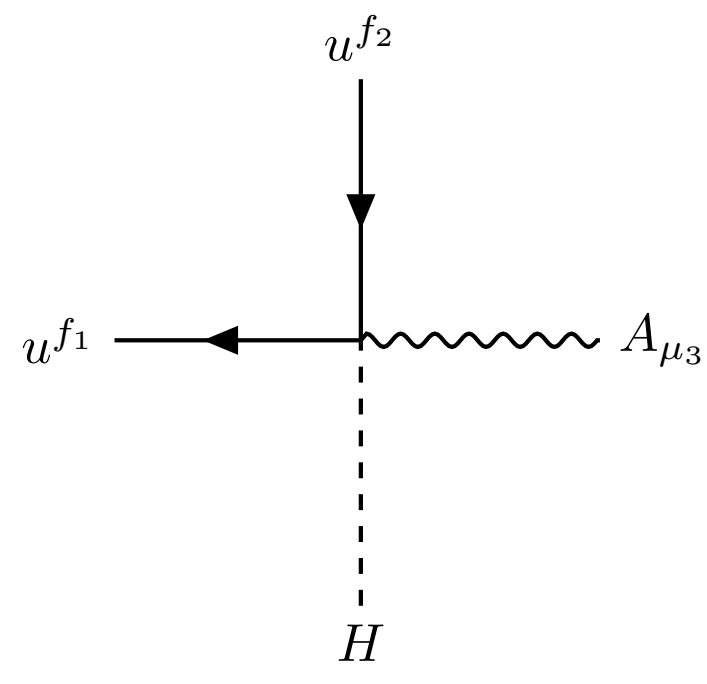

$$
\begin{aligned}
&+\frac{\sqrt{2}c_\beta p_{3\nu}}{\sqrt{\hat{g}'^2+\hat{g}^2}}\left(\hat{g}'\left(\hat{C}^*_{uW\Phi_2,f_2f_1}\sigma^{\mu_3\nu}\mathcal{P}_L+\hat{C}_{uW\Phi_2,f_1f_2}\sigma^{\mu_3\nu}\mathcal{P}_R\right)\right.\\
&\qquad\left.+\hat{g}\left(\hat{C}^*_{uB\Phi_2,f_2f_1}\sigma^{\mu_3\nu}\mathcal{P}_L+\hat{C}_{uB\Phi_2,f_1f_2}\sigma^{\mu_3\nu}\mathcal{P}_R\right)\right)
\end{aligned}
\tag{C.253}
$$

$$
\begin{aligned}
&+\frac{\sqrt{2}c_\beta p_{4\nu}}{\sqrt{\hat{g}'^2+\hat{g}^2}}\left(\hat{g}\left(\hat{C}^{*}_{uW\Phi_2,f_2f_1}\sigma^{\mu_4\nu}\mathcal{P}_L+\hat{C}_{uW\Phi_2,f_1f_2}\sigma^{\mu_4\nu}\mathcal{P}_R\right)\right.\\
&\qquad\left.-\hat{g}'\left(\hat{C}^{*}_{uB\Phi_2,f_2f_1}\sigma^{\mu_4\nu}\mathcal{P}_L+\hat{C}_{uB\Phi_2,f_1f_2}\sigma^{\mu_4\nu}\mathcal{P}_R\right)\right)\\
&+\frac{ivs_{2\beta}\sqrt{\hat{g}'^2+\hat{g}^2}V_{f_1g_1}V_{f_2g_2*}\gamma^{\mu_4}\mathcal{P}_L}{\sqrt{2}}\left(\hat{C}^{(11)[1]}_{\Phi q,g_1g_2}-\hat{C}^{(11)[3]}_{\Phi q,g_1g_2}\right.\\
&\qquad\left.-\hat{C}^{(22)[1]}_{\Phi q,g_1g_2}+\hat{C}^{(22)[3]}_{\Phi q,g_1g_2}\right)\\
&+\frac{ivs_{2\beta}\sqrt{\hat{g}'^2+\hat{g}^2}\gamma^{\mu_4}\mathcal{P}_R}{\sqrt{2}}\left(\hat{C}^{(11)}_{\Phi u,f_1f_2}-\hat{C}^{(22)}_{\Phi u,f_1f_2}\right)
\end{aligned}
\tag{C.254}
$$

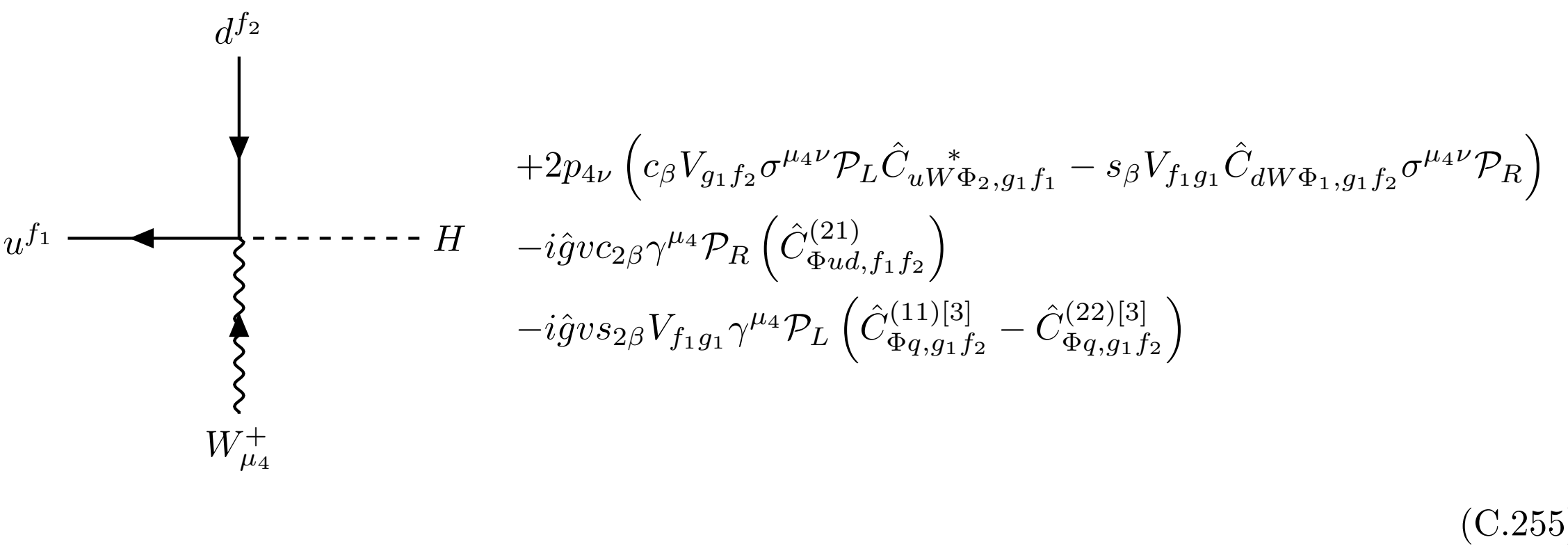

$$
\begin{aligned}
&+2p_{4\nu}\left(c_\beta V_{g_1f_2}\sigma^{\mu_4\nu}\mathcal{P}_L\hat{C}^{*}_{uW\Phi_2,g_1f_1}-s_\beta V_{f_1g_1}\hat{C}_{dW\Phi_1,g_1f_2}\sigma^{\mu_4\nu}\mathcal{P}_R\right)\\
&-i\hat{g}vc_{2\beta}\gamma^{\mu_4}\mathcal{P}_R\left(\hat{C}^{(21)}_{\Phi ud,f_1f_2}\right)\\
&-i\hat{g}vs_{2\beta}V_{f_1g_1}\gamma^{\mu_4}\mathcal{P}_L\left(\hat{C}^{(11)[3]}_{\Phi q,g_1f_2}-\hat{C}^{(22)[3]}_{\Phi q,g_1f_2}\right)
\end{aligned}
\tag{C.255}
$$

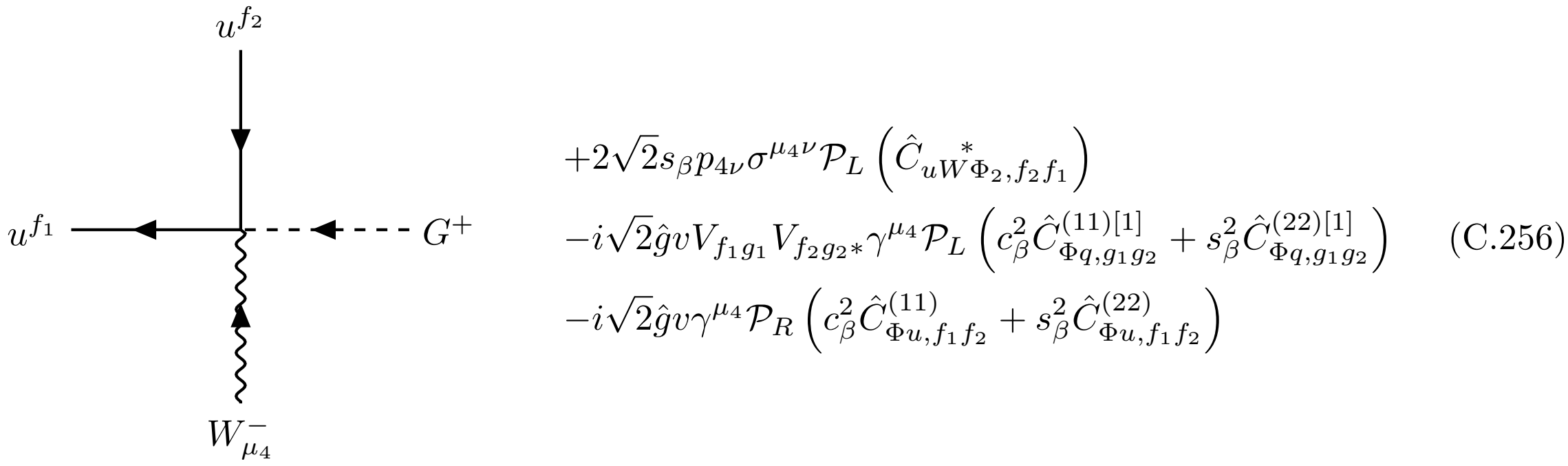

$$
\begin{aligned}
&+2\sqrt{2}s_\beta p_{4\nu}\sigma^{\mu_4\nu}\mathcal{P}_L\left(\hat{C}^{*}_{uW\Phi_2,f_2f_1}\right)\\
&-i\sqrt{2}\hat{g}vV_{f_1g_1}V_{f_2g_2*}\gamma^{\mu_4}\mathcal{P}_L\left(c_\beta^2\hat{C}^{(11)[1]}_{\Phi q,g_1g_2}+s_\beta^2\hat{C}^{(22)[1]}_{\Phi q,g_1g_2}\right)\\
&-i\sqrt{2}\hat{g}v\gamma^{\mu_4}\mathcal{P}_R\left(c_\beta^2\hat{C}^{(11)}_{\Phi u,f_1f_2}+s_\beta^2\hat{C}^{(22)}_{\Phi u,f_1f_2}\right)
\end{aligned}
\tag{C.256}
$$

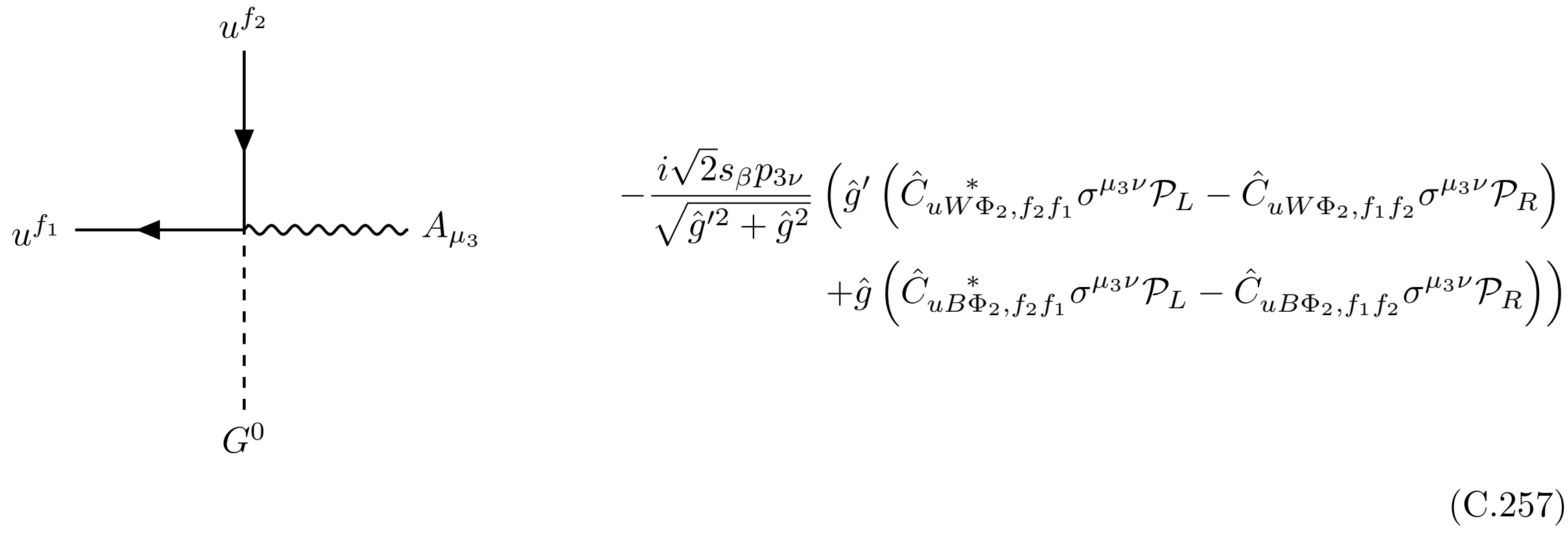

$$-\frac{i\sqrt{2}s_\beta p_{3\nu}}{\sqrt{\hat{g}'^2+\hat{g}^2}}\left(\hat{g}'\left(\hat{C}^*_{uW\Phi_2,f_2f_1}\sigma^{\mu_3\nu}\mathcal{P}_L-\hat{C}_{uW\Phi_2,f_1f_2}\sigma^{\mu_3\nu}\mathcal{P}_R\right)\right.$$
$$\left.+\hat{g}\left(\hat{C}^*_{uB\Phi_2,f_2f_1}\sigma^{\mu_3\nu}\mathcal{P}_L-\hat{C}_{uB\Phi_2,f_1f_2}\sigma^{\mu_3\nu}\mathcal{P}_R\right)\right) \tag{C.257}$$

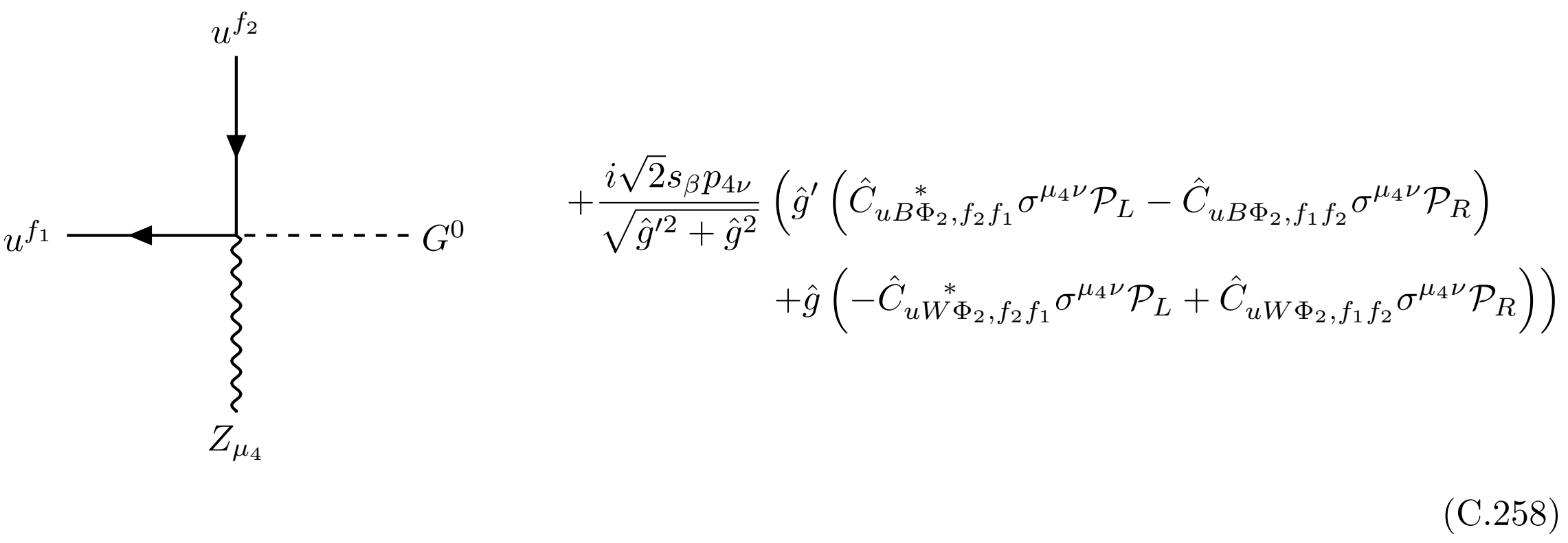

$$+\frac{i\sqrt{2}s_\beta p_{4\nu}}{\sqrt{\hat{g}'^2+\hat{g}^2}}\left(\hat{g}'\left(\hat{C}^*_{uB\Phi_2,f_2f_1}\sigma^{\mu_4\nu}\mathcal{P}_L-\hat{C}_{uB\Phi_2,f_1f_2}\sigma^{\mu_4\nu}\mathcal{P}_R\right)\right.$$
$$\left.+\hat{g}\left(-\hat{C}^*_{uW\Phi_2,f_2f_1}\sigma^{\mu_4\nu}\mathcal{P}_L+\hat{C}_{uW\Phi_2,f_1f_2}\sigma^{\mu_4\nu}\mathcal{P}_R\right)\right) \tag{C.258}$$

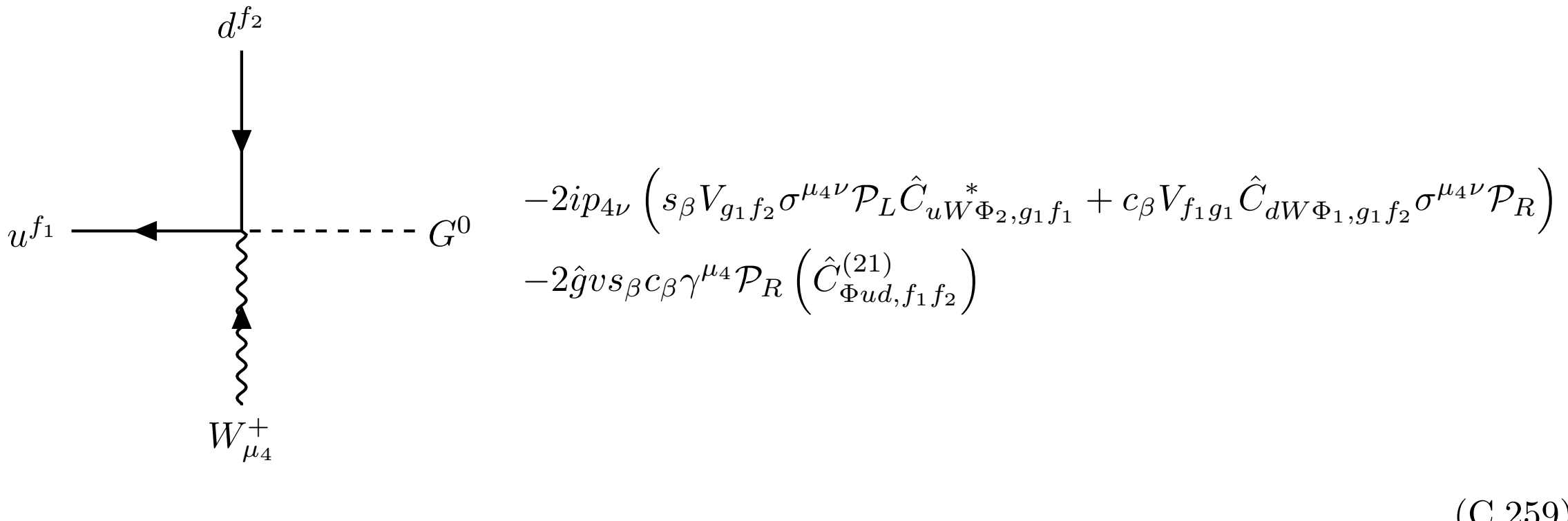

$$-2ip_{4\nu}\left(s_\beta V_{g_1f_2}\sigma^{\mu_4\nu}\mathcal{P}_L\hat{C}^*_{uW\Phi_2,g_1f_1}+c_\beta V_{f_1g_1}\hat{C}_{dW\Phi_1,g_1f_2}\sigma^{\mu_4\nu}\mathcal{P}_R\right)$$
$$-2\hat{g}vs_\beta c_\beta\gamma^{\mu_4}\mathcal{P}_R\left(\hat{C}^{(21)}_{\Phi ud,f_1f_2}\right) \tag{C.259}$$

$d^{f_2}$, $u^{f_1}$, $A_{\mu_3}$, $G^+$

$$\begin{aligned}
&-\frac{2p_{3\nu}}{\sqrt{\hat{g}'^2+\hat{g}^2}}\Big(s_\beta\hat{g}'V_{g_1f_2}\sigma^{\mu_3\nu}\mathcal{P}_L\hat{C}^*_{uW\Phi_2,g_1f_1}\\
&\qquad -\hat{g}s_\beta V_{g_1f_2}\sigma^{\mu_3\nu}\mathcal{P}_L\hat{C}^*_{uB\Phi_2,g_1f_1}\\
&\qquad +\hat{g}c_\beta V_{f_1g_1}\hat{C}_{dB\Phi_1,g_1f_2}\sigma^{\mu_3\nu}\mathcal{P}_R\\
&\qquad +c_\beta\hat{g}'V_{f_1g_1}\hat{C}_{dW\Phi_1,g_1f_2}\sigma^{\mu_3\nu}\mathcal{P}_R\Big)\\
&+\frac{2i\hat{g}vs_\beta c_\beta\hat{g}'\gamma^{\mu_3}\mathcal{P}_R}{\sqrt{\hat{g}'^2+\hat{g}^2}}\Big(\hat{C}^{(21)}_{\Phi ud,f_1f_2}\Big)\\
&-\frac{2i\hat{g}v\hat{g}'V_{f_1g_1}\gamma^{\mu_3}\mathcal{P}_L}{\sqrt{\hat{g}'^2+\hat{g}^2}}\Big(c_\beta^2\hat{C}^{(11)[3]}_{\Phi q,g_1f_2}+s_\beta^2\hat{C}^{(22)[3]}_{\Phi q,g_1f_2}\Big)
\end{aligned}\tag{C.260}$$

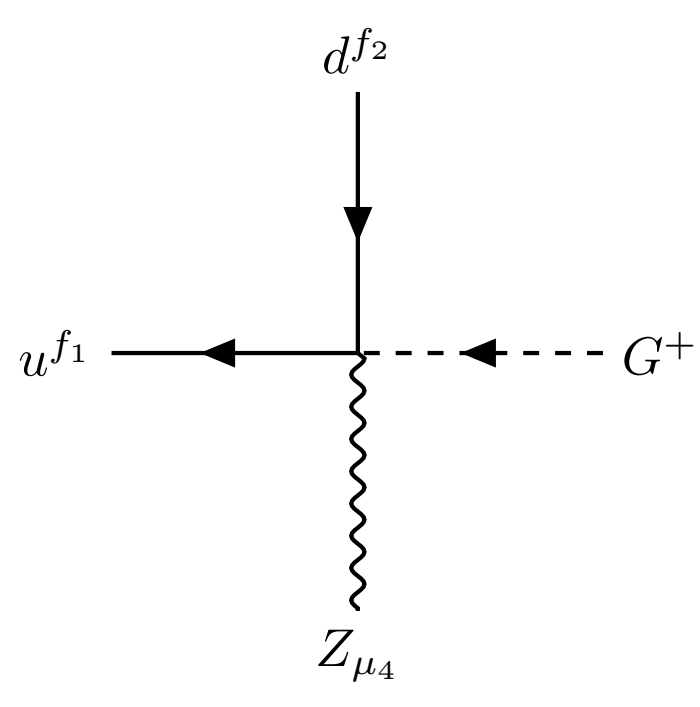

$$\begin{aligned}
&-\frac{2p_{4\nu}}{\sqrt{\hat{g}'^2+\hat{g}^2}}\Big(s_\beta\hat{g}'V_{g_1f_2}\sigma^{\mu_4\nu}\mathcal{P}_L\hat{C}^*_{uB\Phi_2,g_1f_1}\\
&\qquad +\hat{g}s_\beta V_{g_1f_2}\sigma^{\mu_4\nu}\mathcal{P}_L\hat{C}^*_{uW\Phi_2,g_1f_1}\\
&\qquad -c_\beta\hat{g}'V_{f_1g_1}\hat{C}_{dB\Phi_1,g_1f_2}\sigma^{\mu_4\nu}\mathcal{P}_R\\
&\qquad +\hat{g}c_\beta V_{f_1g_1}\hat{C}_{dW\Phi_1,g_1f_2}\sigma^{\mu_4\nu}\mathcal{P}_R\Big)\\
&+\frac{2i\hat{g}^2vs_\beta c_\beta\gamma^{\mu_4}\mathcal{P}_R}{\sqrt{\hat{g}'^2+\hat{g}^2}}\Big(\hat{C}^{(21)}_{\Phi ud,f_1f_2}\Big)\\
&+\frac{2iv\hat{g}'^2V_{f_1g_1}\gamma^{\mu_4}\mathcal{P}_L}{\sqrt{\hat{g}'^2+\hat{g}^2}}\Big(c_\beta^2\hat{C}^{(11)[3]}_{\Phi q,g_1f_2}+s_\beta^2\hat{C}^{(22)[3]}_{\Phi q,g_1f_2}\Big)
\end{aligned}\tag{C.261}$$

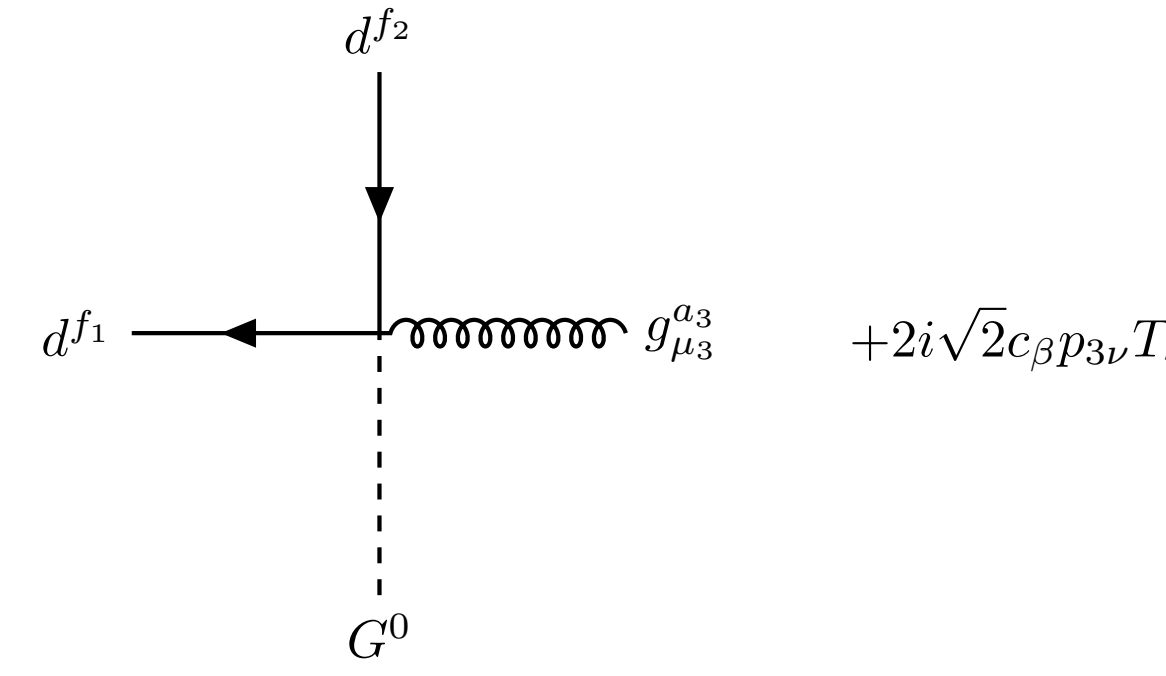

$$+2i\sqrt{2}c_\beta p_{3\nu}T^{a_3}_{m_1m_2}\Big(\hat{C}^*_{dG\Phi_1,f_2f_1}\sigma^{\mu_3\nu}\mathcal{P}_L-\hat{C}_{dG\Phi_1,f_1f_2}\sigma^{\mu_3\nu}\mathcal{P}_R\Big)\tag{C.262}$$

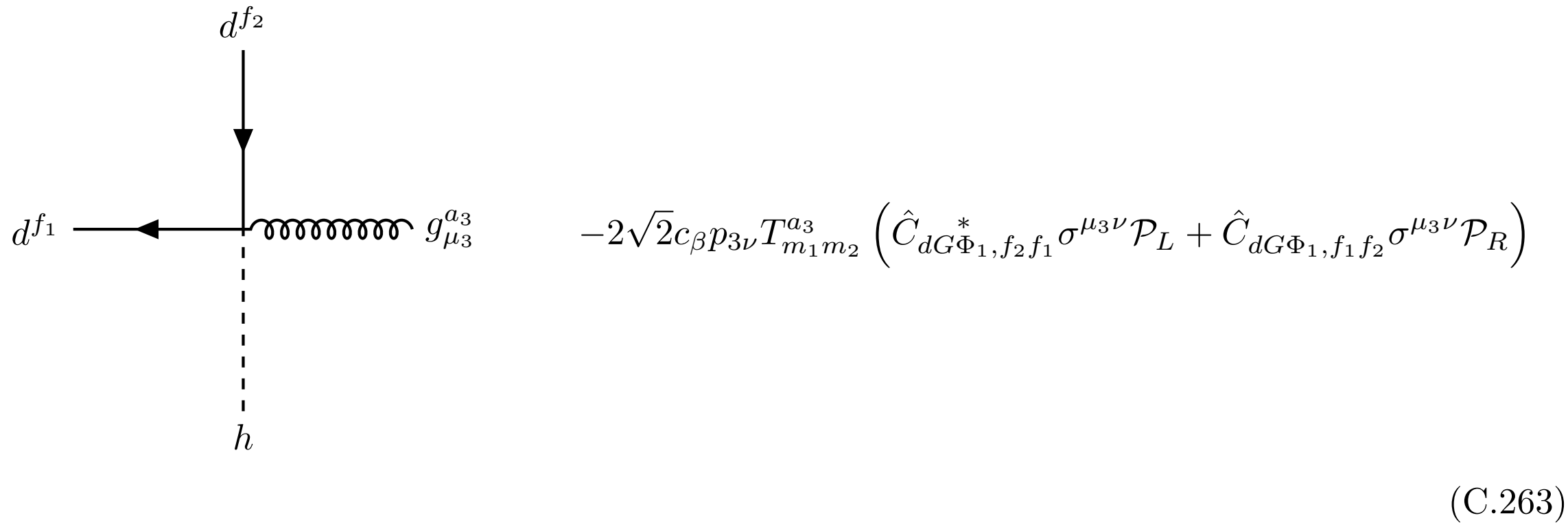

$$-2\sqrt{2}c_\beta p_{3\nu} T^{a_3}_{m_1 m_2} \left( \hat{C}^*_{dG\Phi_1, f_2 f_1} \sigma^{\mu_3 \nu} \mathcal{P}_L + \hat{C}_{dG\Phi_1, f_1 f_2} \sigma^{\mu_3 \nu} \mathcal{P}_R \right) \tag{C.263}$$

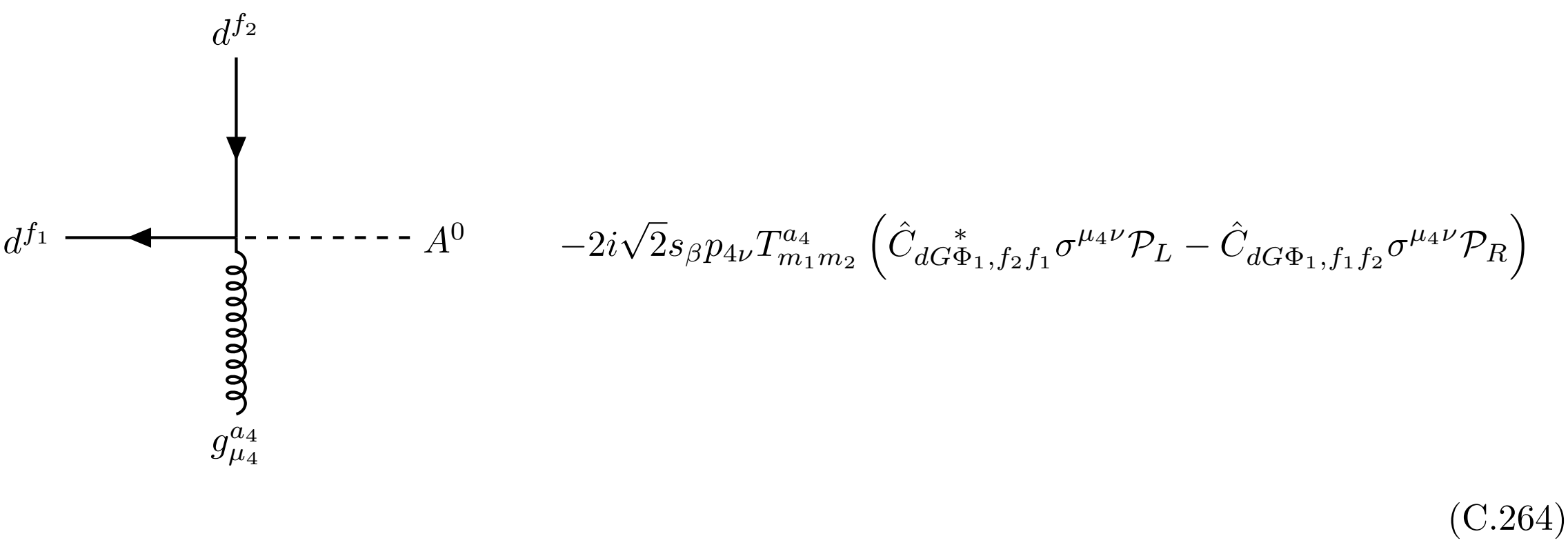

$$-2i\sqrt{2}s_\beta p_{4\nu} T^{a_4}_{m_1 m_2} \left( \hat{C}^*_{dG\Phi_1, f_2 f_1} \sigma^{\mu_4 \nu} \mathcal{P}_L - \hat{C}_{dG\Phi_1, f_1 f_2} \sigma^{\mu_4 \nu} \mathcal{P}_R \right) \tag{C.264}$$

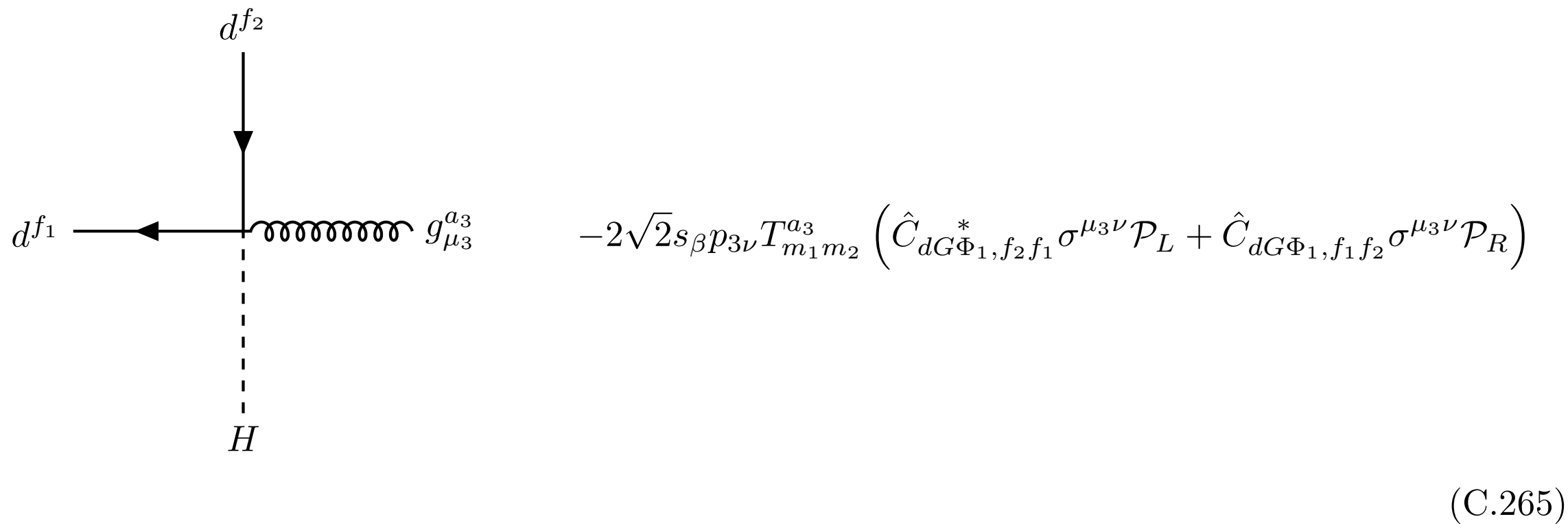

$$-2\sqrt{2}s_\beta p_{3\nu} T^{a_3}_{m_1 m_2} \left( \hat{C}^*_{dG\Phi_1, f_2 f_1} \sigma^{\mu_3 \nu} \mathcal{P}_L + \hat{C}_{dG\Phi_1, f_1 f_2} \sigma^{\mu_3 \nu} \mathcal{P}_R \right) \tag{C.265}$$

$d^{f_2}$, $d^{f_1}$, $g^{a_3}_{\mu_3}$, $g^{a_4}_{\mu_4}$

$$-4ivc_\beta \hat{g}_s f_{a_3 a_4 b_1} T^{b_1}_{m_1 m_2} \left( \sigma^{\mu_3\mu_4} \mathcal{P}_L \hat{C}^*_{dG\Phi_1, f_2 f_1} + \hat{C}_{dG\Phi_1, f_1 f_2} \sigma^{\mu_3\mu_4} \mathcal{P}_R \right) \tag{C.266}$$

$d^{f_2}$, $u^{f_1}$, $g^{a_3}_{\mu_3}$, $H^+$

$$+4p_{3\nu} T^{a_3}_{m_1 m_2} \left( c_\beta V_{g_1 f_2} \sigma^{\mu_3\nu} \mathcal{P}_L \hat{C}^*_{uG\Phi_2, g_1 f_1} + s_\beta V_{f_1 g_1} \hat{C}_{dG\Phi_1, g_1 f_2} \sigma^{\mu_3\nu} \mathcal{P}_R \right) \tag{C.267}$$

$d^{f_2}$, $u^{f_1}$, $g^{a_3}_{\mu_3}$, $G^+$

$$+4p_{3\nu} T^{a_3}_{m_1 m_2} \left( s_\beta V_{g_1 f_2} \sigma^{\mu_3\nu} \mathcal{P}_L \hat{C}^*_{uG\Phi_2, g_1 f_1} - c_\beta V_{f_1 g_1} \hat{C}_{dG\Phi_1, g_1 f_2} \sigma^{\mu_3\nu} \mathcal{P}_R \right) \tag{C.268}$$

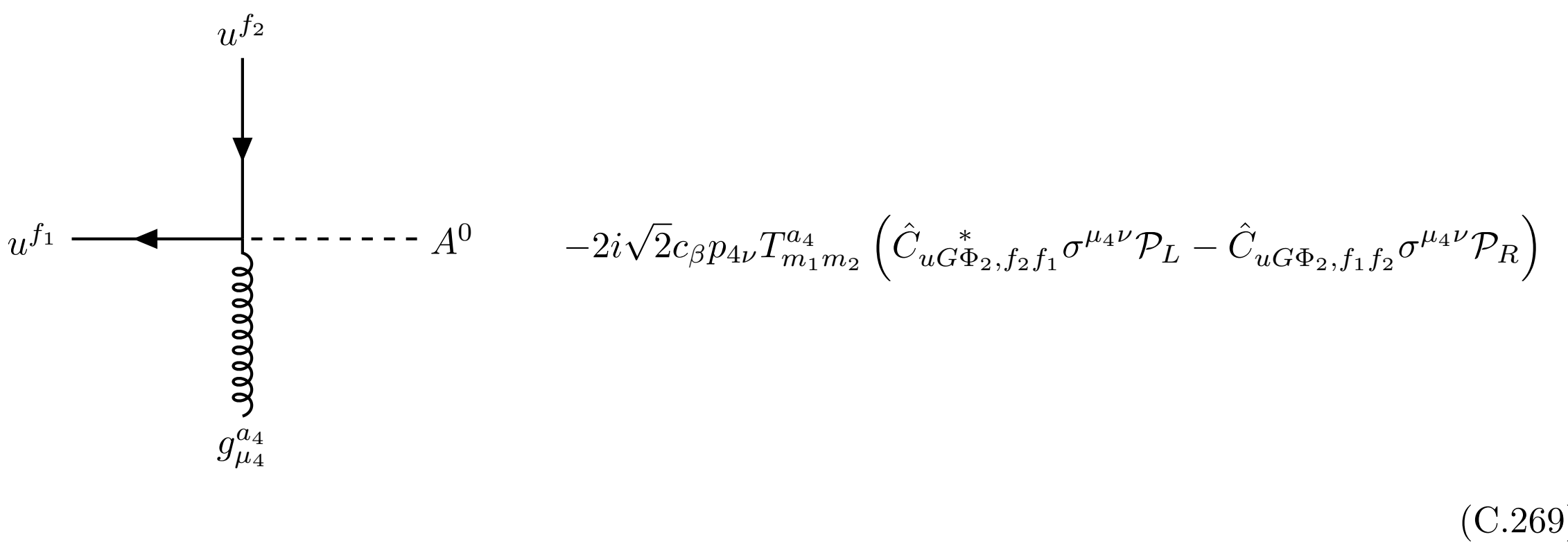

$$-2i\sqrt{2}c_\beta p_{4\nu} T^{a_4}_{m_1 m_2} \left( \hat{C}^*_{uG\Phi_2, f_2 f_1} \sigma^{\mu_4 \nu} \mathcal{P}_L - \hat{C}_{uG\Phi_2, f_1 f_2} \sigma^{\mu_4 \nu} \mathcal{P}_R \right) \tag{C.269}$$

$u^{f_2}$

$u^{f_1}$ $g^{a_3}_{\mu_3}$

$H$

$$+2\sqrt{2}c_\beta p_{3\nu} T^{a_3}_{m_1 m_2} \left( \hat{C}^*_{uG\Phi_2, f_2 f_1} \sigma^{\mu_3 \nu} \mathcal{P}_L + \hat{C}_{uG\Phi_2, f_1 f_2} \sigma^{\mu_3 \nu} \mathcal{P}_R \right) \tag{C.270}$$

$u^{f_2}$

$u^{f_1}$ $g^{a_3}_{\mu_3}$

$G^0$

$$-2i\sqrt{2}s_\beta p_{3\nu} T^{a_3}_{m_1 m_2} \left( \hat{C}^*_{uG\Phi_2, f_2 f_1} \sigma^{\mu_3 \nu} \mathcal{P}_L - \hat{C}_{uG\Phi_2, f_1 f_2} \sigma^{\mu_3 \nu} \mathcal{P}_R \right) \tag{C.271}$$

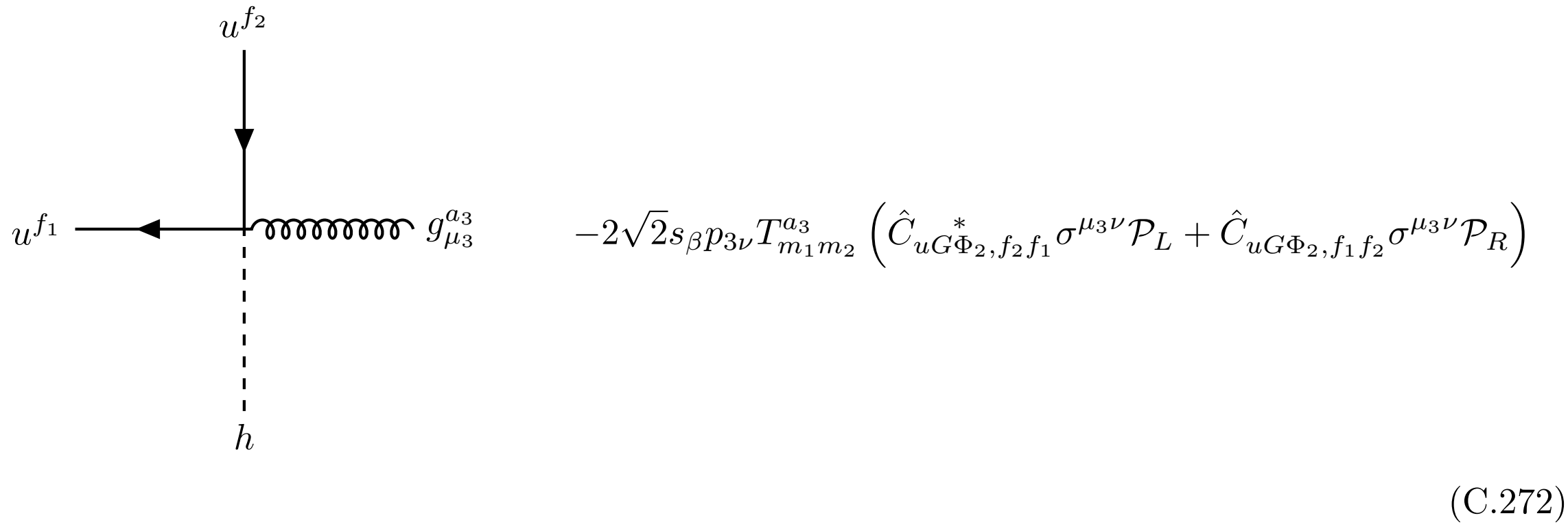

$$-2\sqrt{2}s_\beta p_{3\nu}T^{a_3}_{m_1m_2}\left(\hat{C}^*_{uG\Phi_2,f_2f_1}\sigma^{\mu_3\nu}\mathcal{P}_L+\hat{C}_{uG\Phi_2,f_1f_2}\sigma^{\mu_3\nu}\mathcal{P}_R\right)$$

(C.272)

$u^{f_2}$

$u^{f_1}$ $g^{a_3}_{\mu_3}$

$g^{a_4}_{\mu_4}$

$$-4ivs_\beta\hat{g}_sf_{a_3a_4b_1}T^{b_1}_{m_1m_2}\left(\sigma^{\mu_3\mu_4}\mathcal{P}_L\hat{C}^*_{uG\Phi_2,f_2f_1}+\hat{C}_{uG\Phi_2,f_1f_2}\sigma^{\mu_3\mu_4}\mathcal{P}_R\right)$$

(C.273)

$d^{f_2}$

$u^{f_1}$ $A_{\mu_3}$

$W^+_{\mu_4}$

$$-\frac{2\sqrt{2}\hat{g}v\hat{g}'}{\sqrt{\hat{g}'^2+\hat{g}^2}}\left(s_\beta V_{g_1f_2}\sigma^{\mu_3\mu_4}\mathcal{P}_L\hat{C}^*_{uW\Phi_2,g_1f_1}+c_\beta V_{f_1g_1}\sigma^{\mu_3\mu_4}\mathcal{P}_R\hat{C}_{dW\Phi_1,g_1f_2}\right)$$

(C.274)

$u^{f_2}$, $u^{f_1}$, $W^+_{\mu_3}$, $W^-_{\mu_4}$

$$-2\hat{g}vs_\beta\left(\sigma^{\mu_3\mu_4}\mathcal{P}_L\hat{C}^*_{uW\Phi_2,f_2f_1}+\hat{C}_{uW\Phi_2,f_1f_2}\sigma^{\mu_3\mu_4}\mathcal{P}_R\right) \tag{C.275}$$

$d^{f_2}$, $u^{f_1}$, $W^+_{\mu_3}$, $Z_{\mu_4}$

$$+\frac{2\sqrt{2}\hat{g}^2v}{\sqrt{\hat{g}'^2+\hat{g}^2}}\left(s_\beta V_{g_1f_2}\sigma^{\mu_3\mu_4}\mathcal{P}_L\hat{C}^*_{uW\Phi_2,g_1f_1}+c_\beta V_{f_1g_1}\sigma^{\mu_3\mu_4}\mathcal{P}_R\hat{C}_{dW\Phi_1,g_1f_2}\right) \tag{C.276}$$

$d^{f_2}$, $u^{f_1}$, $G^0$, $G^+$

$$\begin{aligned}&+\sqrt{2}s_\beta c_\beta\left(\not{p}_4\mathcal{P}_R-\not{p}_3\mathcal{P}_R\right)\left(\hat{C}^{(21)}_{\Phi ud,f_1f_2}\right)\\&-\sqrt{2}V_{f_1g_1}\left(\not{p}_3\mathcal{P}_L-\not{p}_4\mathcal{P}_L\right)\left(c_\beta^2\hat{C}^{(11)[3]}_{\Phi q,g_1f_2}+s_\beta^2\hat{C}^{(22)[3]}_{\Phi q,g_1f_2}\right)\end{aligned} \tag{C.277}$$

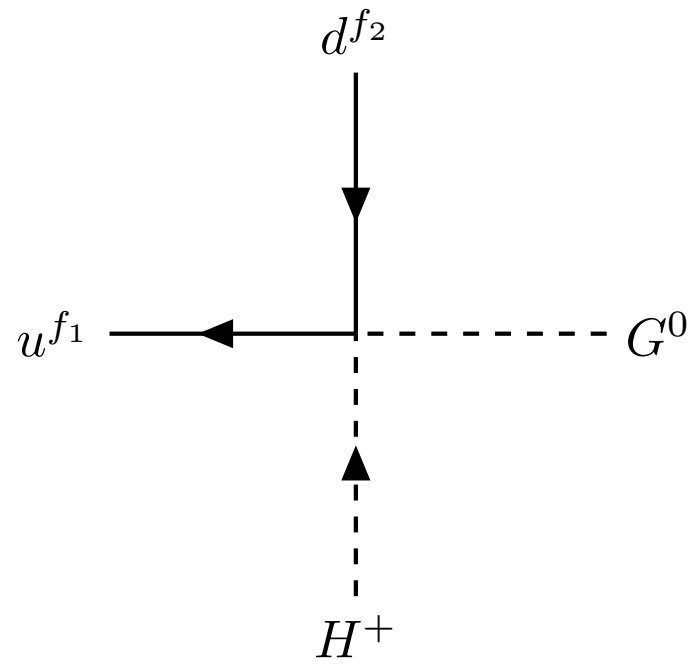

$$
\begin{aligned}
&+\frac{c_{2\beta}\left(\not{p}_4 \mathcal{P}_R - \not{p}_3 \mathcal{P}_R\right)}{\sqrt{2}} \left(\hat{C}^{(21)}_{\Phi ud, f_1 f_2}\right) \\
&+\frac{s_{2\beta} V_{f_1 g_1}\left(\not{p}_3 \mathcal{P}_L - \not{p}_4 \mathcal{P}_L\right)}{\sqrt{2}} \left(\hat{C}^{(11)[3]}_{\Phi q, g_1 f_2} - \hat{C}^{(22)[3]}_{\Phi q, g_1 f_2}\right)
\end{aligned}
\tag{C.278}
$$

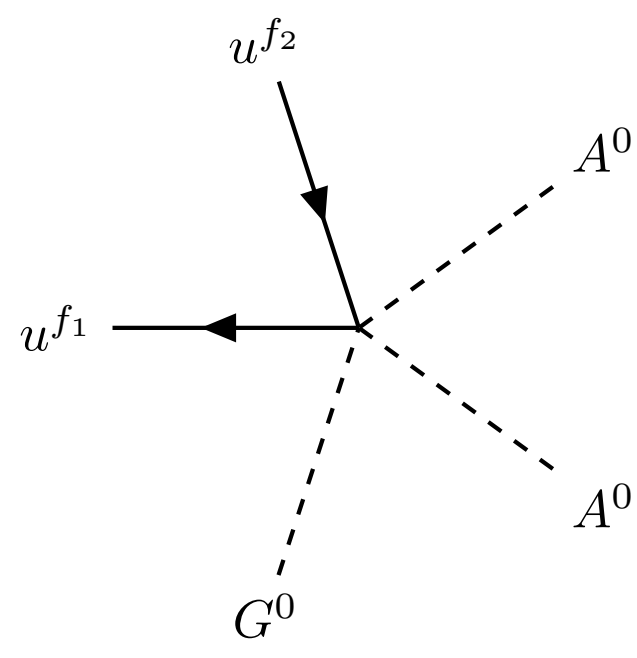

$$
\begin{aligned}
+\frac{s_\beta}{\sqrt{2}} \Big( &\left(2c_\beta^2 - s_\beta^2\right)\left(\mathcal{P}_L \hat{C}^{(11)*}_{u\Phi_2, f_2 f_1} - \mathcal{P}_R \hat{C}^{(11)}_{u\Phi_2, f_1 f_2}\right) \\
&+ \left(2c_\beta^2 - s_\beta^2\right)\left(\mathcal{P}_L \hat{C}^{(12)*}_{u\Phi_1, f_2 f_1} - \mathcal{P}_R \hat{C}^{(12)}_{u\Phi_1, f_1 f_2}\right) \\
&+ \left(2c_\beta^2 - s_\beta^2\right)\left(\mathcal{P}_L \hat{C}^{(21)*}_{u\Phi_1, f_2 f_1} - \mathcal{P}_R \hat{C}^{(21)}_{u\Phi_1, f_1 f_2}\right) \\
&-3c_\beta^2 \left(\mathcal{P}_L \hat{C}^{(22)*}_{u\Phi_2, f_2 f_1} - \mathcal{P}_R \hat{C}^{(22)}_{u\Phi_2, f_1 f_2}\right)\Big)
\end{aligned}
\tag{C.279}
$$

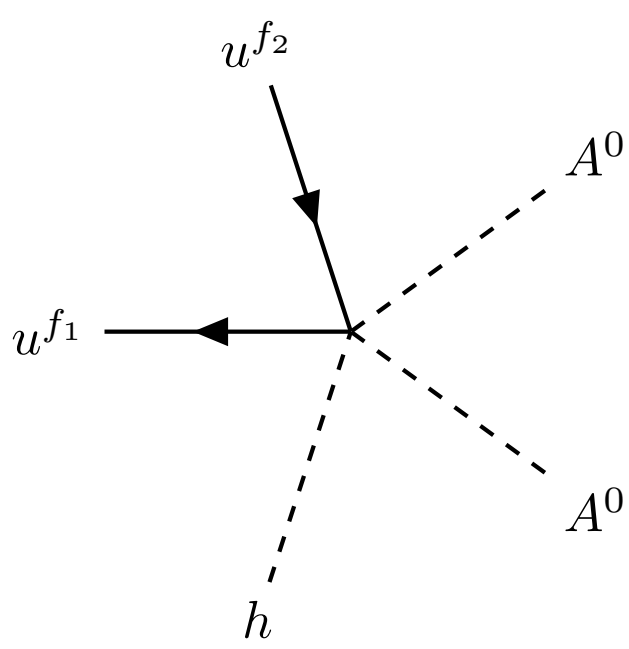

$$
\begin{aligned}
+\frac{is_\beta}{\sqrt{2}} \Big( &s_\beta^2 \left(\mathcal{P}_L \hat{C}^{(11)*}_{u\Phi_2, f_2 f_1} + \mathcal{P}_R \hat{C}^{(11)}_{u\Phi_2, f_1 f_2}\right) \\
&- \left(2c_\beta^2 + s_\beta^2\right)\left(\mathcal{P}_L \hat{C}^{(12)*}_{u\Phi_1, f_2 f_1} + \mathcal{P}_R \hat{C}^{(12)}_{u\Phi_1, f_1 f_2}\right) \\
&+s_\beta^2 \left(\mathcal{P}_L \hat{C}^{(21)*}_{u\Phi_1, f_2 f_1} + \mathcal{P}_R \hat{C}^{(21)}_{u\Phi_1, f_1 f_2}\right) \\
&+c_\beta^2 \left(\mathcal{P}_L \hat{C}^{(22)*}_{u\Phi_2, f_2 f_1} + \mathcal{P}_R \hat{C}^{(22)}_{u\Phi_2, f_1 f_2}\right)\Big)
\end{aligned}
\tag{C.280}
$$

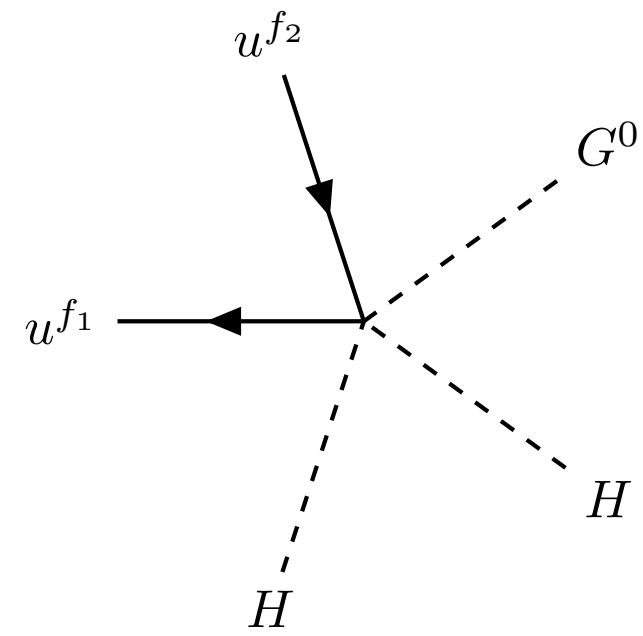

$$
\begin{aligned}
+\frac{s_\beta}{\sqrt{2}} \Big( &-s_\beta^2 \left(\mathcal{P}_L \hat{C}^{(11)*}_{u\Phi_2, f_2 f_1} - \mathcal{P}_R \hat{C}^{(11)}_{u\Phi_2, f_1 f_2}\right) \\
&+ \left(2c_\beta^2 + s_\beta^2\right)\left(\mathcal{P}_L \hat{C}^{(12)*}_{u\Phi_1, f_2 f_1} - \mathcal{P}_R \hat{C}^{(12)}_{u\Phi_1, f_1 f_2}\right) \\
&-s_\beta^2 \left(\mathcal{P}_L \hat{C}^{(21)*}_{u\Phi_1, f_2 f_1} - \mathcal{P}_R \hat{C}^{(21)}_{u\Phi_1, f_1 f_2}\right) \\
&-c_\beta^2 \left(\mathcal{P}_L \hat{C}^{(22)*}_{u\Phi_2, f_2 f_1} - \mathcal{P}_R \hat{C}^{(22)}_{u\Phi_2, f_1 f_2}\right)\Big)
\end{aligned}
\tag{C.281}
$$

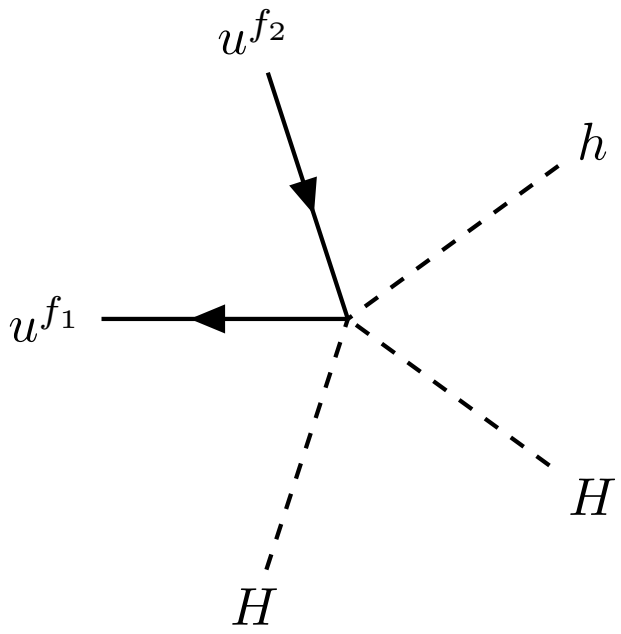

$$+\frac{is_\beta}{\sqrt{2}}\Big(-\left(2c_\beta^2-s_\beta^2\right)\left(\mathcal{P}_L\hat{C}^{(11)*}_{u\Phi_2,f_2f_1}+\mathcal{P}_R\hat{C}^{(11)}_{u\Phi_2,f_1f_2}\right)$$
$$-\left(2c_\beta^2-s_\beta^2\right)\left(\mathcal{P}_L\hat{C}^{(12)*}_{u\Phi_1,f_2f_1}+\mathcal{P}_R\hat{C}^{(12)}_{u\Phi_1,f_1f_2}\right)$$
$$-\left(2c_\beta^2-s_\beta^2\right)\left(\mathcal{P}_L\hat{C}^{(21)*}_{u\Phi_1,f_2f_1}+\mathcal{P}_R\hat{C}^{(21)}_{u\Phi_1,f_1f_2}\right)$$
$$+3c_\beta^2\left(\mathcal{P}_L\hat{C}^{(22)*}_{u\Phi_2,f_2f_1}+\mathcal{P}_R\hat{C}^{(22)}_{u\Phi_2,f_1f_2}\right)\Big) \tag{C.282}$$

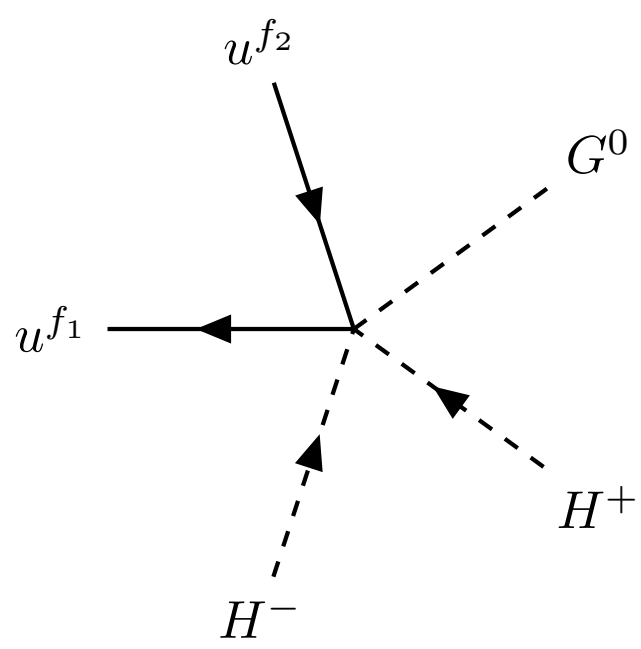

$$+\frac{s_\beta}{\sqrt{2}}\Big(-s_\beta^2\left(\mathcal{P}_L\hat{C}^{(11)*}_{u\Phi_2,f_2f_1}-\mathcal{P}_R\hat{C}^{(11)}_{u\Phi_2,f_1f_2}\right)$$
$$+c_\beta^2\left(\mathcal{P}_L\hat{C}^{(12)*}_{u\Phi_1,f_2f_1}-\mathcal{P}_R\hat{C}^{(12)}_{u\Phi_1,f_1f_2}\right)$$
$$+c_\beta^2\left(\mathcal{P}_L\hat{C}^{(21)*}_{u\Phi_1,f_2f_1}-\mathcal{P}_R\hat{C}^{(21)}_{u\Phi_1,f_1f_2}\right)$$
$$-c_\beta^2\left(\mathcal{P}_L\hat{C}^{(22)*}_{u\Phi_2,f_2f_1}-\mathcal{P}_R\hat{C}^{(22)}_{u\Phi_2,f_1f_2}\right)\Big) \tag{C.283}$$

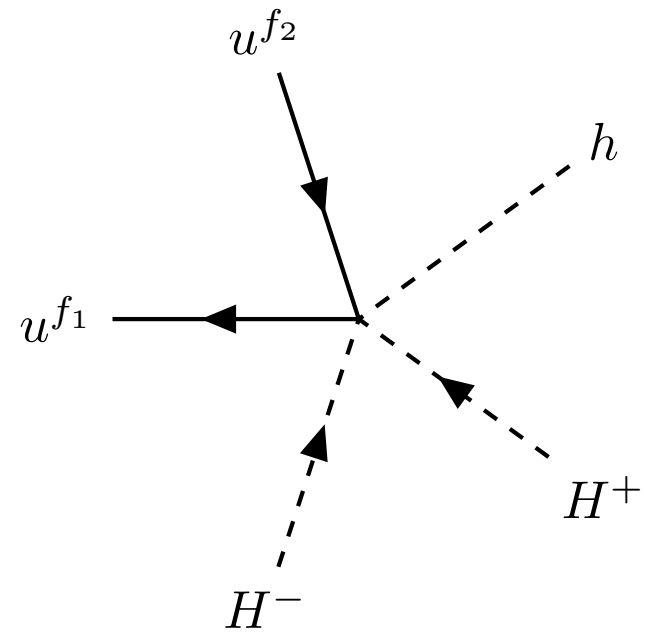

$$+\frac{is_\beta}{\sqrt{2}}\Big(s_\beta^2\left(\mathcal{P}_L\hat{C}^{(11)*}_{u\Phi_2,f_2f_1}+\mathcal{P}_R\hat{C}^{(11)}_{u\Phi_2,f_1f_2}\right)$$
$$-c_\beta^2\left(\mathcal{P}_L\hat{C}^{(12)*}_{u\Phi_1,f_2f_1}+\mathcal{P}_R\hat{C}^{(12)}_{u\Phi_1,f_1f_2}\right)$$
$$-c_\beta^2\left(\mathcal{P}_L\hat{C}^{(21)*}_{u\Phi_1,f_2f_1}+\mathcal{P}_R\hat{C}^{(21)}_{u\Phi_1,f_1f_2}\right)$$
$$+c_\beta^2\left(\mathcal{P}_L\hat{C}^{(22)*}_{u\Phi_2,f_2f_1}+\mathcal{P}_R\hat{C}^{(22)}_{u\Phi_2,f_1f_2}\right)\Big) \tag{C.284}$$

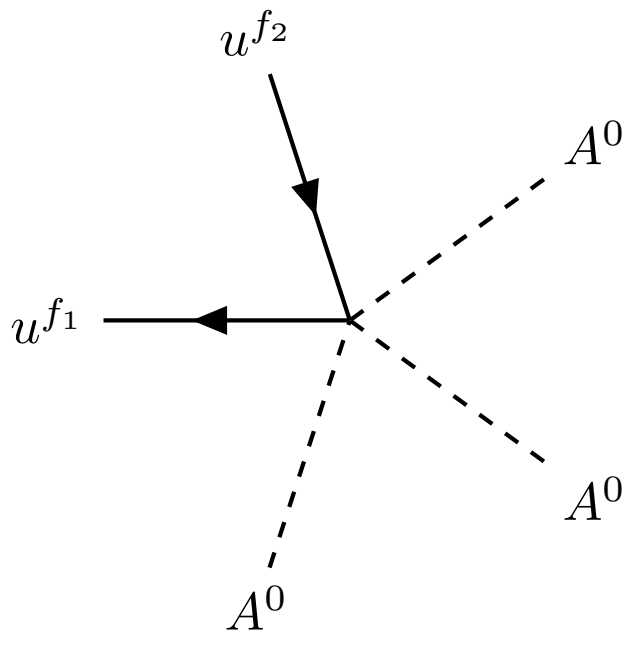

$$\begin{aligned}
&-\frac{3c_\beta}{\sqrt{2}}\left(s_\beta^2\left(\mathcal{P}_L\hat{C}^{(11)*}_{u\Phi_2,f_2f_1}-\mathcal{P}_R\hat{C}^{(11)}_{u\Phi_2,f_1f_2}\right)\right.\\
&\qquad+s_\beta^2\left(\mathcal{P}_L\hat{C}^{(12)*}_{u\Phi_1,f_2f_1}-\mathcal{P}_R\hat{C}^{(12)}_{u\Phi_1,f_1f_2}\right)\\
&\qquad+s_\beta^2\left(\mathcal{P}_L\hat{C}^{(21)*}_{u\Phi_1,f_2f_1}-\mathcal{P}_R\hat{C}^{(21)}_{u\Phi_1,f_1f_2}\right)\\
&\qquad\left.+c_\beta^2\left(\mathcal{P}_L\hat{C}^{(22)*}_{u\Phi_2,f_2f_1}-\mathcal{P}_R\hat{C}^{(22)}_{u\Phi_2,f_1f_2}\right)\right)
\end{aligned}\tag{C.285}$$

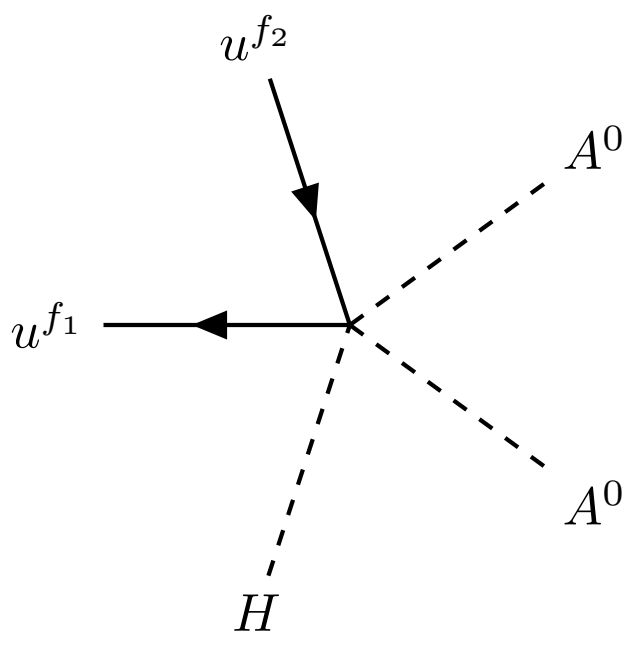

$$\begin{aligned}
&-\frac{ic_\beta}{\sqrt{2}}\left(s_\beta^2\left(\mathcal{P}_L\hat{C}^{(11)*}_{u\Phi_2,f_2f_1}+\mathcal{P}_R\hat{C}^{(11)}_{u\Phi_2,f_1f_2}\right)\right.\\
&\qquad+s_\beta^2\left(\mathcal{P}_L\hat{C}^{(12)*}_{u\Phi_1,f_2f_1}+\mathcal{P}_R\hat{C}^{(12)}_{u\Phi_1,f_1f_2}\right)\\
&\qquad+s_\beta^2\left(\mathcal{P}_L\hat{C}^{(21)*}_{u\Phi_1,f_2f_1}+\mathcal{P}_R\hat{C}^{(21)}_{u\Phi_1,f_1f_2}\right)\\
&\qquad\left.+c_\beta^2\left(\mathcal{P}_L\hat{C}^{(22)*}_{u\Phi_2,f_2f_1}+\mathcal{P}_R\hat{C}^{(22)}_{u\Phi_2,f_1f_2}\right)\right)
\end{aligned}\tag{C.286}$$

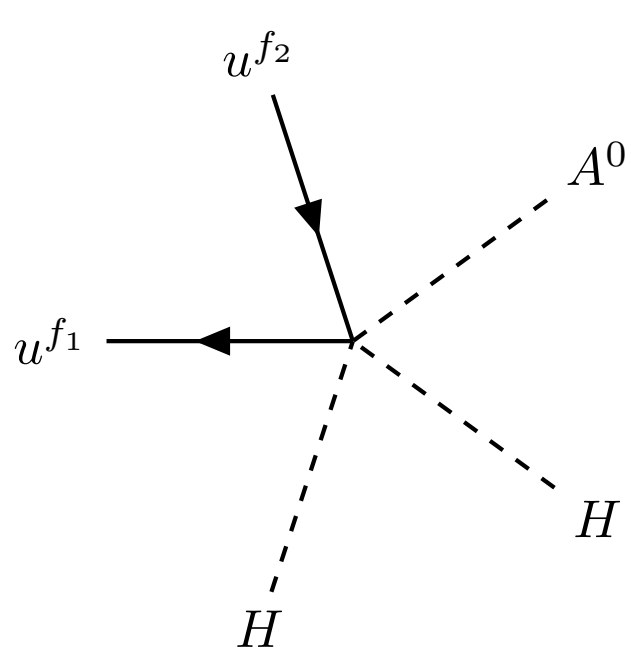

$$\begin{aligned}
&+\frac{c_\beta}{\sqrt{2}}\left(-s_\beta^2\left(\mathcal{P}_L\hat{C}^{(11)*}_{u\Phi_2,f_2f_1}-\mathcal{P}_R\hat{C}^{(11)}_{u\Phi_2,f_1f_2}\right)\right.\\
&\qquad-s_\beta^2\left(\mathcal{P}_L\hat{C}^{(12)*}_{u\Phi_1,f_2f_1}-\mathcal{P}_R\hat{C}^{(12)}_{u\Phi_1,f_1f_2}\right)\\
&\qquad-s_\beta^2\left(\mathcal{P}_L\hat{C}^{(21)*}_{u\Phi_1,f_2f_1}-\mathcal{P}_R\hat{C}^{(21)}_{u\Phi_1,f_1f_2}\right)\\
&\qquad\left.-c_\beta^2\left(\mathcal{P}_L\hat{C}^{(22)*}_{u\Phi_2,f_2f_1}-\mathcal{P}_R\hat{C}^{(22)}_{u\Phi_2,f_1f_2}\right)\right)
\end{aligned}\tag{C.287}$$

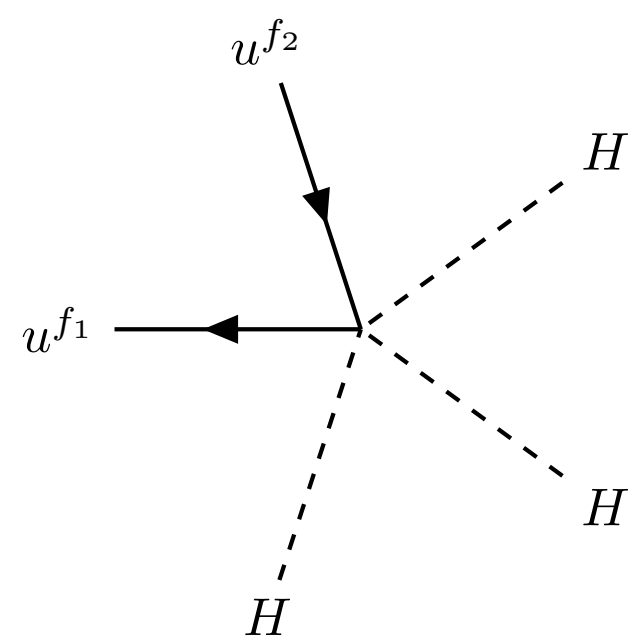

$$\begin{aligned}
&-\frac{3ic_\beta}{\sqrt{2}}\left(s_\beta^2\left(\mathcal{P}_L\hat{C}^{(11)*}_{u\Phi_2,f_2f_1}+\mathcal{P}_R\hat{C}^{(11)}_{u\Phi_2,f_1f_2}\right)\right.\\
&\qquad+s_\beta^2\left(\mathcal{P}_L\hat{C}^{(12)*}_{u\Phi_1,f_2f_1}+\mathcal{P}_R\hat{C}^{(12)}_{u\Phi_1,f_1f_2}\right)\\
&\qquad+s_\beta^2\left(\mathcal{P}_L\hat{C}^{(21)*}_{u\Phi_1,f_2f_1}+\mathcal{P}_R\hat{C}^{(21)}_{u\Phi_1,f_1f_2}\right)\\
&\qquad\left.+c_\beta^2\left(\mathcal{P}_L\hat{C}^{(22)*}_{u\Phi_2,f_2f_1}+\mathcal{P}_R\hat{C}^{(22)}_{u\Phi_2,f_1f_2}\right)\right)
\end{aligned}\tag{C.288}$$

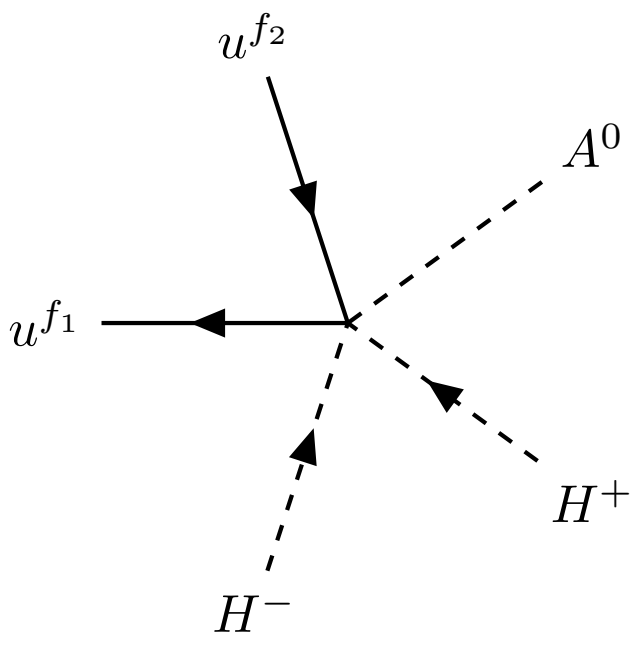

$$
\begin{aligned}
&+\frac{c_\beta}{\sqrt{2}}\left(-s_\beta^2\left(\mathcal{P}_L\hat{C}^{(11)*}_{u\Phi_2,f_2f_1}-\mathcal{P}_R\hat{C}^{(11)}_{u\Phi_2,f_1f_2}\right)\right.\\
&\qquad -s_\beta^2\left(\mathcal{P}_L\hat{C}^{(12)*}_{u\Phi_1,f_2f_1}-\mathcal{P}_R\hat{C}^{(12)}_{u\Phi_1,f_1f_2}\right)\\
&\qquad -s_\beta^2\left(\mathcal{P}_L\hat{C}^{(21)*}_{u\Phi_1,f_2f_1}-\mathcal{P}_R\hat{C}^{(21)}_{u\Phi_1,f_1f_2}\right)\\
&\qquad \left.-c_\beta^2\left(\mathcal{P}_L\hat{C}^{(22)*}_{u\Phi_2,f_2f_1}-\mathcal{P}_R\hat{C}^{(22)}_{u\Phi_2,f_1f_2}\right)\right)
\end{aligned}
\tag{C.289}
$$

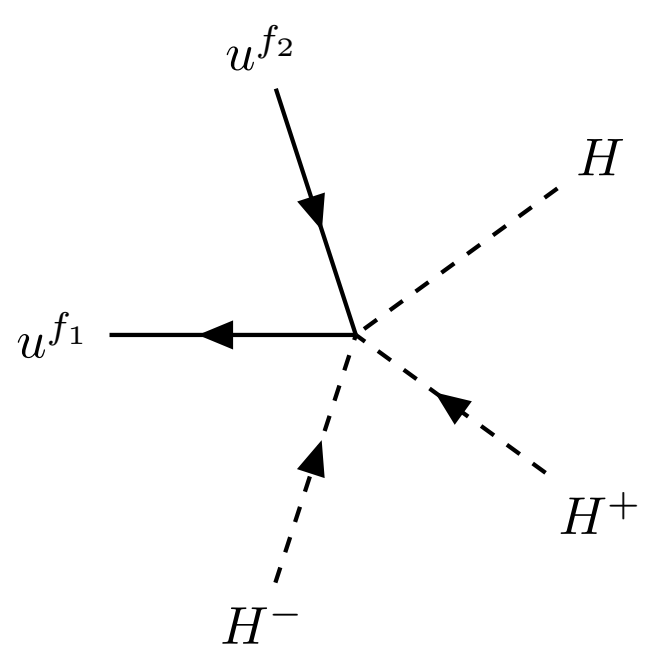

$$
\begin{aligned}
&-\frac{ic_\beta}{\sqrt{2}}\left(s_\beta^2\left(\mathcal{P}_L\hat{C}^{(11)*}_{u\Phi_2,f_2f_1}+\mathcal{P}_R\hat{C}^{(11)}_{u\Phi_2,f_1f_2}\right)\right.\\
&\qquad +s_\beta^2\left(\mathcal{P}_L\hat{C}^{(12)*}_{u\Phi_1,f_2f_1}+\mathcal{P}_R\hat{C}^{(12)}_{u\Phi_1,f_1f_2}\right)\\
&\qquad +s_\beta^2\left(\mathcal{P}_L\hat{C}^{(21)*}_{u\Phi_1,f_2f_1}+\mathcal{P}_R\hat{C}^{(21)}_{u\Phi_1,f_1f_2}\right)\\
&\qquad \left.+c_\beta^2\left(\mathcal{P}_L\hat{C}^{(22)*}_{u\Phi_2,f_2f_1}+\mathcal{P}_R\hat{C}^{(22)}_{u\Phi_2,f_1f_2}\right)\right)
\end{aligned}
\tag{C.290}
$$

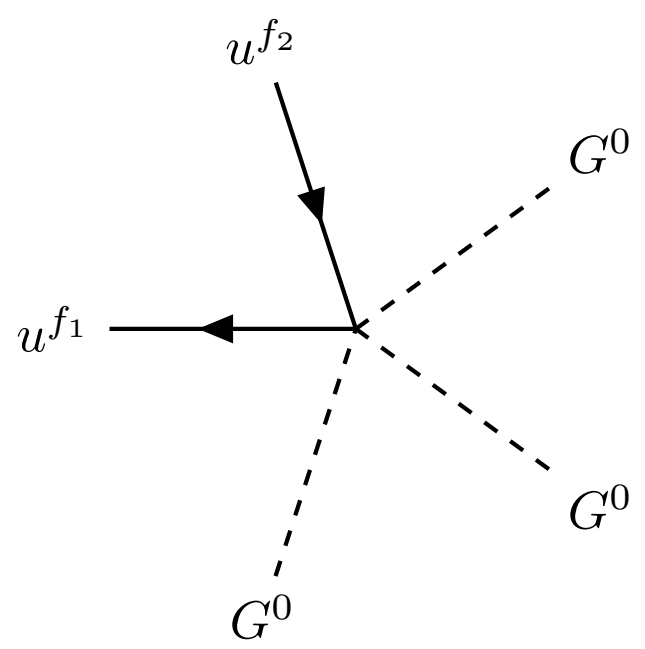

$$
\begin{aligned}
&-\frac{3s_\beta}{\sqrt{2}}\left(c_\beta^2\left(\mathcal{P}_L\hat{C}^{(11)*}_{u\Phi_2,f_2f_1}-\mathcal{P}_R\hat{C}^{(11)}_{u\Phi_2,f_1f_2}\right)\right.\\
&\qquad +c_\beta^2\left(\mathcal{P}_L\hat{C}^{(12)*}_{u\Phi_1,f_2f_1}-\mathcal{P}_R\hat{C}^{(12)}_{u\Phi_1,f_1f_2}\right)\\
&\qquad +c_\beta^2\left(\mathcal{P}_L\hat{C}^{(21)*}_{u\Phi_1,f_2f_1}-\mathcal{P}_R\hat{C}^{(21)}_{u\Phi_1,f_1f_2}\right)\\
&\qquad \left.+s_\beta^2\left(\mathcal{P}_L\hat{C}^{(22)*}_{u\Phi_2,f_2f_1}-\mathcal{P}_R\hat{C}^{(22)}_{u\Phi_2,f_1f_2}\right)\right)
\end{aligned}
\tag{C.291}
$$

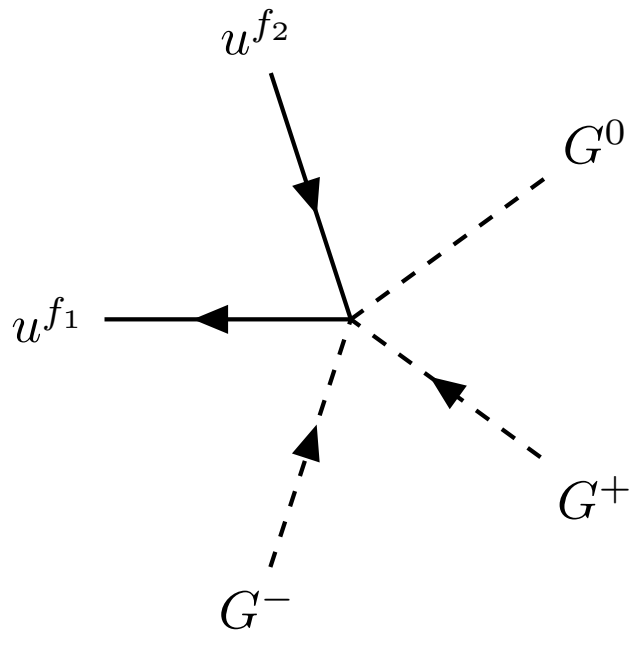

$$\begin{aligned}
&+\frac{s_\beta}{\sqrt{2}}\Big(-c_\beta^2\left(\mathcal{P}_L\hat{C}^{(11)*}_{u\Phi_2,f_2f_1}-\mathcal{P}_R\hat{C}^{(11)}_{u\Phi_2,f_1f_2}\right)\\
&\qquad -c_\beta^2\left(\mathcal{P}_L\hat{C}^{(12)*}_{u\Phi_1,f_2f_1}-\mathcal{P}_R\hat{C}^{(12)}_{u\Phi_1,f_1f_2}\right)\\
&\qquad -c_\beta^2\left(\mathcal{P}_L\hat{C}^{(21)*}_{u\Phi_1,f_2f_1}-\mathcal{P}_R\hat{C}^{(21)}_{u\Phi_1,f_1f_2}\right)\\
&\qquad -s_\beta^2\left(\mathcal{P}_L\hat{C}^{(22)*}_{u\Phi_2,f_2f_1}-\mathcal{P}_R\hat{C}^{(22)}_{u\Phi_2,f_1f_2}\right)\Big)
\end{aligned} \tag{C.292}$$

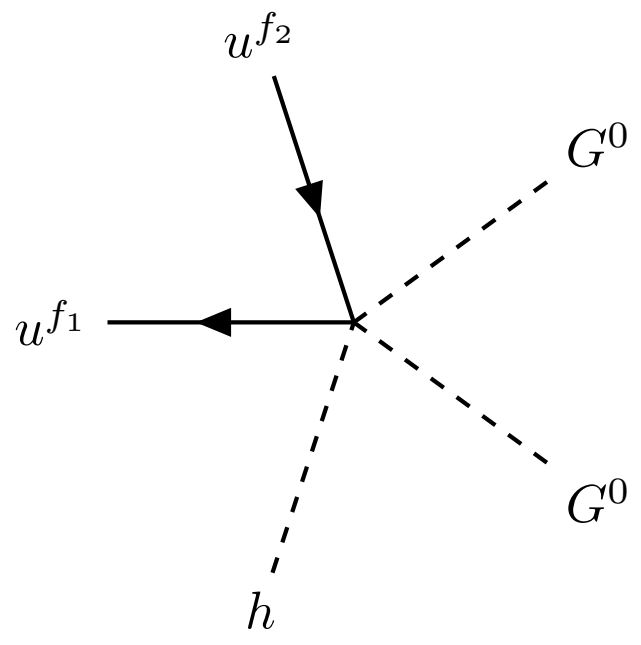

$$\begin{aligned}
&+\frac{is_\beta}{\sqrt{2}}\Big(c_\beta^2\left(\mathcal{P}_L\hat{C}^{(11)*}_{u\Phi_2,f_2f_1}+\mathcal{P}_R\hat{C}^{(11)}_{u\Phi_2,f_1f_2}\right)\\
&\qquad +c_\beta^2\left(\mathcal{P}_L\hat{C}^{(12)*}_{u\Phi_1,f_2f_1}+\mathcal{P}_R\hat{C}^{(12)}_{u\Phi_1,f_1f_2}\right)\\
&\qquad +c_\beta^2\left(\mathcal{P}_L\hat{C}^{(21)*}_{u\Phi_1,f_2f_1}+\mathcal{P}_R\hat{C}^{(21)}_{u\Phi_1,f_1f_2}\right)\\
&\qquad +s_\beta^2\left(\mathcal{P}_L\hat{C}^{(22)*}_{u\Phi_2,f_2f_1}+\mathcal{P}_R\hat{C}^{(22)}_{u\Phi_2,f_1f_2}\right)\Big)
\end{aligned} \tag{C.293}$$

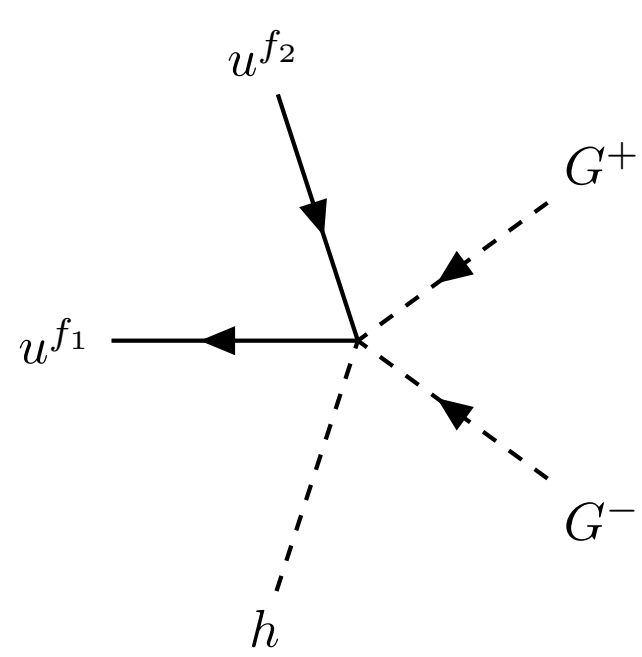

$$\begin{aligned}
&+\frac{is_\beta}{\sqrt{2}}\Big(c_\beta^2\left(\mathcal{P}_L\hat{C}^{(11)*}_{u\Phi_2,f_2f_1}+\mathcal{P}_R\hat{C}^{(11)}_{u\Phi_2,f_1f_2}\right)\\
&\qquad +c_\beta^2\left(\mathcal{P}_L\hat{C}^{(12)*}_{u\Phi_1,f_2f_1}+\mathcal{P}_R\hat{C}^{(12)}_{u\Phi_1,f_1f_2}\right)\\
&\qquad +c_\beta^2\left(\mathcal{P}_L\hat{C}^{(21)*}_{u\Phi_1,f_2f_1}+\mathcal{P}_R\hat{C}^{(21)}_{u\Phi_1,f_1f_2}\right)\\
&\qquad +s_\beta^2\left(\mathcal{P}_L\hat{C}^{(22)*}_{u\Phi_2,f_2f_1}+\mathcal{P}_R\hat{C}^{(22)}_{u\Phi_2,f_1f_2}\right)\Big)
\end{aligned} \tag{C.294}$$

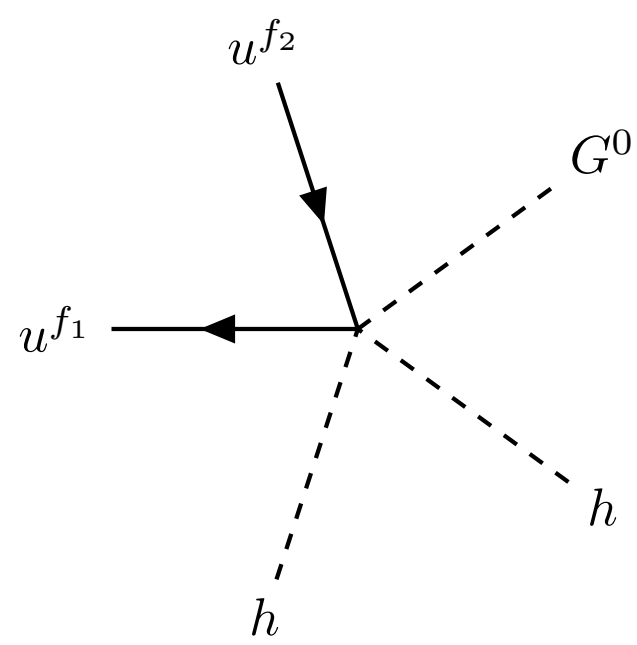

$$\begin{aligned}
&+\frac{s_\beta}{\sqrt{2}}\Big(-c_\beta^2\left(\mathcal{P}_L\hat{C}^{(11)*}_{u\Phi_2,f_2f_1}-\mathcal{P}_R\hat{C}^{(11)}_{u\Phi_2,f_1f_2}\right)\\
&\qquad -c_\beta^2\left(\mathcal{P}_L\hat{C}^{(12)*}_{u\Phi_1,f_2f_1}-\mathcal{P}_R\hat{C}^{(12)}_{u\Phi_1,f_1f_2}\right)\\
&\qquad -c_\beta^2\left(\mathcal{P}_L\hat{C}^{(21)*}_{u\Phi_1,f_2f_1}-\mathcal{P}_R\hat{C}^{(21)}_{u\Phi_1,f_1f_2}\right)\\
&\qquad -s_\beta^2\left(\mathcal{P}_L\hat{C}^{(22)*}_{u\Phi_2,f_2f_1}-\mathcal{P}_R\hat{C}^{(22)}_{u\Phi_2,f_1f_2}\right)\Big)
\end{aligned} \tag{C.295}$$

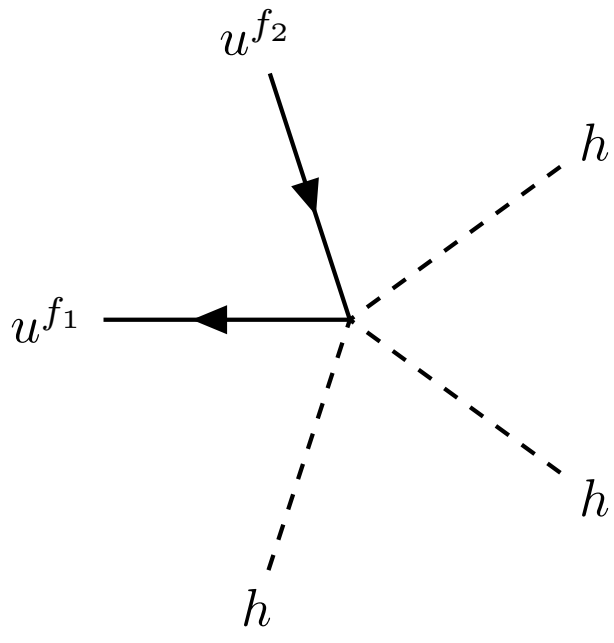

$$
\begin{aligned}
&+\frac{3is_\beta}{\sqrt{2}}\left(c_\beta^2\left(\mathcal{P}_L\hat{C}^{(11)*}_{u\Phi_2,f_2f_1}+\mathcal{P}_R\hat{C}^{(11)}_{u\Phi_2,f_1f_2}\right)\right.\\
&\qquad\quad +c_\beta^2\left(\mathcal{P}_L\hat{C}^{(12)*}_{u\Phi_1,f_2f_1}+\mathcal{P}_R\hat{C}^{(12)}_{u\Phi_1,f_1f_2}\right)\\
&\qquad\quad +c_\beta^2\left(\mathcal{P}_L\hat{C}^{(21)*}_{u\Phi_1,f_2f_1}+\mathcal{P}_R\hat{C}^{(21)}_{u\Phi_1,f_1f_2}\right)\\
&\qquad\quad \left.+s_\beta^2\left(\mathcal{P}_L\hat{C}^{(22)*}_{u\Phi_2,f_2f_1}+\mathcal{P}_R\hat{C}^{(22)}_{u\Phi_2,f_1f_2}\right)\right)
\end{aligned}
\tag{C.296}
$$

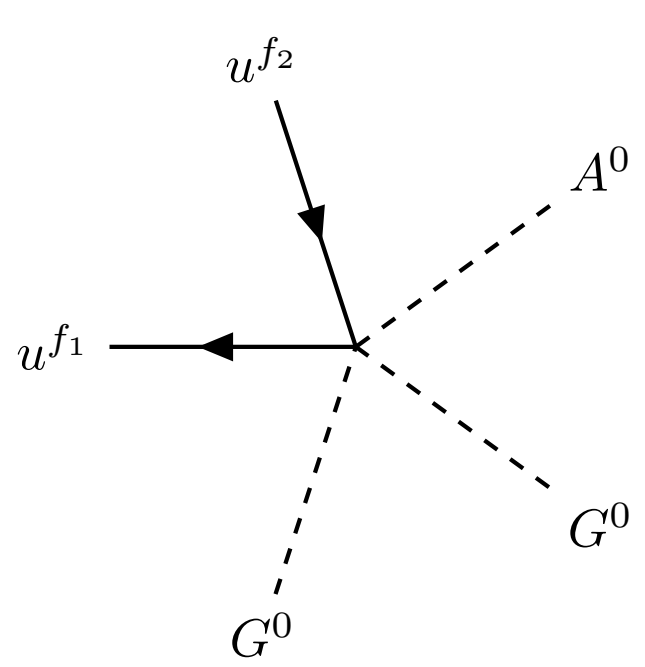

$$
\begin{aligned}
&+\frac{c_\beta}{\sqrt{2}}\left(\left(2s_\beta^2-c_\beta^2\right)\left(\mathcal{P}_L\hat{C}^{(11)*}_{u\Phi_2,f_2f_1}-\mathcal{P}_R\hat{C}^{(11)}_{u\Phi_2,f_1f_2}\right)\right.\\
&\qquad +\left(2s_\beta^2-c_\beta^2\right)\left(\mathcal{P}_L\hat{C}^{(12)*}_{u\Phi_1,f_2f_1}-\mathcal{P}_R\hat{C}^{(12)}_{u\Phi_1,f_1f_2}\right)\\
&\qquad +\left(2s_\beta^2-c_\beta^2\right)\left(\mathcal{P}_L\hat{C}^{(21)*}_{u\Phi_1,f_2f_1}-\mathcal{P}_R\hat{C}^{(21)}_{u\Phi_1,f_1f_2}\right)\\
&\qquad \left.-3s_\beta^2\left(\mathcal{P}_L\hat{C}^{(22)*}_{u\Phi_2,f_2f_1}-\mathcal{P}_R\hat{C}^{(22)}_{u\Phi_2,f_1f_2}\right)\right)
\end{aligned}
\tag{C.297}
$$

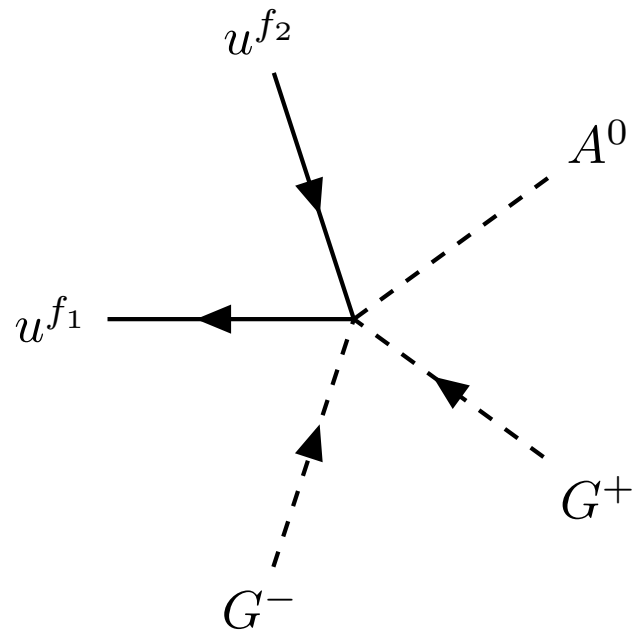

$$
\begin{aligned}
&+\frac{c_\beta}{\sqrt{2}}\left(-c_\beta^2\left(\mathcal{P}_L\hat{C}^{(11)*}_{u\Phi_2,f_2f_1}-\mathcal{P}_R\hat{C}^{(11)}_{u\Phi_2,f_1f_2}\right)\right.\\
&\qquad\quad +s_\beta^2\left(\mathcal{P}_L\hat{C}^{(12)*}_{u\Phi_1,f_2f_1}-\mathcal{P}_R\hat{C}^{(12)}_{u\Phi_1,f_1f_2}\right)\\
&\qquad\quad +s_\beta^2\left(\mathcal{P}_L\hat{C}^{(21)*}_{u\Phi_1,f_2f_1}-\mathcal{P}_R\hat{C}^{(21)}_{u\Phi_1,f_1f_2}\right)\\
&\qquad\quad \left.-s_\beta^2\left(\mathcal{P}_L\hat{C}^{(22)*}_{u\Phi_2,f_2f_1}-\mathcal{P}_R\hat{C}^{(22)}_{u\Phi_2,f_1f_2}\right)\right)
\end{aligned}
\tag{C.298}
$$

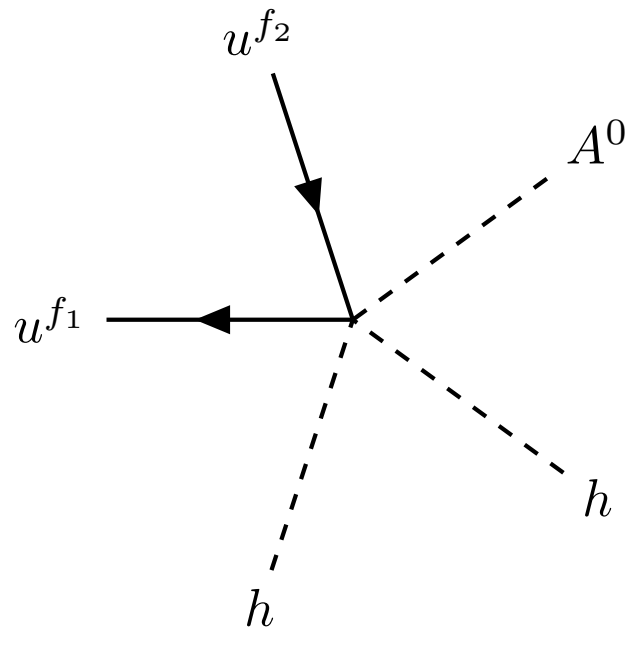

$$\begin{aligned}
&+\frac{c_\beta}{\sqrt{2}}\Big(-c_\beta^2\left(\mathcal{P}_L\hat{C}^{(11)*}_{u\Phi_2,f_2f_1}-\mathcal{P}_R\hat{C}^{(11)}_{u\Phi_2,f_1f_2}\right)\\
&\qquad+\left(2s_\beta^2+c_\beta^2\right)\left(\mathcal{P}_L\hat{C}^{(12)*}_{u\Phi_1,f_2f_1}-\mathcal{P}_R\hat{C}^{(12)}_{u\Phi_1,f_1f_2}\right)\\
&\qquad-c_\beta^2\left(\mathcal{P}_L\hat{C}^{(21)*}_{u\Phi_1,f_2f_1}-\mathcal{P}_R\hat{C}^{(21)}_{u\Phi_1,f_1f_2}\right)\\
&\qquad-s_\beta^2\left(\mathcal{P}_L\hat{C}^{(22)*}_{u\Phi_2,f_2f_1}-\mathcal{P}_R\hat{C}^{(22)}_{u\Phi_2,f_1f_2}\right)\Big)
\end{aligned}\tag{C.299}$$

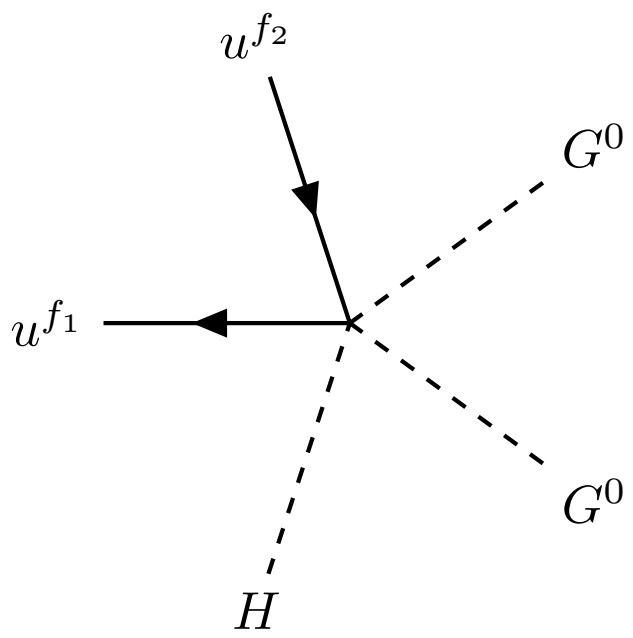

$$\begin{aligned}
&-\frac{ic_\beta}{\sqrt{2}}\Big(c_\beta^2\left(\mathcal{P}_L\hat{C}^{(11)*}_{u\Phi_2,f_2f_1}+\mathcal{P}_R\hat{C}^{(11)}_{u\Phi_2,f_1f_2}\right)\\
&\qquad-\left(2s_\beta^2+c_\beta^2\right)\left(\mathcal{P}_L\hat{C}^{(12)*}_{u\Phi_1,f_2f_1}+\mathcal{P}_R\hat{C}^{(12)}_{u\Phi_1,f_1f_2}\right)\\
&\qquad+c_\beta^2\left(\mathcal{P}_L\hat{C}^{(21)*}_{u\Phi_1,f_2f_1}+\mathcal{P}_R\hat{C}^{(21)}_{u\Phi_1,f_1f_2}\right)\\
&\qquad+s_\beta^2\left(\mathcal{P}_L\hat{C}^{(22)*}_{u\Phi_2,f_2f_1}+\mathcal{P}_R\hat{C}^{(22)}_{u\Phi_2,f_1f_2}\right)\Big)
\end{aligned}\tag{C.300}$$

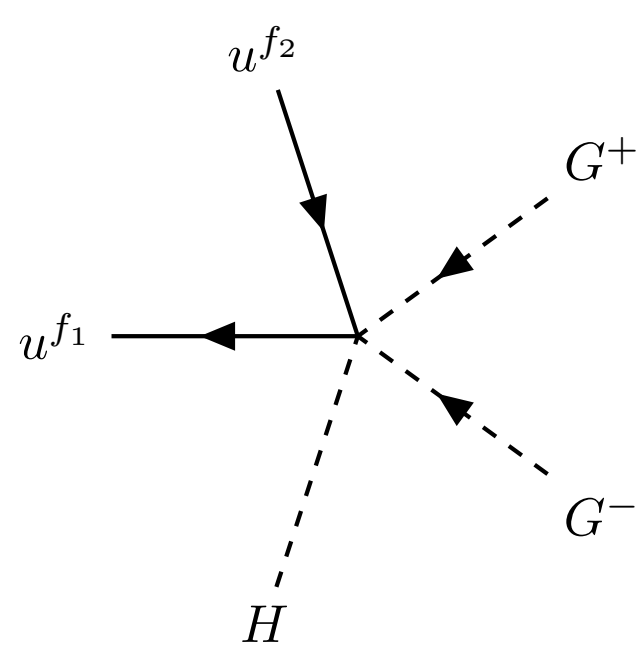

$$\begin{aligned}
&-\frac{ic_\beta}{\sqrt{2}}\Big(c_\beta^2\left(\mathcal{P}_L\hat{C}^{(11)*}_{u\Phi_2,f_2f_1}+\mathcal{P}_R\hat{C}^{(11)}_{u\Phi_2,f_1f_2}\right)\\
&\qquad-s_\beta^2\left(\mathcal{P}_L\hat{C}^{(12)*}_{u\Phi_1,f_2f_1}+\mathcal{P}_R\hat{C}^{(12)}_{u\Phi_1,f_1f_2}\right)\\
&\qquad-s_\beta^2\left(\mathcal{P}_L\hat{C}^{(21)*}_{u\Phi_1,f_2f_1}+\mathcal{P}_R\hat{C}^{(21)}_{u\Phi_1,f_1f_2}\right)\\
&\qquad+s_\beta^2\left(\mathcal{P}_L\hat{C}^{(22)*}_{u\Phi_2,f_2f_1}+\mathcal{P}_R\hat{C}^{(22)}_{u\Phi_2,f_1f_2}\right)\Big)
\end{aligned}\tag{C.301}$$

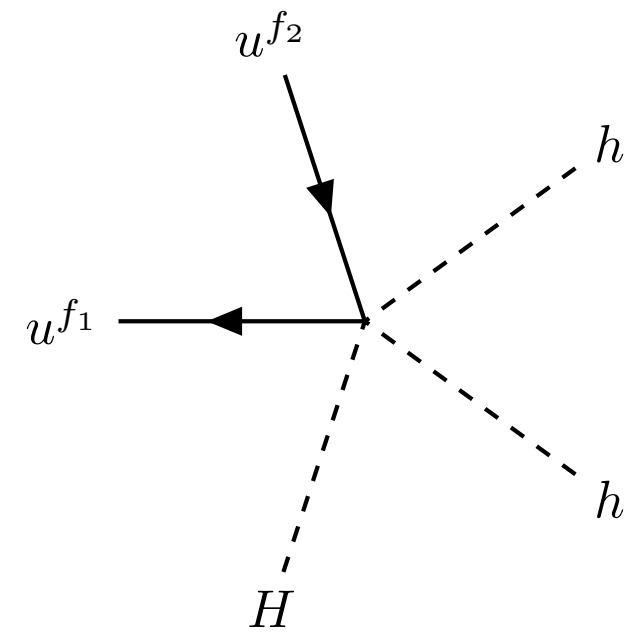

$$\begin{aligned}
&-\frac{ic_\beta}{\sqrt{2}}\Big(-\left(2s_\beta^2-c_\beta^2\right)\left(\mathcal{P}_L\hat{C}^{(11)*}_{u\Phi_2,f_2f_1}+\mathcal{P}_R\hat{C}^{(11)}_{u\Phi_2,f_1f_2}\right)\\
&\qquad-\left(2s_\beta^2-c_\beta^2\right)\left(\mathcal{P}_L\hat{C}^{(12)*}_{u\Phi_1,f_2f_1}+\mathcal{P}_R\hat{C}^{(12)}_{u\Phi_1,f_1f_2}\right)\\
&\qquad-\left(2s_\beta^2-c_\beta^2\right)\left(\mathcal{P}_L\hat{C}^{(21)*}_{u\Phi_1,f_2f_1}+\mathcal{P}_R\hat{C}^{(21)}_{u\Phi_1,f_1f_2}\right)\\
&\qquad+3s_\beta^2\left(\mathcal{P}_L\hat{C}^{(22)*}_{u\Phi_2,f_2f_1}+\mathcal{P}_R\hat{C}^{(22)}_{u\Phi_2,f_1f_2}\right)\Big)
\end{aligned}\tag{C.302}$$

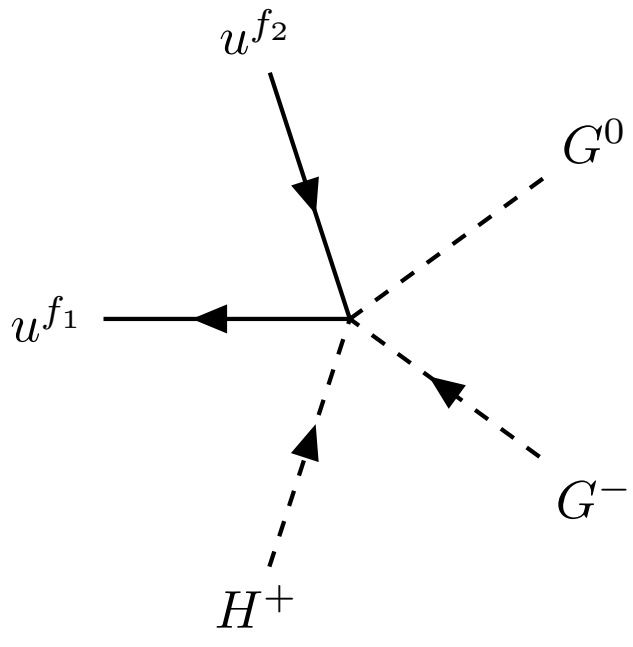

$$\begin{aligned}
&+\frac{c_\beta}{\sqrt{2}}\Big(s_\beta^2\Big(\mathcal{P}_L\hat{C}^{(11)*}_{u\Phi_2,f_2f_1}-\mathcal{P}_R\hat{C}^{(11)}_{u\Phi_2,f_1f_2}\Big)\\
&\qquad -s_\beta^2\Big(\mathcal{P}_L\hat{C}^{(22)*}_{u\Phi_2,f_2f_1}-\mathcal{P}_R\hat{C}^{(22)}_{u\Phi_2,f_1f_2}\Big)\\
&\qquad +s_\beta^2\mathcal{P}_L\hat{C}^{(12)*}_{u\Phi_1,f_2f_1}-c_\beta^2\mathcal{P}_L\hat{C}^{(21)*}_{u\Phi_1,f_2f_1}\\
&\qquad +c_\beta^2\mathcal{P}_R\hat{C}^{(12)}_{u\Phi_1,f_1f_2}-s_\beta^2\mathcal{P}_R\hat{C}^{(21)}_{u\Phi_1,f_1f_2}\Big)
\end{aligned}\tag{C.303}$$

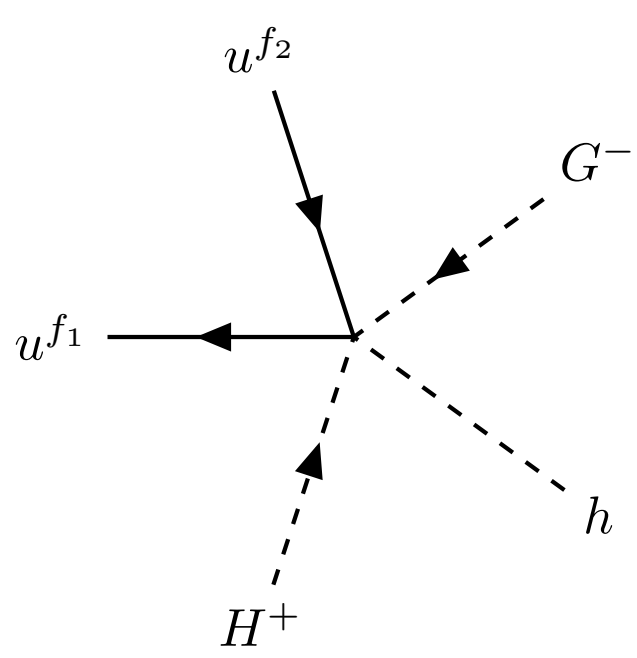

$$\begin{aligned}
&+\frac{ic_\beta}{\sqrt{2}}\Big(-s_\beta^2\Big(\mathcal{P}_L\hat{C}^{(11)*}_{u\Phi_2,f_2f_1}+\mathcal{P}_R\hat{C}^{(11)}_{u\Phi_2,f_1f_2}\Big)\\
&\qquad +s_\beta^2\Big(\mathcal{P}_L\hat{C}^{(22)*}_{u\Phi_2,f_2f_1}+\mathcal{P}_R\hat{C}^{(22)}_{u\Phi_2,f_1f_2}\Big)\\
&\qquad -s_\beta^2\mathcal{P}_L\hat{C}^{(12)*}_{u\Phi_1,f_2f_1}+c_\beta^2\mathcal{P}_L\hat{C}^{(21)*}_{u\Phi_1,f_2f_1}\\
&\qquad +c_\beta^2\mathcal{P}_R\hat{C}^{(12)}_{u\Phi_1,f_1f_2}-s_\beta^2\mathcal{P}_R\hat{C}^{(21)}_{u\Phi_1,f_1f_2}\Big)
\end{aligned}\tag{C.304}$$

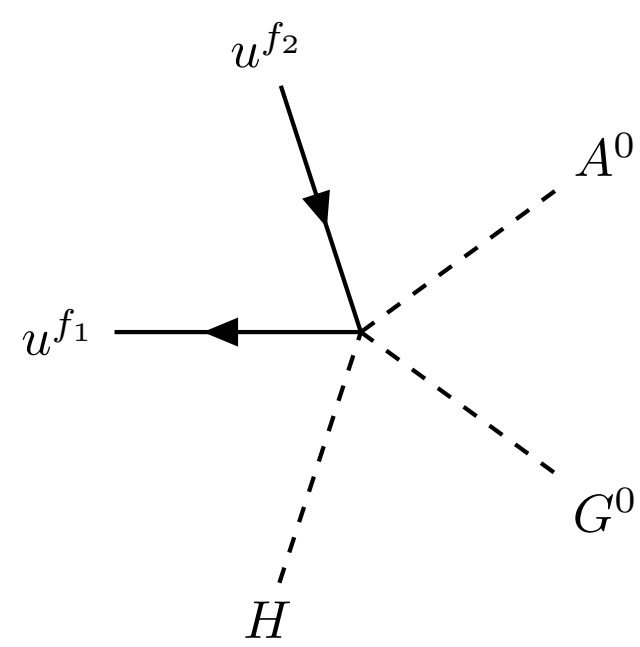

$$\begin{aligned}
&+\frac{is_\beta}{\sqrt{2}}\Big(c_\beta^2\Big(\mathcal{P}_L\hat{C}^{(11)*}_{u\Phi_2,f_2f_1}+\mathcal{P}_R\hat{C}^{(11)}_{u\Phi_2,f_1f_2}\Big)\\
&\qquad -s_\beta^2\Big(\mathcal{P}_L\hat{C}^{(12)*}_{u\Phi_1,f_2f_1}+\mathcal{P}_R\hat{C}^{(12)}_{u\Phi_1,f_1f_2}\Big)\\
&\qquad +c_\beta^2\Big(\mathcal{P}_L\hat{C}^{(21)*}_{u\Phi_1,f_2f_1}+\mathcal{P}_R\hat{C}^{(21)}_{u\Phi_1,f_1f_2}\Big)\\
&\qquad -c_\beta^2\Big(\mathcal{P}_L\hat{C}^{(22)*}_{u\Phi_2,f_2f_1}+\mathcal{P}_R\hat{C}^{(22)}_{u\Phi_2,f_1f_2}\Big)\Big)
\end{aligned}\tag{C.305}$$

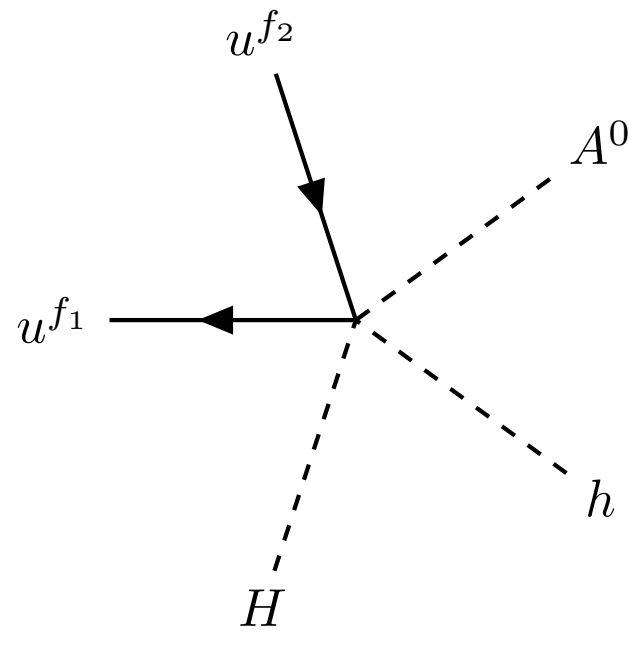

$$
\begin{aligned}
&+\frac{s_\beta}{\sqrt{2}}\Big(-c_\beta^2\left(\mathcal{P}_L\hat{C}^{(11)*}_{u\Phi_2,f_2f_1}-\mathcal{P}_R\hat{C}^{(11)}_{u\Phi_2,f_1f_2}\right)\\
&\qquad+s_\beta^2\left(\mathcal{P}_L\hat{C}^{(12)*}_{u\Phi_1,f_2f_1}-\mathcal{P}_R\hat{C}^{(12)}_{u\Phi_1,f_1f_2}\right)\\
&\qquad-c_\beta^2\left(\mathcal{P}_L\hat{C}^{(21)*}_{u\Phi_1,f_2f_1}-\mathcal{P}_R\hat{C}^{(21)}_{u\Phi_1,f_1f_2}\right)\\
&\qquad+c_\beta^2\left(\mathcal{P}_L\hat{C}^{(22)*}_{u\Phi_2,f_2f_1}-\mathcal{P}_R\hat{C}^{(22)}_{u\Phi_2,f_1f_2}\right)\Big)
\end{aligned}
\tag{C.306}
$$

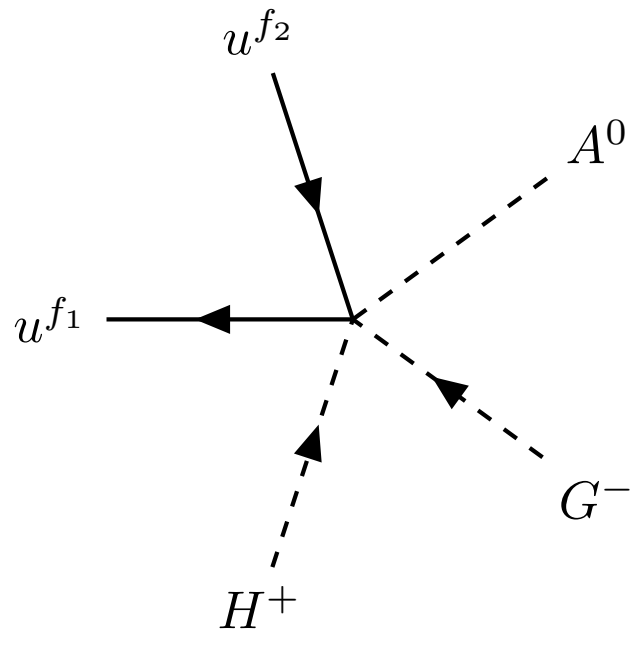

$$
\begin{aligned}
&+\frac{s_\beta}{\sqrt{2}}\Big(c_\beta^2\left(\mathcal{P}_L\hat{C}^{(11)*}_{u\Phi_2,f_2f_1}-\mathcal{P}_R\hat{C}^{(11)}_{u\Phi_2,f_1f_2}\right)\\
&\qquad-c_\beta^2\left(\mathcal{P}_L\hat{C}^{(22)*}_{u\Phi_2,f_2f_1}-\mathcal{P}_R\hat{C}^{(22)}_{u\Phi_2,f_1f_2}\right)\\
&\qquad-s_\beta^2\mathcal{P}_L\hat{C}^{(12)*}_{u\Phi_1,f_2f_1}+c_\beta^2\mathcal{P}_L\hat{C}^{(21)*}_{u\Phi_1,f_2f_1}\\
&\qquad-c_\beta^2\mathcal{P}_R\hat{C}^{(12)}_{u\Phi_1,f_1f_2}+s_\beta^2\mathcal{P}_R\hat{C}^{(21)}_{u\Phi_1,f_1f_2}\Big)
\end{aligned}
\tag{C.307}
$$

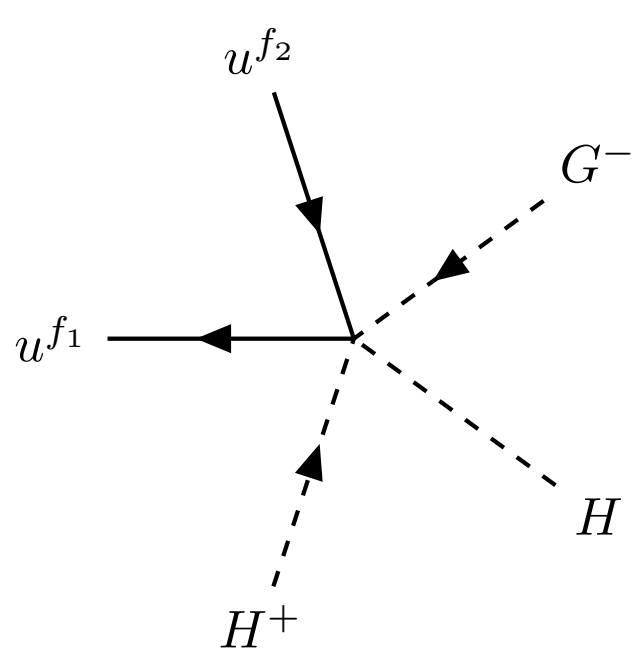

$$
\begin{aligned}
&+\frac{is_\beta}{\sqrt{2}}\Big(c_\beta^2\left(\mathcal{P}_L\hat{C}^{(11)*}_{u\Phi_2,f_2f_1}+\mathcal{P}_R\hat{C}^{(11)}_{u\Phi_2,f_1f_2}\right)\\
&\qquad-c_\beta^2\left(\mathcal{P}_L\hat{C}^{(22)*}_{u\Phi_2,f_2f_1}+\mathcal{P}_R\hat{C}^{(22)}_{u\Phi_2,f_1f_2}\right)\\
&\qquad-s_\beta^2\mathcal{P}_L\hat{C}^{(12)*}_{u\Phi_1,f_2f_1}+c_\beta^2\mathcal{P}_L\hat{C}^{(21)*}_{u\Phi_1,f_2f_1}\\
&\qquad+c_\beta^2\mathcal{P}_R\hat{C}^{(12)}_{u\Phi_1,f_1f_2}-s_\beta^2\mathcal{P}_R\hat{C}^{(21)}_{u\Phi_1,f_1f_2}\Big)
\end{aligned}
\tag{C.308}
$$

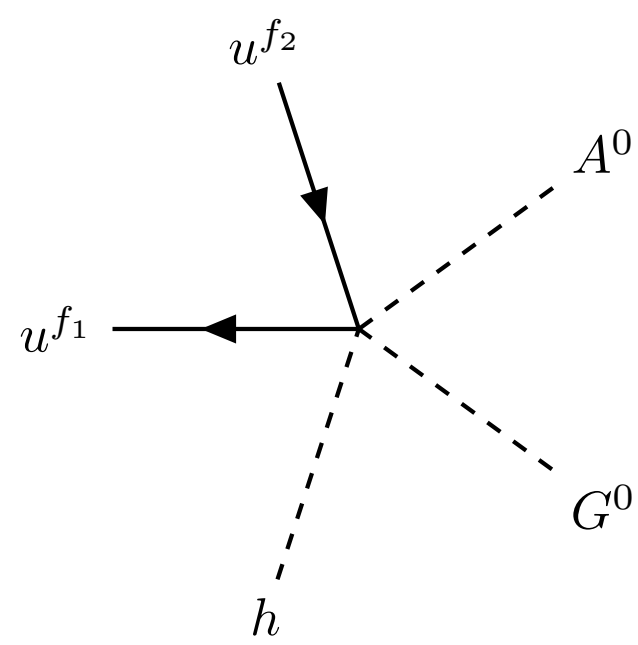

$$
\begin{aligned}
&+\frac{ic_\beta}{\sqrt{2}}\Big(-s_\beta^2\left(\mathcal{P}_L\hat{C}^{(11)*}_{u\Phi_2,f_2f_1}+\mathcal{P}_R\hat{C}^{(11)}_{u\Phi_2,f_1f_2}\right)\\
&\qquad+c_\beta^2\left(\mathcal{P}_L\hat{C}^{(12)*}_{u\Phi_1,f_2f_1}+\mathcal{P}_R\hat{C}^{(12)}_{u\Phi_1,f_1f_2}\right)\\
&\qquad-s_\beta^2\left(\mathcal{P}_L\hat{C}^{(21)*}_{u\Phi_1,f_2f_1}+\mathcal{P}_R\hat{C}^{(21)}_{u\Phi_1,f_1f_2}\right)\\
&\qquad+s_\beta^2\left(\mathcal{P}_L\hat{C}^{(22)*}_{u\Phi_2,f_2f_1}+\mathcal{P}_R\hat{C}^{(22)}_{u\Phi_2,f_1f_2}\right)\Big)
\end{aligned}
\tag{C.309}
$$

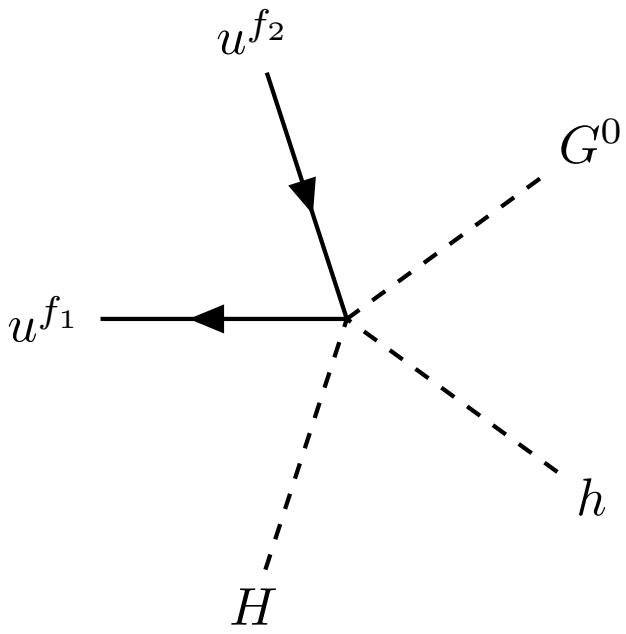

$$
\begin{aligned}
&+\frac{c_\beta}{\sqrt{2}}\left(-s_\beta^2\left(\mathcal{P}_L\hat{C}^{(11)*}_{u\Phi_2,f_2f_1}-\mathcal{P}_R\hat{C}^{(11)}_{u\Phi_2,f_1f_2}\right)\right.\\
&\qquad+c_\beta^2\left(\mathcal{P}_L\hat{C}^{(12)*}_{u\Phi_1,f_2f_1}-\mathcal{P}_R\hat{C}^{(12)}_{u\Phi_1,f_1f_2}\right)\\
&\qquad-s_\beta^2\left(\mathcal{P}_L\hat{C}^{(21)*}_{u\Phi_1,f_2f_1}-\mathcal{P}_R\hat{C}^{(21)}_{u\Phi_1,f_1f_2}\right)\\
&\qquad\left.+s_\beta^2\left(\mathcal{P}_L\hat{C}^{(22)*}_{u\Phi_2,f_2f_1}-\mathcal{P}_R\hat{C}^{(22)}_{u\Phi_2,f_1f_2}\right)\right)
\end{aligned}
\tag{C.310}
$$

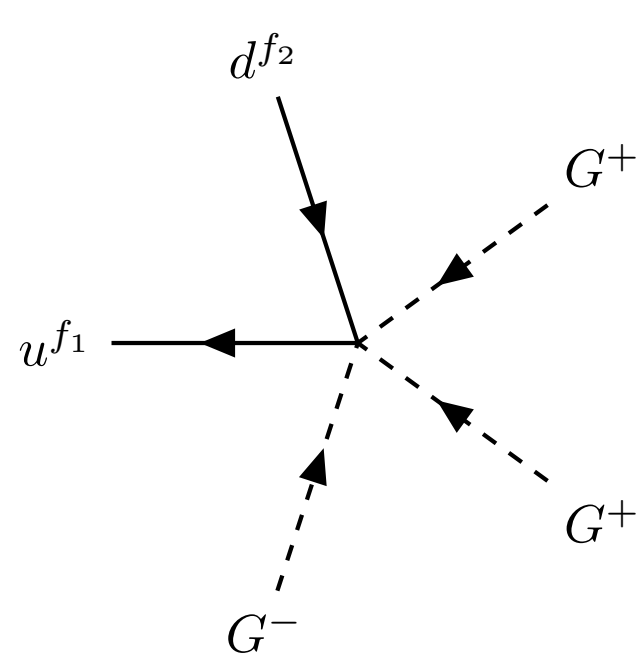

$$
\begin{aligned}
&-2i\left(s_\beta c_\beta^2\mathcal{P}_L V_{g_1f_2}\hat{C}^{(11)*}_{u\Phi_2,g_1f_1}+s_\beta c_\beta^2\mathcal{P}_L V_{g_1f_2}\hat{C}^{(12)*}_{u\Phi_1,g_1f_1}\right.\\
&\qquad+s_\beta c_\beta^2\mathcal{P}_L V_{g_1f_2}\hat{C}^{(21)*}_{u\Phi_1,g_1f_1}+s_\beta^3\mathcal{P}_L V_{g_1f_2}\hat{C}^{(22)*}_{u\Phi_2,g_1f_1}\\
&\qquad-c_\beta^3\mathcal{P}_R V_{f_1g_1}\hat{C}^{(11)}_{d\Phi_1,g_1f_2}-s_\beta^2c_\beta\mathcal{P}_R V_{f_1g_1}\hat{C}^{(12)}_{d\Phi_2,g_1f_2}\\
&\qquad\left.-s_\beta^2c_\beta\mathcal{P}_R V_{f_1g_1}\hat{C}^{(21)}_{d\Phi_2,g_1f_2}-s_\beta^2c_\beta\mathcal{P}_R V_{f_1g_1}\hat{C}^{(22)}_{d\Phi_1,g_1f_2}\right)
\end{aligned}
\tag{C.311}
$$

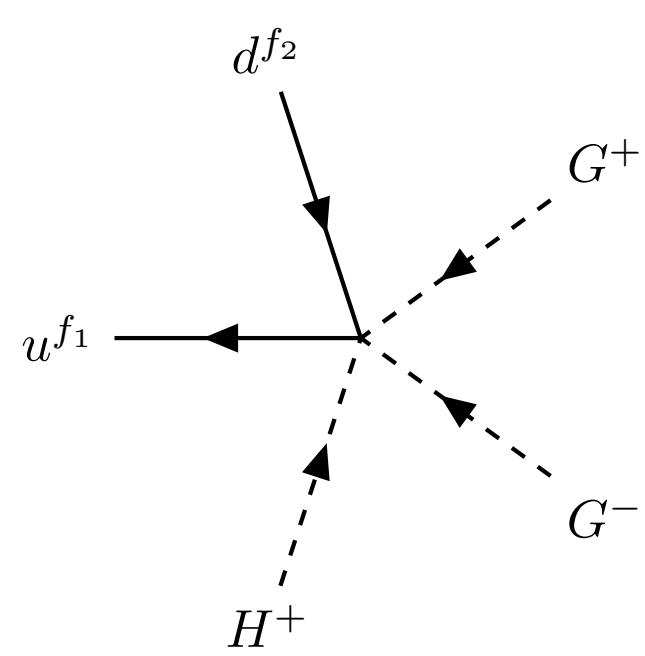

$$
\begin{aligned}
&-i\left(c_\beta^3\mathcal{P}_L V_{g_1f_2}\hat{C}^{(11)*}_{u\Phi_2,g_1f_1}-s_\beta^2c_\beta\mathcal{P}_L V_{g_1f_2}\hat{C}^{(11)*}_{u\Phi_2,g_1f_1}\right.\\
&\qquad-2s_\beta^2c_\beta\mathcal{P}_L V_{g_1f_2}\hat{C}^{(12)*}_{u\Phi_1,g_1f_1}+c_\beta^3\mathcal{P}_L V_{g_1f_2}\hat{C}^{(21)*}_{u\Phi_1,g_1f_1}\\
&\qquad-s_\beta^2c_\beta\mathcal{P}_L V_{g_1f_2}\hat{C}^{(21)*}_{u\Phi_1,g_1f_1}+2s_\beta^2c_\beta\mathcal{P}_L V_{g_1f_2}\hat{C}^{(22)*}_{u\Phi_2,g_1f_1}\\
&\qquad+2s_\beta c_\beta^2\mathcal{P}_R V_{f_1g_1}\hat{C}^{(11)}_{d\Phi_1,g_1f_2}-2s_\beta c_\beta^2\mathcal{P}_R V_{f_1g_1}\hat{C}^{(12)}_{d\Phi_2,g_1f_2}\\
&\qquad+s_\beta^3\mathcal{P}_R V_{f_1g_1}\hat{C}^{(21)}_{d\Phi_2,g_1f_2}-s_\beta c_\beta^2\mathcal{P}_R V_{f_1g_1}\hat{C}^{(21)}_{d\Phi_2,g_1f_2}\\
&\qquad\left.+s_\beta^3\mathcal{P}_R V_{f_1g_1}\hat{C}^{(22)}_{d\Phi_1,g_1f_2}-s_\beta c_\beta^2\mathcal{P}_R V_{f_1g_1}\hat{C}^{(22)}_{d\Phi_1,g_1f_2}\right)
\end{aligned}
\tag{C.312}
$$

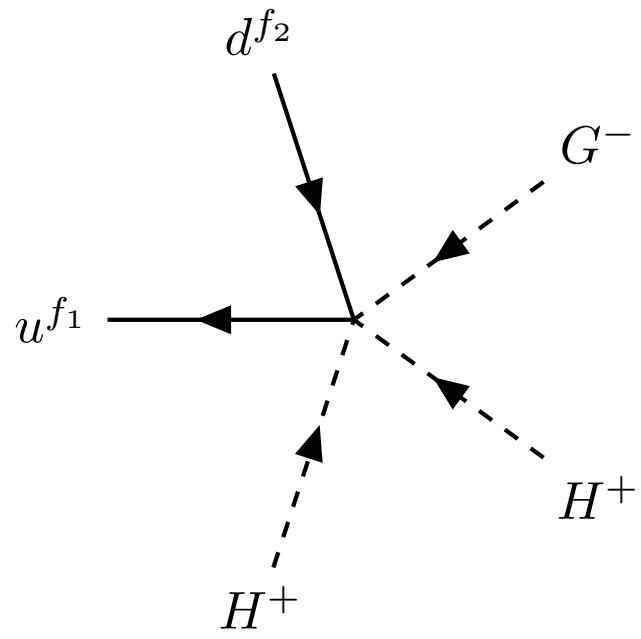

$$
\begin{aligned}
&+2i \Big( s_\beta c_\beta^2 \mathcal{P}_L V_{g_1 f_2} \hat{C}^{(11)*}_{u\Phi_2, g_1 f_1} - s_\beta^3 \mathcal{P}_L V_{g_1 f_2} \hat{C}^{(12)*}_{u\Phi_1, g_1 f_1} \\
&\quad + s_\beta c_\beta^2 \mathcal{P}_L V_{g_1 f_2} \hat{C}^{(21)*}_{u\Phi_1, g_1 f_1} - s_\beta c_\beta^2 \mathcal{P}_L V_{g_1 f_2} \hat{C}^{(22)*}_{u\Phi_2, g_1 f_1} \\
&\quad + s_\beta^2 c_\beta \mathcal{P}_R V_{f_1 g_1} \hat{C}^{(11)}_{d\Phi_1, g_1 f_2} + c_\beta^3 \mathcal{P}_R V_{f_1 g_1} \hat{C}^{(12)}_{d\Phi_2, g_1 f_2} \\
&\quad - s_\beta^2 c_\beta \mathcal{P}_R V_{f_1 g_1} \hat{C}^{(21)}_{d\Phi_2, g_1 f_2} - s_\beta^2 c_\beta \mathcal{P}_R V_{f_1 g_1} \hat{C}^{(22)}_{d\Phi_1, g_1 f_2} \Big)
\end{aligned}
\tag{C.313}
$$

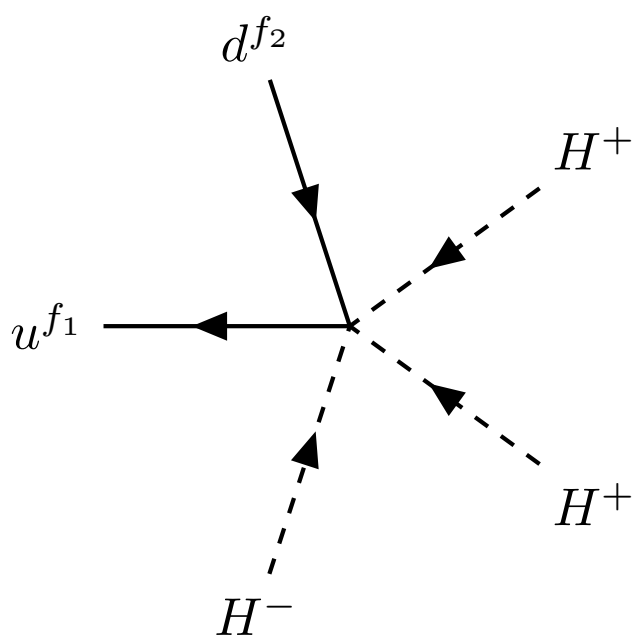

$$
\begin{aligned}
&-2i \Big( s_\beta^2 c_\beta \mathcal{P}_L V_{g_1 f_2} \hat{C}^{(11)*}_{u\Phi_2, g_1 f_1} + s_\beta^2 c_\beta \mathcal{P}_L V_{g_1 f_2} \hat{C}^{(12)*}_{u\Phi_1, g_1 f_1} \\
&\quad + s_\beta^2 c_\beta \mathcal{P}_L V_{g_1 f_2} \hat{C}^{(21)*}_{u\Phi_1, g_1 f_1} + c_\beta^3 \mathcal{P}_L V_{g_1 f_2} \hat{C}^{(22)*}_{u\Phi_2, g_1 f_1} \\
&\quad + s_\beta^3 \mathcal{P}_R V_{f_1 g_1} \hat{C}^{(11)}_{d\Phi_1, g_1 f_2} + s_\beta c_\beta^2 \mathcal{P}_R V_{f_1 g_1} \hat{C}^{(12)}_{d\Phi_2, g_1 f_2} \\
&\quad + s_\beta c_\beta^2 \mathcal{P}_R V_{f_1 g_1} \hat{C}^{(21)}_{d\Phi_2, g_1 f_2} + s_\beta c_\beta^2 \mathcal{P}_R V_{f_1 g_1} \hat{C}^{(22)}_{d\Phi_1, g_1 f_2} \Big)
\end{aligned}
\tag{C.314}
$$

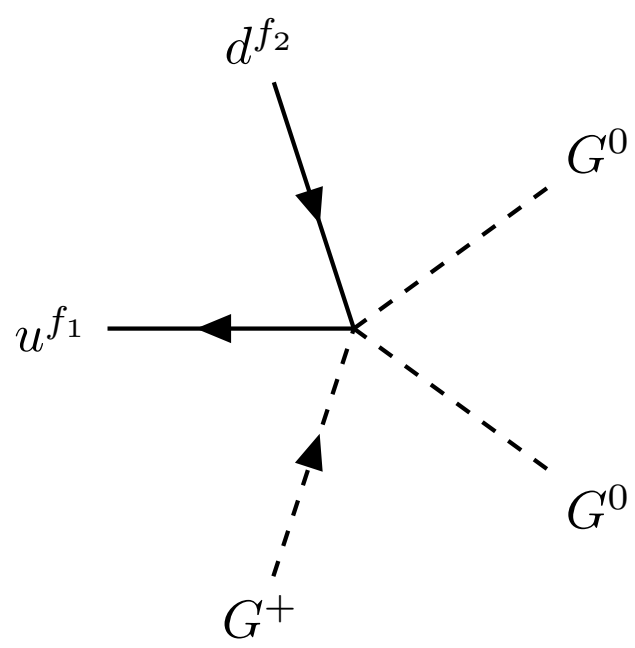

$$
\begin{aligned}
&-i \Big( s_\beta c_\beta^2 \mathcal{P}_L V_{g_1 f_2} \hat{C}^{(11)*}_{u\Phi_2, g_1 f_1} + s_\beta c_\beta^2 \mathcal{P}_L V_{g_1 f_2} \hat{C}^{(12)*}_{u\Phi_1, g_1 f_1} \\
&\quad + s_\beta c_\beta^2 \mathcal{P}_L V_{g_1 f_2} \hat{C}^{(21)*}_{u\Phi_1, g_1 f_1} + s_\beta^3 \mathcal{P}_L V_{g_1 f_2} \hat{C}^{(22)*}_{u\Phi_2, g_1 f_1} \\
&\quad - c_\beta^3 \mathcal{P}_R V_{f_1 g_1} \hat{C}^{(11)}_{d\Phi_1, g_1 f_2} - s_\beta^2 c_\beta \mathcal{P}_R V_{f_1 g_1} \hat{C}^{(12)}_{d\Phi_2, g_1 f_2} \\
&\quad - s_\beta^2 c_\beta \mathcal{P}_R V_{f_1 g_1} \hat{C}^{(21)}_{d\Phi_2, g_1 f_2} - s_\beta^2 c_\beta \mathcal{P}_R V_{f_1 g_1} \hat{C}^{(22)}_{d\Phi_1, g_1 f_2} \Big)
\end{aligned}
\tag{C.315}
$$

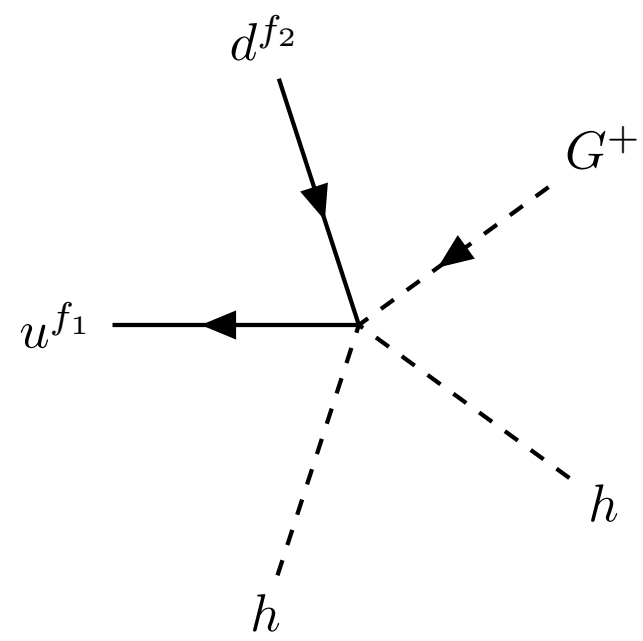

$$
\begin{aligned}
&-i \Big( s_\beta c_\beta^2 \mathcal{P}_L V_{g_1 f_2} \hat{C}^{(11)*}_{u\Phi_2, g_1 f_1} + s_\beta c_\beta^2 \mathcal{P}_L V_{g_1 f_2} \hat{C}^{(12)*}_{u\Phi_1, g_1 f_1} \\
&\quad + s_\beta c_\beta^2 \mathcal{P}_L V_{g_1 f_2} \hat{C}^{(21)*}_{u\Phi_1, g_1 f_1} + s_\beta^3 \mathcal{P}_L V_{g_1 f_2} \hat{C}^{(22)*}_{u\Phi_2, g_1 f_1} \\
&\quad - c_\beta^3 \mathcal{P}_R V_{f_1 g_1} \hat{C}^{(11)}_{d\Phi_1, g_1 f_2} - s_\beta^2 c_\beta \mathcal{P}_R V_{f_1 g_1} \hat{C}^{(12)}_{d\Phi_2, g_1 f_2} \\
&\quad - s_\beta^2 c_\beta \mathcal{P}_R V_{f_1 g_1} \hat{C}^{(21)}_{d\Phi_2, g_1 f_2} - s_\beta^2 c_\beta \mathcal{P}_R V_{f_1 g_1} \hat{C}^{(22)}_{d\Phi_1, g_1 f_2} \Big)
\end{aligned}
\tag{C.316}
$$

$$
\begin{aligned}
&-\frac{1}{2}i\Big(-2s_\beta^2 c_\beta \mathcal{P}_L V_{g_1 f_2}\hat{C}^{(11)*}_{u\Phi_2,g_1 f_1}+c_\beta^3 \mathcal{P}_L V_{g_1 f_2}\hat{C}^{(12)*}_{u\Phi_1,g_1 f_1}\\
&\quad -s_\beta^2 c_\beta \mathcal{P}_L V_{g_1 f_2}\hat{C}^{(12)*}_{u\Phi_1,g_1 f_1}+c_\beta^3 \mathcal{P}_L V_{g_1 f_2}\hat{C}^{(21)*}_{u\Phi_1,g_1 f_1}\\
&\quad -s_\beta^2 c_\beta \mathcal{P}_L V_{g_1 f_2}\hat{C}^{(21)*}_{u\Phi_1,g_1 f_1}+2s_\beta^2 c_\beta \mathcal{P}_L V_{g_1 f_2}\hat{C}^{(22)*}_{u\Phi_2,g_1 f_1}\\
&\quad +2s_\beta c_\beta^2 \mathcal{P}_R V_{f_1 g_1}\hat{C}^{(11)}_{d\Phi_1,g_1 f_2}+s_\beta^3 \mathcal{P}_R V_{f_1 g_1}\hat{C}^{(12)}_{d\Phi_2,g_1 f_2}\\
&\quad -s_\beta c_\beta^2 \mathcal{P}_R V_{f_1 g_1}\hat{C}^{(12)}_{d\Phi_2,g_1 f_2}+s_\beta^3 \mathcal{P}_R V_{f_1 g_1}\hat{C}^{(21)}_{d\Phi_2,g_1 f_2}\\
&\quad -s_\beta c_\beta^2 \mathcal{P}_R V_{f_1 g_1}\hat{C}^{(21)}_{d\Phi_2,g_1 f_2}-2s_\beta c_\beta^2 \mathcal{P}_R V_{f_1 g_1}\hat{C}^{(22)}_{d\Phi_1,g_1 f_2}\Big)
\end{aligned}
\tag{C.317}
$$

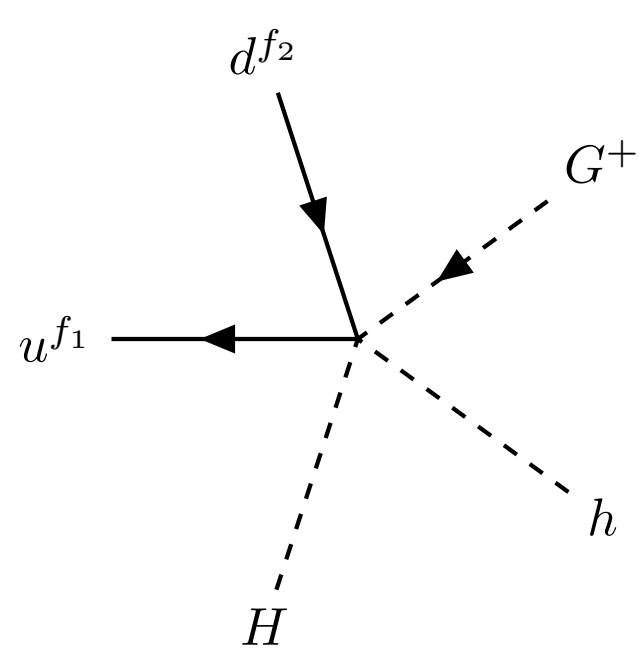

$$
\begin{aligned}
&+\frac{1}{2}i\Big(-2s_\beta^2 c_\beta \mathcal{P}_L V_{g_1 f_2}\hat{C}^{(11)*}_{u\Phi_2,g_1 f_1}+c_\beta^3 \mathcal{P}_L V_{g_1 f_2}\hat{C}^{(12)*}_{u\Phi_1,g_1 f_1}\\
&\quad -s_\beta^2 c_\beta \mathcal{P}_L V_{g_1 f_2}\hat{C}^{(12)*}_{u\Phi_1,g_1 f_1}+c_\beta^3 \mathcal{P}_L V_{g_1 f_2}\hat{C}^{(21)*}_{u\Phi_1,g_1 f_1}\\
&\quad -s_\beta^2 c_\beta \mathcal{P}_L V_{g_1 f_2}\hat{C}^{(21)*}_{u\Phi_1,g_1 f_1}+2s_\beta^2 c_\beta \mathcal{P}_L V_{g_1 f_2}\hat{C}^{(22)*}_{u\Phi_2,g_1 f_1}\\
&\quad +2s_\beta c_\beta^2 \mathcal{P}_R V_{f_1 g_1}\hat{C}^{(11)}_{d\Phi_1,g_1 f_2}+s_\beta^3 \mathcal{P}_R V_{f_1 g_1}\hat{C}^{(12)}_{d\Phi_2,g_1 f_2}\\
&\quad -s_\beta c_\beta^2 \mathcal{P}_R V_{f_1 g_1}\hat{C}^{(12)}_{d\Phi_2,g_1 f_2}+s_\beta^3 \mathcal{P}_R V_{f_1 g_1}\hat{C}^{(21)}_{d\Phi_2,g_1 f_2}\\
&\quad -s_\beta c_\beta^2 \mathcal{P}_R V_{f_1 g_1}\hat{C}^{(21)}_{d\Phi_2,g_1 f_2}-2s_\beta c_\beta^2 \mathcal{P}_R V_{f_1 g_1}\hat{C}^{(22)}_{d\Phi_1,g_1 f_2}\Big)
\end{aligned}
\tag{C.318}
$$

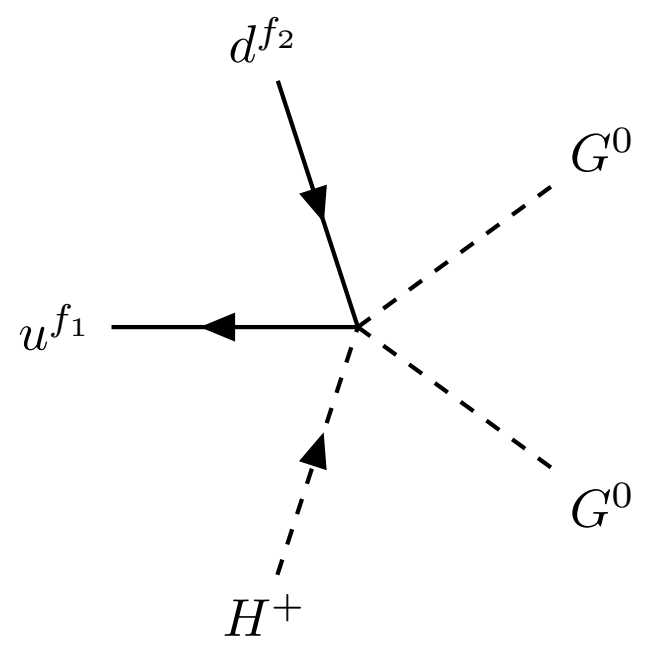

$$
\begin{aligned}
&-i\Big(c_\beta^3 \mathcal{P}_L V_{g_1 f_2}\hat{C}^{(11)*}_{u\Phi_2,g_1 f_1}-s_\beta^2 c_\beta \mathcal{P}_L V_{g_1 f_2}\hat{C}^{(12)*}_{u\Phi_1,g_1 f_1}\\
&\quad -s_\beta^2 c_\beta \mathcal{P}_L V_{g_1 f_2}\hat{C}^{(21)*}_{u\Phi_1,g_1 f_1}+s_\beta^2 c_\beta \mathcal{P}_L V_{g_1 f_2}\hat{C}^{(22)*}_{u\Phi_2,g_1 f_1}\\
&\quad +s_\beta c_\beta^2 \mathcal{P}_R V_{f_1 g_1}\hat{C}^{(11)}_{d\Phi_1,g_1 f_2}-s_\beta c_\beta^2 \mathcal{P}_R V_{f_1 g_1}\hat{C}^{(12)}_{d\Phi_2,g_1 f_2}\\
&\quad -s_\beta c_\beta^2 \mathcal{P}_R V_{f_1 g_1}\hat{C}^{(21)}_{d\Phi_2,g_1 f_2}+s_\beta^3 \mathcal{P}_R V_{f_1 g_1}\hat{C}^{(22)}_{d\Phi_1,g_1 f_2}\Big)
\end{aligned}
\tag{C.319}
$$

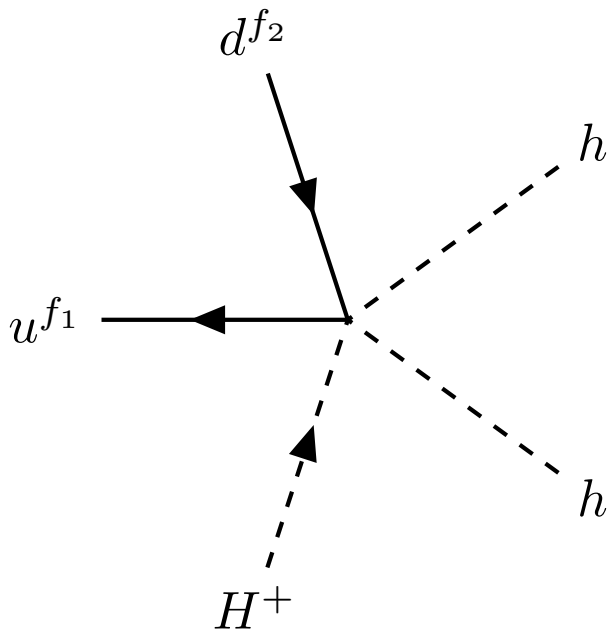

$$\begin{aligned}
&-i\Big(c_\beta^3\mathcal{P}_L V_{g_1f_2}\hat{C}^{(11)*}_{u\Phi_2,g_1f_1} - s_\beta^2 c_\beta\mathcal{P}_L V_{g_1f_2}\hat{C}^{(12)*}_{u\Phi_1,g_1f_1}\\
&\quad -s_\beta^2 c_\beta\mathcal{P}_L V_{g_1f_2}\hat{C}^{(21)*}_{u\Phi_1,g_1f_1} + s_\beta^2 c_\beta\mathcal{P}_L V_{g_1f_2}\hat{C}^{(22)*}_{u\Phi_2,g_1f_1}\\
&\quad +s_\beta c_\beta^2\mathcal{P}_R V_{f_1g_1}\hat{C}^{(11)}_{d\Phi_1,g_1f_2} - s_\beta c_\beta^2\mathcal{P}_R V_{f_1g_1}\hat{C}^{(12)}_{d\Phi_2,g_1f_2}\\
&\quad -s_\beta c_\beta^2\mathcal{P}_R V_{f_1g_1}\hat{C}^{(21)}_{d\Phi_2,g_1f_2} + s_\beta^3\mathcal{P}_R V_{f_1g_1}\hat{C}^{(22)}_{d\Phi_1,g_1f_2}\Big)
\end{aligned}\tag{C.320}$$

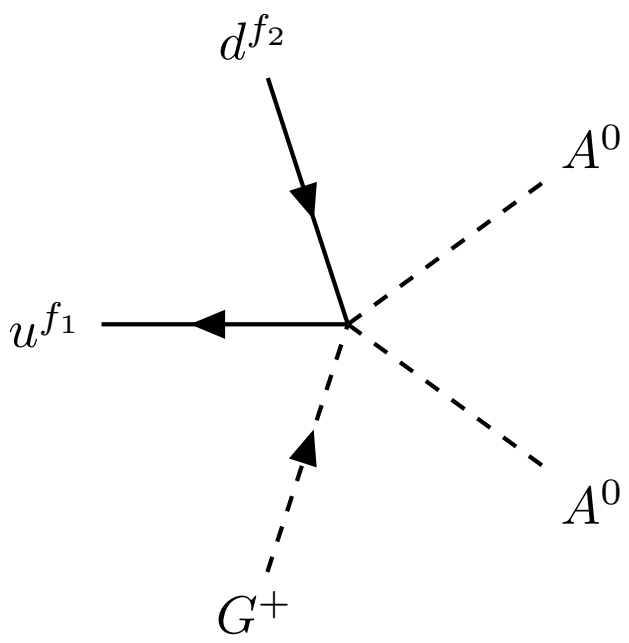

$$\begin{aligned}
&+i\Big(-s_\beta^3\mathcal{P}_L V_{g_1f_2}\hat{C}^{(11)*}_{u\Phi_2,g_1f_1} + s_\beta c_\beta^2\mathcal{P}_L V_{g_1f_2}\hat{C}^{(12)*}_{u\Phi_1,g_1f_1}\\
&\quad +s_\beta c_\beta^2\mathcal{P}_L V_{g_1f_2}\hat{C}^{(21)*}_{u\Phi_1,g_1f_1} - s_\beta c_\beta^2\mathcal{P}_L V_{g_1f_2}\hat{C}^{(22)*}_{u\Phi_2,g_1f_1}\\
&\quad +s_\beta^2 c_\beta\mathcal{P}_R V_{f_1g_1}\hat{C}^{(11)}_{d\Phi_1,g_1f_2} - s_\beta^2 c_\beta\mathcal{P}_R V_{f_1g_1}\hat{C}^{(12)}_{d\Phi_2,g_1f_2}\\
&\quad -s_\beta^2 c_\beta\mathcal{P}_R V_{f_1g_1}\hat{C}^{(21)}_{d\Phi_2,g_1f_2} + c_\beta^3\mathcal{P}_R V_{f_1g_1}\hat{C}^{(22)}_{d\Phi_1,g_1f_2}\Big)
\end{aligned}\tag{C.321}$$

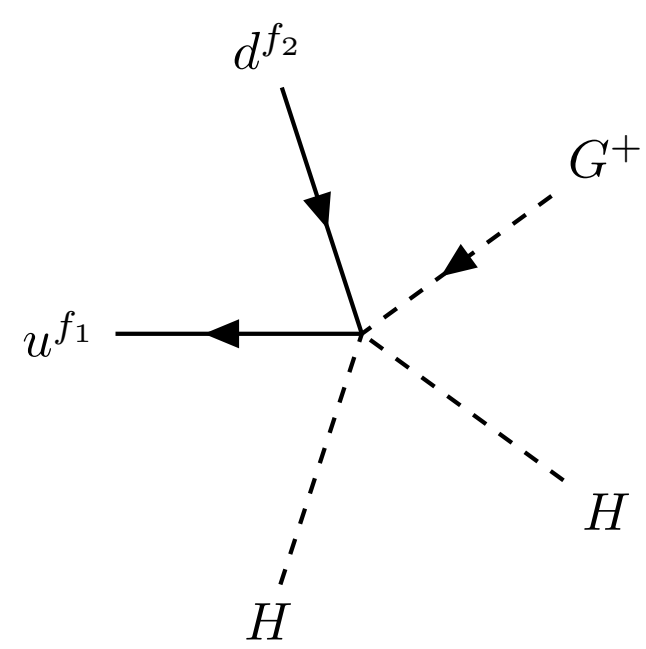

$$\begin{aligned}
&+i\Big(-s_\beta^3\mathcal{P}_L V_{g_1f_2}\hat{C}^{(11)*}_{u\Phi_2,g_1f_1} + s_\beta c_\beta^2\mathcal{P}_L V_{g_1f_2}\hat{C}^{(12)*}_{u\Phi_1,g_1f_1}\\
&\quad +s_\beta c_\beta^2\mathcal{P}_L V_{g_1f_2}\hat{C}^{(21)*}_{u\Phi_1,g_1f_1} - s_\beta c_\beta^2\mathcal{P}_L V_{g_1f_2}\hat{C}^{(22)*}_{u\Phi_2,g_1f_1}\\
&\quad +s_\beta^2 c_\beta\mathcal{P}_R V_{f_1g_1}\hat{C}^{(11)}_{d\Phi_1,g_1f_2} - s_\beta^2 c_\beta\mathcal{P}_R V_{f_1g_1}\hat{C}^{(12)}_{d\Phi_2,g_1f_2}\\
&\quad -s_\beta^2 c_\beta\mathcal{P}_R V_{f_1g_1}\hat{C}^{(21)}_{d\Phi_2,g_1f_2} + c_\beta^3\mathcal{P}_R V_{f_1g_1}\hat{C}^{(22)}_{d\Phi_1,g_1f_2}\Big)
\end{aligned}\tag{C.322}$$

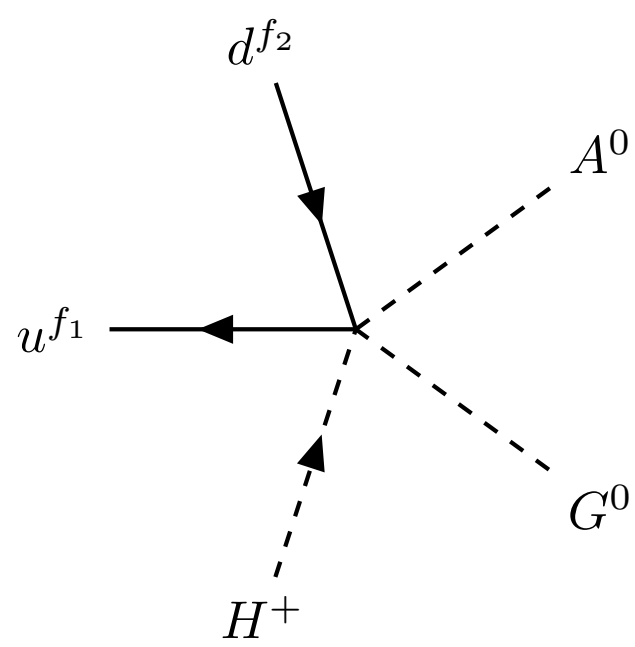

$$\begin{aligned}
&+\frac{1}{2}i\Big(2s_\beta c_\beta^2\mathcal{P}_L V_{g_1f_2}\hat{C}^{(11)*}_{u\Phi_2,g_1f_1} - s_\beta^3\mathcal{P}_L V_{g_1f_2}\hat{C}^{(12)*}_{u\Phi_1,g_1f_1}\\
&\quad +s_\beta c_\beta^2\mathcal{P}_L V_{g_1f_2}\hat{C}^{(12)*}_{u\Phi_1,g_1f_1} - s_\beta^3\mathcal{P}_L V_{g_1f_2}\hat{C}^{(21)*}_{u\Phi_1,g_1f_1}\\
&\quad +s_\beta c_\beta^2\mathcal{P}_L V_{g_1f_2}\hat{C}^{(21)*}_{u\Phi_1,g_1f_1} - 2s_\beta c_\beta^2\mathcal{P}_L V_{g_1f_2}\hat{C}^{(22)*}_{u\Phi_2,g_1f_1}\\
&\quad +2s_\beta^2 c_\beta\mathcal{P}_R V_{f_1g_1}\hat{C}^{(11)}_{d\Phi_1,g_1f_2} + c_\beta^3\mathcal{P}_R V_{f_1g_1}\hat{C}^{(12)}_{d\Phi_2,g_1f_2}\\
&\quad -s_\beta^2 c_\beta\mathcal{P}_R V_{f_1g_1}\hat{C}^{(12)}_{d\Phi_2,g_1f_2} + c_\beta^3\mathcal{P}_R V_{f_1g_1}\hat{C}^{(21)}_{d\Phi_2,g_1f_2}\\
&\quad -s_\beta^2 c_\beta\mathcal{P}_R V_{f_1g_1}\hat{C}^{(21)}_{d\Phi_2,g_1f_2} - 2s_\beta^2 c_\beta\mathcal{P}_R V_{f_1g_1}\hat{C}^{(22)}_{d\Phi_1,g_1f_2}\Big)
\end{aligned}\tag{C.323}$$

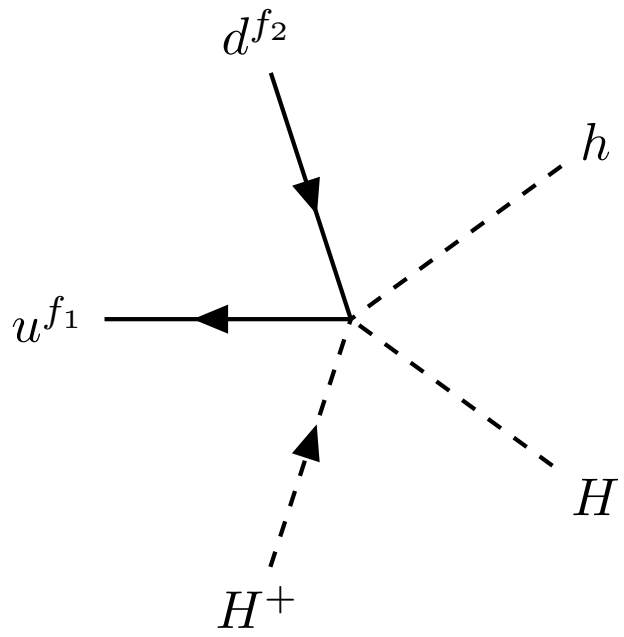

$$
\begin{aligned}
&+\frac{1}{2}i\Big(-2s_\beta c_\beta^2\mathcal{P}_L V_{g_1f_2}\hat{C}^{(11)*}_{u\Phi_2,g_1f_1}+s_\beta^3\mathcal{P}_L V_{g_1f_2}\hat{C}^{(12)*}_{u\Phi_1,g_1f_1}\\
&\quad -s_\beta c_\beta^2\mathcal{P}_L V_{g_1f_2}\hat{C}^{(12)*}_{u\Phi_1,g_1f_1}+s_\beta^3\mathcal{P}_L V_{g_1f_2}\hat{C}^{(21)*}_{u\Phi_1,g_1f_1}\\
&\quad -s_\beta c_\beta^2\mathcal{P}_L V_{g_1f_2}\hat{C}^{(21)*}_{u\Phi_1,g_1f_1}+2s_\beta c_\beta^2\mathcal{P}_L V_{g_1f_2}\hat{C}^{(22)*}_{u\Phi_2,g_1f_1}\\
&\quad -2s_\beta^2 c_\beta\mathcal{P}_R V_{f_1g_1}\hat{C}^{(11)}_{d\Phi_1,g_1f_2}-c_\beta^3\mathcal{P}_R V_{f_1g_1}\hat{C}^{(12)}_{d\Phi_2,g_1f_2}\\
&\quad +s_\beta^2 c_\beta\mathcal{P}_R V_{f_1g_1}\hat{C}^{(12)}_{d\Phi_2,g_1f_2}-c_\beta^3\mathcal{P}_R V_{f_1g_1}\hat{C}^{(21)}_{d\Phi_2,g_1f_2}\\
&\quad +s_\beta^2 c_\beta\mathcal{P}_R V_{f_1g_1}\hat{C}^{(21)}_{d\Phi_2,g_1f_2}+2s_\beta^2 c_\beta\mathcal{P}_R V_{f_1g_1}\hat{C}^{(22)}_{d\Phi_1,g_1f_2}\Big)
\end{aligned}
\tag{C.324}
$$

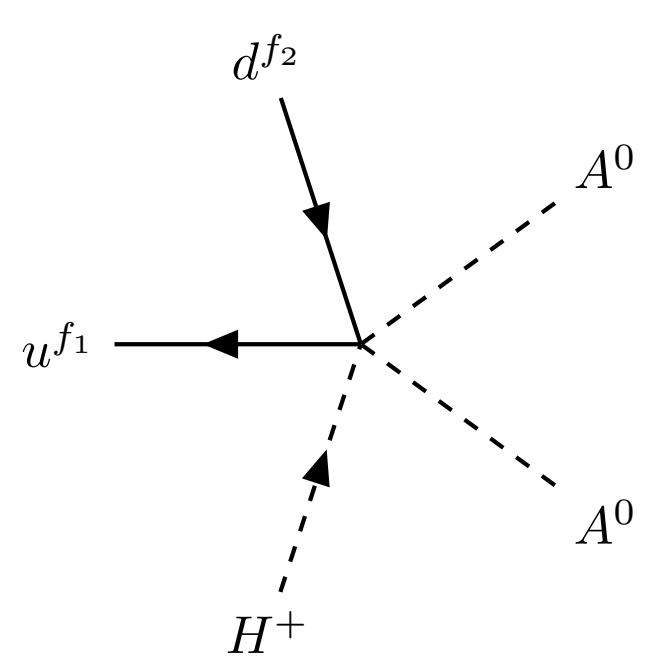

$$
\begin{aligned}
&-i\Big(s_\beta^2 c_\beta\mathcal{P}_L V_{g_1f_2}\hat{C}^{(11)*}_{u\Phi_2,g_1f_1}+s_\beta^2 c_\beta\mathcal{P}_L V_{g_1f_2}\hat{C}^{(12)*}_{u\Phi_1,g_1f_1}\\
&\quad +s_\beta^2 c_\beta\mathcal{P}_L V_{g_1f_2}\hat{C}^{(21)*}_{u\Phi_1,g_1f_1}+c_\beta^3\mathcal{P}_L V_{g_1f_2}\hat{C}^{(22)*}_{u\Phi_2,g_1f_1}\\
&\quad +s_\beta^3\mathcal{P}_R V_{f_1g_1}\hat{C}^{(11)}_{d\Phi_1,g_1f_2}+s_\beta c_\beta^2\mathcal{P}_R V_{f_1g_1}\hat{C}^{(12)}_{d\Phi_2,g_1f_2}\\
&\quad +s_\beta c_\beta^2\mathcal{P}_R V_{f_1g_1}\hat{C}^{(21)}_{d\Phi_2,g_1f_2}+s_\beta c_\beta^2\mathcal{P}_R V_{f_1g_1}\hat{C}^{(22)}_{d\Phi_1,g_1f_2}\Big)
\end{aligned}
\tag{C.325}
$$

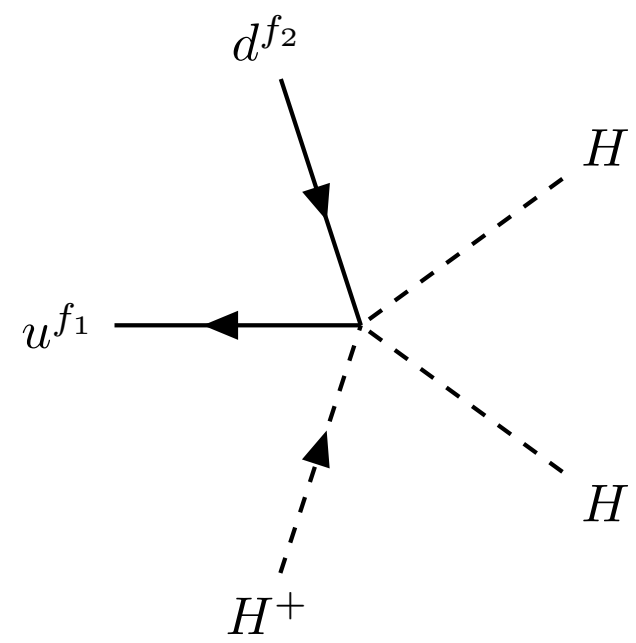

$$
\begin{aligned}
&-i\Big(s_\beta^2 c_\beta\mathcal{P}_L V_{g_1f_2}\hat{C}^{(11)*}_{u\Phi_2,g_1f_1}+s_\beta^2 c_\beta\mathcal{P}_L V_{g_1f_2}\hat{C}^{(12)*}_{u\Phi_1,g_1f_1}\\
&\quad +s_\beta^2 c_\beta\mathcal{P}_L V_{g_1f_2}\hat{C}^{(21)*}_{u\Phi_1,g_1f_1}+c_\beta^3\mathcal{P}_L V_{g_1f_2}\hat{C}^{(22)*}_{u\Phi_2,g_1f_1}\\
&\quad +s_\beta^3\mathcal{P}_R V_{f_1g_1}\hat{C}^{(11)}_{d\Phi_1,g_1f_2}+s_\beta c_\beta^2\mathcal{P}_R V_{f_1g_1}\hat{C}^{(12)}_{d\Phi_2,g_1f_2}\\
&\quad +s_\beta c_\beta^2\mathcal{P}_R V_{f_1g_1}\hat{C}^{(21)}_{d\Phi_2,g_1f_2}+s_\beta c_\beta^2\mathcal{P}_R V_{f_1g_1}\hat{C}^{(22)}_{d\Phi_1,g_1f_2}\Big)
\end{aligned}
\tag{C.326}
$$

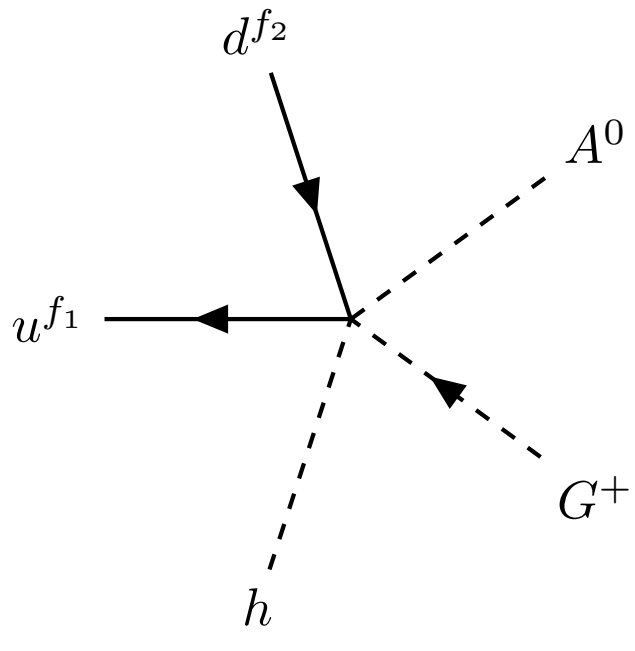

$$
\begin{aligned}
-\frac{1}{2}\Big(&c_\beta^3\mathcal{P}_L V_{g_1f_2}\hat{C}^{(12)*}_{u\Phi_1,g_1f_1} + s_\beta^2 c_\beta\mathcal{P}_L V_{g_1f_2}\hat{C}^{(12)*}_{u\Phi_1,g_1f_1}\\
&-c_\beta^3\mathcal{P}_L V_{g_1f_2}\hat{C}^{(21)*}_{u\Phi_1,g_1f_1} - s_\beta^2 c_\beta\mathcal{P}_L V_{g_1f_2}\hat{C}^{(21)*}_{u\Phi_1,g_1f_1}\\
&+s_\beta^3\mathcal{P}_R V_{f_1g_1}\hat{C}^{(12)}_{d\Phi_2,g_1f_2} + s_\beta c_\beta^2\mathcal{P}_R V_{f_1g_1}\hat{C}^{(12)}_{d\Phi_2,g_1f_2}\\
&-s_\beta^3\mathcal{P}_R V_{f_1g_1}\hat{C}^{(21)}_{d\Phi_2,g_1f_2} - s_\beta c_\beta^2\mathcal{P}_R V_{f_1g_1}\hat{C}^{(21)}_{d\Phi_2,g_1f_2}\Big)
\end{aligned}
\tag{C.327}
$$

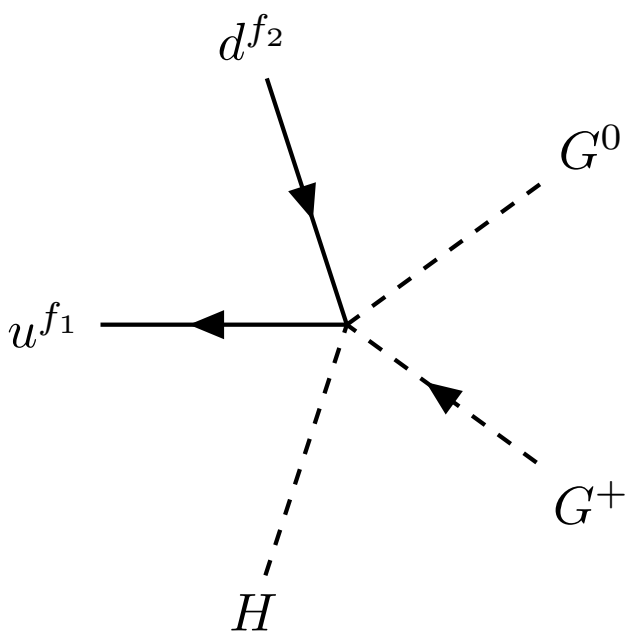

$$
\begin{aligned}
-\frac{1}{2}\Big(&c_\beta^3\mathcal{P}_L V_{g_1f_2}\hat{C}^{(12)*}_{u\Phi_1,g_1f_1} + s_\beta^2 c_\beta\mathcal{P}_L V_{g_1f_2}\hat{C}^{(12)*}_{u\Phi_1,g_1f_1}\\
&-c_\beta^3\mathcal{P}_L V_{g_1f_2}\hat{C}^{(21)*}_{u\Phi_1,g_1f_1} - s_\beta^2 c_\beta\mathcal{P}_L V_{g_1f_2}\hat{C}^{(21)*}_{u\Phi_1,g_1f_1}\\
&+s_\beta^3\mathcal{P}_R V_{f_1g_1}\hat{C}^{(12)}_{d\Phi_2,g_1f_2} + s_\beta c_\beta^2\mathcal{P}_R V_{f_1g_1}\hat{C}^{(12)}_{d\Phi_2,g_1f_2}\\
&-s_\beta^3\mathcal{P}_R V_{f_1g_1}\hat{C}^{(21)}_{d\Phi_2,g_1f_2} - s_\beta c_\beta^2\mathcal{P}_R V_{f_1g_1}\hat{C}^{(21)}_{d\Phi_2,g_1f_2}\Big)
\end{aligned}
\tag{C.328}
$$

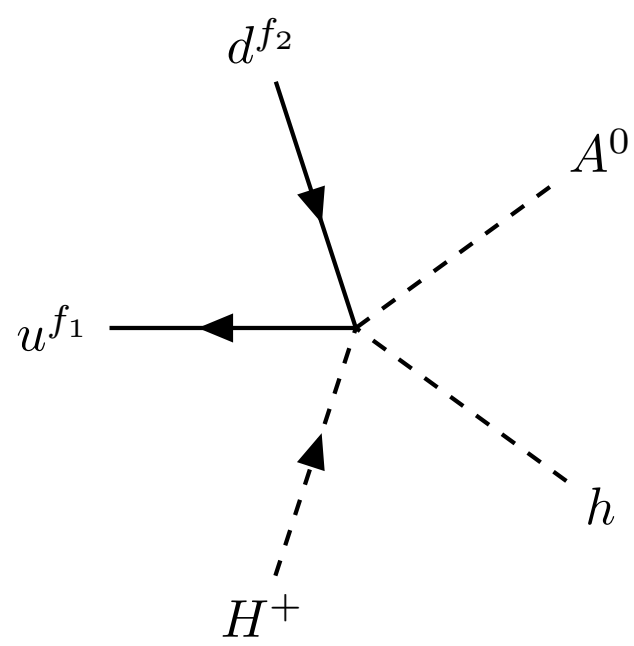

$$
\begin{aligned}
+\frac{1}{2}\Big(&s_\beta^3\mathcal{P}_L V_{g_1f_2}\hat{C}^{(12)*}_{u\Phi_1,g_1f_1} + s_\beta c_\beta^2\mathcal{P}_L V_{g_1f_2}\hat{C}^{(12)*}_{u\Phi_1,g_1f_1}\\
&-s_\beta^3\mathcal{P}_L V_{g_1f_2}\hat{C}^{(21)*}_{u\Phi_1,g_1f_1} - s_\beta c_\beta^2\mathcal{P}_L V_{g_1f_2}\hat{C}^{(21)*}_{u\Phi_1,g_1f_1}\\
&-c_\beta^3\mathcal{P}_R V_{f_1g_1}\hat{C}^{(12)}_{d\Phi_2,g_1f_2} - s_\beta^2 c_\beta\mathcal{P}_R V_{f_1g_1}\hat{C}^{(12)}_{d\Phi_2,g_1f_2}\\
&+c_\beta^3\mathcal{P}_R V_{f_1g_1}\hat{C}^{(21)}_{d\Phi_2,g_1f_2} + s_\beta^2 c_\beta\mathcal{P}_R V_{f_1g_1}\hat{C}^{(21)}_{d\Phi_2,g_1f_2}\Big)
\end{aligned}
\tag{C.329}
$$

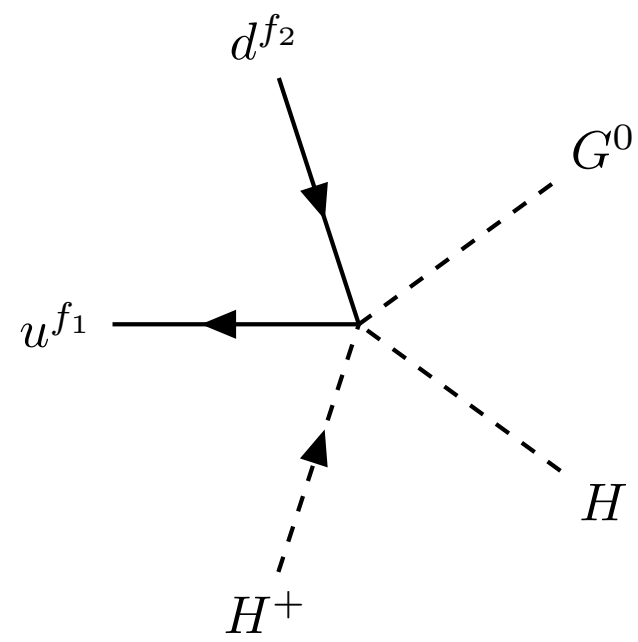

$$
\begin{aligned}
+\frac{1}{2}\Big(&s_\beta^3\mathcal{P}_L V_{g_1f_2}\hat{C}^{(12)*}_{u\Phi_1,g_1f_1} + s_\beta c_\beta^2\mathcal{P}_L V_{g_1f_2}\hat{C}^{(12)*}_{u\Phi_1,g_1f_1}\\
&-s_\beta^3\mathcal{P}_L V_{g_1f_2}\hat{C}^{(21)*}_{u\Phi_1,g_1f_1} - s_\beta c_\beta^2\mathcal{P}_L V_{g_1f_2}\hat{C}^{(21)*}_{u\Phi_1,g_1f_1}\\
&-c_\beta^3\mathcal{P}_R V_{f_1g_1}\hat{C}^{(12)}_{d\Phi_2,g_1f_2} - s_\beta^2 c_\beta\mathcal{P}_R V_{f_1g_1}\hat{C}^{(12)}_{d\Phi_2,g_1f_2}\\
&+c_\beta^3\mathcal{P}_R V_{f_1g_1}\hat{C}^{(21)}_{d\Phi_2,g_1f_2} + s_\beta^2 c_\beta\mathcal{P}_R V_{f_1g_1}\hat{C}^{(21)}_{d\Phi_2,g_1f_2}\Big)
\end{aligned}
\tag{C.330}
$$

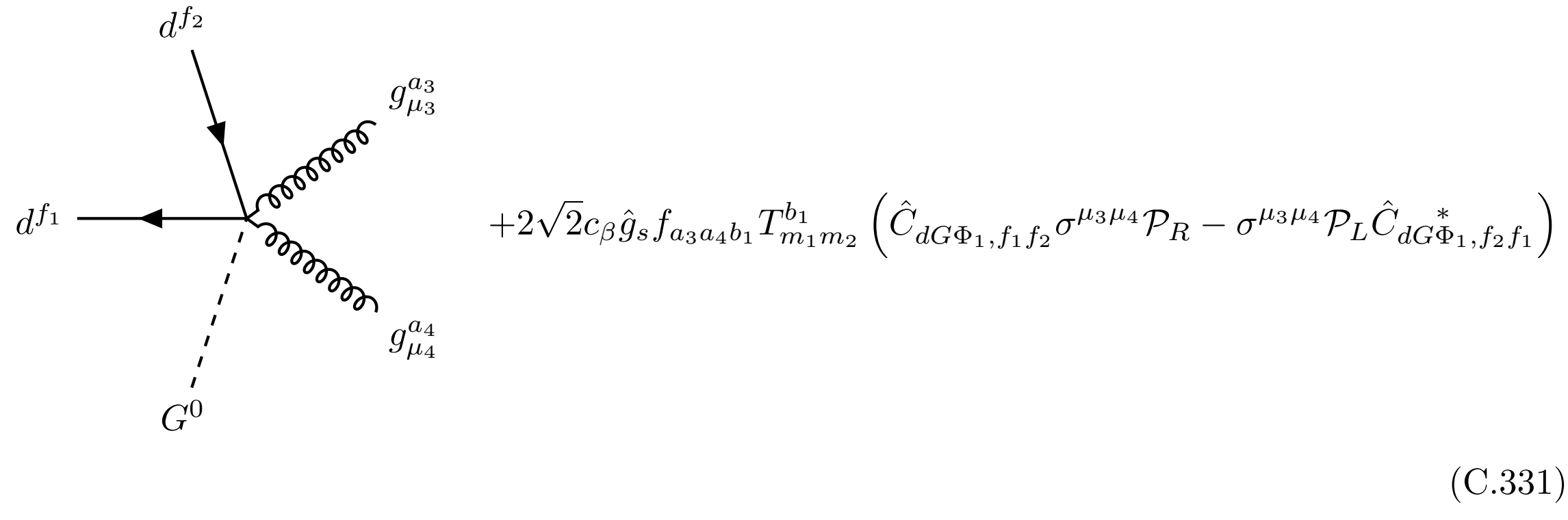

$$+2\sqrt{2}c_\beta \hat{g}_s f_{a_3 a_4 b_1} T^{b_1}_{m_1 m_2} \left( \hat{C}_{dG\Phi_1,f_1 f_2} \sigma^{\mu_3\mu_4} \mathcal{P}_R - \sigma^{\mu_3\mu_4} \mathcal{P}_L \hat{C}^*_{dG\Phi_1,f_2 f_1} \right)$$

(C.331)

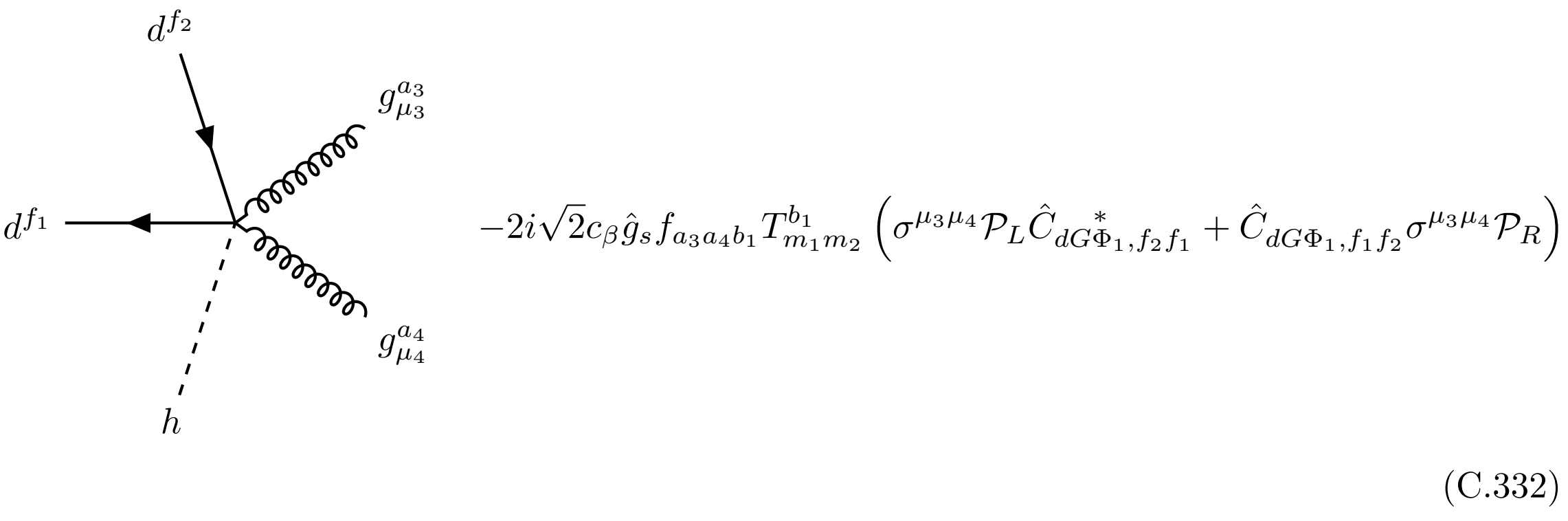

$$-2i\sqrt{2}c_\beta \hat{g}_s f_{a_3 a_4 b_1} T^{b_1}_{m_1 m_2} \left( \sigma^{\mu_3\mu_4} \mathcal{P}_L \hat{C}^*_{dG\Phi_1,f_2 f_1} + \hat{C}_{dG\Phi_1,f_1 f_2} \sigma^{\mu_3\mu_4} \mathcal{P}_R \right)$$

(C.332)

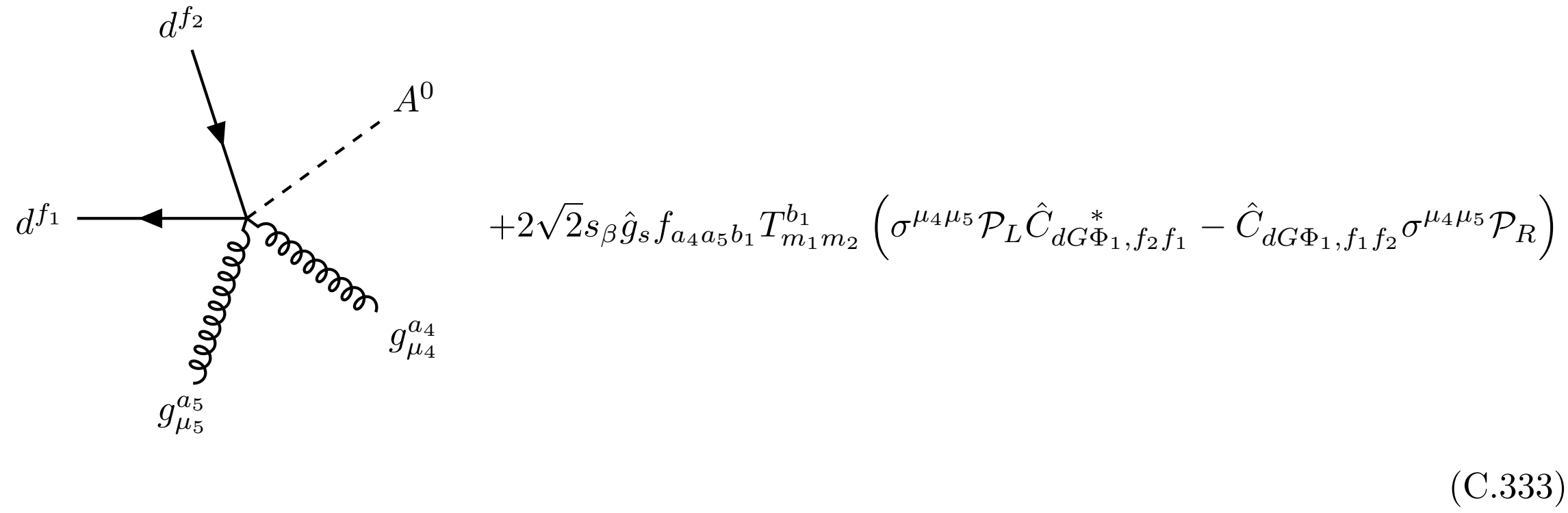

$$+2\sqrt{2}s_\beta \hat{g}_s f_{a_4 a_5 b_1} T^{b_1}_{m_1 m_2} \left( \sigma^{\mu_4\mu_5} \mathcal{P}_L \hat{C}^*_{dG\Phi_1,f_2 f_1} - \hat{C}_{dG\Phi_1,f_1 f_2} \sigma^{\mu_4\mu_5} \mathcal{P}_R \right)$$

(C.333)

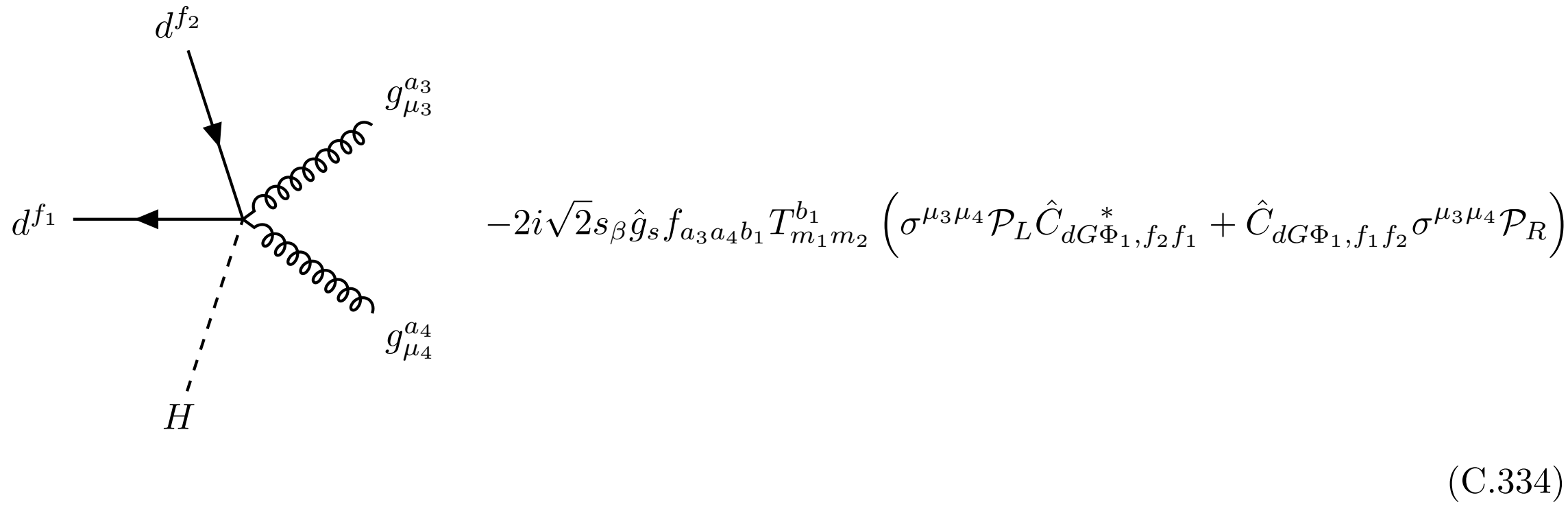

$$-2i\sqrt{2}s_\beta \hat{g}_s f_{a_3 a_4 b_1} T^{b_1}_{m_1 m_2} \left(\sigma^{\mu_3\mu_4}\mathcal{P}_L \hat{C}^*_{dG\Phi_1,f_2 f_1} + \hat{C}_{dG\Phi_1,f_1 f_2}\sigma^{\mu_3\mu_4}\mathcal{P}_R\right) \tag{C.334}$$

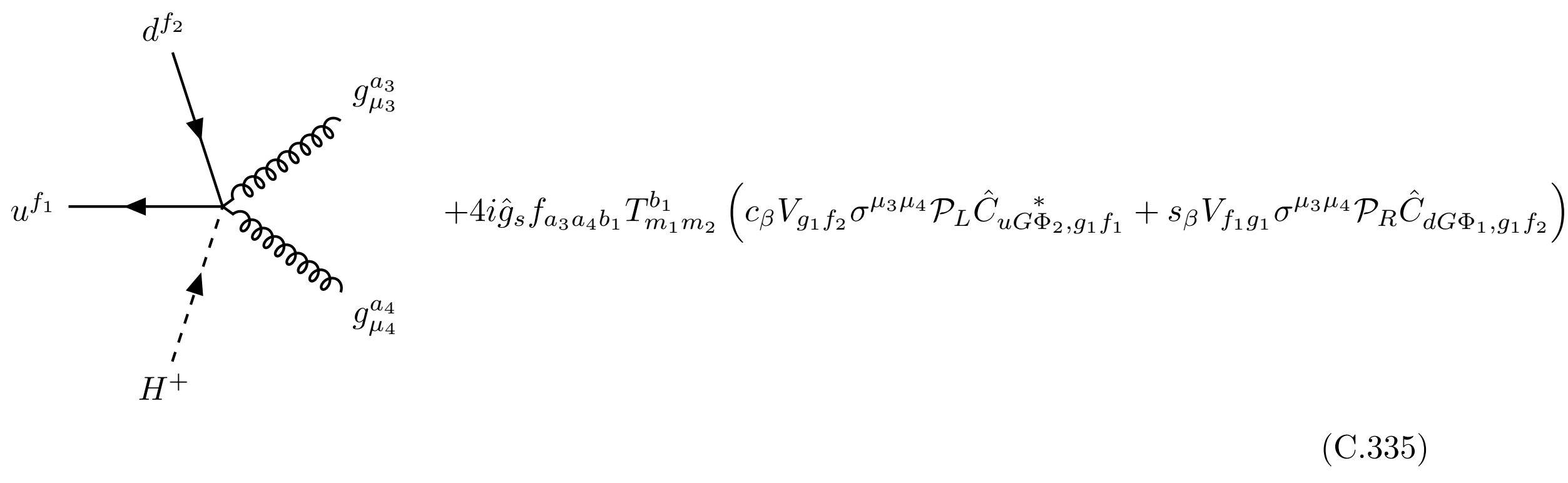

$$+4i\hat{g}_s f_{a_3 a_4 b_1} T^{b_1}_{m_1 m_2} \left(c_\beta V_{g_1 f_2}\sigma^{\mu_3\mu_4}\mathcal{P}_L \hat{C}^*_{uG\Phi_2,g_1 f_1} + s_\beta V_{f_1 g_1}\sigma^{\mu_3\mu_4}\mathcal{P}_R \hat{C}_{dG\Phi_1,g_1 f_2}\right) \tag{C.335}$$

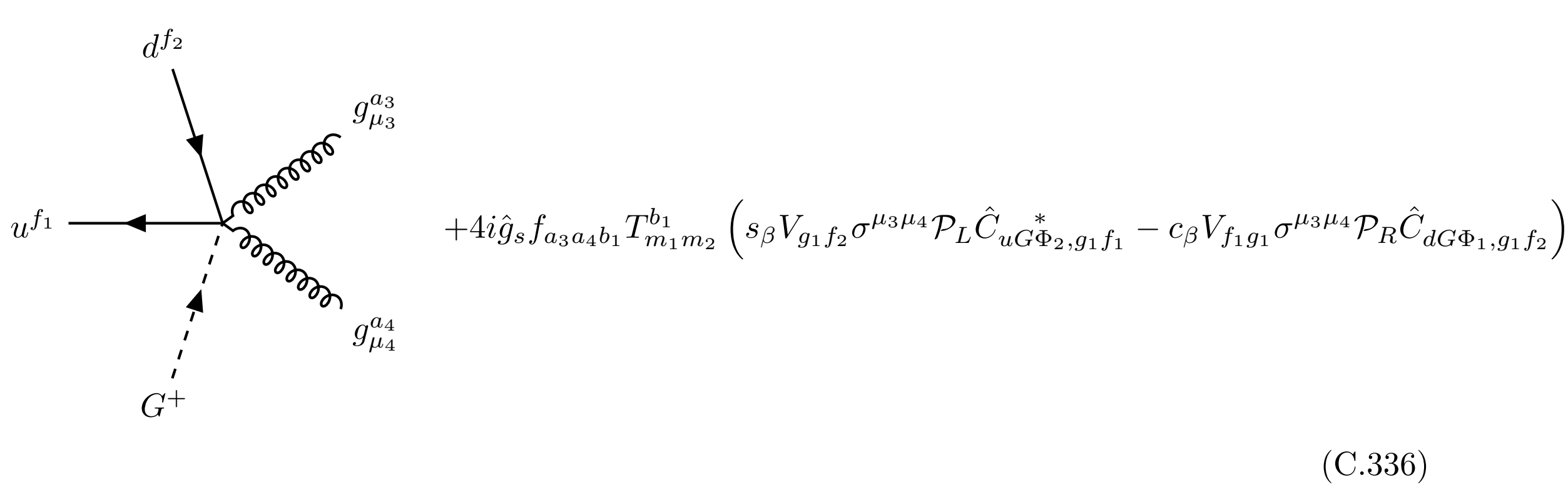

$$+4i\hat{g}_s f_{a_3 a_4 b_1} T^{b_1}_{m_1 m_2} \left(s_\beta V_{g_1 f_2}\sigma^{\mu_3\mu_4}\mathcal{P}_L \hat{C}^*_{uG\Phi_2,g_1 f_1} - c_\beta V_{f_1 g_1}\sigma^{\mu_3\mu_4}\mathcal{P}_R \hat{C}_{dG\Phi_1,g_1 f_2}\right) \tag{C.336}$$

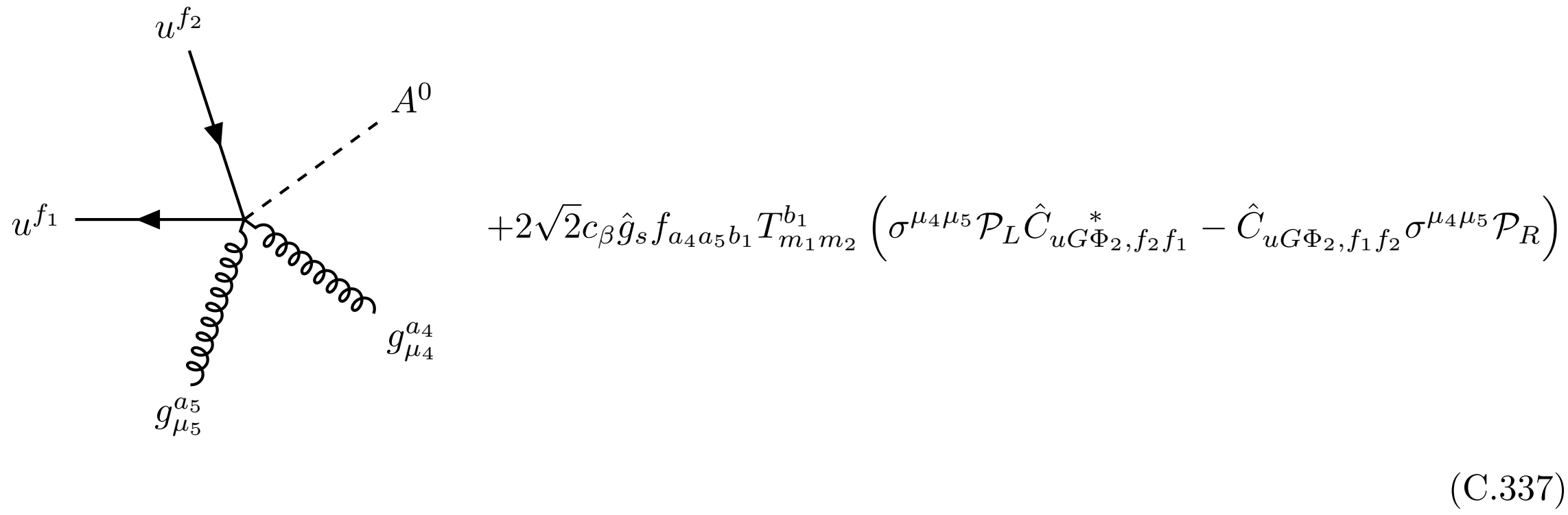

$$+2\sqrt{2}c_\beta\hat{g}_s f_{a_4a_5b_1}T^{b_1}_{m_1m_2}\left(\sigma^{\mu_4\mu_5}\mathcal{P}_L\hat{C}^*_{uG\Phi_2,f_2f_1}-\hat{C}_{uG\Phi_2,f_1f_2}\sigma^{\mu_4\mu_5}\mathcal{P}_R\right) \tag{C.337}$$

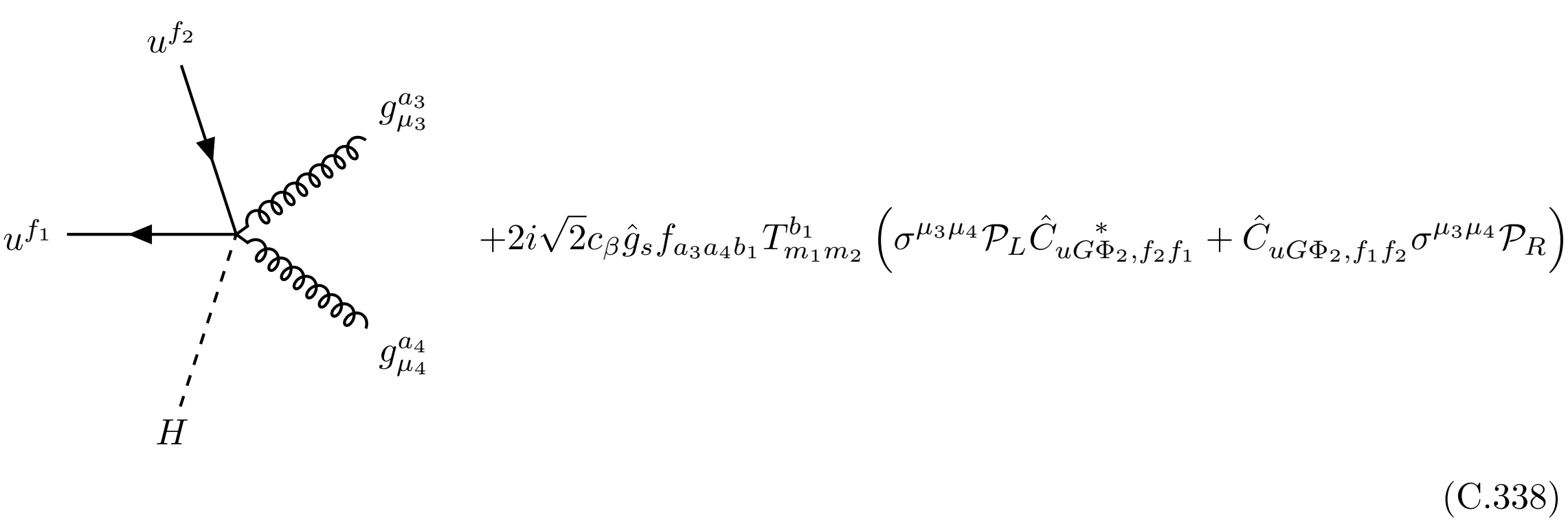

$$+2i\sqrt{2}c_\beta\hat{g}_s f_{a_3a_4b_1}T^{b_1}_{m_1m_2}\left(\sigma^{\mu_3\mu_4}\mathcal{P}_L\hat{C}^*_{uG\Phi_2,f_2f_1}+\hat{C}_{uG\Phi_2,f_1f_2}\sigma^{\mu_3\mu_4}\mathcal{P}_R\right) \tag{C.338}$$

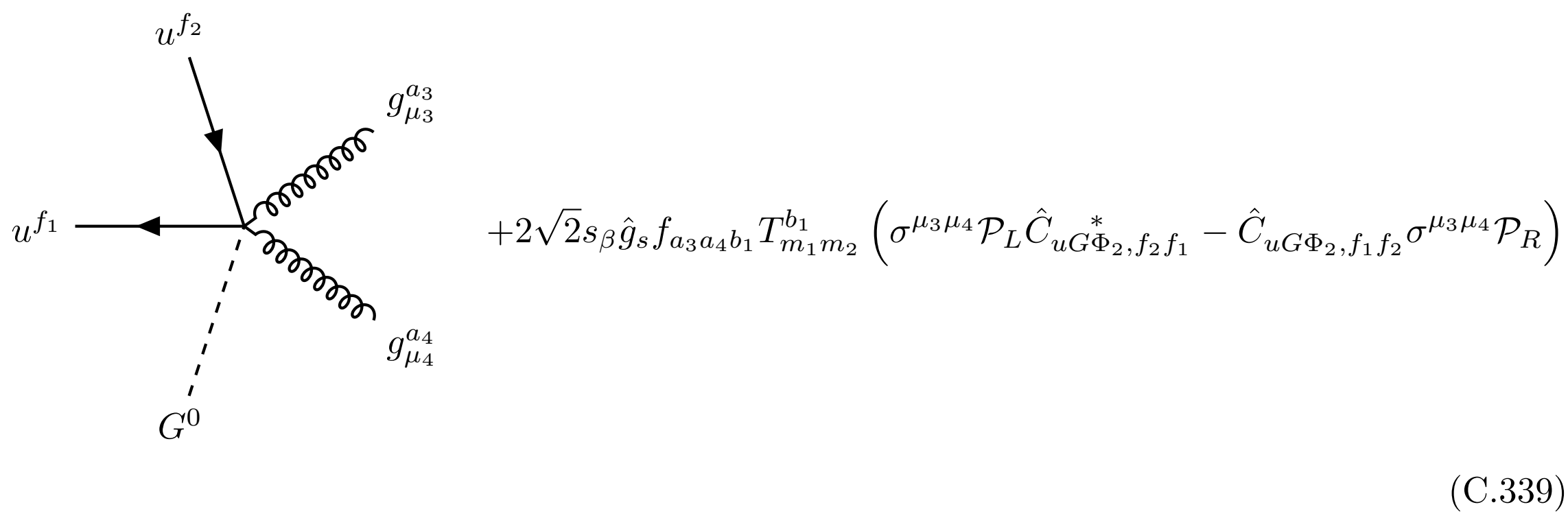

$$+2\sqrt{2}s_\beta\hat{g}_s f_{a_3a_4b_1}T^{b_1}_{m_1m_2}\left(\sigma^{\mu_3\mu_4}\mathcal{P}_L\hat{C}^*_{uG\Phi_2,f_2f_1}-\hat{C}_{uG\Phi_2,f_1f_2}\sigma^{\mu_3\mu_4}\mathcal{P}_R\right) \tag{C.339}$$

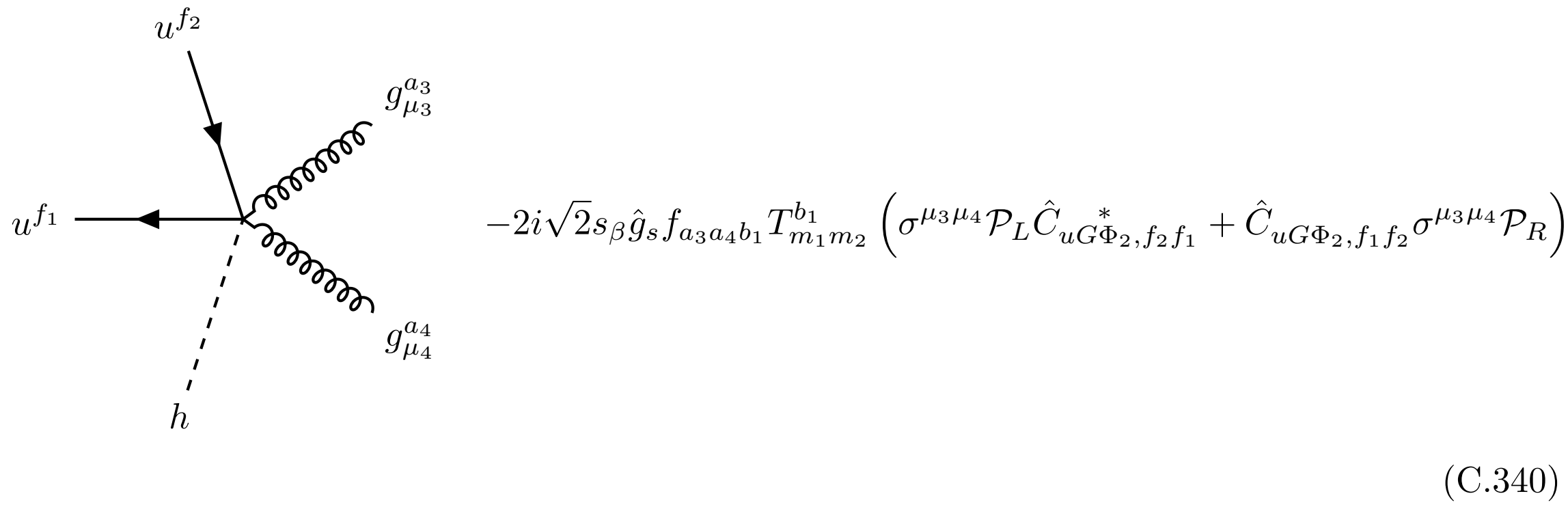

$$-2i\sqrt{2}s_\beta \hat{g}_s f_{a_3 a_4 b_1} T^{b_1}_{m_1 m_2} \left(\sigma^{\mu_3\mu_4}\mathcal{P}_L \hat{C}^{*}_{uG\Phi_2,f_2 f_1} + \hat{C}_{uG\Phi_2,f_1 f_2}\sigma^{\mu_3\mu_4}\mathcal{P}_R\right) \tag{C.340}$$

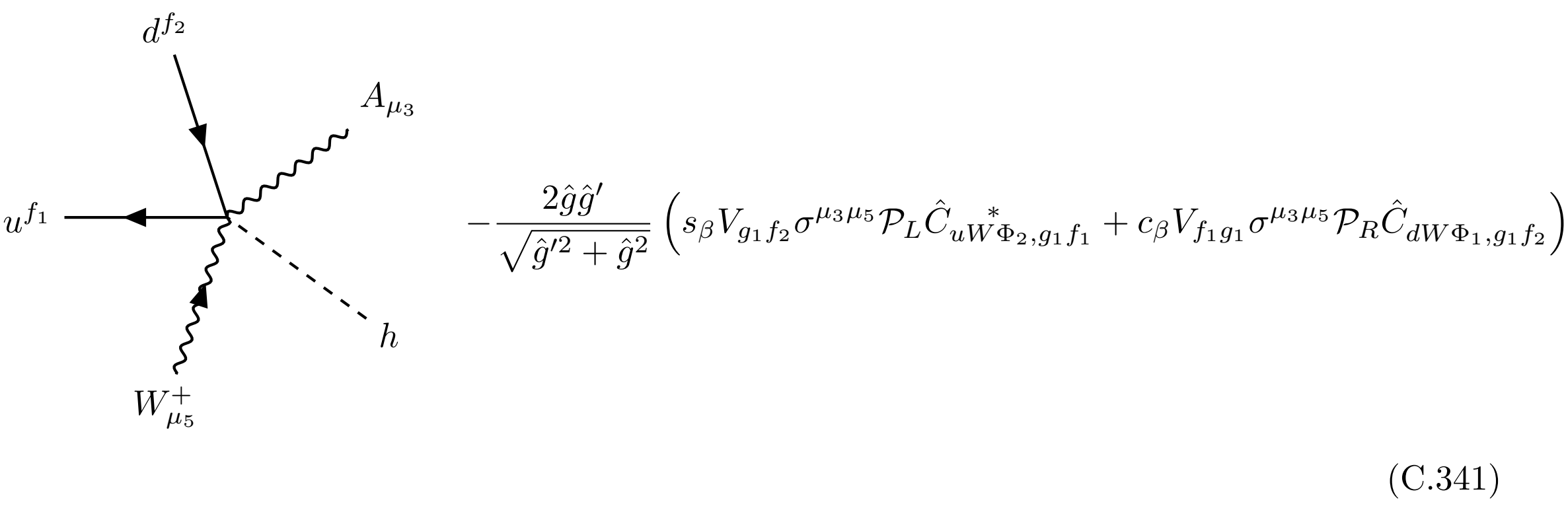

$$-\frac{2\hat{g}\hat{g}'}{\sqrt{\hat{g}'^2+\hat{g}^2}} \left(s_\beta V_{g_1 f_2}\sigma^{\mu_3\mu_5}\mathcal{P}_L \hat{C}^{*}_{uW\Phi_2,g_1 f_1} + c_\beta V_{f_1 g_1}\sigma^{\mu_3\mu_5}\mathcal{P}_R \hat{C}_{dW\Phi_1,g_1 f_2}\right) \tag{C.341}$$

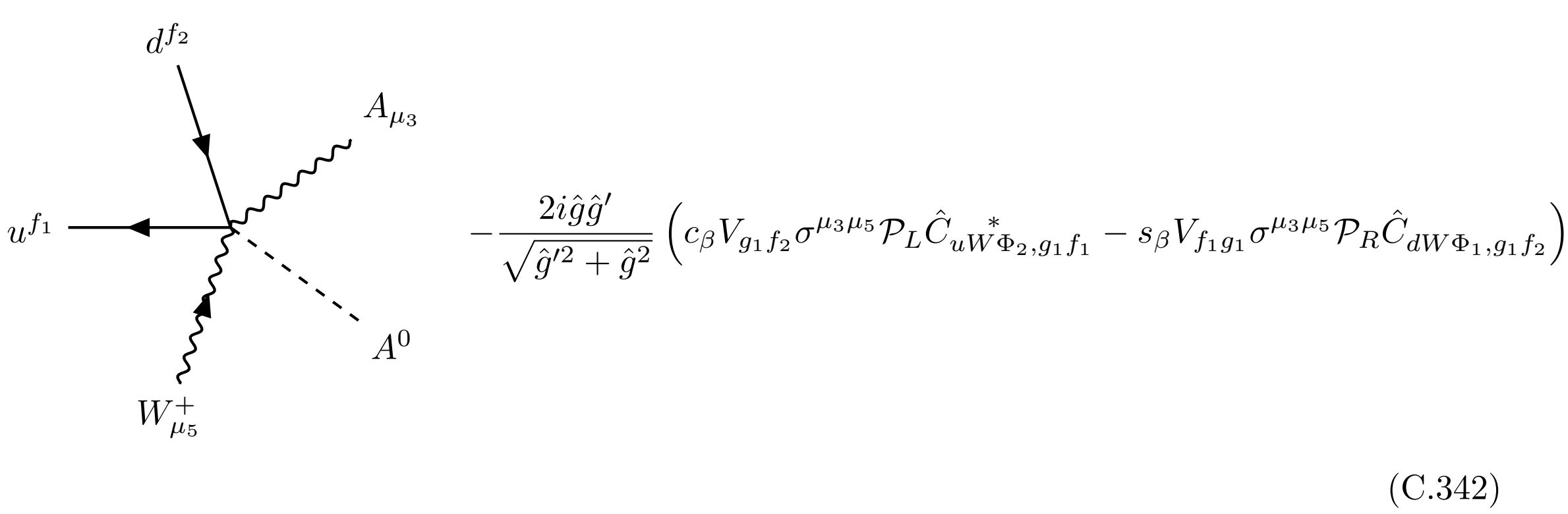

$$-\frac{2i\hat{g}\hat{g}'}{\sqrt{\hat{g}'^2+\hat{g}^2}} \left(c_\beta V_{g_1 f_2}\sigma^{\mu_3\mu_5}\mathcal{P}_L \hat{C}^{*}_{uW\Phi_2,g_1 f_1} - s_\beta V_{f_1 g_1}\sigma^{\mu_3\mu_5}\mathcal{P}_R \hat{C}_{dW\Phi_1,g_1 f_2}\right) \tag{C.342}$$

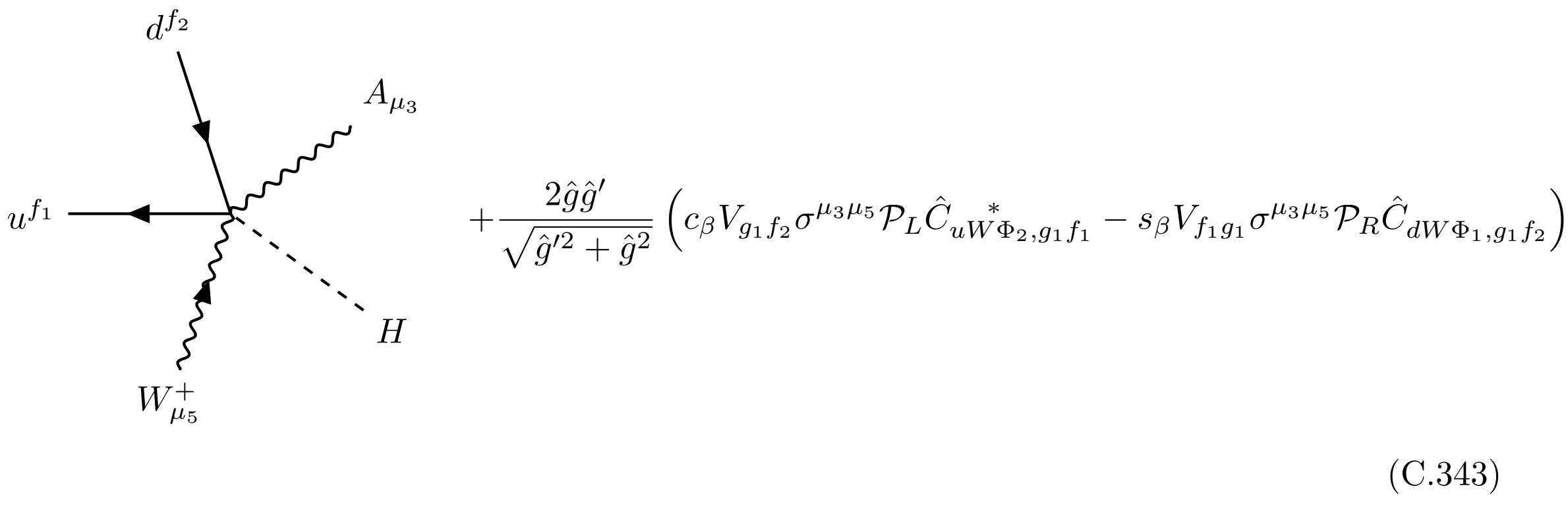

$$+\frac{2\hat{g}\hat{g}'}{\sqrt{\hat{g}'^2+\hat{g}^2}}\left(c_\beta V_{g_1 f_2}\sigma^{\mu_3\mu_5}\mathcal{P}_L\hat{C}^*_{uW\Phi_2,g_1 f_1}-s_\beta V_{f_1 g_1}\sigma^{\mu_3\mu_5}\mathcal{P}_R\hat{C}_{dW\Phi_1,g_1 f_2}\right) \tag{C.343}$$

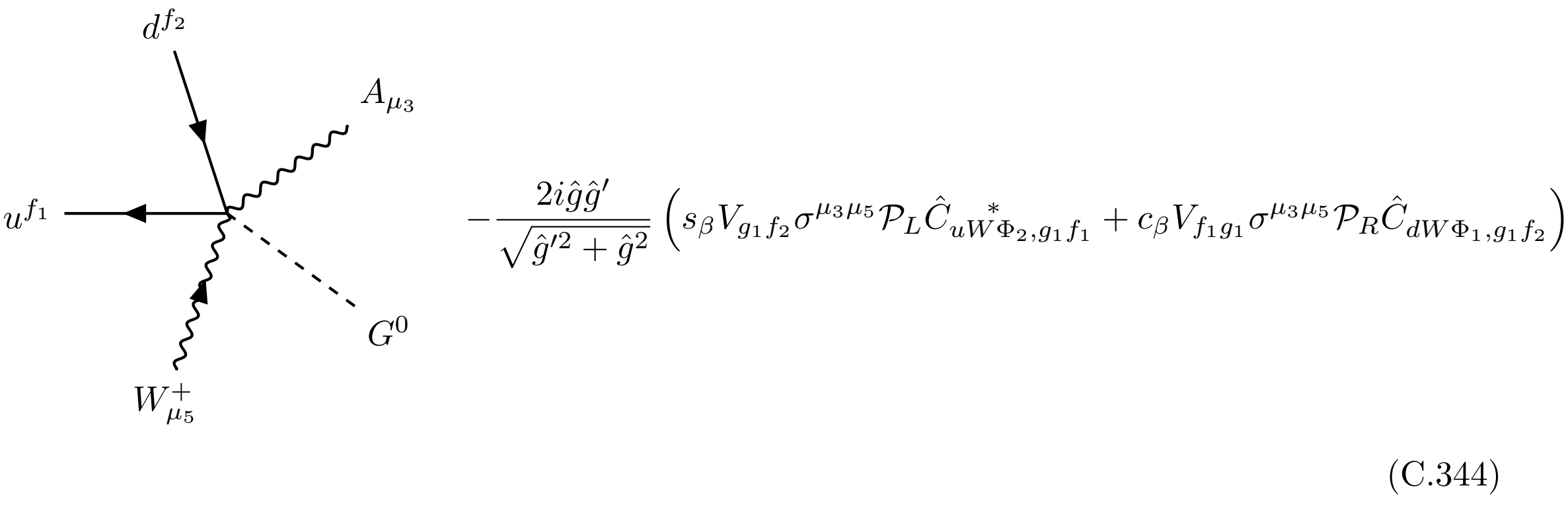

$$-\frac{2i\hat{g}\hat{g}'}{\sqrt{\hat{g}'^2+\hat{g}^2}}\left(s_\beta V_{g_1 f_2}\sigma^{\mu_3\mu_5}\mathcal{P}_L\hat{C}^*_{uW\Phi_2,g_1 f_1}+c_\beta V_{f_1 g_1}\sigma^{\mu_3\mu_5}\mathcal{P}_R\hat{C}_{dW\Phi_1,g_1 f_2}\right) \tag{C.344}$$

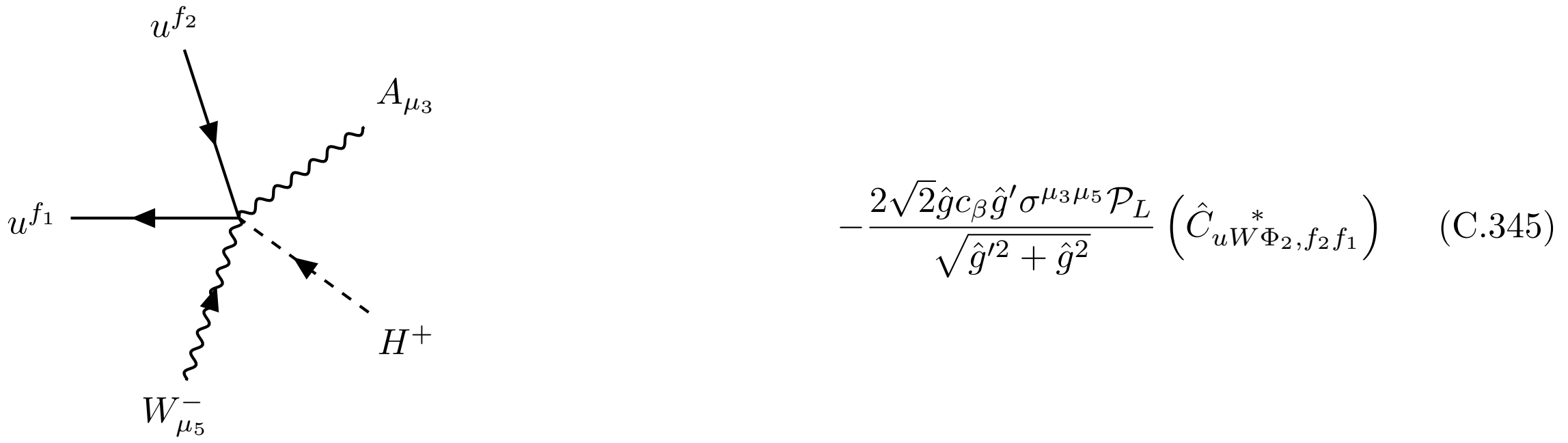

$$-\frac{2\sqrt{2}\hat{g}c_\beta\hat{g}'\sigma^{\mu_3\mu_5}\mathcal{P}_L}{\sqrt{\hat{g}'^2+\hat{g}^2}}\left(\hat{C}^*_{uW\Phi_2,f_2 f_1}\right) \tag{C.345}$$

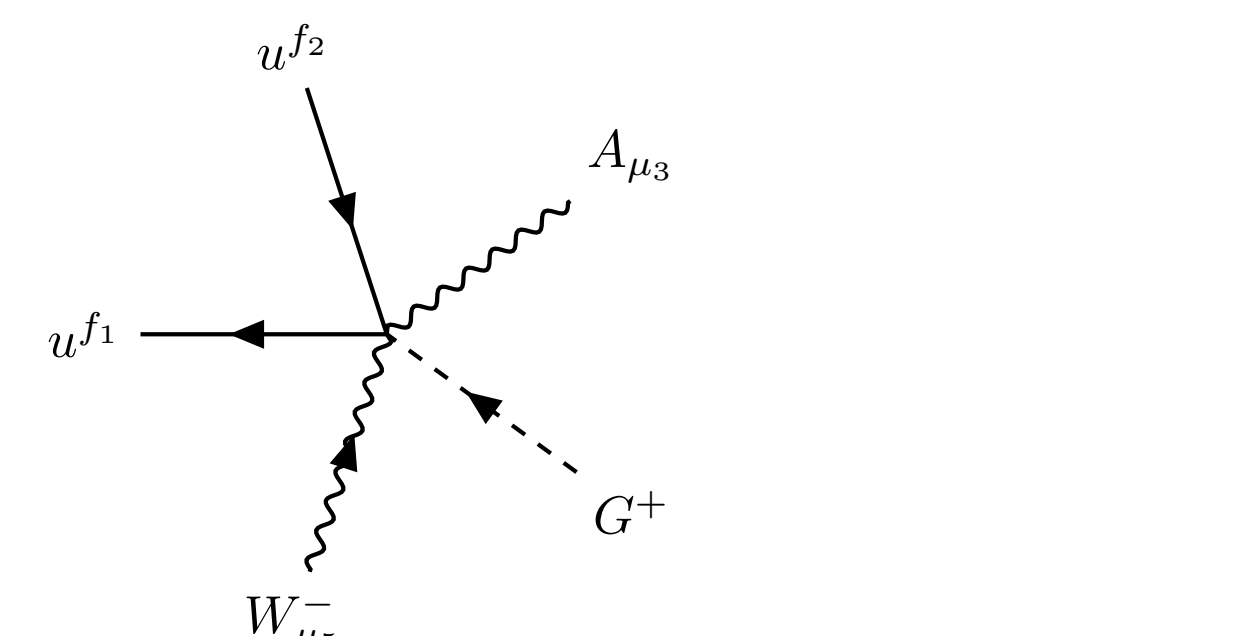

$$-\frac{2\sqrt{2}\hat{g}s_\beta\hat{g}'\sigma^{\mu_3\mu_5}\mathcal{P}_L}{\sqrt{\hat{g}'^2+\hat{g}^2}}\left(\hat{C}^*_{uW\Phi_2,f_2f_1}\right) \tag{C.346}$$

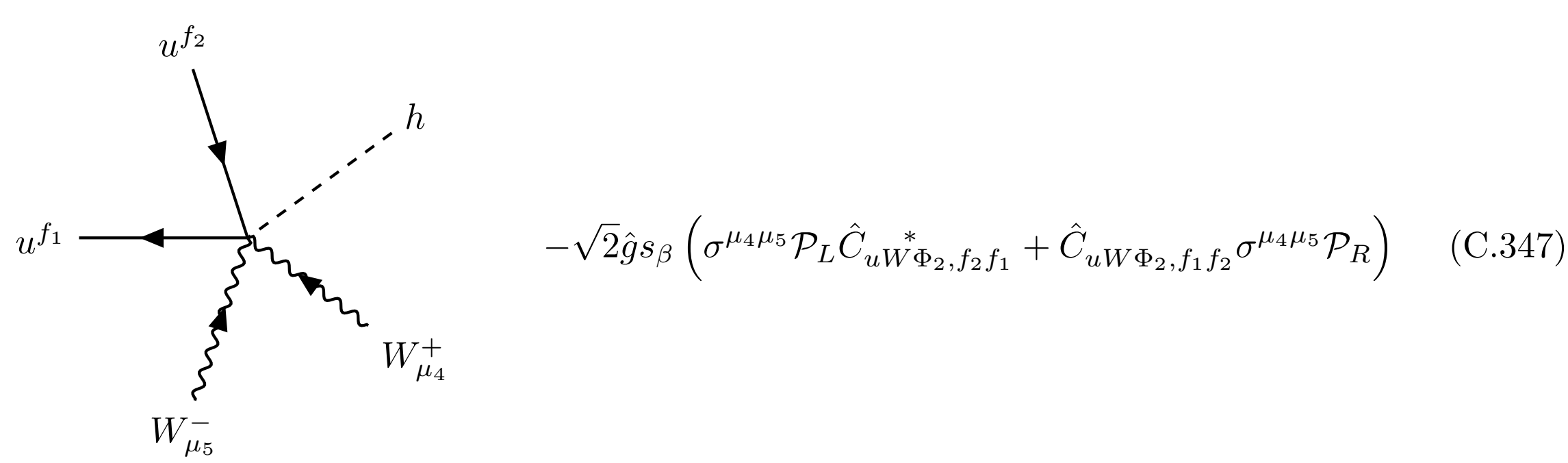

$$-\sqrt{2}\hat{g}s_\beta\left(\sigma^{\mu_4\mu_5}\mathcal{P}_L\hat{C}^*_{uW\Phi_2,f_2f_1}+\hat{C}_{uW\Phi_2,f_1f_2}\sigma^{\mu_4\mu_5}\mathcal{P}_R\right) \tag{C.347}$$

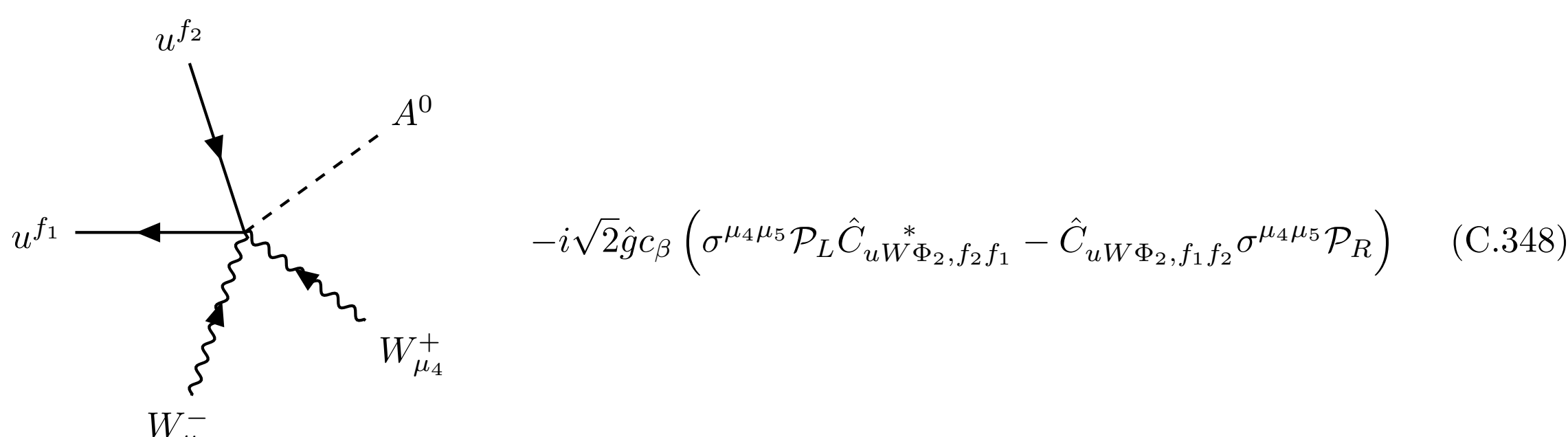

$$-i\sqrt{2}\hat{g}c_\beta\left(\sigma^{\mu_4\mu_5}\mathcal{P}_L\hat{C}^*_{uW\Phi_2,f_2f_1}-\hat{C}_{uW\Phi_2,f_1f_2}\sigma^{\mu_4\mu_5}\mathcal{P}_R\right) \tag{C.348}$$

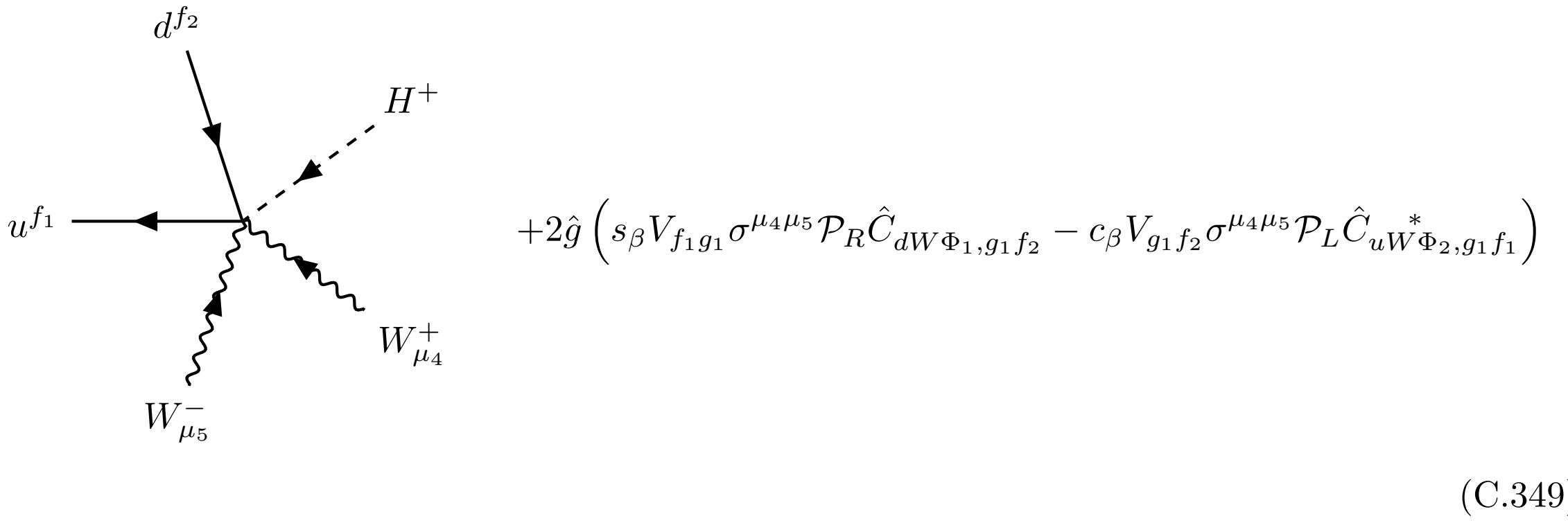

$$+2\hat{g}\left(s_\beta V_{f_1 g_1}\sigma^{\mu_4\mu_5}\mathcal{P}_R\hat{C}_{dW\Phi_1,g_1f_2} - c_\beta V_{g_1f_2}\sigma^{\mu_4\mu_5}\mathcal{P}_L\hat{C}^*_{uW\Phi_2,g_1f_1}\right) \tag{C.349}$$

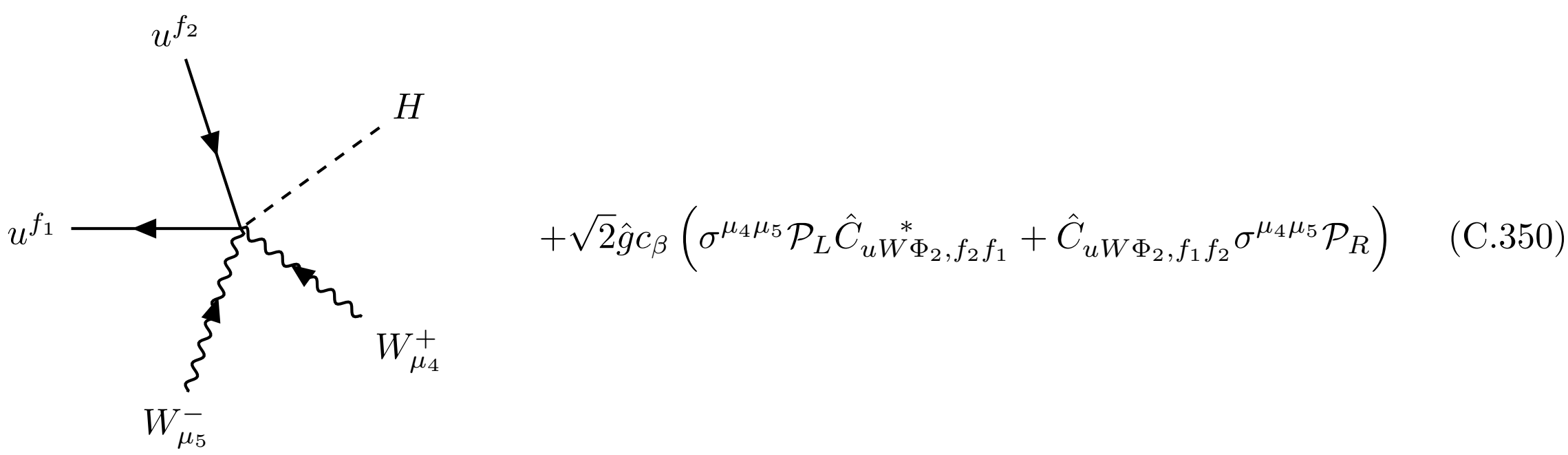

$$+\sqrt{2}\hat{g}c_\beta\left(\sigma^{\mu_4\mu_5}\mathcal{P}_L\hat{C}^*_{uW\Phi_2,f_2f_1} + \hat{C}_{uW\Phi_2,f_1f_2}\sigma^{\mu_4\mu_5}\mathcal{P}_R\right) \tag{C.350}$$

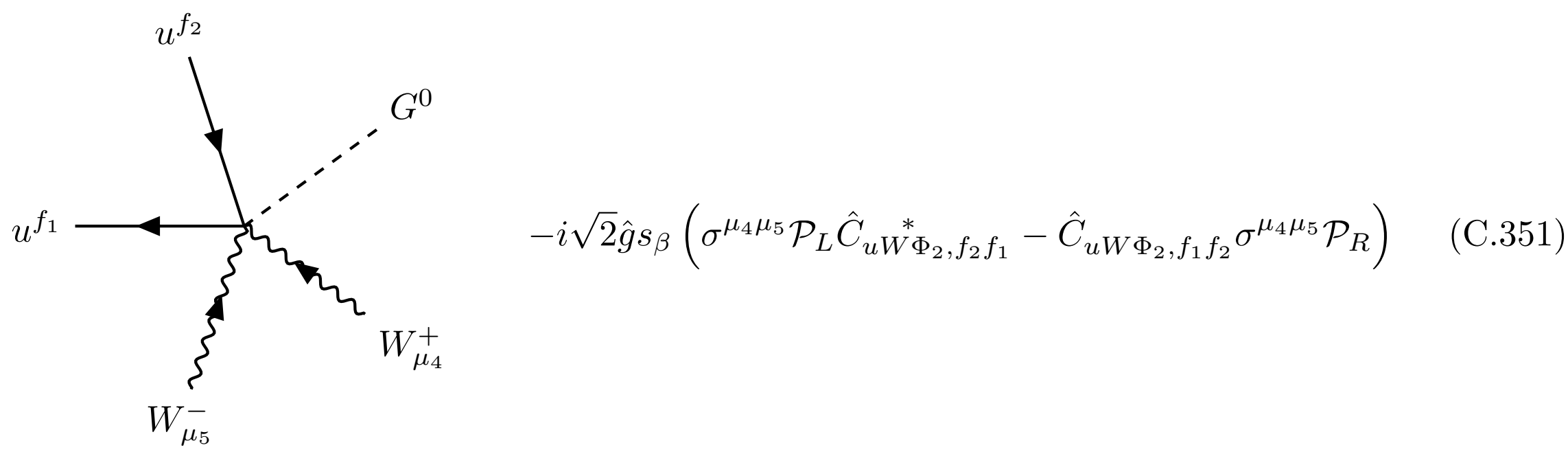

$$-i\sqrt{2}\hat{g}s_\beta\left(\sigma^{\mu_4\mu_5}\mathcal{P}_L\hat{C}^*_{uW\Phi_2,f_2f_1} - \hat{C}_{uW\Phi_2,f_1f_2}\sigma^{\mu_4\mu_5}\mathcal{P}_R\right) \tag{C.351}$$

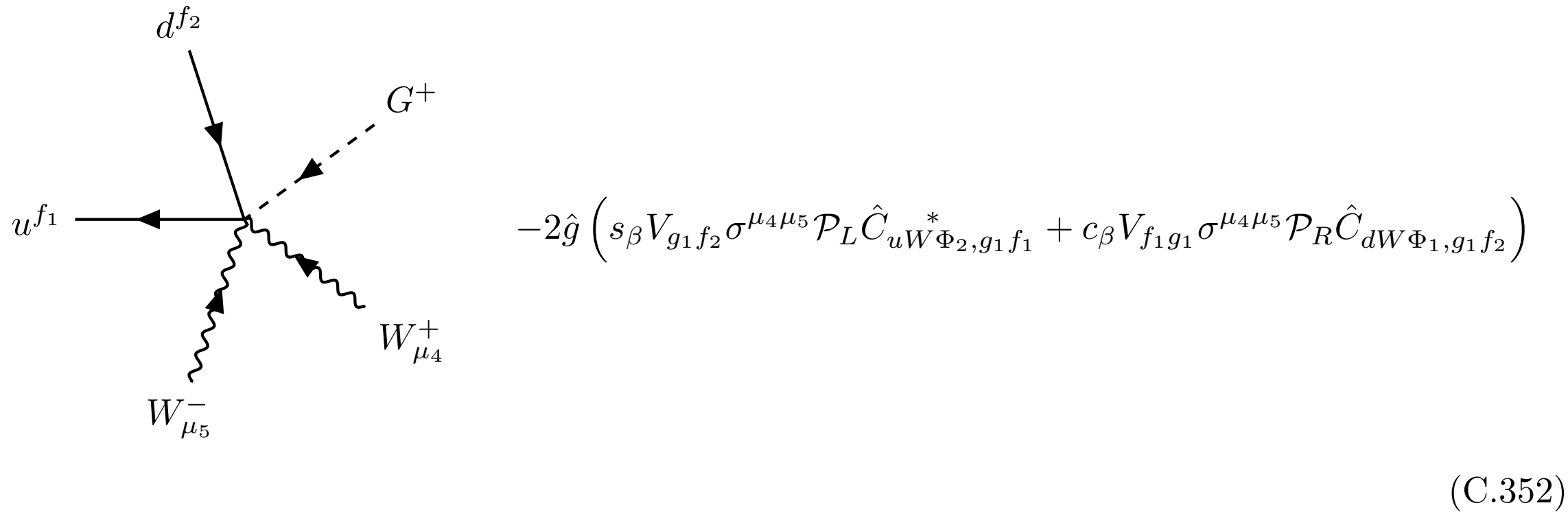

$$-2\hat{g}\left(s_\beta V_{g_1 f_2}\sigma^{\mu_4\mu_5}\mathcal{P}_L\hat{C}^*_{uW\Phi_2,g_1f_1}+c_\beta V_{f_1g_1}\sigma^{\mu_4\mu_5}\mathcal{P}_R\hat{C}_{dW\Phi_1,g_1f_2}\right) \tag{C.352}$$

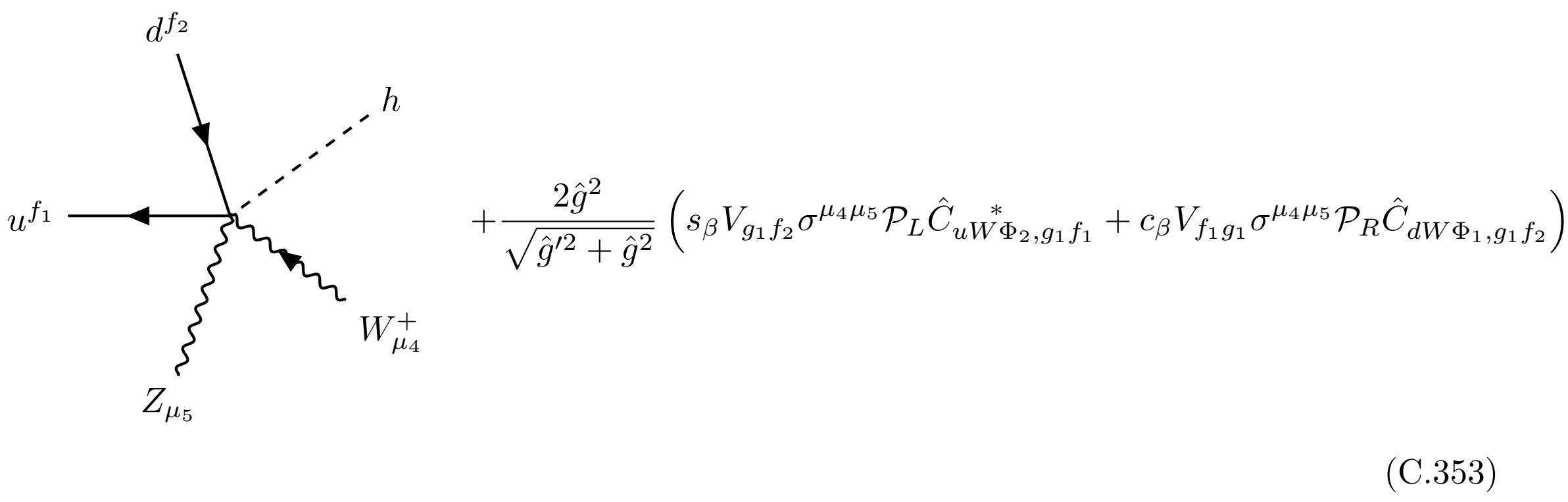

$$+\frac{2\hat{g}^2}{\sqrt{\hat{g}'^2+\hat{g}^2}}\left(s_\beta V_{g_1 f_2}\sigma^{\mu_4\mu_5}\mathcal{P}_L\hat{C}^*_{uW\Phi_2,g_1f_1}+c_\beta V_{f_1g_1}\sigma^{\mu_4\mu_5}\mathcal{P}_R\hat{C}_{dW\Phi_1,g_1f_2}\right) \tag{C.353}$$

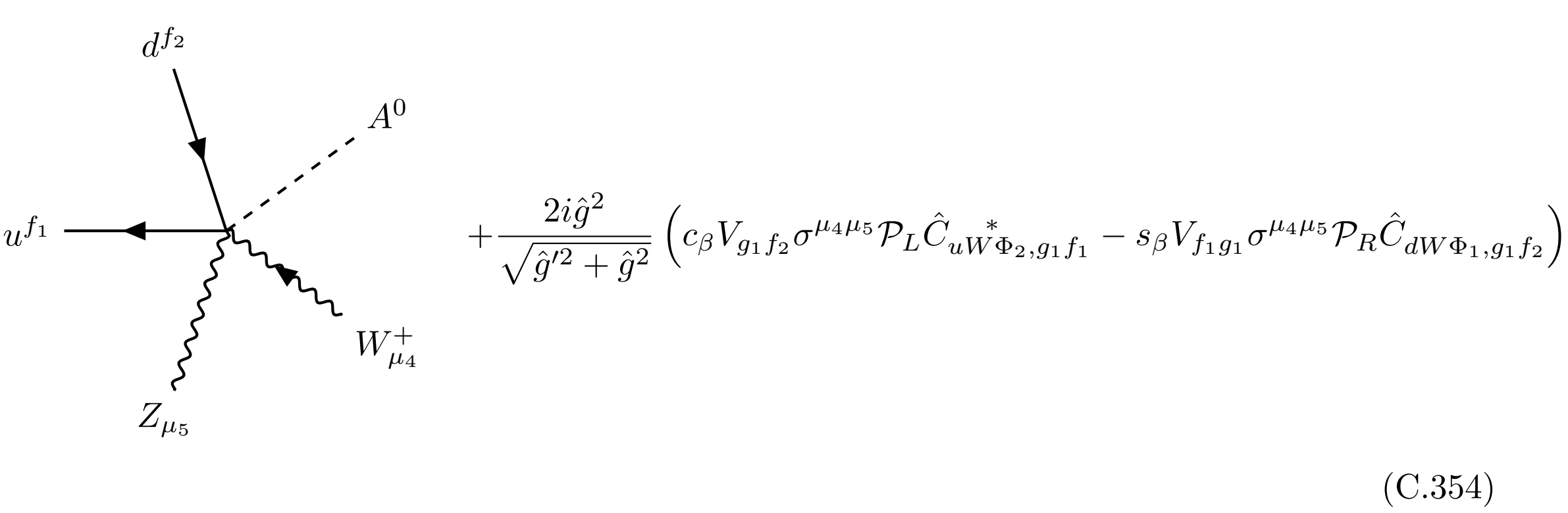

$$+\frac{2i\hat{g}^2}{\sqrt{\hat{g}'^2+\hat{g}^2}}\left(c_\beta V_{g_1 f_2}\sigma^{\mu_4\mu_5}\mathcal{P}_L\hat{C}^*_{uW\Phi_2,g_1f_1}-s_\beta V_{f_1g_1}\sigma^{\mu_4\mu_5}\mathcal{P}_R\hat{C}_{dW\Phi_1,g_1f_2}\right) \tag{C.354}$$

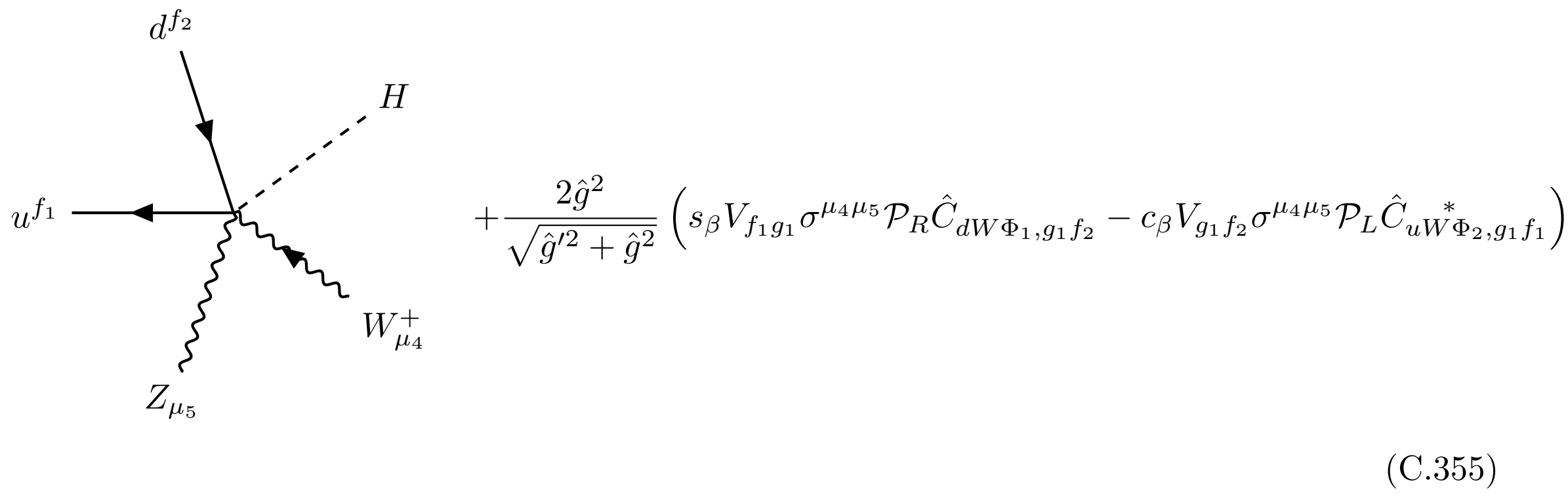

$$+\frac{2\hat{g}^2}{\sqrt{\hat{g}'^2+\hat{g}^2}}\left(s_\beta V_{f_1 g_1}\sigma^{\mu_4\mu_5}\mathcal{P}_R\hat{C}_{dW\Phi_1,g_1f_2}-c_\beta V_{g_1 f_2}\sigma^{\mu_4\mu_5}\mathcal{P}_L\hat{C}^*_{uW\Phi_2,g_1f_1}\right) \tag{C.355}$$

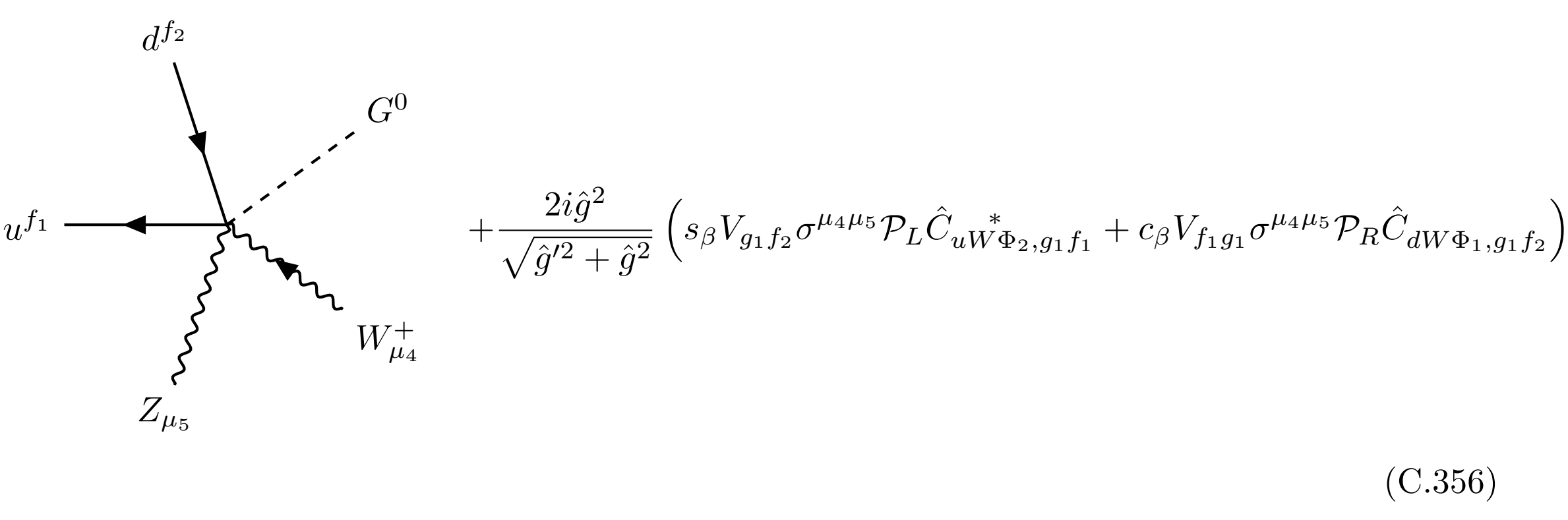

$$+\frac{2i\hat{g}^2}{\sqrt{\hat{g}'^2+\hat{g}^2}}\left(s_\beta V_{g_1 f_2}\sigma^{\mu_4\mu_5}\mathcal{P}_L\hat{C}^*_{uW\Phi_2,g_1f_1}+c_\beta V_{f_1 g_1}\sigma^{\mu_4\mu_5}\mathcal{P}_R\hat{C}_{dW\Phi_1,g_1f_2}\right) \tag{C.356}$$

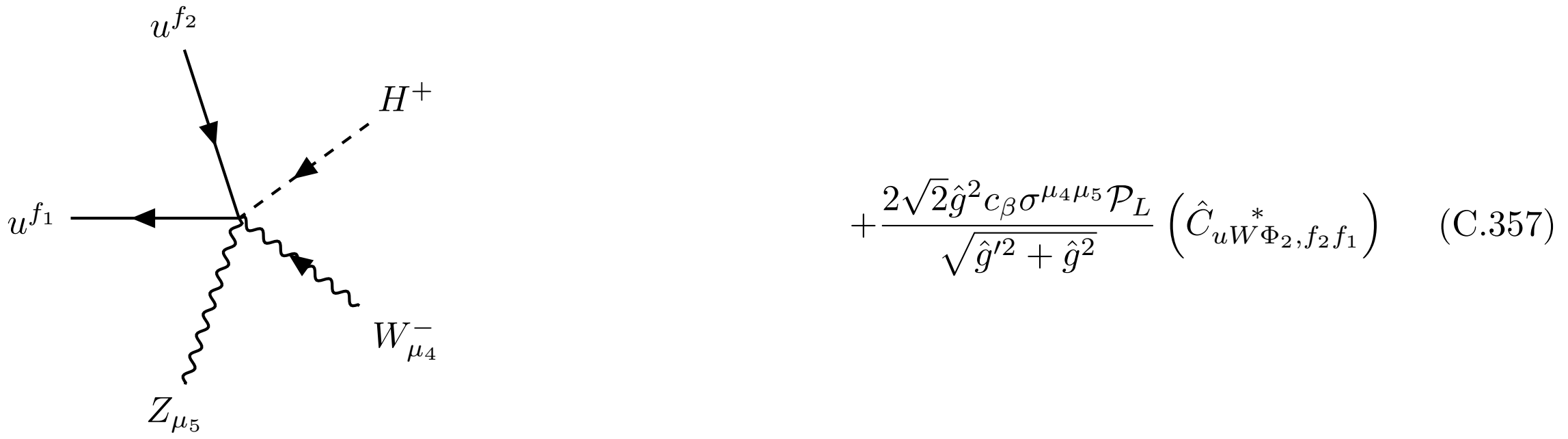

$$+\frac{2\sqrt{2}\hat{g}^2 c_\beta\sigma^{\mu_4\mu_5}\mathcal{P}_L}{\sqrt{\hat{g}'^2+\hat{g}^2}}\left(\hat{C}^*_{uW\Phi_2,f_2f_1}\right) \tag{C.357}$$

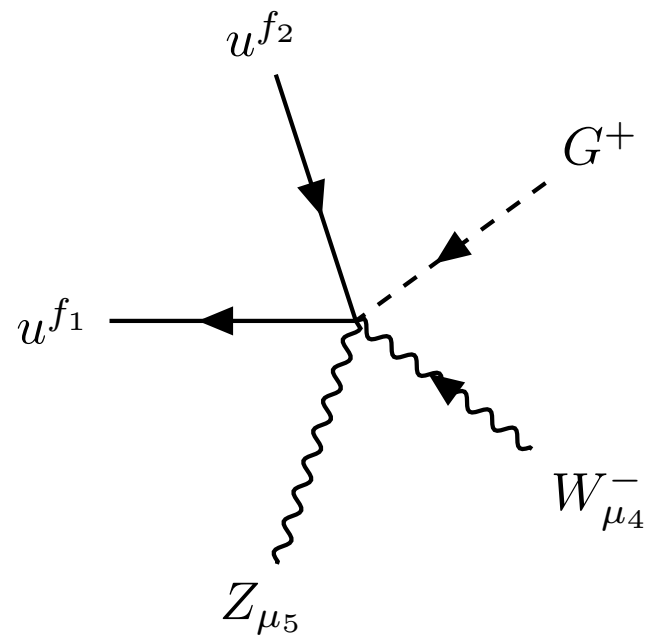

$$+\frac{2\sqrt{2}\hat{g}^2 s_\beta \sigma^{\mu_4\mu_5}\mathcal{P}_L}{\sqrt{\hat{g}'^2+\hat{g}^2}}\left(\hat{C}^*_{uW\Phi_2,f_2f_1}\right) \tag{C.358}$$

$$-\frac{2i\hat{g}\hat{g}' V_{f_1g_1}V^*_{f_2g_2}\gamma^{\mu_3}\mathcal{P}_L}{\sqrt{\hat{g}'^2+\hat{g}^2}}\left(c_\beta^2\hat{C}^{(11)[1]}_{\Phi q,g_1g_2}+c_\beta^2\hat{C}^{(11)[3]}_{\Phi q,g_1g_2}+s_\beta^2\left(\hat{C}^{(22)[1]}_{\Phi q,g_1g_2}+\hat{C}^{(22)[3]}_{\Phi q,g_1g_2}\right)\right)$$
$$-\frac{2i\hat{g}\hat{g}'\gamma^{\mu_3}\mathcal{P}_R}{\sqrt{\hat{g}'^2+\hat{g}^2}}\left(c_\beta^2\hat{C}^{(11)}_{\Phi u,f_1f_2}+s_\beta^2\hat{C}^{(22)}_{\Phi u,f_1f_2}\right) \tag{C.359}$$

$$+\frac{i\hat{g}s_{2\beta}\hat{g}' V_{f_1g_1}V^*_{f_2g_2}\gamma^{\mu_3}\mathcal{P}_L}{\sqrt{\hat{g}'^2+\hat{g}^2}}\left(\hat{C}^{(11)[1]}_{\Phi q,g_1g_2}+\hat{C}^{(11)[3]}_{\Phi q,g_1g_2}-\hat{C}^{(22)[1]}_{\Phi q,g_1g_2}-\hat{C}^{(22)[3]}_{\Phi q,g_1g_2}\right)$$
$$+\frac{i\hat{g}s_{2\beta}\hat{g}'\gamma^{\mu_3}\mathcal{P}_R}{\sqrt{\hat{g}'^2+\hat{g}^2}}\left(\hat{C}^{(11)}_{\Phi u,f_1f_2}-\hat{C}^{(22)}_{\Phi u,f_1f_2}\right) \tag{C.360}$$

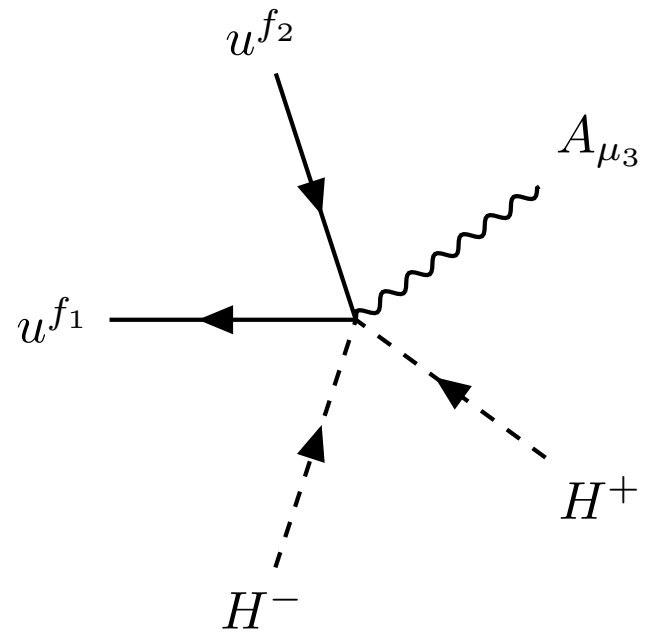

$$\begin{aligned}&-\frac{2i\hat{g}\hat{g}'V_{f_1g_1}V^*_{f_2g_2}\gamma^{\mu_3}\mathcal{P}_L}{\sqrt{\hat{g}'^2+\hat{g}^2}}\left(s_\beta^2\left(\hat{C}^{(11)[1]}_{\Phi q,g_1g_2}+\hat{C}^{(11)[3]}_{\Phi q,g_1g_2}\right)\right.\\&\qquad\qquad\left.+c_\beta^2\hat{C}^{(22)[1]}_{\Phi q,g_1g_2}+c_\beta^2\hat{C}^{(22)[3]}_{\Phi q,g_1g_2}\right)\\&-\frac{2i\hat{g}\hat{g}'\gamma^{\mu_3}\mathcal{P}_R}{\sqrt{\hat{g}'^2+\hat{g}^2}}\left(s_\beta^2\hat{C}^{(11)}_{\Phi u,f_1f_2}+c_\beta^2\hat{C}^{(22)}_{\Phi u,f_1f_2}\right)\end{aligned}\tag{C.361}$$

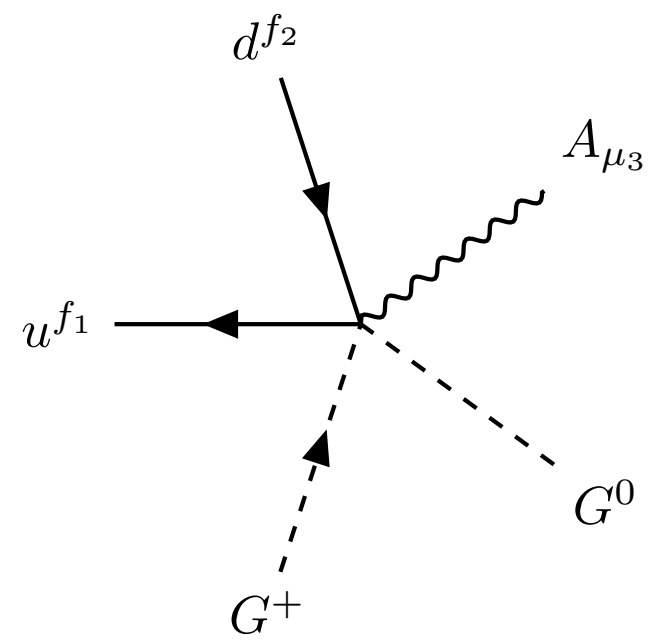

$$\begin{aligned}&-\frac{\sqrt{2}\hat{g}s_\beta c_\beta\hat{g}'\gamma^{\mu_3}\mathcal{P}_R}{\sqrt{\hat{g}'^2+\hat{g}^2}}\left(\hat{C}^{(21)}_{\Phi ud,f_1f_2}\right)\\&-\frac{\sqrt{2}\hat{g}\hat{g}'V_{f_1g_1}\gamma^{\mu_3}\mathcal{P}_L}{\sqrt{\hat{g}'^2+\hat{g}^2}}\left(c_\beta^2\hat{C}^{(11)[3]}_{\Phi q,g_1f_2}+s_\beta^2\hat{C}^{(22)[3]}_{\Phi q,g_1f_2}\right)\end{aligned}\tag{C.362}$$

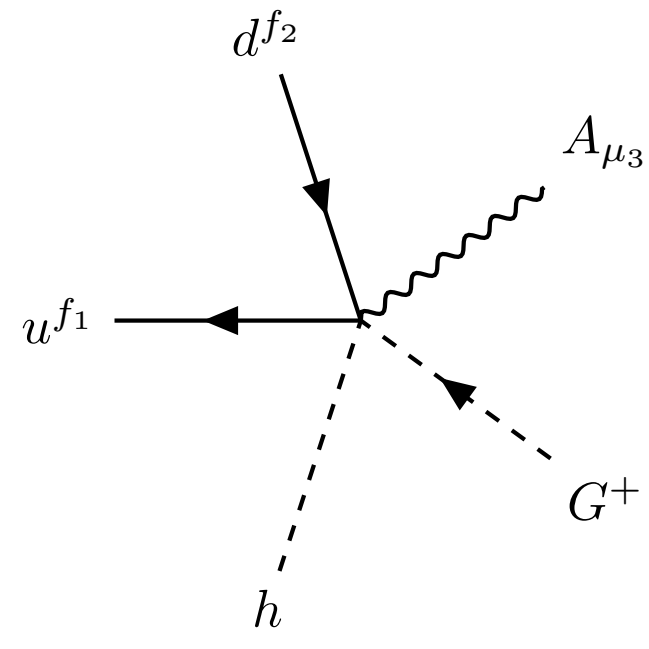

$$\begin{aligned}&+\frac{i\sqrt{2}\hat{g}s_\beta c_\beta\hat{g}'\gamma^{\mu_3}\mathcal{P}_R}{\sqrt{\hat{g}'^2+\hat{g}^2}}\left(\hat{C}^{(21)}_{\Phi ud,f_1f_2}\right)\\&-\frac{i\sqrt{2}\hat{g}\hat{g}'V_{f_1g_1}\gamma^{\mu_3}\mathcal{P}_L}{\sqrt{\hat{g}'^2+\hat{g}^2}}\left(c_\beta^2\hat{C}^{(11)[3]}_{\Phi q,g_1f_2}+s_\beta^2\hat{C}^{(22)[3]}_{\Phi q,g_1f_2}\right)\end{aligned}\tag{C.363}$$

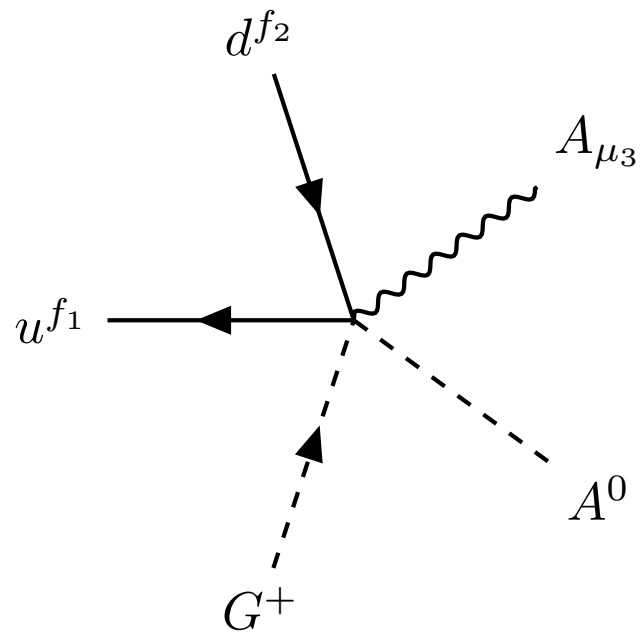

$$\begin{aligned}&-\frac{\hat{g}c_{2\beta}\hat{g}'\gamma^{\mu_3}\mathcal{P}_R}{\sqrt{2}\sqrt{\hat{g}'^2+\hat{g}^2}}\left(\hat{C}^{(21)}_{\Phi ud,f_1f_2}\right)\\&+\frac{\hat{g}s_{2\beta}\hat{g}'V_{f_1g_1}\gamma^{\mu_3}\mathcal{P}_L}{\sqrt{2}\sqrt{\hat{g}'^2+\hat{g}^2}}\left(\hat{C}^{(11)[3]}_{\Phi q,g_1f_2}-\hat{C}^{(22)[3]}_{\Phi q,g_1f_2}\right)\end{aligned}\tag{C.364}$$

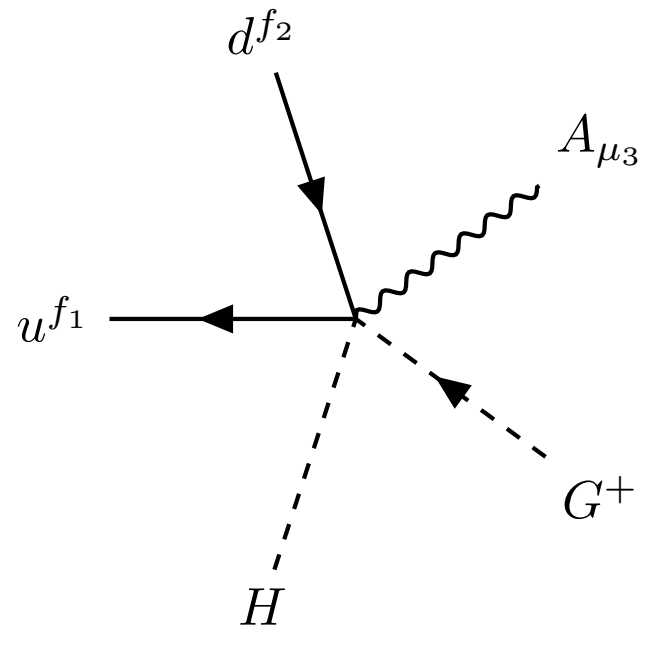

$$\begin{aligned}&-\frac{i\hat{g}c_{2\beta}\hat{g}'\gamma^{\mu_3}\mathcal{P}_R}{\sqrt{2}\sqrt{\hat{g}'^2+\hat{g}^2}}\left(\hat{C}^{(21)}_{\Phi ud,f_1f_2}\right)\\&-\frac{i\hat{g}s_{2\beta}\hat{g}'V_{f_1g_1}\gamma^{\mu_3}\mathcal{P}_L}{\sqrt{2}\sqrt{\hat{g}'^2+\hat{g}^2}}\left(\hat{C}^{(11)[3]}_{\Phi q,g_1f_2}-\hat{C}^{(22)[3]}_{\Phi q,g_1f_2}\right)\end{aligned} \tag{C.365}$$

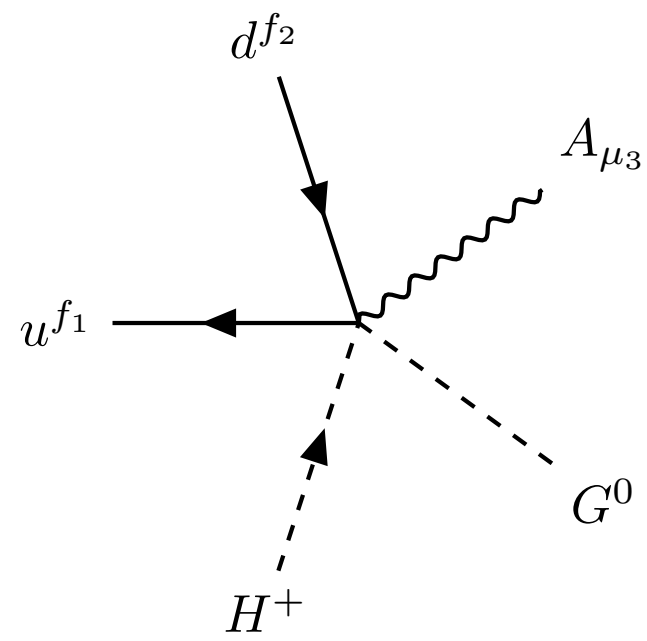

$$\begin{aligned}&-\frac{\hat{g}c_{2\beta}\hat{g}'\gamma^{\mu_3}\mathcal{P}_R}{\sqrt{2}\sqrt{\hat{g}'^2+\hat{g}^2}}\left(\hat{C}^{(21)}_{\Phi ud,f_1f_2}\right)\\&+\frac{\hat{g}s_{2\beta}\hat{g}'V_{f_1g_1}\gamma^{\mu_3}\mathcal{P}_L}{\sqrt{2}\sqrt{\hat{g}'^2+\hat{g}^2}}\left(\hat{C}^{(11)[3]}_{\Phi q,g_1f_2}-\hat{C}^{(22)[3]}_{\Phi q,g_1f_2}\right)\end{aligned} \tag{C.366}$$

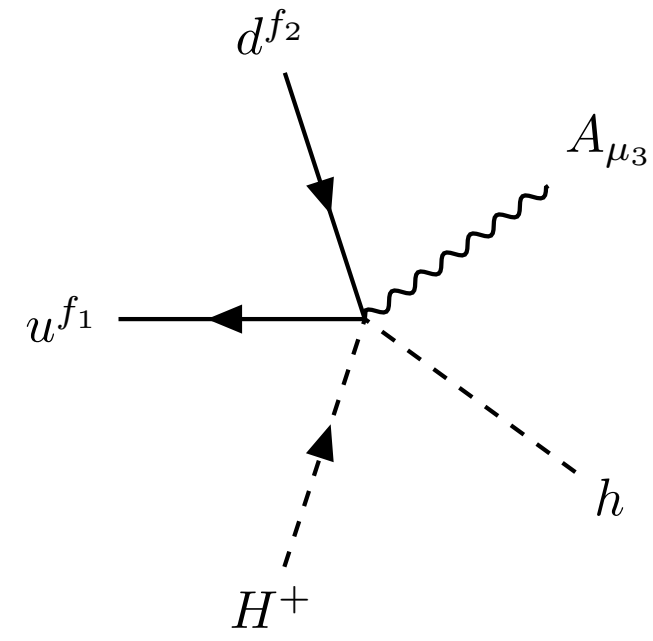

$$\begin{aligned}&+\frac{i\hat{g}c_{2\beta}\hat{g}'\gamma^{\mu_3}\mathcal{P}_R}{\sqrt{2}\sqrt{\hat{g}'^2+\hat{g}^2}}\left(\hat{C}^{(21)}_{\Phi ud,f_1f_2}\right)\\&+\frac{i\hat{g}s_{2\beta}\hat{g}'V_{f_1g_1}\gamma^{\mu_3}\mathcal{P}_L}{\sqrt{2}\sqrt{\hat{g}'^2+\hat{g}^2}}\left(\hat{C}^{(11)[3]}_{\Phi q,g_1f_2}-\hat{C}^{(22)[3]}_{\Phi q,g_1f_2}\right)\end{aligned} \tag{C.367}$$

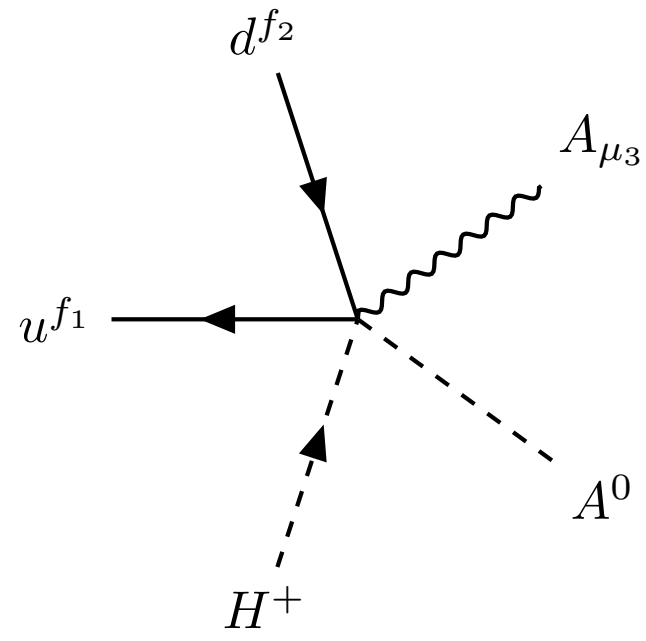

$$
\begin{aligned}
&+\frac{\sqrt{2}\hat{g}s_\beta c_\beta \hat{g}'\gamma^{\mu_3}\mathcal{P}_R}{\sqrt{\hat{g}'^2+\hat{g}^2}}\left(\hat{C}^{(21)}_{\Phi ud,f_1f_2}\right) \\
&-\frac{\sqrt{2}\hat{g}\hat{g}'V_{f_1g_1}\gamma^{\mu_3}\mathcal{P}_L}{\sqrt{\hat{g}'^2+\hat{g}^2}}\left(s_\beta^2\hat{C}^{(11)[3]}_{\Phi q,g_1f_2}+c_\beta^2\hat{C}^{(22)[3]}_{\Phi q,g_1f_2}\right)
\end{aligned}
\tag{C.368}
$$

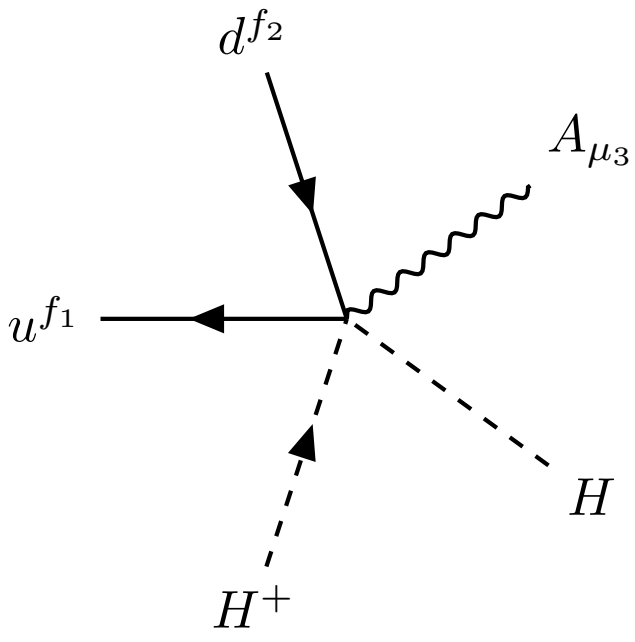

$$
\begin{aligned}
&+\frac{i\sqrt{2}\hat{g}s_\beta c_\beta \hat{g}'\gamma^{\mu_3}\mathcal{P}_R}{\sqrt{\hat{g}'^2+\hat{g}^2}}\left(\hat{C}^{(21)}_{\Phi ud,f_1f_2}\right) \\
&+\frac{i\sqrt{2}\hat{g}\hat{g}'V_{f_1g_1}\gamma^{\mu_3}\mathcal{P}_L}{\sqrt{\hat{g}'^2+\hat{g}^2}}\left(s_\beta^2\hat{C}^{(11)[3]}_{\Phi q,g_1f_2}+c_\beta^2\hat{C}^{(22)[3]}_{\Phi q,g_1f_2}\right)
\end{aligned}
\tag{C.369}
$$

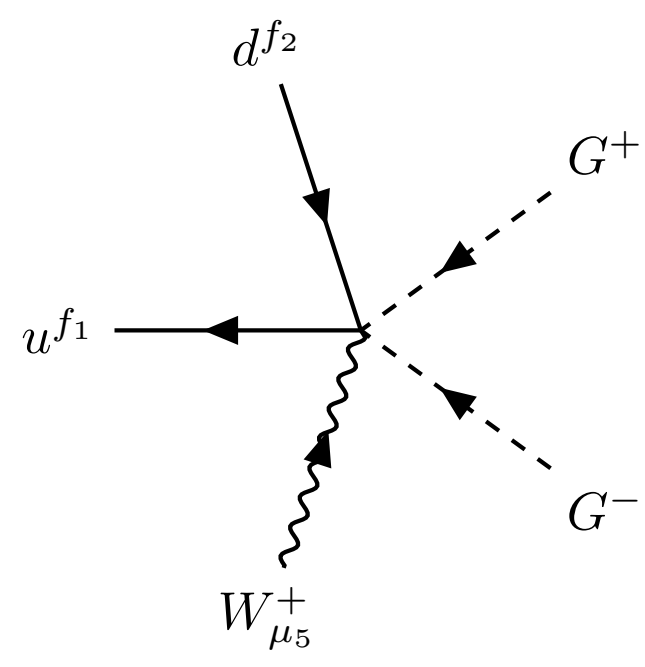

$$
-i\sqrt{2}\hat{g}V_{f_1g_1}\gamma^{\mu_5}\mathcal{P}_L\left(c_\beta^2\hat{C}^{(11)[3]}_{\Phi q,g_1f_2}+s_\beta^2\hat{C}^{(22)[3]}_{\Phi q,g_1f_2}\right)
\tag{C.370}
$$

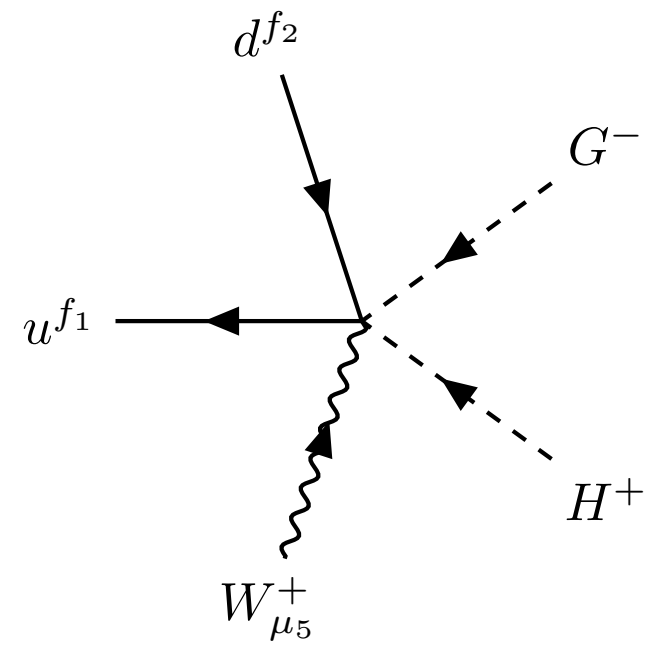

$$
+i\sqrt{2}\hat{g}s_\beta c_\beta V_{f_1g_1}\gamma^{\mu_5}\mathcal{P}_L\left(\hat{C}^{(11)[3]}_{\Phi q,g_1f_2}-\hat{C}^{(22)[3]}_{\Phi q,g_1f_2}\right)
\tag{C.371}
$$

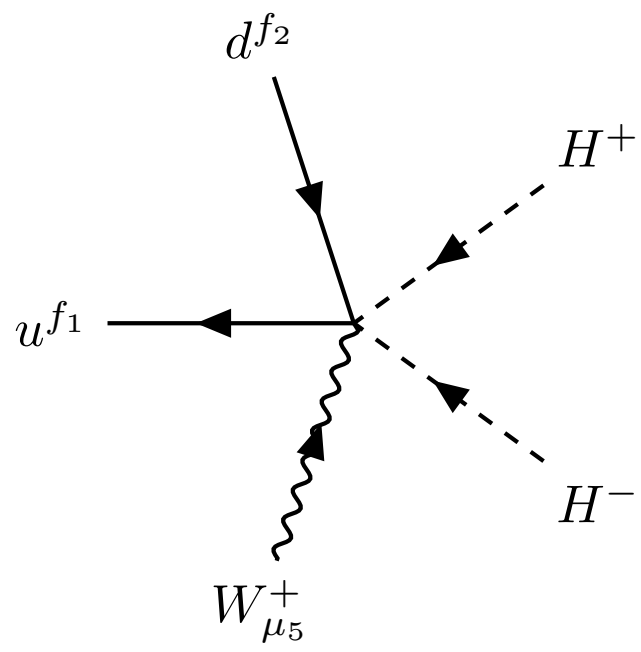

$$-i\sqrt{2}\hat{g}V_{f_1g_1}\gamma^{\mu_5}\mathcal{P}_L\left(s_\beta^2\hat{C}^{(11)[3]}_{\Phi q,g_1f_2}+c_\beta^2\hat{C}^{(22)[3]}_{\Phi q,g_1f_2}\right) \tag{C.372}$$

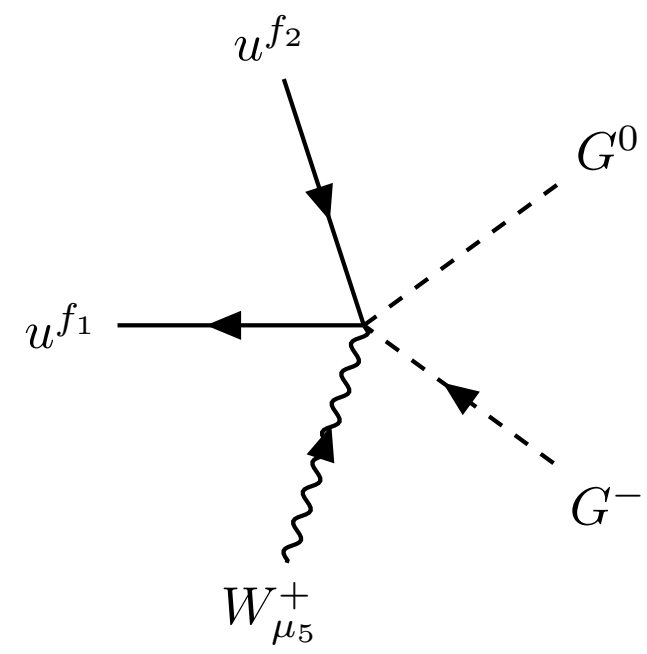

$$\begin{aligned}&+\hat{g}V_{f_1g_1}V_{f_2g_2*}\gamma^{\mu_5}\mathcal{P}_L\left(c_\beta^2\hat{C}^{(11)[1]}_{\Phi q,g_1g_2}+s_\beta^2\hat{C}^{(22)[1]}_{\Phi q,g_1g_2}\right)\\&+\hat{g}\gamma^{\mu_5}\mathcal{P}_R\left(c_\beta^2\hat{C}^{(11)}_{\Phi u,f_1f_2}+s_\beta^2\hat{C}^{(22)}_{\Phi u,f_1f_2}\right)\end{aligned} \tag{C.373}$$

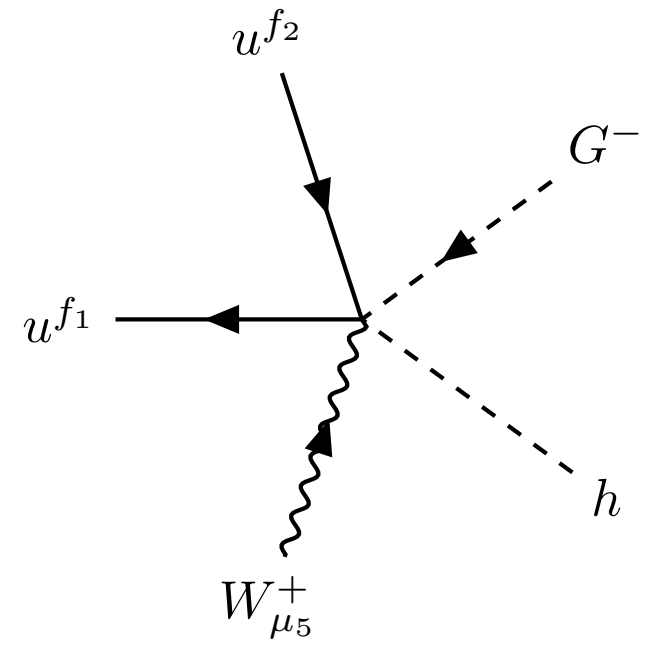

$$\begin{aligned}&-i\hat{g}V_{f_1g_1}V^*_{f_2g_2}\gamma^{\mu_5}\mathcal{P}_L\left(c_\beta^2\hat{C}^{(11)[1]}_{\Phi q,g_1g_2}+s_\beta^2\hat{C}^{(22)[1]}_{\Phi q,g_1g_2}\right)\\&-i\hat{g}\gamma^{\mu_5}\mathcal{P}_R\left(c_\beta^2\hat{C}^{(11)}_{\Phi u,f_1f_2}+s_\beta^2\hat{C}^{(22)}_{\Phi u,f_1f_2}\right)\end{aligned} \tag{C.374}$$

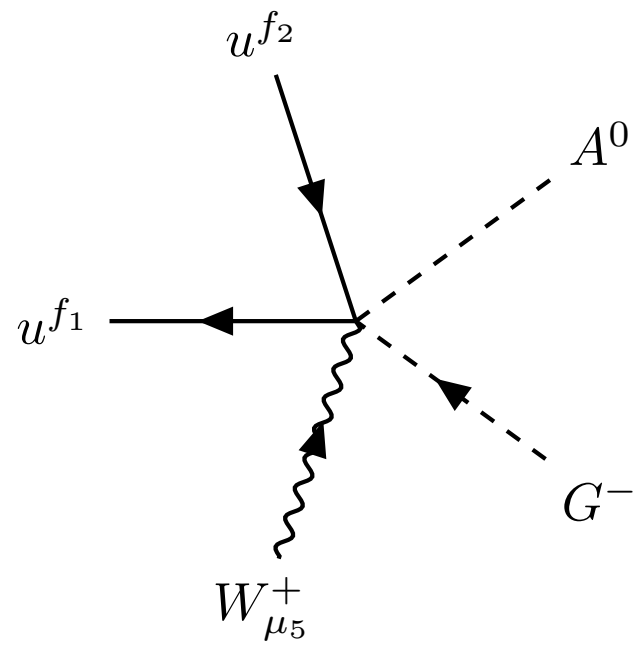

$$
\begin{aligned}
&+\hat{g} s_\beta c_\beta V_{f_1 g_1} V^*_{f_2 g_2} \gamma^{\mu_5} \mathcal{P}_L \left( \hat{C}^{(22)[1]}_{\Phi q, g_1 g_2} - \hat{C}^{(11)[1]}_{\Phi q, g_1 g_2} \right) \\
&+\hat{g} s_\beta c_\beta \gamma^{\mu_5} \mathcal{P}_R \left( \hat{C}^{(22)}_{\Phi u, f_1 f_2} - \hat{C}^{(11)}_{\Phi u, f_1 f_2} \right)
\end{aligned}
\tag{C.375}
$$

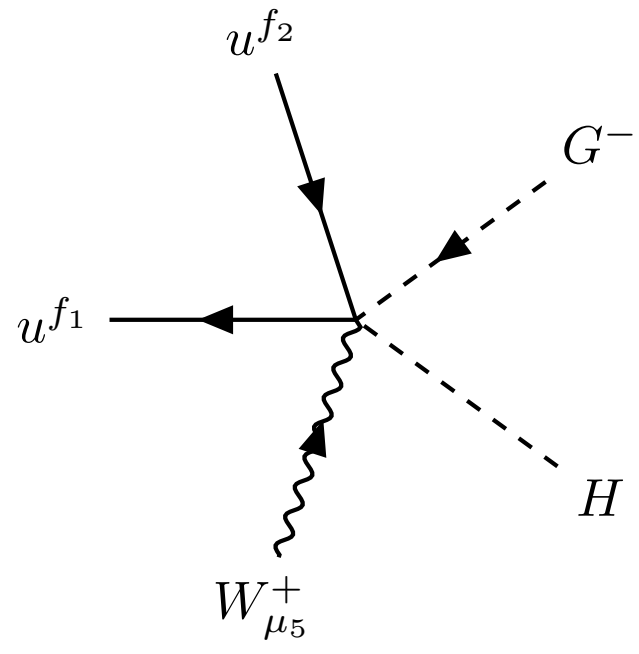

$$
\begin{aligned}
&-\frac{1}{2} i \hat{g} s_{2\beta} V_{f_1 g_1} V^*_{f_2 g_2} \gamma^{\mu_5} \mathcal{P}_L \left( \hat{C}^{(11)[1]}_{\Phi q, g_1 g_2} - \hat{C}^{(22)[1]}_{\Phi q, g_1 g_2} \right) \\
&-\frac{1}{2} i \hat{g} s_{2\beta} \gamma^{\mu_5} \mathcal{P}_R \left( \hat{C}^{(11)}_{\Phi u, f_1 f_2} - \hat{C}^{(22)}_{\Phi u, f_1 f_2} \right)
\end{aligned}
\tag{C.376}
$$

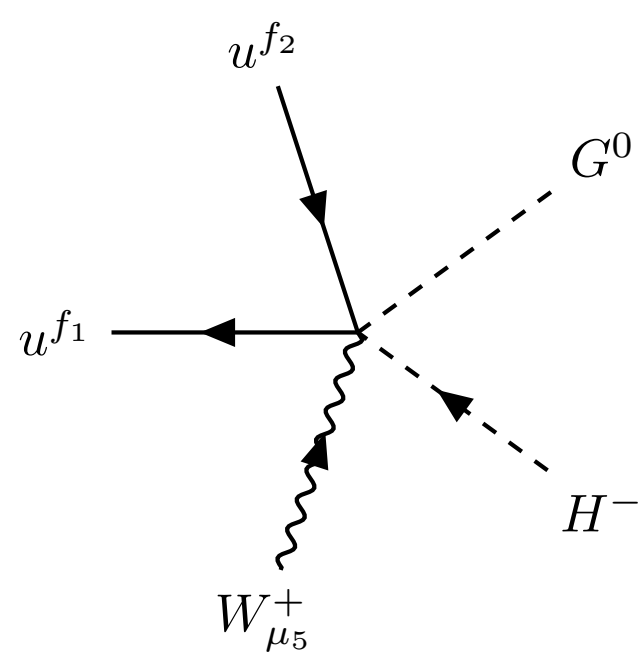

$$
\begin{aligned}
&+\hat{g} s_\beta c_\beta V_{f_1 g_1} V^*_{f_2 g_2} \gamma^{\mu_5} \mathcal{P}_L \left( \hat{C}^{(22)[1]}_{\Phi q, g_1 g_2} - \hat{C}^{(11)[1]}_{\Phi q, g_1 g_2} \right) \\
&+\hat{g} s_\beta c_\beta \gamma^{\mu_5} \mathcal{P}_R \left( \hat{C}^{(22)}_{\Phi u, f_1 f_2} - \hat{C}^{(11)}_{\Phi u, f_1 f_2} \right)
\end{aligned}
\tag{C.377}
$$

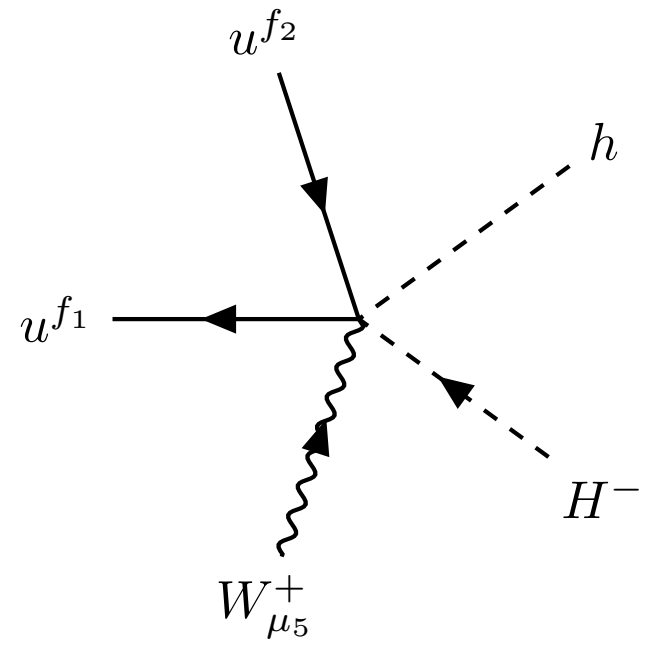

$$
\begin{aligned}
&+\frac{1}{2} i \hat{g} s_{2\beta} V_{f_1 g_1} V^*_{f_2 g_2} \gamma^{\mu_5} \mathcal{P}_L \left( \hat{C}^{(11)[1]}_{\Phi q, g_1 g_2} - \hat{C}^{(22)[1]}_{\Phi q, g_1 g_2} \right) \\
&+\frac{1}{2} i \hat{g} s_{2\beta} \gamma^{\mu_5} \mathcal{P}_R \left( \hat{C}^{(11)}_{\Phi u, f_1 f_2} - \hat{C}^{(22)}_{\Phi u, f_1 f_2} \right)
\end{aligned}
\tag{C.378}
$$

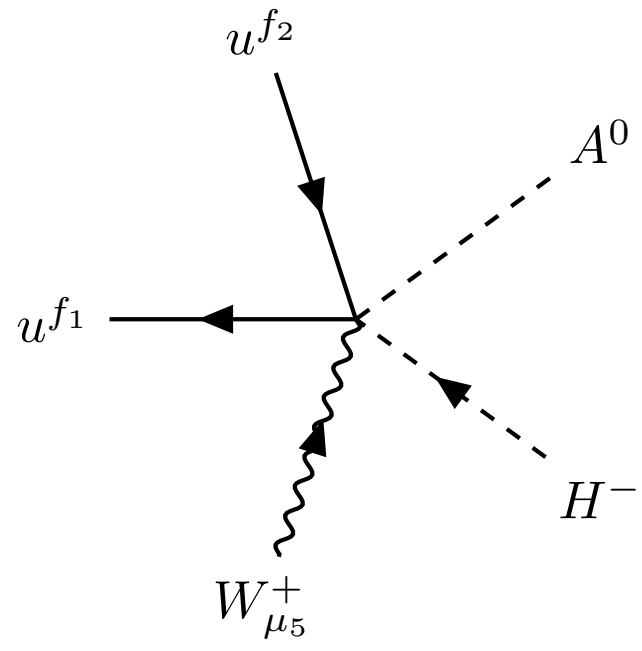

$$
\begin{aligned}
&+\hat{g}V_{f_1g_1}V^*_{f_2g_2}\gamma^{\mu_5}\mathcal{P}_L\left(s^2_\beta\hat{C}^{(11)[1]}_{\Phi q,g_1g_2}+c^2_\beta\hat{C}^{(22)[1]}_{\Phi q,g_1g_2}\right)\\
&+\hat{g}\gamma^{\mu_5}\mathcal{P}_R\left(s^2_\beta\hat{C}^{(11)}_{\Phi u,f_1f_2}+c^2_\beta\hat{C}^{(22)}_{\Phi u,f_1f_2}\right)
\end{aligned}
\tag{C.379}
$$

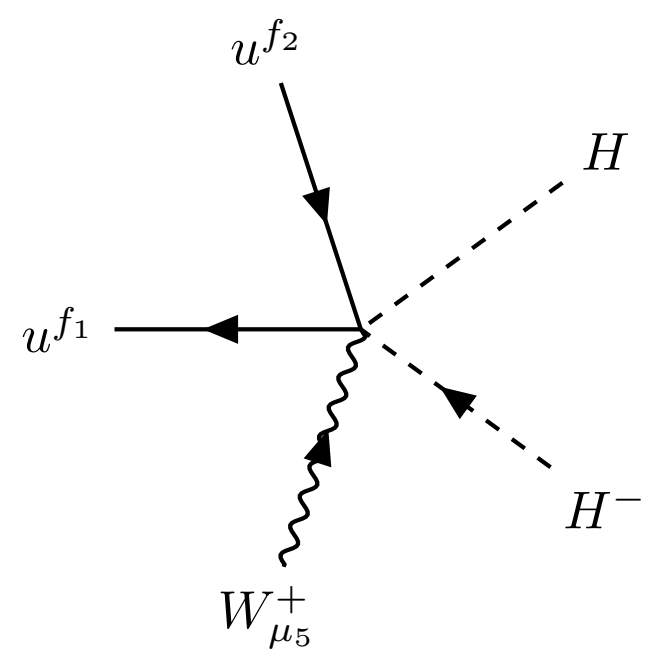

$$
\begin{aligned}
&+i\hat{g}V_{f_1g_1}V^*_{f_2g_2}\gamma^{\mu_5}\mathcal{P}_L\left(s^2_\beta\hat{C}^{(11)[1]}_{\Phi q,g_1g_2}+c^2_\beta\hat{C}^{(22)[1]}_{\Phi q,g_1g_2}\right)\\
&+i\hat{g}\gamma^{\mu_5}\mathcal{P}_R\left(s^2_\beta\hat{C}^{(11)}_{\Phi u,f_1f_2}+c^2_\beta\hat{C}^{(22)}_{\Phi u,f_1f_2}\right)
\end{aligned}
\tag{C.380}
$$

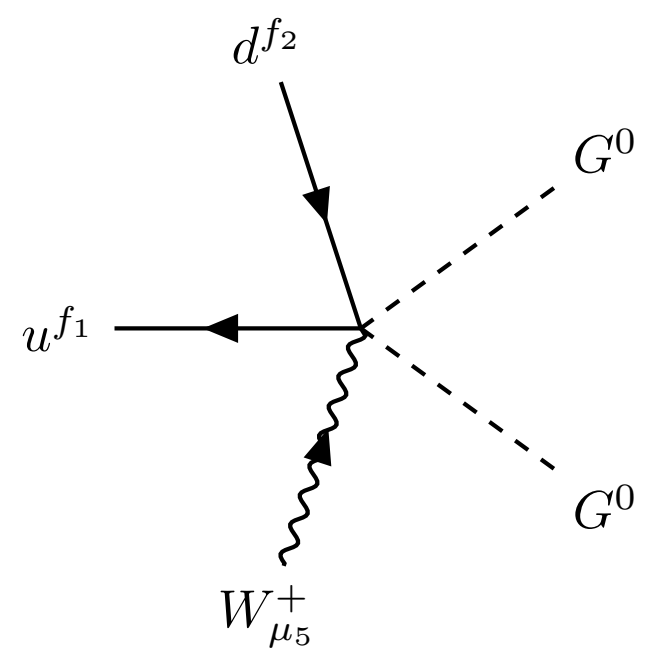

$$
\begin{aligned}
&-i\sqrt{2}\hat{g}s_\beta c_\beta\gamma^{\mu_5}\mathcal{P}_R\left(\hat{C}^{(21)}_{\Phi ud,f_1f_2}\right)\\
&-i\sqrt{2}\hat{g}V_{f_1g_1}\gamma^{\mu_5}\mathcal{P}_L\left(c^2_\beta\hat{C}^{(11)[3]}_{\Phi q,g_1f_2}+s^2_\beta\hat{C}^{(22)[3]}_{\Phi q,g_1f_2}\right)
\end{aligned}
\tag{C.381}
$$

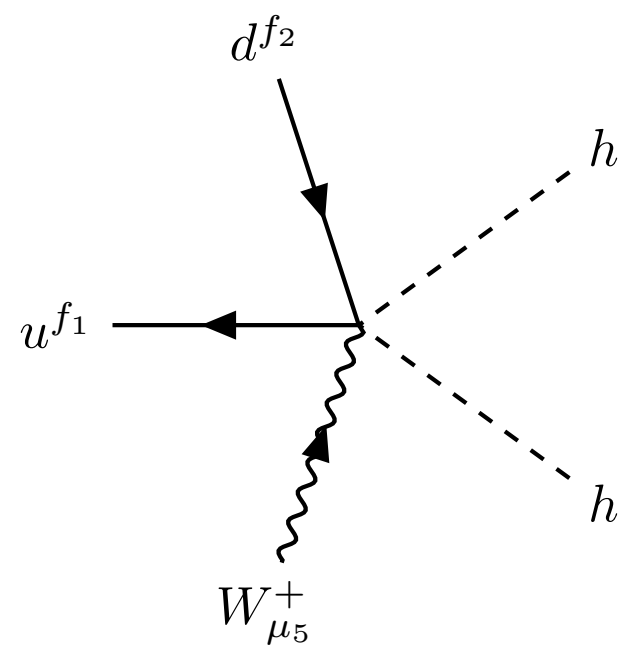

$$+i\sqrt{2}\hat{g}s_\beta c_\beta\gamma^{\mu_5}\mathcal{P}_R\left(\hat{C}^{(21)}_{\Phi ud,f_1f_2}\right)$$
$$-i\sqrt{2}\hat{g}V_{f_1g_1}\gamma^{\mu_5}\mathcal{P}_L\left(c^2_\beta\hat{C}^{(11)[3]}_{\Phi q,g_1f_2}+s^2_\beta\hat{C}^{(22)[3]}_{\Phi q,g_1f_2}\right) \tag{C.382}$$

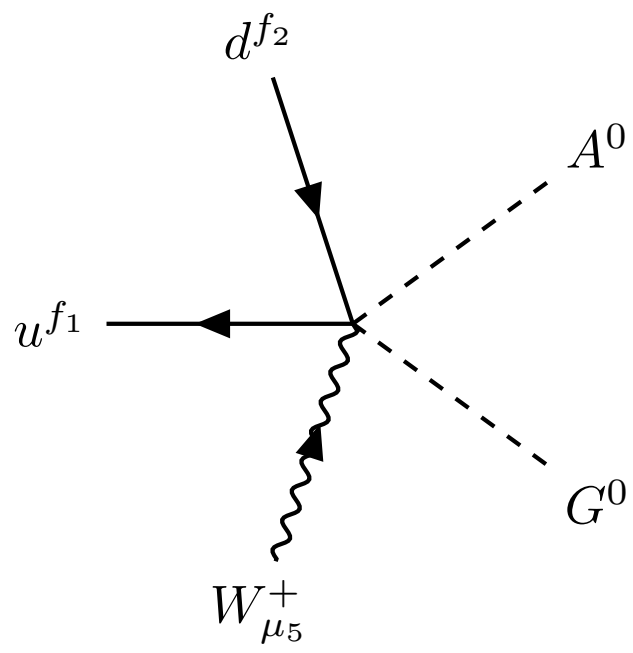

$$-\frac{i\hat{g}c_{2\beta}\gamma^{\mu_5}\mathcal{P}_R}{\sqrt{2}}\left(\hat{C}^{(21)}_{\Phi ud,f_1f_2}\right)$$
$$+\frac{i\hat{g}s_{2\beta}V_{f_1g_1}\gamma^{\mu_5}\mathcal{P}_L}{\sqrt{2}}\left(\hat{C}^{(11)[3]}_{\Phi q,g_1f_2}-\hat{C}^{(22)[3]}_{\Phi q,g_1f_2}\right) \tag{C.383}$$

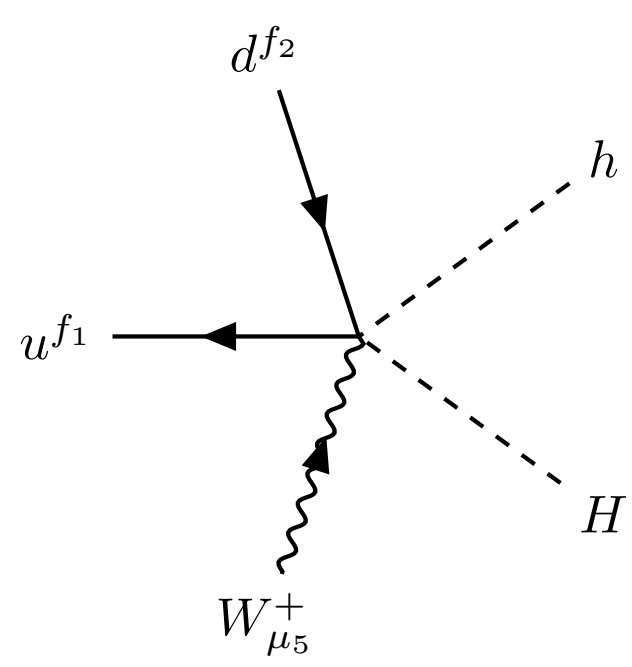

$$-\frac{i\hat{g}c_{2\beta}\gamma^{\mu_5}\mathcal{P}_R}{\sqrt{2}}\left(\hat{C}^{(21)}_{\Phi ud,f_1f_2}\right)$$
$$-\frac{i\hat{g}s_{2\beta}V_{f_1g_1}\gamma^{\mu_5}\mathcal{P}_L}{\sqrt{2}}\left(\hat{C}^{(11)[3]}_{\Phi q,g_1f_2}-\hat{C}^{(22)[3]}_{\Phi q,g_1f_2}\right) \tag{C.384}$$

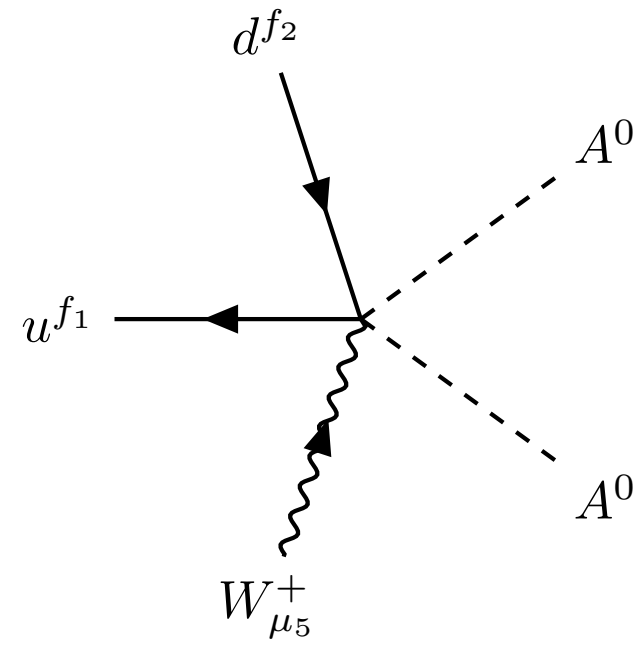

$$+i\sqrt{2}\hat{g}s_\beta c_\beta\gamma^{\mu_5}\mathcal{P}_R\left(\hat{C}^{(21)}_{\Phi ud,f_1f_2}\right)$$
$$-i\sqrt{2}\hat{g}V_{f_1g_1}\gamma^{\mu_5}\mathcal{P}_L\left(s^2_\beta\hat{C}^{(11)[3]}_{\Phi q,g_1f_2}+c^2_\beta\hat{C}^{(22)[3]}_{\Phi q,g_1f_2}\right) \tag{C.385}$$

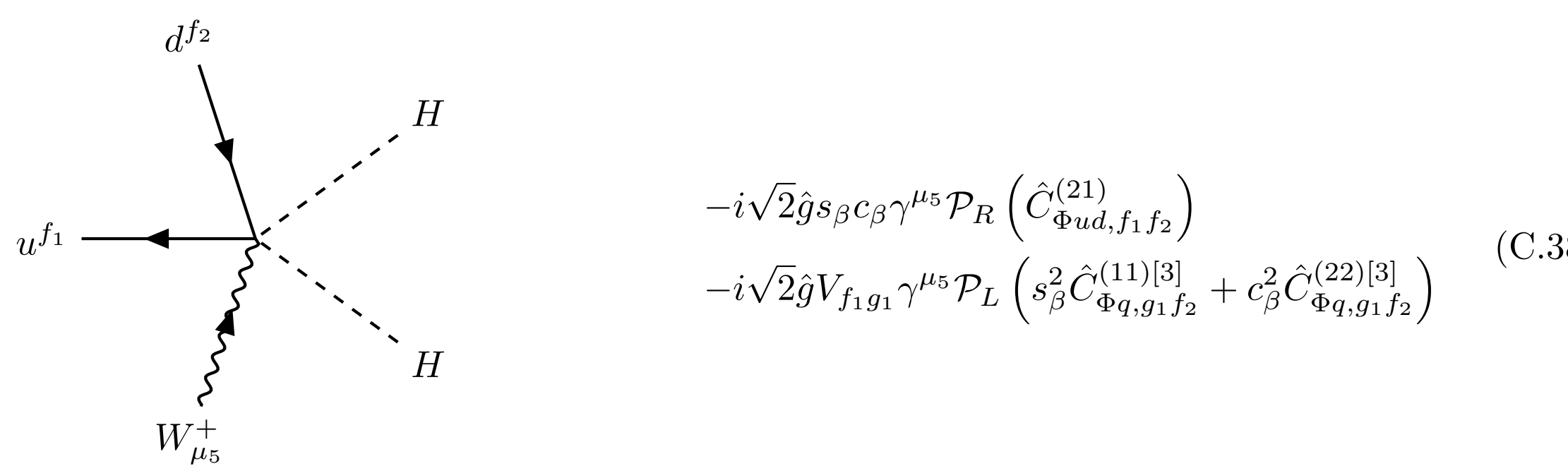

$$
\begin{aligned}
&-i\sqrt{2}\hat{g}s_\beta c_\beta\gamma^{\mu_5}\mathcal{P}_R\left(\hat{C}^{(21)}_{\Phi ud,f_1f_2}\right)\\
&-i\sqrt{2}\hat{g}V_{f_1g_1}\gamma^{\mu_5}\mathcal{P}_L\left(s_\beta^2\hat{C}^{(11)[3]}_{\Phi q,g_1f_2}+c_\beta^2\hat{C}^{(22)[3]}_{\Phi q,g_1f_2}\right)
\end{aligned}
\tag{C.386}
$$

$u^{f_2}$, $G^+$, $u^{f_1}$, $G^-$, $Z_{\mu_5}$

$$
\begin{aligned}
&+\frac{i\left(\hat{g}'^2-\hat{g}^2\right)V_{f_1g_1}V^*_{f_2g_2}\gamma^{\mu_5}\mathcal{P}_L}{\sqrt{\hat{g}'^2+\hat{g}^2}}\left(c_\beta^2\hat{C}^{(11)[1]}_{\Phi q,g_1g_2}+c_\beta^2\hat{C}^{(11)[3]}_{\Phi q,g_1g_2}+s_\beta^2\left(\hat{C}^{(22)[1]}_{\Phi q,g_1g_2}+\hat{C}^{(22)[3]}_{\Phi q,g_1g_2}\right)\right)\\
&+\frac{i\left(\hat{g}'^2-\hat{g}^2\right)\gamma^{\mu_5}\mathcal{P}_R}{\sqrt{\hat{g}'^2+\hat{g}^2}}\left(c_\beta^2\hat{C}^{(11)}_{\Phi u,f_1f_2}+s_\beta^2\hat{C}^{(22)}_{\Phi u,f_1f_2}\right)
\end{aligned}
\tag{C.387}
$$

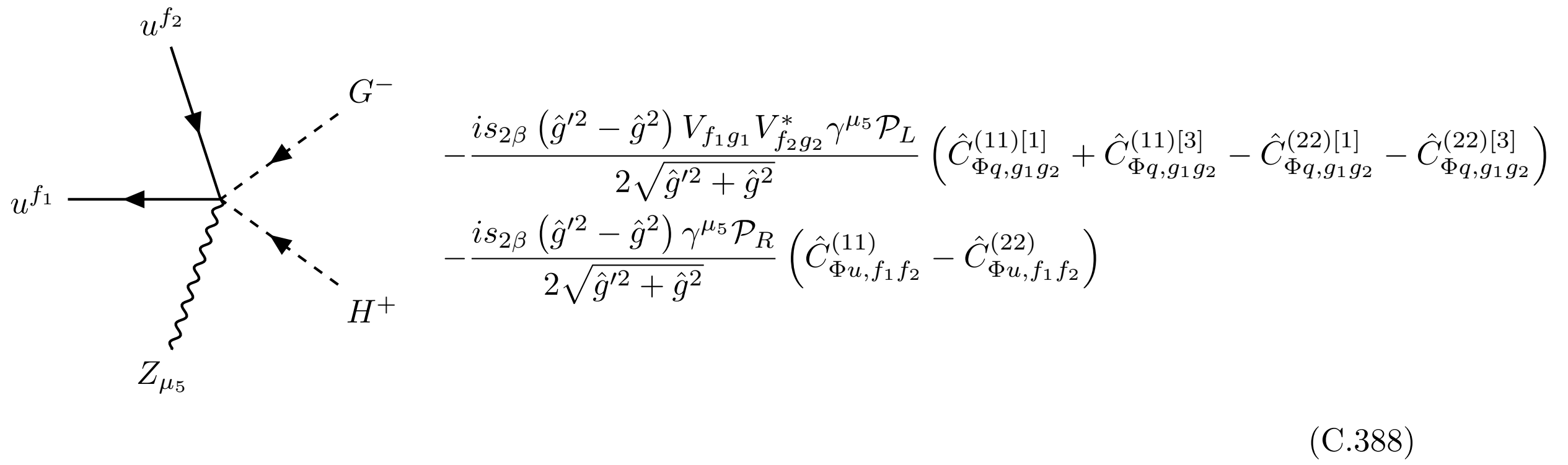

$$
\begin{aligned}
&-\frac{is_{2\beta}\left(\hat{g}'^2-\hat{g}^2\right)V_{f_1g_1}V^*_{f_2g_2}\gamma^{\mu_5}\mathcal{P}_L}{2\sqrt{\hat{g}'^2+\hat{g}^2}}\left(\hat{C}^{(11)[1]}_{\Phi q,g_1g_2}+\hat{C}^{(11)[3]}_{\Phi q,g_1g_2}-\hat{C}^{(22)[1]}_{\Phi q,g_1g_2}-\hat{C}^{(22)[3]}_{\Phi q,g_1g_2}\right)\\
&-\frac{is_{2\beta}\left(\hat{g}'^2-\hat{g}^2\right)\gamma^{\mu_5}\mathcal{P}_R}{2\sqrt{\hat{g}'^2+\hat{g}^2}}\left(\hat{C}^{(11)}_{\Phi u,f_1f_2}-\hat{C}^{(22)}_{\Phi u,f_1f_2}\right)
\end{aligned}
\tag{C.388}
$$

$$
\begin{aligned}
&+\frac{i\left(\hat{g}'^2-\hat{g}^2\right)V_{f_1g_1}V^*_{f_2g_2}\gamma^{\mu_5}\mathcal{P}_L}{\sqrt{\hat{g}'^2+\hat{g}^2}}\left(s_\beta^2\left(\hat{C}^{(11)[1]}_{\Phi q,g_1g_2}+\hat{C}^{(11)[3]}_{\Phi q,g_1g_2}\right)\right.\\
&\qquad\left.+c_\beta^2\hat{C}^{(22)[1]}_{\Phi q,g_1g_2}+c_\beta^2\hat{C}^{(22)[3]}_{\Phi q,g_1g_2}\right)\\
&+\frac{i\left(\hat{g}'^2-\hat{g}^2\right)\gamma^{\mu_5}\mathcal{P}_R}{\sqrt{\hat{g}'^2+\hat{g}^2}}\left(s_\beta^2\hat{C}^{(11)}_{\Phi u,f_1f_2}+c_\beta^2\hat{C}^{(22)}_{\Phi u,f_1f_2}\right)
\end{aligned}
\tag{C.389}
$$

$$
\begin{aligned}
&+i\sqrt{\hat{g}'^2+\hat{g}^2}V_{f_1g_1}V^*_{f_2g_2}\gamma^{\mu_5}\mathcal{P}_L\left(c_\beta^2\hat{C}^{(11)[1]}_{\Phi q,g_1g_2}-c_\beta^2\hat{C}^{(11)[3]}_{\Phi q,g_1g_2}\right.\\
&\qquad\left.+s_\beta^2\left(\hat{C}^{(22)[1]}_{\Phi q,g_1g_2}-\hat{C}^{(22)[3]}_{\Phi q,g_1g_2}\right)\right)\\
&+i\sqrt{\hat{g}'^2+\hat{g}^2}\gamma^{\mu_5}\mathcal{P}_R\left(c_\beta^2\hat{C}^{(11)}_{\Phi u,f_1f_2}+s_\beta^2\hat{C}^{(22)}_{\Phi u,f_1f_2}\right)
\end{aligned}
\tag{C.390}
$$

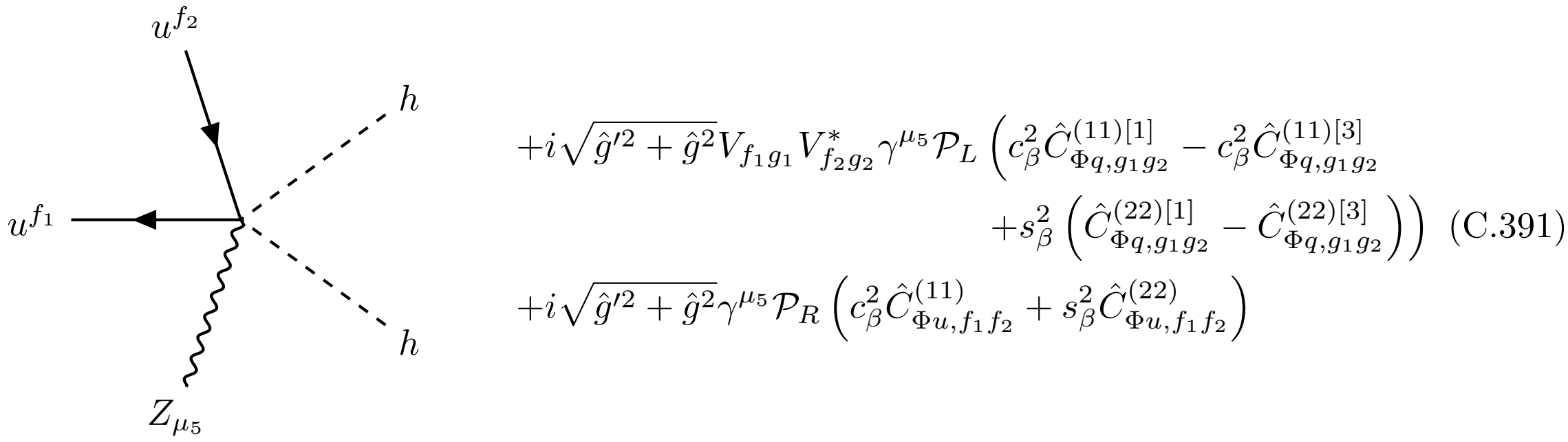

$$
\begin{aligned}
&+i\sqrt{\hat{g}'^2+\hat{g}^2}V_{f_1g_1}V^*_{f_2g_2}\gamma^{\mu_5}\mathcal{P}_L\left(c_\beta^2\hat{C}^{(11)[1]}_{\Phi q,g_1g_2}-c_\beta^2\hat{C}^{(11)[3]}_{\Phi q,g_1g_2}\right.\\
&\qquad\left.+s_\beta^2\left(\hat{C}^{(22)[1]}_{\Phi q,g_1g_2}-\hat{C}^{(22)[3]}_{\Phi q,g_1g_2}\right)\right)\\
&+i\sqrt{\hat{g}'^2+\hat{g}^2}\gamma^{\mu_5}\mathcal{P}_R\left(c_\beta^2\hat{C}^{(11)}_{\Phi u,f_1f_2}+s_\beta^2\hat{C}^{(22)}_{\Phi u,f_1f_2}\right)
\end{aligned}
\tag{C.391}
$$

$$
\begin{aligned}
&-\frac{1}{2}is_{2\beta}\sqrt{\hat{g}'^2+\hat{g}^2}V_{f_1g_1}V^*_{f_2g_2}\gamma^{\mu_5}\mathcal{P}_L\left(\hat{C}^{(11)[1]}_{\Phi q,g_1g_2}-\hat{C}^{(11)[3]}_{\Phi q,g_1g_2}-\hat{C}^{(22)[1]}_{\Phi q,g_1g_2}+\hat{C}^{(22)[3]}_{\Phi q,g_1g_2}\right)\\
&-\frac{1}{2}is_{2\beta}\sqrt{\hat{g}'^2+\hat{g}^2}\gamma^{\mu_5}\mathcal{P}_R\left(\hat{C}^{(11)}_{\Phi u,f_1f_2}-\hat{C}^{(22)}_{\Phi u,f_1f_2}\right)
\end{aligned}
\tag{C.392}
$$

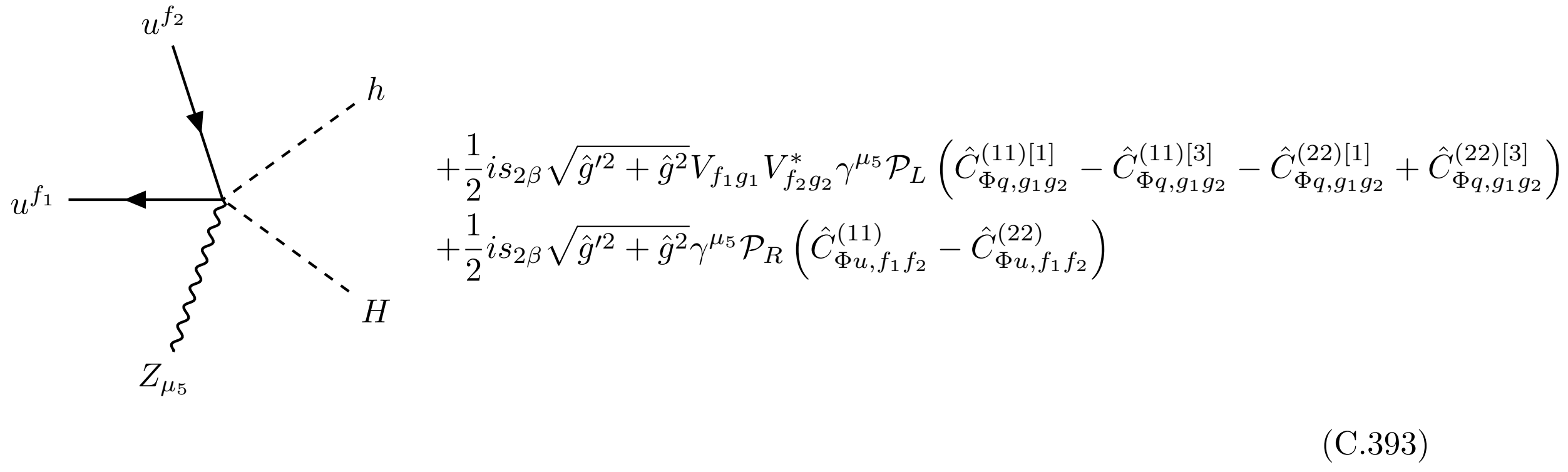

$$\begin{aligned}&+\frac{1}{2} i s_{2\beta}\sqrt{\hat{g}'^2+\hat{g}^2}\, V_{f_1g_1}V^*_{f_2g_2}\gamma^{\mu_5}\mathcal{P}_L\left(\hat{C}^{(11)[1]}_{\Phi q,g_1g_2}-\hat{C}^{(11)[3]}_{\Phi q,g_1g_2}-\hat{C}^{(22)[1]}_{\Phi q,g_1g_2}+\hat{C}^{(22)[3]}_{\Phi q,g_1g_2}\right)\\ &+\frac{1}{2} i s_{2\beta}\sqrt{\hat{g}'^2+\hat{g}^2}\,\gamma^{\mu_5}\mathcal{P}_R\left(\hat{C}^{(11)}_{\Phi u,f_1f_2}-\hat{C}^{(22)}_{\Phi u,f_1f_2}\right)\end{aligned} \tag{C.393}$$

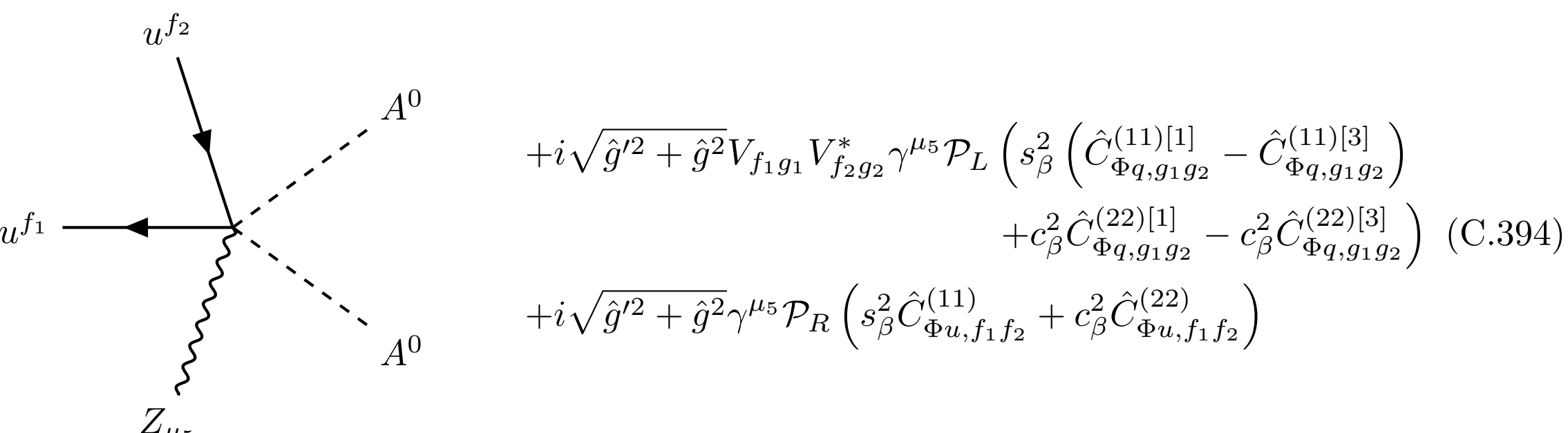

$$\begin{aligned}&+i\sqrt{\hat{g}'^2+\hat{g}^2}\, V_{f_1g_1}V^*_{f_2g_2}\gamma^{\mu_5}\mathcal{P}_L\left(s^2_\beta\left(\hat{C}^{(11)[1]}_{\Phi q,g_1g_2}-\hat{C}^{(11)[3]}_{\Phi q,g_1g_2}\right)\right.\\ &\qquad\qquad\left.+c^2_\beta\hat{C}^{(22)[1]}_{\Phi q,g_1g_2}-c^2_\beta\hat{C}^{(22)[3]}_{\Phi q,g_1g_2}\right)\\ &+i\sqrt{\hat{g}'^2+\hat{g}^2}\,\gamma^{\mu_5}\mathcal{P}_R\left(s^2_\beta\hat{C}^{(11)}_{\Phi u,f_1f_2}+c^2_\beta\hat{C}^{(22)}_{\Phi u,f_1f_2}\right)\end{aligned} \tag{C.394}$$

$u^{f_2}$, $H$, $u^{f_1}$, $H$, $Z_{\mu_5}$

$$\begin{aligned}&+i\sqrt{\hat{g}'^2+\hat{g}^2}\, V_{f_1g_1}V^*_{f_2g_2}\gamma^{\mu_5}\mathcal{P}_L\left(s^2_\beta\left(\hat{C}^{(11)[1]}_{\Phi q,g_1g_2}-\hat{C}^{(11)[3]}_{\Phi q,g_1g_2}\right)\right.\\ &\qquad\qquad\left.+c^2_\beta\hat{C}^{(22)[1]}_{\Phi q,g_1g_2}-c^2_\beta\hat{C}^{(22)[3]}_{\Phi q,g_1g_2}\right)\\ &+i\sqrt{\hat{g}'^2+\hat{g}^2}\,\gamma^{\mu_5}\mathcal{P}_R\left(s^2_\beta\hat{C}^{(11)}_{\Phi u,f_1f_2}+c^2_\beta\hat{C}^{(22)}_{\Phi u,f_1f_2}\right)\end{aligned} \tag{C.395}$$

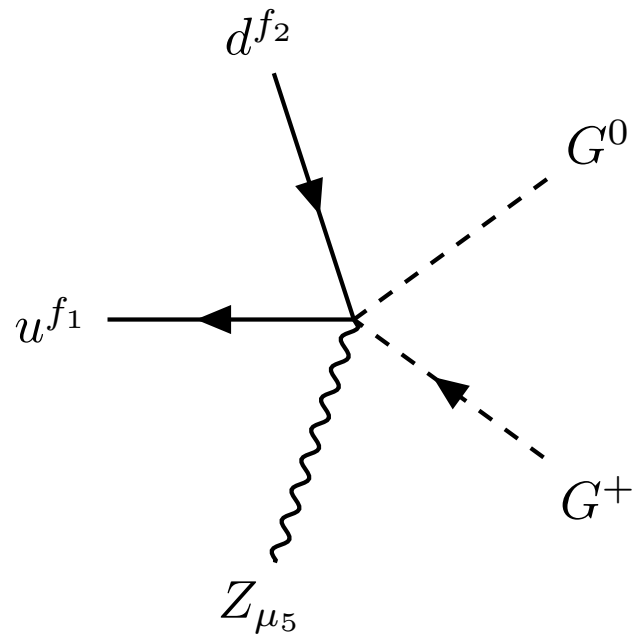

$$\begin{aligned}&-\frac{\sqrt{2}\hat{g}^2 s_\beta c_\beta \gamma^{\mu_5}\mathcal{P}_R}{\sqrt{\hat{g}'^2+\hat{g}^2}}\left(\hat{C}^{(21)}_{\Phi ud,f_1f_2}\right)\\ &+\frac{\sqrt{2}\hat{g}'^2 V_{f_1g_1}\gamma^{\mu_5}\mathcal{P}_L}{\sqrt{\hat{g}'^2+\hat{g}^2}}\left(c_\beta^2\hat{C}^{(11)[3]}_{\Phi q,g_1f_2}+s_\beta^2\hat{C}^{(22)[3]}_{\Phi q,g_1f_2}\right)\end{aligned} \tag{C.396}$$

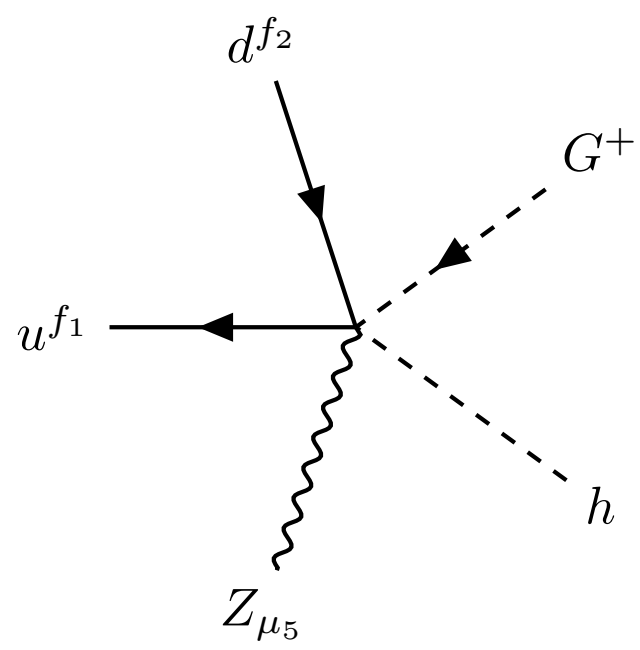

$$\begin{aligned}&+\frac{i\sqrt{2}\hat{g}^2 s_\beta c_\beta \gamma^{\mu_5}\mathcal{P}_R}{\sqrt{\hat{g}'^2+\hat{g}^2}}\left(\hat{C}^{(21)}_{\Phi ud,f_1f_2}\right)\\ &+\frac{i\sqrt{2}\hat{g}'^2 V_{f_1g_1}\gamma^{\mu_5}\mathcal{P}_L}{\sqrt{\hat{g}'^2+\hat{g}^2}}\left(c_\beta^2\hat{C}^{(11)[3]}_{\Phi q,g_1f_2}+s_\beta^2\hat{C}^{(22)[3]}_{\Phi q,g_1f_2}\right)\end{aligned} \tag{C.397}$$

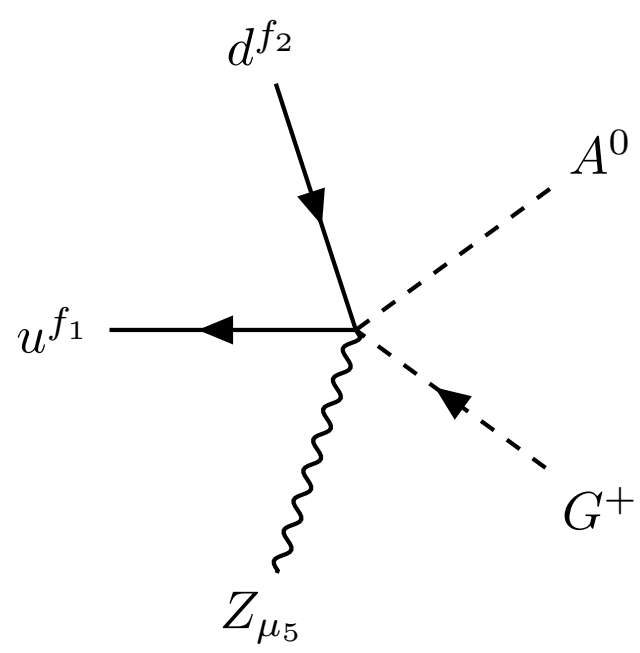

$$\begin{aligned}&-\frac{\hat{g}^2 c_{2\beta} \gamma^{\mu_5}\mathcal{P}_R}{\sqrt{2}\sqrt{\hat{g}'^2+\hat{g}^2}}\left(\hat{C}^{(21)}_{\Phi ud,f_1f_2}\right)\\ &+\frac{s_{2\beta}\hat{g}'^2 V_{f_1g_1}\gamma^{\mu_5}\mathcal{P}_L}{\sqrt{2}\sqrt{\hat{g}'^2+\hat{g}^2}}\left(\hat{C}^{(22)[3]}_{\Phi q,g_1f_2}-\hat{C}^{(11)[3]}_{\Phi q,g_1f_2}\right)\end{aligned} \tag{C.398}$$

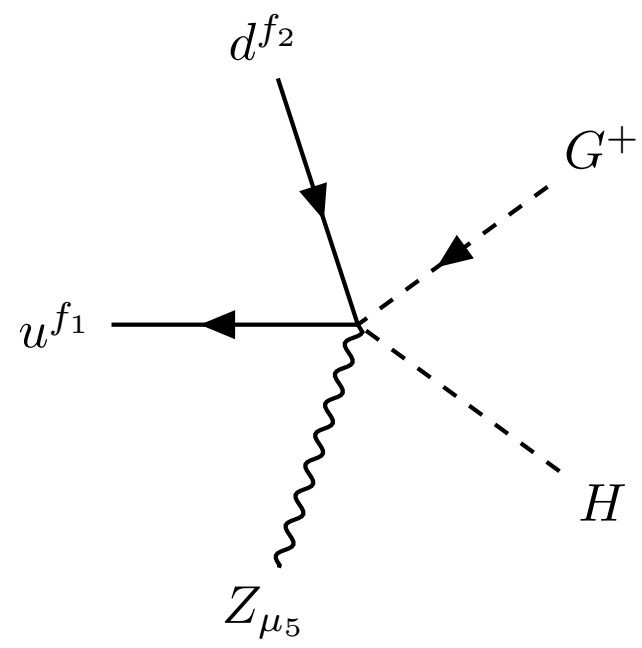

$$\begin{aligned}&-\frac{i\hat{g}^2 c_{2\beta} \gamma^{\mu_5}\mathcal{P}_R}{\sqrt{2}\sqrt{\hat{g}'^2+\hat{g}^2}}\left(\hat{C}^{(21)}_{\Phi ud,f_1f_2}\right)\\ &+\frac{is_{2\beta}\hat{g}'^2 V_{f_1g_1}\gamma^{\mu_5}\mathcal{P}_L}{\sqrt{2}\sqrt{\hat{g}'^2+\hat{g}^2}}\left(\hat{C}^{(11)[3]}_{\Phi q,g_1f_2}-\hat{C}^{(22)[3]}_{\Phi q,g_1f_2}\right)\end{aligned} \tag{C.399}$$

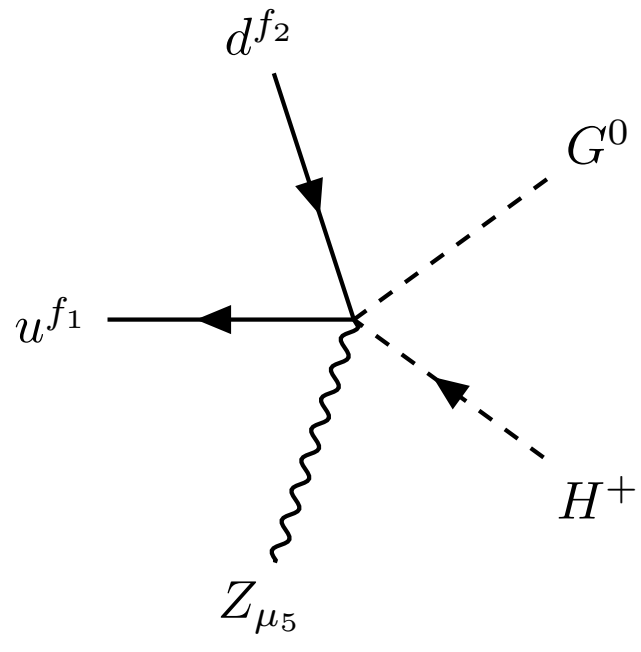

$$-\frac{\hat{g}^2 c_{2\beta}\gamma^{\mu_5}\mathcal{P}_R}{\sqrt{2}\sqrt{\hat{g}'^2+\hat{g}^2}}\left(\hat{C}^{(21)}_{\Phi ud,f_1f_2}\right)$$
$$+\frac{s_{2\beta}\hat{g}'^2 V_{f_1g_1}\gamma^{\mu_5}\mathcal{P}_L}{\sqrt{2}\sqrt{\hat{g}'^2+\hat{g}^2}}\left(\hat{C}^{(22)[3]}_{\Phi q,g_1f_2}-\hat{C}^{(11)[3]}_{\Phi q,g_1f_2}\right) \tag{C.400}$$

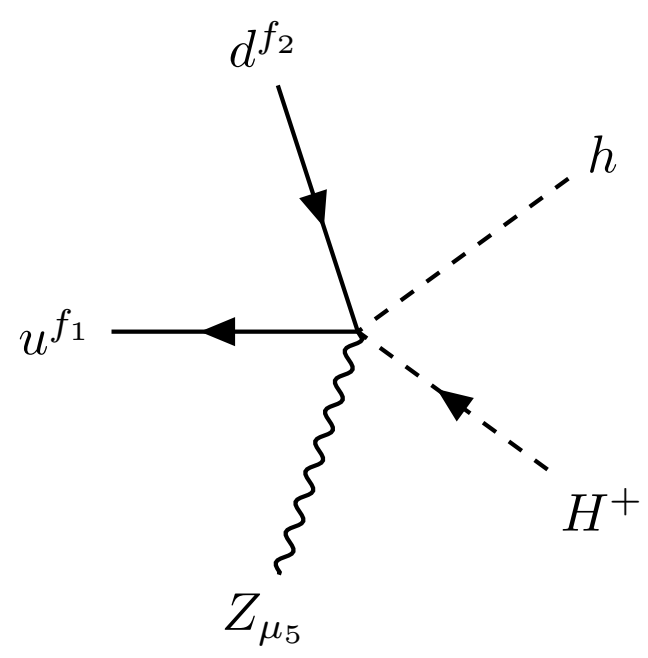

$$+\frac{i\hat{g}^2 c_{2\beta}\gamma^{\mu_5}\mathcal{P}_R}{\sqrt{2}\sqrt{\hat{g}'^2+\hat{g}^2}}\left(\hat{C}^{(21)}_{\Phi ud,f_1f_2}\right)$$
$$-\frac{is_{2\beta}\hat{g}'^2 V_{f_1g_1}\gamma^{\mu_5}\mathcal{P}_L}{\sqrt{2}\sqrt{\hat{g}'^2+\hat{g}^2}}\left(\hat{C}^{(11)[3]}_{\Phi q,g_1f_2}-\hat{C}^{(22)[3]}_{\Phi q,g_1f_2}\right) \tag{C.401}$$

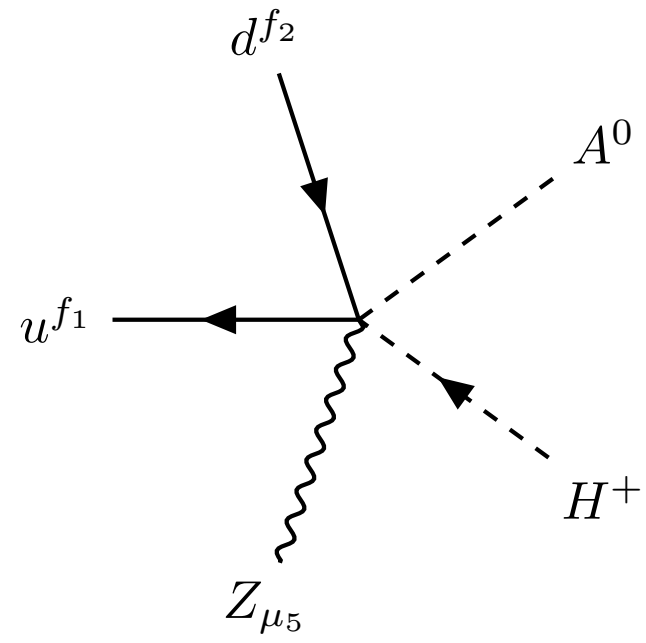

$$+\frac{\sqrt{2}\hat{g}^2 s_\beta c_\beta\gamma^{\mu_5}\mathcal{P}_R}{\sqrt{\hat{g}'^2+\hat{g}^2}}\left(\hat{C}^{(21)}_{\Phi ud,f_1f_2}\right)$$
$$+\frac{\sqrt{2}\hat{g}'^2 V_{f_1g_1}\gamma^{\mu_5}\mathcal{P}_L}{\sqrt{\hat{g}'^2+\hat{g}^2}}\left(s_\beta^2\hat{C}^{(11)[3]}_{\Phi q,g_1f_2}+c_\beta^2\hat{C}^{(22)[3]}_{\Phi q,g_1f_2}\right) \tag{C.402}$$

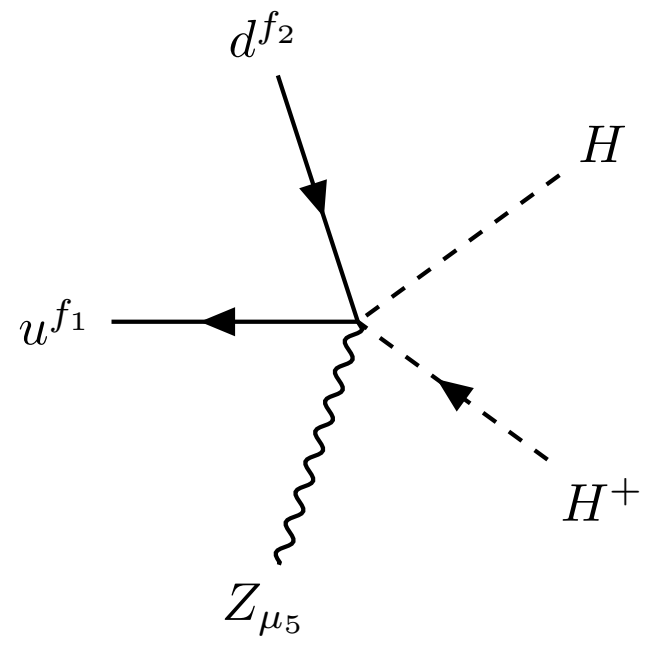

$$\begin{aligned}&+\frac{i\sqrt{2}\hat{g}^2 s_\beta c_\beta \gamma^{\mu_5}\mathcal{P}_R}{\sqrt{\hat{g}'^2+\hat{g}^2}}\left(\hat{C}^{(21)}_{\Phi ud,f_1f_2}\right)\\ &-\frac{i\sqrt{2}\hat{g}'^2 V_{f_1g_1}\gamma^{\mu_5}\mathcal{P}_L}{\sqrt{\hat{g}'^2+\hat{g}^2}}\left(s_\beta^2\hat{C}^{(11)[3]}_{\Phi q,g_1f_2}+c_\beta^2\hat{C}^{(22)[3]}_{\Phi q,g_1f_2}\right)\end{aligned} \tag{C.403}$$

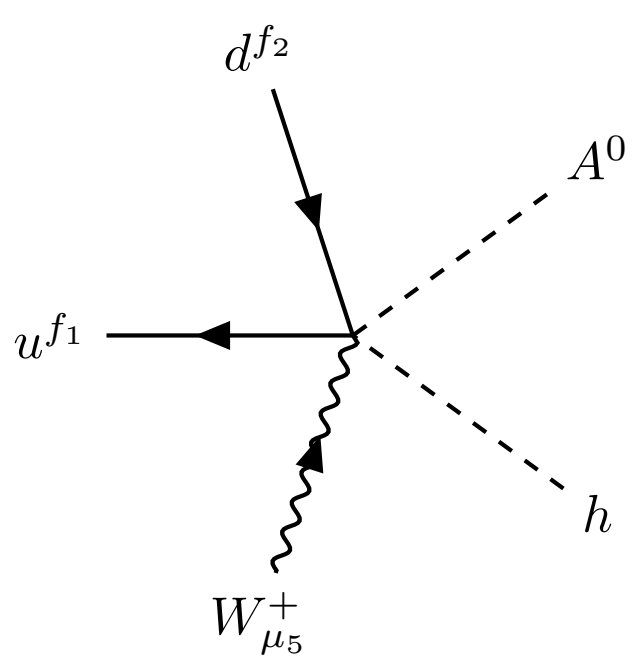

$$-\frac{\hat{g}c_{2\beta}\gamma^{\mu_5}\mathcal{P}_R}{\sqrt{2}}\left(\hat{C}^{(21)}_{\Phi ud,f_1f_2}\right) \tag{C.404}$$

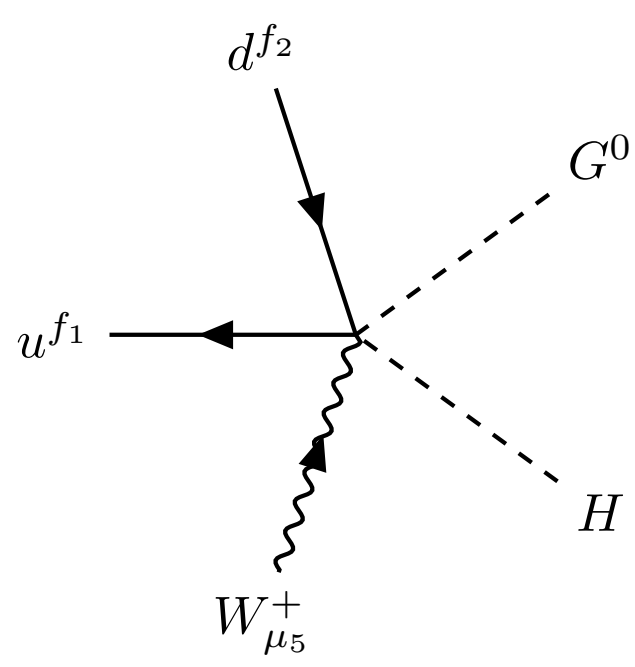

$$+\frac{\hat{g}c_{2\beta}\gamma^{\mu_5}\mathcal{P}_R}{\sqrt{2}}\left(\hat{C}^{(21)}_{\Phi ud,f_1f_2}\right) \tag{C.405}$$

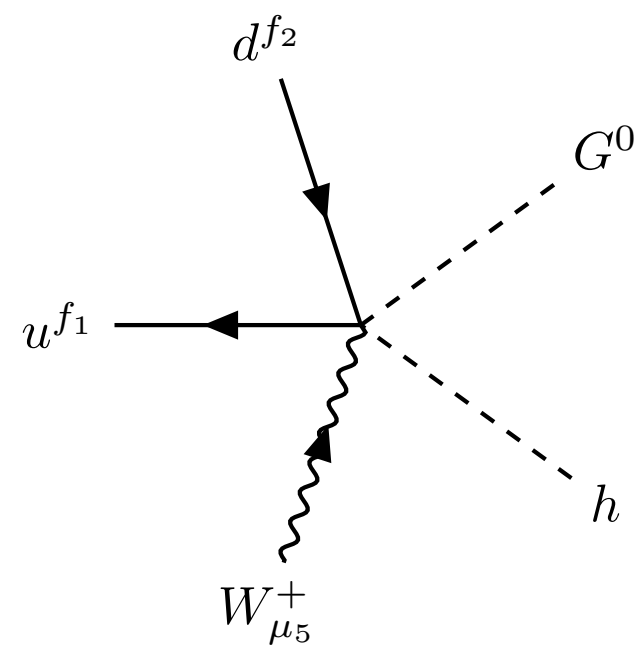

$$-\sqrt{2}\hat{g}s_\beta c_\beta\gamma^{\mu_5}\mathcal{P}_R\left(\hat{C}^{(21)}_{\Phi ud,f_1f_2}\right) \tag{C.406}$$

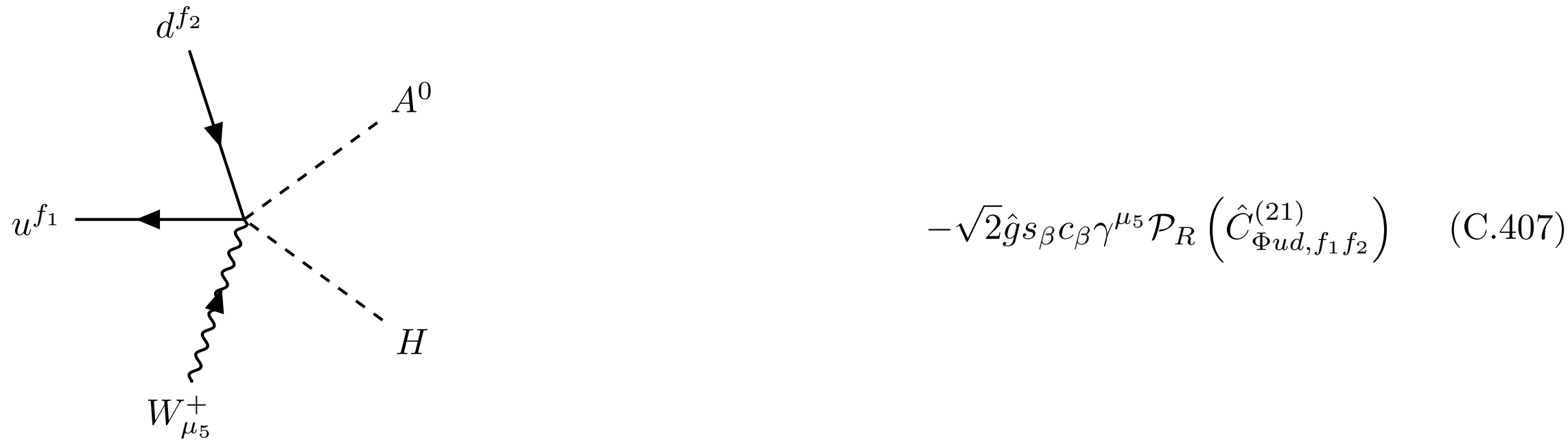

$$-\sqrt{2}\hat{g}s_\beta c_\beta \gamma^{\mu_5}\mathcal{P}_R\left(\hat{C}^{(21)}_{\Phi ud,f_1 f_2}\right) \tag{C.407}$$

## C.3 Scalar-Electroweak Boson Interactions

In this subsection, we present the complete set of Feynman rules for three- and four-point vertices involving electroweak gauge bosons and at least one scalar. We also give an example of a five- and six-point vertex given in Eqs. (C.568) and (C.569).

$$\begin{aligned}
&-\frac{1}{4}\hat{g}\left(p_{1\mu_3}-p_{2\mu_3}\right)\left(s_\beta^2\left(2\left(\delta_{s_{\hat{\beta}\pm}}+\delta_{s_{\hat{\beta}}}-1\right)+A_1'\right)\right.\\
&\qquad\qquad\left.+c_\beta^2\left(A_2'+2\left(\delta_{c_{\hat{\beta}\pm}}+\delta_{c_{\hat{\beta}}}-1\right)\right)-2s_\beta c_\beta B'\right)\\
&-\hat{g}v^2 p_{1\mu_3}\left(2s_\beta^2c_\beta^2\left(2\hat{C}^{(11)(11)}_{D\Phi}-2\hat{C}^{(11)(22)}_{D\Phi}-\hat{C}^{(21)(12)}_{D\Phi}+2\hat{C}^{(22)(22)}_{D\Phi}\right)\right.\\
&\qquad\left.+s_\beta^4\hat{C}^{(21)(12)}_{D\Phi}+c_\beta^4\hat{C}^{(21)(12)}_{D\Phi}\right)\\
&-2\hat{g}v^2c_{2\beta}p_{1\mu_3}\left(c_\beta^2\hat{C}^{(21)(21)}_{D\Phi}-s_\beta^2\hat{C}^{(21)(21)*}_{D\Phi}\right)
\end{aligned} \tag{C.408}$$

$$\begin{aligned}
&-\frac{1}{4}\hat{g}\left(p_{1\mu_3}-p_{2\mu_3}\right)\left(s_\beta^2\left(2\left(\delta_{s_{\hat{\beta}\pm}}+\delta_{s_{\hat{\beta}}}-1\right)+A_2'\right)\right.\\
&\qquad\qquad\left.+c_\beta^2\left(A_1'+2\left(\delta_{c_{\hat{\beta}\pm}}+\delta_{c_{\hat{\beta}}}-1\right)\right)+s_{2\beta}B'\right)\\
&-4\hat{g}v^2 p_{1\mu_3}\left(c_\beta^4\hat{C}^{(11)(11)}_{D\Phi}+s_\beta^2c_\beta^2\left(\hat{C}^{(11)(22)}_{D\Phi}+\hat{C}^{(21)(12)}_{D\Phi}\right)+s_\beta^4\hat{C}^{(22)(22)}_{D\Phi}\right)\\
&-4\hat{g}v^2s_\beta^2c_\beta^2p_{1\mu_3}\left(\hat{C}^{(21)(21)*}_{D\Phi}+\hat{C}^{(21)(21)}_{D\Phi}\right)
\end{aligned} \tag{C.409}$$

$$
\begin{aligned}
&-\frac{1}{4}i\hat{g}\left(p_{1\mu_3}-p_{2\mu_3}\right)\left(s_\beta^2\left(2\delta_{s_{\hat{\beta}\pm}}+A_1-2\right)+c_\beta^2\left(A_2+2\delta_{c_{\hat{\beta}\pm}}-2\right)-2Bs_\beta c_\beta\right)\\
&-i\hat{g}v^2p_{1\mu_3}\left(\hat{C}_{D\Phi}^{(21)(12)}\right)\\
&+2i\hat{g}v^2p_{1\mu_3}\left(s_\beta^4\hat{C}_{D\Phi}^{(21)(21)*}+s_\beta^2c_\beta^2\hat{C}_{D\Phi}^{(21)(21)*}\right.\\
&\qquad\qquad\left.+c_\beta^4\hat{C}_{D\Phi}^{(21)(21)}+s_\beta^2c_\beta^2\hat{C}_{D\Phi}^{(21)(21)}\right)
\end{aligned}
\tag{C.410}
$$

$$
\begin{aligned}
&-\frac{1}{4}\hat{g}\left(p_{1\mu_3}-p_{2\mu_3}\right)\left(s_\beta c_\beta\left(2\left(\delta_{s_{\hat{\beta}\pm}}-\delta_{c_{\hat{\beta}\pm}}+\delta_{c_{\hat{\beta}}}-\delta_{s_{\hat{\beta}}}\right)-A_1'+A_2'\right)-s_\beta^2B'+c_\beta^2B'\right)\\
&+\hat{g}v^2s_{2\beta}p_{1\mu_3}\left(c_\beta^2\left(2\hat{C}_{D\Phi}^{(11)(11)}-\hat{C}_{D\Phi}^{(11)(22)}-\hat{C}_{D\Phi}^{(21)(12)}\right)\right.\\
&\qquad\qquad\left.+s_\beta^2\left(\hat{C}_{D\Phi}^{(11)(22)}+\hat{C}_{D\Phi}^{(21)(12)}-2\hat{C}_{D\Phi}^{(22)(22)}\right)\right)\\
&-2\hat{g}v^2s_\beta c_\beta c_{2\beta}p_{1\mu_3}\left(\hat{C}_{D\Phi}^{(21)(21)*}+\hat{C}_{D\Phi}^{(21)(21)}\right)
\end{aligned}
\tag{C.411}
$$

$$
\begin{aligned}
&-\frac{1}{4}\hat{g}\left(p_{1\mu_3}-p_{2\mu_3}\right)\left(s_\beta c_\beta\left(2\left(-\delta_{s_{\hat{\beta}\pm}}+\delta_{c_{\hat{\beta}\pm}}-\delta_{c_{\hat{\beta}}}+\delta_{s_{\hat{\beta}}}\right)-A_1'+A_2'\right)\right.\\
&\qquad\qquad\left.-s_\beta^2B'+c_\beta^2B'\right)\\
&+\hat{g}v^2s_{2\beta}p_{1\mu_3}\left(c_\beta^2\left(2\hat{C}_{D\Phi}^{(11)(11)}-\hat{C}_{D\Phi}^{(11)(22)}-\hat{C}_{D\Phi}^{(21)(12)}\right)\right.\\
&\qquad\qquad\left.+s_\beta^2\left(\hat{C}_{D\Phi}^{(11)(22)}+\hat{C}_{D\Phi}^{(21)(12)}-2\hat{C}_{D\Phi}^{(22)(22)}\right)\right)\\
&+4\hat{g}v^2s_\beta c_\beta p_{1\mu_3}\left(s_\beta^2\hat{C}_{D\Phi}^{(21)(21)*}-c_\beta^2\hat{C}_{D\Phi}^{(21)(21)}\right)
\end{aligned}
\tag{C.412}
$$

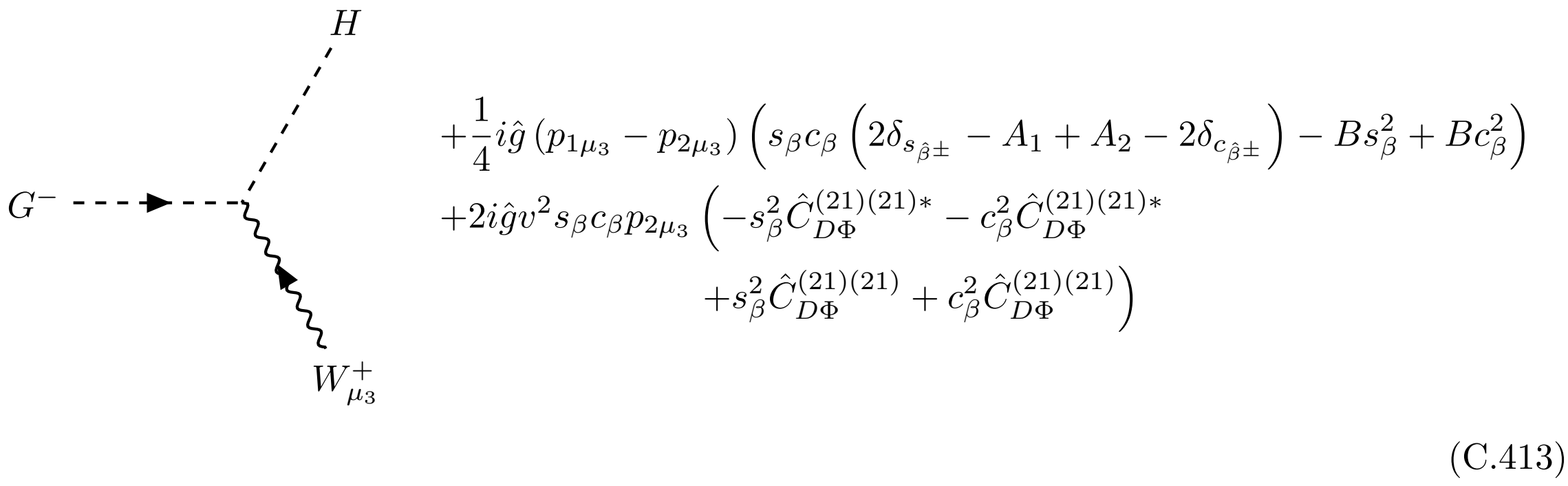

$$
\begin{aligned}
&+\frac{1}{4} i\hat{g}\left(p_{1\mu_3}-p_{2\mu_3}\right)\left(s_\beta c_\beta\left(2\delta_{s_{\hat{\beta}\pm}}-A_1+A_2-2\delta_{c_{\hat{\beta}\pm}}\right)-Bs_\beta^2+Bc_\beta^2\right)\\
&+2i\hat{g}v^2 s_\beta c_\beta p_{2\mu_3}\left(-s_\beta^2\hat{C}_{D\Phi}^{(21)(21)*}-c_\beta^2\hat{C}_{D\Phi}^{(21)(21)*}\right.\\
&\qquad\left.+s_\beta^2\hat{C}_{D\Phi}^{(21)(21)}+c_\beta^2\hat{C}_{D\Phi}^{(21)(21)}\right)
\end{aligned}
\tag{C.413}
$$

$$
-iv^2\left(c_\beta^4-s_\beta^4\right)\sqrt{\hat{g}'^2+\hat{g}^2}\left(p_{1\mu_3}+p_{2\mu_3}\right)\left(\hat{C}_{D\Phi}^{(21)(21)}-\hat{C}_{D\Phi}^{(21)(21)*}\right)
\tag{C.414}
$$

$$
\begin{aligned}
&-\frac{p_{1\mu_3}-p_{2\mu_3}}{4\sqrt{\hat{g}'^2+\hat{g}^2}}\left(c_\beta^2\left(\left(\hat{g}'^2+\hat{g}^2\right)\left(A_2'+A_2+2\delta_{c_{\hat{\beta}}}-2\right)-2\hat{g}X_{WB}\hat{g}'\right)\right.\\
&\qquad+s_\beta^2\left(\left(\hat{g}'^2+\hat{g}^2\right)\left(A_1'+A_1+2\delta_{s_{\hat{\beta}}}-2\right)-2\hat{g}X_{WB}\hat{g}'\right)\\
&\qquad\left.-2s_\beta c_\beta\left(\hat{g}'^2+\hat{g}^2\right)\left(B+B'\right)\right)\\
&-v^2\sqrt{\hat{g}'^2+\hat{g}^2}\left(p_{1\mu_3}\left(2s_\beta^2c_\beta^2\left(3\hat{C}_{D\Phi}^{(11)(11)}-2\hat{C}_{D\Phi}^{(11)(22)}-2\hat{C}_{D\Phi}^{(21)(12)}+3\hat{C}_{D\Phi}^{(22)(22)}\right)\right.\right.\\
&\qquad\left.+s_\beta^4\left(\hat{C}_{D\Phi}^{(11)(22)}+\hat{C}_{D\Phi}^{(21)(12)}\right)+c_\beta^4\left(\hat{C}_{D\Phi}^{(11)(22)}+\hat{C}_{D\Phi}^{(21)(12)}\right)\right)\\
&\qquad-p_{2\mu_3}\left(2s_\beta^2c_\beta^2\left(\hat{C}_{D\Phi}^{(11)(11)}+\hat{C}_{D\Phi}^{(22)(22)}\right)+s_\beta^4\left(\hat{C}_{D\Phi}^{(11)(22)}+\hat{C}_{D\Phi}^{(21)(12)}\right)\right.\\
&\qquad\left.\left.+c_\beta^4\left(\hat{C}_{D\Phi}^{(11)(22)}+\hat{C}_{D\Phi}^{(21)(12)}\right)\right)\right)\\
&-v^2\sqrt{\hat{g}'^2+\hat{g}^2}\left(\left(s_\beta^4+c_\beta^4-4s_\beta^2c_\beta^2\right)p_{1\mu_3}+\left(s_\beta^4+c_\beta^4+4s_\beta^2c_\beta^2\right)p_{2\mu_3}\right)\left(\hat{C}_{D\Phi}^{(21)(21)*}+\hat{C}_{D\Phi}^{(21)(21)}\right)
\end{aligned}
\tag{C.415}
$$

$h$

$G^0$

$Z_{\mu_3}$

$$
\begin{aligned}
&+\frac{p_{1\mu_3}-p_{2\mu_3}}{4\sqrt{\hat{g}'^2+\hat{g}^2}}\left(c_\beta^2\left(\left(\hat{g}'^2+\hat{g}^2\right)\left(A_1'+A_1+2\delta_{c_{\hat{\beta}}}-2\right)-2\hat{g}X_{WB}\hat{g}'\right)\right.\\
&\qquad\qquad +s_\beta^2\left(\left(\hat{g}'^2+\hat{g}^2\right)\left(A_2'+A_2+2\delta_{s_{\hat{\beta}}}-2\right)-2\hat{g}X_{WB}\hat{g}'\right)\\
&\qquad\qquad \left.+s_{2\beta}\left(\hat{g}'^2+\hat{g}^2\right)\left(B+B'\right)\right)\\
&+2v^2\sqrt{\hat{g}'^2+\hat{g}^2}\left(3p_{1\mu_3}-p_{2\mu_3}\right)\left(c_\beta^4\hat{C}_{D\Phi}^{(11)(11)}+s_\beta^2c_\beta^2\left(\hat{C}_{D\Phi}^{(11)(22)}+\hat{C}_{D\Phi}^{(21)(12)}\right)\right.\\
&\qquad\qquad\qquad\qquad \left.+s_\beta^4\hat{C}_{D\Phi}^{(22)(22)}\right)\\
&+2v^2s_\beta^2c_\beta^2\sqrt{\hat{g}'^2+\hat{g}^2}\left(3p_{1\mu_3}-p_{2\mu_3}\right)\left(\hat{C}_{D\Phi}^{(21)(21)*}+\hat{C}_{D\Phi}^{(21)(21)}\right)
\end{aligned}
\tag{C.416}
$$

$G^-$

$G^+$

$Z_{\mu_3}$

$$
\begin{aligned}
&-\frac{i\left(\hat{g}'^2-\hat{g}^2\right)\left(p_{1\mu_3}-p_{2\mu_3}\right)}{2\left(\hat{g}'^2+\hat{g}^2\right)^{3/2}}\left(s_\beta^2\left(\left(\hat{g}'^2+\hat{g}^2\right)\left(2\delta_{s_{\hat{\beta}\pm}}-1\right)+\hat{g}X_{WB}\hat{g}'\right)\right.\\
&\qquad\qquad\qquad \left.+c_\beta^2\left(\left(\hat{g}'^2+\hat{g}^2\right)\left(2\delta_{c_{\hat{\beta}\pm}}-1\right)+\hat{g}X_{WB}\hat{g}'\right)\right)\\
&-2iv^2\sqrt{\hat{g}'^2+\hat{g}^2}\left(p_{1\mu_3}-p_{2\mu_3}\right)\left(c_\beta^4\hat{C}_{D\Phi}^{(11)(11)}+s_\beta^2c_\beta^2\left(\hat{C}_{D\Phi}^{(11)(22)}+\hat{C}_{D\Phi}^{(21)(12)}\right)\right.\\
&\qquad\qquad\qquad\qquad \left.+s_\beta^4\hat{C}_{D\Phi}^{(22)(22)}\right)\\
&-2iv^2s_\beta^2c_\beta^2\sqrt{\hat{g}'^2+\hat{g}^2}\left(p_{1\mu_3}-p_{2\mu_3}\right)\left(\hat{C}_{D\Phi}^{(21)(21)*}+\hat{C}_{D\Phi}^{(21)(21)}\right)
\end{aligned}
\tag{C.417}
$$

$H$

$H$

$Z_{\mu_3}$

$$
+iv^2\left(c_\beta^4-s_\beta^4\right)\sqrt{\hat{g}'^2+\hat{g}^2}\left(p_{1\mu_3}+p_{2\mu_3}\right)\left(\hat{C}_{D\Phi}^{(21)(21)}-\hat{C}_{D\Phi}^{(21)(21)*}\right)
\tag{C.418}
$$

$$
\begin{aligned}
&-\frac{i\left(\hat{g}'^2-\hat{g}^2\right)\left(p_{1\mu_3}-p_{2\mu_3}\right)}{2\left(\hat{g}'^2+\hat{g}^2\right)^{3/2}}\left(s_\beta^2\left(\left(\hat{g}'^2+\hat{g}^2\right)\left(2\delta_{s_{\hat{\beta}\pm}}-1\right)+\hat{g}X_{WB}\hat{g}'\right)\right.\\
&\qquad\left.+c_\beta^2\left(\left(\hat{g}'^2+\hat{g}^2\right)\left(2\delta_{c_{\hat{\beta}\pm}}-1\right)+\hat{g}X_{WB}\hat{g}'\right)\right)\\
&-iv^2\sqrt{\hat{g}'^2+\hat{g}^2}\left(p_{1\mu_3}-p_{2\mu_3}\right)\left(2s_\beta^2c_\beta^2\left(\hat{C}_{D\Phi}^{(11)(11)}-\hat{C}_{D\Phi}^{(21)(12)}+\hat{C}_{D\Phi}^{(22)(22)}\right)\right.\\
&\qquad\left.+s_\beta^4\hat{C}_{D\Phi}^{(11)(22)}+c_\beta^4\hat{C}_{D\Phi}^{(11)(22)}\right)\\
&+2iv^2s_\beta^2c_\beta^2\sqrt{\hat{g}'^2+\hat{g}^2}\left(p_{1\mu_3}-p_{2\mu_3}\right)\left(\hat{C}_{D\Phi}^{(21)(21)*}+\hat{C}_{D\Phi}^{(21)(21)}\right)
\end{aligned}
\tag{C.419}
$$

$$
\begin{aligned}
&+iv^2s_\beta c_\beta\sqrt{\hat{g}'^2+\hat{g}^2}\left(p_{1\mu_3}-3p_{2\mu_3}\right)\left(-s_\beta^2\hat{C}_{D\Phi}^{(21)(21)*}-c_\beta^2\hat{C}_{D\Phi}^{(21)(21)*}\right.\\
&\qquad\left.+s_\beta^2\hat{C}_{D\Phi}^{(21)(21)}+c_\beta^2\hat{C}_{D\Phi}^{(21)(21)}\right)
\end{aligned}
\tag{C.420}
$$

$$
\begin{aligned}
&+\frac{1}{4}\sqrt{\hat{g}'^2+\hat{g}^2}\left(p_{1\mu_3}-p_{2\mu_3}\right)\left(s_\beta c_\beta\left(-A_1'+A_2'-A_1+A_2+2\delta_{c_{\hat{\beta}}}-2\delta_{s_{\hat{\beta}}}\right)\right.\\
&\qquad\left.-s_\beta^2\left(B+B'\right)+c_\beta^2\left(B+B'\right)\right)\\
&-v^2s_\beta c_\beta\sqrt{\hat{g}'^2+\hat{g}^2}\left(3p_{1\mu_3}-p_{2\mu_3}\right)\left(c_\beta^2\left(2\hat{C}_{D\Phi}^{(11)(11)}-\hat{C}_{D\Phi}^{(11)(22)}-\hat{C}_{D\Phi}^{(21)(12)}\right)\right.\\
&\qquad\left.+s_\beta^2\left(\hat{C}_{D\Phi}^{(11)(22)}+\hat{C}_{D\Phi}^{(21)(12)}-2\hat{C}_{D\Phi}^{(22)(22)}\right)\right)\\
&+v^2s_\beta c_\beta c_{2\beta}\sqrt{\hat{g}'^2+\hat{g}^2}\left(3p_{1\mu_3}-p_{2\mu_3}\right)\left(\hat{C}_{D\Phi}^{(21)(21)*}+\hat{C}_{D\Phi}^{(21)(21)}\right)
\end{aligned}
\tag{C.421}
$$

$H$, $G^0$, $Z_{\mu_3}$

$$
\begin{aligned}
&-\frac{1}{4}\sqrt{\hat{g}'^2+\hat{g}^2}\left(p_{1\mu_3}-p_{2\mu_3}\right)\Big(s_\beta c_\beta\left(-A_1'+A_2'-A_1+A_2-2\delta_{c_{\hat{\beta}}}+2\delta_{s_{\hat{\beta}}}\right)\\
&\qquad -s_\beta^2\left(B+B'\right)+c_\beta^2\left(B+B'\right)\Big)\\
&+v^2 s_\beta c_\beta\sqrt{\hat{g}'^2+\hat{g}^2}\left(3p_{1\mu_3}-p_{2\mu_3}\right)\Big(c_\beta^2\left(2\hat{C}_{D\Phi}^{(11)(11)}-\hat{C}_{D\Phi}^{(11)(22)}-\hat{C}_{D\Phi}^{(21)(12)}\right)\\
&\qquad +s_\beta^2\left(\hat{C}_{D\Phi}^{(11)(22)}+\hat{C}_{D\Phi}^{(21)(12)}-2\hat{C}_{D\Phi}^{(22)(22)}\right)\Big)\\
&+v^2 s_\beta c_\beta c_{2\beta}\left(-\sqrt{\hat{g}'^2+\hat{g}^2}\right)\left(3p_{1\mu_3}-p_{2\mu_3}\right)\left(\hat{C}_{D\Phi}^{(21)(21)*}+\hat{C}_{D\Phi}^{(21)(21)}\right)
\end{aligned}
\tag{C.422}
$$

$H^-$, $G^+$, $Z_{\mu_3}$

$$
\begin{aligned}
&+iv^2 s_\beta c_\beta\sqrt{\hat{g}'^2+\hat{g}^2}\left(p_{1\mu_3}-p_{2\mu_3}\right)\Big(c_\beta^2\left(2\hat{C}_{D\Phi}^{(11)(11)}-\hat{C}_{D\Phi}^{(11)(22)}-\hat{C}_{D\Phi}^{(21)(12)}\right)\\
&\qquad +s_\beta^2\left(\hat{C}_{D\Phi}^{(11)(22)}+\hat{C}_{D\Phi}^{(21)(12)}-2\hat{C}_{D\Phi}^{(22)(22)}\right)\Big)\\
&-2iv^2 s_\beta c_\beta\sqrt{\hat{g}'^2+\hat{g}^2}\left(p_{1\mu_3}-p_{2\mu_3}\right)\left(c_\beta^2\hat{C}_{D\Phi}^{(21)(21)}-s_\beta^2\hat{C}_{D\Phi}^{(21)(21)*}\right)
\end{aligned}
\tag{C.423}
$$

$H$, $h$, $Z_{\mu_3}$

$$
\begin{aligned}
&-\frac{1}{4}\left(K-L\right)\sqrt{\hat{g}'^2+\hat{g}^2}\left(s_\beta^2 p_{1\mu_3}-s_\beta^2 p_{2\mu_3}+c_\beta^2 p_{1\mu_3}-c_\beta^2 p_{2\mu_3}\right)\\
&+iv^2 s_\beta c_\beta\sqrt{\hat{g}'^2+\hat{g}^2}\left(p_{1\mu_3}-3p_{2\mu_3}\right)\Big(-s_\beta^2\hat{C}_{D\Phi}^{(21)(21)*}-c_\beta^2\hat{C}_{D\Phi}^{(21)(21)*}\\
&\qquad +s_\beta^2\hat{C}_{D\Phi}^{(21)(21)}+c_\beta^2\hat{C}_{D\Phi}^{(21)(21)}\Big)
\end{aligned}
\tag{C.424}
$$

$G^-$ $W^+_{\mu_2}$ $Z_{\mu_3}$

$$
\begin{aligned}
&+\frac{i\hat{g}vg_{\mu_2\mu_3}\hat{g}'}{\sqrt{2}\left(\hat{g}'^2+\hat{g}^2\right)^{3/2}}\left(s_\beta^2\left(\hat{g}'^3\left(\delta_{s_{\hat{\beta}\pm}}-1\right)+\hat{g}^2\hat{g}'\left(\delta_{s_{\hat{\beta}\pm}}-1\right)-\hat{g}^3X_{WB}\right)\right.\\
&\qquad\qquad\left.+c_\beta^2\left(\left(\delta_{c_{\hat{\beta}\pm}}-1\right)\hat{g}'^3+\hat{g}^2\left(\delta_{c_{\hat{\beta}\pm}}-1\right)\hat{g}'-\hat{g}^3X_{WB}\right)\right)\\
&+2i\sqrt{2}\hat{g}v^3g_{\mu_2\mu_3}\sqrt{\hat{g}'^2+\hat{g}^2}\left(c_\beta^4\hat{C}_{D\Phi}^{(11)(11)}+s_\beta^2c_\beta^2\left(\hat{C}_{D\Phi}^{(11)(22)}+\hat{C}_{D\Phi}^{(21)(12)}\right)+s_\beta^4\hat{C}_{D\Phi}^{(22)(22)}\right)\\
&-\frac{2i\sqrt{2}v\hat{g}'\left(p_{2\mu_3}p_{3\mu_2}-p_2\cdot p_3g_{\mu_2\mu_3}\right)}{\sqrt{\hat{g}'^2+\hat{g}^2}}\left(c_\beta^2\hat{C}_{\Phi WB}^{(11)}+s_\beta^2\hat{C}_{\Phi WB}^{(22)}\right)\\
&-\frac{2i\sqrt{2}v\hat{g}'p_2^\mu p_3^\nu\epsilon_{\mu_2\mu_3\mu\nu}}{\sqrt{\hat{g}'^2+\hat{g}^2}}\left(c_\beta^2\hat{C}_{\Phi B\tilde{W}}^{(11)}+s_\beta^2\hat{C}_{\Phi B\tilde{W}}^{(22)}\right)\\
&+2i\sqrt{2}\hat{g}v^3s_\beta^2c_\beta^2g_{\mu_2\mu_3}\sqrt{\hat{g}'^2+\hat{g}^2}\left(\hat{C}_{D\Phi}^{(21)(21)*}+\hat{C}_{D\Phi}^{(21)(21)}\right)
\end{aligned}
\tag{C.425}
$$

$H^-$ $W^+_{\mu_2}$ $Z_{\mu_3}$

$$
\begin{aligned}
&+\frac{i\hat{g}vc_\beta g_{\mu_2\mu_3}\hat{g}'^2\left(\delta_{c_{\hat{\beta}\pm}}-\delta_{s_{\hat{\beta}\pm}}\right)}{\sqrt{2}\sqrt{\hat{g}'^2+\hat{g}^2}}\left(s_\beta\right)\\
&-i\sqrt{2}\hat{g}v^3s_\beta c_\beta g_{\mu_2\mu_3}\sqrt{\hat{g}'^2+\hat{g}^2}\left(c_\beta^2\left(2\hat{C}_{D\Phi}^{(11)(11)}-\hat{C}_{D\Phi}^{(11)(22)}-\hat{C}_{D\Phi}^{(21)(12)}\right)\right.\\
&\qquad\qquad\left.+s_\beta^2\left(\hat{C}_{D\Phi}^{(11)(22)}+\hat{C}_{D\Phi}^{(21)(12)}-2\hat{C}_{D\Phi}^{(22)(22)}\right)\right)\\
&+\frac{2i\sqrt{2}vs_\beta c_\beta\hat{g}'\left(p_{2\mu_3}p_{3\mu_2}-p_2\cdot p_3g_{\mu_2\mu_3}\right)}{\sqrt{\hat{g}'^2+\hat{g}^2}}\left(\hat{C}_{\Phi WB}^{(11)}-\hat{C}_{\Phi WB}^{(22)}\right)\\
&+\frac{2i\sqrt{2}vs_\beta c_\beta\hat{g}'p_2^\mu p_3^\nu\epsilon_{\mu_2\mu_3\mu\nu}}{\sqrt{\hat{g}'^2+\hat{g}^2}}\left(\hat{C}_{\Phi B\tilde{W}}^{(11)}-\hat{C}_{\Phi B\tilde{W}}^{(22)}\right)\\
&+2i\sqrt{2}\hat{g}v^3s_\beta c_\beta g_{\mu_2\mu_3}\sqrt{\hat{g}'^2+\hat{g}^2}\left(c_\beta^2\hat{C}_{D\Phi}^{(21)(21)}-s_\beta^2\hat{C}_{D\Phi}^{(21)(21)*}\right)
\end{aligned}
\tag{C.426}
$$

$h$ $Z_{\mu_2}$ $Z_{\mu_3}$

$$
\begin{aligned}
&-\frac{ivg_{\mu_2\mu_3}}{2\sqrt{2}}\left(s_\beta^2\left(\left(A_2-2\right)\hat{g}'^2+\left(A_2-2\right)\hat{g}^2-4\hat{g}X_{WB}\hat{g}'\right)\right.\\
&\qquad\qquad+c_\beta^2\left(\left(A_1-2\right)\hat{g}'^2+\left(A_1-2\right)\hat{g}^2-4\hat{g}X_{WB}\hat{g}'\right)\\
&\qquad\qquad\left.+Bs_{2\beta}\left(\hat{g}'^2+\hat{g}^2\right)\right)\\
&-4i\sqrt{2}v^3g_{\mu_2\mu_3}\left(\hat{g}'^2+\hat{g}^2\right)\left(c_\beta^4\hat{C}_{D\Phi}^{(11)(11)}+s_\beta^2c_\beta^2\left(\hat{C}_{D\Phi}^{(11)(22)}+\hat{C}_{D\Phi}^{(21)(12)}\right)+s_\beta^4\hat{C}_{D\Phi}^{(22)(22)}\right)\\
&+\frac{4i\sqrt{2}v\left(p_{2\mu_3}p_{3\mu_2}-p_2\cdot p_3g_{\mu_2\mu_3}\right)}{\hat{g}'^2+\hat{g}^2}\left(c_\beta^2\left(\hat{g}'^2\hat{C}_{\Phi B}^{(11)}+\hat{g}\left(\hat{g}'\hat{C}_{\Phi WB}^{(11)}+\hat{g}\hat{C}_{\Phi W}^{(11)}\right)\right)\right.\\
&\qquad\qquad\left.+s_\beta^2\left(\hat{g}'^2\hat{C}_{\Phi B}^{(22)}+\hat{g}\left(\hat{g}'\hat{C}_{\Phi WB}^{(22)}+\hat{g}\hat{C}_{\Phi W}^{(22)}\right)\right)\right)\\
&+\frac{4i\sqrt{2}vp_2^\mu p_3^\nu\epsilon_{\mu_2\mu_3\mu\nu}}{\hat{g}'^2+\hat{g}^2}\left(c_\beta^2\left(\hat{g}'^2\hat{C}_{\Phi\tilde{B}}^{(11)}+\hat{g}\left(\hat{g}'\hat{C}_{\Phi B\tilde{W}}^{(11)}+\hat{g}\hat{C}_{\Phi\tilde{W}}^{(11)}\right)\right)\right.\\
&\qquad\qquad\left.+s_\beta^2\left(\hat{g}'^2\hat{C}_{\Phi\tilde{B}}^{(22)}+\hat{g}\left(\hat{g}'\hat{C}_{\Phi B\tilde{W}}^{(22)}+\hat{g}\hat{C}_{\Phi\tilde{W}}^{(22)}\right)\right)\right)\\
&-4i\sqrt{2}v^3s_\beta^2c_\beta^2g_{\mu_2\mu_3}\left(\hat{g}'^2+\hat{g}^2\right)\left(\hat{C}_{D\Phi}^{(21)(21)*}+\hat{C}_{D\Phi}^{(21)(21)}\right)
\end{aligned}
\tag{C.427}
$$

$$
\begin{aligned}
&-2\sqrt{2}v^3 s_\beta c_\beta g_{\mu_2\mu_3}\left(\hat{g}'^2+\hat{g}^2\right)\Big(-s_\beta^2\hat{C}_{D\Phi}^{(21)(21)*}-c_\beta^2\hat{C}_{D\Phi}^{(21)(21)*}\\
&\qquad +s_\beta^2\hat{C}_{D\Phi}^{(21)(21)}+c_\beta^2\hat{C}_{D\Phi}^{(21)(21)}\Big)
\end{aligned}
\tag{C.428}
$$

$$
\begin{aligned}
&+\frac{ivg_{\mu_2\mu_3}\left(\hat{g}'^2+\hat{g}^2\right)}{2\sqrt{2}}\left(\left(A_2-A_1\right)s_\beta c_\beta-Bs_\beta^2+Bc_\beta^2\right)\\
&-2i\sqrt{2}v^3 s_\beta c_\beta g_{\mu_2\mu_3}\left(\hat{g}'^2+\hat{g}^2\right)\Big(c_\beta^2\left(2\hat{C}_{D\Phi}^{(11)(11)}-\hat{C}_{D\Phi}^{(11)(22)}-\hat{C}_{D\Phi}^{(21)(12)}\right)\\
&\qquad +s_\beta^2\left(\hat{C}_{D\Phi}^{(11)(22)}+\hat{C}_{D\Phi}^{(21)(12)}-2\hat{C}_{D\Phi}^{(22)(22)}\right)\Big)\\
&+\frac{4i\sqrt{2}vs_\beta c_\beta\left(p_{2\mu_3}p_{3\mu_2}-p_2\cdot p_3 g_{\mu_2\mu_3}\right)}{\hat{g}'^2+\hat{g}^2}\Big(\hat{g}\left(\hat{g}'\hat{C}_{\Phi WB}^{(11)}+\hat{g}\hat{C}_{\Phi W}^{(11)}-\hat{g}'\hat{C}_{\Phi WB}^{(22)}-\hat{g}\hat{C}_{\Phi W}^{(22)}\right)\\
&\qquad +\hat{g}'^2\hat{C}_{\Phi B}^{(11)}-\hat{g}'^2\hat{C}_{\Phi B}^{(22)}\Big)\\
&+\frac{4i\sqrt{2}vs_\beta c_\beta p_2^\mu p_3^\nu\epsilon_{\mu_2\mu_3\mu\nu}}{\hat{g}'^2+\hat{g}^2}\Big(\hat{g}\left(\hat{g}'\hat{C}_{\Phi B\tilde{W}}^{(11)}+\hat{g}\hat{C}_{\Phi\tilde{W}}^{(11)}-\hat{g}'\hat{C}_{\Phi B\tilde{W}}^{(22)}-\hat{g}\hat{C}_{\Phi\tilde{W}}^{(22)}\right)\\
&\qquad +\hat{g}'^2\hat{C}_{\Phi\tilde{B}}^{(11)}-\hat{g}'^2\hat{C}_{\Phi\tilde{B}}^{(22)}\Big)\\
&+2i\sqrt{2}v^3 s_\beta c_\beta c_{2\beta} g_{\mu_2\mu_3}\left(\hat{g}'^2+\hat{g}^2\right)\left(\hat{C}_{D\Phi}^{(21)(21)*}+\hat{C}_{D\Phi}^{(21)(21)}\right)
\end{aligned}
\tag{C.429}
$$

$$
\begin{aligned}
&+\frac{4i\sqrt{2}v\left(p_{1\mu_2}p_{2\mu_1}-p_1\cdot p_2 g_{\mu_1\mu_2}\right)}{\hat{g}'^2+\hat{g}^2}\Big(c_\beta^2\left(\hat{g}\left(\hat{g}\hat{C}_{\Phi B}^{(11)}-\hat{g}'\hat{C}_{\Phi WB}^{(11)}\right)+\hat{g}'^2\hat{C}_{\Phi W}^{(11)}\right)\\
&\qquad +s_\beta^2\left(\hat{g}\left(\hat{g}\hat{C}_{\Phi B}^{(22)}-\hat{g}'\hat{C}_{\Phi WB}^{(22)}\right)+\hat{g}'^2\hat{C}_{\Phi W}^{(22)}\right)\Big)\\
&+\frac{4i\sqrt{2}vp_1^\mu p_2^\nu\epsilon_{\mu_1\mu_2\mu\nu}}{\hat{g}'^2+\hat{g}^2}\Big(c_\beta^2\left(\hat{g}\left(\hat{g}\hat{C}_{\Phi\tilde{B}}^{(11)}-\hat{g}'\hat{C}_{\Phi B\tilde{W}}^{(11)}\right)+\hat{g}'^2\hat{C}_{\Phi\tilde{W}}^{(11)}\right)\\
&\qquad +s_\beta^2\left(\hat{g}\left(\hat{g}\hat{C}_{\Phi\tilde{B}}^{(22)}-\hat{g}'\hat{C}_{\Phi B\tilde{W}}^{(22)}\right)+\hat{g}'^2\hat{C}_{\Phi\tilde{W}}^{(22)}\right)\Big)
\end{aligned}
\tag{C.430}
$$

$$
\begin{aligned}
&-\frac{i\hat{g}^2 v g_{\mu_1\mu_3}\hat{g}'}{\sqrt{2}\sqrt{\hat{g}'^2+\hat{g}^2}}\left(s_\beta^2\left(\delta_{s_{\hat{\beta}\pm}}-1\right)+c_\beta^2\left(\delta_{c_{\hat{\beta}\pm}}-1\right)+\frac{\hat{g}X_{WB}\hat{g}'}{\hat{g}'^2+\hat{g}^2}\right)\\
&+\frac{2i\sqrt{2}\hat{g}v\left(p_{1\mu_3}p_{3\mu_1}-p_1\cdot p_3 g_{\mu_1\mu_3}\right)}{\sqrt{\hat{g}'^2+\hat{g}^2}}\left(c_\beta^2\hat{C}^{(11)}_{\Phi WB}+s_\beta^2\hat{C}^{(22)}_{\Phi WB}\right)\\
&+\frac{2i\sqrt{2}\hat{g}v p_1^\mu p_3^\nu\epsilon_{\mu_1\mu_3\mu\nu}}{\sqrt{\hat{g}'^2+\hat{g}^2}}\left(c_\beta^2\hat{C}^{(11)}_{\Phi B\tilde{W}}+s_\beta^2\hat{C}^{(22)}_{\Phi B\tilde{W}}\right)
\end{aligned}
\tag{C.431}
$$

$$
\begin{aligned}
&-\frac{i\hat{g}^2 v g_{\mu_2\mu_3}}{2\sqrt{2}}\left(\left(A_2-2\right)s_\beta^2+\left(A_1-2\right)c_\beta^2+Bs_{2\beta}\right)\\
&+4i\sqrt{2}v\left(p_{2\mu_3}p_{3\mu_2}-p_2\cdot p_3 g_{\mu_2\mu_3}\right)\left(c_\beta^2\hat{C}^{(11)}_{\Phi W}+s_\beta^2\hat{C}^{(22)}_{\Phi W}\right)\\
&+4i\sqrt{2}v p_2^\mu p_3^\nu\epsilon_{\mu_2\mu_3\mu\nu}\left(c_\beta^2\hat{C}^{(11)}_{\Phi\tilde{W}}+s_\beta^2\hat{C}^{(22)}_{\Phi\tilde{W}}\right)
\end{aligned}
\tag{C.432}
$$

$$
\begin{aligned}
&+\frac{2i\sqrt{2}v\left(p_{1\mu_3}p_{3\mu_1}-p_1\cdot p_3 g_{\mu_1\mu_3}\right)}{\hat{g}'^2+\hat{g}^2}\Bigg(c_\beta^2\left(2\hat{g}\hat{g}'\left(\hat{C}^{(11)}_{\Phi W}-\hat{C}^{(11)}_{\Phi B}\right)+\left(\hat{g}'^2-\hat{g}^2\right)\hat{C}^{(11)}_{\Phi WB}\right)\\
&\qquad\qquad+s_\beta^2\left(2\hat{g}\hat{g}'\left(\hat{C}^{(22)}_{\Phi W}-\hat{C}^{(22)}_{\Phi B}\right)+\left(\hat{g}'^2-\hat{g}^2\right)\hat{C}^{(22)}_{\Phi WB}\right)\Bigg)\\
&+\frac{2i\sqrt{2}v p_1^\mu p_3^\nu\epsilon_{\mu_1\mu_3\mu\nu}}{\hat{g}'^2+\hat{g}^2}\Bigg(c_\beta^2\left(2\hat{g}\hat{g}'\left(\hat{C}^{(11)}_{\Phi\tilde{W}}-\hat{C}^{(11)}_{\Phi\tilde{B}}\right)+\left(\hat{g}'^2-\hat{g}^2\right)\hat{C}^{(11)}_{\Phi B\tilde{W}}\right)\\
&\qquad\qquad+s_\beta^2\left(2\hat{g}\hat{g}'\left(\hat{C}^{(22)}_{\Phi\tilde{W}}-\hat{C}^{(22)}_{\Phi\tilde{B}}\right)+\left(\hat{g}'^2-\hat{g}^2\right)\hat{C}^{(22)}_{\Phi B\tilde{W}}\right)\Bigg)
\end{aligned}
\tag{C.433}
$$

$$
\begin{aligned}
&+\frac{4i\sqrt{2}v s_\beta c_\beta\left(p_{1\mu_2}p_{2\mu_1}-p_1\cdot p_2 g_{\mu_1\mu_2}\right)}{\hat{g}'^2+\hat{g}^2}\Bigg(\hat{g}\left(-\hat{g}'\hat{C}^{(11)}_{\Phi WB}+\hat{g}\hat{C}^{(11)}_{\Phi B}+\hat{g}'\hat{C}^{(22)}_{\Phi WB}-\hat{g}\hat{C}^{(22)}_{\Phi B}\right)\\
&\qquad\qquad+\hat{g}'^2\hat{C}^{(11)}_{\Phi W}-\hat{g}'^2\hat{C}^{(22)}_{\Phi W}\Bigg)\\
&+\frac{4i\sqrt{2}v s_\beta c_\beta p_1^\mu p_2^\nu\epsilon_{\mu_1\mu_2\mu\nu}}{\hat{g}'^2+\hat{g}^2}\Bigg(\hat{g}\left(-\hat{g}'\hat{C}^{(11)}_{\Phi B\tilde{W}}+\hat{g}\hat{C}^{(11)}_{\Phi\tilde{B}}+\hat{g}'\hat{C}^{(22)}_{\Phi B\tilde{W}}-\hat{g}\hat{C}^{(22)}_{\Phi\tilde{B}}\right)\\
&\qquad\qquad+\hat{g}'^2\hat{C}^{(11)}_{\Phi\tilde{W}}-\hat{g}'^2\hat{C}^{(22)}_{\Phi\tilde{W}}\Bigg)
\end{aligned}
\tag{C.434}
$$

$$
\begin{aligned}
&-\frac{i\hat{g}^2 v c_\beta g_{\mu_1\mu_3}\hat{g}'\left(\delta_{c_{\hat{\beta}\pm}}-\delta_{s_{\hat{\beta}\pm}}\right)}{\sqrt{2}\sqrt{\hat{g}'^2+\hat{g}^2}}\left(s_\beta\right)\\
&-\frac{2i\sqrt{2}\hat{g}v s_\beta c_\beta\left(p_{1\mu_3}p_{3\mu_1}-p_1\cdot p_3 g_{\mu_1\mu_3}\right)}{\sqrt{\hat{g}'^2+\hat{g}^2}}\left(\hat{C}^{(11)}_{\Phi WB}-\hat{C}^{(22)}_{\Phi WB}\right)\\
&-\frac{2i\sqrt{2}\hat{g}v s_\beta c_\beta p_1^\mu p_3^\nu\epsilon_{\mu_1\mu_3\mu\nu}}{\sqrt{\hat{g}'^2+\hat{g}^2}}\left(\hat{C}^{(11)}_{\Phi B\tilde{W}}-\hat{C}^{(22)}_{\Phi B\tilde{W}}\right)
\end{aligned}
\tag{C.435}
$$

$$
\begin{aligned}
&+\frac{i\hat{g}^2 v g_{\mu_2\mu_3}}{2\sqrt{2}}\left(\left(A_2-A_1\right)s_\beta c_\beta-Bs_\beta^2+Bc_\beta^2\right)\\
&+4i\sqrt{2}v s_\beta c_\beta\left(p_{2\mu_3}p_{3\mu_2}-p_2\cdot p_3 g_{\mu_2\mu_3}\right)\left(\hat{C}^{(11)}_{\Phi W}-\hat{C}^{(22)}_{\Phi W}\right)\\
&+4i\sqrt{2}v s_\beta c_\beta p_2^\mu p_3^\nu\epsilon_{\mu_2\mu_3\mu\nu}\left(\hat{C}^{(11)}_{\Phi\tilde{W}}-\hat{C}^{(22)}_{\Phi\tilde{W}}\right)
\end{aligned}
\tag{C.436}
$$

$$
\begin{aligned}
&+\frac{2i\sqrt{2}v s_\beta c_\beta\left(p_{1\mu_3}p_{3\mu_1}-p_1\cdot p_3 g_{\mu_1\mu_3}\right)}{\hat{g}'^2+\hat{g}^2}\left(2\hat{g}\hat{g}'\left(-\hat{C}^{(11)}_{\Phi B}+\hat{C}^{(11)}_{\Phi W}+\hat{C}^{(22)}_{\Phi B}-\hat{C}^{(22)}_{\Phi W}\right)\right.\\
&\qquad\qquad\left.+\left(\hat{g}'^2-\hat{g}^2\right)\left(\hat{C}^{(11)}_{\Phi WB}-\hat{C}^{(22)}_{\Phi WB}\right)\right)\\
&+\frac{2i\sqrt{2}v s_\beta c_\beta p_1^\mu p_3^\nu\epsilon_{\mu_1\mu_3\mu\nu}}{\hat{g}'^2+\hat{g}^2}\left(2\hat{g}\hat{g}'\left(-\hat{C}^{(11)}_{\Phi\tilde{B}}+\hat{C}^{(11)}_{\Phi\tilde{W}}+\hat{C}^{(22)}_{\Phi\tilde{B}}-\hat{C}^{(22)}_{\Phi\tilde{W}}\right)\right.\\
&\qquad\qquad\left.+\left(\hat{g}'^2-\hat{g}^2\right)\left(\hat{C}^{(11)}_{\Phi B\tilde{W}}-\hat{C}^{(22)}_{\Phi B\tilde{W}}\right)\right)
\end{aligned}
\tag{C.437}
$$

$G^+$, $A_{\mu_1}$, $G^-$

$$+\frac{i\hat{g}\hat{g}'\left(p_{2\mu_1}-p_{3\mu_1}\right)}{\sqrt{\hat{g}'^2+\hat{g}^2}}\left(s_\beta^2\left(2\delta_{s_{\hat{\beta}\pm}}-1\right)+c_\beta^2\left(2\delta_{c_{\hat{\beta}\pm}}-1\right)+\frac{\hat{g}X_{WB}\hat{g}'}{\hat{g}'^2+\hat{g}^2}\right) \tag{C.438}$$

$H^+$, $A_{\mu_1}$, $H^-$

$$+\frac{i\hat{g}\hat{g}'\left(p_{2\mu_1}-p_{3\mu_1}\right)}{\sqrt{\hat{g}'^2+\hat{g}^2}}\left(s_\beta^2\left(2\delta_{s_{\hat{\beta}\pm}}-1\right)+c_\beta^2\left(2\delta_{c_{\hat{\beta}\pm}}-1\right)+\frac{\hat{g}X_{WB}\hat{g}'}{\hat{g}'^2+\hat{g}^2}\right) \tag{C.439}$$

$h$, $G^-$, $W^+_{\mu_3}$

$$-\frac{1}{4}i\hat{g}\left(p_{1\mu_3}-p_{2\mu_3}\right)\left(s_\beta^2\left(2\delta_{s_{\hat{\beta}\pm}}+A_2-2\right)+c_\beta^2\left(A_1+2\delta_{c_{\hat{\beta}\pm}}-2\right)+Bs_{2\beta}\right) \tag{C.440}$$

$$+\frac{1}{4}i\hat{g}\left(p_{1\mu_3}-p_{2\mu_3}\right)\left(s_\beta c_\beta\left(-2\delta_{s_{\hat{\beta}\pm}}-A_1+A_2+2\delta_{c_{\hat{\beta}\pm}}\right)-Bs_\beta^2+Bc_\beta^2\right) \tag{C.441}$$

$$\begin{aligned}
&-\frac{\sqrt{2}\hat{g}v\hat{g}'p_{2\mu_1}}{\sqrt{\hat{g}'^2+\hat{g}^2}}\left(2s_\beta^2c_\beta^2\left(2\hat{C}_{D\Phi}^{(11)(11)}-2\hat{C}_{D\Phi}^{(11)(22)}-\hat{C}_{D\Phi}^{(21)(12)}+2\hat{C}_{D\Phi}^{(22)(22)}\right)\right.\\
&\qquad\qquad\left.+s_\beta^4\hat{C}_{D\Phi}^{(21)(12)}+c_\beta^4\hat{C}_{D\Phi}^{(21)(12)}\right)\\
&-\frac{2\sqrt{2}\hat{g}vc_{2\beta}\hat{g}'p_{2\mu_1}}{\sqrt{\hat{g}'^2+\hat{g}^2}}\left(c_\beta^2\hat{C}_{D\Phi}^{(21)(21)*}-s_\beta^2\hat{C}_{D\Phi}^{(21)(21)}\right)
\end{aligned} \tag{C.442}$$

$$\begin{aligned}
&-\frac{4\sqrt{2}\hat{g}v\hat{g}'p_{2\mu_1}}{\sqrt{\hat{g}'^2+\hat{g}^2}}\left(c_\beta^4\hat{C}_{D\Phi}^{(11)(11)}+s_\beta^2c_\beta^2\left(\hat{C}_{D\Phi}^{(11)(22)}+\hat{C}_{D\Phi}^{(21)(12)}\right)+s_\beta^4\hat{C}_{D\Phi}^{(22)(22)}\right)\\
&-\frac{4\sqrt{2}\hat{g}vs_\beta^2c_\beta^2\hat{g}'p_{2\mu_1}}{\sqrt{\hat{g}'^2+\hat{g}^2}}\left(\hat{C}_{D\Phi}^{(21)(21)*}+\hat{C}_{D\Phi}^{(21)(21)}\right)
\end{aligned} \tag{C.443}$$

$G^0$ $A_{\mu_1}$ $H^+$ $H^-$

$$
\begin{aligned}
&-\frac{2\sqrt{2}\hat{g}v\hat{g}'p_{2\mu_1}}{\sqrt{\hat{g}'^2+\hat{g}^2}}\left(2s_\beta^2c_\beta^2\left(\hat{C}_{D\Phi}^{(11)(11)}-\hat{C}_{D\Phi}^{(21)(12)}+\hat{C}_{D\Phi}^{(22)(22)}\right)\right.\\
&\qquad\left.+s_\beta^4\hat{C}_{D\Phi}^{(11)(22)}+c_\beta^4\hat{C}_{D\Phi}^{(11)(22)}\right)\\
&+\frac{4\sqrt{2}\hat{g}vs_\beta^2c_\beta^2\hat{g}'p_{2\mu_1}}{\sqrt{\hat{g}'^2+\hat{g}^2}}\left(\hat{C}_{D\Phi}^{(21)(21)*}+\hat{C}_{D\Phi}^{(21)(21)}\right)
\end{aligned}
\tag{C.444}
$$

$G^-$ $A_{\mu_1}$ $H$ $H^+$

$$
\begin{aligned}
&+\frac{i\sqrt{2}\hat{g}v\hat{g}'p_{3\mu_1}}{\sqrt{\hat{g}'^2+\hat{g}^2}}\left(\hat{C}_{D\Phi}^{(21)(12)}\right)\\
&-\frac{2i\sqrt{2}\hat{g}v\hat{g}'p_{3\mu_1}}{\sqrt{\hat{g}'^2+\hat{g}^2}}\left(c_\beta^4\hat{C}_{D\Phi}^{(21)(21)*}+s_\beta^2c_\beta^2\hat{C}_{D\Phi}^{(21)(21)*}\right.\\
&\qquad\left.+s_\beta^4\hat{C}_{D\Phi}^{(21)(21)}+s_\beta^2c_\beta^2\hat{C}_{D\Phi}^{(21)(21)}\right)
\end{aligned}
\tag{C.445}
$$

$A^0$ $A_{\mu_1}$ $G^+$ $G^-$

$$
\begin{aligned}
&+\frac{\sqrt{2}\hat{g}vs_{2\beta}\hat{g}'p_{2\mu_1}}{\sqrt{\hat{g}'^2+\hat{g}^2}}\left(c_\beta^2\left(2\hat{C}_{D\Phi}^{(11)(11)}-\hat{C}_{D\Phi}^{(11)(22)}-\hat{C}_{D\Phi}^{(21)(12)}\right)\right.\\
&\qquad\left.+s_\beta^2\left(\hat{C}_{D\Phi}^{(11)(22)}+\hat{C}_{D\Phi}^{(21)(12)}-2\hat{C}_{D\Phi}^{(22)(22)}\right)\right)\\
&-\frac{2\sqrt{2}\hat{g}vs_\beta c_\beta c_{2\beta}\hat{g}'p_{2\mu_1}}{\sqrt{\hat{g}'^2+\hat{g}^2}}\left(\hat{C}_{D\Phi}^{(21)(21)*}+\hat{C}_{D\Phi}^{(21)(21)}\right)
\end{aligned}
\tag{C.446}
$$

$$
\begin{aligned}
&+\frac{\sqrt{2}\hat{g}vs_{2\beta}\hat{g}'p_{2\mu_1}}{\sqrt{\hat{g}'^2+\hat{g}^2}}\left(s_\beta^2\left(2\hat{C}_{D\Phi}^{(11)(11)}-\hat{C}_{D\Phi}^{(11)(22)}-\hat{C}_{D\Phi}^{(21)(12)}\right)\right.\\
&\qquad\qquad\left.+c_\beta^2\left(\hat{C}_{D\Phi}^{(11)(22)}+\hat{C}_{D\Phi}^{(21)(12)}-2\hat{C}_{D\Phi}^{(22)(22)}\right)\right)\\
&+\frac{\sqrt{2}\hat{g}vs_{2\beta}c_{2\beta}\hat{g}'p_{2\mu_1}}{\sqrt{\hat{g}'^2+\hat{g}^2}}\left(\hat{C}_{D\Phi}^{(21)(21)*}+\hat{C}_{D\Phi}^{(21)(21)}\right)
\end{aligned}
\tag{C.447}
$$

$$
\begin{aligned}
&+\frac{\sqrt{2}\hat{g}vs_{2\beta}\hat{g}'p_{2\mu_1}}{\sqrt{\hat{g}'^2+\hat{g}^2}}\left(c_\beta^2\left(2\hat{C}_{D\Phi}^{(11)(11)}-\hat{C}_{D\Phi}^{(11)(22)}-\hat{C}_{D\Phi}^{(21)(12)}\right)\right.\\
&\qquad\qquad\left.+s_\beta^2\left(\hat{C}_{D\Phi}^{(11)(22)}+\hat{C}_{D\Phi}^{(21)(12)}-2\hat{C}_{D\Phi}^{(22)(22)}\right)\right)\\
&+\frac{4\sqrt{2}\hat{g}vs_\beta c_\beta\hat{g}'p_{2\mu_1}}{\sqrt{\hat{g}'^2+\hat{g}^2}}\left(s_\beta^2\hat{C}_{D\Phi}^{(21)(21)}-c_\beta^2\hat{C}_{D\Phi}^{(21)(21)*}\right)
\end{aligned}
\tag{C.448}
$$

$$
\begin{aligned}
&+\frac{2i\sqrt{2}\hat{g}vs_\beta c_\beta\hat{g}'p_{4\mu_1}}{\sqrt{\hat{g}'^2+\hat{g}^2}}\left(-s_\beta^2\hat{C}_{D\Phi}^{(21)(21)*}-c_\beta^2\hat{C}_{D\Phi}^{(21)(21)*}\right.\\
&\qquad\qquad\left.+s_\beta^2\hat{C}_{D\Phi}^{(21)(21)}+c_\beta^2\hat{C}_{D\Phi}^{(21)(21)}\right)
\end{aligned}
\tag{C.449}
$$

$H$, $A_{\mu_1}$, $H^+$, $H^-$

$$-\frac{2i\sqrt{2}\hat{g}vs_\beta c_\beta\hat{g}'p_{2\mu_1}}{\sqrt{\hat{g}'^2+\hat{g}^2}}\left(-s_\beta^2\hat{C}_{D\Phi}^{(21)(21)*}-c_\beta^2\hat{C}_{D\Phi}^{(21)(21)*}\right.$$
$$\left.+s_\beta^2\hat{C}_{D\Phi}^{(21)(21)}+c_\beta^2\hat{C}_{D\Phi}^{(21)(21)}\right) \tag{C.450}$$

$A^0$, $A^0$, $G^-$, $W^+_{\mu_4}$

$$-\frac{i\hat{g}v\left(p_{1\mu_4}+p_{2\mu_4}\right)}{\sqrt{2}}\left(2s_\beta^2c_\beta^2\left(2\hat{C}_{D\Phi}^{(11)(11)}-2\hat{C}_{D\Phi}^{(11)(22)}-\hat{C}_{D\Phi}^{(21)(12)}+2\hat{C}_{D\Phi}^{(22)(22)}\right)\right.$$
$$\left.+s_\beta^4\hat{C}_{D\Phi}^{(21)(12)}+c_\beta^4\hat{C}_{D\Phi}^{(21)(12)}\right)$$
$$-i\sqrt{2}\hat{g}vc_{2\beta}\left(p_{1\mu_4}+p_{2\mu_4}\right)\left(c_\beta^2\hat{C}_{D\Phi}^{(21)(21)*}-s_\beta^2\hat{C}_{D\Phi}^{(21)(21)}\right) \tag{C.451}$$

$G^-$, $A^0$, $H$, $W^+_{\mu_4}$

$$+\frac{\hat{g}v}{\sqrt{2}}\left(p_{1\mu_4}\left(2s_\beta^2c_\beta^2\left(4\hat{C}_{D\Phi}^{(11)(11)}-2\hat{C}_{D\Phi}^{(11)(22)}-3\hat{C}_{D\Phi}^{(21)(12)}+4\hat{C}_{D\Phi}^{(22)(22)}\right)\right.\right.$$
$$\left.+s_\beta^4\left(2\hat{C}_{D\Phi}^{(11)(22)}+\hat{C}_{D\Phi}^{(21)(12)}\right)+c_\beta^4\left(2\hat{C}_{D\Phi}^{(11)(22)}+\hat{C}_{D\Phi}^{(21)(12)}\right)\right)$$
$$-p_{3\mu_4}\left(2s_\beta^2c_\beta^2\left(2\hat{C}_{D\Phi}^{(11)(11)}-\hat{C}_{D\Phi}^{(21)(12)}+2\hat{C}_{D\Phi}^{(22)(22)}\right)\right.$$
$$\left.\left.+s_\beta^4\left(2\hat{C}_{D\Phi}^{(11)(22)}+\hat{C}_{D\Phi}^{(21)(12)}\right)+c_\beta^4\left(2\hat{C}_{D\Phi}^{(11)(22)}+\hat{C}_{D\Phi}^{(21)(12)}\right)\right)\right)$$
$$+\sqrt{2}\hat{g}v\left(c_\beta^4p_{1\mu_4}\hat{C}_{D\Phi}^{(21)(21)*}+c_\beta^4p_{3\mu_4}\hat{C}_{D\Phi}^{(21)(21)*}\right.$$
$$-3s_\beta^2c_\beta^2p_{1\mu_4}\hat{C}_{D\Phi}^{(21)(21)*}+3s_\beta^2c_\beta^2p_{3\mu_4}\hat{C}_{D\Phi}^{(21)(21)*}$$
$$+s_\beta^4p_{1\mu_4}\hat{C}_{D\Phi}^{(21)(21)}+s_\beta^4p_{3\mu_4}\hat{C}_{D\Phi}^{(21)(21)}$$
$$\left.-3s_\beta^2c_\beta^2p_{1\mu_4}\hat{C}_{D\Phi}^{(21)(21)}+3s_\beta^2c_\beta^2p_{3\mu_4}\hat{C}_{D\Phi}^{(21)(21)}\right) \tag{C.452}$$

$G^0$

$A^0$ $H^-$

$W^+_{\mu_4}$

$$\begin{aligned}
&-\frac{i\hat{g}v}{\sqrt{2}}\left(4s_\beta^2c_\beta^2 p_{1\mu_4}\left(\hat{C}_{D\Phi}^{(11)(11)}-\hat{C}_{D\Phi}^{(11)(22)}-\hat{C}_{D\Phi}^{(21)(12)}+\hat{C}_{D\Phi}^{(22)(22)}\right)\right.\\
&\qquad +p_{2\mu_4}\left(2s_\beta^2c_\beta^2\left(2\hat{C}_{D\Phi}^{(11)(11)}-\hat{C}_{D\Phi}^{(21)(12)}+2\hat{C}_{D\Phi}^{(22)(22)}\right)\right.\\
&\qquad \left.\left.+s_\beta^4\left(2\hat{C}_{D\Phi}^{(11)(22)}+\hat{C}_{D\Phi}^{(21)(12)}\right)+c_\beta^4\left(2\hat{C}_{D\Phi}^{(11)(22)}+\hat{C}_{D\Phi}^{(21)(12)}\right)\right)\right)\\
&-i\sqrt{2}\hat{g}v\left(2s_\beta^4 p_{1\mu_4}\hat{C}_{D\Phi}^{(21)(21)*}-s_\beta^4 p_{2\mu_4}\hat{C}_{D\Phi}^{(21)(21)*}\right.\\
&\qquad -3s_\beta^2c_\beta^2 p_{2\mu_4}\hat{C}_{D\Phi}^{(21)(21)*}+2c_\beta^4 p_{1\mu_4}\hat{C}_{D\Phi}^{(21)(21)}\\
&\qquad \left.-c_\beta^4 p_{2\mu_4}\hat{C}_{D\Phi}^{(21)(21)}-3s_\beta^2c_\beta^2 p_{2\mu_4}\hat{C}_{D\Phi}^{(21)(21)}\right)
\end{aligned} \tag{C.453}$$

$h$

$A^0$ $H^-$

$W^+_{\mu_4}$

$$\begin{aligned}
&-\frac{\hat{g}v\left(2p_{1\mu_4}-p_{2\mu_4}\right)}{\sqrt{2}}\left(2s_\beta^2c_\beta^2\left(2\hat{C}_{D\Phi}^{(11)(11)}-2\hat{C}_{D\Phi}^{(11)(22)}-\hat{C}_{D\Phi}^{(21)(12)}+2\hat{C}_{D\Phi}^{(22)(22)}\right)\right.\\
&\qquad \left.+s_\beta^4\hat{C}_{D\Phi}^{(21)(12)}+c_\beta^4\hat{C}_{D\Phi}^{(21)(12)}\right)\\
&-\sqrt{2}\hat{g}vc_{2\beta}\left(2p_{1\mu_4}-p_{2\mu_4}\right)\left(c_\beta^2\hat{C}_{D\Phi}^{(21)(21)}-s_\beta^2\hat{C}_{D\Phi}^{(21)(21)*}\right)
\end{aligned} \tag{C.454}$$

$G^0$

$G^0$ $G^-$

$W^+_{\mu_4}$

$$\begin{aligned}
&-2i\sqrt{2}\hat{g}v\left(p_{1\mu_4}+p_{2\mu_4}\right)\left(c_\beta^4\hat{C}_{D\Phi}^{(11)(11)}+s_\beta^2c_\beta^2\left(\hat{C}_{D\Phi}^{(11)(22)}+\hat{C}_{D\Phi}^{(21)(12)}\right)\right.\\
&\qquad \left.+s_\beta^4\hat{C}_{D\Phi}^{(22)(22)}\right)\\
&-2i\sqrt{2}\hat{g}vs_\beta^2c_\beta^2\left(p_{1\mu_4}+p_{2\mu_4}\right)\left(\hat{C}_{D\Phi}^{(21)(21)*}+\hat{C}_{D\Phi}^{(21)(21)}\right)
\end{aligned} \tag{C.455}$$

$$
\begin{aligned}
&-2\sqrt{2}\hat{g}v\left(2p_{1\mu_4}-p_{3\mu_4}\right)\left(c_\beta^4\hat{C}_{D\Phi}^{(11)(11)}+s_\beta^2c_\beta^2\left(\hat{C}_{D\Phi}^{(11)(22)}+\hat{C}_{D\Phi}^{(21)(12)}\right)\right.\\
&\qquad\qquad\left.+s_\beta^4\hat{C}_{D\Phi}^{(22)(22)}\right)\\
&-2\sqrt{2}\hat{g}vs_\beta^2c_\beta^2\left(2p_{1\mu_4}-p_{3\mu_4}\right)\left(\hat{C}_{D\Phi}^{(21)(21)*}+\hat{C}_{D\Phi}^{(21)(21)}\right)
\end{aligned}
\tag{C.456}
$$

$$
\begin{aligned}
&+\frac{\hat{g}v}{\sqrt{2}}\left(4s_\beta^2c_\beta^2p_{2\mu_4}\left(-\hat{C}_{D\Phi}^{(11)(11)}+\hat{C}_{D\Phi}^{(11)(22)}+\hat{C}_{D\Phi}^{(21)(12)}-\hat{C}_{D\Phi}^{(22)(22)}\right)\right.\\
&\qquad+p_{1\mu_4}\left(2s_\beta^2c_\beta^2\left(4\hat{C}_{D\Phi}^{(11)(11)}-2\hat{C}_{D\Phi}^{(11)(22)}-3\hat{C}_{D\Phi}^{(21)(12)}+4\hat{C}_{D\Phi}^{(22)(22)}\right)\right.\\
&\qquad\left.\left.+s_\beta^4\left(2\hat{C}_{D\Phi}^{(11)(22)}+\hat{C}_{D\Phi}^{(21)(12)}\right)+c_\beta^4\left(2\hat{C}_{D\Phi}^{(11)(22)}+\hat{C}_{D\Phi}^{(21)(12)}\right)\right)\right)\\
&+\sqrt{2}\hat{g}v\left(s_\beta^4p_{1\mu_4}\hat{C}_{D\Phi}^{(21)(21)*}-2s_\beta^4p_{2\mu_4}\hat{C}_{D\Phi}^{(21)(21)*}\right.\\
&\qquad-3s_\beta^2c_\beta^2p_{1\mu_4}\hat{C}_{D\Phi}^{(21)(21)*}+c_\beta^4p_{1\mu_4}\hat{C}_{D\Phi}^{(21)(21)}\\
&\qquad\left.-2c_\beta^4p_{2\mu_4}\hat{C}_{D\Phi}^{(21)(21)}-3s_\beta^2c_\beta^2p_{1\mu_4}\hat{C}_{D\Phi}^{(21)(21)}\right)
\end{aligned}
\tag{C.457}
$$

$$
\begin{aligned}
&+2i\sqrt{2}\hat{g}v\left(2p_{1\mu_4}-p_{2\mu_4}-p_{3\mu_4}\right)\left(c_\beta^4\hat{C}_{D\Phi}^{(11)(11)}+s_\beta^2c_\beta^2\left(\hat{C}_{D\Phi}^{(11)(22)}+\hat{C}_{D\Phi}^{(21)(12)}\right)\right.\\
&\qquad\qquad\left.+s_\beta^4\hat{C}_{D\Phi}^{(22)(22)}\right)\\
&+2i\sqrt{2}\hat{g}vs_\beta^2c_\beta^2\left(2p_{1\mu_4}-p_{2\mu_4}-p_{3\mu_4}\right)\left(\hat{C}_{D\Phi}^{(21)(21)*}+\hat{C}_{D\Phi}^{(21)(21)}\right)
\end{aligned}
\tag{C.458}
$$

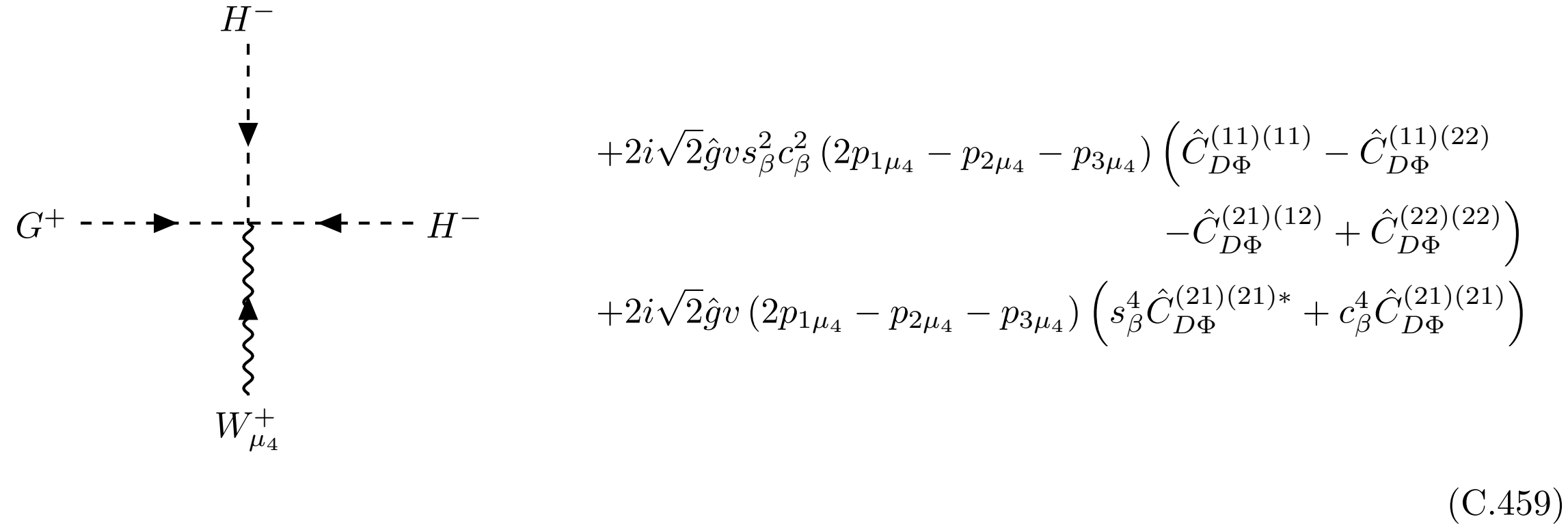

$$
\begin{aligned}
&+2i\sqrt{2}\hat{g}v s_\beta^2 c_\beta^2 \left(2p_{1\mu_4} - p_{2\mu_4} - p_{3\mu_4}\right) \Big(\hat{C}_{D\Phi}^{(11)(11)} - \hat{C}_{D\Phi}^{(11)(22)} \\
&\qquad - \hat{C}_{D\Phi}^{(21)(12)} + \hat{C}_{D\Phi}^{(22)(22)}\Big) \\
&+2i\sqrt{2}\hat{g}v \left(2p_{1\mu_4} - p_{2\mu_4} - p_{3\mu_4}\right) \Big(s_\beta^4 \hat{C}_{D\Phi}^{(21)(21)*} + c_\beta^4 \hat{C}_{D\Phi}^{(21)(21)}\Big)
\end{aligned}
\tag{C.459}
$$

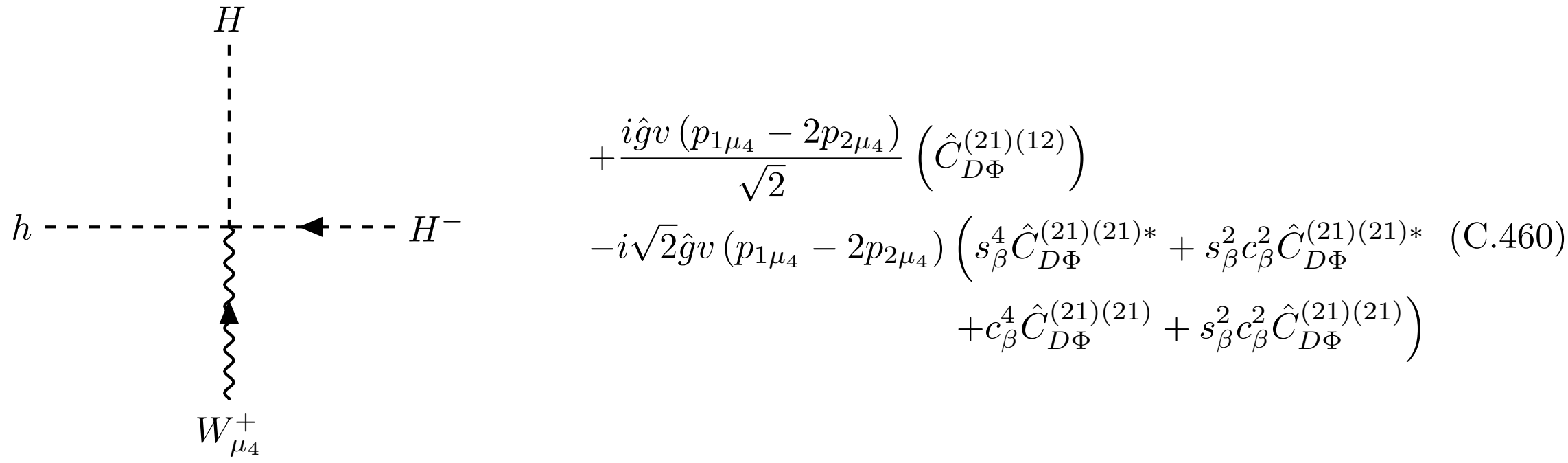

$$
\begin{aligned}
&+\frac{i\hat{g}v \left(p_{1\mu_4} - 2p_{2\mu_4}\right)}{\sqrt{2}} \Big(\hat{C}_{D\Phi}^{(21)(12)}\Big) \\
&-i\sqrt{2}\hat{g}v \left(p_{1\mu_4} - 2p_{2\mu_4}\right) \Big(s_\beta^4 \hat{C}_{D\Phi}^{(21)(21)*} + s_\beta^2 c_\beta^2 \hat{C}_{D\Phi}^{(21)(21)*} \\
&\qquad + c_\beta^4 \hat{C}_{D\Phi}^{(21)(21)} + s_\beta^2 c_\beta^2 \hat{C}_{D\Phi}^{(21)(21)}\Big)
\end{aligned}
\tag{C.460}
$$

$H$

$G^-$ $H$

$W^+_{\mu_4}$

$$
\begin{aligned}
&-\frac{i\hat{g}v \left(p_{2\mu_4} + p_{3\mu_4}\right)}{\sqrt{2}} \Big(\hat{C}_{D\Phi}^{(21)(12)}\Big) \\
&+i\sqrt{2}\hat{g}v \left(p_{2\mu_4} + p_{3\mu_4}\right) \Big(c_\beta^4 \hat{C}_{D\Phi}^{(21)(21)*} + s_\beta^2 c_\beta^2 \hat{C}_{D\Phi}^{(21)(21)*} \\
&\qquad + s_\beta^4 \hat{C}_{D\Phi}^{(21)(21)} + s_\beta^2 c_\beta^2 \hat{C}_{D\Phi}^{(21)(21)}\Big)
\end{aligned}
\tag{C.461}
$$

$$
\begin{aligned}
&+i\sqrt{2}\hat{g}vs_\beta c_\beta \left(p_{1\mu_4}+p_{2\mu_4}\right)\Big(c_\beta^2\left(2\hat{C}_{D\Phi}^{(11)(11)}-\hat{C}_{D\Phi}^{(11)(22)}-\hat{C}_{D\Phi}^{(21)(12)}\right)\\
&\qquad\qquad +s_\beta^2\left(\hat{C}_{D\Phi}^{(11)(22)}+\hat{C}_{D\Phi}^{(21)(12)}-2\hat{C}_{D\Phi}^{(22)(22)}\right)\Big)\\
&-i\sqrt{2}\hat{g}vs_\beta c_\beta\Big(-2s_\beta^2 p_{1\mu_4}\hat{C}_{D\Phi}^{(21)(21)*}+s_\beta^2 p_{2\mu_4}\hat{C}_{D\Phi}^{(21)(21)*}\\
&\qquad\qquad +3c_\beta^2 p_{2\mu_4}\hat{C}_{D\Phi}^{(21)(21)*}-3s_\beta^2 p_{2\mu_4}\hat{C}_{D\Phi}^{(21)(21)}\\
&\qquad\qquad +2c_\beta^2 p_{1\mu_4}\hat{C}_{D\Phi}^{(21)(21)}-c_\beta^2 p_{2\mu_4}\hat{C}_{D\Phi}^{(21)(21)}\Big)
\end{aligned}
\tag{C.462}
$$

$$
\begin{aligned}
&+\sqrt{2}\hat{g}vs_\beta c_\beta \left(2p_{1\mu_4}-p_{3\mu_4}\right)\Big(c_\beta^2\left(2\hat{C}_{D\Phi}^{(11)(11)}-\hat{C}_{D\Phi}^{(11)(22)}-\hat{C}_{D\Phi}^{(21)(12)}\right)\\
&\qquad\qquad +s_\beta^2\left(\hat{C}_{D\Phi}^{(11)(22)}+\hat{C}_{D\Phi}^{(21)(12)}-2\hat{C}_{D\Phi}^{(22)(22)}\right)\Big)\\
&-\sqrt{2}\hat{g}vs_\beta c_\beta c_{2\beta}\left(2p_{1\mu_4}-p_{3\mu_4}\right)\left(\hat{C}_{D\Phi}^{(21)(21)*}+\hat{C}_{D\Phi}^{(21)(21)}\right)
\end{aligned}
\tag{C.463}
$$

$$
\begin{aligned}
&+i\sqrt{2}\hat{g}vs_\beta c_\beta \left(p_{1\mu_4}+p_{2\mu_4}\right)\Big(s_\beta^2\left(2\hat{C}_{D\Phi}^{(11)(11)}-\hat{C}_{D\Phi}^{(11)(22)}-\hat{C}_{D\Phi}^{(21)(12)}\right)\\
&\qquad\qquad +c_\beta^2\left(\hat{C}_{D\Phi}^{(11)(22)}+\hat{C}_{D\Phi}^{(21)(12)}-2\hat{C}_{D\Phi}^{(22)(22)}\right)\Big)\\
&+i\sqrt{2}\hat{g}vs_\beta c_\beta c_{2\beta}\left(p_{1\mu_4}+p_{2\mu_4}\right)\left(\hat{C}_{D\Phi}^{(21)(21)*}+\hat{C}_{D\Phi}^{(21)(21)}\right)
\end{aligned}
\tag{C.464}
$$

$H$, $A^0$, $H^-$, $W^+_{\mu_4}$

$$\begin{aligned}
&-\sqrt{2}\hat{g}vs_\beta c_\beta\left(2p_{1\mu_4}-p_{2\mu_4}\right)\Big(s_\beta^2\left(2\hat{C}_{D\Phi}^{(11)(11)}-\hat{C}_{D\Phi}^{(11)(22)}-\hat{C}_{D\Phi}^{(21)(12)}\right)\\
&\qquad+c_\beta^2\left(\hat{C}_{D\Phi}^{(11)(22)}+\hat{C}_{D\Phi}^{(21)(12)}-2\hat{C}_{D\Phi}^{(22)(22)}\right)\Big)\\
&+\sqrt{2}\hat{g}vs_\beta c_\beta\Big(3s_\beta^2p_{1\mu_4}\hat{C}_{D\Phi}^{(21)(21)*}-3s_\beta^2p_{2\mu_4}\hat{C}_{D\Phi}^{(21)(21)*}\\
&\qquad-c_\beta^2p_{1\mu_4}\hat{C}_{D\Phi}^{(21)(21)*}-c_\beta^2p_{2\mu_4}\hat{C}_{D\Phi}^{(21)(21)*}\\
&\qquad+s_\beta^2p_{1\mu_4}\hat{C}_{D\Phi}^{(21)(21)}+s_\beta^2p_{2\mu_4}\hat{C}_{D\Phi}^{(21)(21)}\\
&\qquad-3c_\beta^2p_{1\mu_4}\hat{C}_{D\Phi}^{(21)(21)}+3c_\beta^2p_{2\mu_4}\hat{C}_{D\Phi}^{(21)(21)}\Big)
\end{aligned} \tag{C.465}$$

$G^-$, $G^0$, $H$, $W^+_{\mu_4}$

$$\begin{aligned}
&-\sqrt{2}\hat{g}vs_\beta c_\beta\left(2p_{1\mu_4}-p_{3\mu_4}\right)\Big(c_\beta^2\left(2\hat{C}_{D\Phi}^{(11)(11)}-\hat{C}_{D\Phi}^{(11)(22)}-\hat{C}_{D\Phi}^{(21)(12)}\right)\\
&\qquad+s_\beta^2\left(\hat{C}_{D\Phi}^{(11)(22)}+\hat{C}_{D\Phi}^{(21)(12)}-2\hat{C}_{D\Phi}^{(22)(22)}\right)\Big)\\
&+\sqrt{2}\hat{g}vs_\beta c_\beta\Big(-s_\beta^2p_{1\mu_4}\hat{C}_{D\Phi}^{(21)(21)*}+2s_\beta^2p_{3\mu_4}\hat{C}_{D\Phi}^{(21)(21)*}\\
&\qquad+3c_\beta^2p_{1\mu_4}\hat{C}_{D\Phi}^{(21)(21)*}-3s_\beta^2p_{1\mu_4}\hat{C}_{D\Phi}^{(21)(21)}\\
&\qquad+c_\beta^2p_{1\mu_4}\hat{C}_{D\Phi}^{(21)(21)}-2c_\beta^2p_{3\mu_4}\hat{C}_{D\Phi}^{(21)(21)}\Big)
\end{aligned} \tag{C.466}$$

$G^0$, $G^0$, $H^-$, $W^+_{\mu_4}$

$$\begin{aligned}
&+i\sqrt{2}\hat{g}vs_\beta c_\beta\left(p_{1\mu_4}+p_{2\mu_4}\right)\Big(c_\beta^2\left(2\hat{C}_{D\Phi}^{(11)(11)}-\hat{C}_{D\Phi}^{(11)(22)}-\hat{C}_{D\Phi}^{(21)(12)}\right)\\
&\qquad+s_\beta^2\left(\hat{C}_{D\Phi}^{(11)(22)}+\hat{C}_{D\Phi}^{(21)(12)}-2\hat{C}_{D\Phi}^{(22)(22)}\right)\Big)\\
&-2i\sqrt{2}\hat{g}vs_\beta c_\beta\left(p_{1\mu_4}+p_{2\mu_4}\right)\left(c_\beta^2\hat{C}_{D\Phi}^{(21)(21)}-s_\beta^2\hat{C}_{D\Phi}^{(21)(21)*}\right)
\end{aligned} \tag{C.467}$$

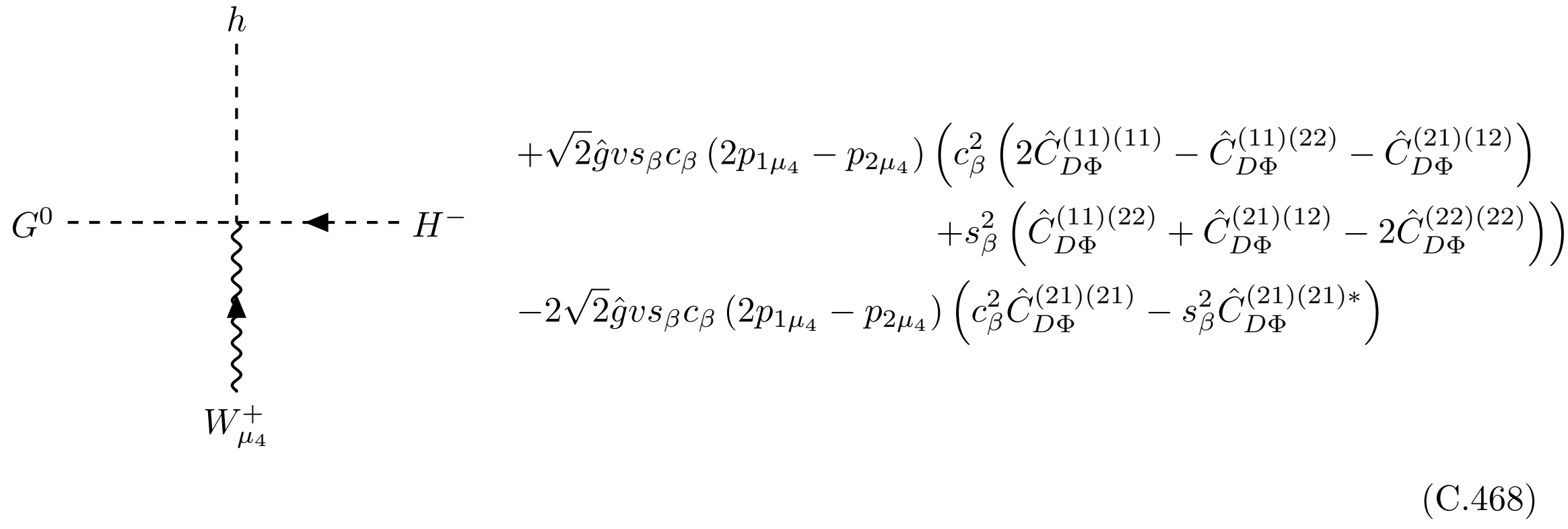

$$+\sqrt{2}\hat{g}vs_\beta c_\beta\left(2p_{1\mu_4}-p_{2\mu_4}\right)\left(c_\beta^2\left(2\hat{C}_{D\Phi}^{(11)(11)}-\hat{C}_{D\Phi}^{(11)(22)}-\hat{C}_{D\Phi}^{(21)(12)}\right)+s_\beta^2\left(\hat{C}_{D\Phi}^{(11)(22)}+\hat{C}_{D\Phi}^{(21)(12)}-2\hat{C}_{D\Phi}^{(22)(22)}\right)\right)-2\sqrt{2}\hat{g}vs_\beta c_\beta\left(2p_{1\mu_4}-p_{2\mu_4}\right)\left(c_\beta^2\hat{C}_{D\Phi}^{(21)(21)}-s_\beta^2\hat{C}_{D\Phi}^{(21)(21)*}\right) \tag{C.468}$$

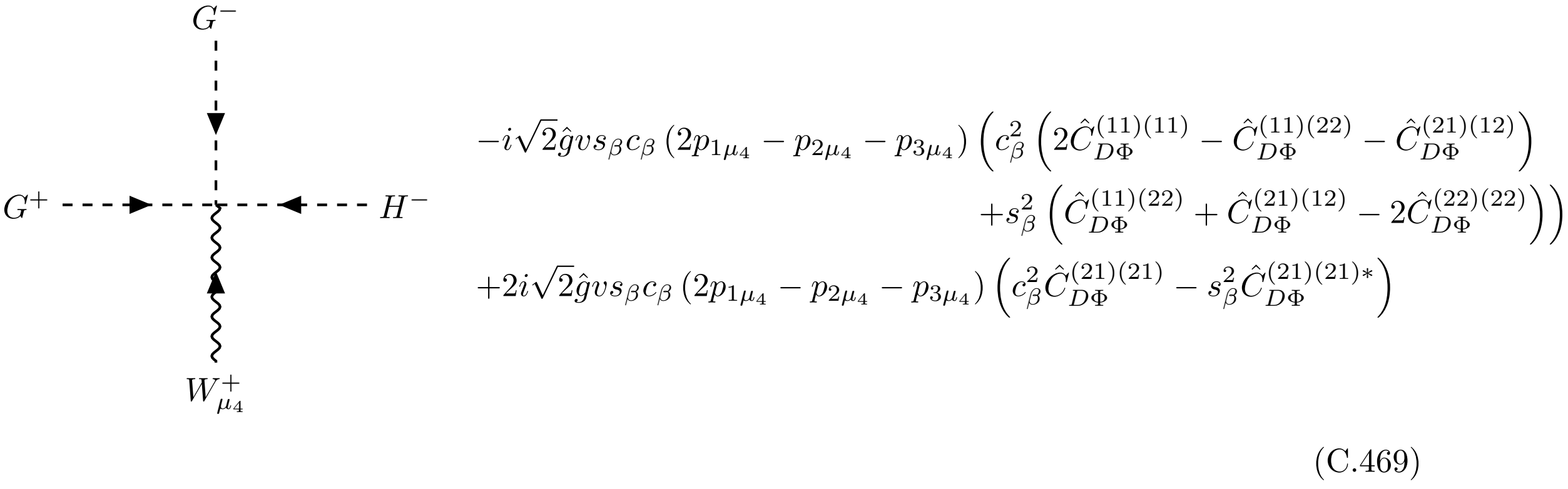

$$-i\sqrt{2}\hat{g}vs_\beta c_\beta\left(2p_{1\mu_4}-p_{2\mu_4}-p_{3\mu_4}\right)\left(c_\beta^2\left(2\hat{C}_{D\Phi}^{(11)(11)}-\hat{C}_{D\Phi}^{(11)(22)}-\hat{C}_{D\Phi}^{(21)(12)}\right)+s_\beta^2\left(\hat{C}_{D\Phi}^{(11)(22)}+\hat{C}_{D\Phi}^{(21)(12)}-2\hat{C}_{D\Phi}^{(22)(22)}\right)\right)+2i\sqrt{2}\hat{g}vs_\beta c_\beta\left(2p_{1\mu_4}-p_{2\mu_4}-p_{3\mu_4}\right)\left(c_\beta^2\hat{C}_{D\Phi}^{(21)(21)}-s_\beta^2\hat{C}_{D\Phi}^{(21)(21)*}\right) \tag{C.469}$$

$$-i\sqrt{2}\hat{g}vs_\beta c_\beta\left(p_{2\mu_4}-2p_{3\mu_4}\right)\left(-s_\beta^2\hat{C}_{D\Phi}^{(21)(21)*}-c_\beta^2\hat{C}_{D\Phi}^{(21)(21)*}+s_\beta^2\hat{C}_{D\Phi}^{(21)(21)}+c_\beta^2\hat{C}_{D\Phi}^{(21)(21)}\right) \tag{C.470}$$

$$+i\sqrt{2}\hat{g}vs_\beta c_\beta \left(p_{1\mu_4}+p_{2\mu_4}\right)\Big(-s_\beta^2\hat{C}_{D\Phi}^{(21)(21)*} - c_\beta^2 \hat{C}_{D\Phi}^{(21)(21)*} + s_\beta^2 \hat{C}^{(21)(21)}_{D\Phi} + c_\beta^2\hat{C}^{(21)(21)}_{D\Phi}\Big) \tag{C.471}$$

$$\begin{aligned}&-i\sqrt{2}\hat{g}vs_\beta c_\beta \left(2p_{1\mu_4}-p_{2\mu_4}-p_{3\mu_4}\right)\Big(s_\beta^2\left(2\hat{C}_{D\Phi}^{(11)(11)}-\hat{C}_{D\Phi}^{(11)(22)}-\hat{C}_{D\Phi}^{(21)(12)}\right)\\&\qquad\qquad +c_\beta^2\left(\hat{C}_{D\Phi}^{(11)(22)}+\hat{C}_{D\Phi}^{(21)(12)}-2\hat{C}_{D\Phi}^{(22)(22)}\right)\Big)\\&-2i\sqrt{2}\hat{g}vs_\beta c_\beta \left(2p_{1\mu_4}-p_{2\mu_4}-p_{3\mu_4}\right)\left(c_\beta^2\hat{C}_{D\Phi}^{(21)(21)}-s_\beta^2\hat{C}_{D\Phi}^{(21)(21)*}\right)\end{aligned} \tag{C.472}$$

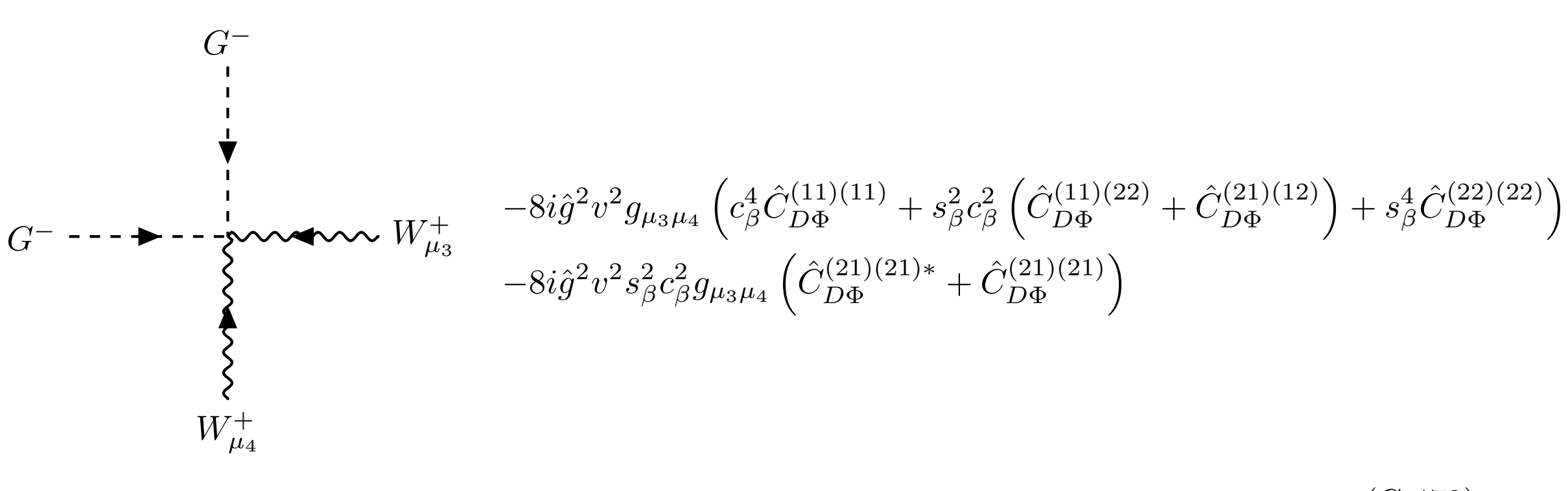

$$\begin{aligned}&-8i\hat{g}^2v^2g_{\mu_3\mu_4}\left(c_\beta^4\hat{C}_{D\Phi}^{(11)(11)}+s_\beta^2c_\beta^2\left(\hat{C}_{D\Phi}^{(11)(22)}+\hat{C}_{D\Phi}^{(21)(12)}\right)+s_\beta^4\hat{C}_{D\Phi}^{(22)(22)}\right)\\&-8i\hat{g}^2v^2s_\beta^2c_\beta^2g_{\mu_3\mu_4}\left(\hat{C}_{D\Phi}^{(21)(21)*}+\hat{C}_{D\Phi}^{(21)(21)}\right)\end{aligned} \tag{C.473}$$

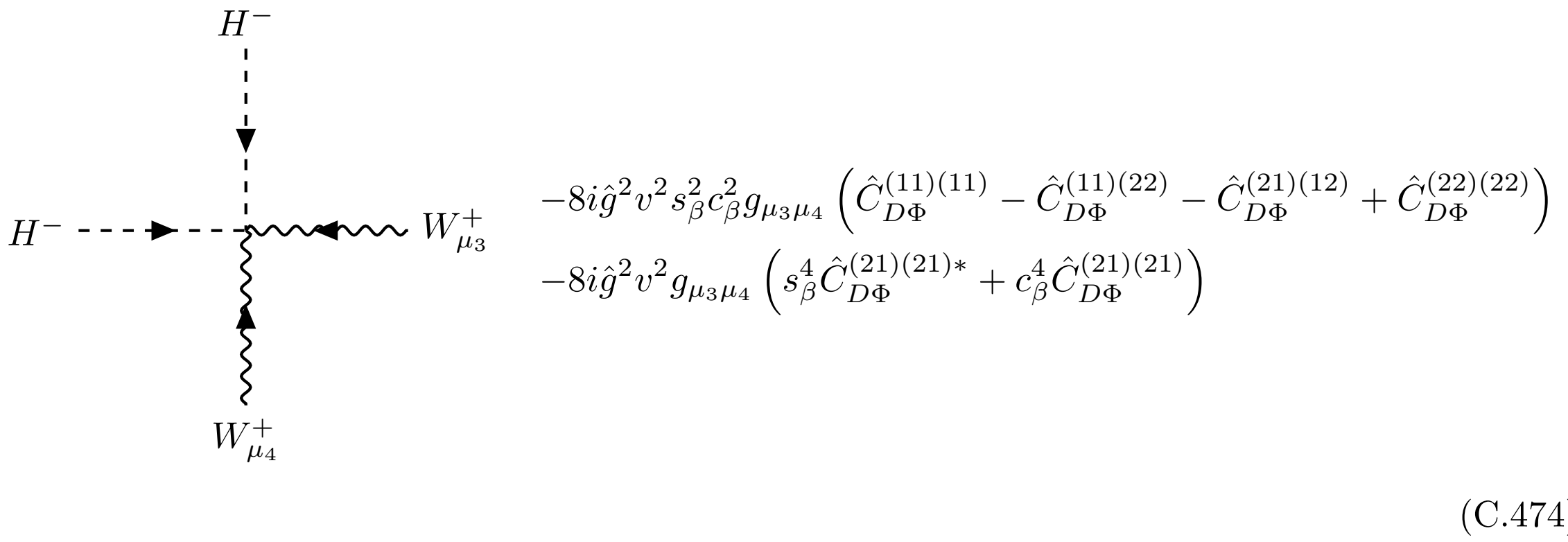

$$
\begin{aligned}
&-8i\hat{g}^2 v^2 s_\beta^2 c_\beta^2 g_{\mu_3\mu_4}\left(\hat{C}_{D\Phi}^{(11)(11)}-\hat{C}_{D\Phi}^{(11)(22)}-\hat{C}_{D\Phi}^{(21)(12)}+\hat{C}_{D\Phi}^{(22)(22)}\right)\\
&-8i\hat{g}^2 v^2 g_{\mu_3\mu_4}\left(s_\beta^4\hat{C}_{D\Phi}^{(21)(21)*}+c_\beta^4\hat{C}_{D\Phi}^{(21)(21)}\right)
\end{aligned}
\tag{C.474}
$$

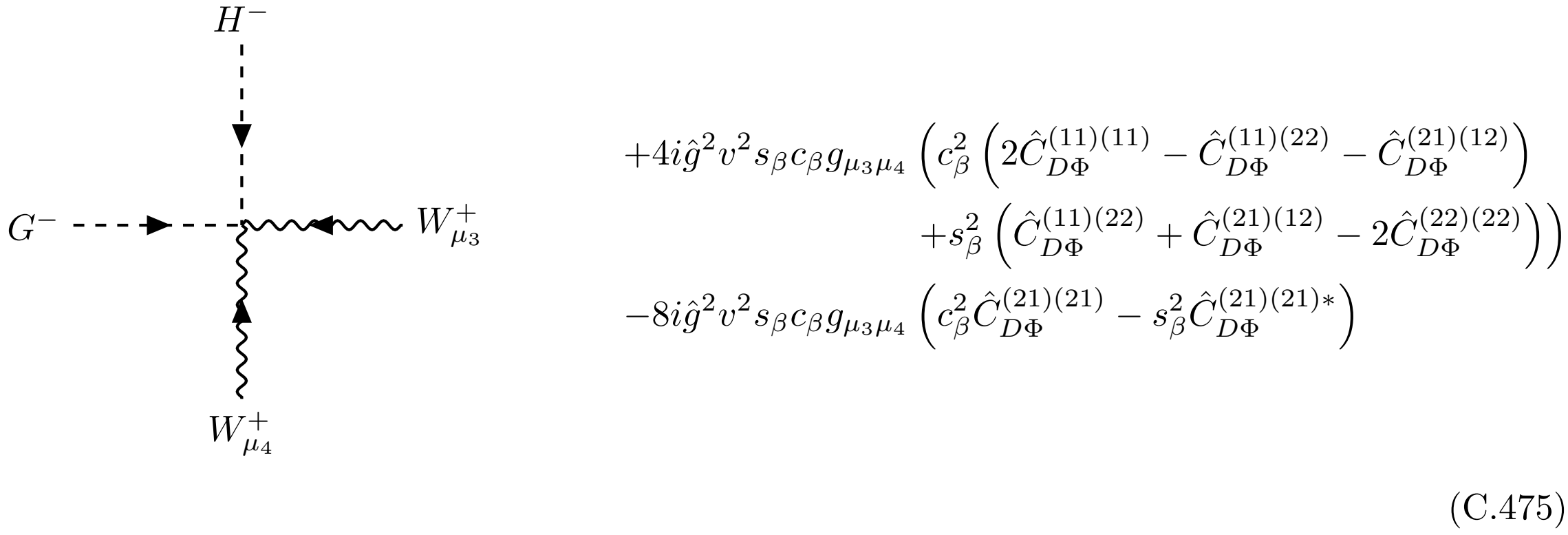

$$
\begin{aligned}
&+4i\hat{g}^2 v^2 s_\beta c_\beta g_{\mu_3\mu_4}\left(c_\beta^2\left(2\hat{C}_{D\Phi}^{(11)(11)}-\hat{C}_{D\Phi}^{(11)(22)}-\hat{C}_{D\Phi}^{(21)(12)}\right)\right.\\
&\qquad\qquad\left.+s_\beta^2\left(\hat{C}_{D\Phi}^{(11)(22)}+\hat{C}_{D\Phi}^{(21)(12)}-2\hat{C}_{D\Phi}^{(22)(22)}\right)\right)\\
&-8i\hat{g}^2 v^2 s_\beta c_\beta g_{\mu_3\mu_4}\left(c_\beta^2\hat{C}_{D\Phi}^{(21)(21)}-s_\beta^2\hat{C}_{D\Phi}^{(21)(21)*}\right)
\end{aligned}
\tag{C.475}
$$

$G^+$, $G^-$, $A_{\mu_1}$, $Z_{\mu_4}$

$$
\begin{aligned}
&+\frac{i\hat{g}g_{\mu_1\mu_4}\hat{g}'\left(\hat{g}'^2-\hat{g}^2\right)}{\left(\hat{g}'^2+\hat{g}^2\right)^2}\Big(s_\beta^2\Big(\hat{g}'^2\Big(2\delta_{s_{\hat{\beta}\pm}}-1\Big)+\hat{g}^2\Big(2\delta_{s_{\hat{\beta}\pm}}-1\Big)\Big)\\
&\qquad\qquad +c_\beta^2\Big(\hat{g}^2\Big(2\delta_{c_{\hat{\beta}\pm}}-1\Big)+\Big(2\delta_{c_{\hat{\beta}\pm}}-1\Big)\hat{g}'^2\Big)\\
&\qquad\qquad +2\hat{g}X_{WB}\hat{g}'\Big)\\
&+4i\hat{g}v^2g_{\mu_1\mu_4}\hat{g}'\Big(c_\beta^4\hat{C}_{D\Phi}^{(11)(11)}+s_\beta^2c_\beta^2\Big(\hat{C}_{D\Phi}^{(11)(22)}+\hat{C}_{D\Phi}^{(21)(12)}\Big)+s_\beta^4\hat{C}_{D\Phi}^{(22)(22)}\Big)\\
&-\frac{2i\left(p_{1\mu_4}p_{4\mu_1}-p_1\cdot p_4g_{\mu_1\mu_4}\right)}{\hat{g}'^2+\hat{g}^2}\Big(c_\beta^2\Big(2\hat{g}\hat{g}'\Big(\hat{C}_{\Phi B}^{(11)}-\hat{C}_{\Phi W}^{(11)}\Big)\\
&\qquad\qquad +\left(\hat{g}'^2-\hat{g}^2\right)\hat{C}_{\Phi WB}^{(11)}\Big)\\
&\qquad\qquad +s_\beta^2\Big(2\hat{g}\hat{g}'\Big(\hat{C}_{\Phi B}^{(22)}-\hat{C}_{\Phi W}^{(22)}\Big)\\
&\qquad\qquad +\left(\hat{g}'^2-\hat{g}^2\right)\hat{C}_{\Phi WB}^{(22)}\Big)\Big)\\
&-\frac{2ip_1^\mu p_4^\nu\epsilon_{\mu_1\mu_4\mu\nu}}{\hat{g}'^2+\hat{g}^2}\Big(c_\beta^2\Big(2\hat{g}\hat{g}'\Big(\hat{C}_{\Phi\tilde{B}}^{(11)}-\hat{C}_{\Phi\tilde{W}}^{(11)}\Big)+\left(\hat{g}'^2-\hat{g}^2\right)\hat{C}_{\Phi B\tilde{W}}^{(11)}\Big)\\
&\qquad\qquad +s_\beta^2\Big(2\hat{g}\hat{g}'\Big(\hat{C}_{\Phi\tilde{B}}^{(22)}-\hat{C}_{\Phi\tilde{W}}^{(22)}\Big)+\left(\hat{g}'^2-\hat{g}^2\right)\hat{C}_{\Phi B\tilde{W}}^{(22)}\Big)\Big)\\
&+4i\hat{g}v^2s_\beta^2c_\beta^2g_{\mu_1\mu_4}\hat{g}'\Big(\hat{C}_{D\Phi}^{(21)(21)*}+\hat{C}_{D\Phi}^{(21)(21)}\Big)
\end{aligned}
\tag{C.476}
$$

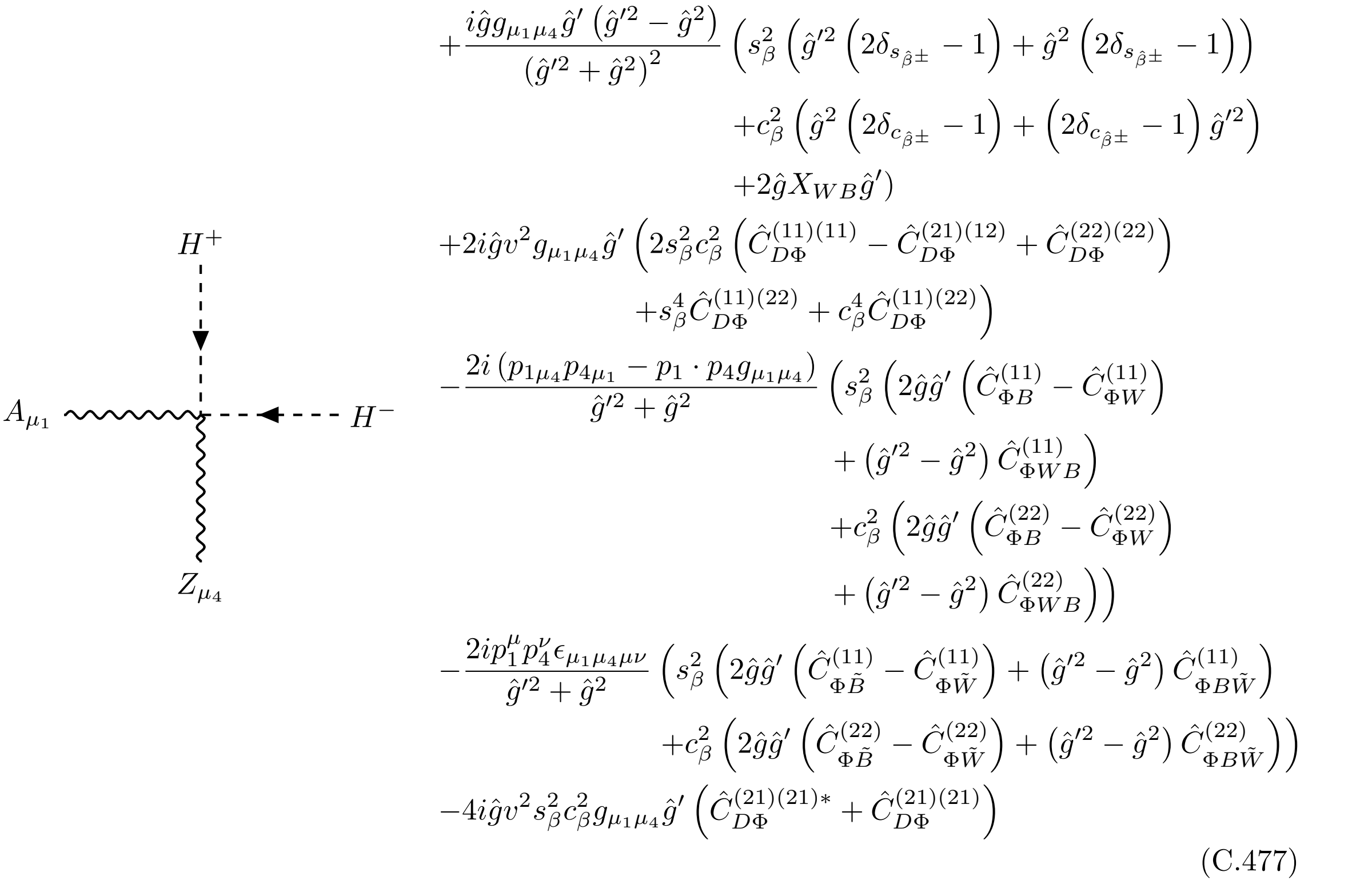

$$
\begin{aligned}
&+\frac{i\hat{g}g_{\mu_1\mu_4}\hat{g}'\left(\hat{g}'^2-\hat{g}^2\right)}{\left(\hat{g}'^2+\hat{g}^2\right)^2}\Big(s_\beta^2\Big(\hat{g}'^2\Big(2\delta_{s_{\hat{\beta}\pm}}-1\Big)+\hat{g}^2\Big(2\delta_{s_{\hat{\beta}\pm}}-1\Big)\Big)\\
&\qquad\qquad +c_\beta^2\Big(\hat{g}^2\Big(2\delta_{c_{\hat{\beta}\pm}}-1\Big)+\Big(2\delta_{c_{\hat{\beta}\pm}}-1\Big)\hat{g}'^2\Big)\\
&\qquad\qquad +2\hat{g}X_{WB}\hat{g}'\Big)\\
&+2i\hat{g}v^2g_{\mu_1\mu_4}\hat{g}'\Big(2s_\beta^2c_\beta^2\Big(\hat{C}_{D\Phi}^{(11)(11)}-\hat{C}_{D\Phi}^{(21)(12)}+\hat{C}_{D\Phi}^{(22)(22)}\Big)\\
&\qquad\qquad +s_\beta^4\hat{C}_{D\Phi}^{(11)(22)}+c_\beta^4\hat{C}_{D\Phi}^{(11)(22)}\Big)\\
&-\frac{2i\left(p_{1\mu_4}p_{4\mu_1}-p_1\cdot p_4g_{\mu_1\mu_4}\right)}{\hat{g}'^2+\hat{g}^2}\Big(s_\beta^2\Big(2\hat{g}\hat{g}'\Big(\hat{C}_{\Phi B}^{(11)}-\hat{C}_{\Phi W}^{(11)}\Big)\\
&\qquad\qquad +\left(\hat{g}'^2-\hat{g}^2\right)\hat{C}_{\Phi WB}^{(11)}\Big)\\
&\qquad\qquad +c_\beta^2\Big(2\hat{g}\hat{g}'\Big(\hat{C}_{\Phi B}^{(22)}-\hat{C}_{\Phi W}^{(22)}\Big)\\
&\qquad\qquad +\left(\hat{g}'^2-\hat{g}^2\right)\hat{C}_{\Phi WB}^{(22)}\Big)\Big)\\
&-\frac{2ip_1^\mu p_4^\nu\epsilon_{\mu_1\mu_4\mu\nu}}{\hat{g}'^2+\hat{g}^2}\Big(s_\beta^2\Big(2\hat{g}\hat{g}'\Big(\hat{C}_{\Phi\tilde{B}}^{(11)}-\hat{C}_{\Phi\tilde{W}}^{(11)}\Big)+\left(\hat{g}'^2-\hat{g}^2\right)\hat{C}_{\Phi B\tilde{W}}^{(11)}\Big)\\
&\qquad\qquad +c_\beta^2\Big(2\hat{g}\hat{g}'\Big(\hat{C}_{\Phi\tilde{B}}^{(22)}-\hat{C}_{\Phi\tilde{W}}^{(22)}\Big)+\left(\hat{g}'^2-\hat{g}^2\right)\hat{C}_{\Phi B\tilde{W}}^{(22)}\Big)\Big)\\
&-4i\hat{g}v^2s_\beta^2c_\beta^2g_{\mu_1\mu_4}\hat{g}'\Big(\hat{C}_{D\Phi}^{(21)(21)*}+\hat{C}_{D\Phi}^{(21)(21)}\Big)
\end{aligned}
\tag{C.477}
$$

$A^0$, $A^0$, $G^0$, $Z_{\mu_4}$

$$
\begin{aligned}
&+\sqrt{2}v\sqrt{\hat{g}'^2+\hat{g}^2}\left(2s_\beta^2c_\beta^2\left(p_{1\mu_4}+p_{2\mu_4}\right)\left(\hat{C}_{D\Phi}^{(11)(11)}-\hat{C}_{D\Phi}^{(11)(22)}\right.\right.\\
&\qquad\left.-\hat{C}_{D\Phi}^{(21)(12)}+\hat{C}_{D\Phi}^{(22)(22)}\right)\\
&\qquad+p_{3\mu_4}\left(2s_\beta^2c_\beta^2\left(\hat{C}_{D\Phi}^{(11)(11)}+\hat{C}_{D\Phi}^{(22)(22)}\right)\right.\\
&\qquad\left.\left.+s_\beta^4\left(\hat{C}_{D\Phi}^{(11)(22)}+\hat{C}_{D\Phi}^{(21)(12)}\right)+c_\beta^4\left(\hat{C}_{D\Phi}^{(11)(22)}+\hat{C}_{D\Phi}^{(21)(12)}\right)\right)\right)\\
&+\sqrt{2}v\sqrt{\hat{g}'^2+\hat{g}^2}\left(\left(s_\beta^4+c_\beta^4\right)\left(p_{1\mu_4}+p_{2\mu_4}\right)\right.\\
&\qquad\left.-\left(s_\beta^4+c_\beta^4+4s_\beta^2c_\beta^2\right)p_{3\mu_4}\right)\left(\hat{C}_{D\Phi}^{(21)(21)*}+\hat{C}_{D\Phi}^{(21)(21)}\right)
\end{aligned}
\tag{C.478}
$$

$A^0$, $A^0$, $h$, $Z_{\mu_4}$

$$
-i\sqrt{2}v\left(c_\beta^4-s_\beta^4\right)\sqrt{\hat{g}'^2+\hat{g}^2}\left(p_{1\mu_4}+p_{2\mu_4}-p_{3\mu_4}\right)\left(\hat{C}_{D\Phi}^{(21)(21)}-\hat{C}_{D\Phi}^{(21)(21)*}\right)
\tag{C.479}
$$

$G^0$, $A^0$, $H$, $Z_{\mu_4}$

$$
-i\sqrt{2}v\left(c_\beta^4-s_\beta^4\right)\sqrt{\hat{g}'^2+\hat{g}^2}\left(p_{1\mu_4}-p_{2\mu_4}+p_{3\mu_4}\right)\left(\hat{C}_{D\Phi}^{(21)(21)}-\hat{C}_{D\Phi}^{(21)(21)*}\right)
\tag{C.480}
$$

$$
\begin{aligned}
&-\sqrt{2}v\sqrt{\hat{g}'^2+\hat{g}^2}\left(2s_\beta^2c_\beta^2p_{2\mu_4}\left(-\hat{C}_{D\Phi}^{(11)(11)}+\hat{C}_{D\Phi}^{(11)(22)}+\hat{C}_{D\Phi}^{(21)(12)}-\hat{C}_{D\Phi}^{(22)(22)}\right)\right.\\
&\qquad -p_{3\mu_4}\left(2s_\beta^2c_\beta^2\left(\hat{C}_{D\Phi}^{(11)(11)}+\hat{C}_{D\Phi}^{(22)(22)}\right)\right.\\
&\qquad \left.+s_\beta^4\left(\hat{C}_{D\Phi}^{(11)(22)}+\hat{C}_{D\Phi}^{(21)(12)}\right)+c_\beta^4\left(\hat{C}_{D\Phi}^{(11)(22)}+\hat{C}_{D\Phi}^{(21)(12)}\right)\right)\\
&\qquad +p_{1\mu_4}\left(2s_\beta^2c_\beta^2\left(3\hat{C}_{D\Phi}^{(11)(11)}-2\hat{C}_{D\Phi}^{(11)(22)}\right.\right.\\
&\qquad \left.-2\hat{C}_{D\Phi}^{(21)(12)}+3\hat{C}_{D\Phi}^{(22)(22)}\right)\\
&\qquad \left.\left.+s_\beta^4\left(\hat{C}_{D\Phi}^{(11)(22)}+\hat{C}_{D\Phi}^{(21)(12)}\right)+c_\beta^4\left(\hat{C}_{D\Phi}^{(11)(22)}+\hat{C}_{D\Phi}^{(21)(12)}\right)\right)\right)\\
&-\sqrt{2}v\sqrt{\hat{g}'^2+\hat{g}^2}\left(\left(s_\beta^4+c_\beta^4\right)\left(-p_{2\mu_4}+p_{1\mu_4}+p_{3\mu_4}\right)\right.\\
&\qquad \left.-4s_\beta^2c_\beta^2\left(p_{1\mu_4}-p_{3\mu_4}\right)\right)\left(\hat{C}_{D\Phi}^{(21)(21)*}+\hat{C}_{D\Phi}^{(21)(21)}\right)
\end{aligned}
\tag{C.481}
$$

$$
\begin{aligned}
&+\frac{v}{\sqrt{2}\sqrt{\hat{g}'^2+\hat{g}^2}}\left(\left(\hat{g}'^2-\hat{g}^2\right)p_{1\mu_4}\left(2s_\beta^2c_\beta^2\left(2\hat{C}_{D\Phi}^{(11)(11)}-2\hat{C}_{D\Phi}^{(11)(22)}\right.\right.\right.\\
&\qquad \left.\left.-\hat{C}_{D\Phi}^{(21)(12)}+2\hat{C}_{D\Phi}^{(22)(22)}\right)+s_\beta^4\hat{C}_{D\Phi}^{(21)(12)}+c_\beta^4\hat{C}_{D\Phi}^{(21)(12)}\right)\\
&\qquad \left.+\left(\hat{g}'^2+\hat{g}^2\right)\left(p_{2\mu_4}-p_{3\mu_4}\right)\hat{C}_{D\Phi}^{(21)(12)}\right)\\
&+\frac{\sqrt{2}v}{\sqrt{\hat{g}'^2+\hat{g}^2}}\left(-\hat{g}^2c_\beta^4p_{1\mu_4}\hat{C}_{D\Phi}^{(21)(21)*}-\hat{g}^2c_\beta^4p_{2\mu_4}\hat{C}_{D\Phi}^{(21)(21)*}\right.\\
&\qquad +\hat{g}^2c_\beta^4p_{3\mu_4}\hat{C}_{D\Phi}^{(21)(21)*}+\hat{g}^2s_\beta^2c_\beta^2p_{1\mu_4}\hat{C}_{D\Phi}^{(21)(21)*}\\
&\qquad -\hat{g}^2s_\beta^2c_\beta^2p_{2\mu_4}\hat{C}_{D\Phi}^{(21)(21)*}+\hat{g}^2s_\beta^2c_\beta^2p_{3\mu_4}\hat{C}_{D\Phi}^{(21)(21)*}\\
&\qquad +c_\beta^4\hat{g}'^2p_{1\mu_4}\hat{C}_{D\Phi}^{(21)(21)*}-c_\beta^4\hat{g}'^2p_{2\mu_4}\hat{C}_{D\Phi}^{(21)(21)*}\\
&\qquad +c_\beta^4\hat{g}'^2p_{3\mu_4}\hat{C}_{D\Phi}^{(21)(21)*}-s_\beta^2c_\beta^2\hat{g}'^2p_{1\mu_4}\hat{C}_{D\Phi}^{(21)(21)*}\\
&\qquad -s_\beta^2c_\beta^2\hat{g}'^2p_{2\mu_4}\hat{C}_{D\Phi}^{(21)(21)*}+s_\beta^2c_\beta^2\hat{g}'^2p_{3\mu_4}\hat{C}_{D\Phi}^{(21)(21)*}\\
&\qquad -\hat{g}^2s_\beta^4p_{1\mu_4}\hat{C}_{D\Phi}^{(21)(21)}-\hat{g}^2s_\beta^4p_{2\mu_4}\hat{C}_{D\Phi}^{(21)(21)}\\
&\qquad +\hat{g}^2s_\beta^4p_{3\mu_4}\hat{C}_{D\Phi}^{(21)(21)}+\hat{g}^2s_\beta^2c_\beta^2p_{1\mu_4}\hat{C}_{D\Phi}^{(21)(21)}\\
&\qquad -\hat{g}^2s_\beta^2c_\beta^2p_{2\mu_4}\hat{C}_{D\Phi}^{(21)(21)}+\hat{g}^2s_\beta^2c_\beta^2p_{3\mu_4}\hat{C}_{D\Phi}^{(21)(21)}\\
&\qquad +s_\beta^4\hat{g}'^2p_{1\mu_4}\hat{C}_{D\Phi}^{(21)(21)}-s_\beta^4\hat{g}'^2p_{2\mu_4}\hat{C}_{D\Phi}^{(21)(21)}\\
&\qquad +s_\beta^4\hat{g}'^2p_{3\mu_4}\hat{C}_{D\Phi}^{(21)(21)}-s_\beta^2c_\beta^2\hat{g}'^2p_{1\mu_4}\hat{C}_{D\Phi}^{(21)(21)}\\
&\qquad \left.-s_\beta^2c_\beta^2\hat{g}'^2p_{2\mu_4}\hat{C}_{D\Phi}^{(21)(21)}+s_\beta^2c_\beta^2\hat{g}'^2p_{3\mu_4}\hat{C}_{D\Phi}^{(21)(21)}\right)
\end{aligned}
\tag{C.482}
$$

$$
\begin{aligned}
&+2\sqrt{2}v\sqrt{\hat{g}'^2+\hat{g}^2}\left(p_{1\mu_4}+p_{2\mu_4}+p_{3\mu_4}\right)\Big(c_\beta^4\hat{C}_{D\Phi}^{(11)(11)}\\
&\qquad +s_\beta^2c_\beta^2\left(\hat{C}_{D\Phi}^{(11)(22)}+\hat{C}_{D\Phi}^{(21)(12)}\right)+s_\beta^4\hat{C}_{D\Phi}^{(22)(22)}\Big)\\
&+2\sqrt{2}vs_\beta^2c_\beta^2\sqrt{\hat{g}'^2+\hat{g}^2}\left(p_{1\mu_4}+p_{2\mu_4}+p_{3\mu_4}\right)\left(\hat{C}_{D\Phi}^{(21)(21)*}+\hat{C}_{D\Phi}^{(21)(21)}\right)
\end{aligned}
\tag{C.483}
$$

$$
\begin{aligned}
&+\frac{2\sqrt{2}v\left(\hat{g}'^2-\hat{g}^2\right)p_{1\mu_4}}{\sqrt{\hat{g}'^2+\hat{g}^2}}\Big(c_\beta^4\hat{C}_{D\Phi}^{(11)(11)}+s_\beta^2c_\beta^2\left(\hat{C}_{D\Phi}^{(11)(22)}+\hat{C}_{D\Phi}^{(21)(12)}\right)\\
&\qquad +s_\beta^4\hat{C}_{D\Phi}^{(22)(22)}\Big)\\
&+\frac{2\sqrt{2}vs_\beta^2c_\beta^2\left(\hat{g}'^2-\hat{g}^2\right)p_{1\mu_4}}{\sqrt{\hat{g}'^2+\hat{g}^2}}\left(\hat{C}_{D\Phi}^{(21)(21)*}+\hat{C}_{D\Phi}^{(21)(21)}\right)
\end{aligned}
\tag{C.484}
$$

$$
\begin{aligned}
&+2\sqrt{2}v\sqrt{\hat{g}'^2+\hat{g}^2}\left(3p_{1\mu_4}-p_{2\mu_4}-p_{3\mu_4}\right)\Big(c_\beta^4\hat{C}_{D\Phi}^{(11)(11)}\\
&\qquad +s_\beta^2c_\beta^2\left(\hat{C}_{D\Phi}^{(11)(22)}+\hat{C}_{D\Phi}^{(21)(12)}\right)+s_\beta^4\hat{C}_{D\Phi}^{(22)(22)}\Big)\\
&+2\sqrt{2}vs_\beta^2c_\beta^2\sqrt{\hat{g}'^2+\hat{g}^2}\left(3p_{1\mu_4}-p_{2\mu_4}-p_{3\mu_4}\right)\left(\hat{C}_{D\Phi}^{(21)(21)*}+\hat{C}_{D\Phi}^{(21)(21)}\right)
\end{aligned}
\tag{C.485}
$$

$$
\begin{aligned}
&+\sqrt{2}v\sqrt{\hat{g}'^2+\hat{g}^2}\Big(p_{1\mu_4}\Big(2s_\beta^2c_\beta^2\Big(3\hat{C}_{D\Phi}^{(11)(11)}-2\hat{C}_{D\Phi}^{(11)(22)}-2\hat{C}_{D\Phi}^{(21)(12)}+3\hat{C}_{D\Phi}^{(22)(22)}\Big)\\
&\qquad+s_\beta^4\Big(\hat{C}_{D\Phi}^{(11)(22)}+\hat{C}_{D\Phi}^{(21)(12)}\Big)+c_\beta^4\Big(\hat{C}_{D\Phi}^{(11)(22)}+\hat{C}_{D\Phi}^{(21)(12)}\Big)\Big)\\
&\qquad-2s_\beta^2c_\beta^2\left(p_{2\mu_4}+p_{3\mu_4}\right)\Big(\hat{C}_{D\Phi}^{(11)(11)}-\hat{C}_{D\Phi}^{(11)(22)}\\
&\qquad-\hat{C}_{D\Phi}^{(21)(12)}+\hat{C}_{D\Phi}^{(22)(22)}\Big)\Big)\\
&+\sqrt{2}v\sqrt{\hat{g}'^2+\hat{g}^2}\left(\left(s_\beta^4+c_\beta^4-4s_\beta^2c_\beta^2\right)p_{1\mu_4}\right.\\
&\qquad\left.-\left(s_\beta^4+c_\beta^4\right)\left(p_{2\mu_4}+p_{3\mu_4}\right)\right)\Big(\hat{C}_{D\Phi}^{(21)(21)*}+\hat{C}_{D\Phi}^{(21)(21)}\Big)
\end{aligned}
\tag{C.486}
$$

$$
\begin{aligned}
&+\frac{\sqrt{2}v\left(\hat{g}'^2-\hat{g}^2\right)p_{1\mu_4}}{\sqrt{\hat{g}'^2+\hat{g}^2}}\Big(2s_\beta^2c_\beta^2\Big(\hat{C}_{D\Phi}^{(11)(11)}-\hat{C}_{D\Phi}^{(21)(12)}+\hat{C}_{D\Phi}^{(22)(22)}\Big)\\
&\qquad+s_\beta^4\hat{C}_{D\Phi}^{(11)(22)}+c_\beta^4\hat{C}_{D\Phi}^{(11)(22)}\Big)\\
&-\frac{2\sqrt{2}vs_\beta^2c_\beta^2\left(\hat{g}'^2-\hat{g}^2\right)p_{1\mu_4}}{\sqrt{\hat{g}'^2+\hat{g}^2}}\Big(\hat{C}_{D\Phi}^{(21)(21)*}+\hat{C}_{D\Phi}^{(21)(21)}\Big)
\end{aligned}
\tag{C.487}
$$

$$
\begin{aligned}
&-2i\sqrt{2}v\sqrt{\hat{g}'^2+\hat{g}^2}\left(p_{1\mu_4}-p_{2\mu_4}\right)\Big(c_\beta^4\hat{C}_{D\Phi}^{(11)(11)}+s_\beta^2c_\beta^2\Big(\hat{C}_{D\Phi}^{(11)(22)}+\hat{C}_{D\Phi}^{(21)(12)}\Big)\\
&\qquad+s_\beta^4\hat{C}_{D\Phi}^{(22)(22)}\Big)\\
&-2i\sqrt{2}vs_\beta^2c_\beta^2\sqrt{\hat{g}'^2+\hat{g}^2}\left(p_{1\mu_4}-p_{2\mu_4}\right)\Big(\hat{C}_{D\Phi}^{(21)(21)*}+\hat{C}_{D\Phi}^{(21)(21)}\Big)
\end{aligned}
\tag{C.488}
$$

$$
\begin{aligned}
&+\frac{iv}{\sqrt{2}\sqrt{\hat{g}'^2+\hat{g}^2}}\Bigg(\left(\hat{g}'^2+\hat{g}^2\right)p_{1\mu_4}\Bigg(2s_\beta^2c_\beta^2\Bigg(2\hat{C}_{D\Phi}^{(11)(11)}-2\hat{C}_{D\Phi}^{(11)(22)}\\
&\qquad -\hat{C}_{D\Phi}^{(21)(12)}+2\hat{C}_{D\Phi}^{(22)(22)}\Bigg)+s_\beta^4\hat{C}_{D\Phi}^{(21)(12)}+c_\beta^4\hat{C}_{D\Phi}^{(21)(12)}\Bigg)\\
&\qquad -\left(\hat{g}'^2+\hat{g}^2\right)p_{3\mu_4}\Bigg(2s_\beta^2c_\beta^2\Bigg(2\hat{C}_{D\Phi}^{(11)(11)}-2\hat{C}_{D\Phi}^{(11)(22)}\\
&\qquad -\hat{C}_{D\Phi}^{(21)(12)}+2\hat{C}_{D\Phi}^{(22)(22)}\Bigg)+s_\beta^4\hat{C}_{D\Phi}^{(21)(12)}+c_\beta^4\hat{C}_{D\Phi}^{(21)(12)}\Bigg)\\
&\qquad +\left(\hat{g}'^2-\hat{g}^2\right)p_{2\mu_4}\hat{C}_{D\Phi}^{(21)(12)}\Bigg)\\
&+\frac{i\sqrt{2}v}{\sqrt{\hat{g}'^2+\hat{g}^2}}\Bigg(\hat{g}^2s_\beta^4p_{1\mu_4}\hat{C}_{D\Phi}^{(21)(21)*}+\hat{g}^2s_\beta^4p_{2\mu_4}\hat{C}_{D\Phi}^{(21)(21)*}\\
&\quad -\hat{g}^2s_\beta^4p_{3\mu_4}\hat{C}_{D\Phi}^{(21)(21)*}-\hat{g}^2s_\beta^2c_\beta^2p_{1\mu_4}\hat{C}_{D\Phi}^{(21)(21)*}\\
&\quad +\hat{g}^2s_\beta^2c_\beta^2p_{2\mu_4}\hat{C}_{D\Phi}^{(21)(21)*}+\hat{g}^2s_\beta^2c_\beta^2p_{3\mu_4}\hat{C}_{D\Phi}^{(21)(21)*}\\
&\quad +s_\beta^4\hat{g}'^2p_{1\mu_4}\hat{C}_{D\Phi}^{(21)(21)*}-s_\beta^4\hat{g}'^2p_{2\mu_4}\hat{C}_{D\Phi}^{(21)(21)*}\\
&\quad -s_\beta^4\hat{g}'^2p_{3\mu_4}\hat{C}_{D\Phi}^{(21)(21)*}-s_\beta^2c_\beta^2\hat{g}'^2p_{1\mu_4}\hat{C}_{D\Phi}^{(21)(21)*}\\
&\quad -s_\beta^2c_\beta^2\hat{g}'^2p_{2\mu_4}\hat{C}_{D\Phi}^{(21)(21)*}+s_\beta^2c_\beta^2\hat{g}'^2p_{3\mu_4}\hat{C}_{D\Phi}^{(21)(21)*}\\
&\quad +\hat{g}^2c_\beta^4p_{1\mu_4}\hat{C}_{D\Phi}^{(21)(21)}+\hat{g}^2c_\beta^4p_{2\mu_4}\hat{C}_{D\Phi}^{(21)(21)}\\
&\quad -\hat{g}^2c_\beta^4p_{3\mu_4}\hat{C}_{D\Phi}^{(21)(21)}-\hat{g}^2s_\beta^2c_\beta^2p_{1\mu_4}\hat{C}_{D\Phi}^{(21)(21)}\\
&\quad +\hat{g}^2s_\beta^2c_\beta^2p_{2\mu_4}\hat{C}_{D\Phi}^{(21)(21)}+\hat{g}^2s_\beta^2c_\beta^2p_{3\mu_4}\hat{C}_{D\Phi}^{(21)(21)}\\
&\quad +c_\beta^4\hat{g}'^2p_{1\mu_4}\hat{C}_{D\Phi}^{(21)(21)}-c_\beta^4\hat{g}'^2p_{2\mu_4}\hat{C}_{D\Phi}^{(21)(21)}\\
&\quad -c_\beta^4\hat{g}'^2p_{3\mu_4}\hat{C}_{D\Phi}^{(21)(21)}-s_\beta^2c_\beta^2\hat{g}'^2p_{1\mu_4}\hat{C}_{D\Phi}^{(21)(21)}\\
&\quad -s_\beta^2c_\beta^2\hat{g}'^2p_{2\mu_4}\hat{C}_{D\Phi}^{(21)(21)}+s_\beta^2c_\beta^2\hat{g}'^2p_{3\mu_4}\hat{C}_{D\Phi}^{(21)(21)}\Bigg)
\end{aligned}
\tag{C.489}
$$

$H$

$G^+$ $H^-$

$Z_{\mu_4}$

$H$

$h$ $H$

$Z_{\mu_4}$

$$
-i\sqrt{2}v\left(c_\beta^4-s_\beta^4\right)\sqrt{\hat{g}'^2+\hat{g}^2}\left(p_{1\mu_4}-p_{2\mu_4}-p_{3\mu_4}\right)\left(\hat{C}_{D\Phi}^{(21)(21)}-\hat{C}_{D\Phi}^{(21)(21)*}\right)
\tag{C.490}
$$

$H^+$, $h$, $H^-$, $Z_{\mu_4}$

$$
\begin{aligned}
&-i\sqrt{2}v\sqrt{\hat{g}'^2+\hat{g}^2}\left(p_{2\mu_4}-p_{3\mu_4}\right)\Big(2s_\beta^2c_\beta^2\left(\hat{C}_{D\Phi}^{(11)(11)}-\hat{C}_{D\Phi}^{(21)(12)}+\hat{C}_{D\Phi}^{(22)(22)}\right)\\
&\qquad\qquad +s_\beta^4\hat{C}_{D\Phi}^{(11)(22)}+c_\beta^4\hat{C}_{D\Phi}^{(11)(22)}\Big)\\
&+2i\sqrt{2}vs_\beta^2c_\beta^2\sqrt{\hat{g}'^2+\hat{g}^2}\left(p_{2\mu_4}-p_{3\mu_4}\right)\left(\hat{C}_{D\Phi}^{(21)(21)*}+\hat{C}_{D\Phi}^{(21)(21)}\right)
\end{aligned}
\tag{C.491}
$$

$G^-$, $A_{\mu_1}$, $H^+$, $Z_{\mu_4}$

$$
\begin{aligned}
&-2i\hat{g}v^2s_\beta c_\beta g_{\mu_1\mu_4}\hat{g}'\Big(c_\beta^2\left(2\hat{C}_{D\Phi}^{(11)(11)}-\hat{C}_{D\Phi}^{(11)(22)}-\hat{C}_{D\Phi}^{(21)(12)}\right)\\
&\qquad\qquad +s_\beta^2\left(\hat{C}_{D\Phi}^{(11)(22)}+\hat{C}_{D\Phi}^{(21)(12)}-2\hat{C}_{D\Phi}^{(22)(22)}\right)\Big)\\
&+\frac{2is_\beta c_\beta\left(p_{1\mu_4}p_{4\mu_1}-p_1\cdot p_4g_{\mu_1\mu_4}\right)}{\hat{g}'^2+\hat{g}^2}\Big(2\hat{g}\hat{g}'\left(\hat{C}_{\Phi B}^{(11)}-\hat{C}_{\Phi W}^{(11)}-\hat{C}_{\Phi B}^{(22)}+\hat{C}_{\Phi W}^{(22)}\right)\\
&\qquad\qquad +\left(\hat{g}'^2-\hat{g}^2\right)\left(\hat{C}_{\Phi WB}^{(11)}-\hat{C}_{\Phi WB}^{(22)}\right)\Big)\\
&+\frac{2is_\beta c_\beta p_1^\mu p_4^\nu\epsilon_{\mu_1\mu_4\mu\nu}}{\hat{g}'^2+\hat{g}^2}\Big(2\hat{g}\hat{g}'\left(\hat{C}_{\Phi\tilde{B}}^{(11)}-\hat{C}_{\Phi\tilde{W}}^{(11)}-\hat{C}_{\Phi\tilde{B}}^{(22)}+\hat{C}_{\Phi\tilde{W}}^{(22)}\right)\\
&\qquad\qquad +\left(\hat{g}'^2-\hat{g}^2\right)\left(\hat{C}_{\Phi B\tilde{W}}^{(11)}-\hat{C}_{\Phi B\tilde{W}}^{(22)}\right)\Big)\\
&+4i\hat{g}v^2s_\beta c_\beta g_{\mu_1\mu_4}\hat{g}'\left(c_\beta^2\hat{C}_{D\Phi}^{(21)(21)*}-s_\beta^2\hat{C}_{D\Phi}^{(21)(21)}\right)
\end{aligned}
\tag{C.492}
$$

$A^0$, $A^0$, $A^0$, $Z_{\mu_4}$

$$
\begin{aligned}
&-\sqrt{2}vs_\beta c_\beta\sqrt{\hat{g}'^2+\hat{g}^2}\left(p_{1\mu_4}+p_{2\mu_4}+p_{3\mu_4}\right)\\
&\qquad\Big(s_\beta^2\left(2\hat{C}_{D\Phi}^{(11)(11)}-\hat{C}_{D\Phi}^{(11)(22)}-\hat{C}_{D\Phi}^{(21)(12)}\right)\\
&\qquad\qquad +c_\beta^2\left(\hat{C}_{D\Phi}^{(11)(22)}+\hat{C}_{D\Phi}^{(21)(12)}-2\hat{C}_{D\Phi}^{(22)(22)}\right)\Big)\\
&-\sqrt{2}vs_\beta c_\beta c_{2\beta}\sqrt{\hat{g}'^2+\hat{g}^2}\left(p_{1\mu_4}+p_{2\mu_4}+p_{3\mu_4}\right)\left(\hat{C}_{D\Phi}^{(21)(21)*}+\hat{C}_{D\Phi}^{(21)(21)}\right)
\end{aligned}
\tag{C.493}
$$

$G^0$

$A^0$ $G^0$

$Z_{\mu_4}$

$$
\begin{aligned}
&-\sqrt{2}vs_\beta c_\beta\sqrt{\hat{g}'^2+\hat{g}^2}\left(p_{1\mu_4}+p_{2\mu_4}+p_{3\mu_4}\right)\\
&\qquad\left(c_\beta^2\left(2\hat{C}_{D\Phi}^{(11)(11)}-\hat{C}_{D\Phi}^{(11)(22)}-\hat{C}_{D\Phi}^{(21)(12)}\right)\right.\\
&\qquad\left.+s_\beta^2\left(\hat{C}_{D\Phi}^{(11)(22)}+\hat{C}_{D\Phi}^{(21)(12)}-2\hat{C}_{D\Phi}^{(22)(22)}\right)\right)\\
&+\sqrt{2}vs_\beta c_\beta c_{2\beta}\sqrt{\hat{g}'^2+\hat{g}^2}\left(p_{1\mu_4}+p_{2\mu_4}+p_{3\mu_4}\right)\left(\hat{C}_{D\Phi}^{(21)(21)*}+\hat{C}_{D\Phi}^{(21)(21)}\right)
\end{aligned}
\tag{C.494}
$$

$G^+$

$A^0$ $G^-$

$Z_{\mu_4}$

$$
\begin{aligned}
&-\frac{\sqrt{2}vs_\beta c_\beta\left(\hat{g}'^2-\hat{g}^2\right)p_{1\mu_4}}{\sqrt{\hat{g}'^2+\hat{g}^2}}\left(c_\beta^2\left(2\hat{C}_{D\Phi}^{(11)(11)}-\hat{C}_{D\Phi}^{(11)(22)}-\hat{C}_{D\Phi}^{(21)(12)}\right)\right.\\
&\qquad\left.+s_\beta^2\left(\hat{C}_{D\Phi}^{(11)(22)}+\hat{C}_{D\Phi}^{(21)(12)}-2\hat{C}_{D\Phi}^{(22)(22)}\right)\right)\\
&+\frac{\sqrt{2}vs_\beta c_\beta}{\sqrt{\hat{g}'^2+\hat{g}^2}}\left(\hat{g}^2s_\beta^2p_{1\mu_4}\hat{C}_{D\Phi}^{(21)(21)*}+\hat{g}^2s_\beta^2p_{2\mu_4}\hat{C}_{D\Phi}^{(21)(21)*}\right.\\
&\qquad-\hat{g}^2s_\beta^2p_{3\mu_4}\hat{C}_{D\Phi}^{(21)(21)*}-\hat{g}^2c_\beta^2p_{1\mu_4}\hat{C}_{D\Phi}^{(21)(21)*}\\
&\qquad+\hat{g}^2c_\beta^2p_{2\mu_4}\hat{C}_{D\Phi}^{(21)(21)*}-\hat{g}^2c_\beta^2p_{3\mu_4}\hat{C}_{D\Phi}^{(21)(21)*}\\
&\qquad-s_\beta^2\hat{g}'^2p_{1\mu_4}\hat{C}_{D\Phi}^{(21)(21)*}+s_\beta^2\hat{g}'^2p_{2\mu_4}\hat{C}_{D\Phi}^{(21)(21)*}\\
&\qquad-s_\beta^2\hat{g}'^2p_{3\mu_4}\hat{C}_{D\Phi}^{(21)(21)*}+c_\beta^2\hat{g}'^2p_{1\mu_4}\hat{C}_{D\Phi}^{(21)(21)*}\\
&\qquad+c_\beta^2\hat{g}'^2p_{2\mu_4}\hat{C}_{D\Phi}^{(21)(21)*}-c_\beta^2\hat{g}'^2p_{3\mu_4}\hat{C}_{D\Phi}^{(21)(21)*}\\
&\qquad+\hat{g}^2s_\beta^2p_{1\mu_4}\hat{C}_{D\Phi}^{(21)(21)}-\hat{g}^2s_\beta^2p_{2\mu_4}\hat{C}_{D\Phi}^{(21)(21)}\\
&\qquad+\hat{g}^2s_\beta^2p_{3\mu_4}\hat{C}_{D\Phi}^{(21)(21)}-\hat{g}^2c_\beta^2p_{1\mu_4}\hat{C}_{D\Phi}^{(21)(21)}\\
&\qquad-\hat{g}^2c_\beta^2p_{2\mu_4}\hat{C}_{D\Phi}^{(21)(21)}+\hat{g}^2c_\beta^2p_{3\mu_4}\hat{C}_{D\Phi}^{(21)(21)}\\
&\qquad-s_\beta^2\hat{g}'^2p_{1\mu_4}\hat{C}_{D\Phi}^{(21)(21)}-s_\beta^2\hat{g}'^2p_{2\mu_4}\hat{C}_{D\Phi}^{(21)(21)}\\
&\qquad+s_\beta^2\hat{g}'^2p_{3\mu_4}\hat{C}_{D\Phi}^{(21)(21)}+c_\beta^2\hat{g}'^2p_{1\mu_4}\hat{C}_{D\Phi}^{(21)(21)}\\
&\qquad\left.-c_\beta^2\hat{g}'^2p_{2\mu_4}\hat{C}_{D\Phi}^{(21)(21)}+c_\beta^2\hat{g}'^2p_{3\mu_4}\hat{C}_{D\Phi}^{(21)(21)}\right)
\end{aligned}
\tag{C.495}
$$

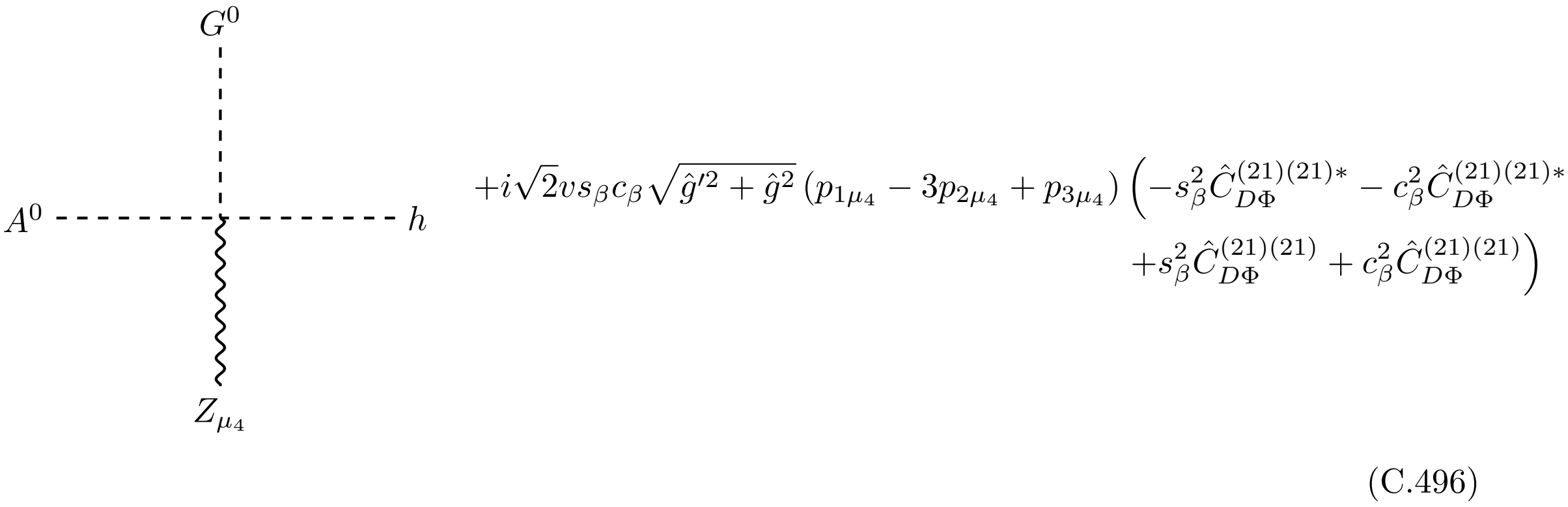

$$+i\sqrt{2}vs_\beta c_\beta\sqrt{\hat{g}'^2+\hat{g}^2}\left(p_{1\mu_4}-3p_{2\mu_4}+p_{3\mu_4}\right)\Big(-s_\beta^2\hat{C}_{D\Phi}^{(21)(21)*}-c_\beta^2\hat{C}_{D\Phi}^{(21)(21)*}$$
$$+s_\beta^2\hat{C}_{D\Phi}^{(21)(21)}+c_\beta^2\hat{C}_{D\Phi}^{(21)(21)}\Big) \tag{C.496}$$

$$-\sqrt{2}vs_\beta c_\beta\sqrt{\hat{g}'^2+\hat{g}^2}\left(3p_{1\mu_4}-p_{2\mu_4}-p_{3\mu_4}\right)$$
$$\Big(c_\beta^2\Big(2\hat{C}_{D\Phi}^{(11)(11)}-\hat{C}_{D\Phi}^{(11)(22)}-\hat{C}_{D\Phi}^{(21)(12)}\Big)$$
$$+s_\beta^2\Big(\hat{C}_{D\Phi}^{(11)(22)}+\hat{C}_{D\Phi}^{(21)(12)}-2\hat{C}_{D\Phi}^{(22)(22)}\Big)\Big)$$
$$+\sqrt{2}vs_\beta c_\beta c_{2\beta}\sqrt{\hat{g}'^2+\hat{g}^2}\left(3p_{1\mu_4}-p_{2\mu_4}-p_{3\mu_4}\right)\Big(\hat{C}_{D\Phi}^{(21)(21)*}+\hat{C}_{D\Phi}^{(21)(21)}\Big) \tag{C.497}$$

$$-i\sqrt{2}vs_\beta c_\beta\sqrt{\hat{g}'^2+\hat{g}^2}\left(p_{1\mu_4}+p_{2\mu_4}-3p_{3\mu_4}\right)\Big(-s_\beta^2\hat{C}_{D\Phi}^{(21)(21)*}-c_\beta^2\hat{C}_{D\Phi}^{(21)(21)*}$$
$$+s_\beta^2\hat{C}_{D\Phi}^{(21)(21)}+c_\beta^2\hat{C}_{D\Phi}^{(21)(21)}\Big) \tag{C.498}$$

H

A⁰ H H

$Z_{\mu_4}$

$$
\begin{aligned}
&-\sqrt{2}vs_\beta c_\beta\sqrt{\hat{g}'^2+\hat{g}^2}\left(3p_{1_{\mu_4}}-p_{2_{\mu_4}}-p_{3_{\mu_4}}\right)\\
&\qquad\left(s_\beta^2\left(2\hat{C}_{D\Phi}^{(11)(11)}-\hat{C}_{D\Phi}^{(11)(22)}-\hat{C}_{D\Phi}^{(21)(12)}\right)\right.\\
&\qquad\left.+c_\beta^2\left(\hat{C}_{D\Phi}^{(11)(22)}+\hat{C}_{D\Phi}^{(21)(12)}-2\hat{C}_{D\Phi}^{(22)(22)}\right)\right)\\
&-\sqrt{2}vs_\beta c_\beta c_{2\beta}\sqrt{\hat{g}'^2+\hat{g}^2}\left(3p_{1_{\mu_4}}-p_{2_{\mu_4}}-p_{3_{\mu_4}}\right)\left(\hat{C}_{D\Phi}^{(21)(21)*}+\hat{C}_{D\Phi}^{(21)(21)}\right)
\end{aligned}
\tag{C.499}
$$

$H^+$

$A^0$ $H^-$

$Z_{\mu_4}$

$$
\begin{aligned}
&-\frac{\sqrt{2}vs_\beta c_\beta\left(\hat{g}'^2-\hat{g}^2\right)p_{1_{\mu_4}}}{\sqrt{\hat{g}'^2+\hat{g}^2}}\left(s_\beta^2\left(2\hat{C}_{D\Phi}^{(11)(11)}-\hat{C}_{D\Phi}^{(11)(22)}-\hat{C}_{D\Phi}^{(21)(12)}\right)\right.\\
&\qquad\left.+c_\beta^2\left(\hat{C}_{D\Phi}^{(11)(22)}+\hat{C}_{D\Phi}^{(21)(12)}-2\hat{C}_{D\Phi}^{(22)(22)}\right)\right)\\
&+\frac{\sqrt{2}vs_\beta c_\beta}{\sqrt{\hat{g}'^2+\hat{g}^2}}\left(-\hat{g}^2s_\beta^2p_{1_{\mu_4}}\hat{C}_{D\Phi}^{(21)(21)*}-\hat{g}^2s_\beta^2p_{2_{\mu_4}}\hat{C}_{D\Phi}^{(21)(21)*}\right.\\
&\qquad+\hat{g}^2s_\beta^2p_{3_{\mu_4}}\hat{C}_{D\Phi}^{(21)(21)*}+\hat{g}^2c_\beta^2p_{1_{\mu_4}}\hat{C}_{D\Phi}^{(21)(21)*}\\
&\qquad-\hat{g}^2c_\beta^2p_{2_{\mu_4}}\hat{C}_{D\Phi}^{(21)(21)*}+\hat{g}^2c_\beta^2p_{3_{\mu_4}}\hat{C}_{D\Phi}^{(21)(21)*}\\
&\qquad+s_\beta^2\hat{g}'^2p_{1_{\mu_4}}\hat{C}_{D\Phi}^{(21)(21)*}-s_\beta^2\hat{g}'^2p_{2_{\mu_4}}\hat{C}_{D\Phi}^{(21)(21)*}\\
&\qquad+s_\beta^2\hat{g}'^2p_{3_{\mu_4}}\hat{C}_{D\Phi}^{(21)(21)*}-c_\beta^2\hat{g}'^2p_{1_{\mu_4}}\hat{C}_{D\Phi}^{(21)(21)*}\\
&\qquad-c_\beta^2\hat{g}'^2p_{2_{\mu_4}}\hat{C}_{D\Phi}^{(21)(21)*}+c_\beta^2\hat{g}'^2p_{3_{\mu_4}}\hat{C}_{D\Phi}^{(21)(21)*}\\
&\qquad-\hat{g}^2s_\beta^2p_{1_{\mu_4}}\hat{C}_{D\Phi}^{(21)(21)}+\hat{g}^2s_\beta^2p_{2_{\mu_4}}\hat{C}_{D\Phi}^{(21)(21)}\\
&\qquad-\hat{g}^2s_\beta^2p_{3_{\mu_4}}\hat{C}_{D\Phi}^{(21)(21)}+\hat{g}^2c_\beta^2p_{1_{\mu_4}}\hat{C}_{D\Phi}^{(21)(21)}\\
&\qquad+\hat{g}^2c_\beta^2p_{2_{\mu_4}}\hat{C}_{D\Phi}^{(21)(21)}-\hat{g}^2c_\beta^2p_{3_{\mu_4}}\hat{C}_{D\Phi}^{(21)(21)}\\
&\qquad+s_\beta^2\hat{g}'^2p_{1_{\mu_4}}\hat{C}_{D\Phi}^{(21)(21)}+s_\beta^2\hat{g}'^2p_{2_{\mu_4}}\hat{C}_{D\Phi}^{(21)(21)}\\
&\qquad-s_\beta^2\hat{g}'^2p_{3_{\mu_4}}\hat{C}_{D\Phi}^{(21)(21)}-c_\beta^2\hat{g}'^2p_{1_{\mu_4}}-\hat{C}_{D\Phi}^{(21)(21)}\\
&\qquad\left.+c_\beta^2\hat{g}'^2p_{2_{\mu_4}}\hat{C}_{D\Phi}^{(21)(21)}-c_\beta^2\hat{g}'^2p_{3_{\mu_4}}\hat{C}_{D\Phi}^{(21)(21)}\right)
\end{aligned}
\tag{C.500}
$$

$$-i\sqrt{2}vs_\beta c_\beta\sqrt{\hat{g}'^2+\hat{g}^2}\left(p_{1\mu_4}+p_{2\mu_4}+p_{3\mu_4}\right)\Big(-s_\beta^2\hat{C}_{D\Phi}^{(21)(21)*}-c_\beta^2\hat{C}_{D\Phi}^{(21)(21)*} +s_\beta^2\hat{C}_{D\Phi}^{(21)(21)}+c_\beta^2\hat{C}_{D\Phi}^{(21)(21)}\Big) \tag{C.501}$$

$$\begin{aligned}&+\sqrt{2}vs_\beta c_\beta\sqrt{\hat{g}'^2+\hat{g}^2}\left(3p_{1\mu_4}-p_{2\mu_4}-p_{3\mu_4}\right)\\ &\qquad\Big(c_\beta^2\Big(2\hat{C}_{D\Phi}^{(11)(11)}-\hat{C}_{D\Phi}^{(11)(22)}-\hat{C}_{D\Phi}^{(21)(12)}\Big)\\ &\qquad+s_\beta^2\Big(\hat{C}_{D\Phi}^{(11)(22)}+\hat{C}_{D\Phi}^{(21)(12)}-2\hat{C}_{D\Phi}^{(22)(22)}\Big)\Big)\\ &-\sqrt{2}vs_\beta c_\beta c_{2\beta}\sqrt{\hat{g}'^2+\hat{g}^2}\left(3p_{1\mu_4}-p_{2\mu_4}-p_{3\mu_4}\right)\Big(\hat{C}_{D\Phi}^{(21)(21)*}+\hat{C}_{D\Phi}^{(21)(21)}\Big)\end{aligned} \tag{C.502}$$

$$\begin{aligned}&-\frac{\sqrt{2}vs_\beta c_\beta\left(\hat{g}'^2-\hat{g}^2\right)p_{1\mu_4}}{\sqrt{\hat{g}'^2+\hat{g}^2}}\Big(c_\beta^2\Big(2\hat{C}_{D\Phi}^{(11)(11)}-\hat{C}_{D\Phi}^{(11)(22)}-\hat{C}_{D\Phi}^{(21)(12)}\Big)\\ &\qquad+s_\beta^2\Big(\hat{C}_{D\Phi}^{(11)(22)}+\hat{C}_{D\Phi}^{(21)(12)}-2\hat{C}_{D\Phi}^{(22)(22)}\Big)\Big)\\ &+\frac{\sqrt{2}vs_{2\beta}\left(\hat{g}'^2-\hat{g}^2\right)p_{1\mu_4}}{\sqrt{\hat{g}'^2+\hat{g}^2}}\Big(c_\beta^2\hat{C}_{D\Phi}^{(21)(21)*}-s_\beta^2\hat{C}_{D\Phi}^{(21)(21)}\Big)\end{aligned} \tag{C.503}$$

$G^-$, $G^+$, $H$, $Z_{\mu_4}$

$$
\begin{aligned}
&-i\sqrt{2}vs_\beta c_\beta\sqrt{\hat{g}'^2+\hat{g}^2}\left(p_{1\mu_4}-p_{2\mu_4}\right)\Big(c_\beta^2\left(2\hat{C}_{D\Phi}^{(11)(11)}-\hat{C}_{D\Phi}^{(11)(22)}-\hat{C}_{D\Phi}^{(21)(12)}\right)\\
&\qquad\qquad\qquad\qquad\qquad\qquad +s_\beta^2\left(\hat{C}_{D\Phi}^{(11)(22)}+\hat{C}_{D\Phi}^{(21)(12)}-2\hat{C}_{D\Phi}^{(22)(22)}\right)\Big)\\
&+\frac{i\sqrt{2}vs_\beta c_\beta}{\sqrt{\hat{g}'^2+\hat{g}^2}}\Big(-\hat{g}^2s_\beta^2p_{1\mu_4}\hat{C}_{D\Phi}^{(21)(21)*}+\hat{g}^2s_\beta^2p_{2\mu_4}\hat{C}_{D\Phi}^{(21)(21)*}\\
&\qquad -\hat{g}^2s_\beta^2p_{3\mu_4}\hat{C}_{D\Phi}^{(21)(21)*}+\hat{g}^2c_\beta^2p_{1\mu_4}\hat{C}_{D\Phi}^{(21)(21)*}\\
&\qquad -\hat{g}^2c_\beta^2p_{2\mu_4}\hat{C}_{D\Phi}^{(21)(21)*}-\hat{g}^2c_\beta^2p_{3\mu_4}\hat{C}_{D\Phi}^{(21)(21)*}\\
&\qquad -s_\beta^2\hat{g}'^2p_{1\mu_4}\hat{C}_{D\Phi}^{(21)(21)*}+s_\beta^2\hat{g}'^2p_{2\mu_4}\hat{C}_{D\Phi}^{(21)(21)*}\\
&\qquad +s_\beta^2\hat{g}'^2p_{3\mu_4}\hat{C}_{D\Phi}^{(21)(21)*}+c_\beta^2\hat{g}'^2p_{1\mu_4}\hat{C}_{D\Phi}^{(21)(21)*}\\
&\qquad -c_\beta^2\hat{g}'^2p_{2\mu_4}\hat{C}_{D\Phi}^{(21)(21)*}+c_\beta^2\hat{g}'^2p_{3\mu_4}\hat{C}_{D\Phi}^{(21)(21)*}\\
&\qquad -\hat{g}^2s_\beta^2p_{1\mu_4}\hat{C}_{D\Phi}^{(21)(21)}+\hat{g}^2s_\beta^2p_{2\mu_4}\hat{C}_{D\Phi}^{(21)(21)}\\
&\qquad +\hat{g}^2s_\beta^2p_{3\mu_4}\hat{C}_{D\Phi}^{(21)(21)}+\hat{g}^2c_\beta^2p_{1\mu_4}\hat{C}_{D\Phi}^{(21)(21)}\\
&\qquad -\hat{g}^2c_\beta^2p_{2\mu_4}\hat{C}_{D\Phi}^{(21)(21)}+\hat{g}^2c_\beta^2p_{3\mu_4}\hat{C}_{D\Phi}^{(21)(21)}\\
&\qquad -s_\beta^2\hat{g}'^2p_{1\mu_4}\hat{C}_{D\Phi}^{(21)(21)}+s_\beta^2\hat{g}'^2p_{2\mu_4}\hat{C}_{D\Phi}^{(21)(21)}\\
&\qquad -s_\beta^2\hat{g}'^2p_{3\mu_4}\hat{C}_{D\Phi}^{(21)(21)}+c_\beta^2\hat{g}'^2p_{1\mu_4}\hat{C}_{D\Phi}^{(21)(21)}\\
&\qquad -c_\beta^2\hat{g}'^2p_{2\mu_4}\hat{C}_{D\Phi}^{(21)(21)}-c_\beta^2\hat{g}'^2p_{3\mu_4}\hat{C}_{D\Phi}^{(21)(21)}\Big)
\end{aligned}
\tag{C.504}
$$

$h$, $G^+$, $H^-$, $Z_{\mu_4}$

$$
\begin{aligned}
&+i\sqrt{2}vs_\beta c_\beta\sqrt{\hat{g}'^2+\hat{g}^2}\left(p_{1\mu_4}-p_{3\mu_4}\right)\Big(c_\beta^2\left(2\hat{C}_{D\Phi}^{(11)(11)}-\hat{C}_{D\Phi}^{(11)(22)}-\hat{C}_{D\Phi}^{(21)(12)}\right)\\
&\qquad\qquad\qquad\qquad\qquad\qquad +s_\beta^2\left(\hat{C}_{D\Phi}^{(11)(22)}+\hat{C}_{D\Phi}^{(21)(12)}-2\hat{C}_{D\Phi}^{(22)(22)}\right)\Big)\\
&-2i\sqrt{2}vs_\beta c_\beta\sqrt{\hat{g}'^2+\hat{g}^2}\left(p_{1\mu_4}-p_{3\mu_4}\right)\left(c_\beta^2\hat{C}_{D\Phi}^{(21)(21)}-s_\beta^2\hat{C}_{D\Phi}^{(21)(21)*}\right)
\end{aligned}
\tag{C.505}
$$

$h$, $h$, $H$, $Z_{\mu_4}$

$$+i\sqrt{2}vs_\beta c_\beta\sqrt{\hat{g}'^2+\hat{g}^2}\left(p_{1\mu_4}+p_{2\mu_4}-3p_{3\mu_4}\right)\Big(-s_\beta^2\hat{C}_{D\Phi}^{(21)(21)*}-c_\beta^2\hat{C}_{D\Phi}^{(21)(21)*}+s_\beta^2\hat{C}_{D\Phi}^{(21)(21)}+c_\beta^2\hat{C}_{D\Phi}^{(21)(21)}\Big) \tag{C.506}$$

$H$, $H$, $H$, $Z_{\mu_4}$

$$+i\sqrt{2}vs_\beta c_\beta\sqrt{\hat{g}'^2+\hat{g}^2}\left(p_{1\mu_4}+p_{2\mu_4}+p_{3\mu_4}\right)\Big(-s_\beta^2\hat{C}_{D\Phi}^{(21)(21)*}-c_\beta^2\hat{C}_{D\Phi}^{(21)(21)*}+s_\beta^2\hat{C}_{D\Phi}^{(21)(21)}+c_\beta^2\hat{C}_{D\Phi}^{(21)(21)}\Big) \tag{C.507}$$

$$
\begin{aligned}
&-i\sqrt{2}vs_\beta c_\beta\sqrt{\hat{g}'^2+\hat{g}^2}\left(p_{2\mu_4}-p_{3\mu_4}\right)\Big(s_\beta^2\Big(2\hat{C}_{D\Phi}^{(11)(11)}-\hat{C}_{D\Phi}^{(11)(22)}-\hat{C}_{D\Phi}^{(21)(12)}\Big)\\
&\qquad\qquad+c_\beta^2\Big(\hat{C}_{D\Phi}^{(11)(22)}+\hat{C}_{D\Phi}^{(21)(12)}-2\hat{C}_{D\Phi}^{(22)(22)}\Big)\Big)\\
&+\frac{i\sqrt{2}vs_\beta c_\beta}{\sqrt{\hat{g}'^2+\hat{g}^2}}\Big(\hat{g}^2s_\beta^2p_{1\mu_4}\hat{C}_{D\Phi}^{(21)(21)*}+\hat{g}^2s_\beta^2p_{2\mu_4}\hat{C}_{D\Phi}^{(21)(21)*}\\
&\qquad-\hat{g}^2s_\beta^2p_{3\mu_4}\hat{C}_{D\Phi}^{(21)(21)*}+\hat{g}^2c_\beta^2p_{1\mu_4}\hat{C}_{D\Phi}^{(21)(21)*}\\
&\qquad-\hat{g}^2c_\beta^2p_{2\mu_4}\hat{C}_{D\Phi}^{(21)(21)*}+\hat{g}^2c_\beta^2p_{3\mu_4}\hat{C}_{D\Phi}^{(21)(21)*}\\
&\qquad-s_\beta^2\hat{g}'^2p_{1\mu_4}\hat{C}_{D\Phi}^{(21)(21)*}+s_\beta^2\hat{g}'^2p_{2\mu_4}\hat{C}_{D\Phi}^{(21)(21)*}\\
&\qquad-s_\beta^2\hat{g}'^2p_{3\mu_4}\hat{C}_{D\Phi}^{(21)(21)*}-c_\beta^2\hat{g}'^2p_{1\mu_4}\hat{C}_{D\Phi}^{(21)(21)*}\\
&\qquad-c_\beta^2\hat{g}'^2p_{2\mu_4}\hat{C}_{D\Phi}^{(21)(21)*}+c_\beta^2\hat{g}'^2p_{3\mu_4}\hat{C}_{D\Phi}^{(21)(21)*}\\
&\qquad-\hat{g}^2s_\beta^2p_{1\mu_4}\hat{C}_{D\Phi}^{(21)(21)}+\hat{g}^2s_\beta^2p_{2\mu_4}\hat{C}_{D\Phi}^{(21)(21)}\\
&\qquad-\hat{g}^2s_\beta^2p_{3\mu_4}\hat{C}_{D\Phi}^{(21)(21)}-\hat{g}^2c_\beta^2p_{1\mu_4}\hat{C}_{D\Phi}^{(21)(21)}\\
&\qquad-\hat{g}^2c_\beta^2p_{2\mu_4}\hat{C}_{D\Phi}^{(21)(21)}+\hat{g}^2c_\beta^2p_{3\mu_4}\hat{C}_{D\Phi}^{(21)(21)}\\
&\qquad+s_\beta^2\hat{g}'^2p_{1\mu_4}\hat{C}_{D\Phi}^{(21)(21)}+s_\beta^2\hat{g}'^2p_{2\mu_4}\hat{C}_{D\Phi}^{(21)(21)}\\
&\qquad-s_\beta^2\hat{g}'^2p_{3\mu_4}\hat{C}_{D\Phi}^{(21)(21)}+c_\beta^2\hat{g}'^2p_{1\mu_4}\hat{C}_{D\Phi}^{(21)(21)}\\
&\qquad-c_\beta^2\hat{g}'^2p_{2\mu_4}\hat{C}_{D\Phi}^{(21)(21)}+c_\beta^2\hat{g}'^2p_{3\mu_4}\hat{C}_{D\Phi}^{(21)(21)}\Big)
\end{aligned}
\tag{C.508}
$$

$$
\begin{aligned}
&-\frac{\hat{g}g_{\mu_3\mu_4}\hat{g}'}{4\left(\hat{g}'^2+\hat{g}^2\right)^{3/2}}\Big(s_\beta^2\Big(\hat{g}'^3\Big(2\Big(\delta_{s_{\hat{\beta}\pm}}+\delta_{s_{\hat{\beta}}}-1\Big)+A_2'\Big)\\
&\qquad+\hat{g}^2\hat{g}'\Big(2\Big(\delta_{s_{\hat{\beta}\pm}}+\delta_{s_{\hat{\beta}}}-1\Big)+A_2'\Big)-2\hat{g}^3X_{WB}\Big)\\
&\qquad+c_\beta^2\Big(\hat{g}'^3\Big(A_1'+2\Big(\delta_{c_{\hat{\beta}\pm}}+\delta_{c_{\hat{\beta}}}-1\Big)\Big)\\
&\qquad+\hat{g}^2\hat{g}'\Big(A_1'+2\Big(\delta_{c_{\hat{\beta}\pm}}+\delta_{c_{\hat{\beta}}}-1\Big)\Big)-2\hat{g}^3X_{WB}\Big)\\
&\qquad+s_{2\beta}B'\hat{g}'\left(\hat{g}'^2+\hat{g}^2\right)\Big)\\
&-2\hat{g}v^2g_{\mu_3\mu_4}\sqrt{\hat{g}'^2+\hat{g}^2}\Big(c_\beta^4\hat{C}_{D\Phi}^{(11)(11)}+s_\beta^2c_\beta^2\Big(\hat{C}_{D\Phi}^{(11)(22)}+\hat{C}_{D\Phi}^{(21)(12)}\Big)\\
&\qquad+s_\beta^4\hat{C}_{D\Phi}^{(22)(22)}\Big)\\
&+\frac{2\hat{g}'\left(p_{3\mu_4}p_{4\mu_3}-p_3\cdot p_4g_{\mu_3\mu_4}\right)}{\sqrt{\hat{g}'^2+\hat{g}^2}}\Big(c_\beta^2\hat{C}_{\Phi WB}^{(11)}+s_\beta^2\hat{C}_{\Phi WB}^{(22)}\Big)\\
&+\frac{2\hat{g}'p_3^\mu p_4^\nu\epsilon_{\mu_3\mu_4\mu\nu}}{\sqrt{\hat{g}'^2+\hat{g}^2}}\Big(c_\beta^2\hat{C}_{\Phi B\tilde{W}}^{(11)}+s_\beta^2\hat{C}_{\Phi B\tilde{W}}^{(22)}\Big)\\
&-2\hat{g}v^2s_\beta^2c_\beta^2g_{\mu_3\mu_4}\sqrt{\hat{g}'^2+\hat{g}^2}\Big(\hat{C}_{D\Phi}^{(21)(21)*}+\hat{C}_{D\Phi}^{(21)(21)}\Big)
\end{aligned}
\tag{C.509}
$$

$$
\begin{aligned}
&+\frac{i\hat{g}g_{\mu_3\mu_4}\hat{g}'}{4\left(\hat{g}'^2+\hat{g}^2\right)^{3/2}}\left(s_\beta^2\left(\hat{g}'^3\left(2\delta_{s_{\hat{\beta}\pm}}+A_2-2\right)\right.\right.\\
&\qquad\left.+\hat{g}^2\hat{g}'\left(2\delta_{s_{\hat{\beta}\pm}}+A_2-2\right)-2\hat{g}^3X_{WB}\right)\\
&\qquad+c_\beta^2\left(\hat{g}'^3\left(A_1+2\delta_{c_{\hat{\beta}\pm}}-2\right)\right.\\
&\qquad\left.+\hat{g}^2\hat{g}'\left(A_1+2\delta_{c_{\hat{\beta}\pm}}-2\right)-2\hat{g}^3X_{WB}\right)\\
&\qquad\left.+Bs_{2\beta}\hat{g}'\left(\hat{g}'^2+\hat{g}^2\right)\right)\\
&+6i\hat{g}v^2g_{\mu_3\mu_4}\sqrt{\hat{g}'^2+\hat{g}^2}\left(c_\beta^4\hat{C}_{D\Phi}^{(11)(11)}+s_\beta^2c_\beta^2\left(\hat{C}_{D\Phi}^{(11)(22)}+\hat{C}_{D\Phi}^{(21)(12)}\right)\right.\\
&\qquad\left.+s_\beta^4\hat{C}_{D\Phi}^{(22)(22)}\right)\\
&-\frac{2i\hat{g}'\left(p_{3\mu_4}p_{4\mu_3}-p_3\cdot p_4g_{\mu_3\mu_4}\right)}{\sqrt{\hat{g}'^2+\hat{g}^2}}\left(c_\beta^2\hat{C}_{\Phi WB}^{(11)}+s_\beta^2\hat{C}_{\Phi WB}^{(22)}\right)\\
&-\frac{2i\hat{g}'p_3^\mu p_4^\nu\epsilon_{\mu_3\mu_4\mu\nu}}{\sqrt{\hat{g}'^2+\hat{g}^2}}\left(c_\beta^2\hat{C}_{\Phi B\tilde{W}}^{(11)}+s_\beta^2\hat{C}_{\Phi B\tilde{W}}^{(22)}\right)\\
&+6i\hat{g}v^2s_\beta^2c_\beta^2g_{\mu_3\mu_4}\sqrt{\hat{g}'^2+\hat{g}^2}\left(\hat{C}_{D\Phi}^{(21)(21)*}+\hat{C}_{D\Phi}^{(21)(21)}\right)
\end{aligned}
\tag{C.510}
$$

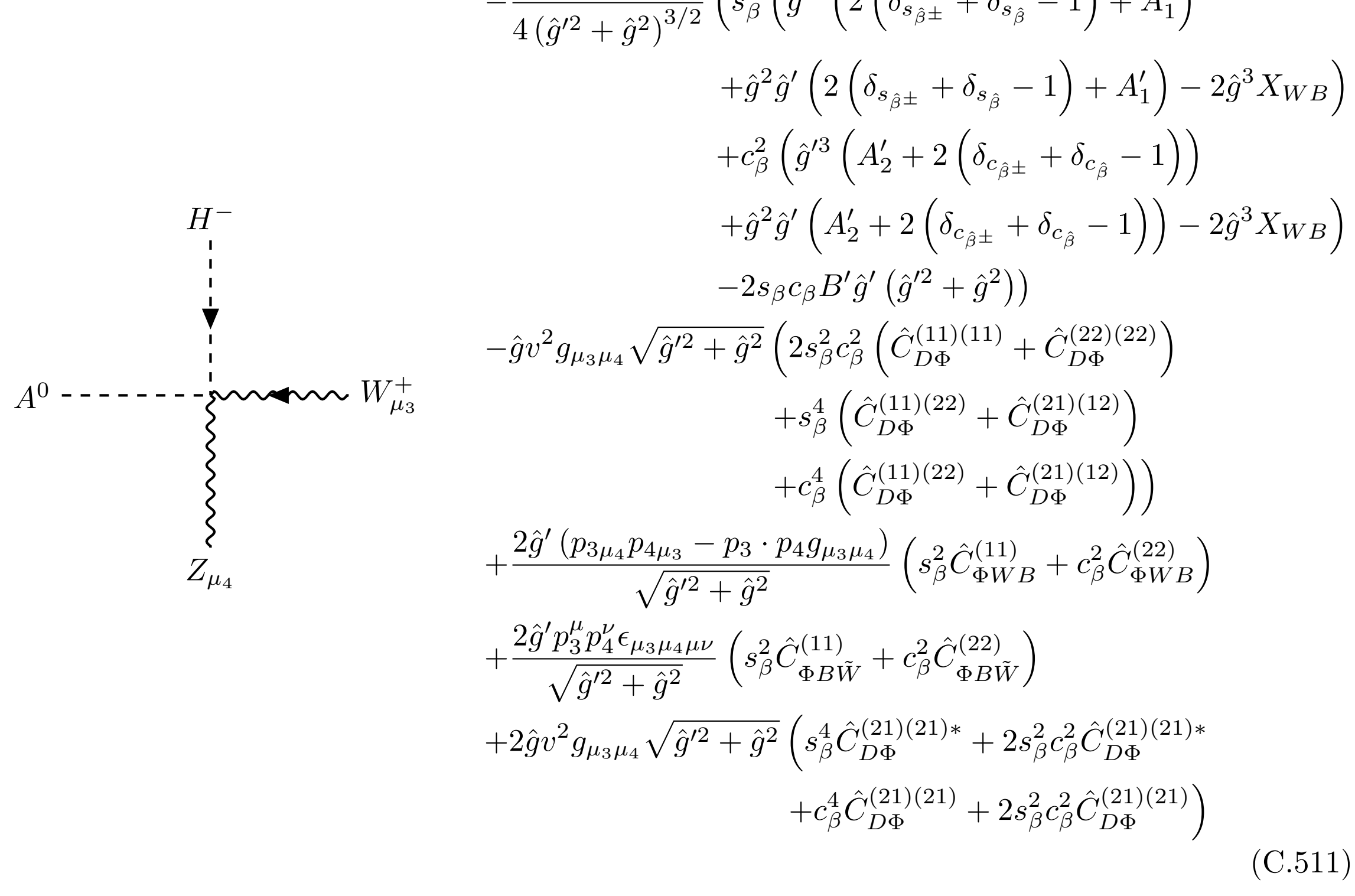

$$
\begin{aligned}
&-\frac{\hat{g}g_{\mu_3\mu_4}\hat{g}'}{4\left(\hat{g}'^2+\hat{g}^2\right)^{3/2}}\left(s_\beta^2\left(\hat{g}'^3\left(2\left(\delta_{s_{\hat{\beta}\pm}}+\delta_{s_{\hat{\beta}}}-1\right)+A_1'\right)\right.\right.\\
&\qquad\left.+\hat{g}^2\hat{g}'\left(2\left(\delta_{s_{\hat{\beta}\pm}}+\delta_{s_{\hat{\beta}}}-1\right)+A_1'\right)-2\hat{g}^3X_{WB}\right)\\
&\qquad+c_\beta^2\left(\hat{g}'^3\left(A_2'+2\left(\delta_{c_{\hat{\beta}\pm}}+\delta_{c_{\hat{\beta}}}-1\right)\right)\right.\\
&\qquad\left.+\hat{g}^2\hat{g}'\left(A_2'+2\left(\delta_{c_{\hat{\beta}\pm}}+\delta_{c_{\hat{\beta}}}-1\right)\right)-2\hat{g}^3X_{WB}\right)\\
&\qquad\left.-2s_\beta c_\beta B'\hat{g}'\left(\hat{g}'^2+\hat{g}^2\right)\right)\\
&-\hat{g}v^2g_{\mu_3\mu_4}\sqrt{\hat{g}'^2+\hat{g}^2}\left(2s_\beta^2c_\beta^2\left(\hat{C}_{D\Phi}^{(11)(11)}+\hat{C}_{D\Phi}^{(22)(22)}\right)\right.\\
&\qquad+s_\beta^4\left(\hat{C}_{D\Phi}^{(11)(22)}+\hat{C}_{D\Phi}^{(21)(12)}\right)\\
&\qquad\left.+c_\beta^4\left(\hat{C}_{D\Phi}^{(11)(22)}+\hat{C}_{D\Phi}^{(21)(12)}\right)\right)\\
&+\frac{2\hat{g}'\left(p_{3\mu_4}p_{4\mu_3}-p_3\cdot p_4g_{\mu_3\mu_4}\right)}{\sqrt{\hat{g}'^2+\hat{g}^2}}\left(s_\beta^2\hat{C}_{\Phi WB}^{(11)}+c_\beta^2\hat{C}_{\Phi WB}^{(22)}\right)\\
&+\frac{2\hat{g}'p_3^\mu p_4^\nu\epsilon_{\mu_3\mu_4\mu\nu}}{\sqrt{\hat{g}'^2+\hat{g}^2}}\left(s_\beta^2\hat{C}_{\Phi B\tilde{W}}^{(11)}+c_\beta^2\hat{C}_{\Phi B\tilde{W}}^{(22)}\right)\\
&+2\hat{g}v^2g_{\mu_3\mu_4}\sqrt{\hat{g}'^2+\hat{g}^2}\left(s_\beta^4\hat{C}_{D\Phi}^{(21)(21)*}+2s_\beta^2c_\beta^2\hat{C}_{D\Phi}^{(21)(21)*}\right.\\
&\qquad\left.+c_\beta^4\hat{C}_{D\Phi}^{(21)(21)}+2s_\beta^2c_\beta^2\hat{C}_{D\Phi}^{(21)(21)}\right)
\end{aligned}
\tag{C.511}
$$

$$
\begin{aligned}
&-\frac{i\hat{g}g_{\mu_3\mu_4}\hat{g}'}{4\left(\hat{g}'^2+\hat{g}^2\right)^{3/2}}\left(s_\beta^2\left(\hat{g}'^3\left(2\delta_{s_{\hat{\beta}\pm}}+A_1-2\right)\right.\right.\\
&\qquad\left.+\hat{g}^2\hat{g}'\left(2\delta_{s_{\hat{\beta}\pm}}+A_1-2\right)-2\hat{g}^3X_{WB}\right)\\
&\qquad+c_\beta^2\left(\hat{g}'^3\left(A_2+2\delta_{c_{\hat{\beta}\pm}}-2\right)\right.\\
&\qquad\left.+\hat{g}^2\hat{g}'\left(A_2+2\delta_{c_{\hat{\beta}\pm}}-2\right)-2\hat{g}^3X_{WB}\right)\\
&\qquad\left.-2Bs_\beta c_\beta\hat{g}'\left(\hat{g}'^2+\hat{g}^2\right)\right)\\
&-i\hat{g}v^2g_{\mu_3\mu_4}\sqrt{\hat{g}'^2+\hat{g}^2}\left(2s_\beta^2c_\beta^2\left(3\hat{C}_{D\Phi}^{(11)(11)}-2\hat{C}_{D\Phi}^{(11)(22)}\right.\right.\\
&\qquad\left.-2\hat{C}_{D\Phi}^{(21)(12)}+3\hat{C}_{D\Phi}^{(22)(22)}\right)\\
&\qquad+s_\beta^4\left(\hat{C}_{D\Phi}^{(11)(22)}+\hat{C}_{D\Phi}^{(21)(12)}\right)\\
&\qquad\left.+c_\beta^4\left(\hat{C}_{D\Phi}^{(11)(22)}+\hat{C}_{D\Phi}^{(21)(12)}\right)\right)\\
&+\frac{2i\hat{g}'\left(p_{3\mu_4}p_{4\mu_3}-p_3\cdot p_4g_{\mu_3\mu_4}\right)}{\sqrt{\hat{g}'^2+\hat{g}^2}}\left(s_\beta^2\hat{C}_{\Phi WB}^{(11)}+c_\beta^2\hat{C}_{\Phi WB}^{(22)}\right)\\
&+\frac{2i\hat{g}'p_3^\mu p_4^\nu\epsilon_{\mu_3\mu_4\mu\nu}}{\sqrt{\hat{g}'^2+\hat{g}^2}}\left(s_\beta^2\hat{C}_{\Phi B\tilde{W}}^{(11)}+c_\beta^2\hat{C}_{\Phi B\tilde{W}}^{(22)}\right)\\
&-2i\hat{g}v^2g_{\mu_3\mu_4}\sqrt{\hat{g}'^2+\hat{g}^2}\left(s_\beta^4\hat{C}_{D\Phi}^{(21)(21)*}-2s_\beta^2c_\beta^2\hat{C}_{D\Phi}^{(21)(21)*}\right.\\
&\qquad\left.+c_\beta^4\hat{C}_{D\Phi}^{(21)(21)}-2s_\beta^2c_\beta^2\hat{C}_{D\Phi}^{(21)(21)}\right)
\end{aligned}
\tag{C.512}
$$

$$
\begin{aligned}
&+\frac{\hat{g}g_{\mu_3\mu_4}\hat{g}'^2}{4\sqrt{\hat{g}'^2+\hat{g}^2}}\left(s_\beta c_\beta\left(2\left(-\delta_{s_{\hat{\beta}\pm}}+\delta_{c_{\hat{\beta}\pm}}-\delta_{c_{\hat{\beta}}}+\delta_{s_{\hat{\beta}}}\right)+A_1'-A_2'\right)+s_\beta^2B'-c_\beta^2B'\right)\\
&+\hat{g}v^2s_\beta c_\beta g_{\mu_3\mu_4}\sqrt{\hat{g}'^2+\hat{g}^2}\left(c_\beta^2\left(2\hat{C}_{D\Phi}^{(11)(11)}-\hat{C}_{D\Phi}^{(11)(22)}-\hat{C}_{D\Phi}^{(21)(12)}\right)\right.\\
&\qquad\left.+s_\beta^2\left(\hat{C}_{D\Phi}^{(11)(22)}+\hat{C}_{D\Phi}^{(21)(12)}-2\hat{C}_{D\Phi}^{(22)(22)}\right)\right)\\
&-\frac{2s_\beta c_\beta\hat{g}'\left(p_{3\mu_4}p_{4\mu_3}-p_3\cdot p_4g_{\mu_3\mu_4}\right)}{\sqrt{\hat{g}'^2+\hat{g}^2}}\left(\hat{C}_{\Phi WB}^{(11)}-\hat{C}_{\Phi WB}^{(22)}\right)\\
&+\frac{s_{2\beta}\hat{g}'p_3^\mu p_4^\nu\epsilon_{\mu_3\mu_4\mu\nu}}{\sqrt{\hat{g}'^2+\hat{g}^2}}\left(\hat{C}_{\Phi B\tilde{W}}^{(22)}-\hat{C}_{\Phi B\tilde{W}}^{(11)}\right)\\
&+\hat{g}v^2s_{2\beta}g_{\mu_3\mu_4}\sqrt{\hat{g}'^2+\hat{g}^2}\left(-s_\beta^2\hat{C}_{D\Phi}^{(21)(21)*}-2c_\beta^2\hat{C}_{D\Phi}^{(21)(21)*}\right.\\
&\qquad\left.+2s_\beta^2\hat{C}_{D\Phi}^{(21)(21)}+c_\beta^2\hat{C}_{D\Phi}^{(21)(21)}\right)
\end{aligned}
\tag{C.513}
$$

$H$, $G^-$, $W^+_{\mu_3}$, $Z_{\mu_4}$

$$
\begin{aligned}
&-\frac{i\hat{g}g_{\mu_3\mu_4}\hat{g}'^2}{4\sqrt{\hat{g}'^2+\hat{g}^2}}\left(s_\beta c_\beta\left(2\delta_{s_{\hat\beta\pm}}-A_1+A_2-2\delta_{c_{\hat\beta\pm}}\right)-Bs_\beta^2+Bc_\beta^2\right)\\
&+3i\hat{g}v^2 s_\beta c_\beta g_{\mu_3\mu_4}\sqrt{\hat{g}'^2+\hat{g}^2}\left(c_\beta^2\left(2\hat{C}_{D\Phi}^{(11)(11)}-\hat{C}_{D\Phi}^{(11)(22)}-\hat{C}_{D\Phi}^{(21)(12)}\right)\right.\\
&\qquad\left.+s_\beta^2\left(\hat{C}_{D\Phi}^{(11)(22)}+\hat{C}_{D\Phi}^{(21)(12)}-2\hat{C}_{D\Phi}^{(22)(22)}\right)\right)\\
&-\frac{2is_\beta c_\beta\hat{g}'\left(p_{3\mu_4}p_{4\mu_3}-p_3\cdot p_4 g_{\mu_3\mu_4}\right)}{\sqrt{\hat{g}'^2+\hat{g}^2}}\left(\hat{C}_{\Phi WB}^{(11)}-\hat{C}_{\Phi WB}^{(22)}\right)\\
&-\frac{2is_\beta c_\beta\hat{g}'p_3^\mu p_4^\nu\epsilon_{\mu_3\mu_4\mu\nu}}{\sqrt{\hat{g}'^2+\hat{g}^2}}\left(\hat{C}_{\Phi B\tilde{W}}^{(11)}-\hat{C}_{\Phi B\tilde{W}}^{(22)}\right)\\
&-2i\hat{g}v^2 s_\beta c_\beta g_{\mu_3\mu_4}\sqrt{\hat{g}'^2+\hat{g}^2}\left(-s_\beta^2\hat{C}_{D\Phi}^{(21)(21)*}+2c_\beta^2\hat{C}_{D\Phi}^{(21)(21)*}\right.\\
&\qquad\left.-2s_\beta^2\hat{C}_{D\Phi}^{(21)(21)}+c_\beta^2\hat{C}_{D\Phi}^{(21)(21)}\right)
\end{aligned}
\tag{C.514}
$$

$H^-$, $G^0$, $W^+_{\mu_3}$, $Z_{\mu_4}$

$$
\begin{aligned}
&+\frac{\hat{g}g_{\mu_3\mu_4}\hat{g}'^2}{4\sqrt{\hat{g}'^2+\hat{g}^2}}\left(s_\beta c_\beta\left(2\left(\delta_{s_{\hat\beta\pm}}-\delta_{c_{\hat\beta\pm}}+\delta_{c_{\hat\beta}}-\delta_{s_{\hat\beta}}\right)+A_1'-A_2'\right)+s_\beta^2B'-c_\beta^2B'\right)\\
&+\hat{g}v^2 s_\beta c_\beta g_{\mu_3\mu_4}\sqrt{\hat{g}'^2+\hat{g}^2}\left(c_\beta^2\left(2\hat{C}_{D\Phi}^{(11)(11)}-\hat{C}_{D\Phi}^{(11)(22)}-\hat{C}_{D\Phi}^{(21)(12)}\right)\right.\\
&\qquad\left.+s_\beta^2\left(\hat{C}_{D\Phi}^{(11)(22)}+\hat{C}_{D\Phi}^{(21)(12)}-2\hat{C}_{D\Phi}^{(22)(22)}\right)\right)\\
&-\frac{2s_\beta c_\beta\hat{g}'\left(p_{3\mu_4}p_{4\mu_3}-p_3\cdot p_4 g_{\mu_3\mu_4}\right)}{\sqrt{\hat{g}'^2+\hat{g}^2}}\left(\hat{C}_{\Phi WB}^{(11)}-\hat{C}_{\Phi WB}^{(22)}\right)\\
&+\frac{s_{2\beta}\hat{g}'p_3^\mu p_4^\nu\epsilon_{\mu_3\mu_4\mu\nu}}{\sqrt{\hat{g}'^2+\hat{g}^2}}\left(\hat{C}_{\Phi B\tilde{W}}^{(22)}-\hat{C}_{\Phi B\tilde{W}}^{(11)}\right)\\
&-2\hat{g}v^2 s_\beta c_\beta g_{\mu_3\mu_4}\sqrt{\hat{g}'^2+\hat{g}^2}\left(c_\beta^2\hat{C}_{D\Phi}^{(21)(21)}-s_\beta^2\hat{C}_{D\Phi}^{(21)(21)*}\right)
\end{aligned}
\tag{C.515}
$$

$H^-$, $h$, $W^+_{\mu_3}$, $Z_{\mu_4}$

$$
\begin{aligned}
&+\frac{i\hat{g}g_{\mu_3\mu_4}\hat{g}'^2}{4\sqrt{\hat{g}'^2+\hat{g}^2}}\left(s_\beta c_\beta\left(-2\delta_{s_{\hat\beta\pm}}-A_1+A_2+2\delta_{c_{\hat\beta\pm}}\right)-Bs_\beta^2+Bc_\beta^2\right)\\
&-3i\hat{g}v^2 s_\beta c_\beta g_{\mu_3\mu_4}\sqrt{\hat{g}'^2+\hat{g}^2}\left(c_\beta^2\left(2\hat{C}_{D\Phi}^{(11)(11)}-\hat{C}_{D\Phi}^{(11)(22)}-\hat{C}_{D\Phi}^{(21)(12)}\right)\right.\\
&\qquad\left.+s_\beta^2\left(\hat{C}_{D\Phi}^{(11)(22)}+\hat{C}_{D\Phi}^{(21)(12)}-2\hat{C}_{D\Phi}^{(22)(22)}\right)\right)\\
&+\frac{2is_\beta c_\beta\hat{g}'\left(p_{3\mu_4}p_{4\mu_3}-p_3\cdot p_4 g_{\mu_3\mu_4}\right)}{\sqrt{\hat{g}'^2+\hat{g}^2}}\left(\hat{C}_{\Phi WB}^{(11)}-\hat{C}_{\Phi WB}^{(22)}\right)\\
&+\frac{2is_\beta c_\beta\hat{g}'p_3^\mu p_4^\nu\epsilon_{\mu_3\mu_4\mu\nu}}{\sqrt{\hat{g}'^2+\hat{g}^2}}\left(\hat{C}_{\Phi B\tilde{W}}^{(11)}-\hat{C}_{\Phi B\tilde{W}}^{(22)}\right)\\
&+6i\hat{g}v^2 s_\beta c_\beta g_{\mu_3\mu_4}\sqrt{\hat{g}'^2+\hat{g}^2}\left(c_\beta^2\hat{C}_{D\Phi}^{(21)(21)}-s_\beta^2\hat{C}_{D\Phi}^{(21)(21)*}\right)
\end{aligned}
\tag{C.516}
$$

$A^0$

$A^0$ $Z_{\mu_3}$

$Z_{\mu_4}$

$$
\begin{aligned}
&-\frac{1}{2} i g_{\mu_3\mu_4} \Big( c_\beta^2 \Big( \hat{g}^2 \Big( A_2' + 2\delta_{c_{\hat{\beta}}} - 1 \Big) + \hat{g}'^2 \Big( A_2' + 2\delta_{c_{\hat{\beta}}} - 1 \Big) - 2\hat{g} X_{WB} \hat{g}' \Big) \\
&\qquad + s_\beta^2 \Big( \hat{g}^2 \Big( A_1' + 2\delta_{s_{\hat{\beta}}} - 1 \Big) + \hat{g}'^2 \Big( A_1' + 2\delta_{s_{\hat{\beta}}} - 1 \Big) - 2\hat{g} X_{WB} \hat{g}' \Big) \\
&\qquad - 2 s_\beta c_\beta B' \left( \hat{g}'^2 + \hat{g}^2 \right) \Big) \\
&-2iv^2 g_{\mu_3\mu_4} \left( \hat{g}'^2 + \hat{g}^2 \right) \Big( 2 s_\beta^2 c_\beta^2 \Big( \hat{C}_{D\Phi}^{(11)(11)} + \hat{C}_{D\Phi}^{(22)(22)} \Big) \\
&\qquad + s_\beta^4 \Big( \hat{C}_{D\Phi}^{(11)(22)} + \hat{C}_{D\Phi}^{(21)(12)} \Big) \\
&\qquad + c_\beta^4 \Big( \hat{C}_{D\Phi}^{(11)(22)} + \hat{C}_{D\Phi}^{(21)(12)} \Big) \Big) \\
&+ \frac{4i \left( p_{3\mu_4} p_{4\mu_3} - p_3 \cdot p_4 g_{\mu_3\mu_4} \right)}{\hat{g}'^2 + \hat{g}^2} \Big( s_\beta^2 \Big( \hat{g}'^2 \hat{C}_{\Phi B}^{(11)} + \hat{g} \Big( \hat{g}' \hat{C}_{\Phi WB}^{(11)} + \hat{g} \hat{C}_{\Phi W}^{(11)} \Big) \Big) \\
&\qquad + c_\beta^2 \Big( \hat{g}'^2 \hat{C}_{\Phi B}^{(22)} + \hat{g} \Big( \hat{g}' \hat{C}_{\Phi WB}^{(22)} + \hat{g} \hat{C}_{\Phi W}^{(22)} \Big) \Big) \Big) \\
&+ \frac{4i p_3^\mu p_4^\nu \epsilon_{\mu_3\mu_4\mu\nu}}{\hat{g}'^2 + \hat{g}^2} \Big( s_\beta^2 \Big( \hat{g}'^2 \hat{C}_{\Phi\tilde{B}}^{(11)} + \hat{g} \Big( \hat{g}' \hat{C}_{\Phi B\tilde{W}}^{(11)} + \hat{g} \hat{C}_{\Phi\tilde{W}}^{(11)} \Big) \Big) \\
&\qquad + c_\beta^2 \Big( \hat{g}'^2 \hat{C}_{\Phi\tilde{B}}^{(22)} + \hat{g} \Big( \hat{g}' \hat{C}_{\Phi B\tilde{W}}^{(22)} + \hat{g} \hat{C}_{\Phi\tilde{W}}^{(22)} \Big) \Big) \Big) \\
&+ 2iv^2 g_{\mu_3\mu_4} \left( s_\beta^4 + c_\beta^4 + 4 s_\beta^2 c_\beta^2 \right) \left( \hat{g}'^2 + \hat{g}^2 \right) \Big( \hat{C}_{D\Phi}^{(21)(21)*} + \hat{C}_{D\Phi}^{(21)(21)} \Big)
\end{aligned}
\tag{C.517}
$$

$G^0$

$G^0$ $Z_{\mu_3}$

$Z_{\mu_4}$

$$
\begin{aligned}
&-\frac{1}{2} i g_{\mu_3\mu_4} \Big( c_\beta^2 \Big( \hat{g}^2 \Big( A_1' + 2\delta_{c_{\hat{\beta}}} - 1 \Big) + \hat{g}'^2 \Big( A_1' + 2\delta_{c_{\hat{\beta}}} - 1 \Big) - 2\hat{g} X_{WB} \hat{g}' \Big) \\
&\qquad + s_\beta^2 \Big( \hat{g}^2 \Big( A_2' + 2\delta_{s_{\hat{\beta}}} - 1 \Big) + \hat{g}'^2 \Big( A_2' + 2\delta_{s_{\hat{\beta}}} - 1 \Big) - 2\hat{g} X_{WB} \hat{g}' \Big) \\
&\qquad + s_{2\beta} B' \left( \hat{g}'^2 + \hat{g}^2 \right) \Big) \\
&-4iv^2 g_{\mu_3\mu_4} \left( \hat{g}'^2 + \hat{g}^2 \right) \Big( c_\beta^4 \hat{C}_{D\Phi}^{(11)(11)} + s_\beta^2 c_\beta^2 \Big( \hat{C}_{D\Phi}^{(11)(22)} + \hat{C}_{D\Phi}^{(21)(12)} \Big) \\
&\qquad + s_\beta^4 \hat{C}_{D\Phi}^{(22)(22)} \Big) \\
&+ \frac{4i \left( p_{3\mu_4} p_{4\mu_3} - p_3 \cdot p_4 g_{\mu_3\mu_4} \right)}{\hat{g}'^2 + \hat{g}^2} \Big( c_\beta^2 \Big( \hat{g}'^2 \hat{C}_{\Phi B}^{(11)} + \hat{g} \Big( \hat{g}' \hat{C}_{\Phi WB}^{(11)} + \hat{g} \hat{C}_{\Phi W}^{(11)} \Big) \Big) \\
&\qquad + s_\beta^2 \Big( \hat{g}'^2 \hat{C}_{\Phi B}^{(22)} + \hat{g} \Big( \hat{g}' \hat{C}_{\Phi WB}^{(22)} + \hat{g} \hat{C}_{\Phi W}^{(22)} \Big) \Big) \Big) \\
&+ \frac{4i p_3^\mu p_4^\nu \epsilon_{\mu_3\mu_4\mu\nu}}{\hat{g}'^2 + \hat{g}^2} \Big( c_\beta^2 \Big( \hat{g}'^2 \hat{C}_{\Phi\tilde{B}}^{(11)} + \hat{g} \Big( \hat{g}' \hat{C}_{\Phi B\tilde{W}}^{(11)} + \hat{g} \hat{C}_{\Phi\tilde{W}}^{(11)} \Big) \Big) \\
&\qquad + s_\beta^2 \Big( \hat{g}'^2 \hat{C}_{\Phi\tilde{B}}^{(22)} + \hat{g} \Big( \hat{g}' \hat{C}_{\Phi B\tilde{W}}^{(22)} + \hat{g} \hat{C}_{\Phi\tilde{W}}^{(22)} \Big) \Big) \Big) \\
&- 4iv^2 s_\beta^2 c_\beta^2 g_{\mu_3\mu_4} \left( \hat{g}'^2 + \hat{g}^2 \right) \Big( \hat{C}_{D\Phi}^{(21)(21)*} + \hat{C}_{D\Phi}^{(21)(21)} \Big)
\end{aligned}
\tag{C.518}
$$

$G^-$, $G^+$, $Z_{\mu_3}$, $Z_{\mu_4}$

$$
\begin{aligned}
&-\frac{i g_{\mu_3\mu_4}\left(\hat{g}'^2-\hat{g}^2\right)^2}{2\left(\hat{g}'^2+\hat{g}^2\right)^2}\left(s_\beta^2\left(\hat{g}'^2\left(2\delta_{s_{\hat{\beta}\pm}}-1\right)+\hat{g}^2\left(2\delta_{s_{\hat{\beta}\pm}}-1\right)+2\hat{g}X_{WB}\hat{g}'\right)\right.\\
&\qquad\left.+c_\beta^2\left(\hat{g}^2\left(2\delta_{c_{\hat{\beta}\pm}}-1\right)+\left(2\delta_{c_{\hat{\beta}\pm}}-1\right)\hat{g}'^2+2\hat{g}X_{WB}\hat{g}'\right)\right)\\
&-4iv^2 g_{\mu_3\mu_4}\left(\hat{g}'^2-\hat{g}^2\right)\left(c_\beta^4\hat{C}_{D\Phi}^{(11)(11)}+s_\beta^2c_\beta^2\left(\hat{C}_{D\Phi}^{(11)(22)}+\hat{C}_{D\Phi}^{(21)(12)}\right)\right.\\
&\qquad\left.+s_\beta^4\hat{C}_{D\Phi}^{(22)(22)}\right)\\
&+\frac{4i\left(p_{3\mu_4}p_{4\mu_3}-p_3\cdot p_4 g_{\mu_3\mu_4}\right)}{\hat{g}'^2+\hat{g}^2}\left(c_\beta^2\left(\hat{g}'^2\hat{C}_{\Phi B}^{(11)}+\hat{g}\left(\hat{g}\hat{C}_{\Phi W}^{(11)}-\hat{g}'\hat{C}_{\Phi WB}^{(11)}\right)\right)\right.\\
&\qquad\left.+s_\beta^2\left(\hat{g}'^2\hat{C}_{\Phi B}^{(22)}+\hat{g}\left(\hat{g}\hat{C}_{\Phi W}^{(22)}-\hat{g}'\hat{C}_{\Phi WB}^{(22)}\right)\right)\right)\\
&+\frac{4ip_3^\mu p_4^\nu\epsilon_{\mu_3\mu_4\mu\nu}}{\hat{g}'^2+\hat{g}^2}\left(c_\beta^2\left(\hat{g}'^2\hat{C}_{\Phi\tilde{B}}^{(11)}+\hat{g}\left(\hat{g}\hat{C}_{\Phi\tilde{W}}^{(11)}-\hat{g}'\hat{C}_{\Phi B\tilde{W}}^{(11)}\right)\right)\right.\\
&\qquad\left.+s_\beta^2\left(\hat{g}'^2\hat{C}_{\Phi\tilde{B}}^{(22)}+\hat{g}\left(\hat{g}\hat{C}_{\Phi\tilde{W}}^{(22)}-\hat{g}'\hat{C}_{\Phi B\tilde{W}}^{(22)}\right)\right)\right)\\
&-4iv^2s_\beta^2c_\beta^2 g_{\mu_3\mu_4}\left(\hat{g}'^2-\hat{g}^2\right)\left(\hat{C}_{D\Phi}^{(21)(21)*}+\hat{C}_{D\Phi}^{(21)(21)}\right)
\end{aligned}
\tag{C.519}
$$

$h$, $h$, $Z_{\mu_3}$, $Z_{\mu_4}$

$$
\begin{aligned}
&-\frac{1}{2}ig_{\mu_3\mu_4}\left(s_\beta^2\left(\left(A_2-1\right)\hat{g}'^2+\left(A_2-1\right)\hat{g}^2-2\hat{g}X_{WB}\hat{g}'\right)\right.\\
&\qquad+c_\beta^2\left(\left(A_1-1\right)\hat{g}'^2+\left(A_1-1\right)\hat{g}^2-2\hat{g}X_{WB}\hat{g}'\right)\\
&\qquad\left.+Bs_{2\beta}\left(\hat{g}'^2+\hat{g}^2\right)\right)\\
&-2iv^2 g_{\mu_3\mu_4}\left(\hat{g}'^2+\hat{g}^2\right)\left(c_\beta^4\hat{C}_{D\Phi}^{(11)(11)}+s_\beta^2c_\beta^2\left(\hat{C}_{D\Phi}^{(11)(22)}+\hat{C}_{D\Phi}^{(21)(12)}\right)\right.\\
&\qquad\left.+s_\beta^4\hat{C}_{D\Phi}^{(22)(22)}\right)\\
&+\frac{4i\left(p_{3\mu_4}p_{4\mu_3}-p_3\cdot p_4 g_{\mu_3\mu_4}\right)}{\hat{g}'^2+\hat{g}^2}\left(c_\beta^2\left(\hat{g}'^2\hat{C}_{\Phi B}^{(11)}+\hat{g}\left(\hat{g}'\hat{C}_{\Phi WB}^{(11)}+\hat{g}\hat{C}_{\Phi W}^{(11)}\right)\right)\right.\\
&\qquad\left.+s_\beta^2\left(\hat{g}'^2\hat{C}_{\Phi B}^{(22)}+\hat{g}\left(\hat{g}'\hat{C}_{\Phi WB}^{(22)}+\hat{g}\hat{C}_{\Phi W}^{(22)}\right)\right)\right)\\
&+\frac{4ip_3^\mu p_4^\nu\epsilon_{\mu_3\mu_4\mu\nu}}{\hat{g}'^2+\hat{g}^2}\left(c_\beta^2\left(\hat{g}'^2\hat{C}_{\Phi\tilde{B}}^{(11)}+\hat{g}\left(\hat{g}'\hat{C}_{\Phi B\tilde{W}}^{(11)}+\hat{g}\hat{C}_{\Phi\tilde{W}}^{(11)}\right)\right)\right.\\
&\qquad\left.+s_\beta^2\left(\hat{g}'^2\hat{C}_{\Phi\tilde{B}}^{(22)}+\hat{g}\left(\hat{g}'\hat{C}_{\Phi B\tilde{W}}^{(22)}+\hat{g}\hat{C}_{\Phi\tilde{W}}^{(22)}\right)\right)\right)\\
&-2iv^2s_\beta^2c_\beta^2 g_{\mu_3\mu_4}\left(\hat{g}'^2+\hat{g}^2\right)\left(\hat{C}_{D\Phi}^{(21)(21)*}+\hat{C}_{D\Phi}^{(21)(21)}\right)
\end{aligned}
\tag{C.520}
$$

$H$

$A^0$ $Z_{\mu_3}$

$Z_{\mu_4}$

$$+2v^2 g_{\mu_3\mu_4}\left(c_\beta^4 - s_\beta^4\right)\left(\hat{g}'^2+\hat{g}^2\right)\left(\hat{C}_{D\Phi}^{(21)(21)} - \hat{C}_{D\Phi}^{(21)(21)*}\right) \tag{C.521}$$

$H$

$H$ $Z_{\mu_3}$

$Z_{\mu_4}$

$$\begin{aligned}
&-\frac{1}{2} i g_{\mu_3\mu_4}\left(s_\beta^2\left(\left(A_1-1\right)\hat{g}'^2+\left(A_1-1\right)\hat{g}^2-2\hat{g}X_{WB}\hat{g}'\right)\right.\\
&\qquad\qquad +c_\beta^2\left(\left(A_2-1\right)\hat{g}'^2+\left(A_2-1\right)\hat{g}^2-2\hat{g}X_{WB}\hat{g}'\right)\\
&\qquad\qquad \left.-2Bs_\beta c_\beta\left(\hat{g}'^2+\hat{g}^2\right)\right)\\
&-2iv^2 g_{\mu_3\mu_4}\left(\hat{g}'^2+\hat{g}^2\right)\left(2s_\beta^2c_\beta^2\left(3\hat{C}_{D\Phi}^{(11)(11)}-2\hat{C}_{D\Phi}^{(11)(22)}\right.\right.\\
&\qquad\qquad\qquad\qquad \left.-2\hat{C}_{D\Phi}^{(21)(12)}+3\hat{C}_{D\Phi}^{(22)(22)}\right)\\
&\qquad\qquad\qquad\qquad +s_\beta^4\left(\hat{C}_{D\Phi}^{(11)(22)}+\hat{C}_{D\Phi}^{(21)(12)}\right)\\
&\qquad\qquad\qquad\qquad \left.+c_\beta^4\left(\hat{C}_{D\Phi}^{(11)(22)}+\hat{C}_{D\Phi}^{(21)(12)}\right)\right)\\
&+\frac{4i\left(p_{3\mu_4}p_{4\mu_3}-p_3\cdot p_4 g_{\mu_3\mu_4}\right)}{\hat{g}'^2+\hat{g}^2}\left(s_\beta^2\left(\hat{g}'^2\hat{C}_{\Phi B}^{(11)}+\hat{g}\left(\hat{g}'\hat{C}_{\Phi WB}^{(11)}+\hat{g}\hat{C}_{\Phi W}^{(11)}\right)\right)\right.\\
&\qquad\qquad\qquad\qquad\qquad \left.+c_\beta^2\left(\hat{g}'^2\hat{C}_{\Phi B}^{(22)}+\hat{g}\left(\hat{g}'\hat{C}_{\Phi WB}^{(22)}+\hat{g}\hat{C}_{\Phi W}^{(22)}\right)\right)\right)\\
&+\frac{4ip_3^\mu p_4^\nu\epsilon_{\mu_3\mu_4\mu\nu}}{\hat{g}'^2+\hat{g}^2}\left(s_\beta^2\left(\hat{g}'^2\hat{C}_{\Phi\tilde{B}}^{(11)}+\hat{g}\left(\hat{g}'\hat{C}_{\Phi B\tilde{W}}^{(11)}+\hat{g}\hat{C}_{\Phi\tilde{W}}^{(11)}\right)\right)\right.\\
&\qquad\qquad\qquad \left.+c_\beta^2\left(\hat{g}'^2\hat{C}_{\Phi\tilde{B}}^{(22)}+\hat{g}\left(\hat{g}'\hat{C}_{\Phi B\tilde{W}}^{(22)}+\hat{g}\hat{C}_{\Phi\tilde{W}}^{(22)}\right)\right)\right)\\
&-2iv^2 g_{\mu_3\mu_4}\left(s_\beta^4+c_\beta^4-4s_\beta^2c_\beta^2\right)\left(\hat{g}'^2+\hat{g}^2\right)\left(\hat{C}_{D\Phi}^{(21)(21)*}+\hat{C}_{D\Phi}^{(21)(21)}\right)
\end{aligned} \tag{C.522}$$

$H^-$, $H^+$, $Z_{\mu_3}$, $Z_{\mu_4}$

$$
\begin{aligned}
&-\frac{ig_{\mu_3\mu_4}\left(\hat{g}'^2-\hat{g}^2\right)^2}{2\left(\hat{g}'^2+\hat{g}^2\right)^2}\left(s_\beta^2\left(\hat{g}'^2\left(2\delta_{s_{\hat{\beta}\pm}}-1\right)+\hat{g}^2\left(2\delta_{s_{\hat{\beta}\pm}}-1\right)+2\hat{g}X_{WB}\hat{g}'\right)\right.\\
&\qquad\left.+c_\beta^2\left(\hat{g}^2\left(2\delta_{c_{\hat{\beta}\pm}}-1\right)+\left(2\delta_{c_{\hat{\beta}\pm}}-1\right)\hat{g}'^2+2\hat{g}X_{WB}\hat{g}'\right)\right)\\
&-2iv^2g_{\mu_3\mu_4}\left(\hat{g}'^2-\hat{g}^2\right)\left(2s_\beta^2c_\beta^2\left(\hat{C}_{D\Phi}^{(11)(11)}-\hat{C}_{D\Phi}^{(21)(12)}+\hat{C}_{D\Phi}^{(22)(22)}\right)\right.\\
&\qquad\left.+s_\beta^4\hat{C}_{D\Phi}^{(11)(22)}+c_\beta^4\hat{C}_{D\Phi}^{(11)(22)}\right)\\
&+\frac{4i\left(p_{3\mu_4}p_{4\mu_3}-p_3\cdot p_4g_{\mu_3\mu_4}\right)}{\hat{g}'^2+\hat{g}^2}\left(s_\beta^2\left(\hat{g}'^2\hat{C}_{\Phi B}^{(11)}+\hat{g}\left(\hat{g}\hat{C}_{\Phi W}^{(11)}-\hat{g}'\hat{C}_{\Phi WB}^{(11)}\right)\right)\right.\\
&\qquad\left.+c_\beta^2\left(\hat{g}'^2\hat{C}_{\Phi B}^{(22)}+\hat{g}\left(\hat{g}\hat{C}_{\Phi W}^{(22)}-\hat{g}'\hat{C}_{\Phi WB}^{(22)}\right)\right)\right)\\
&+\frac{4ip_3^\mu p_4^\nu\epsilon_{\mu_3\mu_4\mu\nu}}{\hat{g}'^2+\hat{g}^2}\left(s_\beta^2\left(\hat{g}'^2\hat{C}_{\Phi\tilde{B}}^{(11)}+\hat{g}\left(\hat{g}\hat{C}_{\Phi\tilde{W}}^{(11)}-\hat{g}'\hat{C}_{\Phi B\tilde{W}}^{(11)}\right)\right)\right.\\
&\qquad\left.+c_\beta^2\left(\hat{g}'^2\hat{C}_{\Phi\tilde{B}}^{(22)}+\hat{g}\left(\hat{g}\hat{C}_{\Phi\tilde{W}}^{(22)}-\hat{g}'\hat{C}_{\Phi B\tilde{W}}^{(22)}\right)\right)\right)\\
&+4iv^2s_\beta^2c_\beta^2g_{\mu_3\mu_4}\left(\hat{g}'^2-\hat{g}^2\right)\left(\hat{C}_{D\Phi}^{(21)(21)*}+\hat{C}_{D\Phi}^{(21)(21)}\right)
\end{aligned}
\tag{C.523}
$$

$G^0$, $A^0$, $Z_{\mu_3}$, $Z_{\mu_4}$

$$
\begin{aligned}
&-\frac{1}{2}ig_{\mu_3\mu_4}\left(\hat{g}'^2+\hat{g}^2\right)\left(\left(A_2'-A_1'\right)s_\beta c_\beta-s_\beta^2B'+c_\beta^2B'\right)\\
&+2iv^2s_\beta c_\beta g_{\mu_3\mu_4}\left(\hat{g}'^2+\hat{g}^2\right)\left(c_\beta^2\left(2\hat{C}_{D\Phi}^{(11)(11)}-\hat{C}_{D\Phi}^{(11)(22)}-\hat{C}_{D\Phi}^{(21)(12)}\right)\right.\\
&\qquad\left.+s_\beta^2\left(\hat{C}_{D\Phi}^{(11)(22)}+\hat{C}_{D\Phi}^{(21)(12)}-2\hat{C}_{D\Phi}^{(22)(22)}\right)\right)\\
&-\frac{4is_\beta c_\beta\left(p_{3\mu_4}p_{4\mu_3}-p_3\cdot p_4g_{\mu_3\mu_4}\right)}{\hat{g}'^2+\hat{g}^2}\left(\hat{g}\left(\hat{g}'\hat{C}_{\Phi WB}^{(11)}+\hat{g}\hat{C}_{\Phi W}^{(11)}\right.\right.\\
&\qquad\left.-\hat{g}'\hat{C}_{\Phi WB}^{(22)}-\hat{g}\hat{C}_{\Phi W}^{(22)}\right)\\
&\qquad\left.+\hat{g}'^2\hat{C}_{\Phi B}^{(11)}-\hat{g}'^2\hat{C}_{\Phi B}^{(22)}\right)\\
&-\frac{4is_\beta c_\beta p_3^\mu p_4^\nu\epsilon_{\mu_3\mu_4\mu\nu}}{\hat{g}'^2+\hat{g}^2}\left(\hat{g}\left(\hat{g}'\hat{C}_{\Phi B\tilde{W}}^{(11)}+\hat{g}\hat{C}_{\Phi\tilde{W}}^{(11)}-\hat{g}'\hat{C}_{\Phi B\tilde{W}}^{(22)}-\hat{g}\hat{C}_{\Phi\tilde{W}}^{(22)}\right)\right.\\
&\qquad\left.+\hat{g}'^2\hat{C}_{\Phi\tilde{B}}^{(11)}-\hat{g}'^2\hat{C}_{\Phi\tilde{B}}^{(22)}\right)\\
&-2iv^2s_\beta c_\beta c_{2\beta}g_{\mu_3\mu_4}\left(\hat{g}'^2+\hat{g}^2\right)\left(\hat{C}_{D\Phi}^{(21)(21)*}+\hat{C}_{D\Phi}^{(21)(21)}\right)
\end{aligned}
\tag{C.524}
$$

$h$, $A^0$, $Z_{\mu_3}$, $Z_{\mu_4}$

$$-6v^2 s_\beta c_\beta g_{\mu_3\mu_4}\left(\hat{g}'^2+\hat{g}^2\right)\Big(-s_\beta^2\hat{C}_{D\Phi}^{(21)(21)*}-c_\beta^2\hat{C}_{D\Phi}^{(21)(21)*} +s_\beta^2\hat{C}_{D\Phi}^{(21)(21)}+c_\beta^2\hat{C}_{D\Phi}^{(21)(21)}\Big) \tag{C.525}$$

$H$, $G^0$, $Z_{\mu_3}$, $Z_{\mu_4}$

$$-2v^2 s_\beta c_\beta g_{\mu_3\mu_4}\left(\hat{g}'^2+\hat{g}^2\right)\Big(-s_\beta^2\hat{C}_{D\Phi}^{(21)(21)*}-c_\beta^2\hat{C}_{D\Phi}^{(21)(21)*} +s_\beta^2\hat{C}_{D\Phi}^{(21)(21)}+c_\beta^2\hat{C}_{D\Phi}^{(21)(21)}\Big) \tag{C.526}$$

$H$, $h$, $Z_{\mu_3}$, $Z_{\mu_4}$

$$\begin{aligned}
&+\frac{1}{2}ig_{\mu_3\mu_4}\left(\hat{g}'^2+\hat{g}^2\right)\left(\left(A_2-A_1\right)s_\beta c_\beta-Bs_\beta^2+Bc_\beta^2\right)\\
&-6iv^2 s_\beta c_\beta g_{\mu_3\mu_4}\left(\hat{g}'^2+\hat{g}^2\right)\Big(c_\beta^2\left(2\hat{C}_{D\Phi}^{(11)(11)}-\hat{C}_{D\Phi}^{(11)(22)}-\hat{C}_{D\Phi}^{(21)(12)}\right)\\
&\qquad +s_\beta^2\left(\hat{C}_{D\Phi}^{(11)(22)}+\hat{C}_{D\Phi}^{(21)(12)}-2\hat{C}_{D\Phi}^{(22)(22)}\right)\Big)\\
&+\frac{4is_\beta c_\beta\left(p_{3\mu_4}p_{4\mu_3}-p_3\cdot p_4 g_{\mu_3\mu_4}\right)}{\hat{g}'^2+\hat{g}^2}\Big(\hat{g}\left(\hat{g}'\hat{C}_{\Phi WB}^{(11)}+\hat{g}\hat{C}_{\Phi W}^{(11)}-\hat{g}'\hat{C}_{\Phi WB}^{(22)}-\hat{g}\hat{C}_{\Phi W}^{(22)}\right)\\
&\qquad +\hat{g}'^2\hat{C}_{\Phi B}^{(11)}-\hat{g}'^2\hat{C}_{\Phi B}^{(22)}\Big)\\
&+\frac{4is_\beta c_\beta p_3^\mu p_4^\nu\epsilon_{\mu_3\mu_4\mu\nu}}{\hat{g}'^2+\hat{g}^2}\Big(\hat{g}\left(\hat{g}'\hat{C}_{\Phi B\tilde{W}}^{(11)}+\hat{g}\hat{C}_{\Phi\tilde{W}}^{(11)}-\hat{g}'\hat{C}_{\Phi B\tilde{W}}^{(22)}-\hat{g}\hat{C}_{\Phi\tilde{W}}^{(22)}\right)\\
&\qquad +\hat{g}'^2\hat{C}_{\Phi\tilde{B}}^{(11)}-\hat{g}'^2\hat{C}_{\Phi\tilde{B}}^{(22)}\Big)\\
&+6iv^2 s_\beta c_\beta c_{2\beta} g_{\mu_3\mu_4}\left(\hat{g}'^2+\hat{g}^2\right)\left(\hat{C}_{D\Phi}^{(21)(21)*}+\hat{C}_{D\Phi}^{(21)(21)}\right)
\end{aligned} \tag{C.527}$$

$H^+$, $G^-$, $Z_{\mu_3}$, $Z_{\mu_4}$

$$
\begin{aligned}
&+2iv^2 s_\beta c_\beta g_{\mu_3\mu_4}\left(\hat{g}'^2-\hat{g}^2\right)\left(c_\beta^2\left(2\hat{C}_{D\Phi}^{(11)(11)}-\hat{C}_{D\Phi}^{(11)(22)}-\hat{C}_{D\Phi}^{(21)(12)}\right)\right.\\
&\qquad\left.+s_\beta^2\left(\hat{C}_{D\Phi}^{(11)(22)}+\hat{C}_{D\Phi}^{(21)(12)}-2\hat{C}_{D\Phi}^{(22)(22)}\right)\right)\\
&-\frac{4is_\beta c_\beta\left(p_{3\mu_4}p_{4\mu_3}-p_3\cdot p_4 g_{\mu_3\mu_4}\right)}{\hat{g}'^2+\hat{g}^2}\left(-\hat{g}\left(\hat{g}'\hat{C}_{\Phi WB}^{(11)}+\hat{g}\hat{C}_{\Phi W}^{(11)}\right.\right.\\
&\qquad\left.+\hat{g}'\hat{C}_{\Phi WB}^{(22)}-\hat{g}\hat{C}_{\Phi W}^{(22)}\right)\\
&\qquad\left.+\hat{g}'^2\hat{C}_{\Phi B}^{(11)}-\hat{g}'^2\hat{C}_{\Phi B}^{(22)}\right)\\
&-\frac{4is_\beta c_\beta p_3^\mu p_4^\nu \epsilon_{\mu_3\mu_4\mu\nu}}{\hat{g}'^2+\hat{g}^2}\left(-\hat{g}\left(\hat{g}\hat{C}_{\Phi B\tilde{W}}^{(11)}+\hat{g}\hat{C}_{\Phi\tilde{W}}^{(11)}\right.\right.\\
&\qquad\left.+\hat{g}'\hat{C}_{\Phi B\tilde{W}}^{(22)}-\hat{g}\hat{C}_{\Phi\tilde{W}}^{(22)}\right)\\
&\qquad\left.+\hat{g}'^2\hat{C}_{\Phi\tilde{B}}^{(11)}-\hat{g}'^2\hat{C}_{\Phi\tilde{B}}^{(22)}\right)\\
&-4iv^2 s_\beta c_\beta g_{\mu_3\mu_4}\left(\hat{g}'^2-\hat{g}^2\right)\left(c_\beta^2\hat{C}_{D\Phi}^{(21)(21)*}-s_\beta^2\hat{C}_{D\Phi}^{(21)(21)}\right)
\end{aligned}
\tag{C.528}
$$

$A_{\mu_2}$, $A_{\mu_1}$, $A^0$, $A^0$

$$
\begin{aligned}
&+\frac{4i\left(p_{1\mu_2}p_{2\mu_1}-p_1\cdot p_2 g_{\mu_1\mu_2}\right)}{\hat{g}'^2+\hat{g}^2}\left(s_\beta^2\left(\hat{g}\left(\hat{g}\hat{C}_{\Phi B}^{(11)}-\hat{g}'\hat{C}_{\Phi WB}^{(11)}\right)+\hat{g}'^2\hat{C}_{\Phi W}^{(11)}\right)\right.\\
&\qquad\left.+c_\beta^2\left(\hat{g}\left(\hat{g}\hat{C}_{\Phi B}^{(22)}-\hat{g}'\hat{C}_{\Phi WB}^{(22)}\right)+\hat{g}'^2\hat{C}_{\Phi W}^{(22)}\right)\right)\\
&+\frac{4ip_1^\mu p_2^\nu \epsilon_{\mu_1\mu_2\mu\nu}}{\hat{g}'^2+\hat{g}^2}\left(s_\beta^2\left(\hat{g}\left(\hat{g}\hat{C}_{\Phi\tilde{B}}^{(11)}-\hat{g}'\hat{C}_{\Phi B\tilde{W}}^{(11)}\right)+\hat{g}'^2\hat{C}_{\Phi\tilde{W}}^{(11)}\right)\right.\\
&\qquad\left.+c_\beta^2\left(\hat{g}\left(\hat{g}\hat{C}_{\Phi\tilde{B}}^{(22)}-\hat{g}'\hat{C}_{\Phi B\tilde{W}}^{(22)}\right)+\hat{g}'^2\hat{C}_{\Phi\tilde{W}}^{(22)}\right)\right)
\end{aligned}
\tag{C.529}
$$

$A_{\mu_2}$, $A_{\mu_1}$, $G^0$, $G^0$

$$
\begin{aligned}
&+\frac{4i\left(p_{1\mu_2}p_{2\mu_1}-p_1\cdot p_2 g_{\mu_1\mu_2}\right)}{\hat{g}'^2+\hat{g}^2}\left(c_\beta^2\left(\hat{g}\left(\hat{g}\hat{C}_{\Phi B}^{(11)}-\hat{g}'\hat{C}_{\Phi WB}^{(11)}\right)+\hat{g}'^2\hat{C}_{\Phi W}^{(11)}\right)\right.\\
&\qquad\left.+s_\beta^2\left(\hat{g}\left(\hat{g}\hat{C}_{\Phi B}^{(22)}-\hat{g}'\hat{C}_{\Phi WB}^{(22)}\right)+\hat{g}'^2\hat{C}_{\Phi W}^{(22)}\right)\right)\\
&+\frac{4ip_1^\mu p_2^\nu \epsilon_{\mu_1\mu_2\mu\nu}}{\hat{g}'^2+\hat{g}^2}\left(c_\beta^2\left(\hat{g}\left(\hat{g}\hat{C}_{\Phi\tilde{B}}^{(11)}-\hat{g}'\hat{C}_{\Phi B\tilde{W}}^{(11)}\right)+\hat{g}'^2\hat{C}_{\Phi\tilde{W}}^{(11)}\right)\right.\\
&\qquad\left.+s_\beta^2\left(\hat{g}\left(\hat{g}\hat{C}_{\Phi\tilde{B}}^{(22)}-\hat{g}'\hat{C}_{\Phi B\tilde{W}}^{(22)}\right)+\hat{g}'^2\hat{C}_{\Phi\tilde{W}}^{(22)}\right)\right)
\end{aligned}
\tag{C.530}
$$

$$
\begin{aligned}
&-\frac{2i\hat{g}^2 g_{\mu_1\mu_2}\hat{g}'^2}{\hat{g}'^2+\hat{g}^2}\left(s_\beta^2\left(2\delta_{s_{\hat{\beta}\pm}}-1\right)+c_\beta^2\left(2\delta_{c_{\hat{\beta}\pm}}-1\right)+\frac{2\hat{g}X_{WB}\hat{g}'}{\hat{g}'^2+\hat{g}^2}\right)\\
&+\frac{4i\left(p_{1\mu_2}p_{2\mu_1}-p_1\cdot p_2 g_{\mu_1\mu_2}\right)}{\hat{g}'^2+\hat{g}^2}\left(c_\beta^2\left(\hat{g}\left(\hat{g}'\hat{C}^{(11)}_{\Phi WB}+\hat{g}\hat{C}^{(11)}_{\Phi B}\right)+\hat{g}'^2\hat{C}^{(11)}_{\Phi W}\right)\right.\\
&\qquad\left.+s_\beta^2\left(\hat{g}\left(\hat{g}'\hat{C}^{(22)}_{\Phi WB}+\hat{g}\hat{C}^{(22)}_{\Phi B}\right)+\hat{g}'^2\hat{C}^{(22)}_{\Phi W}\right)\right)\\
&+\frac{4ip_1^{\mu}p_2^{\nu}\epsilon_{\mu_1\mu_2\mu\nu}}{\hat{g}'^2+\hat{g}^2}\left(c_\beta^2\left(\hat{g}\left(\hat{g}'\hat{C}^{(11)}_{\Phi B\tilde{W}}+\hat{g}\hat{C}^{(11)}_{\Phi\tilde{B}}\right)+\hat{g}'^2\hat{C}^{(11)}_{\Phi\tilde{W}}\right)\right.\\
&\qquad\left.+s_\beta^2\left(\hat{g}\left(\hat{g}'\hat{C}^{(22)}_{\Phi B\tilde{W}}+\hat{g}\hat{C}^{(22)}_{\Phi\tilde{B}}\right)+\hat{g}'^2\hat{C}^{(22)}_{\Phi\tilde{W}}\right)\right)
\end{aligned}
\tag{C.531}
$$

$$
\begin{aligned}
&+\frac{4i\left(p_{1\mu_2}p_{2\mu_1}-p_1\cdot p_2 g_{\mu_1\mu_2}\right)}{\hat{g}'^2+\hat{g}^2}\left(c_\beta^2\left(\hat{g}\left(\hat{g}\hat{C}^{(11)}_{\Phi B}-\hat{g}'\hat{C}^{(11)}_{\Phi WB}\right)+\hat{g}'^2\hat{C}^{(11)}_{\Phi W}\right)\right.\\
&\qquad\left.+s_\beta^2\left(\hat{g}\left(\hat{g}\hat{C}^{(22)}_{\Phi B}-\hat{g}'\hat{C}^{(22)}_{\Phi WB}\right)+\hat{g}'^2\hat{C}^{(22)}_{\Phi W}\right)\right)\\
&+\frac{4ip_1^{\mu}p_2^{\nu}\epsilon_{\mu_1\mu_2\mu\nu}}{\hat{g}'^2+\hat{g}^2}\left(c_\beta^2\left(\hat{g}\left(\hat{g}\hat{C}^{(11)}_{\Phi\tilde{B}}-\hat{g}'\hat{C}^{(11)}_{\Phi B\tilde{W}}\right)+\hat{g}'^2\hat{C}^{(11)}_{\Phi\tilde{W}}\right)\right.\\
&\qquad\left.+s_\beta^2\left(\hat{g}\left(\hat{g}\hat{C}^{(22)}_{\Phi\tilde{B}}-\hat{g}'\hat{C}^{(22)}_{\Phi B\tilde{W}}\right)+\hat{g}'^2\hat{C}^{(22)}_{\Phi\tilde{W}}\right)\right)
\end{aligned}
\tag{C.532}
$$

$$
\begin{aligned}
&+\frac{4i\left(p_{1\mu_2}p_{2\mu_1}-p_1\cdot p_2 g_{\mu_1\mu_2}\right)}{\hat{g}'^2+\hat{g}^2}\left(s_\beta^2\left(\hat{g}\left(\hat{g}\hat{C}^{(11)}_{\Phi B}-\hat{g}'\hat{C}^{(11)}_{\Phi WB}\right)+\hat{g}'^2\hat{C}^{(11)}_{\Phi W}\right)\right.\\
&\qquad\left.+c_\beta^2\left(\hat{g}\left(\hat{g}\hat{C}^{(22)}_{\Phi B}-\hat{g}'\hat{C}^{(22)}_{\Phi WB}\right)+\hat{g}'^2\hat{C}^{(22)}_{\Phi W}\right)\right)\\
&+\frac{4ip_1^{\mu}p_2^{\nu}\epsilon_{\mu_1\mu_2\mu\nu}}{\hat{g}'^2+\hat{g}^2}\left(s_\beta^2\left(\hat{g}\left(\hat{g}\hat{C}^{(11)}_{\Phi\tilde{B}}-\hat{g}'\hat{C}^{(11)}_{\Phi B\tilde{W}}\right)+\hat{g}'^2\hat{C}^{(11)}_{\Phi\tilde{W}}\right)\right.\\
&\qquad\left.+c_\beta^2\left(\hat{g}\left(\hat{g}\hat{C}^{(22)}_{\Phi\tilde{B}}-\hat{g}'\hat{C}^{(22)}_{\Phi B\tilde{W}}\right)+\hat{g}'^2\hat{C}^{(22)}_{\Phi\tilde{W}}\right)\right)
\end{aligned}
\tag{C.533}
$$

$$
\begin{aligned}
&-\frac{2i\hat{g}^2 g_{\mu_1\mu_2}\hat{g}'^2}{\hat{g}'^2+\hat{g}^2}\left(s_\beta^2\left(2\delta_{s_{\hat{\beta}\pm}}-1\right)+c_\beta^2\left(2\delta_{c_{\hat{\beta}\pm}}-1\right)+\frac{2\hat{g}X_{WB}\hat{g}'}{\hat{g}'^2+\hat{g}^2}\right)\\
&+\frac{4i\left(p_{1\mu_2}p_{2\mu_1}-p_1\cdot p_2 g_{\mu_1\mu_2}\right)}{\hat{g}'^2+\hat{g}^2}\left(s_\beta^2\left(\hat{g}\left(\hat{g}'\hat{C}^{(11)}_{\Phi WB}+\hat{g}\hat{C}^{(11)}_{\Phi B}\right)+\hat{g}'^2\hat{C}^{(11)}_{\Phi W}\right)\right.\\
&\qquad\qquad\left.+c_\beta^2\left(\hat{g}\left(\hat{g}'\hat{C}^{(22)}_{\Phi WB}+\hat{g}\hat{C}^{(22)}_{\Phi B}\right)+\hat{g}'^2\hat{C}^{(22)}_{\Phi W}\right)\right)\\
&+\frac{4ip_1^\mu p_2^\nu\epsilon_{\mu_1\mu_2\mu\nu}}{\hat{g}'^2+\hat{g}^2}\left(s_\beta^2\left(\hat{g}\left(\hat{g}'\hat{C}^{(11)}_{\Phi B\tilde{W}}+\hat{g}\hat{C}^{(11)}_{\Phi\tilde{B}}\right)+\hat{g}'^2\hat{C}^{(11)}_{\Phi\tilde{W}}\right)\right.\\
&\qquad\qquad\left.+c_\beta^2\left(\hat{g}\left(\hat{g}'\hat{C}^{(22)}_{\Phi B\tilde{W}}+\hat{g}\hat{C}^{(22)}_{\Phi\tilde{B}}\right)+\hat{g}'^2\hat{C}^{(22)}_{\Phi\tilde{W}}\right)\right)
\end{aligned}
\tag{C.534}
$$

$$
\begin{aligned}
&+\frac{\hat{g}^2 g_{\mu_1\mu_4}\hat{g}'}{4\sqrt{\hat{g}'^2+\hat{g}^2}}\left(s_\beta^2\left(2\left(\delta_{s_{\hat{\beta}\pm}}+\delta_{s_{\hat{\beta}}}-1\right)+A_2'\right)\right.\\
&\qquad\qquad+c_\beta^2\left(A_1'+2\left(\delta_{c_{\hat{\beta}\pm}}+\delta_{c_{\hat{\beta}}}-1\right)\right)\\
&\qquad\qquad\left.+s_{2\beta}B'+\frac{2\hat{g}X_{WB}\hat{g}'}{\hat{g}'^2+\hat{g}^2}\right)\\
&-\frac{2\hat{g}\left(p_{1\mu_4}p_{4\mu_1}-p_1\cdot p_4 g_{\mu_1\mu_4}\right)}{\sqrt{\hat{g}'^2+\hat{g}^2}}\left(c_\beta^2\hat{C}^{(11)}_{\Phi WB}+s_\beta^2\hat{C}^{(22)}_{\Phi WB}\right)\\
&-\frac{2\hat{g}p_1^\mu p_4^\nu\epsilon_{\mu_1\mu_4\mu\nu}}{\sqrt{\hat{g}'^2+\hat{g}^2}}\left(c_\beta^2\hat{C}^{(11)}_{\Phi B\tilde{W}}+s_\beta^2\hat{C}^{(22)}_{\Phi B\tilde{W}}\right)
\end{aligned}
\tag{C.535}
$$

$$
\begin{aligned}
&-\frac{i\hat{g}^2 g_{\mu_1\mu_4}\hat{g}'}{4\sqrt{\hat{g}'^2+\hat{g}^2}}\left(s_\beta^2\left(2\delta_{s_{\hat{\beta}\pm}}+A_2-2\right)+c_\beta^2\left(A_1+2\delta_{c_{\hat{\beta}\pm}}-2\right)\right.\\
&\qquad\qquad\left.+Bs_{2\beta}+\frac{2\hat{g}X_{WB}\hat{g}'}{\hat{g}'^2+\hat{g}^2}\right)\\
&+\frac{2i\hat{g}\left(p_{1\mu_4}p_{4\mu_1}-p_1\cdot p_4 g_{\mu_1\mu_4}\right)}{\sqrt{\hat{g}'^2+\hat{g}^2}}\left(c_\beta^2\hat{C}^{(11)}_{\Phi WB}+s_\beta^2\hat{C}^{(22)}_{\Phi WB}\right)\\
&+\frac{2i\hat{g}p_1^\mu p_4^\nu\epsilon_{\mu_1\mu_4\mu\nu}}{\sqrt{\hat{g}'^2+\hat{g}^2}}\left(c_\beta^2\hat{C}^{(11)}_{\Phi B\tilde{W}}+s_\beta^2\hat{C}^{(22)}_{\Phi B\tilde{W}}\right)
\end{aligned}
\tag{C.536}
$$

$A^0$ $A_{\mu_1}$ $H^-$ $W^+_{\mu_4}$

$$
\begin{aligned}
&+\frac{\hat{g}^2 g_{\mu_1\mu_4}\hat{g}'}{4\sqrt{\hat{g}'^2+\hat{g}^2}}\left(s_\beta^2\left(2\left(\delta_{s_{\hat{\beta}\pm}}+\delta_{s_{\hat{\beta}}}-1\right)+A_1'\right)\right.\\
&\qquad\qquad +c_\beta^2\left(A_2'+2\left(\delta_{c_{\hat{\beta}\pm}}+\delta_{c_{\hat{\beta}}}-1\right)\right)\\
&\qquad\qquad \left.-2s_\beta c_\beta B'+\frac{2\hat{g}X_{WB}\hat{g}'}{\hat{g}'^2+\hat{g}^2}\right)\\
&-\frac{2\hat{g}\left(p_{1\mu_4}p_{4\mu_1}-p_1\cdot p_4 g_{\mu_1\mu_4}\right)}{\sqrt{\hat{g}'^2+\hat{g}^2}}\left(s_\beta^2\hat{C}^{(11)}_{\Phi WB}+c_\beta^2\hat{C}^{(22)}_{\Phi WB}\right)\\
&-\frac{2\hat{g}p_1^\mu p_4^\nu\epsilon_{\mu_1\mu_4\mu\nu}}{\sqrt{\hat{g}'^2+\hat{g}^2}}\left(s_\beta^2\hat{C}^{(11)}_{\Phi B\tilde{W}}+c_\beta^2\hat{C}^{(22)}_{\Phi B\tilde{W}}\right)
\end{aligned}
\tag{C.537}
$$

$H$ $A_{\mu_1}$ $H^-$ $W^+_{\mu_4}$

$$
\begin{aligned}
&+\frac{i\hat{g}^2 g_{\mu_1\mu_4}\hat{g}'}{4\sqrt{\hat{g}'^2+\hat{g}^2}}\left(s_\beta^2\left(2\delta_{s_{\hat{\beta}\pm}}+A_1-2\right)+c_\beta^2\left(A_2+2\delta_{c_{\hat{\beta}\pm}}-2\right)\right.\\
&\qquad\qquad \left.-2Bs_\beta c_\beta+\frac{2\hat{g}X_{WB}\hat{g}'}{\hat{g}'^2+\hat{g}^2}\right)\\
&-\frac{2i\hat{g}\left(p_{1\mu_4}p_{4\mu_1}-p_1\cdot p_4 g_{\mu_1\mu_4}\right)}{\sqrt{\hat{g}'^2+\hat{g}^2}}\left(s_\beta^2\hat{C}^{(11)}_{\Phi WB}+c_\beta^2\hat{C}^{(22)}_{\Phi WB}\right)\\
&-\frac{2i\hat{g}p_1^\mu p_4^\nu\epsilon_{\mu_1\mu_4\mu\nu}}{\sqrt{\hat{g}'^2+\hat{g}^2}}\left(s_\beta^2\hat{C}^{(11)}_{\Phi B\tilde{W}}+c_\beta^2\hat{C}^{(22)}_{\Phi B\tilde{W}}\right)
\end{aligned}
\tag{C.538}
$$

$A^0$ $A^0$ $W^+_{\mu_3}$ $W^-_{\mu_4}$

$$
\begin{aligned}
&-\frac{1}{2}i\hat{g}^2 g_{\mu_3\mu_4}\left(c_\beta^2\left(A_2'+2\delta_{c_{\hat{\beta}}}-1\right)+s_\beta^2\left(A_1'+2\delta_{s_{\hat{\beta}}}-1\right)-2s_\beta c_\beta B'\right)\\
&+4i\left(p_{3\mu_4}p_{4\mu_3}-p_3\cdot p_4 g_{\mu_3\mu_4}\right)\left(s_\beta^2\hat{C}^{(11)}_{\Phi W}+c_\beta^2\hat{C}^{(22)}_{\Phi W}\right)\\
&+4ip_3^\mu p_4^\nu\epsilon_{\mu_3\mu_4\mu\nu}\left(s_\beta^2\hat{C}^{(11)}_{\Phi\tilde{W}}+c_\beta^2\hat{C}^{(22)}_{\Phi\tilde{W}}\right)
\end{aligned}
\tag{C.539}
$$

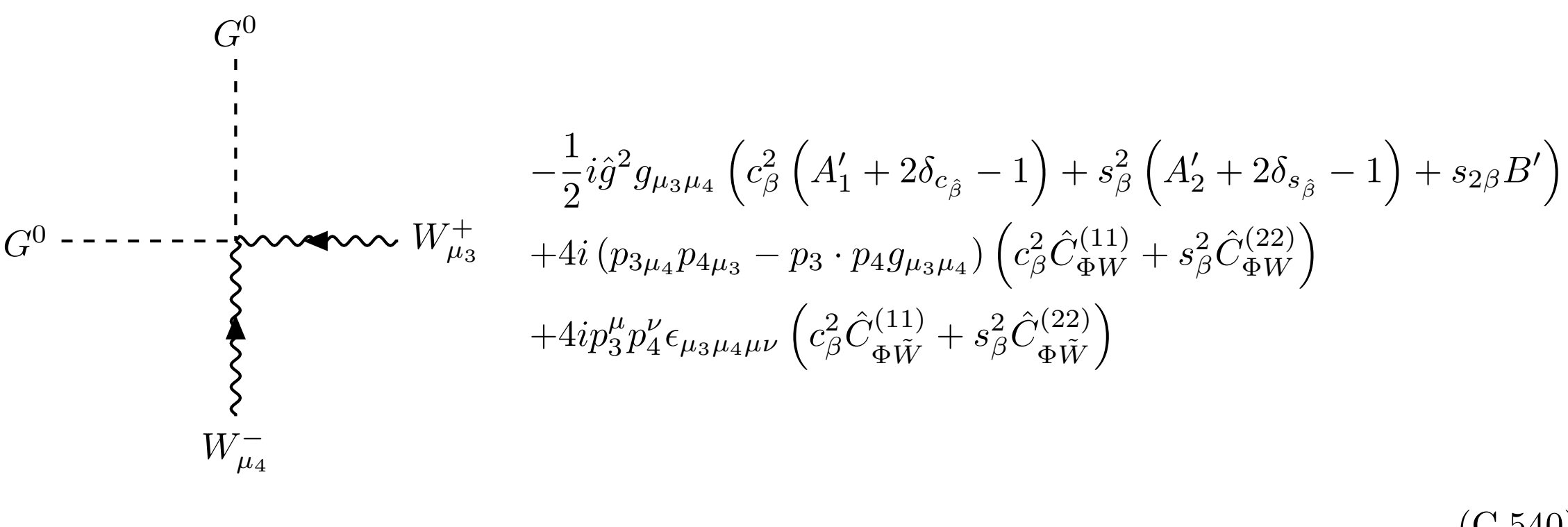

$$
\begin{aligned}
&-\frac{1}{2} i \hat{g}^2 g_{\mu_3\mu_4} \left( c_\beta^2 \left( A_1' + 2\delta_{c_{\hat{\beta}}} - 1 \right) + s_\beta^2 \left( A_2' + 2\delta_{s_{\hat{\beta}}} - 1 \right) + s_{2\beta} B' \right) \\
&+4i \left( p_{3\mu_4} p_{4\mu_3} - p_3 \cdot p_4 g_{\mu_3\mu_4} \right) \left( c_\beta^2 \hat{C}^{(11)}_{\Phi W} + s_\beta^2 \hat{C}^{(22)}_{\Phi W} \right) \\
&+4i p_3^\mu p_4^\nu \epsilon_{\mu_3\mu_4\mu\nu} \left( c_\beta^2 \hat{C}^{(11)}_{\Phi \tilde{W}} + s_\beta^2 \hat{C}^{(22)}_{\Phi \tilde{W}} \right)
\end{aligned}
\tag{C.540}
$$

$$
\begin{aligned}
&-\frac{1}{2} i \hat{g}^2 g_{\mu_3\mu_4} \left( \left( A_2 - 1 \right) s_\beta^2 + \left( A_1 - 1 \right) c_\beta^2 + B s_{2\beta} \right) \\
&+4i \left( p_{3\mu_4} p_{4\mu_3} - p_3 \cdot p_4 g_{\mu_3\mu_4} \right) \left( c_\beta^2 \hat{C}^{(11)}_{\Phi W} + s_\beta^2 \hat{C}^{(22)}_{\Phi W} \right) \\
&+4i p_3^\mu p_4^\nu \epsilon_{\mu_3\mu_4\mu\nu} \left( c_\beta^2 \hat{C}^{(11)}_{\Phi \tilde{W}} + s_\beta^2 \hat{C}^{(22)}_{\Phi \tilde{W}} \right)
\end{aligned}
\tag{C.541}
$$

$$
\begin{aligned}
&-\frac{1}{2} i \hat{g}^2 g_{\mu_3\mu_4} \left( \left( A_1 - 1 \right) s_\beta^2 + \left( A_2 - 1 \right) c_\beta^2 - 2 B s_\beta c_\beta \right) \\
&+4i \left( p_{3\mu_4} p_{4\mu_3} - p_3 \cdot p_4 g_{\mu_3\mu_4} \right) \left( s_\beta^2 \hat{C}^{(11)}_{\Phi W} + c_\beta^2 \hat{C}^{(22)}_{\Phi W} \right) \\
&+4i p_3^\mu p_4^\nu \epsilon_{\mu_3\mu_4\mu\nu} \left( s_\beta^2 \hat{C}^{(11)}_{\Phi \tilde{W}} + c_\beta^2 \hat{C}^{(22)}_{\Phi \tilde{W}} \right)
\end{aligned}
\tag{C.542}
$$

$A^0$

$A_{\mu_1}$ $A^0$

$Z_{\mu_4}$

$$
\begin{aligned}
&+\frac{2i\left(p_{1\mu_4}p_{4\mu_1}-p_1\cdot p_4 g_{\mu_1\mu_4}\right)}{\hat{g}'^2+\hat{g}^2}\left(s_\beta^2\left(2\hat{g}\hat{g}'\left(\hat{C}^{(11)}_{\Phi W}-\hat{C}^{(11)}_{\Phi B}\right)\right.\right.\\
&\qquad\qquad\left.+\left(\hat{g}'^2-\hat{g}^2\right)\hat{C}^{(11)}_{\Phi WB}\right)\\
&\qquad\qquad+c_\beta^2\left(2\hat{g}\hat{g}'\left(\hat{C}^{(22)}_{\Phi W}-\hat{C}^{(22)}_{\Phi B}\right)\right.\\
&\qquad\qquad\left.\left.+\left(\hat{g}'^2-\hat{g}^2\right)\hat{C}^{(22)}_{\Phi WB}\right)\right)\\
&+\frac{2ip_1^\mu p_4^\nu\epsilon_{\mu_1\mu_4\mu\nu}}{\hat{g}'^2+\hat{g}^2}\left(s_\beta^2\left(2\hat{g}\hat{g}'\left(\hat{C}^{(11)}_{\Phi\tilde{W}}-\hat{C}^{(11)}_{\Phi\tilde{B}}\right)+\left(\hat{g}'^2-\hat{g}^2\right)\hat{C}^{(11)}_{\Phi B\tilde{W}}\right)\right.\\
&\qquad\left.+c_\beta^2\left(2\hat{g}\hat{g}'\left(\hat{C}^{(22)}_{\Phi\tilde{W}}-\hat{C}^{(22)}_{\Phi\tilde{B}}\right)+\left(\hat{g}'^2-\hat{g}^2\right)\hat{C}^{(22)}_{\Phi B\tilde{W}}\right)\right)
\end{aligned}
\tag{C.543}
$$

$G^0$

$A_{\mu_1}$ $G^0$

$Z_{\mu_4}$

$$
\begin{aligned}
&+\frac{2i\left(p_{1\mu_4}p_{4\mu_1}-p_1\cdot p_4 g_{\mu_1\mu_4}\right)}{\hat{g}'^2+\hat{g}^2}\left(c_\beta^2\left(2\hat{g}\hat{g}'\left(\hat{C}^{(11)}_{\Phi W}-\hat{C}^{(11)}_{\Phi B}\right)\right.\right.\\
&\qquad\qquad\left.+\left(\hat{g}'^2-\hat{g}^2\right)\hat{C}^{(11)}_{\Phi WB}\right)\\
&\qquad\qquad+s_\beta^2\left(2\hat{g}\hat{g}'\left(\hat{C}^{(22)}_{\Phi W}-\hat{C}^{(22)}_{\Phi B}\right)\right.\\
&\qquad\qquad\left.\left.+\left(\hat{g}'^2-\hat{g}^2\right)\hat{C}^{(22)}_{\Phi WB}\right)\right)\\
&+\frac{2ip_1^\mu p_4^\nu\epsilon_{\mu_1\mu_4\mu\nu}}{\hat{g}'^2+\hat{g}^2}\left(c_\beta^2\left(2\hat{g}\hat{g}'\left(\hat{C}^{(11)}_{\Phi\tilde{W}}-\hat{C}^{(11)}_{\Phi\tilde{B}}\right)+\left(\hat{g}'^2-\hat{g}^2\right)\hat{C}^{(11)}_{\Phi B\tilde{W}}\right)\right.\\
&\qquad\left.+s_\beta^2\left(2\hat{g}\hat{g}'\left(\hat{C}^{(22)}_{\Phi\tilde{W}}-\hat{C}^{(22)}_{\Phi\tilde{B}}\right)+\left(\hat{g}'^2-\hat{g}^2\right)\hat{C}^{(22)}_{\Phi B\tilde{W}}\right)\right)
\end{aligned}
\tag{C.544}
$$

$h$

$A_{\mu_1}$ $h$

$Z_{\mu_4}$

$$
\begin{aligned}
&+\frac{2i\left(p_{1\mu_4}p_{4\mu_1}-p_1\cdot p_4 g_{\mu_1\mu_4}\right)}{\hat{g}'^2+\hat{g}^2}\left(c_\beta^2\left(2\hat{g}\hat{g}'\left(\hat{C}^{(11)}_{\Phi W}-\hat{C}^{(11)}_{\Phi B}\right)\right.\right.\\
&\qquad\qquad\left.+\left(\hat{g}'^2-\hat{g}^2\right)\hat{C}^{(11)}_{\Phi WB}\right)\\
&\qquad\qquad+s_\beta^2\left(2\hat{g}\hat{g}'\left(\hat{C}^{(22)}_{\Phi W}-\hat{C}^{(22)}_{\Phi B}\right)\right.\\
&\qquad\qquad\left.\left.+\left(\hat{g}'^2-\hat{g}^2\right)\hat{C}^{(22)}_{\Phi WB}\right)\right)\\
&+\frac{2ip_1^\mu p_4^\nu\epsilon_{\mu_1\mu_4\mu\nu}}{\hat{g}'^2+\hat{g}^2}\left(c_\beta^2\left(2\hat{g}\hat{g}'\left(\hat{C}^{(11)}_{\Phi\tilde{W}}-\hat{C}^{(11)}_{\Phi\tilde{B}}\right)+\left(\hat{g}'^2-\hat{g}^2\right)\hat{C}^{(11)}_{\Phi B\tilde{W}}\right)\right.\\
&\qquad\left.+s_\beta^2\left(2\hat{g}\hat{g}'\left(\hat{C}^{(22)}_{\Phi\tilde{W}}-\hat{C}^{(22)}_{\Phi\tilde{B}}\right)+\left(\hat{g}'^2-\hat{g}^2\right)\hat{C}^{(22)}_{\Phi B\tilde{W}}\right)\right)
\end{aligned}
\tag{C.545}
$$

$H$, $A_{\mu_1}$, $H$, $Z_{\mu_4}$

$$
\begin{aligned}
&+\frac{2i\left(p_{1\mu_4}p_{4\mu_1}-p_1\cdot p_4 g_{\mu_1\mu_4}\right)}{\hat{g}'^2+\hat{g}^2}\left(s_\beta^2\left(2\hat{g}\hat{g}'\left(\hat{C}^{(11)}_{\Phi W}-\hat{C}^{(11)}_{\Phi B}\right)\right.\right.\\
&\qquad\left.+\left(\hat{g}'^2-\hat{g}^2\right)\hat{C}^{(11)}_{\Phi WB}\right)\\
&\qquad+c_\beta^2\left(2\hat{g}\hat{g}'\left(\hat{C}^{(22)}_{\Phi W}-\hat{C}^{(22)}_{\Phi B}\right)\right.\\
&\qquad\left.\left.+\left(\hat{g}'^2-\hat{g}^2\right)\hat{C}^{(22)}_{\Phi WB}\right)\right)\\
&+\frac{2ip_1^\mu p_4^\nu\epsilon_{\mu_1\mu_4\mu\nu}}{\hat{g}'^2+\hat{g}^2}\left(s_\beta^2\left(2\hat{g}\hat{g}'\left(\hat{C}^{(11)}_{\Phi\tilde{W}}-\hat{C}^{(11)}_{\Phi\tilde{B}}\right)+\left(\hat{g}'^2-\hat{g}^2\right)\hat{C}^{(11)}_{\Phi B\tilde{W}}\right)\right.\\
&\qquad\left.+c_\beta^2\left(2\hat{g}\hat{g}'\left(\hat{C}^{(22)}_{\Phi\tilde{W}}-\hat{C}^{(22)}_{\Phi\tilde{B}}\right)+\left(\hat{g}'^2-\hat{g}^2\right)\hat{C}^{(22)}_{\Phi B\tilde{W}}\right)\right)
\end{aligned}
\tag{C.546}
$$

$A_{\mu_2}$, $A_{\mu_1}$, $A^0$, $G^0$

$$
\begin{aligned}
&-\frac{4is_\beta c_\beta\left(p_{1\mu_2}p_{2\mu_1}-p_1\cdot p_2 g_{\mu_1\mu_2}\right)}{\hat{g}'^2+\hat{g}^2}\left(\hat{g}\left(-\hat{g}'\hat{C}^{(11)}_{\Phi WB}+\hat{g}\hat{C}^{(11)}_{\Phi B}\right.\right.\\
&\qquad\left.+\hat{g}'\hat{C}^{(22)}_{\Phi WB}-\hat{g}\hat{C}^{(22)}_{\Phi B}\right)\\
&\qquad\left.+\hat{g}'^2\hat{C}^{(11)}_{\Phi W}-\hat{g}'^2\hat{C}^{(22)}_{\Phi W}\right)\\
&-\frac{4is_\beta c_\beta p_1^\mu p_2^\nu\epsilon_{\mu_1\mu_2\mu\nu}}{\hat{g}'^2+\hat{g}^2}\left(\hat{g}\left(-\hat{g}'\hat{C}^{(11)}_{\Phi B\tilde{W}}+\hat{g}\hat{C}^{(11)}_{\Phi\tilde{B}}\right.\right.\\
&\qquad\left.+\hat{g}'\hat{C}^{(22)}_{\Phi B\tilde{W}}-\hat{g}\hat{C}^{(22)}_{\Phi\tilde{B}}\right)\\
&\qquad\left.+\hat{g}'^2\hat{C}^{(11)}_{\Phi\tilde{W}}-\hat{g}'^2\hat{C}^{(22)}_{\Phi\tilde{W}}\right)
\end{aligned}
\tag{C.547}
$$

$A_{\mu_2}$, $A_{\mu_1}$, $h$, $H$

$$
\begin{aligned}
&+\frac{4is_\beta c_\beta\left(p_{1\mu_2}p_{2\mu_1}-p_1\cdot p_2 g_{\mu_1\mu_2}\right)}{\hat{g}'^2+\hat{g}^2}\left(\hat{g}\left(-\hat{g}'\hat{C}^{(11)}_{\Phi WB}+\hat{g}\hat{C}^{(11)}_{\Phi B}\right.\right.\\
&\qquad\left.+\hat{g}'\hat{C}^{(22)}_{\Phi WB}-\hat{g}\hat{C}^{(22)}_{\Phi B}\right)\\
&\qquad\left.+\hat{g}'^2\hat{C}^{(11)}_{\Phi W}-\hat{g}'^2\hat{C}^{(22)}_{\Phi W}\right)\\
&+\frac{4is_\beta c_\beta p_1^\mu p_2^\nu\epsilon_{\mu_1\mu_2\mu\nu}}{\hat{g}'^2+\hat{g}^2}\left(\hat{g}\left(-\hat{g}'\hat{C}^{(11)}_{\Phi B\tilde{W}}+\hat{g}\hat{C}^{(11)}_{\Phi\tilde{B}}\right.\right.\\
&\qquad\left.+\hat{g}'\hat{C}^{(22)}_{\Phi B\tilde{W}}-\hat{g}\hat{C}^{(22)}_{\Phi\tilde{B}}\right)\\
&\qquad\left.+\hat{g}'^2\hat{C}^{(11)}_{\Phi\tilde{W}}-\hat{g}'^2\hat{C}^{(22)}_{\Phi\tilde{W}}\right)
\end{aligned}
\tag{C.548}
$$

$A_{\mu_2}$, $A_{\mu_1}$, $G^-$, $H^+$

$$-\frac{4is_\beta c_\beta \left(p_{1\mu_2}p_{2\mu_1} - p_1 \cdot p_2 g_{\mu_1\mu_2}\right)}{\hat{g}'^2+\hat{g}^2}\left(\hat{g}\left(\hat{g}'\hat{C}^{(11)}_{\Phi WB} + \hat{g}\hat{C}^{(11)}_{\Phi B} - \hat{g}'\hat{C}^{(22)}_{\Phi WB} - \hat{g}\hat{C}^{(22)}_{\Phi B}\right) + \hat{g}'^2\hat{C}^{(11)}_{\Phi W} - \hat{g}'^2\hat{C}^{(22)}_{\Phi W}\right) - \frac{4is_\beta c_\beta p_1^\mu p_2^\nu \epsilon_{\mu_1\mu_2\mu\nu}}{\hat{g}'^2+\hat{g}^2}\left(\hat{g}\left(\hat{g}'\hat{C}^{(11)}_{\Phi B\tilde{W}} + \hat{g}\hat{C}^{(11)}_{\Phi\tilde{B}} - \hat{g}'\hat{C}^{(22)}_{\Phi B\tilde{W}} - \hat{g}\hat{C}^{(22)}_{\Phi\tilde{B}}\right) + \hat{g}'^2\hat{C}^{(11)}_{\Phi\tilde{W}} - \hat{g}'^2\hat{C}^{(22)}_{\Phi\tilde{W}}\right) \tag{C.549}$$

$A^0$, $A_{\mu_1}$, $G^-$, $W^+_{\mu_4}$

$$+\frac{\hat{g}^2 g_{\mu_1\mu_4}\hat{g}'}{4\sqrt{\hat{g}'^2+\hat{g}^2}}\left(s_\beta c_\beta\left(2\left(\delta_{s_{\hat{\beta}\pm}} - \delta_{c_{\hat{\beta}\pm}} + \delta_{c_{\hat{\beta}}} - \delta_{s_{\hat{\beta}}}\right) - A'_1 + A'_2\right) - s_\beta^2 B' + c_\beta^2 B'\right) + \frac{\hat{g}s_{2\beta}\left(p_{1\mu_4}p_{4\mu_1} - p_1 \cdot p_4 g_{\mu_1\mu_4}\right)}{\sqrt{\hat{g}'^2+\hat{g}^2}}\left(\hat{C}^{(11)}_{\Phi WB} - \hat{C}^{(22)}_{\Phi WB}\right) + \frac{\hat{g}s_{2\beta}p_1^\mu p_4^\nu \epsilon_{\mu_1\mu_4\mu\nu}}{\sqrt{\hat{g}'^2+\hat{g}^2}}\left(\hat{C}^{(11)}_{\Phi B\tilde{W}} - \hat{C}^{(22)}_{\Phi B\tilde{W}}\right) \tag{C.550}$$

$G^-$, $A_{\mu_1}$, $H$, $W^+_{\mu_4}$

$$+\frac{i\hat{g}^2 g_{\mu_1\mu_4}\hat{g}'}{4\sqrt{\hat{g}'^2+\hat{g}^2}}\left(s_\beta c_\beta\left(2\delta_{s_{\hat{\beta}\pm}} - A_1 + A_2 - 2\delta_{c_{\hat{\beta}\pm}}\right) - Bs_\beta^2 + Bc_\beta^2\right) + \frac{2i\hat{g}s_\beta c_\beta\left(p_{1\mu_4}p_{4\mu_1} - p_1 \cdot p_4 g_{\mu_1\mu_4}\right)}{\sqrt{\hat{g}'^2+\hat{g}^2}}\left(\hat{C}^{(11)}_{\Phi WB} - \hat{C}^{(22)}_{\Phi WB}\right) + \frac{2i\hat{g}s_\beta c_\beta p_1^\mu p_4^\nu \epsilon_{\mu_1\mu_4\mu\nu}}{\sqrt{\hat{g}'^2+\hat{g}^2}}\left(\hat{C}^{(11)}_{\Phi B\tilde{W}} - \hat{C}^{(22)}_{\Phi B\tilde{W}}\right) \tag{C.551}$$

$G^0$, $A_{\mu_1}$, $H^-$, $W^+_{\mu_4}$

$$
\begin{aligned}
&+\frac{\hat{g}^2 g_{\mu_1\mu_4}\hat{g}'}{4\sqrt{\hat{g}'^2+\hat{g}^2}}\left(s_\beta c_\beta\left(2\left(-\delta_{s_{\hat{\beta}\pm}}+\delta_{c_{\hat{\beta}\pm}}-\delta_{c_{\hat{\beta}}}+\delta_{s_{\hat{\beta}}}\right)-A'_1+A'_2\right)\right.\\
&\qquad\qquad\left.-s_\beta^2 B'+c_\beta^2 B'\right)\\
&+\frac{\hat{g}s_{2\beta}\left(p_{1\mu_4}p_{4\mu_1}-p_1\cdot p_4 g_{\mu_1\mu_4}\right)}{\sqrt{\hat{g}'^2+\hat{g}^2}}\left(\hat{C}^{(11)}_{\Phi WB}-\hat{C}^{(22)}_{\Phi WB}\right)\\
&+\frac{\hat{g}s_{2\beta}p_1^\mu p_4^\nu\epsilon_{\mu_1\mu_4\mu\nu}}{\sqrt{\hat{g}'^2+\hat{g}^2}}\left(\hat{C}^{(11)}_{\Phi B\tilde{W}}-\hat{C}^{(22)}_{\Phi B\tilde{W}}\right)
\end{aligned}
\tag{C.552}
$$

$h$, $A_{\mu_1}$, $H^-$, $W^+_{\mu_4}$

$$
\begin{aligned}
&-\frac{i\hat{g}^2 g_{\mu_1\mu_4}\hat{g}'}{4\sqrt{\hat{g}'^2+\hat{g}^2}}\left(s_\beta c_\beta\left(-2\delta_{s_{\hat{\beta}\pm}}-A_1+A_2+2\delta_{c_{\hat{\beta}\pm}}\right)-Bs_\beta^2+Bc_\beta^2\right)\\
&-\frac{2i\hat{g}s_\beta c_\beta\left(p_{1\mu_4}p_{4\mu_1}-p_1\cdot p_4 g_{\mu_1\mu_4}\right)}{\sqrt{\hat{g}'^2+\hat{g}^2}}\left(\hat{C}^{(11)}_{\Phi WB}-\hat{C}^{(22)}_{\Phi WB}\right)\\
&-\frac{2i\hat{g}s_\beta c_\beta p_1^\mu p_4^\nu\epsilon_{\mu_1\mu_4\mu\nu}}{\sqrt{\hat{g}'^2+\hat{g}^2}}\left(\hat{C}^{(11)}_{\Phi B\tilde{W}}-\hat{C}^{(22)}_{\Phi B\tilde{W}}\right)
\end{aligned}
\tag{C.553}
$$

$G^0$, $A^0$, $W^+_{\mu_3}$, $W^-_{\mu_4}$

$$
\begin{aligned}
&-\frac{1}{2}i\hat{g}^2 g_{\mu_3\mu_4}\left(\left(A'_2-A'_1\right)s_\beta c_\beta-s_\beta^2 B'+c_\beta^2 B'\right)\\
&-4is_\beta c_\beta\left(p_{3\mu_4}p_{4\mu_3}-p_3\cdot p_4 g_{\mu_3\mu_4}\right)\left(\hat{C}^{(11)}_{\Phi W}-\hat{C}^{(22)}_{\Phi W}\right)\\
&-4is_\beta c_\beta p_3^\mu p_4^\nu\epsilon_{\mu_3\mu_4\mu\nu}\left(\hat{C}^{(11)}_{\Phi\tilde{W}}-\hat{C}^{(22)}_{\Phi\tilde{W}}\right)
\end{aligned}
\tag{C.554}
$$

$$
\begin{aligned}
&+\frac{1}{2} i\hat{g}^2 g_{\mu_3\mu_4}\left(\left(A_2-A_1\right) s_\beta c_\beta - B s_\beta^2 + B c_\beta^2\right)\\
&+4 i s_\beta c_\beta\left(p_{3\mu_4} p_{4\mu_3} - p_3\cdot p_4 g_{\mu_3\mu_4}\right)\left(\hat{C}_{\Phi W}^{(11)}-\hat{C}_{\Phi W}^{(22)}\right)\\
&+4 i s_\beta c_\beta p_3^\mu p_4^\nu \epsilon_{\mu_3\mu_4\mu\nu}\left(\hat{C}_{\Phi\tilde{W}}^{(11)}-\hat{C}_{\Phi\tilde{W}}^{(22)}\right)
\end{aligned}
\tag{C.555}
$$

$$
\begin{aligned}
&-\frac{2 i s_\beta c_\beta\left(p_{1\mu_4} p_{4\mu_1} - p_1\cdot p_4 g_{\mu_1\mu_4}\right)}{\hat{g}'^2+\hat{g}^2}\left(2\hat{g}\hat{g}'\left(-\hat{C}_{\Phi B}^{(11)}+\hat{C}_{\Phi W}^{(11)}\right.\right.\\
&\qquad\left.+\hat{C}_{\Phi B}^{(22)}-\hat{C}_{\Phi W}^{(22)}\right)\\
&\qquad\left.+\left(\hat{g}'^2-\hat{g}^2\right)\left(\hat{C}_{\Phi WB}^{(11)}-\hat{C}_{\Phi WB}^{(22)}\right)\right)\\
&-\frac{2 i s_\beta c_\beta p_1^\mu p_4^\nu \epsilon_{\mu_1\mu_4\mu\nu}}{\hat{g}'^2+\hat{g}^2}\left(2\hat{g}\hat{g}'\left(-\hat{C}_{\Phi\tilde{B}}^{(11)}+\hat{C}_{\Phi\tilde{W}}^{(11)}\right.\right.\\
&\qquad\left.+\hat{C}_{\Phi\tilde{B}}^{(22)}-\hat{C}_{\Phi\tilde{W}}^{(22)}\right)\\
&\qquad\left.+\left(\hat{g}'^2-\hat{g}^2\right)\left(\hat{C}_{\Phi B\tilde{W}}^{(11)}-\hat{C}_{\Phi B\tilde{W}}^{(22)}\right)\right)
\end{aligned}
\tag{C.556}
$$

$$
\begin{aligned}
&+\frac{2 i s_\beta c_\beta\left(p_{1\mu_4} p_{4\mu_1} - p_1\cdot p_4 g_{\mu_1\mu_4}\right)}{\hat{g}'^2+\hat{g}^2}\left(2\hat{g}\hat{g}'\left(-\hat{C}_{\Phi B}^{(11)}+\hat{C}_{\Phi W}^{(11)}\right.\right.\\
&\qquad\left.+\hat{C}_{\Phi B}^{(22)}-\hat{C}_{\Phi W}^{(22)}\right)\\
&\qquad\left.+\left(\hat{g}'^2-\hat{g}^2\right)\left(\hat{C}_{\Phi WB}^{(11)}-\hat{C}_{\Phi WB}^{(22)}\right)\right)\\
&+\frac{2 i s_\beta c_\beta p_1^\mu p_4^\nu \epsilon_{\mu_1\mu_4\mu\nu}}{\hat{g}'^2+\hat{g}^2}\left(2\hat{g}\hat{g}'\left(-\hat{C}_{\Phi\tilde{B}}^{(11)}+\hat{C}_{\Phi\tilde{W}}^{(11)}\right.\right.\\
&\qquad\left.+\hat{C}_{\Phi\tilde{B}}^{(22)}-\hat{C}_{\Phi\tilde{W}}^{(22)}\right)\\
&\qquad\left.+\left(\hat{g}'^2-\hat{g}^2\right)\left(\hat{C}_{\Phi B\tilde{W}}^{(11)}-\hat{C}_{\Phi B\tilde{W}}^{(22)}\right)\right)
\end{aligned}
\tag{C.557}
$$

$A_{\mu_2}$ $A_{\mu_1}$ $G^-$ $W^+_{\mu_4}$

$$
\begin{aligned}
&-\frac{2i\sqrt{2}\hat{g}^2 v\hat{g}'\left(g_{\mu_1\mu_2}p_{1\mu_4}+g_{\mu_1\mu_2}p_{2\mu_4}-g_{\mu_1\mu_4}p_{1\mu_2}-g_{\mu_2\mu_4}p_{2\mu_1}\right)}{\hat{g}'^2+\hat{g}^2} \\
&\times\left(c_\beta^2\hat{C}^{(11)}_{\Phi WB}+s_\beta^2\hat{C}^{(22)}_{\Phi WB}\right) \\
&-\frac{2i\sqrt{2}\hat{g}^2 v\hat{g}'\left(p_1^\mu-p_2^\mu\right)\epsilon_{\mu_1\mu_2\mu_4\mu}}{\hat{g}'^2+\hat{g}^2}\left(c_\beta^2\hat{C}^{(11)}_{\Phi B\tilde{W}}+s_\beta^2\hat{C}^{(22)}_{\Phi B\tilde{W}}\right)
\end{aligned}
\tag{C.558}
$$

$h$ $A_{\mu_1}$ $W^+_{\mu_3}$ $W^-_{\mu_4}$

$$
\begin{aligned}
&-\frac{2i\sqrt{2}\hat{g}v}{\sqrt{\hat{g}'^2+\hat{g}^2}}\Bigg(g_{\mu_1\mu_3}p_{1\mu_4}\left(c_\beta^2\left(2\hat{g}'\hat{C}^{(11)}_{\Phi W}-\hat{g}\hat{C}^{(11)}_{\Phi WB}\right)+s_\beta^2\left(2\hat{g}'\hat{C}^{(22)}_{\Phi W}-\hat{g}\hat{C}^{(22)}_{\Phi WB}\right)\right) \\
&\qquad -2g_{\mu_1\mu_3}\hat{g}'p_{3\mu_4}\left(c_\beta^2\hat{C}^{(11)}_{\Phi W}+s_\beta^2\hat{C}^{(22)}_{\Phi W}\right) \\
&\qquad -2g_{\mu_1\mu_4}\hat{g}'\left(p_{1\mu_3}-p_{4\mu_3}\right)\left(c_\beta^2\hat{C}^{(11)}_{\Phi W}+s_\beta^2\hat{C}^{(22)}_{\Phi W}\right) \\
&\qquad +2g_{\mu_3\mu_4}\hat{g}'\left(p_{3\mu_1}-p_{4\mu_1}\right)\left(c_\beta^2\hat{C}^{(11)}_{\Phi W}+s_\beta^2\hat{C}^{(11)}_{\Phi W}\right) \\
&\qquad +\hat{g}g_{\mu_1\mu_4}p_{1\mu_3}\left(c_\beta^2\hat{C}^{(11)}_{\Phi WB}+c_\beta^2\hat{C}^{(22)}_{\Phi WB}\right)\Bigg) \\
&+\frac{2i\sqrt{2}\hat{g}v\epsilon^\mu_{\mu_1\mu_4\mu_3}}{\sqrt{\hat{g}'^2+\hat{g}^2}}\Bigg(p_1^\mu\left(c_\beta^2\left(2\hat{g}'\hat{C}^{(11)}_{\Phi\tilde{W}}-\hat{g}\hat{C}^{(11)}_{\Phi B\tilde{W}}\right)+s_\beta^2\left(2\hat{g}'\hat{C}^{(22)}_{\Phi\tilde{W}}-\hat{g}\hat{C}^{(22)}_{\Phi B\tilde{W}}\right)\right) \\
&\qquad +2\hat{g}'\left(p_3^\mu+p_4^\mu\right)\left(c_\beta^2\hat{C}^{(11)}_{\Phi\tilde{W}}+s_\beta^2\hat{C}^{(22)}_{\Phi\tilde{W}}\right)\Bigg)
\end{aligned}
\tag{C.559}
$$

$G^-$ $A_{\mu_1}$ $W^+_{\mu_3}$ $Z_{\mu_4}$

$$
\begin{aligned}
&+\frac{2i\sqrt{2}\hat{g}v\left(\hat{g}^2 g_{\mu_1\mu_3}p_{1\mu_4}-\hat{g}^2 g_{\mu_1\mu_4}p_{1\mu_3}+\hat{g}'^2\left(g_{\mu_1\mu_4}p_{4\mu_3}-g_{\mu_3\mu_4}p_{4\mu_1}\right)\right)}{\hat{g}'^2+\hat{g}^2} \\
&\times\left(c_\beta^2\hat{C}^{(11)}_{\Phi WB}+s_\beta^2\hat{C}^{(22)}_{\Phi WB}\right) \\
&-\frac{2i\sqrt{2}\hat{g}v\epsilon^\mu_{\mu_1\mu_4\mu_3}\left(\hat{g}'^2 p_4^\mu+\hat{g}^2 p_1^\mu\right)}{\hat{g}'^2+\hat{g}^2}\left(c_\beta^2\hat{C}^{(11)}_{\Phi B\tilde{W}}+s_\beta^2\hat{C}^{(22)}_{\Phi B\tilde{W}}\right)
\end{aligned}
\tag{C.560}
$$

$A_{\mu_1}$, $A_{\mu_2}$, $H^-$, $W^+_{\mu_4}$

$$
\begin{aligned}
&+\frac{2i\sqrt{2}\hat{g}^2 v s_\beta c_\beta \hat{g}' \left(g_{\mu_1\mu_2}p_{1\mu_4} + g_{\mu_1\mu_2}p_{2\mu_4} - g_{\mu_1\mu_4}p_{1\mu_2} - g_{\mu_2\mu_4}p_{2\mu_1}\right)}{\hat{g}'^2+\hat{g}^2} \\
&\times \left(\hat{C}^{(11)}_{\Phi WB} - \hat{C}^{(22)}_{\Phi WB}\right) \\
&+\frac{2i\sqrt{2}\hat{g}^2 v s_\beta c_\beta \hat{g}' \left(p_1^\mu - p_2^\mu\right)\epsilon_{\mu_1\mu_2\mu_4\mu}}{\hat{g}'^2+\hat{g}^2}\left(\hat{C}^{(11)}_{\Phi B\tilde{W}} - \hat{C}^{(22)}_{\Phi B\tilde{W}}\right)
\end{aligned}
\tag{C.561}
$$

$A_{\mu_1}$, $H$, $W^+_{\mu_3}$, $W^-_{\mu_4}$

$$
\begin{aligned}
&+\frac{2i\sqrt{2}\hat{g} v s_\beta c_\beta}{\sqrt{\hat{g}'^2+\hat{g}^2}}\Big(g_{\mu_1\mu_3}p_{1\mu_4}\Big(-2\hat{g}'\hat{C}^{(11)}_{\Phi W} + \hat{g}\hat{C}^{(11)}_{\Phi WB} \\
&\qquad +2\hat{g}'\hat{C}^{(22)}_{\Phi W} - \hat{g}\hat{C}^{(22)}_{\Phi WB}\Big) \\
&\qquad +2g_{\mu_1\mu_3}\hat{g}'p_{3\mu_4}\left(\hat{C}^{(11)}_{\Phi W} - \hat{C}^{(22)}_{\Phi W}\right) \\
&\qquad +2g_{\mu_1\mu_4}\hat{g}'\left(p_{1\mu_3} - p_{4\mu_3}\right)\left(\hat{C}^{(11)}_{\Phi W} - \hat{C}^{(22)}_{\Phi W}\right) \\
&\qquad -2g_{\mu_3\mu_4}\hat{g}'\left(p_{3\mu_1} - p_{4\mu_1}\right)\left(\hat{C}^{(11)}_{\Phi W} - \hat{C}^{(22)}_{\Phi W}\right) \\
&\qquad -\hat{g}g_{\mu_1\mu_4}p_{1\mu_3}\left(\hat{C}^{(11)}_{\Phi WB} - \hat{C}^{(22)}_{\Phi WB}\right)\Big) \\
&+\frac{2i\sqrt{2}\hat{g} v s_\beta c_\beta \epsilon^{\mu}_{\mu_1\mu_4\mu_3}}{\sqrt{\hat{g}'^2+\hat{g}^2}}\Big(p_1^\mu\Big(-\hat{g}\hat{C}^{(11)}_{\Phi B\tilde{W}} + 2\hat{g}'\hat{C}^{(11)}_{\Phi\tilde{W}} \\
&\qquad +\hat{g}\hat{C}^{(22)}_{\Phi B\tilde{W}} - 2\hat{g}'\hat{C}^{(22)}_{\Phi\tilde{W}}\Big) \\
&\qquad +2\hat{g}'\left(p_3^\mu + p_4^\mu\right)\left(\hat{C}^{(11)}_{\Phi\tilde{W}} - \hat{C}^{(22)}_{\Phi\tilde{W}}\right)\Big)
\end{aligned}
\tag{C.562}
$$

$A_{\mu_1}$, $H^-$, $W^+_{\mu_3}$, $Z_{\mu_4}$

$$
\begin{aligned}
&-\frac{2i\sqrt{2}\hat{g} v s_\beta c_\beta \left(\hat{g}^2 g_{\mu_1\mu_3}p_{1\mu_4} - \hat{g}^2 g_{\mu_1\mu_4}p_{1\mu_3} + \hat{g}'^2\left(g_{\mu_1\mu_4}p_{4\mu_3} - g_{\mu_3\mu_4}p_{4\mu_1}\right)\right)}{\hat{g}'^2+\hat{g}^2} \\
&\times \left(\hat{C}^{(11)}_{\Phi WB} - \hat{C}^{(22)}_{\Phi WB}\right) \\
&+\frac{2i\sqrt{2}\hat{g} v s_\beta c_\beta \epsilon^{\mu}_{\mu_1\mu_4\mu_3}\left(\hat{g}'^2 p_4^\mu + \hat{g}^2 p_1^\mu\right)}{\hat{g}'^2+\hat{g}^2}\left(\hat{C}^{(11)}_{\Phi B\tilde{W}} - \hat{C}^{(22)}_{\Phi B\tilde{W}}\right)
\end{aligned}
\tag{C.563}
$$

$$
\begin{aligned}
&+\frac{2i\sqrt{2}\hat{g}v}{\sqrt{\hat{g}'^2+\hat{g}^2}}\Big(-2\hat{g}g_{\mu_2\mu_3}\left(p_{2\mu_4}-p_{3\mu_4}\right)\Big(c_\beta^2\hat{C}^{(11)}_{\Phi W}+s_\beta^2\hat{C}^{(22)}_{\Phi W}\Big)\\
&\qquad+2\hat{g}g_{\mu_2\mu_4}\left(p_{2\mu_3}-p_{4\mu_3}\right)\Big(c_\beta^2\hat{C}^{(11)}_{\Phi W}+s_\beta^2\hat{C}^{(22)}_{\Phi W}\Big)\\
&\qquad-2\hat{g}g_{\mu_3\mu_4}\left(p_{3\mu_2}-p_{4\mu_2}\right)\Big(c_\beta^2\hat{C}^{(11)}_{\Phi W}+s_\beta^2\hat{C}^{(22)}_{\Phi W}\Big)\\
&\qquad-g_{\mu_2\mu_4}\hat{g}'p_{4\mu_3}\Big(c_\beta^2\hat{C}^{(11)}_{\Phi WB}+s_\beta^2\hat{C}^{(22)}_{\Phi WB}\Big)\\
&\qquad+g_{\mu_3\mu_4}\hat{g}'p_{4\mu_2}\Big(c_\beta^2\hat{C}^{(11)}_{\Phi WB}+s_\beta^2\hat{C}^{(22)}_{\Phi WB}\Big)\Big)\\
&-\frac{2i\sqrt{2}\hat{g}v\epsilon^{\mu}_{\mu_4\mu_2\mu_3}}{\sqrt{\hat{g}'^2+\hat{g}^2}}\Big(p_4^\mu\Big(c_\beta^2\Big(\hat{g}'\hat{C}^{(11)}_{\Phi B\tilde{W}}+2\hat{g}\hat{C}^{(11)}_{\Phi\tilde{W}}\Big)\\
&\qquad+s_\beta^2\Big(\hat{g}'\hat{C}^{(22)}_{\Phi B\tilde{W}}+2\hat{g}\hat{C}^{(22)}_{\Phi\tilde{W}}\Big)\Big)\\
&\qquad+2\hat{g}p_2^\mu\Big(c_\beta^2\hat{C}^{(11)}_{\Phi\tilde{W}}+s_\beta^2\hat{C}^{(22)}_{\Phi\tilde{W}}\Big)\\
&\qquad+2\hat{g}p_3^\mu\Big(c_\beta^2\hat{C}^{(11)}_{\Phi\tilde{W}}+s_\beta^2\hat{C}^{(22)}_{\Phi\tilde{W}}\Big)\Big)
\end{aligned}
\tag{C.564}
$$

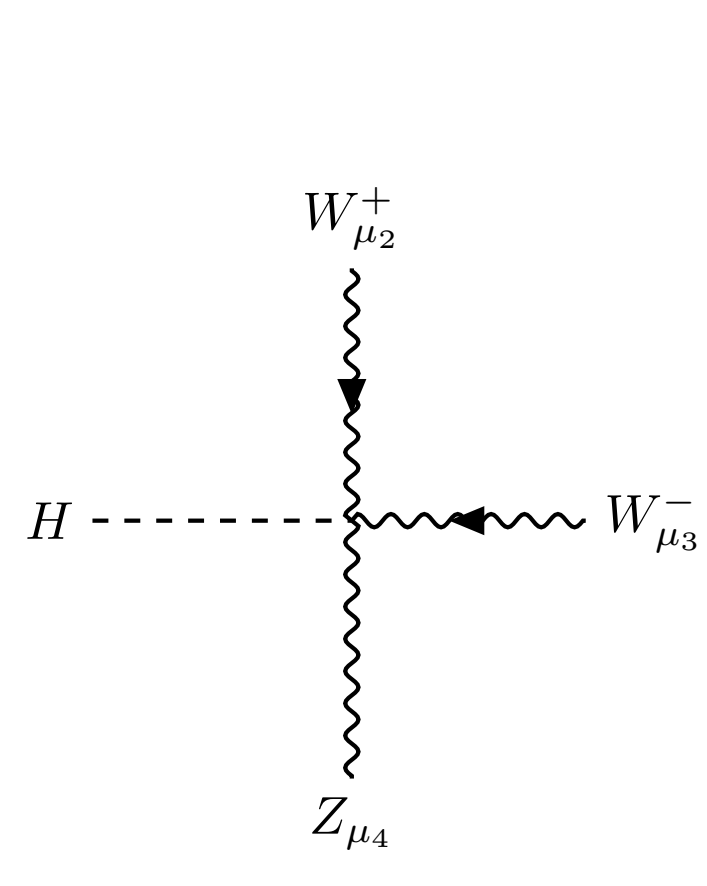

$$
\begin{aligned}
&-\frac{2i\sqrt{2}\hat{g}vs_\beta c_\beta}{\sqrt{\hat{g}'^2+\hat{g}^2}}\Big(2\hat{g}g_{\mu_2\mu_3}p_{2\mu_4}\Big(\hat{C}^{(11)}_{\Phi W}-\hat{C}^{(22)}_{\Phi W}\Big)\\
&\qquad+2\hat{g}g_{\mu_2\mu_3}p_{3\mu_4}\Big(\hat{C}^{(22)}_{\Phi W}-\hat{C}^{(11)}_{\Phi W}\Big)\\
&\qquad-2\hat{g}g_{\mu_2\mu_4}\left(p_{2\mu_3}-p_{4\mu_3}\right)\Big(\hat{C}^{(11)}_{\Phi W}-\hat{C}^{(22)}_{\Phi W}\Big)\\
&\qquad+2\hat{g}g_{\mu_3\mu_4}\left(p_{3\mu_2}-p_{4\mu_2}\right)\Big(\hat{C}^{(11)}_{\Phi W}-\hat{C}^{(22)}_{\Phi W}\Big)\\
&\qquad+g_{\mu_2\mu_4}\hat{g}'p_{4\mu_3}\Big(\hat{C}^{(11)}_{\Phi WB}-\hat{C}^{(22)}_{\Phi WB}\Big)\\
&\qquad-g_{\mu_3\mu_4}\hat{g}'p_{4\mu_2}\Big(\hat{C}^{(11)}_{\Phi WB}-\hat{C}^{(22)}_{\Phi WB}\Big)\Big)\\
&-\frac{2i\sqrt{2}\hat{g}vs_\beta c_\beta\epsilon^{\mu}_{\mu_4\mu_2\mu_3}}{\sqrt{\hat{g}'^2+\hat{g}^2}}\Big(p_4^\mu\Big(\hat{g}'\hat{C}^{(11)}_{\Phi B\tilde{W}}+2\hat{g}\hat{C}^{(11)}_{\Phi\tilde{W}}\\
&\qquad-\hat{g}'\hat{C}^{(22)}_{\Phi B\tilde{W}}-2\hat{g}\hat{C}^{(22)}_{\Phi\tilde{W}}\Big)\\
&\qquad+2\hat{g}p_2^\mu\Big(\hat{C}^{(11)}_{\Phi\tilde{W}}-\hat{C}^{(22)}_{\Phi\tilde{W}}\Big)\\
&\qquad+2\hat{g}p_3^\mu\Big(\hat{C}^{(11)}_{\Phi\tilde{W}}-\hat{C}^{(22)}_{\Phi\tilde{W}}\Big)\Big)
\end{aligned}
\tag{C.565}
$$

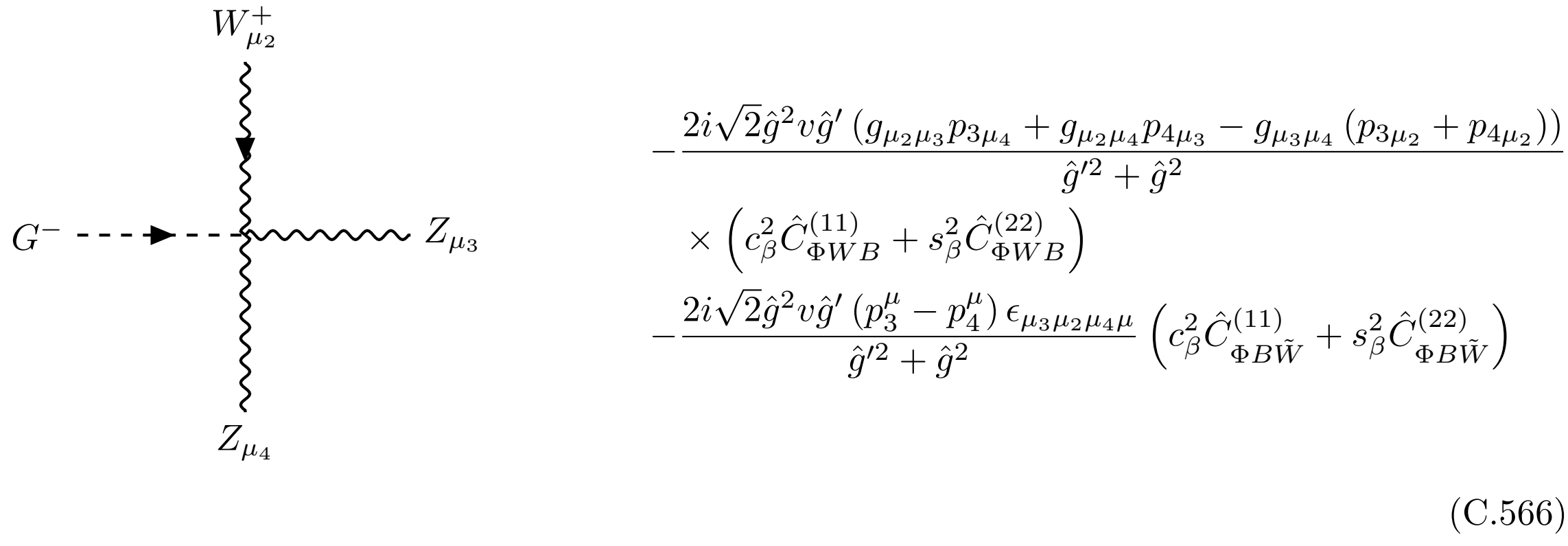

$$
\begin{aligned}
&-\frac{2i\sqrt{2}\hat{g}^2 v\hat{g}'\left(g_{\mu_2\mu_3}p_{3\mu_4}+g_{\mu_2\mu_4}p_{4\mu_3}-g_{\mu_3\mu_4}\left(p_{3\mu_2}+p_{4\mu_2}\right)\right)}{\hat{g}'^2+\hat{g}^2}\\
&\times\left(c_\beta^2\hat{C}^{(11)}_{\Phi WB}+s_\beta^2\hat{C}^{(22)}_{\Phi WB}\right)\\
&-\frac{2i\sqrt{2}\hat{g}^2 v\hat{g}'\left(p_3^\mu-p_4^\mu\right)\epsilon_{\mu_3\mu_2\mu_4\mu}}{\hat{g}'^2+\hat{g}^2}\left(c_\beta^2\hat{C}^{(11)}_{\Phi B\tilde{W}}+s_\beta^2\hat{C}^{(22)}_{\Phi B\tilde{W}}\right)
\end{aligned}
\tag{C.566}
$$

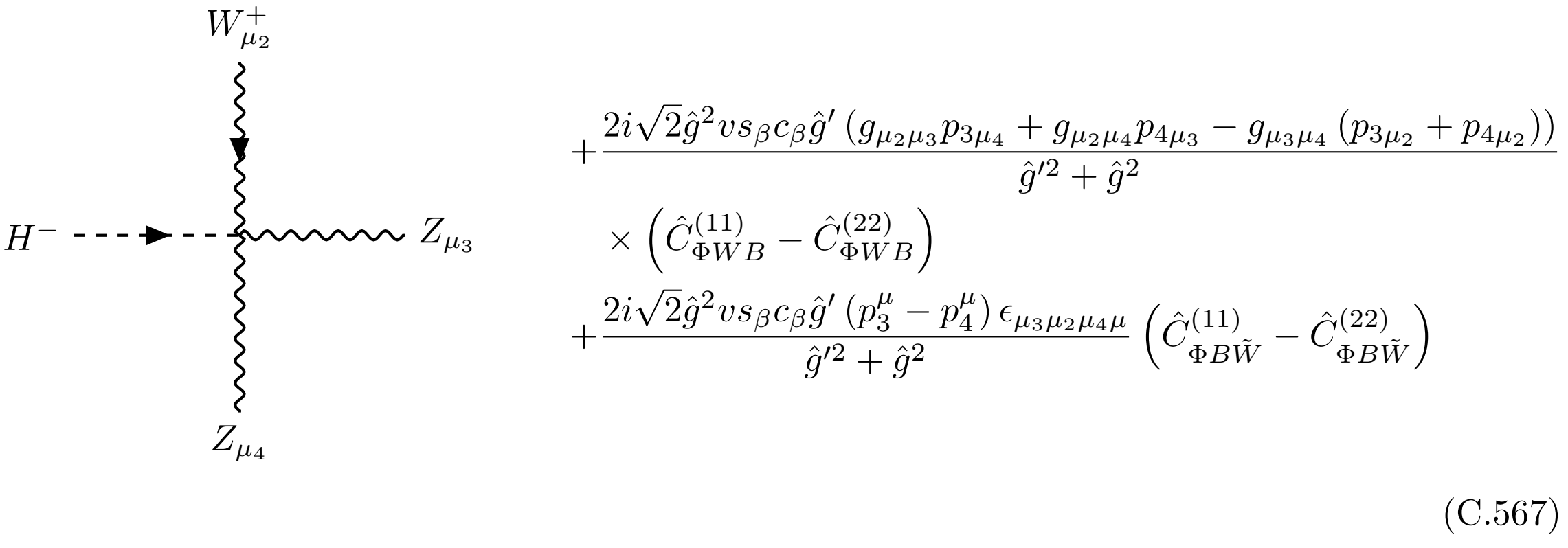

$$
\begin{aligned}
&+\frac{2i\sqrt{2}\hat{g}^2 v s_\beta c_\beta\hat{g}'\left(g_{\mu_2\mu_3}p_{3\mu_4}+g_{\mu_2\mu_4}p_{4\mu_3}-g_{\mu_3\mu_4}\left(p_{3\mu_2}+p_{4\mu_2}\right)\right)}{\hat{g}'^2+\hat{g}^2}\\
&\times\left(\hat{C}^{(11)}_{\Phi WB}-\hat{C}^{(22)}_{\Phi WB}\right)\\
&+\frac{2i\sqrt{2}\hat{g}^2 v s_\beta c_\beta\hat{g}'\left(p_3^\mu-p_4^\mu\right)\epsilon_{\mu_3\mu_2\mu_4\mu}}{\hat{g}'^2+\hat{g}^2}\left(\hat{C}^{(11)}_{\Phi B\tilde{W}}-\hat{C}^{(22)}_{\Phi B\tilde{W}}\right)
\end{aligned}
\tag{C.567}
$$

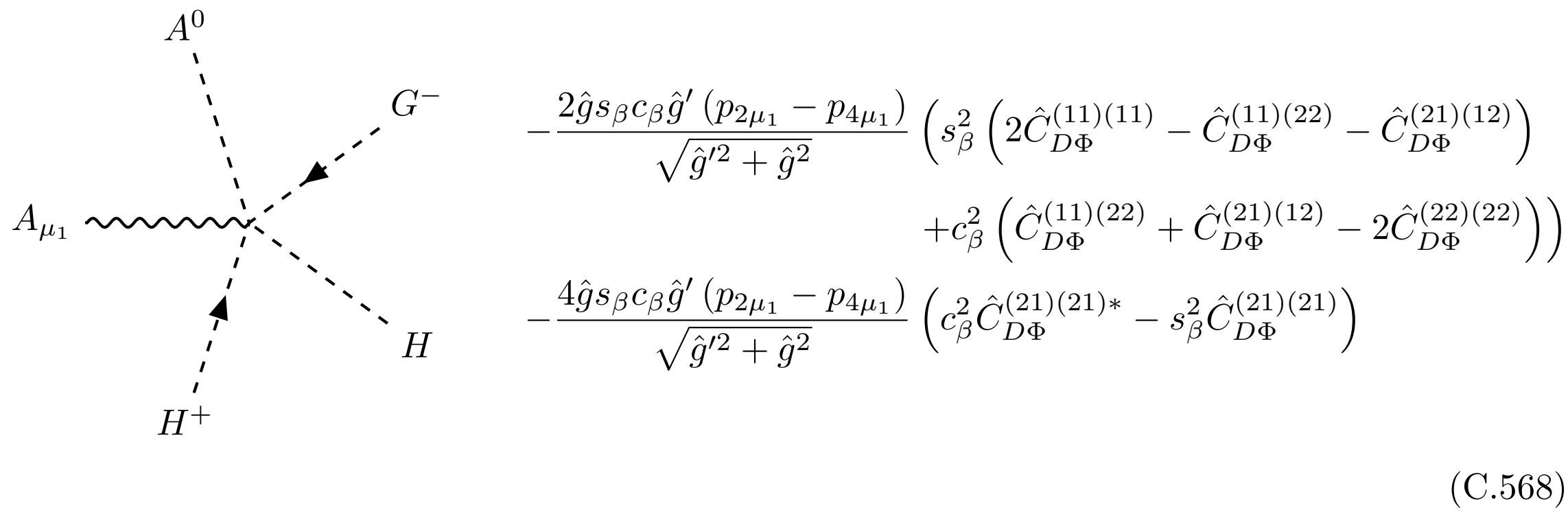

$$
\begin{aligned}
&-\frac{2\hat{g}s_\beta c_\beta\hat{g}'\left(p_{2\mu_1}-p_{4\mu_1}\right)}{\sqrt{\hat{g}'^2+\hat{g}^2}}\left(s_\beta^2\left(2\hat{C}^{(11)(11)}_{D\Phi}-\hat{C}^{(11)(22)}_{D\Phi}-\hat{C}^{(21)(12)}_{D\Phi}\right)\right.\\
&\qquad\left.+c_\beta^2\left(\hat{C}^{(11)(22)}_{D\Phi}+\hat{C}^{(21)(12)}_{D\Phi}-2\hat{C}^{(22)(22)}_{D\Phi}\right)\right)\\
&-\frac{4\hat{g}s_\beta c_\beta\hat{g}'\left(p_{2\mu_1}-p_{4\mu_1}\right)}{\sqrt{\hat{g}'^2+\hat{g}^2}}\left(c_\beta^2\hat{C}^{(21)(21)*}_{D\Phi}-s_\beta^2\hat{C}^{(21)(21)}_{D\Phi}\right)
\end{aligned}
\tag{C.568}
$$

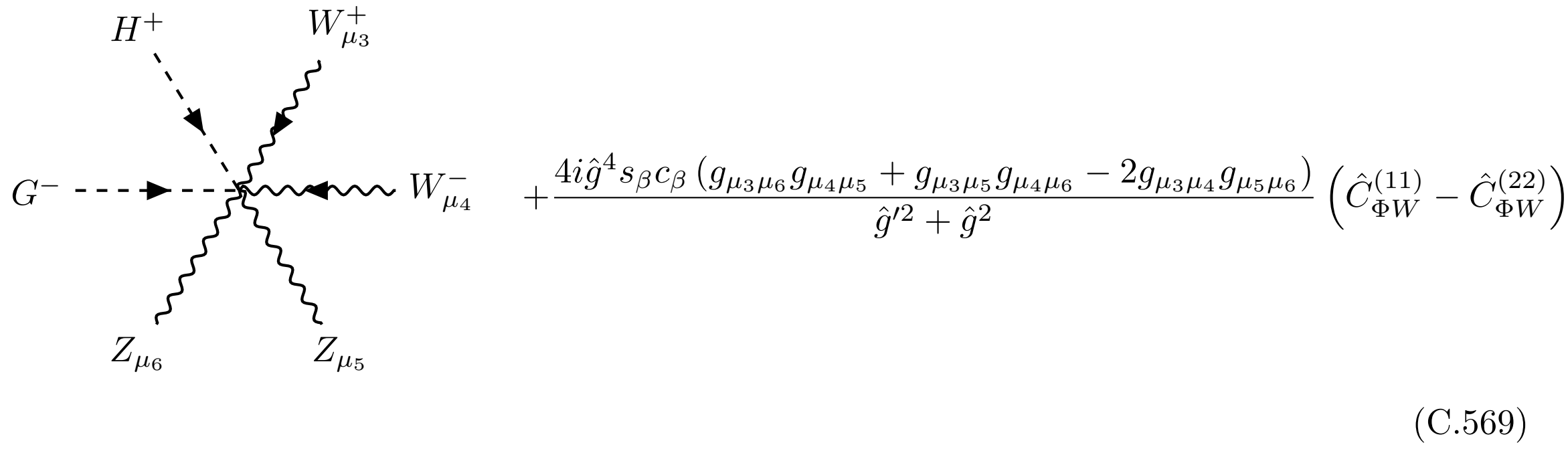

$$+\frac{4i\hat{g}^4 s_\beta c_\beta \left(g_{\mu_3\mu_6}g_{\mu_4\mu_5}+g_{\mu_3\mu_5}g_{\mu_4\mu_6}-2g_{\mu_3\mu_4}g_{\mu_5\mu_6}\right)}{\hat{g}'^2+\hat{g}^2}\left(\hat{C}^{(11)}_{\Phi W}-\hat{C}^{(22)}_{\Phi W}\right) \tag{C.569}$$

## C.4 Scalar-Gluon Interactions

This last subsection includes Feynman rules for vertices involving two gluons and either one or two scalars.

$$\begin{aligned}&+4i\sqrt{2}v\delta_{a_1a_2}\left(p_1^\lambda p_2^\rho \epsilon_{\mu_1\mu_2\lambda\rho}\left(c_\beta^2\hat{C}^{(11)}_{\Phi\tilde{G}}+s_\beta^2\hat{C}^{(22)}_{\Phi\tilde{G}}\right)\right.\\&\qquad\left.+\left(p_{1\mu_2}p_{2\mu_1}-p_1\cdot p_2 g_{\mu_1\mu_2}\right)\left(c_\beta^2\hat{C}^{(11)}_{\Phi G}+s_\beta^2\hat{C}^{(22)}_{\Phi G}\right)\right)\end{aligned} \tag{C.570}$$

$$\begin{aligned}&+4i\sqrt{2}v s_\beta c_\beta\delta_{a_1a_2}\left(p_1^\lambda p_2^\rho \epsilon_{\mu_1\mu_2\lambda\rho}\left(\hat{C}^{(11)}_{\Phi\tilde{G}}-\hat{C}^{(22)}_{\Phi\tilde{G}}\right)\right.\\&\qquad\left.+\left(\hat{C}^{(11)}_{\Phi G}-\hat{C}^{(22)}_{\Phi G}\right)\left(p_{1\mu_2}p_{2\mu_1}-p_1\cdot p_2 g_{\mu_1\mu_2}\right)\right)\end{aligned} \tag{C.571}$$

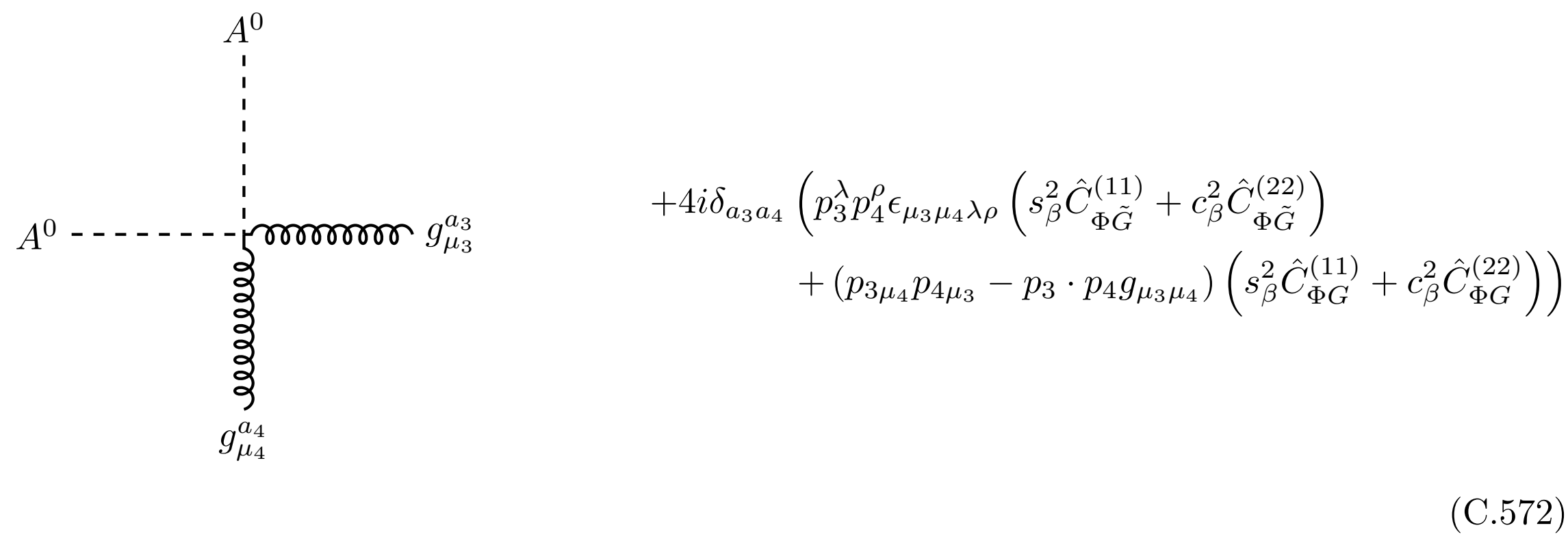

$$+4i\delta_{a_3a_4}\left(p_3^\lambda p_4^\rho \epsilon_{\mu_3\mu_4\lambda\rho}\left(s_\beta^2 \hat{C}^{(11)}_{\Phi\tilde{G}} + c_\beta^2 \hat{C}^{(22)}_{\Phi\tilde{G}}\right) + \left(p_{3\mu_4}p_{4\mu_3} - p_3\cdot p_4 g_{\mu_3\mu_4}\right)\left(s_\beta^2 \hat{C}^{(11)}_{\Phi G} + c_\beta^2 \hat{C}^{(22)}_{\Phi G}\right)\right) \tag{C.572}$$

$$+4i\delta_{a_1a_2}\left(p_1^\lambda p_2^\rho \epsilon_{\mu_1\mu_2\lambda\rho}\left(c_\beta^2 \hat{C}^{(11)}_{\Phi\tilde{G}} + s_\beta^2 \hat{C}^{(22)}_{\Phi\tilde{G}}\right) + \left(p_{1\mu_2}p_{2\mu_1} - p_1\cdot p_2 g_{\mu_1\mu_2}\right)\left(c_\beta^2 \hat{C}^{(11)}_{\Phi G} + s_\beta^2 \hat{C}^{(22)}_{\Phi G}\right)\right) \tag{C.573}$$

$$+4i\delta_{a_1a_2}\left(p_1^\lambda p_2^\rho \epsilon_{\mu_1\mu_2\lambda\rho}\left(c_\beta^2 \hat{C}^{(11)}_{\Phi\tilde{G}} + s_\beta^2 \hat{C}^{(22)}_{\Phi\tilde{G}}\right) + \left(p_{1\mu_2}p_{2\mu_1} - p_1\cdot p_2 g_{\mu_1\mu_2}\right)\left(c_\beta^2 \hat{C}^{(11)}_{\Phi G} + s_\beta^2 \hat{C}^{(22)}_{\Phi G}\right)\right) \tag{C.574}$$

$g^{a_2}_{\mu_2}$, $g^{a_1}_{\mu_1}$, $h$, $h$

$$
\begin{aligned}
&+4i\delta_{a_1a_2}\Big(p_1^\lambda p_2^\rho \epsilon_{\mu_1\mu_2\lambda\rho}\left(c_\beta^2 \hat{C}^{(11)}_{\Phi\tilde{G}} + s_\beta^2 \hat{C}^{(22)}_{\Phi\tilde{G}}\right)\\
&\qquad + \left(p_{1\mu_2}p_{2\mu_1} - p_1\cdot p_2 g_{\mu_1\mu_2}\right)\left(c_\beta^2 \hat{C}^{(11)}_{\Phi G} + s_\beta^2 \hat{C}^{(22)}_{\Phi G}\right)\Big)
\end{aligned}
\tag{C.575}
$$

$g^{a_2}_{\mu_2}$, $g^{a_1}_{\mu_1}$, $H$, $H$

$$
\begin{aligned}
&+4i\delta_{a_1a_2}\Big(p_1^\lambda p_2^\rho \epsilon_{\mu_1\mu_2\lambda\rho}\left(s_\beta^2 \hat{C}^{(11)}_{\Phi\tilde{G}} + c_\beta^2 \hat{C}^{(22)}_{\Phi\tilde{G}}\right)\\
&\qquad + \left(p_{1\mu_2}p_{2\mu_1} - p_1\cdot p_2 g_{\mu_1\mu_2}\right)\left(s_\beta^2 \hat{C}^{(11)}_{\Phi G} + c_\beta^2 \hat{C}^{(22)}_{\Phi G}\right)\Big)
\end{aligned}
\tag{C.576}
$$

$g^{a_2}_{\mu_2}$, $g^{a_1}_{\mu_1}$, $H^+$, $H^-$

$$
\begin{aligned}
&+4i\delta_{a_1a_2}\Big(p_1^\lambda p_2^\rho \epsilon_{\mu_1\mu_2\lambda\rho}\left(s_\beta^2 \hat{C}^{(11)}_{\Phi\tilde{G}} + c_\beta^2 \hat{C}^{(22)}_{\Phi\tilde{G}}\right)\\
&\qquad + \left(p_{1\mu_2}p_{2\mu_1} - p_1\cdot p_2 g_{\mu_1\mu_2}\right)\left(s_\beta^2 \hat{C}^{(11)}_{\Phi G} + c_\beta^2 \hat{C}^{(22)}_{\Phi G}\right)\Big)
\end{aligned}
\tag{C.577}
$$

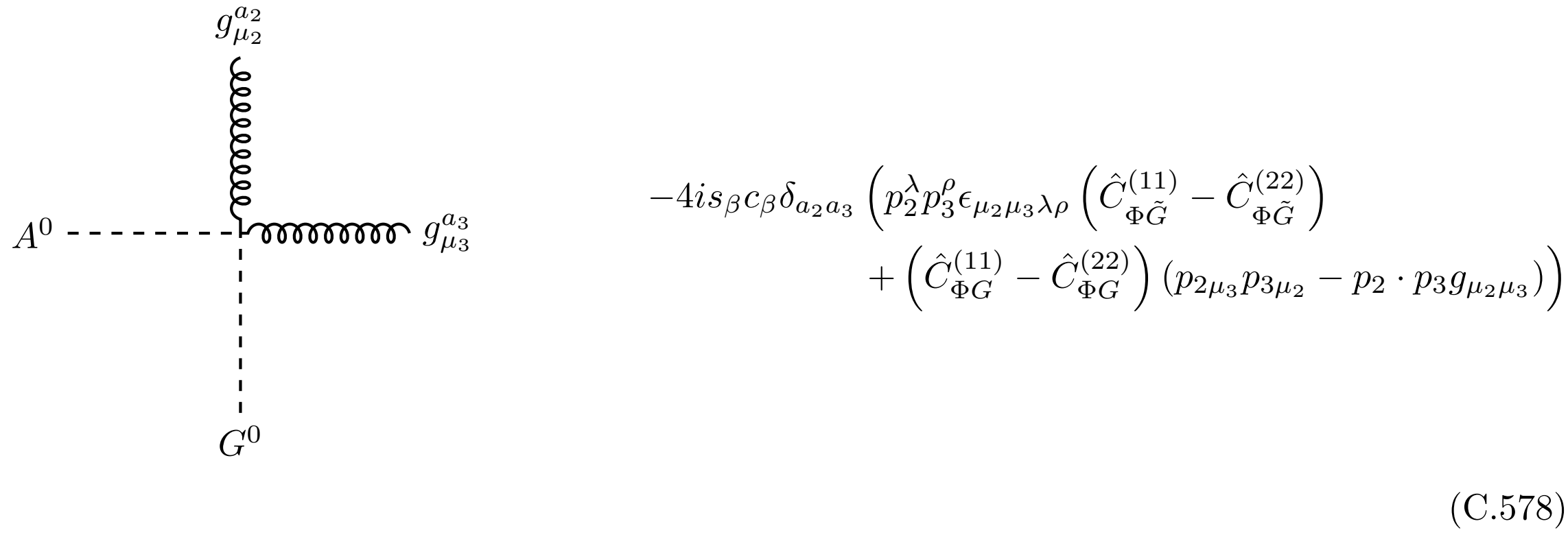

$$-4is_\beta c_\beta \delta_{a_2 a_3} \left( p_2^\lambda p_3^\rho \epsilon_{\mu_2 \mu_3 \lambda \rho} \left( \hat{C}^{(11)}_{\Phi \tilde{G}} - \hat{C}^{(22)}_{\Phi \tilde{G}} \right) + \left( \hat{C}^{(11)}_{\Phi G} - \hat{C}^{(22)}_{\Phi G} \right) \left( p_{2\mu_3} p_{3\mu_2} - p_2 \cdot p_3 g_{\mu_2 \mu_3} \right) \right) \tag{C.578}$$

$g^{a_2}_{\mu_2}$, $g^{a_1}_{\mu_1}$, $h$, $H$

$$+4is_\beta c_\beta \delta_{a_1 a_2} \left( p_1^\lambda p_2^\rho \epsilon_{\mu_1 \mu_2 \lambda \rho} \left( \hat{C}^{(11)}_{\Phi \tilde{G}} - \hat{C}^{(22)}_{\Phi \tilde{G}} \right) + \left( \hat{C}^{(11)}_{\Phi G} - \hat{C}^{(22)}_{\Phi G} \right) \left( p_{1\mu_2} p_{2\mu_1} - p_1 \cdot p_2 g_{\mu_1 \mu_2} \right) \right) \tag{C.579}$$

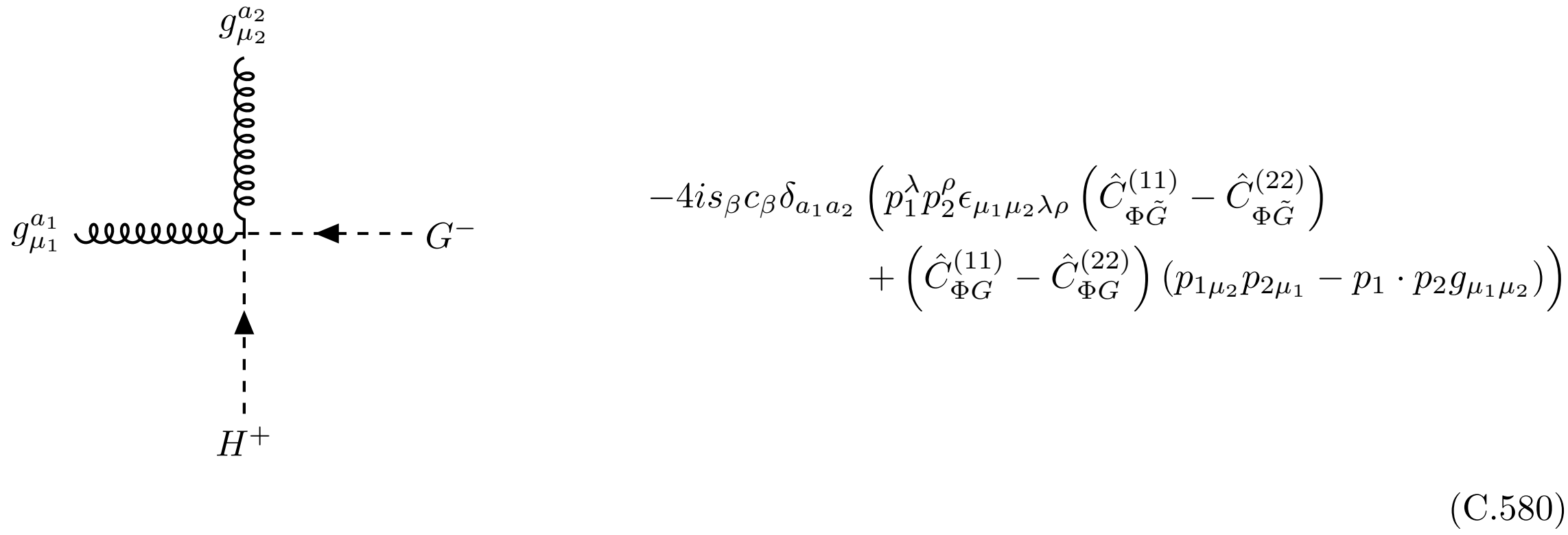

$$-4is_\beta c_\beta \delta_{a_1 a_2} \left( p_1^\lambda p_2^\rho \epsilon_{\mu_1 \mu_2 \lambda \rho} \left( \hat{C}^{(11)}_{\Phi \tilde{G}} - \hat{C}^{(22)}_{\Phi \tilde{G}} \right) + \left( \hat{C}^{(11)}_{\Phi G} - \hat{C}^{(22)}_{\Phi G} \right) \left( p_{1\mu_2} p_{2\mu_1} - p_1 \cdot p_2 g_{\mu_1 \mu_2} \right) \right) \tag{C.580}$$